\documentclass[10pt,a4paper,onecolumn,twoside,
               openright,titlepage,fleqn,final,
               headsepline,headinclude,%
               liststotoc,bibtotoc,idxtotoc,%
               normalheadings,abstracton]{scrreprt}

\typearea{12}                                    
\usepackage[latin1]{inputenc}                    
\usepackage[dvips]{graphicx}                     
\usepackage{color}                               
\usepackage{calc}                                
\usepackage{latexsym}                            
\usepackage{amsfonts}                            
\usepackage{amssymb}                             
\usepackage[mathscr]{eucal}                      
\usepackage{fancyhdr}                            
\usepackage{pictex}                              
\usepackage{epsf}                                
\usepackage{rotating}                            
\usepackage{theorem}                             
\usepackage{makeidx}                             
\usepackage{hyperref}                            

\definecolor{light-gray}{gray}{0.9}              

\graphicspath{{.}{Graphics/}}                    

\theoremstyle{plain}
\theorembodyfont{\slshape}
\theoremheaderfont{\scshape}

\newtheorem{theorem}{Theorem}
\newtheorem{postulate}{Postulate}
\newtheorem{definition}{Definition}

\newcommand{\ALiCE}{\textbf{ALiCE}}
\newcommand{\SESAM}{\textbf{SESAM}}
\newcommand{\TXL}{\textbf{T$\chi$L}}
\newcommand{\APE}{\textbf{APE}}
\newcommand{\DESY}{\textbf{DESY}}
\newcommand{\NIC}{\textbf{NIC}}
\newcommand{\Nicse}{\textbf{Nicse}}
\newcommand{\GNU}{\textbf{GNU}}
\newcommand{\TAO}{\texttt{TAO}}
\newcommand{\Zz}{\texttt{Zz}}
\newcommand{\CRAY}{\textbf{CRAY}}
\newcommand{\ZAM}{\textbf{ZAM}}
\newcommand{\lagrangian}{\mathscr{L}}
\newcommand{\hamiltonian}{\hat{H}}
\newcommand{\Hilbert}{{\cal H}}
\newcommand{\wightm}{\mathfrak{W}}
\newcommand{\schwinger}{\mathfrak{S}}
\newcommand{\feynm}{\mathfrak{F}}

\newcommand{\Ensemble}[1]{\lbrace{#1}\rbrace}
\newcommand{\EnsembleDens}[1]{\rho\lbrace{#1}\rbrace}
\newcommand{\EnsembleMeas}[1]{[\mbox{d}#1]}
\newcommand{\MarkovChain}[1]{[#1]}
\newcommand{\MarkovDens}[1]{\tilde{\rho}[#1]}
\newcommand{\MarkovDensVar}[2]{\tilde{\rho}_{#2}[#1]}
\newcommand{\MarkovDensEq}[1]{\MarkovDensVar{#1}{\mbox{\tiny Eq}}}
\newcommand{\MarkovDensElemVar}[2]{\tilde{\rho}_{#2}(#1)}
\newcommand{\MarkovDensEqElem}[1]%
{\MarkovDensElemVar{#1}{\mbox{\tiny Eq}}}
\newcommand{\MarkovProb}[2]{{\cal P}\left(\MarkovChain{#1}%
    \rightarrow\MarkovChain{#2}\right)}
\newcommand{\MarkovTrans}{{\cal P}}

\newcommand{\Exp}[2]{#1\times10^{#2}}

\newcommand{\sumat}[4]%
{\left(\begin{array}{cc}#1&#2\\#3&#4\end{array}\right)}

\newcommand{\sqlint}{\texttt{INTEGER}}
\newcommand{\sqldou}{\texttt{DOUBLE PRECISION}}
\newcommand{\sqlvch}{\texttt{VARCHAR(255)}}
\newcommand{\sqltxt}{\texttt{TEXT}}
\newcommand{\sqlenm}{\texttt{ENUM}}
\newcommand{\sqldat}{\texttt{DATETIME}}

\def\highstrut{\hbox{\vrule height12pt depth5pt width0pt}}

\newcounter{saveeqn}%
\newcommand{\eqcite}[1]{%
  \setcounter{saveeqn}{\value{equation}}%
  \let\savete\theequation%
  \renewcommand{\theequation}{%
    \mbox{\ref{#1}}}}
\newcommand{\fineqcite}{%
  \setcounter{equation}{\value{saveeqn}}%
  \let\theequation\savete}

\makeindex

\epsfclipon

\hypersetup{%
  debug=false,                                    
  a4paper=true,                                   
  dvips,                                          
  hyperindex=true,                                
  pdfpagemode={UseOutline},                       
  pdftitle={Advanced Algorithms for
    the Simulation of Gauge Theories
    with Dynamical Fermionic Degrees of Freedom}, 
  pdfauthor={Wolfram Schroers},                   
  pdfsubject={PhD Thesis},                        
  pdfcreator={Wolfram Schroers},                  
  pdfproducer={Wolfram Schroers},                 
  pdfkeywords={Physics,Quantum Field Theory,
    Quantum Chromodynamics,Lattice,Algorithms,
    Quantum Spin,Multiboson,
    Three Dynamical Flavors}                      
  }

\begin{document}

\parindent=0pt                                   
\pagenumbering{roman}
\pagestyle{headings}


\titlehead{%
  \href{http://www.physik.uni-wuppertal.de}{Fachbereich Physik} \hfill
  \textbf{WUB-DIS 2001-7}                                       \\
  \href{http://www.uni-wuppertal.de}{BUGH Wuppertal}            \\
  \href{http://www.wuppertal.de}{D-42097 Wuppertal}             \\
  Germany}
\title{%
  Advanced Algorithms for the Simulation of Gauge Theories with
  Dynamical Fermionic Degrees of Freedom }
\subject{Ph.D. Thesis}
\date{November 2, 2001}
\publishers{}
\author{%
  \begin{minipage}{0.75\textwidth}
    \begin{center}
      {\includegraphics[scale=1.0]{Loewe.eps}}
    \end{center}\vspace{2cm}
    \begin{center}Wolfram Schroers\end{center}
  \end{minipage}
  }
\lowertitleback{%
}

\maketitle


\begin{abstract}
  The topic of this thesis is the numerical simulation of quantum
  chromodynamics including dynamical fermions. Two major problems of
  most simulation algorithms that deal with dynamical fermions are (i)
  their restriction to only two mass-degenerate quarks, and (ii) their
  limitation to relatively heavy masses. Realistic simulations of
  quantum chromodynamics, however, require the inclusion of three
  light dynamical fermion flavors. It is therefore highly important to
  develop algorithms which are efficient in this situation.
  
  This thesis is focused on the implementation and the application of
  a novel kind of algorithm which is expected to overcome the
  limitations of older schemes. This new algorithm is named Multiboson
  Method. It allows to simulate an arbitrary number of dynamical
  fermion flavors, which can in principle have different masses. It
  will be shown that it exhibits better scaling properties for light
  fermions than other methods. Therefore, it has the potential to
  become the method of choice.
  
  An explorative investigation of the parameter space of quantum
  chromodynamics with three flavors finishes this work. The results
  may serve as a starting point for future realistic simulations.
\end{abstract}

\tableofcontents

\listoffigures

\listoftables


\pagenumbering{arabic}

\chapter{Introduction}
\label{sec:introduction}
\begin{verse}
  \begin{flushright}
    \textsf{The works of the LORD are great, \\
      sought out by all them \\
      that have pleasure in them. \\
    \textsl{Psalm 111, Verse 2.}}
  \end{flushright}
\end{verse}

Throughout history one of the fundamental driving forces of man has
been the desire to understand nature. In the past few centuries,
natural sciences have paved the way to several revolutionary insights
to the structures underlying our world. Some of them founded
fundamental benefits for the quality of life and the advance of
civilization. In particular, in the recent decades, computer
technology has set the base for a major leap of several important
facets of human existence.

A major challenge is posed by fundamental research, which does not
directly aim towards developing industrial applications, but instead
examines structures and relations relevant for future technologies.
The goal of fundamental research is to formulate theories which
comprise as many different phenomena as possible and which are, at the
same time, as simple as they can be.

The branch of natural sciences which concentrates on the structures
underlying matter and energy is termed particle physics. This field
lives on contributions from experiments, providing insights of how the
particles which constitute our world interact. A further driving force
behind particle physics is the desire to find a simple description of
the mechanisms underlying these experiments. The field of particle
physics has benefitted from the evolution in computer science, but on
the other hand theoretical physicists have triggered many pivotal
developments in technology we have today.

\index{Interactions} Current physical theories categorize the
interactions between matter and energy into four different types of
fundamental forces. \index{Forces in nature} These forces are
gravitation, electromagnetism, and finally the weak and the strong
interactions. The strong interaction is responsible for the forces
acting between \index{Hadrons} hadrons, i.e.~between neutrons, protons
and nuclei built up from these particles. Its name originates from the
fact that it is the strongest among the other forces on the energy
scale of hadronic interactions. The strengths of the four forces can
be stated in terms of their coupling constants \cite{Groom:2000in}:
\begin{eqnarray}
  \label{eq:fundamental-couplings}
  \alpha_{\mbox{\tiny em}}(Q^2=0) &\sim& 1/137.035\,999\,76(50)\,,
  \nonumber \\
  \alpha_{\mbox{\tiny strong}}(Q^2=m_Z^2) &\sim& 0.1185(20)\,, 
  \nonumber \\
  G_{\mbox{\tiny weak}}     &\sim& \Exp{1.166\,39(1)}{-5}\,
  \mbox{GeV}^{-2}\,, \nonumber \\
  G_{\mbox{\tiny Newton}}   &\sim& \Exp{6.707(10)}{-39}\,
  \mbox{GeV}^{-2}\,.
\end{eqnarray}
It is apparent, that gravitation is many orders of magnitude weaker
than the other forces and thus is expected to play no role on the
energy scales important for particle physics so far
\cite{SciAmi:1989xx}.

In contrast to electromagnetic interactions and gravity, strong and
weak forces do not have infinite ranges. They exhibit finite ranges up
to at most the size of a nucleus. Thus, they only play a role in
nuclear interactions, but are almost completely negligible on the
level of atoms and molecules. Any particular process can be dissected
into a multitude of processes acting at a smaller scale.  Hence, a
process with a given action $S\lbrace\mbox{Process}\rbrace$ may be
described by a set of subprocesses $\sum_i
S\lbrace\mbox{Subprocess}_i\rbrace$ with smaller actions
$S\lbrace\mbox{Subprocess}_i\rbrace<S\lbrace\mbox{Process}\rbrace$.
We can tell from observations that processes are essentially
deterministic if the action of the process in question
$S\lbrace\mbox{Process}\rbrace$ is far larger than some action $\hbar$
which is known as ``Planck's constant'' \cite{Landau:1979pt}. However,
if the action of the process is of the order of $\hbar$, then the
system cannot be described by a deterministic theory any longer and a
non-deterministic theory known as quantum mechanics must be employed.
Today, all interactions can be described within a quantum-mechanical
framework, with the exception of gravity, which has so far not been
successfully formulated as a consistent quantum theory. For a
discussion of these topics the reader may consult
\cite{Rovelli:2000aw} and references therein.

\index{WKB-expansion} A necessary requirement for the above iteration
to make sense is that it must be possible to recover a classical
(non-probabilistic) theory in a certain limit (naturally the $\hbar\to
0$ limit) from a quantum theory. This limit is given by the WKB
approximation \cite{Messiah:1991aa}. One finds that an expansion
exists, where the amplitude of a process can be formulated as a
power-series in $\hbar$:
\begin{equation}
  \label{eq:wkb-amplitude-expansion}
  A\lbrace\mbox{Process}\rbrace = A\lbrace\mbox{Classical}\rbrace +
  \hbar A_1 + \hbar^2 A_2 + \cdots\;.
\end{equation}
In the limit $\hbar\to 0$ the classical amplitude is recovered. But
Eq.~(\ref{eq:wkb-amplitude-expansion}) also shows that the quantum
theory contains more information than the classical theory: Different
choices for the series coefficients $\lbrace A_1, A_2, \dots\rbrace$
obviously lead to the same classical theory. Thus, if one intends to
construct a quantum theory starting from the classical theory, there
is always an ambiguity of how to proceed \cite{Messiah:1991aa}. The
correct prescription can only be found by experimental means.

There exist several quantization prescriptions to build a quantum
theory. All of these have in common that they describe a system which
exhibits a probabilistic interpretation of physical observables and a
deterministic evolution equation of some underlying degrees of
freedom. The quantum theory which is believed to constitute the
correct quantum theory of the strong interaction is called quantum
chromodynamics (QCD).

In general, a quantum mechanical system can not be solved exactly, but
only by use of certain approximations. The approximation most commonly
employed is known as perturbation theory and is applicable in a large
variety of cases. It is known to fail, however, when applied to the
low-energy regime of QCD, where a diversity of interesting phenomena
occurs. Hence, different techniques commonly called
``non-perturbative'' methods must be employed. One of these
non-perturbative methods is the simulation of the system in a
large-scale numerical calculation, an approach which is referred to as
``Lattice QCD''. This approach is in the focus of this thesis.

When performing numerical simulations within Lattice QCD, one finds
that the simulation of the bosonic constituents of QCD, the
``gluons'', stand at the basis of research efforts, see
\cite{Creutz:1980zw} for a pioneering publication.

The inclusion of dynamical fermions poses a serious problem. Although
it has become clear in \cite{Aoki:1999yr,Kanaya:1998sd} that without
dynamical fermions the low-energy hadron spectrum is reproduced with
$10\%$ accuracy, several important aspects of low-energy QCD require
the inclusion of dynamical fermions. One case where the inclusion of
dynamical fermions is phenomenologically vital is given by the mass of
the $\eta'$ meson, cf.~\cite{Struckmann:2000ts}.

The numerical simulation of dynamical fermions is plagued by severe
difficulties. In particular, the requirement that the fermions must be
light is to be met, since only in this case the chiral behavior of
QCD, i.e.~the behavior at light fermion masses, is reproduced
correctly. But in this particular limit, the algorithms suffer from a
phenomenon known as critical slowing down, i.e.~a polynomial decrease
in efficiency as the chiral point is approached.

Furthermore, the majority of algorithms in use today can only treat
two mass-degenerate dynamical fermion flavors, a situation not present
in strong interactions as observed in nature. In fact, one has to use
three dynamical fermion flavors \cite{Sharpe:2000bc}.

Thus, the demands on an algorithm for the simulation of dynamical
fermion flavors must consist of (i) the suitability for the simulation
of three dynamical fermion flavors, and (ii) the efficiency of the
algorithm with regard to critical slowing down. These two requirements
are not met by the commonly used algorithm in Lattice QCD, the hybrid
Monte-Carlo (HMC) algorithm.

The topic of this thesis is the exploration of a new type of
algorithm, known as the multiboson algorithm. This algorithm is
expected to be superior to the HMC algorithm with regard to the above
properties. In particular, it will be examined how to tune and
optimize this class of algorithms and if these algorithms are suitable
for the simulation of three light, dynamical, and mass-degenerate
fermion flavors.

The thesis is organized as follows: the theoretical background of the
strong interaction, the quantization of field theories, and the
definition of lattice gauge theories is given in
chapter~\ref{sec:quant-field-theor}. The tools required to perform
numerical simulations in lattice theories and the analysis of time
series are formulated in chapter~\ref{sec:numerical-methods}. The
optimization and tuning of the algorithm is discussed in
chapter~\ref{sec:tuning-mult-algor}.

A direct comparison of the multiboson algorithm
with the hybrid Monte-Carlo method is performed in
chapter~\ref{sec:comp-dynam-ferm}. In particular, the scaling of the
algorithms with the quark mass has been focused at.

A particularly useful application of the multiboson algorithm appears
to be the simulation of QCD with three dynamical fermion flavors. Such
a simulation allows to assess the suitability of multiboson algorithms
for future simulations aimed at obtaining physically relevant results.
A first, explorative investigation of the parameter space which might
be relevant for future simulations is presented in
chapter~\ref{sec:expl-invest-param}.

Finally the conclusions are summarized in
chapter~\ref{sec:summary-outlook-1}.

Appendix~\ref{sec:notat-conv} contains a short overview of the
notation used in this thesis. An introduction to group theory and the
corresponding algebras is given in App.~\ref{sec:groups-algebras}. The
explicit expressions used for the local actions required for the
implementation of multiboson algorithms are listed in
App.~\ref{sec:local-forms-actions}. At last,
App.~\ref{sec:logist-runn-large} explains the concepts required for
running large production runs, where a huge amount of data is
typically generated.

I have to thank many colleagues and friends who have accompanied me
during the completion of this thesis and my scientific work. I am
indebted to my parents, Astrid B{\"o}rger, Claus Gebert, Ivan Hip,
Boris Postler, and Zbygniew Sroczynski for the time they invested to
proof-read my thesis. For the interesting scientific collaborations
and many useful discussions I express my gratitude to Guido Arnold,
Sabrina Casanova, Massimo D'Elia, Norbert Eicker, Federico Farchioni,
Philippe de Forcrand, Christoph Gattringer, Rainer Jacob, Peter Kroll,
Thomas Moschny, Hartmut Neff, Boris Orth, Pavel Pobylitza, Nicos
Stefanis, and in particular to Istv{\'a}n Montvay, Thomas Lippert, and
Klaus Schilling.

\chapter{Quantum Field Theories and Hadronic Physics}
\label{sec:quant-field-theor}
This chapter provides a general introduction into the topic of
particle physics. It covers both the phenomenological aspects, the
mathematical structures commonly used to describe these systems, and
the particular methods to obtain results from the basic principles.

Section~\ref{sec:phen-strong-inter} gives a general overview of the
phenomenology of the strong interaction without making direct
reference to a particular model.

A short overview of classical (i.e.~non-quantum) field theories is
given in Sec.~\ref{sec:class-field-theor}. With this basis, the
general principles of constructing a quantum field theory starting
from a classical field theory are presented in
Sec.~\ref{sec:quantization}. The case of non-relativistic theories is
covered in Sec.~\ref{sec:non-relat-quant}, while the generalization to
relativistic quantum field theories requires far more effort. This is
described in Sec.~\ref{sec:axioms-relat-quant}, where the basic
axiomatic frameworks of relativistic quantum field theories are
stated.

Particular emphasis will be placed on the path integral quantization
which allows for a rigorous and efficient construction of a quantum
theory. Section~\ref{sec:path-integral} provides a detailed treatise
of this method and also contains a discussion of how one can perform
computations in practice. An important tool for the evaluation of path
integrals is the concept of ensembles, which is introduced in
Sec.~\ref{sec:ensembles}. It will turn out to be essential in
numerical simulations of quantum field theories.

Section~\ref{sec:gauge-theories} introduces an important class of
quantum field theories, namely the class of gauge theories. These
models will be of central importance in the following.

With all necessary tools prepared, Sec.~\ref{sec:quant-chrom} will
introduce a gauge theory which is expected to be able to describe the
whole phenomenology of the strong interaction. This theory is known as
quantum chromodynamics (QCD) and it is the main scope of this thesis.
After a general introduction to the properties of QCD in
Sec.~\ref{sec:runn-coupl-energy}, a method known as factorization is
discussed in Sec.~\ref{sec:fact-proc}. This method allows to combine
information from the different energy scales and thus provides an
essential tool for actual predictions in QCD calculations. Finally,
the method of Lattice QCD is discussed in Sec.~\ref{sec:lattice-qcd}.
Lattice simulations exploit numerical integration schemes to gain
information about the structure of QCD, and represent the major tool
for the purposes of this thesis.

The construction of a quantum field theory based on path integrals
requires a certain discretization scheme. This scheme is particularly
important in lattice simulations. Therefore,
Sec.~\ref{sec:discretization} covers the common discretizations for
the different types of fields one encounters in quantum field
theories. The case of scalar fields allows for a simple and efficient
construction as will be shown in Sec.~\ref{sec:scalar-fields}. The
case of gauge fields is more involved since there exist several
proposals how this implementation should be done. Contemporary
simulations focus mainly on the Wilson discretization, although
recently a new and probably superior method has been proposed. This
method, known as D-theory, is reviewed in Sec.~\ref{sec:d-theory}.

The discretization of fermion fields is even more involved. The
necessary conditions such a scheme has to fulfill are given in
Sec.~\ref{sec:fermion-fields} and the scheme used in this thesis,
namely the Wilson-fermion scheme, is constructed. In contrast to the
cases of scalar and gauge fields, a large number of different fermion
discretization schemes are used in actual simulations today, and each
has its particular advantages and disadvantages.

This chapter is concluded by the application of the previously
discussed discretization schemes to a gauge theory containing both
fermions and gauge fields in Sec.~\ref{sec:yang-mills-theory}. Such a
model is expected to be the lattice version of gauge theories with
fermions, and in particular of QCD.

\section{Phenomenology of Strong Interactions}
\label{sec:phen-strong-inter}
Until 1932 only the electron $e$, the photon $\gamma$ and the proton
$p$ have been known as elementary particles (for overviews of the
history of particles physics see
\cite{Nachtmann:1990ta,Leader:1996hk,Sutton:1992xx}).  The only strong
process known was the $\alpha$-decay of a nucleus. The milestones in
this period were the detection of the neutron by \textsc{Chadwick} in
1932 and the prediction of the $\pi$-meson (today it is customary to
call it simply ``pion'') by \textsc{Yukawa} in 1935 as the mediator of
the strong force. However, it took until 1948 before the charged pion
was actually detected by \textsc{Lattes}. In 1947 particles carrying a
new type of quantum number called ``strangeness'' have been detected
by \textsc{Rochester}.  In 1950 the neutral pion was detected by
\textsc{Carlson} and \textsc{Bjorkland}. It was soon realized that the
hadrons were not point-like objects like the leptons, but had an
internal structure and accordingly a finite spatial extent.

The experiments to observe the structure of hadrons usually consist of
scattering two incoming particles off each other, producing several
outgoing particles of possibly different type. If one considers a
particular subset of processes where all outgoing particles are of a
determined type, one speaks of \textit{exclusive reactions}. A
sub-class of exclusive reactions are the \textit{elastic scattering}
processes, where the incoming and outgoing particles are identical.

The \textit{inclusive reactions} are obtained by summing over all
possible exclusive reactions for given incoming particles. Inclusive
electron-nucleon scattering at very large energies is called
\textit{deep-inelastic scattering} (DIS) and played an important role
in the understanding of the structure of hadrons. The prediction of
scaling by \textsc{Bjorken} in 1969 was confirmed experimentally and
led to the insight that the hadrons consist of point-like
sub-particles. In 1968 \textsc{Feynman} proposed a model which
exhibited this feature, the \index{Partons}\textit{parton model}.

A collection of hadrons, as known in the early 60's, is given in
Tab.~\ref{tab:particlelist} together with their properties in the form
of quantum numbers. These quantum numbers are known as spin, parity,
electric charge $Q$, baryon number $\mathcal{B}$, and strangeness $S$.
They are conserved by the strong interaction\footnote{Note, however,
  that the weak interaction violates both the baryon number and the
  strangeness.  While the latter phenomenon has been observed in
  experiment so far \cite{Nachtmann:1990ta}, the former violation may
  never be observed directly in earth-bound experiments
  \cite{Rubakov:1996vz}}.
\begin{table}[htb]
  \begin{center}
    \begin{tabular}[h]{l|c|ccc|r}\hline\hline
      {\bf Hadron} & {\bf Spin}$^{\mbox{\bf Parity\strut}}$ &
      {$\mathbf{Q}$} & {$\mathbf{\cal B}$} & {$\mathbf{S}$} &
      \textbf{$\mathbf{m/}$MeV} \\
      \hline
      $\pi^{\pm}$ & $0^-$   & $\pm 1$ & $0$ & $0$ & $140$  \\ 
      $\pi^0$     & $0^-$   & $0$  & $0$  & $0$  & $135$  \\ 
      $K^0$       & $0^-$   & $0$  & $0$  & $+1$ & $498$  \\
      $\bar{K}^0$ & $0^-$   & $0$  & $0$  & $-1$ & $498$  \\
      $K^{\pm}$   & $0^-$   & $\pm 1$ & $0$  & $\pm 1$ & $494$  \\
      $\eta$      & $0^-$   & $0$  & $0$  & $0$  & $547$  \\
      $\eta^{\prime}$ & $0^-$ & $0$ & $0$ & $0$  & $958$ \\
      $\rho^{\pm}$& $1^-$   & $\pm 1$ & $0$ & $0$ & $767$ \\
      $\rho^0$    & $1^-$   & $0$  & $0$  & $0$  & $769$ \\
      $K^{*0}$    & $1^-$   & $0$  & $0$  & $+1$ & $896$ \\
      $\bar{K}^{*0}$ & $1^-$ & $0$ & $0$  & $-1$ & $896$ \\
      $K^{*\pm}$  & $1^-$   & $\pm 1$ & $0$ & $\pm 1$ & $892$ \\
      $\omega$    & $1^-$   & $0$  & $0$  & $0$  & $783$ \\
      $\phi$      & $1^-$   & $0$  & $0$  & $0$  & $1019$ \\
      $p$         & $1/2^+$ & $+1$ & $+1$ & $0$  & $938$ \\
      $n$         & $1/2^+$ & $0$  & $+1$ & $0$  & $940$ \\
      $\Sigma^+$  & $1/2^+$ & $+1$ & $+1$ & $-1$ & $1189$ \\
      $\Sigma^-$  & $1/2^+$ & $-1$ & $+1$ & $-1$ & $1197$ \\
      $\Sigma^0$  & $1/2^+$ & $0$  & $+1$ & $-1$ & $1193$ \\
      $\Lambda$   & $1/2^+$ & $0$  & $+1$ & $-1$ & $1116$ \\
      $\Xi^0$     & $1/2^+$ & $0$  & $+1$ & $-2$ & $1315$ \\
      $\Xi^-$     & $1/2^+$ & $-1$ & $+1$ & $-2$ & $1321$ \\
      \hline\hline
    \end{tabular}
    \caption{List of selected hadrons with their quantum numbers and
      their masses in MeV.}
    \label{tab:particlelist}
  \end{center}
\end{table}
The particles may be divided into several groups according to their
spin and parity: the particles with even spin are called
\textit{mesons} and the particles with odd spin \textit{baryons}.
Because of their parity and spin, the particles $\pi^0$, $\pi^\pm$,
$K^\pm$, $K^0$, $\bar{K}^0$, $\eta$ and $\eta^{'}$ are usually called
\index{Hadrons!mesons} ``pseudoscalar mesons''.  Similarly, the
particles $\rho^0$, $\rho^\pm$, $\omega$, $K^{*0}$, $K^{*\pm}$,
$\bar{K}^{*0}$ and $\varphi$ are named ``vector mesons''. The group
$p$, $n$, $\Lambda$, $\Sigma^0$, $\Sigma^\pm$, $\Xi^-$ and $\Xi^0$ is
simply called \index{Hadrons!baryons} ``baryons''. In each group, the
members have roughly similar masses (with the exception of the
$\eta^{'}$ in the group of the pseudoscalar mesons).

In 1963 \textsc{Gell-Mann} and \textsc{Zweig} independently proposed a
scheme to classify the known particles as multiplets of the Lie group
SU$(3)_F$. It turned out that the classification is indeed possible
with the exception that there were no particles corresponding to the
fundamental triplets of the group. This would imply that the
corresponding particles carry fractional charges; particles with such
a property have never been seen in any experiment. However,
experiments at SLAC in 1971 involving neutrino-nucleon scattering
clearly indicated that the data could be accounted for if the parton
inside the nucleon had the properties of the particles in the
fundamental triplet of the SU$(3)_F$ group. This led finally to the
identification of the (charged) partons from Feynman's model with the
particles from the classification scheme of \textsc{Gell-Mann} and
\textsc{Zweig}. \index{Quarks} These particles are known today as
\textit{quarks}.

Today a huge number of particles subject to the strong interaction is
known from different kinds of experiments. For a complete overview see
\cite{Groom:2000in}. To classify them the quark model had to be
extended to six kinds of quarks. They are referred to as having
``different flavors'', which are new quantum numbers. Thus, they are
conserved by the strong interaction. The quarks cover a huge range of
masses. Their properties are listed in Tab.~\ref{tab:quarks}.
\begin{table}[htbp]
  \begin{center}
    \begin{tabular}[h]{*{10}{c}} \hline\hline
      & $\mathbf{Y}$   & $\mathbf{T_3}$  & $\mathbf{Q}$ &
      $\mathbf{\cal B}$& $\mathbf{S}$    & $\mathbf{C}$ & 
      $\mathbf{B}$     & $\mathbf{T}$    & $\mathbf{m/\mbox{MeV}}$ \\
      \hline
      $u$ & $1/2$ & $1/2$  & $2/3$  & $1/3$ & $0$ & $0$ & $0$ & $0$ &
      $1-5$   \\
      $c$ & $0$   & $0$    & $2/3$  & $1/3$ & $0$ & $1$ & $0$ & $0$ &
      $1150-1350$    \\
      $t$ & $0$   & $0$    & $2/3$  & $1/3$ & $0$ & $0$ & $0$ & $1$ &
      $174300\pm 5100$  \\
      $d$ & $1/2$ & $-1/2$ & $-1/3$ & $1/3$ & $0$ & $0$ & $0$ & $0$ &
      $3-9$   \\
      $s$ & $0$   & $0$    & $-1/3$ & $1/3$ & $-1$ & $0$ & $0$ & $0$ &
      $75-170$ \\
      $b$ & $0$   & $0$    & $-1/3$ & $1/3$ & $0$ & $0$ & $-1$ & $0$ &
      $4000-4400$    \\ \hline\hline
    \end{tabular}
    \caption{Different flavors of quarks together with their
      associated quantum numbers.}
    \label{tab:quarks}
  \end{center}
\end{table}
To classify the multiplets of the flavor group SU$(3)_F$, two numbers
are required which are usually called $Y$ (the strong hyper-charge)
and $T_3$ (the isospin). They are defined by
\[ Y = S + B, \]
and
\[ T_3 = Q - \frac{1}{2}\left( S + B \right). \]
The three light quarks together with the corresponding anti-particles
are shown in Fig.~\ref{fig:quark-triplet}. The classifications for the
pseudoscalar mesons, the vector mesons and the baryons are given in
Figs.~\ref{fig:pseudoscalar-octet},~\ref{fig:vector-octet},
and~\ref{fig:baryon-octet}.
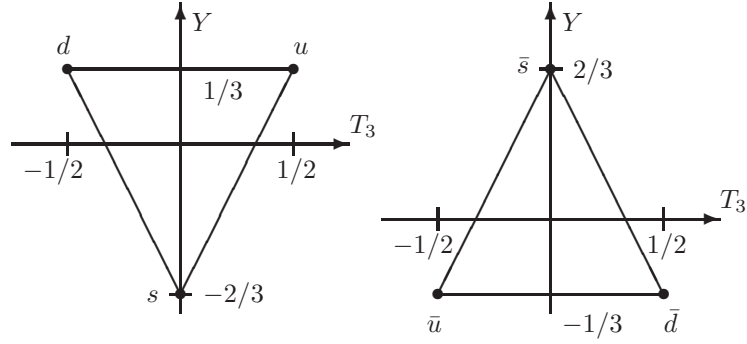
\begin{figure}[htb]
  \begin{center}
    \setlength{\unitlength}{1.5cm}
    \begin{tabular}[h]{cc}
      \begin{picture}(3,2.75)(-1.5,-1.55)
        \thicklines
        \put(-1.5,0){\vector(1,0){3}}
        \put(0,-1.5){\vector(0,1){2.75}}
        \put(1.5,0.1){$T_3$}   \put(0.1,1.0){$Y$}
        \put(0.85,-0.3){$1/2$} \put(-1.4,-0.3){$-1/2$}
        \put(0.2,0.4){$1/3$}   \put(0.2,-1.4){$-2/3$}
        \put(1,-0.1){\line(0,1){0.2}}  \put(-1,-0.1){\line(0,1){0.2}}
        \put(-0.1,-1.33333333333){\line(1,0){0.2}}
        \put(-0.1,0.66666666666){\line(1,0){0.2}}
        \put(0,-1.33333333333){\line(1,2){1}}
        \put(1,0.66666666666){\line(-1,0){2}}
        \put(0,-1.33333333333){\line(-1,2){1}}
        \put(0,-1.33333333333){\circle*{0.1}}
        \put(1,0.66666666666){\circle*{0.1}}
        \put(-1,0.66666666666){\circle*{0.1}}
        \put(-1.1,0.8){$d$} \put(1.0,0.8){$u$} \put(-0.3,-1.4){$s$}
      \end{picture} &
      \begin{picture}(3,2.75)(-1.5,-0.8833333333)
        \thicklines
        \put(-1.5,0){\vector(1,0){3}}
        \put(0,-0.83333333333){\vector(0,1){2.75}}
        \put(1.5,0.1){$T_3$}   \put(0.1,1.666666666){$Y$}
        \put(0.85,-0.3){$1/2$} \put(-1.4,-0.3){$-1/2$}
        \put(0.1,-1.0){$-1/3$} \put(0.2,1.25){$2/3$}
        \put(1,-0.1){\line(0,1){0.2}}  \put(-1,-0.1){\line(0,1){0.2}}
        \put(-0.1,1.33333333333){\line(1,0){0.2}}
        \put(-0.1,-0.66666666666){\line(1,0){0.2}}
        \put(0,1.33333333333){\line(1,-2){1}}
        \put(1,-0.66666666666){\line(-1,0){2}}
        \put(0,1.33333333333){\line(-1,-2){1}}
        \put(0,1.33333333333){\circle*{0.1}}
        \put(1,-0.66666666666){\circle*{0.1}}
        \put(-1,-0.66666666666){\circle*{0.1}}
        \put(-1.1,-1.0){$\bar{u}$} \put(1.0,-1.0){$\bar{d}$}
        \put(-0.3,1.3){$\bar{s}$}
      \end{picture}
    \end{tabular}
    \caption{Fundamental representations of the SU$(3)_F$
      flavor group. The left graph shows the quark triplet
      ($u$,$d$,$s$) and the right graph shows the anti-quark triplet
      ($\bar{u}$,$\bar{d}$,$\bar{s}$).}
    \label{fig:quark-triplet}
  \end{center}
\end{figure}
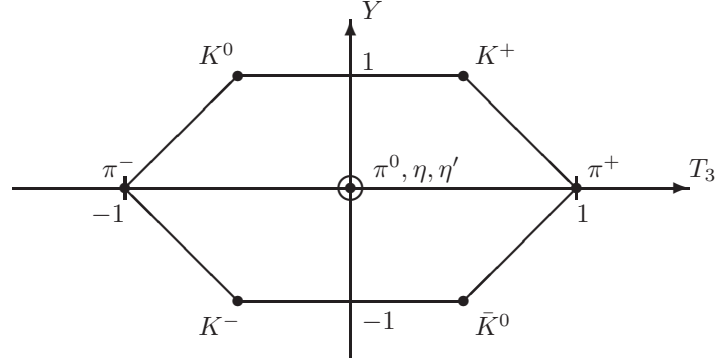
\begin{figure}[htb]
  \begin{center}
    \setlength{\unitlength}{1.5cm}
    \begin{picture}(6,3)(-3,-1.5)
      \thicklines
      \put(-3,0){\vector(1,0){6}}
      \put(0,-1.5){\vector(0,1){3}}
      \put(3,0.1){$T_3$}  \put(0.1,1.5){$Y$}
      \put(2.0,-0.3){$1$} \put(-2.3,-0.3){$-1$}
      \put(0.1,1.05){$1$} \put(0.1,-1.25){$-1$}
      \put(2,-0.1){\line(0,1){0.2}}   \put(-2,-0.1){\line(0,1){0.2}}
      \put(-0.1,-1){\line(1,0){0.2}}  \put(-0.1,1){\line(1,0){0.2}}
      \put(-2,0){\line(1,1){1}}       \put(2,0){\line(-1,-1){1}}
      \put(-1,1){\line(1,0){2}}       \put(1,-1){\line(-1,0){2}}
      \put(1,1){\line(1,-1){1}}       \put(-1,-1){\line(-1,1){1}}
      \put(-2,0){\circle*{0.1}}       \put(2,0){\circle*{0.1}}
      \put(-1,1){\circle*{0.1}}       \put(1,1){\circle*{0.1}}
      \put(-1,-1){\circle*{0.1}}      \put(1,-1){\circle*{0.1}}
      \put(0,0){\circle*{0.1}}        \put(0,0){\circle{0.2}}
      \put(0.2,0.1){$\pi^0, \eta, \eta^{\prime}$}
      \put(-1.35,1.1){$K^0$}
      \put(1.1,1.1){$K^+$}
      \put(-2.2,0.1){$\pi^-$}
      \put(2.1,0.1){$\pi^+$}
      \put(-1.35,-1.3){$K^-$}
      \put(1.1,-1.3){$\bar{K}^0$}
    \end{picture}
    \caption{Pseudoscalar meson octet together with the singlet (the
      $\eta'$ state) as classified by the parameters of the SU$(3)_F$
      group.}
    \label{fig:pseudoscalar-octet}
  \end{center}
\end{figure}
\begin{figure}[htb]
  \begin{center}
    \setlength{\unitlength}{1.5cm}
    \begin{picture}(6,3)(-3,-1.5)
      \thicklines
      \put(-3,0){\vector(1,0){6}}
      \put(0,-1.5){\vector(0,1){3}}
      \put(3,0.1){$T_3$}  \put(0.1,1.5){$Y$}
      \put(2.0,-0.3){$1$} \put(-2.3,-0.3){$-1$}
      \put(0.1,1.05){$1$}  \put(0.1,-1.25){$-1$}
      \put(2,-0.1){\line(0,1){0.2}}   \put(-2,-0.1){\line(0,1){0.2}}
      \put(-0.1,-1){\line(1,0){0.2}}  \put(-0.1,1){\line(1,0){0.2}}
      \put(-2,0){\line(1,1){1}}       \put(2,0){\line(-1,-1){1}}
      \put(-1,1){\line(1,0){2}}       \put(1,-1){\line(-1,0){2}}
      \put(1,1){\line(1,-1){1}}       \put(-1,-1){\line(-1,1){1}}
      \put(-2,0){\circle*{0.1}}       \put(2,0){\circle*{0.1}}
      \put(-1,1){\circle*{0.1}}       \put(1,1){\circle*{0.1}}
      \put(-1,-1){\circle*{0.1}}      \put(1,-1){\circle*{0.1}}
      \put(0,0){\circle*{0.1}}        \put(0,0){\circle{0.2}}
      \put(0.2,0.1){$\rho^0, \omega, \phi$}
      \put(-1.35,1.1){$K_0^*$}
      \put(1.1,1.1){$K^{+*}$}
      \put(-2.2,0.1){$\rho^-$}
      \put(2.1,0.1){$\rho^+$}
      \put(-1.35,-1.3){$K^{-*}$}
      \put(1.1,-1.3){$\bar{K}_0^*$}
    \end{picture}
    \caption{Vector meson octet together with the singlet (the $\phi$
      state) as classified by the parameters of the SU$(3)_F$ group.}
    \label{fig:vector-octet}
  \end{center}
\end{figure}
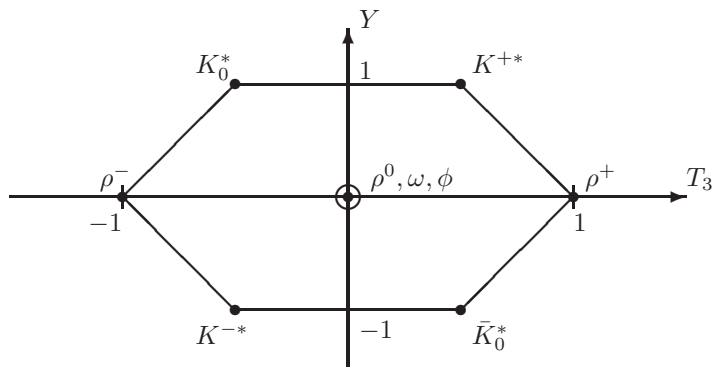
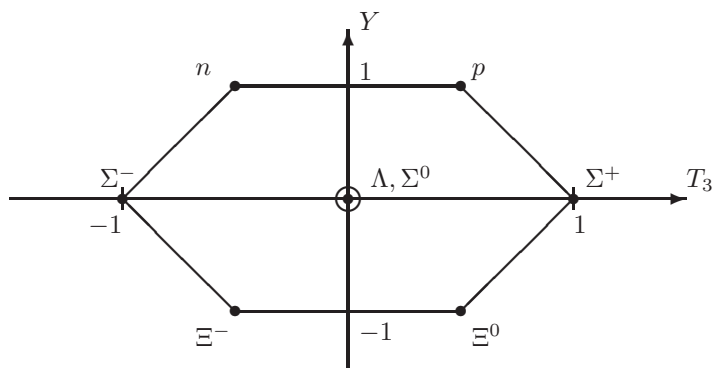
\begin{figure}[htb]
  \begin{center}
    \setlength{\unitlength}{1.5cm}
    \begin{picture}(6,3)(-3,-1.5)
      \thicklines
      \put(-3,0){\vector(1,0){6}}
      \put(0,-1.5){\vector(0,1){3}}
      \put(3,0.1){$T_3$}  \put(0.1,1.5){$Y$}
      \put(2.0,-0.3){$1$} \put(-2.3,-0.3){$-1$}
      \put(0.1,1.05){$1$}  \put(0.1,-1.25){$-1$}
      \put(2,-0.1){\line(0,1){0.2}}   \put(-2,-0.1){\line(0,1){0.2}}
      \put(-0.1,-1){\line(1,0){0.2}}  \put(-0.1,1){\line(1,0){0.2}}
      \put(-2,0){\line(1,1){1}}       \put(2,0){\line(-1,-1){1}}
      \put(-1,1){\line(1,0){2}}       \put(1,-1){\line(-1,0){2}}
      \put(1,1){\line(1,-1){1}}       \put(-1,-1){\line(-1,1){1}}
      \put(-2,0){\circle*{0.1}}       \put(2,0){\circle*{0.1}}
      \put(-1,1){\circle*{0.1}}       \put(1,1){\circle*{0.1}}
      \put(-1,-1){\circle*{0.1}}      \put(1,-1){\circle*{0.1}}
      \put(0,0){\circle*{0.1}}        \put(0,0){\circle{0.2}}
      \put(0.2,0.1){$\Lambda,\Sigma^0$}
      \put(-1.35,1.1){$n$}
      \put(1.1,1.1){$p$}
      \put(-2.2,0.1){$\Sigma^-$}
      \put(2.1,0.1){$\Sigma^+$}
      \put(-1.35,-1.3){$\Xi^-$}
      \put(1.1,-1.3){$\Xi^0$}
    \end{picture}
    \caption{Baryon octet as classified by the parameters of the
      SU$(3)_F$ group.}
    \label{fig:baryon-octet}
  \end{center}
\end{figure}
The multiplets containing the particles are \textit{irreducible}
representations of the SU$(3)_F$ group; they may be built from tensor
products of the fundamental triplet in the following way:
\begin{eqnarray}
  \label{eq:tensor-multiplets}
  3\otimes\bar{3} &=& 8\oplus 1, \nonumber\\
  3\otimes 3\otimes 3 &=& 10\oplus 8\oplus 8\oplus 1\,.
\end{eqnarray}
This explains why there are always nine particles in each of the meson
groups: the first eight belong to an octet and the remaining one is
the singlet state. In the group of the pseudoscalar mesons, the
singlet state $\eta'$ deserves special attention because its mass is
extremely heavy.

The particle content to lowest order of the pseudoscalar mesons
(Fig.~\ref{fig:pseudoscalar-octet}) in terms of the different quark
flavors is given by
\begin{eqnarray}
  \label{eq:pseudoscal-mes-wavefunctions}
  \pi^+ &=& d\bar{u}\,, \quad\pi^- = u\bar{d}\,, \quad K^0 =
  d\bar{s}\,, \quad \bar{K}^0 = s\bar{d}\,, \quad K^+ = u\bar{s}\,,
  \quad K^- = s\bar{u}\,, \nonumber \\
  \pi^0 &=& \frac{1}{\sqrt{2}}\left(u\bar{u}-d\bar{d}\right)\,,
  \quad\eta =
  \frac{1}{\sqrt{6}}\left(u\bar{u}+d\bar{d}-2s\bar{s}\right)\,,
  \quad\eta^{\prime} = \frac{1}{\sqrt{2}}\left(u\bar{u}+d\bar{d}+
    s\bar{s}\right)\,.
  \nonumber \\
\end{eqnarray}
Similarly the lowest-order particle content of the vector mesons
(Fig.~\ref{fig:vector-octet}) is given by
\begin{eqnarray}
  \label{eq:vect-mes-wavefunctions}
  \rho^+ &=& u\bar{d}\,, \quad\rho^- = d\bar{u}\,, \quad K_0^* =
  d\bar{s}\,, \quad \bar{K}_0^* = s\bar{d}\,, \quad K^{+*} =
  u\bar{s}\,, \quad K^{-*} = s\bar{u}\,, \nonumber \\
  \rho^0 &=& \frac{1}{\sqrt{2}}\left(u\bar{u}-d\bar{d}\right)\,,
  \quad\omega = \frac{1}{\sqrt{2}}\left(u\bar{u}+d\bar{d}\right)\,,
  \quad\phi=s\bar{s}\,.
\end{eqnarray}
\index{Quarks!color} A very important step was the discovery of the
``color'' degree of freedom. A first indication towards this feature
was the observation that the $\Delta^{++}$ baryon is a particle with
flavor content of three $u$-type quarks in the ground state and spin
pointing in the same direction. Without an additional quantum number
this would imply that all quarks building up these particles are in
the same quantum state which is not possible with Fermi-Dirac
particles since their wave-functions should anti-commute.
Consequently, a further quantum number should exist which indeed has
been found in experiments \cite{Nachtmann:1990ta}. This quantum number
has been termed \textit{color} and can take on three values. If this
property is described again in terms of an SU$(3)$ group, then the
quarks must transform as the representation of the fundamental
multiplet. The hadrons, however, are color singlets since they display
no color charge. This and the observation that the quarks have never
been seen as free particles outside of hadrons led to the hypothesis
of \index{Confinement}\textit{confinement} which will be examined more
closely in Sec.~\ref{sec:runn-coupl-energy}. According to this
hypothesis, free quarks could never be observed in nature directly
since the force between them grows infinitely.

\section{Classical Field Theories}
\label{sec:class-field-theor}
There are quite a few frameworks for the description of a
classical\footnote{The meaning of the word ``classical'' in the title
  of this section and in the context of this paragraph is referring to
  any system described by a finite or infinite number of degrees of
  freedom with deterministic dynamics regardless of the symmetries of
  the underlying space-time or the system itself. It should be noted,
  that the term ``classical'' is perhaps the word with the largest
  variety of meanings in the literature of physics --- it is used for
  non-relativistic, non-quantum systems, in a different context for
  relativistic quantum field theories and also for anything in
  between.} physical system. For a general review of such frameworks,
the reader may consult \cite{Scheck:1992me}, for the generalization to
field theories \cite{Thirring:1978bo}.  For the later generalization
to quantum mechanical systems, the
\textit{Lagrangian}\index{Lagrangian} method will attract our
attention.

This approach has the advantage that it may be formulated in a
coordinate-invariant manner, i.e.~it does not depend on any specific
geometry of the physical system. The basic postulate underlying the
dynamics of this formulation is (see e.g.~\cite{Masujima:2000uq} for
textbook overview):
\begin{description}
\item[Principle of extremal action:] A set of classical fields
  $\lbrace\varphi_i(x)\rbrace$, $i=1,\dots,N$, is described by a local
  $C^2$ function
  $\lagrangian\left(\varphi_i(x),\partial\varphi_i(x)\right)$ which is
  called the Lagrangian density. The integral $\lagrangian$ over a
  region in Minkowski-space $\mathbb{M}^4$ is defined as the
  \index{Action} action of the system: \[
  \mathscr{S}[\varphi(x)]=\int_{\mathbb{M}^4} d^4x
  \lagrangian\left(\varphi_i(x),\partial\varphi_i(x)\right). \] The
  equations of motion follow from the requirement that the action
  becomes minimal. A necessary condition is \[ \frac{\delta
    \mathscr{S}[\varphi_i(x)]}{\delta\varphi_i} = 0\,. \]
\end{description}
From this postulate, we obtain a necessary condition which the
functions $\varphi_k(x)$ have to fulfill in order to describe the
motion of the physical system
\begin{equation}
  \label{eq:euler-lagrange}
  \frac{\delta\lagrangian}{\delta\varphi_i} - \frac{d}{dt}\left(\frac
    {\delta\lagrangian}{\delta\left(\partial\varphi_i\right)}\right)
   = 0\,.
\end{equation}
This set of equations is called the \textit{Euler-Lagrange equations}.
Using $\lagrangian$, we can define the momentum conjugate $\pi_i(x)$
of $\varphi_i(x)$ via \[ \pi_i(x) =
\frac{\delta\lagrangian\left(\varphi_i(x),\partial\varphi_i(x)\right)}
{\delta\left(\partial_0\varphi_i(x)\right)}\,. \] Then the
\index{Hamiltonian}\textit{Hamiltonian} $\hamiltonian$ is defined by a
Legendre-transformation
\[ \hamiltonian\left(\varphi_i(x),\pi_i(x)\right) = \int d^3x \left(
  \pi_i(x)\partial_0\varphi_i(x) -
  \lagrangian\left(\varphi_i(x),\partial\varphi_i(x)\right) \right)\,.
\] From Eq.~(\ref{eq:euler-lagrange}) the canonical equations of
motion follow:
\begin{eqnarray}
  \label{eq:canonical-motion}
  \partial_0\varphi_i(x) &=&
  \frac{\delta\hamiltonian\left(\varphi_i(x),\pi_i(x)\right)}
  {\delta\pi_i(x)}\,, \nonumber \\
  \partial_0\pi_i(x)     &=& - \frac{
  \delta\hamiltonian\left(\varphi_i(x),\pi_i(x)\right)}
  {\delta\varphi_i(x)}\,.
\end{eqnarray}
Similarly to the situation in classical mechanics, the Lagrangian
framework and the Hamiltonian framework provide the bases of two
different formulations for the quantum mechanical description of a
system.

Since the particles of a quantum theory should transform as
representations of multiplets of certain symmetry groups, the
representations of the Poincar\'{e} group (see
App.~\ref{sec:poincare-group}) will require special attention. The
lowest representations are given by
\begin{enumerate}
\item \index{Fields} the singlet representation, described by a scalar
  field $\phi(x):x\mapsto\phi(x)\in\mathbb{M}^4$. The field transforms
  as
  \[ \left(\Lambda,a\right):x\mapsto x'\Rightarrow
  \phi(x)\mapsto\phi(x')\,. \]
\item the doublet representation, described by a Weyl spinor field
  $\xi_\alpha$, transforming as the fundamental representation under
  $\mathcal{L}^\uparrow_+$. In this thesis we deal with Dirac spinors
  composed of two Weyl spinor fields \[ \psi=\pmatrix{ \xi \\
    \bar{\chi} }\,. \]
\item the vector representation, described by a four-vector field
  $A^\mu(x)$, transforming as \[ A^\mu = \Lambda^\mu_\nu A^\nu\,. \]
\end{enumerate}
The Lagrangians describing the corresponding particles should obey
Lorentz- and $CPT$-invariance and possibly transform according to an
internal symmetry group. As an excellent introduction see
\cite{Weinberg:1995mt}. So far these requirements fix at least the
free field Lagrangians. Examples for Lagrangians obeying these
principles are listed in the following:
\begin{enumerate}
\item The Lagrangian for a complex scalar field $\phi(x)$ is given by
  \begin{equation}
    \label{eq:scalar-lagr}
    \lagrangian\left(\phi\right) =
    (\partial_\mu\phi^\dagger)(\partial^\mu\phi) + m^2
    \phi^\dagger\phi + V(\phi^\dagger,\phi)\,,
  \end{equation}
  where the free field (i.e.~the Lagrangian describing a field
  propagating without an external force) is given by
  $V(\phi^\dagger,\phi)=0$.
\item The spin-statistics theorem (see App.~\ref{sec:spin-stat-theor})
  suggests that a quantum mechanical spin-$1/2$ particle should be
  described by anticommuting field variables. Thus, the fields should
  be Grassmann variables (cf.~Appendix~\ref{sec:grassmann-algebras}).
  The free field Lagrangian is given by
  \begin{equation}
    \label{eq:free-dirac-lagr}
    \lagrangian\left(\bar{\psi},\psi \right) = \bar{\psi}\left(
      \mbox{i}\slash\!\!\!\partial - m\right)\psi\,.
  \end{equation}
  A generalization is the free $N$-component \index{Fields!Yang-Mills}
  Yang-Mills field described by an $N$-component vector
  $\Psi_N=(\psi_1,\dots,\psi_N)$ of independent fields
  $\lbrace\psi_i\rbrace$. Its Lagrangian is the sum of the
  single-field Lagrangians and thus given by
  \begin{equation}
    \label{eq:free-ym-lagr}
    \lagrangian\left(\bar{\Psi}_N,\Psi_N \right) =
    \bar{\Psi}_N\left( \mbox{i}\slash\!\!\!\partial -
      m\right)\Psi_N\,.
  \end{equation}
\item The vector field $A^\mu$ with a field strength
  $F^{\mu\nu}=\partial^\mu A^\nu - \partial^\nu A^\mu$ is described in
  the non-interacting case by
  \begin{equation}
    \label{eq:free-vector-lagr}
    \lagrangian\left(A_\mu\right) = -
    \frac{1}{4}F^{\mu\nu}F_{\mu\nu}\,.
  \end{equation}
  The generalization to a vector field with several components will be
  discussed in Sec.~\ref{sec:gauge-theories}.
\end{enumerate}

\section{Quantization}
\label{sec:quantization}
As it has been pointed out in Chapter~\ref{sec:introduction}, the
world of elementary particle physics is essentially of quantum
mechanical nature. However, any theory of particle physics should also
obey the principle of Lorentz invariance. Therefore the need arises to
find a quantum mechanical model which is at least globally invariant
under the Poincar\'{e}-group. This is hard to do within the framework
of quantum mechanics for pointlike-particles. In fact, the
single-particle interpretation of the relativistic Dirac equation is
subjected to several paradoxes (see e.g.~\cite{Itzykson:1980rh}) which
can only be resolved if one considers instead \textit{fields} (or
rather generalized concepts called \textit{operator-valued
  distributions}) \cite{Weinberg:1995mt}.

\subsection{Non-relativistic Quantum Mechanics}
\label{sec:non-relat-quant}
Before embarking on the definition of a relativistic quantum field
theory, we should recall the concepts of a non-relativistic quantum
theory. In general, a quantum theory has the following general
structure \cite{Peres:1993bo,Fernandez:1992jh}:
\begin{description}
\item[Hilbert space $\Hilbert$:] \index{Hilbert space} The discussion
  will now be limited to \textit{pure states}, which are given by unit
  rays of a complex Hilbert space $\Hilbert$ with scalar product
  $\langle\cdot\vert\cdot\rangle$.
\item[Observables:] \index{Observable} An operator $\hat{A}$ on
  $\Hilbert$ is called an \textit{observable} if it is a self-adjoint
  operator on $\Hilbert$. Thus, its eigenvalues are real. Observables
  correspond to quantities which can be measured in an experiment. If
  the system is in a state $\vert\psi\rangle\in\Hilbert$, the
  \textit{expectation value} $\langle\hat{A}\rangle$ of the observable
  $\hat{A}$ is given by
  \[ \langle\hat{A}\rangle = \langle\psi|\hat{A}\psi\rangle\,. \]
\item[Symmetries:] The \textit{symmetries} of the system are
  represented by unitary (or anti-unitary) operators on $\Hilbert$.
\item[Evolution:] The evolution equation of the states, the
  \textit{Schr{\"o}dinger equation}, is given by
  \begin{equation}
    \label{eq:schrodinger}
    \hamiltonian\vert\psi\rangle = -\mbox{i}\frac{\partial}{\partial
      t} \vert\psi\rangle\,.
  \end{equation}
  The Hamiltonian operator $\hamiltonian$ is a Hermitian operator
  acting in $\Hilbert$.
\end{description}
The state space $\lbrace\vert \psi_i\rangle\rbrace$ may be finite,
infinite or even uncountably infinite. In the case that it is finite,
the solution of Eq.~(\ref{eq:schrodinger}) is well-defined and can be
found by diagonalizing the operator $\hat{H}$.

But already in the case of an infinite but countable state space,
there may be physically equivalent observables which are not unitarily
equivalent \cite{Itzykson:1980rh,Haag:1992hx}. However, we will see
below that information about the quantum field theory can be extracted
even without knowing the complete state space. In fact, only few cases
are known, where the state space has been constructed in a
mathematically rigorous manner. So far, this does not include any
interacting quantum field theory in four space-time dimensions.

\subsection{The Axioms of Relativistic Quantum Field Theories}
\label{sec:axioms-relat-quant}
The mathematically rigorous formulation of relativistic quantum field
theories requires the introduction of operator-like objects replacing
the classical fields; however, it turns out to be impossible to use
operator-valued \textit{functions} to define $\phi(x)$ since the
relativistic quantum field is too singular at short distances
\cite{Fernandez:1992jh}. Rather, a quantum field must be defined as an
operator-valued \textit{distribution} $\phi\left[f(x)\right]$; the
only objects with physical meaning are thus given by the
\textit{smeared fields} $\phi[f]$, where $f(x)$ is a smooth test
function in the Schwartz space ${\cal S}\left(\mathbb{M}^4\right)$ and
\begin{equation}
  \label{eq:smeared-distribution}
  \phi[f] = \int dx\; \phi\left[f(x)\right]\,.
\end{equation}
The fact that the $\phi$-operators only get a meaning in conjunction
with the functions $f(x)$ already displays the need to regularize a
quantum field theory --- a requirement that must be implemented by all
methods striving to compute observables. With this notation, a
relativistic quantum field theory can be postulated using the
\textit{G{\aa}rding-Wightman axioms} \cite{Fernandez:1992jh} (where
the discussion is now limited to the case of a single scalar field in
four dimensions):

\begin{description}
\item[States:] \index{Vacuum} The states of the system are the unit
  rays of a separable Hilbert space $\Hilbert$. There is a
  distinguished state $\vert\Omega\rangle$, called the
  \textit{vacuum}.
\item[Fields:] \index{Fields} There exists a dense subspace
  $D\subset\Hilbert$, and for each test function $f$ in ${\cal
    S}\left(\mathbb{R}^4\right)$ there exists an operator $\phi[f]$
  with domain $D$, such that:
  \begin{enumerate}
  \item The map $f\mapsto\langle\psi_1|\phi[f]\psi_2\rangle$ is a
    tempered distribution $\forall\; \left(\vert\psi_1\rangle,
    \vert\psi_2\rangle\right)\in D$.
  \item For all $f(x)\in\mathbb{R}$, the operator $\phi[f]$ is
    Hermitian.
  \item The vacuum $\vert\Omega\rangle$ belongs to $D$.
  \item $\phi[f]$ leaves $D$ invariant: given an arbitrary
    $\vert\psi\rangle\in D$ implies that $\phi[f]\vert\psi\rangle\in
    D$.
  \item The set $D_0$ of finite linear combinations of vectors of the
    form $\phi[f_1]\dots\phi[f_n]\vert\Omega\rangle$ with $n\geq 0$
    and $f_1,\dots,f_n\in{\cal S}\left(\mathbb{R}^4\right)$ is dense
    in $\Hilbert$.
  \end{enumerate}
\item[Relativistic covariance:] There is a continuous unitary
  representation $U\left(a,\Lambda\right)$ of the proper
  orthochroneous Poincar\'{e} group ${\cal P}_+^\uparrow$ such that
  \begin{enumerate}
  \item $\vert\psi\rangle\in D$ implies
    $U\left(a,\Lambda\right)\vert\psi\rangle\in D$.
  \item $U\left(a,\Lambda\right)\vert\Omega\rangle =
    \vert\Omega\rangle\qquad\forall\left(a,\Lambda\right) \in{\cal
      P}_+^\uparrow$.
  \item $U\left(a,\Lambda\right)\phi[f(x)]U
    \left(a,\Lambda\right)^{-1} =
    \phi\left[f\left(\Lambda^{-1}(x-a)\right)\right]$.
  \end{enumerate}
\item[Spectral condition:] The joint spectrum of the infinitesimal
  generators of the translation subgroup $U\left(a,\mathbf{1}\right)$
  is contained in the forward light cone \[ V_+ = \lbrace
  p=\left(p^0,\mathbf{p}\right)\in\mathbb{R}^4|
  p^0\geq\vert\mathbf{p}\vert\rbrace\,. \]
\item[Locality:] If $f$ and $g$ have spacelike-separated supports,
  then $\phi[f]$ and $\phi[g]$ commute:
  \[ \left(\phi[f]\phi[g]  - \phi[g]\phi[f]\right)\vert\psi\rangle =
  0\qquad\forall \vert\psi\rangle\in D\,. \]
\end{description}

The quantities of major interest are the vacuum expectation values of
products of field operators $\wightm_n\left(f_1,\dots,f_n\right)$.
These objects are called
\index{Distributions!Wightman}\textit{Wightman distributions}:
\begin{equation}
  \label{eq:wightman-distribs}
  \wightm_n\left(f_1,\dots,f_n\right) = \langle\Omega|
  \phi\left[f_1\right] \dots \phi\left[f_n\right]\Omega\rangle\,.
\end{equation}
It can be shown \cite{Streater:1989vi} using the so-called
\index{Reconstruction theorem} ``reconstruction theorem'' that all
information of the quantum theory can be obtained from these vacuum
expectation values. Essentially it allows to construct the state space
as well as the field operators.  Since the $\wightm_n$ are
numerical-valued quantities, they are much easier to work with than
the operator-valued fields $\phi$. This allows for a simpler treatment
of the problem. Assuming that some smearing functions $\lbrace
f_i(x)\rbrace$, $1\leq i\leq n$, peaked around the points $\lbrace
x_i\rbrace$ have been chosen, one can speak of \textit{Wightman
  functions} and introduce the more convenient notation
\begin{equation}
  \label{eq:wightman-functions}
  \wightm_n\left( x_1, \dots, x_n \right) = \wightm_n\left(
  f_1(x),\dots,f_n(x)\right)\,.
\end{equation}
A very powerful observation is the non-trivial fact that the Wightman
functions may be analytically continued from the Minkowski space
$\mathbb{M}^4$ to Euclidean space $\mathbb{R}^4$. This can be done by
applying the following transformation of a four-vector $x$ in
Minkowski-space
\begin{equation}
  \label{eq:Wick-rotation}
  x \mapsto x^{\prime}: x^\prime =
  \left(x^\prime_0,\mathbf{x}^\prime\right) = \left(-\mbox{i}x_0,
  \mathbf{x}\right)\,.
\end{equation}
The Minkowski-metric $g_{\mu\nu}$ is changed to $\delta_{\mu\nu}$.
This transformation is also known as the \index{Wick rotation}
\textit{Wick rotation}. This allows for the definition of the
\index{Distributions!Schwinger}\textit{Schwinger
  functions}\footnote{Again one should not forget that the objects
  under consideration are in fact distributions.}
\begin{equation}
  \label{eq:schwinger-def}
  \schwinger_n\left( x_1, \dots, x_n \right) \equiv \wightm_n\left(
  x^\prime_1,\dots, x^\prime_n \right)\,.
\end{equation}
As discussed in \cite{Montvay:1994cy} this analytic continuation is
possible in the whole complex plane, i.e.~the Schwinger distributions
exist if all $x_i$ are distinct. It can be shown using the
G{\aa}rding-Wightman axioms \cite{Fernandez:1992jh} that the
$\schwinger_n\left(x_1,\dots,x_n\right)$ have the following
properties:
\begin{description}
\item[Reflection positivity:] Let $\theta x =
  \theta\left(x^0,\mathbf{x}\right) = \left(-x^0,\mathbf{x}\right)$
  denote reflection on the real axis. The $\schwinger$ satisfy the
  condition
  \[ \schwinger_n\left(\theta x_1,\dots,\theta x_n\right) =
  \schwinger^*_n\left(x_1,\dots,x_n\right)\,. \]
\item[Euclidean invariance:] The $\schwinger_n$ are invariant under
  all Euclidean transformations
  $\left(a,R\right)\in\mbox{SO}\left(4\right)$:
  \[ \schwinger_n\left( R x_1+a,\dots, R x_n+a\right) =
  \schwinger_n\left(x_1,\dots,x_n\right)\,. \]
\item[Positive definiteness:] Define the composition of test functions
  $f_i\in{\cal S}\left(\mathbb{R}^{4n}\right)$, $i=0,\dots,n$, by
  (with $k+l<n$)
  \[ \left(f_k\otimes f_l\right)\left(x_1,\dots,x_{k+l}\right) =
  f_k\left(x_1,\dots,x_k\right) f_l\left(x_{k+1},\dots,x_{k+l}\right)\,.
  \] The $\schwinger_n$ then obey the following condition:
  \[ \sum_{k,l=0}^n \schwinger_{k+l}\left(f_k^*\left(\theta x_n,\dots,
      \theta x_1\right)\right)\otimes
  f_l\left(x_1,\dots,x_l\right)\geq 0\,. \]
\item[Permutation symmetry:] The $\schwinger_n$ are symmetric in their
  arguments.
\end{description}
Now there is another important theorem due to \textsc{Osterwalder} and
\textsc{Schrader}: given the Schwinger distributions
(\ref{eq:schwinger-def}) satisfying the above conditions, one can
reconstruct the whole quantum field theory in Minkowski space
\cite{Osterwalder:1973ar,Osterwalder:1975ar,Osterwalder:1973bo}. So
the axioms due to G{\aa}rding-Wightman and Osterwalder-Schrader are
equivalent and one can use the Osterwalder-Schrader framework to
actually \textit{define} a relativistic quantum field theory.

Consequently, it is sufficient to compute the Schwinger functions for
a Euclidean quantum field theory and then reconstruct the Minkowski
theory from them. As it will be discussed later on, the Schwinger
functions are easier to handle than the Wightman distributions.
However, this has to be taken with a grain of salt: physical
observables calculated in the Euclidean theory must afterwards be
analytically continued back to Minkowski space to allow for a
comparison with experiments. After all, the physical quantities are
defined in the Minkowski theory and not in the Euclidean domain. In
some cases (e.g.~for the correlation lengths of the two-point function
which is the inverse of the particle mass), the results are identical,
i.e.~the inverse Wick-rotation does \textit{not} change the value
obtained in the calculation. However, there exist a lot of cases,
where the analytic continuation is non-trivial. For details the reader
is encouraged to consult \cite{Montvay:1994cy}.

\begin{figure}[htb]
  \begin{center}
    \includegraphics[scale=0.75]{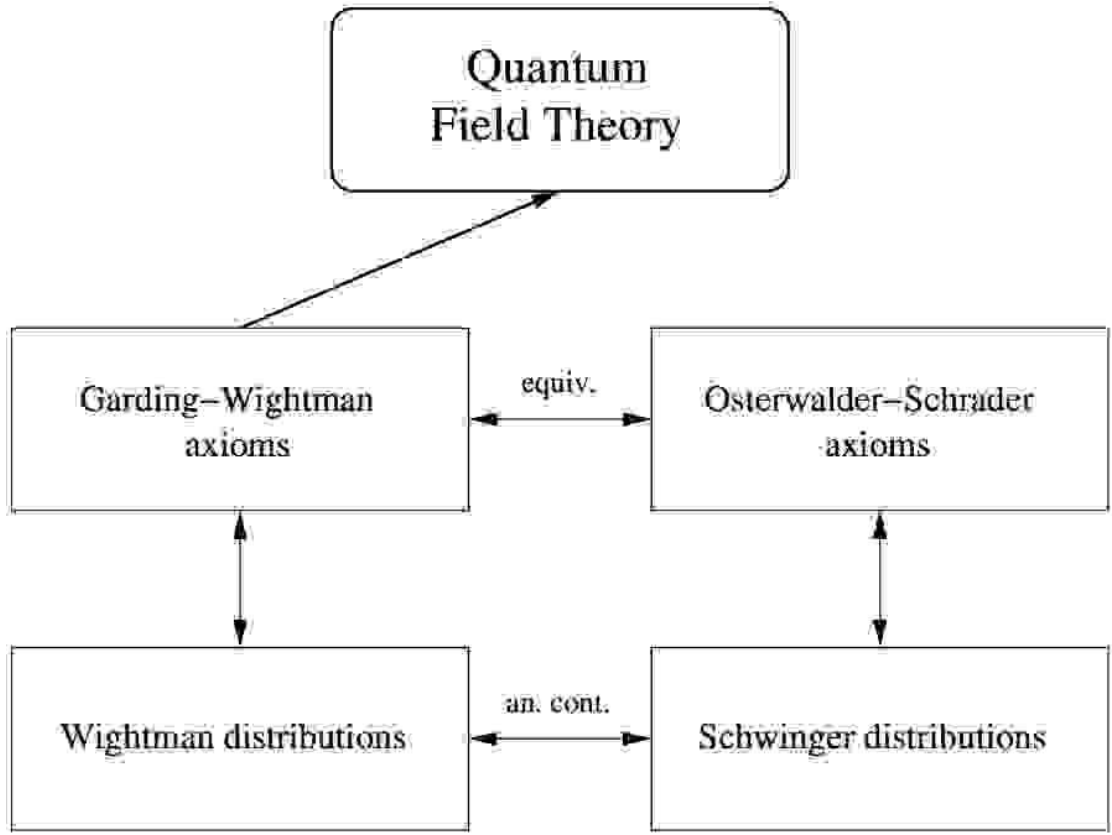}
    \caption{Relations of different axiomatic frameworks for quantum
      field theory.}
    \label{fig:qft-axioms}
  \end{center}
\end{figure}

The relations between the different axiomatic settings discussed so
far are given in Fig.~\ref{fig:qft-axioms}. From the Wightman
distributions, the whole QFT can be constructed. However, the
Osterwalder-Schrader axioms are an equivalent formulation. The
Schwinger distributions and the Wightman distributions are related by
analytic continuation.

\begin{figure}[htb]
  \begin{center}
    \includegraphics[scale=0.75]{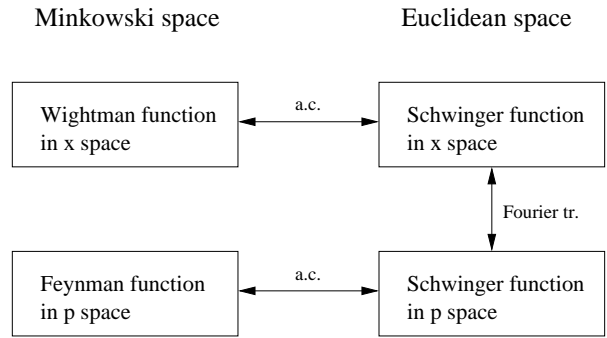}
    \caption{Relations between the different kinds of $n$-point
      functions consisting of vacuum expectation values of products of
      field operators. The figure has been taken from
      \cite{Roepstorff:1991xb}.}
    \label{fig:npoint-funcs}
  \end{center}
\end{figure}

The permutation symmetry of the Schwinger functions allows for the
construction of a generating functional. In contrast, the Wightman
functions are only symmetric for spacelike-separated arguments. Thus,
they can not be computed in terms of a generating functional. Another
elegant way to define generating functions in Minkowski space is the
introduction of \index{Distributions!Feynman}Feynman functions which
can be defined as ``time-ordered'' products of field operators:
\begin{equation}
  \label{eq:feynman-distrib}
  \feynm_n\left(x_1, \dots, x_n\right) =
  \langle\Omega\vert\mathscr{T}\left\lbrace\phi\left[x_1\right]
  \dots\phi\left[x_n\right]\right\rbrace\vert\Omega\rangle\,,
\end{equation}
where the time-ordering is defined as the product with factors
arranged so that the one with the last time-argument is placed
leftmost, the next-latest next to the leftmost
etc.~\cite{Weinberg:1995mt}. There is also an alternative definition
in \cite{Roepstorff:1991xb}. It is given by applying a Fourier
transform to the Schwinger function and performing an analytic
continuation back to Minkowski space afterwards. This is displayed in
Fig.~\ref{fig:npoint-funcs}.

By construction, the Feynman functions are also symmetric for
timelike-separated arguments and thus they are symmetric for arbitrary
arguments.
Hence, both the Schwinger and the Feynman functions allow for the
construction of a generating functional $W_{\schwinger}\left[J\right]$
and $W_{\feynm}\left[J\right]$. Only the Schwinger functions will be
considered here --- formally the generating functional (sometimes it
is also called the vacuum-vacuum functional) is given by
\begin{eqnarray}
  \label{eq:schwinger-generator}
  W_{\schwinger}\left[J\right] &=& \sum_{n=0}^\infty
  \frac{\mbox{i}^n}{n!} \int dx_1 \dots dx_n \schwinger_n\left(
    x_1,\dots,x_n \right) J(x_1)\dots J(x_n) \nonumber \\
  &=& \langle\Omega\vert\exp\left[ \mbox{i}\int d^4x
  \phi[f]J(x)\right]\vert\Omega\rangle\,,
\end{eqnarray}
where the functions $J(x_i)$ are taken from the Schwartz space ${\cal
  S}\left(\mathbb{R}^4\right)$. Using $W_{\schwinger}[J]$, the
Schwinger functions can be recovered by a functional derivative
\begin{equation}
  \label{eq:schwinger-recover}
  \schwinger_n\left(x_1,\dots,x_n\right) = \frac{\delta^n
  W_{\schwinger}\left[J\right]}{\delta J_1\dots\delta J_n}\,.
\end{equation}
Knowledge of $W_{\schwinger}$ is thus equivalent to solving the
quantum field theory.

\subsection{The Path Integral}
\label{sec:path-integral}
Having now discussed what a quantum field theory \textit{is}, one
needs a recipe of how to construct it. In fact, there exist several
prescriptions of how to build a quantum theory if the Lagrangian of
the classical field theory to which the quantum theory should reduce
is known. The two most commonly used quantization schemes are the
\index{Quantization!canonical} \textit{canonical quantization scheme}
(which is described in detail in standard textbooks like
\cite{Messiah:1991aa,Peres:1993bo,Haag:1992hx,Weinberg:1995mt}) and
the \textit{path-integral quantization} which will be used in this
thesis. Both of these schemes break the general covariance of the
classical theory discussed in Sec.~\ref{sec:class-field-theor}. The
quantum theory still stays invariant under global Lorentz
transformations (and there even exist generalizations to curved, but
fixed spacetimes, see e.g.~\cite{Sexl:1983im} and references therein),
but the quantization prescriptions implicitly assume the existence of
a global, canonical basis. However, local Lorentz invariance is only
of importance for a quantum theory of gravity, so all to be said in
the following can be applied to any quantum theory of strong
interactions discussed in Sec.~\ref{sec:phen-strong-inter}.

\subsubsection{Construction Principle}
\label{sec:constr-princ}
Before attempting to define a prescription for a quantum field theory,
let us go back to the case of non-relativistic quantum mechanics. The
notion of a path integral is closely related to the notion of a random
walk. To make this relation obvious, consider the expectation value
$\langle E\rangle$ of the evolution operator applied to a single
particle in one dimension between two states $\vert x\rangle, \vert
y\rangle\in\Hilbert$:
\begin{equation}
  \label{eq:expect-nonrel}
  \langle E\rangle =
  \langle x\vert\exp\left[-\mbox{i}\hamiltonian t\right]\vert
  y\rangle\,.
\end{equation}
$\langle E\rangle$ is the probability amplitude for the particle to
move from position $x$ to position $y$ in time $t$. If the Hamiltonian
corresponds to a free particle, \[ \hamiltonian =
\frac{\hat{p}^2}{2m}, \] then the solution to (\ref{eq:expect-nonrel})
can be given immediately \cite{Montvay:1994cy}:
\begin{equation}
  \label{eq:expect-gauss-sol}
  \langle E\rangle = \left(\frac{m}{\mbox{i}2\pi t}\right)^{1/2}
  \exp\left[\mbox{i}\frac{m}{2t} (x-y)^2\right]\,.
\end{equation}
On the other hand, the probability for a one-dimensional random walk
to go from position $x$ to position $y$ in time $t$ is given by
\cite{Roepstorff:1991xb}:
\begin{equation}
  \label{eq:expect-random-walk}
  P_0(x-y,t) = \left(\frac{1}{4\pi Dt}\right)^{1/2}
  \exp\left[-\frac{1}{4Dt}(x-y)^2\right]\,,
\end{equation}
with $D$ being the diffusion constant. The quantum mechanical
expectation value is obtained by analytic continuation of
Eq.~(\ref{eq:expect-random-walk}) to imaginary time and the
identification $D=1/2m$. Thus, the quantum mechanical amplitude may be
computed by considering a classical random walk and analytically
continuing the result to imaginary time. If one adopts this
interpretation, the amplitude can be computed via a
\index{Quantization!path-integral}\textit{path-integral} using a
\index{Measure!Wiener}\textit{conditional Wiener measure}, see
\cite{Roepstorff:1991xb} for a rigorous mathematical treatment.

To extend Eq.~(\ref{eq:expect-gauss-sol}) also to the case of
non-Gaussian Hamiltonians, we decompose the full Hamiltonian
$\hamiltonian$ into a Gaussian and a non-Gaussian part:
\begin{equation}
  \label{eq:full-hamiltonian}
  \hamiltonian = \hamiltonian_0 + \hat{V}(x)\,,
\end{equation}
and perform a time-slicing procedure. Consider the evolution operator
$\hat{U}(t)=\exp[-\mbox{i}\hamiltonian t]$ for small imaginary times
$t\rightarrow \mbox{i}\epsilon$. In the leading order,
$\hat{U}\left(\mbox{i}\epsilon\right)$ coincides with the operator
$\hat{W}\left(\epsilon\right)$ which is defined in the following way:
\begin{equation}
  \label{eq:transfer-matrix}
  \hat{U}\left(\mbox{i}\epsilon\right) =
  \hat{W}\left(\epsilon\right) + {\cal O}\left(\epsilon^3\right) =
  \exp\left( \hat{V} \frac{\epsilon}{2}\right)\exp\left(
    \hamiltonian_0\epsilon \right)\exp\left( \hat{V}\frac{\epsilon}{2}
  \right) + {\cal O}\left(\epsilon^3\right)\,.
\end{equation}
The operator $\hat{W}\left(\epsilon\right)$ is known as the
\textit{transfer matrix}; its matrix elements can be computed to
yield:
\begin{equation}
  \label{eq:smalleps-opera}
  \langle x\vert\hat{W}\left(\epsilon\right)\vert y\rangle = \left(
  \frac{m}{2\pi\epsilon} \right)^{1/2} \exp\left[ \frac{m}{2\epsilon}
  (x-y)^2 + \frac{\epsilon}{2}\left(V(x)+V(y)\right) \right]\,.
\end{equation}
Using the Lie-Trotter formula,
one gets:
\begin{equation}
  \label{eq:finite-evol}
  \exp\left[-\left(\hamiltonian_0 + \hat{V}\right) t\right] =
  \lim_{N\rightarrow\infty}\hat{W}^N\left(\epsilon\right)\,.
\end{equation}
Inserting $N-1$ times the identity $\mathbf{1}=\int dx_i \vert
x_i\rangle\langle x_i\vert$ into Eq.~(\ref{eq:finite-evol}) finally
yields the expression:
\begin{eqnarray}
  \label{eq:pathint-expand}
  \langle x\vert\exp\left[-\hamiltonian t\right]\vert y\rangle &=&
  \lim_{N\rightarrow\infty} \left(\frac{m}{2\pi\epsilon}\right)^{N/2}
  \int dx_1\dots dx_{N-1} \nonumber\\
  && \quad\times\exp\Biggl[ -\frac{m}{2\epsilon}\left(\left(
      x-x_1\right)^2+\dots+\left(x_{N-1}-y\right)^2\right) \nonumber\\
  && \quad-\epsilon\left(\frac{1}{2}V(x)+V(x_1)+\dots+V(x_{N-1})
    +\frac{1}{2}V(y)\right) \Biggr] \nonumber \\
  &=& \lim_{N\rightarrow\infty} \left(\frac{m}
    {2\pi\epsilon}\right)^{N/2} \int dx_1\dots dx_{N-1} \nonumber \\
  && \qquad\qquad\qquad\times
  \exp\left[-S_N\left(x,x_1,\dots,x_{N-1},y\right) \right]\,.
\end{eqnarray}
In the continuum limit this equation can now be interpreted as a path
integral over a set of random walks with the weight function given in
the exponential. If we denote all paths $\omega(\tau)$ with fixed
end-points from $\omega(\tau=0)=x_0=x$ to $\omega(\tau=t)=x_N=y$, then
we can write (using the Wiener measure $\EnsembleMeas{\omega}$):
\begin{eqnarray}
  \langle E\rangle &=& \int\EnsembleMeas{\omega}
  \exp\left(-V[\omega]\right) \nonumber \\
  &=& \lim_{N\rightarrow\infty} \int dx_1 \dots dx_{N-1}
  P\left(x_0,\dots,x_N\right)
  \exp\left(-V\left(x_0,\dots,x_N\right)\right)\,,
  \label{eq:pathint-def}
\end{eqnarray}
where \[ P\left(x_0,\dots,x_N\right)=\prod_{i=0}^N
P_0\left(x_{i+1}-x_i, t_{i+1}-t_i\right)\,. \] Now there is an
important difference between the forms of
Eqs.~(\ref{eq:pathint-expand}) and (\ref{eq:pathint-def}): In the
latter, the exponential weight which connects neighbor points of the
paths is already a part of the measure $\EnsembleMeas{\omega}$, while
in the former the exponential weight is contained in the expression
for $S_N\left(x_0,\dots,x_N\right)$. What is then the interpretation
of this weight factor? From Eq.~(\ref{eq:wkb-amplitude-expansion}) we
know, that any amplitude can be expanded in a power series of
$\hbar^n$ with the amplitude for the classical process being the
leading amplitude. Thus, in the limit $\hbar\rightarrow 0$, only the
classical (leading) contribution should contribute to the expression
(\ref{eq:pathint-expand}). The exponential weight factor should thus
be peaked around the classical solution, i.e.~the exponential factor
will become minimal for the classical trajectory, just like minimizing
the \index{Action} action $\mathscr{S}\left[\omega(\tau)\right]$
yields the classical path. Thus, the exponential weight factor in the
limit $N\rightarrow\infty$ should coincide with the classical action
if one inserts a differentiable trajectory. However, there is an
important difference: the classical action
$\mathscr{S}\left[\omega(\tau)\right]$ is only defined for
differentiable paths, while the exponential factor
$\lim_{N\rightarrow\infty}S_N(x_0,\dots,x_N)$ in
(\ref{eq:pathint-expand}) is defined for any continuous path (which is
a superset of the set of all differential paths). This gives rise to a
certain freedom in the choice of $S_N(x_0,\dots,x_N)$. The actual
choice should thus be guided by the desire to simplify the problem at
hand. Especially in the case of chiral fermions, a wide class of
possible actions has been proposed, see Sec.~\ref{sec:fermion-fields}.

Symbolically we can thus introduce a functional
$S\left[\omega(\tau)\right]$ which projects any continuous path to a
real number and write
\begin{equation}
  \label{eq:pathint-nonrel}
  \langle x\vert\exp\left[-\hamiltonian t\right]\vert y\rangle =
  \int{\cal D}x \exp\left[-S\left[x\right]\right]\,,
\end{equation}
where the integral measure is given by
\begin{equation}
  \label{eq:nonrel-intmeas}
  {\cal D}x = \lim_{N\rightarrow\infty} \left(\frac{m}{2\pi\epsilon}
  \right)^{N/2} dx_1\dots dx_{N-1}\,.
\end{equation}
The resulting expression Eq.~(\ref{eq:pathint-nonrel}) can be
analytically continued back to imaginary times using
$t\rightarrow\mbox{i}t$ which yields the desired transition amplitude
Eq.~(\ref{eq:expect-gauss-sol}). Such a \index{Wick rotation}
Wick-rotated form of Eq.~(\ref{eq:pathint-nonrel}) is known as the
\textit{Feynman-Kac} formula. Sometimes the derivation is directly
carried out in Minkowski space, but the problem is that the integrand
is highly oscillatory and not well-defined, for further reading
cf.~\cite{Albeverio:1978ew}.

As already pointed out, the difference between (\ref{eq:pathint-def})
and (\ref{eq:pathint-nonrel}) lies in the interpretation of the
measure. One can perform substitutions ${\cal D}x\rightarrow{\cal
  D}x^\prime$ to (\ref{eq:nonrel-intmeas}), giving rise to different
integral measures. Since the path integral has close resemblance to a
system of statistical mechanics (via its affinity to the random walk),
we will classify the different classes of paths which can be used in
(\ref{eq:pathint-nonrel}) by the means of \index{Ensembles}
\textit{ensembles}. This topic is discussed in detail in
Sec.~\ref{sec:ensembles}. For the time being, we want to interpret
(\ref{eq:pathint-expand}) as an integral over \textit{random paths}
with a weight given by the entire exponential. This amounts to
choosing the measure \[ {\cal D}x \propto\lim_{N\rightarrow\infty}
dx_1\dots dx_{N-1}\,., \] Later in Sec.~\ref{sec:ensembles} it will be
argued that these paths are taken from the \textit{random ensemble}.
In contrast, in the expression (\ref{eq:pathint-def}) using the Wiener
measure $\EnsembleMeas{\omega}$, the paths are taken from the
\textit{canonical ensemble}. This integral measure already contains
the kinetic term, but not the potential term. In this way, the problem
of assigning a meaning to the derivative from the classical action is
circumvented. This procedure is not possible in the case for quantum
field theories which will be discussed below since in that case there
is no such thing as a Wiener measure.

\subsubsection{Computing Observables}
\label{sec:comp-observ}
As discussed in Sec.~\ref{sec:axioms-relat-quant}, one is interested
in ground-state expectation values of certain operators, $\langle0
|\hat{A}|0\rangle$. Consider a (countable) Hilbert space $\Hilbert$
with Hamiltonian $\hamiltonian$. Let $\lbrace E_i\rbrace$, $i\geq 0$,
be the eigenvalues and $\lbrace|i\rangle\rbrace$ be the corresponding
eigenvectors of $\hamiltonian$ in ascending order. Taking the trace of
the evolution operator and $\hat{A}$ provides us with \[
\mbox{Tr}\,\left( \hat{A}\exp\left[-\hamiltonian\tau\right]\right) =
\sum_{i=0}^\infty \exp\left[-E_i\tau\right] \langle
i|\hat{A}|i\rangle\,. \] In the limit $\tau\rightarrow\infty$ only the
term with $E_0$ in the exponential survives and we are left with
\begin{equation}
  \label{eq:vacuum-expect}
  \langle 0|\hat{A}|0\rangle = \lim_{\tau\rightarrow\infty}
  \frac{1}{Z(\tau)} \mbox{Tr}\,\left(\hat{A}\exp\left[
      -\hamiltonian\tau\right]\right)\,,
\end{equation}
with the \textit{partition function} \index{Partition function}
\index{State sum|see{Partition function}}
\begin{equation}
  \label{eq:state-sum-nofields}
  Z(\tau) = \mbox{Tr}\,\left(\exp\left[ -\hamiltonian\tau\right]\right)
  = \sum_{i=0}^\infty \exp\left[-E_i\tau\right]\,.
\end{equation}
For the application to field theory, the operators $\hat{x}(t_n)$ will
require special attention, since they are analogous to the Schwinger
functions encountered in Euclidean quantum field theories in
Sec.~\ref{sec:axioms-relat-quant}). Using
$\hat{x}(t)=\exp[\hamiltonian t]\hat{x}\exp[-\hamiltonian t]$, we
consider the $n$-point correlation function $\langle x(\tau_1)\dots
x(\tau_n)\rangle$, with $\tau_1<\dots<\tau_n$. It is straightforward
\cite{Montvay:1994cy} to show that
\begin{eqnarray}
  \label{eq:nonrel-schwinger}
  \langle 0|\hat{x}(\tau_1)\dots\hat{x}(\tau_n)|0\rangle &=& \nonumber
  \lim_{\tau\rightarrow\infty}\frac{1}{Z(\tau)} 
  \mbox{Tr}\,\left( e^{-\hamiltonian(\tau/2-\tau_1)}\hat{x}
    e^{-\hamiltonian(\tau_1-\tau_2)} \dots \hat{x}
    e^{-\hamiltonian(\tau_n+\tau/2)} \right) \\
  &=& \lim_{\tau\rightarrow\infty}\frac{1}{Z(\tau)} \int {\cal
  D}x\, x(\tau_1)\dots
  x(\tau_n)\exp\left[-S\left[x(\tau)\right]\right]\,,
\end{eqnarray}
where the paths obey periodic boundary conditions, \[
x\left(-\tau/2\right) = x\left(\tau/2\right), \] and the partition
function $Z(\tau)$ can be written as
\begin{equation}
  \label{eq:state-sum}
  Z(\tau) = \int {\cal D}x
  \exp\left[-S\left[x(\tau)\right]\right]\,.
\end{equation}
Hence, the $n$-point correlation function $\langle x(\tau_1)\dots
x(\tau_n)\rangle$ is written in (\ref{eq:nonrel-schwinger}) as the
moment of the measure ${\cal D}x$. There is another possibility to
obtain the correlation function from a generating functional
$Z_J[J(\tau)]$ with $J(\tau)$ being a continuous path by means of the
following definition:
\begin{eqnarray}
  \label{eq:state-sum-pert}
  Z_J[J(\tau)] &=& \lim_{N\rightarrow\infty} \left(\frac{m}
    {2\pi\epsilon}\right)^{N/2} dx_1\dots dx_{N-1} \nonumber \\
  && \qquad\times\exp\left[ -S(x_0,\dots,x_N) + \sum_{i=1}^{n}x_i
    J(\tau_i)\right]\,.
\end{eqnarray}
Using
\begin{eqnarray}
  \label{eq:gen-func-diff}
  && \lim_{N\rightarrow\infty}\left(
    \frac{m}{2\pi\epsilon}\right)^{N/2} \int dx_1 \dots dx_{N-1}
  \frac{d}{dJ(\tau_1)}\dots \frac{d}{dJ(\tau_n)} \nonumber \\
  && \qquad\times\exp\left[-S(x_0,\dots,x_N) + \sum_{i=1}^{n}x_i
    J(\tau_i)\right] \nonumber \\
  &=& \lim_{N\rightarrow\infty} \left(
    \frac{m}{2\pi\epsilon}\right)^{N/2} \int dx_1 \dots dx_{N-1}\,
  x(\tau_1) \dots x(\tau_n) \nonumber \\
  && \qquad\times\exp\left[-S(x_0,\dots,x_N) + \sum_{i=1}^{n}x_i
    J(\tau_i)\right]\,,
\end{eqnarray}
one recovers Eq.~(\ref{eq:nonrel-schwinger}). The meaning of the
derivatives can be understood by considering again the subset of
differentiable paths. The expression then reduces to the
\textit{functional derivative}. Thus, $Z_J[J(\tau)]$ can be considered
to be the generating functional for the $n$-point correlation
functions and we can write symbolically:
\begin{equation}
  \label{eq:nonrel-schwinger2}
  \langle x(\tau_1)\dots x(\tau_n)\rangle =
  \lim_{\tau\rightarrow\infty} \frac{\delta^n Z_J[J(\tau)]} {\delta
  J(\tau_1) \dots \delta J(\tau_2)}\,.
\end{equation}
Here the notation of the conventional functional derivative has been
employed, but with a meaning corresponding to
Eq.~(\ref{eq:gen-func-diff}. This expression is similar to the
generating functional of the Schwinger functions
(\ref{eq:schwinger-recover}), implying that generalizing
$Z_J[J(\tau)]$ to the case of Euclidean fields is the key to find a
quantization prescription for quantum field theories.

\subsubsection{Euclidean Field Theory}
\label{sec:eucl-field-theory}
The generalization of Eq.~(\ref{eq:nonrel-schwinger2}) to the case of
Euclidean fields is very difficult, however. As a starting point one
can expect that the expectation values for the Schwinger function in
(\ref{eq:schwinger-def}) can also be written using a path integral
just like the $n$-point functions in (\ref{eq:nonrel-schwinger}).
\index{Distributions!Schwinger} They would then be moments of some
suitably defined measure
\begin{equation}
  \label{eq:pathint-field}
  \EnsembleMeas{\phi} =
  \mathscr{N}\exp\left[-S\left[\phi\right]\right] {\cal D}\phi\,.
\end{equation}
The Schwinger functions $\schwinger_n(x_1,\dots,x_n)$ can hence be
written
\begin{equation}
  \label{eq:schwinger-pathint}
  \schwinger_n(x_1,\dots,x_n) = \frac{1}{W_\schwinger[0]}
  \int\EnsembleMeas{\phi} \phi(x_1)\dots\phi(x_n)\,,
\end{equation}
where the generating functional $W_\schwinger[0]$
\index{Partition function} is the field theory analogue of
Eq.~(\ref{eq:state-sum}). It can symbolically be written as
\begin{equation}
  \label{eq:gen-func-euclid}
  W_\schwinger[0] = \int\EnsembleMeas{\phi} \equiv Z\,.
\end{equation}
The functional $S[\phi]$ appearing in Eq.~(\ref{eq:pathint-field}) is
again a suitable generalization of the Euclidean action to a superset
of continuous, but non-differentiable fields. The vacuum expectation
value $\langle\hat{A}\rangle= \langle\Omega|\hat{A}|\Omega\rangle$ of
a general operator, $\hat{A}\left[\phi(x)\right]$, is then defined by
the path integral
\begin{equation}
  \label{eq:operator-def}
  \langle\hat{A}\rangle = Z^{-1} \int\EnsembleMeas{\phi}\,
  \hat{A}\left[\phi\right]\,,
\end{equation}
where the partition function $Z$ is given by
(\ref{eq:gen-func-euclid}). However, this definition encounters severe
difficulties because of the fact that the $\phi[f]$ are not
pointwise-defined objects.

By inverting the logic which led to the path-integral formula
Eq.~(\ref{eq:pathint-def}), one can define a prescription to formulate
a quantum field theory starting from a classical action $\mathscr{S}$.
This procedure which gives meaning to Eq.~(\ref{eq:pathint-field}) is
called \index{Renormalization|(}\textit{renormalization theory} and
consists of the following steps \cite{Fernandez:1992jh}:
\begin{enumerate}
\item Regularize the theory by imposing an ultraviolet cutoff
  $\Lambda=a^{-1}$ (where $a$ is a distance short compared to the
  intrinsic scales of the theory) so that (\ref{eq:pathint-field}) is
  a well-defined measure. This can e.g.~be done by discretizing the
  Euclidean space $\mathbb{R}^4$ to describe the system using a
  (finite) lattice in $\mathbb{Z}^4_\Omega$ such that all $\lbrace
  x_i\rbrace\in\mathbb{Z}^4_\Omega$. Find a functional $S_{\lbrace
    g_i\rbrace}[\phi(x)]$ with parameters $\lbrace g_i\rbrace$ on the
  lattice which reduces to the classical action $\mathscr{S}[\phi(x)]$
  for differentiable continuum fields.  This prescription is not
  unique. In any case, however, either Euclidean invariance or
  Osterwalder-Schrader positivity or both are broken. Let
  $\langle\phi(x_1)\dots\phi(x_n)\rangle_{\lbrace g_i\rbrace}$ be the
  $n$-point functions of the discrete theory.
\item Perform the infinite volume limit $\Omega\rightarrow\infty$ for
  the system with $\lbrace g_i\rbrace$ held fixed. This limit must
  exist and be unique.
\item Allow the parameters $\lbrace g_i\rbrace$ of $S_{\lbrace
    g_i\rbrace}\left[\phi\right]$ to be functions of $\Lambda$:
  $\lbrace g_i\rbrace\mapsto\lbrace g_i\left(\Lambda\right)\rbrace$.
  The parameters occurring in (\ref{eq:pathint-field}) are then called
  the \textit{bare parameters}.
\item Perform the \textit{continuum limit}
  $\Lambda=a^{-1}\rightarrow\infty$. A continuum quantum field theory
  is obtained from the sequence of lattice theories by rescaling the
  lengths by a factor $\Lambda$ and rescaling the fields by a factor
  $Z(\Lambda)$:
  \begin{equation}
    \label{eq:cont-scale}
    \schwinger_n(x_1,\dots,x_n) = \lim_{\Lambda\rightarrow\infty}
    Z(\Lambda)^n \langle \phi(x_1)\dots\phi(x_n)\rangle_{\lbrace
      g_i(\Lambda)\rbrace}\,.
  \end{equation}
  For each choice $\lbrace g_i\left(\Lambda\right)\rbrace$ check the
  convergence properties of $\schwinger_n$ and if they satisfy the
  Osterwalder-Schrader axioms.
\item Consider all possible choices of $\lbrace g_i(a^{-1})\rbrace$
  and $Z(a^{-1})$; classify all limiting theories
  $\schwinger_n(x_1,\dots,x_n)$ and study their properties.
\end{enumerate}
This procedure may give rise to continuum theories which can be
categorized as follows:
\begin{description}
\item[No limit:] For at least one $n$, the limit (\ref{eq:cont-scale})
  does not exist.
\item[Unimportant limit:] All resulting $\lbrace\schwinger_n\rbrace$
  exist, but are devoid of information (like $\schwinger_n=0\ \forall
  n$ etc.)
\item[Gaussian limit:] The limiting theory
  $\lbrace\schwinger_n\rbrace$ is Gaussian, i.e.~a generalized free
  field. This situation is commonly referred to as
  \textit{triviality}.
\item[Non-Gaussian limit:] The limiting theory is non-Gaussian giving
  rise to a nontrivial theory. This may, however, still imply that the
  scattering matrix is the identity.
\end{description}
For a non-trivial limit to exist, the lattice theories should have
\index{Correlation length}\textit{correlation lengths}
$\xi(\Lambda)\simeq \Lambda^{-1}\xi\lbrace g_i(\Lambda)\rbrace$ as
$\Lambda\rightarrow\infty$ (otherwise the physical lengths would get
rescaled to $0$). Thus, the parameters $\lbrace g_i(a^{-1})\rbrace$
should approach or sit on the critical surface and the theory must
undergo a phase transition of second order where the correlation
lengths diverge. This is expected to be the case for most interesting
quantum field theories whose critical behavior can be handled using
the renormalization group of \textsc{Wilson}, see \cite{Wilson:1974jj}
for the historical paper and \cite{Montvay:1994cy,Rothe:1992nt} for
standard textbooks. There is also a second very interesting case where
$\xi\lbrace g_i(\Lambda)\rbrace=\infty$ for all $\Lambda$, i.e.~the
parameters $\lbrace g_i\rbrace$ already sit on the critical surface
for finite lattice spacings. This is e.g.~the case in non-compact
U$(1)$ pure gauge theories. For compact U$(1)$ the situation is less
clear so far, consult for a description of simulation results the work
of \textsc{Arnold} \cite{Arnold:1998vn} and references
therein.\index{Renormalization|)}.

Despite the huge phenomenological successes of quantum field theories
in practice, a rigorous proof that the resulting theory exists in the
sense defined above, has been stated so far only for a few special
cases. In four dimensions, so far only free fields have been proven
with mathematical rigor to give rise to a relativistic quantum field
theory.

\subsubsection{Evaluation of Path Integrals}
\label{sec:eval-path-integr}
Having now a definition for the path integral, we also need a way to
evaluate it. In principle there are two different ways to compute
expressions of the form (\ref{eq:pathint-nonrel}) and
(\ref{eq:operator-def}):
\begin{itemize}
\item Consider a series of weight factors $\left\lbrace \exp\left[
      -S_i\right]\right\rbrace$, $i=1,\dots,N$, which converges to the
  desired weight factor
  $\lim_{i\rightarrow\infty}\exp\left[-S_i\right] =
  \exp\left[-S\right]$. The path integrals (\ref{eq:pathint-nonrel})
  should be computable for each $\exp\left[ -S_i\right]$.
\item Compute an approximation to (\ref{eq:pathint-nonrel}) for finite
  $N$ in the measure (\ref{eq:nonrel-intmeas}). The resulting
  approximation will depend on $N$. Then perform the limit
  $N\rightarrow\infty$.
\end{itemize}
As has already been mentioned, taking the limit in
(\ref{eq:cont-scale}) is only possible in some simple models, or in
the case that the resulting integrals have Gaussian shape. One way to
also extend the applicability to non-Gaussian models is thus to
approximate the ``true'' function $W_{\schwinger}$ by integrable
Gaussian models which reduce to the $W_{\schwinger}$ in some suitable
limit.

\index{Perturbation theory} The most popular form to do this is to
expand the exponential into a Gaussian part and a small, non-Gaussian
part:
\begin{equation}
  \label{eq:perturbative-expansion}
  \exp\left[-S[\phi]\right] = \exp\left[-S_{\mbox{\tiny
  Gaussian}}\right] \sum_{n=0}^\infty \frac{1}{n!} \frac{\delta^n
  \left(\exp\left[-(S-S_{\mbox{\tiny
  Gaussian}})[\phi]\right]\right)}{\delta\phi^n}\,.
\end{equation}
The idea is then to form the path-integral of the r.h.s.~of
Eq.~(\ref{eq:perturbative-expansion}) and take the result to be the
sum of all contributions. The problem behind the series obtained this
way is that in several cases the sum fails to converge. This is the
case of the common four-dimensional models, as has first been noted by
\textsc{Dyson} in \cite{Dyson:1952tj}.

As an example consider the ``field theory'' at a single site with
``partition function'' \cite{Cheng:1984xx}
\begin{equation}
  \label{eq:state-sum-example}
  Z = \frac{1}{\sqrt{2\pi}}\int_{-\infty}^{\infty} d\phi\;
  \exp\left(-\phi^2/2 + g\phi^4\right)\,.
\end{equation}
The function $Z=Z(g)$ contains an essential singularity at the origin.
Performing the perturbative expansion
(\ref{eq:perturbative-expansion}) yields \[ Z = \sum_{k=0}^\infty g^k
Z_k,\quad Z_k=\frac{(-1)^k}{k!}\sqrt{\frac{2}{\pi}}\int_0^\infty d\phi\;
\exp\left(-\phi^2/2 + 4k\ln\phi\right)\,, \] which has a convergence
radius of $g=0$. Performing a semi-classical expansion around the
saddle point $\phi_c=2\sqrt{k}$ and integrating over the quadratic
deviations yields \[ Z_k \approx \frac{(-16)^k}{\sqrt{\pi}} \exp\left(
  (k-1/2)\ln k-k\right)\,. \] Obviously the $Z_k$ are divergent (and
the divergence is in fact logarithmic), but the power series is at
least asymptotic in the complex $g$ plane cut along the negative real
axis since
\begin{eqnarray*}
  \left\vert Z(g) - \sum_{k=0}^\infty g^k Z_k\right\vert &<&
  \frac{4^{n+1}\Gamma(2n+3/2)} {\sqrt{\pi}(n+1)!} \frac{|g|^{n+1}}
  {(\cos(1/2\mbox{Arg}\, g))^{2n+3/2}}\,.
\end{eqnarray*}
This means that for fixed $n$ the right hand side can be made
arbitrarily small by choosing $g$ small enough. It may even be
possible to recover the full partition function $Z$ from the series
expansion $\lbrace Z_k\rbrace$ using resummation. For recent reviews
of the application of resummation techniques consult
\cite{Beneke:1998ui,Beneke:2000kc}.

Despite these conceptional difficulties, perturbation theory turns out
to be the most effective approach to treat many problems in quantum
field theory provided the expansion parameter is sufficiently small.
However, in several situations of interest, the latter condition is
not fulfilled and the perturbative expansion is not even asymptotic,
or the expansion parameter is too large, causing it to diverge already
in the lowest orders. In these situations, one has to resort to
different ways to approximate the Schwinger functions. One possibility
is a numerical simulation of Euclidean QFT on the finite, discrete
lattice $\mathbb{Z}^4_\Omega$. It has some very intriguing advantages:
it does not resort to any assumptions of the dynamics of the model one
is examining other than the information underlying the regularized
action and it is directly based on the definition of the quantities
under consideration. In essence, any operator
$\hat{A}\left[\phi(x)\right]$ corresponding to a physical observable
can be written via (\ref{eq:operator-def}) as the corresponding moment
of a measure $\EnsembleMeas{\phi}$ on the underlying space. The
ensemble of field configurations $\phi(x)$ is distributed according to
the partition function (\ref{eq:gen-func-euclid}). Consequently, the
latter is the quantity which one tries to access in numerical
simulations.

However, this approach has the shortcoming that the actual continuum
limit can never be performed and at best one has to resort to
extrapolation techniques giving rise to further uncertainties. Since
the actual shape of the Schwinger functions is not recovered, an
analytic continuation to Minkowski-space is not possible either and
objects like distribution amplitudes are not directly accessible.
Nonetheless it is possible to compute integrals over these functions
and their moments, which help to shed light on their behavior. This
approach has been used in
e.g.~\cite{Gockeler:1996wg,Gockeler:1997jk,Best:1997ab,Dolgov:2000ca,%
  Detmold:2001jb,Detmold:2001dv} to extract information about form
factors and structure functions from lattice simulations.

One important question is if the theory is renormalizable if one uses
a perturbative expansion. There are models which are renormalizable
non-perturbatively, but are non-renormalizable when employing a
perturbative expansion. This is the case for the Gross-Neveau model at
large $N$ in three dimensions \cite{Fernandez:1992jh}.

However, due to the great importance of perturbative methods, the
models which are perturbatively renormalizable are considered in most
practical applications. This means that one has to choose Lagrangians
with mass dimension $d(\lagrangian)\leq 4$ \cite{Cheng:1984xx}, where
the mass dimension for the scalars $\phi$, Dirac spinors $\psi$ and
vector fields $A^\mu$ and their derivatives are given by:
\begin{eqnarray}
  \label{eq:mass-dims}
  d(\phi) = 1\,,\qquad d(\partial^n\phi) = 1+n\,, && \nonumber \\
  d(\psi) = 3/2\,,\qquad d(\partial^n\psi) = 3/2+n\,, && \nonumber \\
  d(A^\mu) = 1\,,\qquad d(\partial^n A^\mu) = 1+n\,.
\end{eqnarray}
The mass dimension of a composite term in the Lagrangian is given by
adding the mass dimensions of its factors. A dimension-four term then
corresponds to a renormalizable interaction, less than four is
super-renormalizable and greater than four is non-renormalizable.

\subsection{Ensembles}
\label{sec:ensembles}
Following its definition, Eq.~(\ref{eq:operator-def}), the quantum
mechanical vacuum expectation value $\langle \hat{A}\rangle$ of some
functional $\hat{A}\left[\phi\right]$ of the fundamental fields in the
theory, $\phi(x)$, can be written as the moment of the measure
(\ref{eq:pathint-field}). \index{Operators} As discussed in
Sec.~\ref{sec:eval-path-integr}, the analytic treatment of equation
(\ref{eq:operator-def}) is only possible in case the path integral has
the shape of a Gaussian or in some toy models. If one does not want to
recourse to expansion techniques or simplifying assumption at this
stage, the only alternative method known today is the numerical
treatment of (\ref{eq:operator-def}). However, a straightforward
integration does not appear to be feasible, since the dimensionality
of the integral in simulations as they are run today is easily
exceeding $10^6$ \cite{Sokal:1989ea}. The only alternative is
therefore a \textit{Monte-Carlo integration}. To define possible
techniques for treating this problem, the concept of
\index{Ensembles}\textit{ensembles of configurations} has turned out
to be extremely useful \cite{Montvay:1994cy}:
\begin{description}
\item[Ensembles:] An ensemble $\left(\Ensemble{\phi},
    \EnsembleDens{\phi}, \EnsembleMeas{\phi}\right)$ consists of an
  infinite number of field configurations $\Ensemble{\phi}$ with a
  density $\EnsembleDens{\phi}$ defined on the measure
  $\EnsembleMeas{\phi}$.
\end{description}
A simple example is the \index{Ensembles!micro-canonical}
\textit{micro-canonical} ensemble, which is defined by
\begin{equation}
  \label{eq:micro-can-ensemble}
  \EnsembleDens{\phi}_{\mbox{\tiny $\mu$-can}} \propto
  \delta\left(S\left[\phi\right] - C\right)\,,
\end{equation}
with a constant $C\in\mathbb{R}^1$. Thus, this ensemble only consists
of configurations with a fixed action. Obviously, this ensemble cannot
be used for the evaluation of (\ref{eq:operator-def}), since the
majority of configurations appearing in the path-integral are not
members of $\Ensemble{\phi}_{\mbox{\tiny $\mu$-can}}$. To take account
of the need to include \textit{any} possible configuration in the
ensemble, we also have to introduce the notion of \index{Ergodicity}
\textit{ergodicity}:
\begin{description}
\item[Ergodicity:] An ensemble $\left(\Ensemble{\phi},
    \EnsembleDens{\phi}, \EnsembleMeas{\phi}\right)$ is called ergodic
  if \[ \EnsembleDens{\phi} > 0 \quad\forall\phi\in\mathbb{E}^4\,.
  \]
\end{description}
\index{Ensembles!random} An example of an ergodic ensemble is given by
the random ensemble, where each possible field configuration enters
with equal probability:
\begin{equation}
  \label{eq:random-ensemble}
  \EnsembleDens{\phi}_{\mbox{\tiny rand}} \propto 1\,.
\end{equation}
With the measure $\EnsembleMeas{\phi}_{\mbox{\tiny rand}}$ from the
random ensemble, the expression (\ref{eq:operator-def}) becomes
\begin{equation}
  \label{eq:random-ens-opdef}
  \langle A\rangle = Z^{-1} \int\EnsembleMeas{\phi}_{\mbox{\tiny
      rand}}\; e^{-S[\phi]} A\left[\phi\right]\,, \qquad Z = \int
  \EnsembleMeas{\phi}_{\mbox{\tiny rand}}\; e^{-S[\phi]}\,.
\end{equation}
Switching to different ensembles in path integrals consists of a
re-parameterization of the measure. It is therefore equivalent to the
substitution rule in ordinary integrals.

\index{Ensembles!canonical}Another example of an ergodic ensemble is
given by the \textit{canonical ensemble} (also known as the
``equilibrium ensemble'') which is defined by
\begin{equation}
  \label{eq:can-ensemble}
  \EnsembleDens{\phi}_{\mbox{\tiny can}} \propto
  e^{-S\left[\phi\right]}\,.
\end{equation}
The measure in (\ref{eq:pathint-field}) is corresponding to the
canonical ensemble and therefore underlying the path integral
definition in Eq.~(\ref{eq:operator-def}). Due to this simple form of
the operator expectation value, the canonical ensemble
(\ref{eq:can-ensemble}) plays a huge role in numerical simulations of
quantum field theories.

\index{Ensembles!multi-canonical}Finally an important generalization
of the canonical ensemble is given by the \textit{multi-canonical
  ensemble}. Suppose the underlying action in
Eq.~(\ref{eq:can-ensemble}) is replaced by an action
$S\left[\phi\right]\rightarrow S^{'}\left[\phi\right] =
S\left[\phi\right] + \gamma\tilde{S}\left[\phi\right]$, with some
parameter $\gamma$. The ensemble
$\lbrace\phi\rbrace^\gamma_{\mbox{\tiny multi-can}}$ with density
\begin{equation}
  \label{eq:multi-can-ensemble}
  \EnsembleDens{\phi}^\gamma_{\mbox{\tiny multi-can}} \propto
  e^{-S^{'}\left[\phi\right]}
\end{equation}
leads to the following shape of (\ref{eq:operator-def}):
\begin{equation}
  \label{eq:multi-can-operator}
  \langle\hat{A}\rangle = Z_{\gamma}^{-1}
  \int\EnsembleMeas{\phi}^\gamma_{\mbox{\tiny multi-can}}\;
  e^{\gamma\tilde{S}\left[\phi\right]} A\left[\phi\right]\,,
  \qquad Z_{\gamma} = \int\EnsembleMeas{\phi}^{\gamma}_{\mbox{\tiny
      multi-can}}\; e^{\gamma\tilde{S}\left[\phi\right]}\,.
\end{equation}
The reason why (\ref{eq:multi-can-ensemble}) is useful is that it is
often possible to find an action $S'\left[\phi\right]$ which is
numerically simpler to handle and simulate than the original action
$S\left[\phi\right]$ and with the ensembles (\ref{eq:can-ensemble})
and (\ref{eq:multi-can-ensemble}) being close enough to each other
such that the \index{Reweighting} ``reweighting correction'' in
(\ref{eq:multi-can-operator}) is small. A situation where this is the
case is given in this thesis in the framework of the TSMB algorithm to
be discussed in Sec.~\ref{sec:mult-algor}.

The ensemble is given by an infinite set of field configurations
$\Ensemble{\phi}$. The introduction of ensembles thus apparently made
the problem of integrating a complicated multi-dimensional system even
worse instead of simplifying it. However, the re-formulation of the
problem allows for a solution by a different integration technique,
the Monte-Carlo integration
\cite{Montvay:1994cy,Rothe:1992nt,numerical-recipes}. This numerical
method is going to be discussed in Sec.~\ref{sec:monte-carlo-algor}.

\section{Gauge Theories}
\label{sec:gauge-theories}
The guiding principle of the construction of quantum field theories in
Sec.~\ref{sec:axioms-relat-quant} was the idea of locality. For a
start, consider the $N$-component ($N\geq 2$)
\index{Fields!Yang-Mills} Yang-Mills theory described by the
Lagrangian:
\begin{equation}
  \label{eq:yang-mills}
  \lagrangian\left(\bar{\Psi}_N,\Psi_N\right) =
  \bar{\Psi}_N\left(\mbox{i}\slash\!\!\!\partial -
  m\right) \Psi_N\,,
\end{equation}
which is invariant under global transformations $U\in\mbox{SU}(N)$:
\begin{equation}
  \label{eq:glob-gauge-trans}
  U:\left\lbrace\begin{array}{l}
      \Psi_N\mapsto\Psi_N^\prime = U\,\Psi_N, \\
      \bar{\Psi}_N\mapsto\bar{\Psi}_N^\prime = \bar{\Psi}_N\,
      U^\dagger\,. \end{array}\right.
\end{equation}
However, a global transformation on the fields living in
$\mathbb{R}^4$ is not consistent with the idea of locality. Rather we
want a theory which is invariant under \textit{local} gauge
transformations $U(x)\in\mbox{SU}(N)$:
\begin{equation}
  \label{eq:local-gauge-trans}
  U(x):\left\lbrace\begin{array}{l}
      \Psi_N\mapsto\Psi_N^\prime = U(x)\,\Psi_N, \\
      \bar{\Psi}_N\mapsto\bar{\Psi}_N^\prime = \bar{\Psi}_N\,
      U^\dagger(x)\,. \end{array}\right.
\end{equation}
A theory invariant under these transformations is called a
\textit{gauge theory}. It is possible to add to
Eq.~(\ref{eq:yang-mills}) a term containing a new set of fields
$A^\mu_a(x)$ such that it stays invariant under the transformation
(\ref{eq:local-gauge-trans}). The simplest way to do this is to choose
\begin{equation}
  \label{eq:yang-mills-extent}
  \lagrangian(\Psi_N,\bar{\Psi}_N,A^\mu_a) =
  \bar{\Psi}_N\left(\mbox{i}\ \slash\!\!\!\!D - m\right) \Psi_N\,,
\end{equation}
where the covariant derivative $\slash\!\!\!\! D$ is given by
\begin{equation}
  \label{eq:covar-deriv}
  \slash\!\!\!\! D =
  \gamma_\mu\left(\partial^\mu+\mbox{i}gA^\mu_a\right)\,,
\end{equation}
and the transformation of $A^\mu_a(x)$ must be given by
\begin{equation}
  \label{eq:adjoint-transform}
  U(x):A^\mu_a(x)\mapsto{A^\mu_a}^\prime(x) =
  U^\dagger(x)\left(\partial^\mu + A^\mu_a(x)\right)U(x)\,,
\end{equation}
meaning that the $A^\mu_a$ lie in the adjoint representation of
SU$(N)$ and that $1\leq a\leq\mbox{dim}\left(\mbox{SU}(N)\right)$.
Thus, the resulting theory will now contain the fields $\bar{\Psi}_N$,
$\Psi_N$, and $A^\mu_a(x)$. The new fields $A^\mu_a(x)$ are termed
\index{Fields!gauge} \textit{gauge fields} and their coupling to the
fields $\bar{\Psi}_N$, $\Psi_N$ is given by the dimensionless coupling
strength $g$.

By postulating the fields to be invariant under the transformations
(\ref{eq:local-gauge-trans}) and (\ref{eq:adjoint-transform}) and
requiring that the Lagrangian only contains perturbatively
renormalizable terms (see Sec.~\ref{sec:eval-path-integr}), one is
finally led to the general form
\begin{equation}
  \label{eq:yang-mills-total}
  \lagrangian(\bar{\Psi}_N,\Psi_N,A^\mu) = -\frac{1}{4}
  \sum_{a=1}^{N^2-1} F_{\mu\nu}^a F^{\mu\nu a} + \bar{\Psi}_N\left(
  \mbox{i}\ \slash\!\!\!\!D - m\right)\Psi_N\,,
\end{equation}
with the \textit{field strength}
\begin{equation}
  \label{eq:ym-field-strength}
  F_{\mu\nu}^a = \partial_\mu A_\nu^a - \partial_\nu A_\mu^a + g
  \sum_{b,c=1}^{N^2-1} f_{abc} A_\mu^b A_\nu^c\,.
\end{equation}
There is an important difference between the pure gauge part in the
SU$(N)$ Lagrangian (\ref{eq:yang-mills-total}) and the single gauge
field Lagrangian (\ref{eq:free-vector-lagr}) corresponding to an
Abelian gauge group: The former contains interactions between
different components of the gauge field $A^\mu_a$, while the latter
describes a true free field. Thus, the $N$-component vector theory
contains interactions even in the case of a purely gauge theory
without coupling to a matter field. It is argued below, that this
phenomenon leads to the dynamical generation of a mass scale in the
case of the quantized theory. This phenomenon is also known as
\index{Dimensional transmutation} \textit{dimensional transmutation}.

\index{Fields!non-Abelian} Since the group SU$(N)$ is non-Abelian ---
their elements don't commute --- Eq.~(\ref{eq:yang-mills-total}) is
referred to as a \textit{non-Abelian gauge theory}. For alternative
ways to define a gauge theory cf.~\cite{Creutz:1983bo,Montvay:1994cy}
and references therein.

In addition to the SU$(N)$ symmetry, the Lagrangian
(\ref{eq:yang-mills-total}) is also invariant under axial rotations of
the fermion fields, provided, the Dirac part is massless ($m=0$):
\begin{eqnarray}
  \label{eq:axial-transforms}
  \Psi\mapsto\Psi'             &=& \exp\left[\gamma_5\alpha\right]
  \Psi\,, \nonumber \\
  \bar{\Psi}\mapsto\bar{\Psi}' &=& \bar{\Psi}\exp\left[
    -\gamma_5\alpha\right]\,.
\end{eqnarray}
The question arises, whether this symmetry exists also on the quantum
level, or if it is broken by an anomaly. As has been realized by
\textsc{Adler} \cite{Adler:1969gk} and \textsc{Bell} and
\textsc{Jackiw} \cite{Bell:1969ts}, for an Abelian gauge theory this
is indeed the case. The anomaly responsible for breaking the axial
current corresponding to the symmetry (\ref{eq:axial-transforms}) is
known as the \textit{Abelian anomaly} or
ABJ-anomaly\index{Abelian anomaly}.  It is present once the theory
contains fermions and is independent of the fermion masses. This
result has also been derived non-perturbatively by \textsc{Fujikawa}
\cite{Fujikawa:1979ay}. An extension to non-Abelian theories has been
given in \cite{Bardeen:1969md}. For a textbook containing a rigorous
mathematical treatment consult \cite{Bertlmann:1996xk}.

\section{Quantum Chromodynamics}
\label{sec:quant-chrom}
Now the ground has been prepared to formulate quantum chromodynamics
(QCD) as the theory underlying the strong interaction. It is a
Yang-Mills gauge theory (see Sec.~\ref{sec:gauge-theories}) symmetric
under the SU$(3)$ group (as discussed in
Sec.~\ref{sec:phen-strong-inter}), where the latter symmetry group
refers to the color degree of freedom of the quarks. It contains six
flavors of quarks with masses $\lbrace m_k\rbrace$, with each flavor
of quarks transforming as the fundamental triplet representation of
the color group. The accompanying vector bosons, the ``gluons''
\index{Gluons} transform according to the adjoint representation.
Furthermore we require the theory to be perturbatively renormalizable.
Thence, the resulting Lagrangian $\lagrangian_{\mbox{\tiny QCD}}$ in
Minkowski-space is given by (the number of colors is denoted by
$N_{c}=3$)
\begin{equation}
  \label{eq:qcd-lagrangian}
  \lagrangian_{\mbox{\tiny QCD}} = -\frac{1}{4}\sum_{a=1}^{8}
  G_{\mu\nu}^a G^{\mu\nu a} + \sum_{k=1}^6 \sum_{\alpha=1}^{N_{c}}
  \bar{\Psi}_{k\alpha}\left( \mbox{i}\ \slash\!\!\!\!D - m_k
  \right)\Psi_{k\alpha}\,.
\end{equation}
It is also possible (without violating perturbative renormalizability)
to add a term of the form \[ \lagrangian_\theta \propto \frac{\theta
  g^2}{64\pi^2} \sum_{a=1}^{8} \varepsilon_{\alpha\beta\mu\nu}
G^{\alpha\beta a} G^{\mu\nu a} \] to (\ref{eq:qcd-lagrangian}). This
term is known as the ``$\theta$-term'' and would be a source of $CP$
violation \cite{Nachtmann:1990ta}. The experimental limit for $\theta$
is ${\cal O}(10^{-9})$.  Thus, this term will not be considered in this
thesis.

Several important properties of QCD can be learned by considering the
symmetries of (\ref{eq:qcd-lagrangian}) \cite{Cheng:1984xx}. For
$N_{f}$ massless quark flavors, $\lagrangian_{\mbox{\tiny QCD}}$ is
invariant under several global symmetry transformations. In
particular, one can decompose the Dirac spinors into left- and
right-handed quark fields and perform independent rotations on the
resulting Weyl spinors. This yields a global
$\mbox{SU}(N_{f})_{L}\otimes \mbox{SU}(N_{f})_{R}$ symmetry (this
symmetry is also known as \textit{chiral}). Furthermore one can make
independent global vector and axial rotations on the full Dirac
spinors resulting in a global $\mbox{U}(1)_{V}\otimes \mbox{U}(1)_{A}$
symmetry.  When looking at the masses of the different quark flavors,
one can indeed consider the masses of the $u$- and $d$- quark flavors
to be almost zero compared to the typical scales of hadronic
resonances. To a lesser extent this is also valid for the $s$-quark
flavor. Thus, QCD contains three almost massless fermion flavors and
should consequently have a global
$\mbox{SU}(3)_{L}\otimes\mbox{SU}(3)_{R}\otimes \mbox{U}(1)_{V}\otimes
\mbox{U}(1)_{A}$ symmetry.

According to the Noether theorem, there should be conserved charges
corresponding to each symmetry of the Lagrangian. The
U$(1)_{V}$-symmetry is indeed associated with a conserved quantum
number, namely the baryon number which is conserved exactly by the
strong interaction. The current corresponding to the axial
U$(1)_{A}$-symmetry is, however, explicitly broken by the ABJ anomaly
(cf.~Sec.~\ref{sec:gauge-theories}) if the theory is quantized.
Nonetheless, one can find a modified, conserved current albeit it will
be gauge-dependent and thus not represent a physical current.

From the remaining chiral symmetry, one half is indeed present in the
hadron spectrum, namely as the flavor SU$(3)_{F}$ symmetry discussed
in Sec.~\ref{sec:phen-strong-inter}. This half corresponds to a vector
symmetry transformation of the Dirac spinors.  The other half,
however, which corresponds to an axial vector transformation would
result in a parity degeneracy of the particles which is clearly not
observed. To be specific, there are no parity degeneracies present in
the hadron spectrum at all. Thus, the quantization of QCD must break
this symmetry.  Since there is no anomaly which could attribute for
this symmetry breaking, it must be broken in a spontaneous manner,
i.e.~the ground state of the theory will not be invariant.  Due to the
Goldstone theorem \cite{Cheng:1984xx}, consequently there exist
massless particles corresponding to the pseudoscalar mesons whose
masses are much smaller than those of the other hadrons. The fact that
they are not zero can be attributed to the explicit breaking of chiral
symmetry due to the small masses of the light quarks. Within the
framework of \textit{Chiral Perturbation Theory} ($\chi$PT) (see
Sec.~\ref{sec:lattice-qcd}), it can indeed be shown that for small
quark masses, the effect can be treated perturbatively.

But there does not seem to exist any Goldstone boson corresponding to
the breaking of the axial U$(1)_{A}$ charge. The only particle with
the correct symmetries is the $\eta'$-meson whose mass is far too
large (see Tab.~\ref{tab:particlelist}). The solution of this problem
is related to the topology of the gauge field. Topological transitions
can produce the $\eta'$-mass via the axial anomaly. A possible
explanation is that \index{Instantons} \textit{instanton transitions}
(see below) are responsible for these topological charge fluctuations.

\subsection{Running Coupling and Energy Scales}
\label{sec:runn-coupl-energy}
The quantum theory build upon (\ref{eq:qcd-lagrangian}) is
characterized by a running coupling (for details see
e.g.~\cite{Cheng:1984xx}). Performing a leading order perturbative
analysis and renormalizing the theory, the behavior of the running
coupling ``constant'' is found to be\index{Coupling!strong}
\begin{equation}
  \label{eq:running-alphas}
  \alpha_S(Q^2) =
  \frac{4\pi}{\beta_0}\frac{1}{\ln\left(Q^2/\Lambda_{\mbox{\tiny
          QCD}}^2\right)}\,,
\end{equation}
with $\beta_0=11-2/3N_f$, where $N_f$ is the number of active flavors
\cite{Nachtmann:1990ta}. This defines the coupling at an energy scale
$Q^2$. There are two important lessons to be learned from
(\ref{eq:running-alphas}):
\begin{itemize}
\item The coupling $\alpha_S(Q^2)$ \textit{decreases} for increasing
  values of $Q^2$. The interaction vanishes for $Q^2\rightarrow\infty$
  and the particles becomes free in this limit. This property is
  referred to as \index{Asymptotic freedom} \textit{asymptotic
    freedom}.
\item The coupling becomes infinite for a certain finite value of
  $Q^2$, $\Lambda^2_{\mbox{\tiny QCD}}$. This happens also in case of
  an Abelian gauge theory (where the underlying group is U$(1)$) and
  shows an intrinsic inconsistency under which
  (\ref{eq:running-alphas}) has been derived: The assumption that
  $\alpha(Q^2)$ is small becomes invalid for increasing
  $\alpha_S(Q^2)$ at some point and the series starts to diverge
  already at the first order beyond tree level. This singularity is
  called the \index{Landau pole} \textit{Landau pole} and is
  considered to be an unphysical remnant only present due to the fact
  that perturbation theory cannot be applied for too large expansion
  parameters.  The appearance of the Landau pole thus sets a limit to
  the applicability of perturbative calculations. On the other hand
  one can expect the calculation to be valid at energies far larger
  than $\Lambda_{\mbox{\tiny QCD}}$.
\end{itemize}
It is usually assumed that, when ``solving'' full QCD by the methods
sketched in Sec.~\ref{sec:path-integral}, one also obtains the whole
low-energy phenomenology with minimal input. There is no reason why
the failure of a single method, namely the perturbative expansion
around the free field, should imply that QCD is not valid at low
energy scales. However, a concise solution of interacting quantum
field theories is not in sight, so one has to stick with a number of
models parameterizing the low-energy behavior. One of these
parameterizations is \index{Chiral perturbation theory}
\index{CPT@$\chi$PT|see{Chiral perturbation theory}} $\chi$PT
\cite{Gasser:1982ap}. Besides the latter, there are also different
effective theories which parameterize the behavior of the strong
interaction at low energies: models like the Nambu-Jona-Lasignio model
\cite{Nambu:1961tp}, the skyrmion model \cite{Klebanov:1997ny}, or
models based on instantons (see below) are different attempts to
describe the properties of low energy strong interactions. The hadrons
built up from one of the three heavy quarks can be described using
\textit{Heavy-Quark Effective Theory} (HQET), see
\cite{Wise:1993wa,Casalbuoni:1997pg} for introductions.

However, all these theories are only able to predict the low-energy
properties of the strong interaction; they do not incorporate an
adequate mechanism for the description of the parton content of
hadrons. For the high energy regime, the perturbative treatment of QCD
has to be used, which describes the interaction using the color group
with the gluons being the mediators of the strong force.  However, if
the strong interaction is described using the flavor group as an
interaction between the baryons (the octet multiplet in the flavor
SU$(3)$ group), then the mediating particles are the pseudoscalar
mesons.

One particularly important concept in the development of QCD is the
hypothesis of \index{Confinement} \textit{confinement}. The common
understanding of confinement is that in a world without sea quarks,
the static potential of two quarks would be linear growing without
limit.  This leads to bound quarks not being separable and thus free
quarks being unobservable. One consequence of this picture of
confinement could be that the classical limit,
Eq.~(\ref{eq:wkb-amplitude-expansion}), may not exist. Thus, the
consequences of confinement could be wide-reaching. The best tools
which have so far been used to address this particular issue are
lattice simulations.  For a recent discussion of lattice simulations
regarding confinement, see \cite{Diakonov:1998rk} and references
therein.

There is another very important property of QCD shared with other
non-Abelian gauge theories: Consider (\ref{eq:yang-mills-total})
without fermions. Then it can be shown \cite{'tHooft:1976up} that
there exist gauge field configurations which vanish at spatial
infinity, but fall into different topological classes. They may be
characterized by the winding number $n(U)$, which is given by the
Chern-Simons three form on the gauge fields
\cite{Jackiw:1997xx,Bertlmann:1996xk}. The transition between the
different topological sectors may be performed using the
\index{Instantons}\textit{instanton solutions}. These are solutions of
the classical equations of motion and they may also contribute
significantly in the quantized theory. For recent overviews consult
\cite{Schafer:1998wv,Diakonov:1997sj,Diakonov:2000pa}. The importance
of instantons for hadron physics has also been demonstrated on the
lattice in \cite{Negele:1997na,vanBaal:1998vi}. Recently, a method to
examine a prediction of the instanton model with lattice simulations
has been proposed in \cite{Horvath:2001ir}. This method has been
applied in \cite{Hip:2001hc}, confirming the predictions of the
instanton model. Indications for this picture have also been found in
an earlier publication \cite{DeGrand:2001pj} and in later works
\cite{Hasenfratz:2001wd,Edwards:2001nd}.

One particularly important point is that the quantum field theory
built upon (\ref{eq:yang-mills-total}) puts a lower limit to the
magnitude of the instanton actions resulting in a certain mass scale
of the theory.  Thus, a mass-scale is generated although the classical
theory is scale-free (and has no free parameters except for the
coupling $g$ which can be rescaled to any value). As has already been
mentioned in Sec.~\ref{sec:gauge-theories}, this phenomenon is known
as dimensional transmutation. A different widely discussed
manifestation of dimensional transmutation is the existence of
glue-balls (see e.g.~\cite{Michael:2001qz} for a recent overview).

\subsection{Factorizable Processes}
\label{sec:fact-proc}
Several observables in QCD (like structure functions and form factors
etc., see e.g.~\cite{Lepage:1982gd,Mueller:1989hs} and for a more
recent review \cite{Brodsky:2000dr} and references therein) depend on
input from both regimes. For several interesting processes involving
these observables, a method known as \textit{factorization} is
applicable. The formal framework of factorization is the
operator-product expansion, whenever it applies.  Consider two local
operators $\hat{A}(x)$, $\hat{B}(y)$. The Wilson expansion of the time
ordered product of the composite operator for short distances
$(x-y)\rightarrow 0$ can then be performed as \cite{Yndurain:1993bo}
\begin{equation}
  \label{eq:ope-def}
  \mathscr{T}\left\lbrace\hat{A}(x)\hat{B}(y)\right\rbrace = \sum_i
  C_i(x-y) \hat{N}_i(x)\,.
\end{equation}
This relation is only established perturbatively, however. The
singularities of the composite operator
$T\left\lbrace\hat{A}(x)\hat{B}(y)\right\rbrace$ are then contained in
the $\lbrace C_i\rbrace$ which are $C$-numbers. They are called Wilson
coefficients and contain the high-energy physics. Consequently, they
can be computed perturbatively. The operators
$\lbrace\hat{N}_i\rbrace$ are local operators containing information
about the low-energy regime and hence are usually not accessible by
perturbative methods. The individual terms in the sum
(\ref{eq:ope-def}) can be arranged in such an order that the single
terms behave as a power series in $Q^{-2N}$, where $N$ characterizes
the order of the associated term.  This is done by ascribing a certain
``twist'' to each term. The first term (which vanishes slowest) is
called the ``leading twist contribution'' and the higher terms are
consequently ``higher twist contributions''. The series then takes a
form reminiscent of the perturbative expansion,
Eq.~(\ref{eq:perturbative-expansion}).

In the form of (\ref{eq:ope-def}), the high energy regime and the low
energy part can be treated separately, and the object under
consideration factorizes in the two separate contributions. The major
ingredient to a factorization scheme is the \textit{factorization
  scale}, i.e.~the scale describing which contributions belong to the
low-energy regime and thus, to the operators $\hat{N}_i(x)$, and which
contributions belong to the high energy part, i.e.~the functions
$C_i(x-y)$. This leaves a certain freedom in the application of the
factorization approach. This freedom should be exploited to keep
higher-order corrections in the perturbative series as small as
possible, shifting the majority of contributions into the leading
order.

Naturally the question arises to what extent it is possible to ascribe
any meaning to a series like (\ref{eq:ope-def}) if it involves a
running coupling (\ref{eq:running-alphas}) which is singular at some
point in the physical parameter space. This question has been
addressed in e.g.~\cite{Dokshitzer:1998nz}. From a pragmatic point of
view one can adopt the series despite the conceptional problems.
However, one has to circumvent the Landau singularity; to achieve
this, a number of proposals have been made: one is to apply a
``freezing'' prescription, i.e.~simply hold the coupling constant
fixed below a certain point \cite{Parisi:1980jy}. Another consists of
introducing an effective gluon mass \cite{Cornwall:1982zr}.  A
different approach relies on the application of an analytization
procedure (first applied to QED by \textsc{Lehmann} and then
\textsc{Bogoliubov}, see
\cite{Bogolyubov:1955or,Bogolyubov:1956or,Bogolyubov:1959or} and
references therein), which was originally invented to extent
(\ref{eq:running-alphas}) also to the regime where $Q^2$ is a timelike
momentum transfer \cite{Radyushkin:1996kg,Krasnikov:1982fx}.  Later a
framework of analytic perturbation theory has been founded on this
bases by \textsc{Shirkov} and \textsc{Solovtsov} in
\cite{Shirkov:1997wi,Shirkov:1998sb,Solovtsov:1999in,Shirkov:2000db}.
In essence, the Landau singularity in Eq.~(\ref{eq:running-alphas})
can be compensated in a minimal way by adding a unique power-term
replacing the running coupling by
\begin{equation}
  \label{eq:analytic-alphas}
  \alpha_S(Q^2) =
  \frac{4\pi}{\beta_0}\left(
    \frac{1}{\ln\left(Q^2/\Lambda_{\mbox{\tiny QCD}}^2\right)} +
    \frac{\Lambda_{\mbox{\tiny QCD}}^2} {\Lambda_{\mbox{\tiny
          QCD}}^2-Q^2}\right)\,.
\end{equation}
In contrast to the conventional expansion, the contribution of higher
terms appears to be suppressed (cf.~\cite{Shirkov:1998sb}). This
observation together with a renormalization and factorization scheme
optimized for putting most higher order contributions into the leading
order should allow for a consistent and efficient description of
factorizable processes. Indeed, it has been found that this program
works for the cases of the electromagnetic form factor of the pion and
the $\pi\rightarrow\gamma^*\gamma$ transition form factor
\cite{Stefanis:1998dg,Stefanis:1998ud,Stefanis:2000vd} and yields an
excellent agreement with the experimental data while providing a
consistent framework for the computation of hadronic observables.

\subsection{Lattice QCD}
\label{sec:lattice-qcd}
\index{Lattice QCD} The approach to perform a numerical simulation on
a finite lattice yielding an approximation to the Schwinger functions
$\schwinger_n$ in discrete Euclidean space $\mathbb{Z}^4_\Omega$ is
referred to as \textit{lattice gauge theory} and provides in principle
the only means known so far to access the complete structure of both
the low and high energy regime of QCD\@. Anyhow, due to the technical
difficulties inherent to this method, the quality of results is poor
when compared to perturbation theory (whenever the latter is
applicable). Thus, contemporary lattice investigations always
concentrate on the non-perturbative regime of QCD calculating the
properties of the low-energy parameterizations.

As will be shown in Sec.~\ref{sec:fermion-fields}, there are problems
concerning the formulation of massless fermions on the lattice.
\index{Chiral perturbation theory} On the other hand, $\chi$PT, as a
low energy model of QCD, performs an expansion in the quark mass
around the point $m_q=0$ and thus allows for a systematic treatment of
near-massless fermions; for this reason, it is of particular interest
for lattice investigations, since one is usually interested in
performing extrapolations in the quark mass (see
e.g.~\cite{Detmold:2001jb} for a recent proposal of how to do this).
$\chi$PT is, however, limited to the continuum theory. Thus, the
continuum extrapolation should precede the application of $\chi$PT.

Since in lattice simulations one often chooses quark masses occurring
in virtual quark loops \index{Quarks!sea} (the so-called
``sea-quarks'') different from the quark masses appearing in hadrons
\index{Quarks!valence} (the so-called ``valence-quarks''), an
extension of the original $\chi$PT-formulation is necessary to handle
also these models. The first extension was to set the sea-quark mass
equal to zero \index{Quenched approximation} (the \textit{quenched
  approximation}) yielding ``quenched chiral perturbation theory''
(for a short discussion and the references, see
\cite{Bernard:1994sv}). This model allows for the extraction of
phenomenology from lattice simulations if one completely disregards
dynamical fermion contributions.

With the advent of dynamical fermion simulations, a further extension
of this model introducing different masses for sea and valence quarks
was proposed by \textsc{Bernard} and \textsc{Golterman} in
\cite{Bernard:1994sv} resulting in the ``partially quenched chiral
perturbation theory''. In principle, partially quenched chiral
perturbation theory should allow for the first time to gain direct
access to phenomenological quantities from lattice simulations
provided a number of conditions is met \cite{Sharpe:2000bc}. In
essence, one has to perform simulations with three dynamical quark
flavors (which may be even mass-degenerate) at rather small masses of
about $1/4m_{s}$. This goal is out of reach with the resources
available to the lattice community today, but it may pave the way for
future lattice simulations aiming at precise measurements of hadron
properties.

While quenched simulations already allow for a rather precise
determination of many phenomena in QCD
\cite{Aoki:1999yr,Kanaya:1998sd}, there are observables which depend
also on dynamical fermion contributions. For example, the mass of the
$\eta'$ meson (see above) is only properly accessible in unquenched
simulations (see \cite{Struckmann:2000ts} for a discussion).

\begin{figure}[htb]
  \begin{center}
    \includegraphics[scale=0.7]{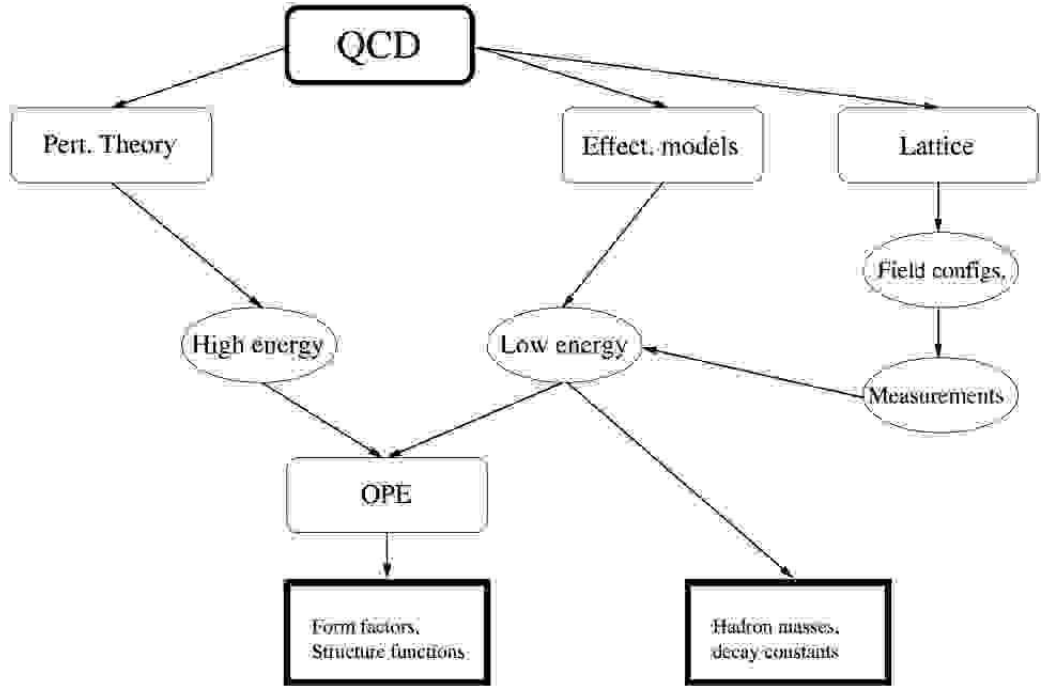}
    \caption{Different methods for obtaining prediction in QCD.}
    \label{fig:qcd-methods}
  \end{center}
\end{figure}

The different methods for computations in QCD are visualized in
Fig.~\ref{fig:qcd-methods}.

\section{Discretization}
\label{sec:discretization}
As discussed in~\ref{sec:eucl-field-theory}, for the construction of a
quantum field theory on a lattice, the functional $S[\varphi(x)]$ is
required. This functional should reduce to the Euclidean action
$\mathscr{S}[\varphi(x)]$ in the continuum limit and for
differentiable paths. Before applying the limit prescription, it will
thus differ by ${\cal O}(a)$-effects from the continuum expression ---
meaning that in general the choice of $S[\varphi(x)]$ is not unique
but still leaves freedom to choose all terms of order ${\cal O}(a^n)$
with $n\geq 1$. This freedom should be used to find the form best
suited for numerical calculations.

\subsection{Scalar Fields}
\label{sec:scalar-fields}
\index{Fields} Consider the complex field $\phi(x)$ defined on the
sites $x\in \mathbb{Z}^4_\Omega$. The continuum Lagrangian
corresponding to this situation is given by
Eq.~(\ref{eq:scalar-lagr}). One candidate for the lattice version of
the action is then given by \cite{Montvay:1994cy}:
\begin{equation}
  \label{eq:lattice-scafi}
  S[\phi(x)] = \sum_x\left( \sum_{\mu=0}^3 \phi(x)^\dagger
  \phi\left(x+\hat{\mu}\right) + m^2 \phi(x)^\dagger\phi(x) +
  V[\phi(x)] \right)\,.
\end{equation}
There are certainly other ways to replace the derivative, but the
present choice is the simplest way to incorporate neighbor fields.
Consequently, this choice is suitable for numerical investigations and
will be used in this thesis.

A particularly interesting model is the so-called $\phi^4$-model,
where one sets \[ V[\phi] = \frac{\lambda}{4!}\phi^4\,. \] This model
appears to be an interacting, nontrivial field theory at first sight,
but already early it has been conjectured \cite{Wilson:1974jj} that it
might only give rise to a non-interacting theory of free particles. In
later investigations this surmise has been corroborated
\cite{Baker:1979gc,Luscher:1988gk}. However, a rigorous proof is still
missing.

\subsection{Gauge Fields}
\label{sec:gauge-fields}
In the continuum form (\ref{eq:yang-mills-total}), the gauge field is
given in terms of parallel transporters along infinitesimal distances.
By putting the system on a lattice, the shortest (non-zero) distance
is the lattice spacing $a$. The parallel transporter connecting a
point $x\in\mathbb{Z}^4_\Omega$ with its neighbor $x+\hat{\mu}$ is
denoted by $U(x,x+\hat{\mu})$. It is an element of SU$(N)$. The
simplest gauge-invariant object one can construct is a closed loop
with a side length of one lattice unit usually called the
\textit{plaquette}. Starting from the point $x\in\mathbb{Z}^4_\Omega$,
one can construct the plaquette lying in the $\mu\nu$-plane by
considering
\begin{equation}
  \label{eq:old-plaquette}
  U_{\mu\nu}(x) = U(x,x+\hat{\mu})
  U(x+\hat{\mu},x+\hat{\mu}+\hat{\nu})
  U(x+\hat{\mu}+\hat{\nu},x+\hat{\nu}) U(x+\hat{\nu},x)\,.
\end{equation}
Due to the fact that $U(x,x+\hat{\mu})=U^\dagger(x+\hat{\mu},x)$ one
can rewrite (\ref{eq:old-plaquette}):
\begin{equation}
  \label{eq:plaquette}
  U_{\mu\nu}(x) = U(x,x+\hat{\mu})
  U(x+\hat{\mu},x+\hat{\mu}+\hat{\nu})
  U^\dagger(x+\hat{\nu},x+\hat{\mu}+\hat{\nu})
  U^\dagger(x,x+\hat{\nu})\,.
\end{equation}
The suggestion of Wilson \cite{Wilson:1974sk} was to use real part of
the trace of $U_{\mu\nu}$ summed over all plaquettes as the action of
the system,
\begin{eqnarray}
  \label{eq:wilson-gauge-action}
  S_{\mbox{\tiny g}}[U(x)] &=& -\beta \frac{1}{2\mbox{Tr}\,\mathbf{1}}
  \sum_{x}\sum_{\mu<\nu} \left(\mbox{Tr}\, U_{\mu\nu}(x) + \mbox{Tr}\,
    U_{\mu\nu}^{-1} - 1\right) \nonumber \\
  &=& \beta\sum_x\sum_{\mu\nu} \left(1-\frac{1}{N}\mbox{Re}\,\mbox{Tr}\,
  U_{\mu\nu}(x)\right)\,,
\end{eqnarray}
which is the discretized form of the non-Abelian gauge field part in
Eq.~(\ref{eq:yang-mills-total}). This form, however, is also
applicable to the case of Abelian gauge fields with the Lagrangian
given by Eq.~(\ref{eq:free-vector-lagr}).

For later applications, the more convenient notation $U_\mu(x)\equiv
U(x,x+\hat{\mu})$ will be used from now on; hence, the plaquette,
Eq.~(\ref{eq:plaquette}), is written as
\begin{equation}
  \label{eq:new-plaquette}
  U_{\mu\nu}(x) = U_\mu(x) U_\nu(x+\hat{\mu})
  U^\dagger_\mu(x+\hat{\nu}) U^\dagger_\nu(x)\,.
\end{equation}

The path integral form of the partition function,
Eq.~(\ref{eq:gen-func-euclid}), on a finite lattice
$\mathbb{Z}^4_\Omega$ is given by
\begin{equation}
  \label{eq:gauge-field-gen-func}
  Z = \int \prod_x dU(x) \exp\left[-S[U(x)]\right]\,,
\end{equation}
where the measure $dU(x)$ is the Haar measure on the gauge group (see
App.~\ref{sec:groups-repr}).

One of the properties of the Haar measure is that the total integral
over the group space on a single lattice point $x$ is finite. Thus,
any gauge-fixing procedure is unnecessary \cite{Montvay:1994cy}.
However, if one attempts to apply a saddle-point approximation to
(\ref{eq:gauge-field-gen-func}) (see Sec.~\ref{sec:eval-path-integr}),
the presence of zero modes will spoil the inverse of the two-point
function \cite{Cheng:1984xx}. This requires the introduction of a
gauge fixing procedure and auxiliary fields known as
\textit{Faddeev-Popov} ghosts \cite{Cheng:1984xx}.

It is important to point out that the entire partition function
(\ref{eq:gauge-field-gen-func}) is manifestly gauge invariant since it
is composed of gauge invariant loops $U_{\mu\nu}(x)$ only. It can be
shown \cite{Creutz:1983bo,Montvay:1994cy} that it is impossible to
break gauge invariance spontaneously. This fact implies that the
expectation value of the $U_\mu(x)$ will always vanish, a fact which
is also known as the \textsc{Elitzur} theorem: \[ \langle
U_\mu(x)\rangle =0\,. \]

Eq.~(\ref{eq:wilson-gauge-action}) has two limits, where the form of
the gauge fields can be written down explicitly: If one considers
$\beta\rightarrow 0$, the situation resembles the hot temperature
limit in thermodynamics. Thus, the resulting gauge field
configurations are called \textit{hot configurations} --- the values of
the gauge field variables are arbitrary and they can take any random
values from their domain of definition. The limit
$\beta\rightarrow\infty$, is referred to as the zero-temperature
limit. The correlation between neighbor points increases, therefore
the total correlations length will increase. Finally, the values of
the field variables are given by $U_\mu(x)=1\ \forall x\in\Omega$.  If
the system undergoes a second order phase transition, i.e.~if the
correlation length diverges, this describes the continuum quantum
field theory.

\subsection{D-Theory}
\label{sec:d-theory}
\index{D-theory} There exists also a discretization technique entirely
different from the methods discussed in Sec.~\ref{sec:gauge-fields}.
It is based on a quantum link model, which is constructed in such a
way that it is still locally gauge invariant and should thus reduce to
the correct continuum form of the Yang-Mills theory, just as
(\ref{eq:wilson-gauge-action}) is expected to do. Quantum link models
with local gauge invariance have been formulated by \textsc{Horn} in
\cite{Horn:1981kk} for the first time, where a model with a local
SU$(2)\otimes$U$(1)$ gauge invariance has been formulated. They have
been extended to the case of SU$(N)$ by \textsc{Orland} and
\textsc{Rohrlich} in \cite{Orland:1990st}. These models, however, did
not yet relate to the quantum field theories with continuous
symmetries as discussed in this thesis. Only recently it has been
realized by \textsc{Chandrasekharan} and \textsc{Wiese} in
\cite{Chandrasekharan:1997ih}, how one can relate the discrete quantum
link models with continuum field theories. For a review of the
ingredients of quantum link models see \cite{Wiese:1998nh}.

\subsubsection{Globally Symmetric Models}
\label{sec:glob-symm-models}
The basic ideas behind the construction of D-theory become clear if
one considers a spin model, namely the O$(3)$-model in two space
dimensions. The action is given by
(cf.~(\ref{eq:wilson-gauge-action}))
\begin{equation}
  \label{eq:o3-model}
  S[\vec{s}] = -\frac{1}{g}\sum_x\sum_\mu
  \vec{s}(x)\cdot\vec{s}(x+\hat{\mu})\,,
\end{equation}
with $\vec{s}(x)$, $x\in\mathbb{Z}^2_\Omega$, being three-component unit
vectors. The coupling constant is given by $g$. After quantizing this
spin system (cf.~Sec.~\ref{sec:path-integral}) by considering the
partition sum \index{Partition function}
\begin{equation}
  \label{eq:o3-quant-part}
  Z = \int\EnsembleMeas{\vec{s}} \exp\left[-S[\vec{s}]\right]\,,
\end{equation}
one arrives at a model which is asymptotically free and has a
non-perturbatively generated mass-gap. The question arises, whether it
is possible to find a different lattice system (with no resemblance to
Eq.~(\ref{eq:o3-model})) which still reduces to the same continuum
field theory in the sense discussed in Sec.~\ref{sec:path-integral}.
This construction is indeed possible and can be done as follows:
\begin{enumerate}
\item Replace the classical vectors $\vec{s}(x)$ by quantum spin
  operators $\hat{S}(x)$ which are elements of the algebra
  $\hat{S}(x)\in\mbox{su}(2)$, i.e.~they are the generators of the
  SU$(2)$-group (cf.~Appendix~\ref{sec:sun-groups}).
\item Replace the classical action (\ref{eq:o3-model}) by the
  Hamiltonian 
  \begin{equation}
    \label{eq:o3-hamiltonian}
    \hamiltonian = J\sum_x\sum_\mu
    \hat{S}(x)\cdot\hat{S}(x+\hat{\mu})\,,
  \end{equation}
  which yields the quantum Heisenberg model. The partition sum
  (\ref{eq:o3-quant-part}) is therefore replaced by the state sum
  \index{State sum}
  \begin{equation}
    \label{eq:o3-quant-ham-part}
    Z = \mbox{Tr}\,\exp\left[-\beta H\right]\,.
  \end{equation}
  It is important to point out that the particular representation of
  the group is not important --- the trace can be taken over any
  representation, although in practice one usually adopts the
  fundamental representation \cite{Wiese:1998nh}\footnote{This choice
    allows one to restrict to the smallest possible Hilbert space}. In
  the following, the discussion is restricted to the case $J>0$,
  i.e.~the anti-ferromagnetic Heisenberg system.
\item By using a Suzuki-Trotter discretization, the state sum
  (\ref{eq:o3-quant-ham-part}) becomes a partition function of a
  three-dimensional model with continuous symmetry with a certain
  lattice spacing $a_{3d}$. This model is invariant under a global
  SO$(3)$-symmetry since the Hamiltonian (\ref{eq:o3-hamiltonian}) is
  also invariant.  The low-energy properties of the resulting model
  can be described using chiral perturbation theory
  \cite{Hasenfratz:1990pk}. The symmetry is spontaneously broken in
  the ground state, resulting in two Goldstone bosons which are
  represented by fields in the coset SO$(3)$/SO$(2)=S^2$. Thus, they
  describe the same kind of three-component unit vectors which appear
  in the original action, Eq.~(\ref{eq:o3-model}). The low-energy
  effective action of the Goldstone bosons can be formulated using
  chiral perturbation theory:
  \begin{equation}
    \label{eq:chpt-spinwaves}
    S[\vec{s}] = \int_0^{L_0}\int d^2x\;
    \frac{\rho_2}{2}\left(\partial_\mu\vec{s}
      \cdot\partial^{\mu}\vec{s} + \frac{1}{c^2} d_0\vec{s}\cdot
    d^0\vec{s}\right)\,,
  \end{equation}
  with $L_0$ being the extend of the third dimension which has been
  introduced by the Suzuki-Trotter discretization. The parameters $c$
  and $\rho_s$ constitute the spin-wave velocity and the stiffness,
  respectively.
\item Finally, there exists a mapping of the two systems, which has
  been suggested by \textsc{Hasenfratz} and \textsc{Niedermayer}
  \cite{Hasenfratz:1991jw}. This is achieved by a block spin
  transformation, which maps subvolumes of size $\Omega_{\mbox{\tiny
      sub}}=L_0\times (L_0 c)^2$ to a new lattice system. The new
  lattice will then have a lattice spacing given by $a_{2d}=L_0 c$ and
  the coupling constant $g$ of the transformed system is given by
  \begin{equation}
    \label{eq:blockspin-coupling}
    1/g = L_0 \rho_s + {\cal O}(1/(L_0\rho_s))\,.
  \end{equation}
  Thus, the continuum limit of the new lattice model is obtained in
  the limit $L_0\rightarrow\infty$. The correlation length (and thus
  the inverse mass scale of the system) is given in terms of $L_0$ by
  \begin{equation}
    \label{eq:corr-length}
    \xi = \frac{e c}{16\pi\rho_s}\exp\left(2\pi L_0\rho_s\right)
    \left( 1-\frac{1}{4\pi L_0\rho_s} + {\cal
        O}(1/(L_0\rho_s)^2)\right)\,.
  \end{equation}
  In the limit $L_0\rightarrow\infty$, the correlation length thus
  diverges exponentially and the extent $L_0\ll\xi$ becomes negligible
  and hence the system undergoes dimensional reduction.
\end{enumerate}

In conclusion, one can say that D-theory introduces a substructure to
the original system. The lattice spacing of this substructure is much
smaller than the corresponding lattice spacing of the original theory.
However, the resulting lattice action is obtained from exact blocking
of the continuum fields, implying that the lattice artifacts are of
order ${\cal O}(a_{3d})$. This means that in practical simulations,
one can use lattice spacings of the same order of magnitude as with
the Wilson discretization and the resulting theory has a lattice
spacing $a_{2d}\gg a_{3d}$.

\subsubsection{Models with Local Gauge Symmetries}
\label{sec:models-with-local}
The construction principle underlying D-theory can be applied to other
models as well. The important cases of U$(1)$ and SU$(2)$ gauge
theories have been discussed in \cite{Chandrasekharan:1997ih}. The
application to the case of QCD has been considered in
\cite{Brower:1997ha}. For a review consult \cite{Beard:1998ic}.

In \cite{Chandrasekharan:1998ck}, it has been conjectured how the
parameter space of the D-theory formulation is related the coupling of
the conventional theory. Also the principal chiral model could have
been formulated in this way and has been shown to reduce to the
conventional discretization formulation \cite{Schlittgen:1999ay}.
From these discussions it becomes clear that in fact D-theory is an
alternative formulation of the discretization of quantum field
theories with local gauge symmetries.

\subsubsection{Simulation Algorithms}
\label{sec:simul-algor}
For the simulation of quantum spin systems, a particular efficient
class of algorithms is available, known as \textit{cluster
  algorithms}. While the most efficient algorithms to be discussed in
Chapter~\ref{sec:numerical-methods} which are applicable to the Wilson
action Eq.~(\ref{eq:wilson-gauge-action}) are all local, the cluster
algorithms are global.

\index{Algorithms!cluster} Cluster algorithms have first been
introduced to quantum spin systems by \textsc{Swendsen} and
\textsc{Wang} in \cite{Swendsen:1987ce}.  These algorithms exploit the
mapping introduced by \textsc{Fortuin} and \textsc{Kasteleyn}
\cite{Fortuin:1972ar} to rewrite the partition function and to
formulate a global algorithm which is able to flip a large cluster of
spins at once. In this way, critical slowing down which will be
discussed in Sec.~\ref{sec:scaling-behavior} is effectively reduced,
provided the average cluster size scales proportional to the
correlation length of the system. For a general overview of cluster
algorithms see \cite{Niedermayer:1996ea}. A useful generalization of
cluster algorithms which might be applicable to D-theory is given by
the world-line Monte-Carlo algorithms, see
\cite{Evertz:1993rb,Kawashima:1995ar,Evertz:1997ar} and
\cite{Beard:1996pr} for a new implementation.

If indeed locally gauge symmetric models can be simulated efficiently
using a quantum spin system, the inclusion of dynamical fermions would
be straightforward \cite{Wiese:1998nh}. Thus, full Yang-Mills theory
might be efficiently simulated. There is furthermore reason to
believe, that the fermionic sign problem for a discussion) may be
handled better in the framework of quantum spin systems. For an
overview see \cite{Chandrasekharan:2000dj}. For further readings
consult \cite{Galli:1996ye}.

This benefit could then be used to overcome the limitations of current
algorithms regarding the sign of the fermionic determinant. This
problem occurs whenever an odd number of dynamical fermion flavors is
being simulated very close to massless fermion flavors. This point
will be discussed in Secs.~\ref{sec:fermion-fields}
and~\ref{sec:summary-5}.

\subsection{Fermion Fields}
\label{sec:fermion-fields}
The Euclidean space version of (\ref{eq:yang-mills-extent}) is given
by
\begin{equation}
  \label{eq:yang-mills-euclid}
  \lagrangian(\Psi_N,\bar{\Psi}_N,A^\mu_a) =
  \bar{\Psi}_N\left(\slash\!\!\!\!D - m\right) \Psi_N\,,
\end{equation}
where the $\gamma_\mu$-matrices in Euclidean space must be employed,
cf.~App.~\ref{sec:dirac-matrices}. The representation of
Eq.~(\ref{eq:yang-mills-euclid}) on the lattice is a very complicated
task. As shown in App.~\ref{sec:grassmann-algebras}, the basic fields
$\bar{\Psi}(x), \Psi(x)$ are elements of a Grassmann algebra.  These
fields admit a representation as four-component vectors with the
choice of $\lbrace\gamma_\mu\rbrace$ as given in
App.~\ref{sec:dirac-matrices}. Thus, the task of putting an
$N$-component Yang-Mills field in Euclidean space, on the lattice is
equivalent to finding a matrix $Q_{ab,\mu\nu}(y,x)$ with
$a,b=1,\dots,N$, $\mu,\nu=0,\dots,3$, and $x,y\in\mathbb{Z}^4_\Omega$,
giving rise to the action \begin{equation}
  \label{eq:sferm-lattice-def}
  S_{\mbox{\tiny f}} = -\sum_{xy} \sum_{ab,\mu\nu}
  \bar{\Psi}^{a\mu}(y) Q_{ab,\mu\nu}(y,x) \Psi^{b\nu}(x)\,.
\end{equation}
To simplify the notation, the indices $a,b$, and $\mu,\nu$ will be
suppressed from now on. The corresponding path integral defining the
quantum partition function, Eq.~(\ref{eq:gen-func-euclid}), is then
given by (cf.~Eq.~(\ref{eq:grassmann-exp-int}) in
App.~\ref{sec:integration})
\begin{equation}
  \label{eq:state-sum-fermions}
  Z = \int\EnsembleMeas{\bar{\psi}}\EnsembleMeas{\psi} \exp\left[
    -\sum_{xy} \bar{\Psi}(y) Q(y,x) \Psi(x) \right] =
  \det {Q}\,.
\end{equation}
For the discretization of the fermionic action, a number of choices is
available. However, the Nielsen-Ninomiya theorem
\cite{Nielsen:1981rz,Nielsen:1981xu} puts a general limit on any
lattice fermion action; under some natural assumptions on the lattice
action, it follows that there is an equal number of left- and
right-handed particles for every set of quantum numbers.

This implies that on the lattice the fermion spectrum consists of
pairs of fermions and fermion-mirrors. Thus, apparently it appears to
be impossible to implement the structure of Dirac fermions on a
discrete space-time. However, one can evade the physical consequences
by decoupling the superfluous fermion states. In QCD this can be
achieved, for instance, by giving the fermion doublers a mass
proportional to the cut-off $a^{-1}$. This procedure, however, does
not work in a chirally symmetric model; in fact, it is a general
consequence of the topological character of lattice theory that there
does not exist a regularized chiral fermion theory that has the
following properties (see for a proof of this no-go theorem
\cite{Nielsen:1981hk}):
\begin{enumerate}
\item global invariance under the gauge group,
\item a different number of left- and right-handed species for given
  charge combinations,
\item the (correct) Adler-Bell-Jackiw anomaly,
\item and an action bilinear in the Weyl field.
\end{enumerate}
The absence of the Adler-Bell-Jackiw anomaly displays the fact that
the axial U$_A(1)$ current is conserved because of the cancellation of
opposite-handed species.

Of course, in the continuum formulation any gauge invariant
regularization scheme yields the same expression for the axial
anomaly. Thus, this should also be valid for the lattice
regularization, too. Consequently, any candidate for the lattice
discretization of gauge theories should reproduce the axial anomaly in
the continuum limit. Indeed it has been shown in \cite{Karsten:1981wd}
that the Wilson discretization \cite{Wilson:1975id} does reproduce the
chiral anomaly in the continuum limit. The Wilson action breaks chiral
symmetry on the lattice explicitly thus removing the unwanted
doublers from the propagators. The chiral symmetry breaking term is
actually an irrelevant contribution to the lattice Ward identity,
i.e.~it is proportional to the lattice spacing, $a$. However, it does
not disappear in the limit $a\rightarrow 0$, but rather accounts
precisely for the anomaly. For a discussion of the phase structure
associated with Wilson fermions on the lattice consult
\cite{Aoki:1984qi}.

A theorem showing that, under the rather general conditions of
locality, gauge covariance and the absence of species doubling, the
lattice action gives rise to the axial anomaly has been given in
\cite{Reisz:1999cm,Reisz:1999cn} for Abelian gauge theories and
generalized to the case of QCD (which can in principle be generalized
to any non-Abelian gauge theory) in \cite{Frewer:2000ee}. However, the
axial flavor mixing current should be non-anomalous. That this is
indeed the case has been shown in \cite{Reisz:2000dt}. The proofs have
all been done perturbatively on the lattice using the expansion from
\cite{Wohlert:1985hk}.

The problem of representation of chiral symmetry on the lattice has
been resolved only recently, when it was realized that a solution of
the Ginsparg-Wilson relation (GWR) introduced in
\cite{Ginsparg:1982bj} has an exact chiral symmetry on the lattice, as
has first been discussed in \cite{Luscher:1998pq}. The first fermionic
action which actually satisfies the GWR was the perfect action of
\cite{Hasenfratz:1998ft}. For practical purposes, the solution of
Neuberger \cite{Neuberger:1998fp,Neuberger:1998wv} is the most widely
used today (for a historical overview of the development leading to
the Neuberger representation, see \cite{Neuberger:2001nb}). Finally it
is important to point out that the theorem in \cite{Frewer:2000ee}
also applies to Ginsparg-Wilson fermions thus ensuring that they
reduce to the correct fermionic action in the continuum limit.

However, since the numerical effort for the evaluation of Neuberger
fermions increases by $1-2$ orders of magnitude compared to Wilson
fermions, the calculation with dynamical Neuberger fermions is still
prohibitively expensive.

As argued above, the Wilson action breaks chiral symmetry on the
lattice with a term of order ${\cal O}(a)$. Thus, the action depends
linearly on the cutoff and physical observables might show sizable
lattice artifacts when approaching the continuum limit. As has been
put forward by \textsc{Sheikholeslami} and \textsc{Wohlert} in
\cite{Sheikholeslami:1985ij} the cancellation of the ${\cal O}(a)$
dependence can be calculated perturbatively up to a prefactor, the
parameter $c_{\mbox{\tiny SW}}$. Observables computed using this
fermionic action with a non-perturbatively calculated $c_{\mbox{\tiny
    SW}}$ indeed show weaker artifacts with an ${\cal
  O}(a^2)$-dependence as has been demonstrated in
e.g.~\cite{Allton:1998gi,Aoki:1999ff}. This program is also called
\textit{clover-improvement}, since the perturbative correction has the
shape of a four-leaf clover. Clover-improvement turned out to be
useful in a number of studies employing the \textit{hybrid
  Monte-Carlo} (HMC) algorithm (see
Sec.~\ref{sec:hybrid-monte-carlo}). When applying it to
\textit{multiboson} (MB) algorithms (cf.~Sec.~\ref{sec:mult-algor}),
however, the required local staples (see
App.~\ref{sec:local-forms-actions}) would soon become extremely
complicated and the merits of the improvement might become obscured by
the increased algorithmic demands.

Since the major focus of this thesis lies on Wilson fermions, a few
words about its explicit breaking of chiral symmetry are in order.
Having no chiral symmetry means that there is explicit symmetry
breaking by the non-chiral fermion mass. Thus, the physics of
spontaneous chiral symmetry breaking may be shadowed. In fact, it
turns out to be extremely difficult to perform lattice calculations
with light quarks since the numerical effort increases polynomially in
the inverse quark mass \cite{Lippert:2001ha}. However, when performing
the continuum limit at sufficiently small quark masses (where the
precise meaning of ``sufficient'' can only be given very roughly
within $\chi$PT \cite{Sharpe:2000bc}), one can afterwards extrapolate
to the desired quark mass and still be able to extract correct
continuum physics from numerical lattice simulations. This is the
method usually adopted in actual calculations employing light
fermions.

\index{Action!Wilson} With the conventions used in this work, the
Wilson action for a single fermion flavor reads:
\begin{equation}
  \label{eq:wilsonaction}
  S_{\mbox{\tiny f}} = \sum_{xy} \Psi(y)^\dagger Q(y,x)\Psi(x)\,,
\end{equation}
\index{Wilson matrix} where the Wilson matrix $Q(y,x)$ is defined to
be
\begin{eqnarray}
  \label{eq:wilson-matrix}
  Q(y,x) &=& \nonumber\displaystyle\delta(y,x) -
  \kappa\sum_{\rho=0}^3\Bigl( 
  U_\rho\left(y-\hat{\rho}\right) \left(1+\gamma_\rho\right)
  \delta\left(y, x+\hat{\rho}\right) \\
  && \displaystyle\phantom{\delta(y,x) -
    \kappa\sum} + U_\rho^\dagger(y)
  \left(1-\gamma_\rho\right) \delta\left(y, x-\hat{\rho}\right)
  \Bigr)\,,
\end{eqnarray}
with $\kappa$ being a function of the bare mass parameter which is
called \textit{hopping parameter}\index{Hopping parameter}. Due to the
anticommutivity of the fermion field, one also has to include
antiperiodic boundary conditions in the coupling to the gauge field,
see \cite{Montvay:1994cy} for a thorough discussion. This usually
proceeds by choosing all $U_0(x)\rightarrow -U_0(x)$, with $x$
restricted to a single timeslice when applying the matrix
multiplication with (\ref{eq:wilson-matrix}). This sign is not
explicitly written here.  For the local form to be discussed in
App.~\ref{sec:local-forms-actions}, however, it is necessary to treat
this factor separately.

The matrix $Q(y,x)$ in (\ref{eq:wilson-matrix}) consists of the local
$\delta$-function contribution and a ``derivative'' term containing
nearest-neighbor interactions. This is often called the
\textit{hopping matrix}, $D(y,x)$, and can be considered to be the
lattice version of the covariant derivative in the continuum Dirac
matrix Eq.~(\ref{eq:yang-mills-extent}), $\slash\!\!\!\!D$. The
``mass'' has been taken to unity and the hopping parameter $\kappa$
has been written in front of the lattice derivative term which can be
achieved by a redefinition of the fields $\Psi$.  Thus, the Wilson
matrix explicitly breaks chiral symmetry on the lattice. As will soon
become clear, one can nonetheless recover the correct chiral behavior
by fine-tuning the $\kappa$ parameter. In terms of the hopping matrix,
Eq.~(\ref{eq:wilson-matrix}) can be written as
\begin{eqnarray}
  \label{eq:wilson-and-hopping-matrix}
  Q(y,x) &=& \nonumber\displaystyle\delta(y,x) - \kappa D(y,x)\,, \\
  D(y,x) &=& \nonumber\displaystyle \sum_{\rho=0}^3\Bigl(
  U_\rho\left(y-\hat{\rho}\right) \left(1-\gamma_\rho\right)
  \delta\left(y, x+\hat{\rho}\right) \\
  && \displaystyle\phantom{\delta(y,x) -
    \kappa\sum} + U_\rho^\dagger(y)
  \left(1-\gamma_\rho\right) \delta\left(y, x-\hat{\rho}\right)
  \Bigr)\,.
\end{eqnarray}
The Wilson matrix, $Q(y,x)$, fulfills the $\gamma_5$-hermiticity
property
\begin{equation}
  \label{eq:gamma5-hermiticity}
  Q^\dagger(y,x) = \gamma_5 Q(y,x) \gamma_5\,,
\end{equation}
as can be seen from inspection. This leads to the following
properties: the eigenvalues are either real or come in complex
conjugate pairs. If one takes the determinant of $Q(y,x)$, it can
therefore only change sign if an odd number of purely real eigenvalues
becomes negative. At this point it should also be remarked that the
total number of real eigenvalues is in the continuum related to the
topological charge via the Atiyah-Singer-index theorem
\cite{Bertlmann:1996xk}. For an investigation of the validity of the
index theorem on the lattice see
\cite{Gattringer:1998ab,Gattringer:1998ci}.

\begin{figure}[htb]
  \begin{center}
    \includegraphics[scale=0.4,clip=true]%
    {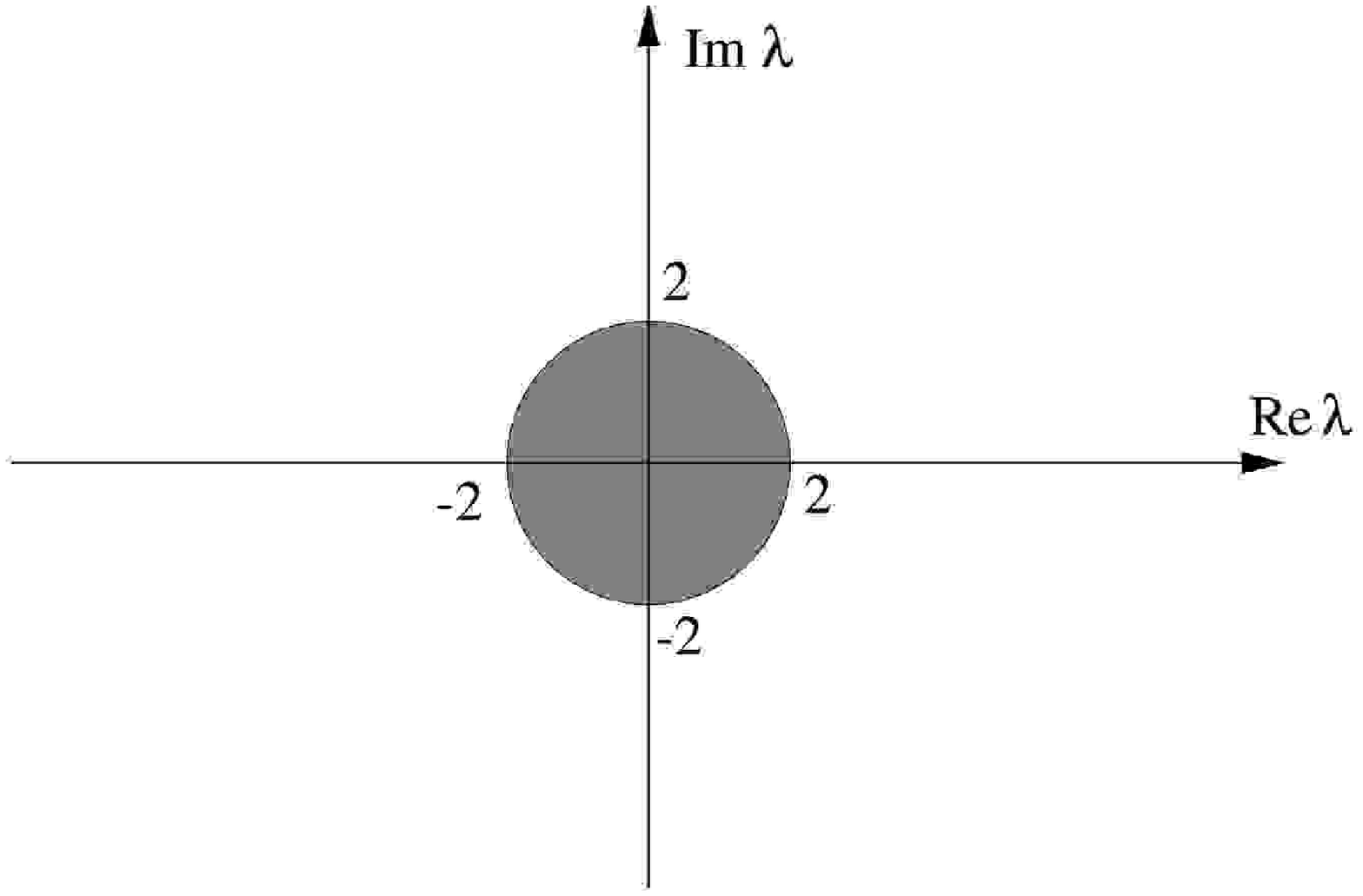}
    \caption{Spectrum of the hopping matrix, $D(y,x)$ in
      Eq.~(\ref{eq:wilson-and-hopping-matrix}), in the limit
      $\beta\rightarrow0$.}
    \label{fig:strong-coupling-hopping-spectrum}
  \end{center}
\end{figure}

\begin{figure}[htb]
  \begin{center}
    \includegraphics[scale=0.4,clip=true]%
    {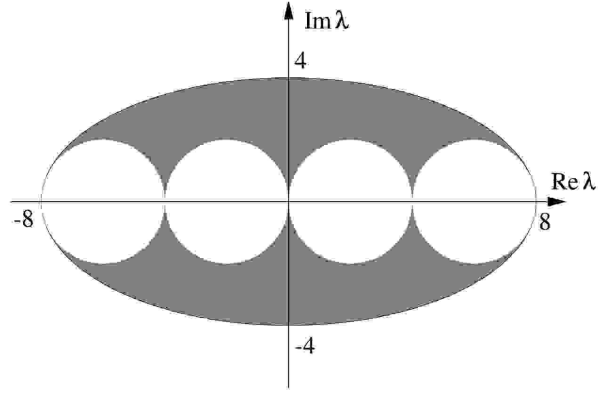}
    \caption{Spectrum of the hopping matrix, $D(y,x)$ in
      Eq.~(\ref{eq:wilson-and-hopping-matrix}), in the limit
      $\beta\rightarrow\infty$.}
    \label{fig:weak-coupling-hopping-spectrum}
  \end{center}
\end{figure}

The spectrum of the hopping matrix $D(y,x)$ in
Eq.~(\ref{eq:wilson-and-hopping-matrix}) has been examined in
\cite{Setoodeh:1988ds}. For recent overviews and results obtained from
eigenvalue methods, see \cite{Lippert:1993ph,Hip:1999ph,Neff:2001ph}.
In general, the following picture emerges: For a configuration with
$\beta=0$ (cf.~Eq.~(\ref{eq:wilson-gauge-action})), the spectrum fills
a disc centered at the origin with radius two (see
Fig.~\ref{fig:strong-coupling-hopping-spectrum}). In the small
coupling regime, the structure is more complicated (consult
Fig.~\ref{fig:weak-coupling-hopping-spectrum}): The outer shape of the
eigenvalues forms an ellipse which has a large radius of eight and a
small radius of four. However, four circles with radius two each,
centered on the real axis, are left out. At intermediate values of
$\beta$, one finds spectra interpolating between these two situations:
the spectrum starts to spread and the holes start to form, but the
eigenvalue density is not yet completely zero in the holes.
Especially, the real eigenvalues tend to populate the bulks for a
rather long time compared to the imaginary ones (see
\cite{Gattringer:1998ab}). When measuring the lattice spacing in
physical units, $a$, one finds that ${\cal O}(a)$ effects manifest
themselves prominently in the real eigenvalues still lying in the
holes \cite{Gattringer:1998ab}.

Considering then the complete Wilson matrix, $Q(y,x)$, one finds that
the lower bound of the spectrum becomes zero if (in the free case)
$\kappa_{\mbox{\tiny free}}=1/8$. A derivation of this result for free
configurations can also be found in \cite{Montvay:1994cy}. In this
case, the Wilson matrix describes massless Dirac fermions. This point
is called the \index{Chiral point} \textit{chiral point} and the
associated value of $\kappa$ is called the \textit{critical value}
$\kappa_{\mbox{\tiny crit}}$.

Hence, if $\beta$ increases from zero to $\infty$, the values of
$\kappa_{\mbox{\tiny crit}}$ decrease from $\kappa_{\mbox{\tiny
    crit}}=1/4$ down to $\kappa_{\mbox{\tiny
    crit}}=\kappa_{\mbox{\tiny free}}$. For practical determinations
of $\kappa_{\mbox{\tiny crit}}$, see Sec.~\ref{sec:chiral-limit}.

The Wilson matrix given in equation
(\ref{eq:wilson-and-hopping-matrix}) is not only non-Hermitian, it is
even non-normal, i.e. \[ [Q(y,x),Q^\dagger(y,x)]\neq 0\,. \] Thus, it
cannot be diagonalized by a unitary matrix; however, it is possible to
diagonalize the Wilson matrix by a similarity transformation with
non-unitary matrices,
\begin{equation}
  \label{eq:wilson-diagonalize}
  \mathbf{Q} = S^{-1} \cdot Q \cdot S\,.
\end{equation}
A further consequence of non-normality is that $Q(y,x)$ will in
general have different left- and right eigenvectors
\cite{Smit:1987fn}, a property which should be respected in the
definition of matrix elements in terms of the eigenvectors
\cite{Hip:2001hc,Hip:2001mh}. In several cases (as it is the case for
the sampling algorithms to be discussed in
Chapter~\ref{sec:numerical-methods}), one only needs the determinant
$\det{Q}$. Therefore it is often convenient to use the Hermitian
variant of the Wilson action which can be obtained by replacing the
matrix $Q(y,x)$ by the Hermitian Wilson matrix $\tilde{Q}(y,x)$:
\begin{equation}
  \label{eq:wilson-hermitian}
  \tilde{Q}(y,x) = \gamma_5 Q(y,x)\,.
\end{equation}
It is easy to show that $\tilde{Q}(y,x)$ is in fact self-adjoint:
\begin{eqnarray*}
  \tilde{Q}^\dagger(y,x) &=& Q^\dagger(y,x) \gamma_5
  \nonumber\\
  &=& \gamma_5 Q(y,x) = \tilde{Q}(y,x)\,.
\end{eqnarray*}
The spectrum of $\tilde{Q}(y,x)$ is more complicated than the spectrum
of $Q(y,x)$ and determining the sign of the determinant is a
non-trivial task. Exploiting the fact that $\det\gamma_5=1$, one can,
however, always use \[ \det\tilde{Q} = \det Q\,. \]

\subsubsection{Even-Odd Preconditioning}
\label{sec:even-odd-prec}
\index{Even-odd preconditioning} A simple transformation allows the
Wilson action to be rewritten \cite{DeGrand:1988vx,DeGrand:1990dk}
such that the condition number is reduced. To do this we divide the
lattice into two distinct subsets of ``even'' and ``odd'' coordinates:
\begin{description}
\item[Even-odd splitting:] If the coordinates of a given lattice site
  are given by $\left(t,x,y,z\right)\in\Omega$ then a point belongs to
  the ``odd'' subset iff \[ \left(t+x+y+z\right)\ \mathop{mod}\ 2 = 1.
  \] Otherwise they belong to the ``even'' subset.
\end{description}
If we rearrange the components of the vector in
(\ref{eq:wilson-matrix}) in such a way that the color spinor is given
by $\left(\phi_{\mbox{\tiny even}},\phi_{\mbox{\tiny odd}}\right)$
with the first half being ``even'' sites and the second half ``odd''
sites, then the Wilson matrix (\ref{eq:wilson-matrix}) takes the
following shape:
\begin{equation}
  \label{eq:wilson-eo-coords}
  \tilde{Q}(y,x) = \gamma_5\left( \begin{array}{cc} 1 & -\kappa D_{eo}
      \\ -\kappa D_{oe} & 1 \end{array} \right)\,.
\end{equation}
Using the Schur decomposition \cite{golub-vanloan}
\begin{equation}
  \label{eq:schur-decomp}
  \det\left( \begin{array}{cc} A & B \\ C & D \end{array} \right) =
  \det A\; \det \left( D-CA^{-1}B \right)\,,
\end{equation}
one arrives at the preconditioned action
\begin{equation}
  \label{eq:wilson-eo-precond}
  \hat{Q} = \gamma_5 \left( 1 - \kappa^2 D_{oe}D_{eo} \right)\,.
\end{equation}
Since this matrix has the same determinant as (\ref{eq:wilson-matrix})
it yields the same action (\ref{eq:wilsonaction}). However, the
smallest eigenvalue is about a factor of two larger, making the
inversion simpler. On the other hand, (\ref{eq:wilson-eo-precond}) has
a more complicated shape (it now contains next-to-nearest neighbor
interactions). Therefore the total effort for a matrix multiplication
stays the same, but the memory requirement for a color spinor has been
reduced.

\subsection{Yang-Mills Theory}
\label{sec:yang-mills-theory}
Finally, one can write down the total discretized form of the
continuum Yang-Mills action whose Lagrangian is given in
Eq.~(\ref{eq:yang-mills-total}):
\begin{eqnarray}
  \label{eq:yang-mills-lattice}
  S_{\mbox{\tiny g+f}} &=& \beta\sum_x\sum_{\mu\nu}
  \left(1-\frac{1}{N}\mbox{Re}\, \mbox{Tr}\, U_{\mu\nu}(x)\right) +
  \sum_{xy} \Psi(y)^\dagger Q(y,x)\Psi(x)
  \nonumber \\
  &=& S_{\mbox{\tiny g}} + \ln\det Q(y,x)\,,
\end{eqnarray}
with $Q(y,x)$ being the Wilson matrix (\ref{eq:wilson-matrix}). The
bare mass parameter $\kappa$ appearing in $Q(y,x)$ refers to the
contribution of dynamical sea quarks (i.e.~the virtual quark loops).
It is therefore termed $\kappa_{\mbox{\tiny sea}}$. The evaluation of
the determinant becomes increasingly difficult as $\kappa_{\mbox{\tiny
    sea}}$ approaches $\kappa_{\mbox{\tiny crit}}$, whose precise
value can only be determined non-perturbatively (see
Sec.~\ref{sec:chiral-limit}). Since the evaluation of the determinant
of such a huge matrix is highly difficult, it is sometimes being set
equal to one (which corresponds to $\kappa_{\mbox{\tiny sea}}=0$),
resulting in the fermionic contribution to
(\ref{eq:yang-mills-lattice}) being totally absent. If this is only
being done in the generation of configurations (i.e.~the ensemble is
sampled with the pure gauge action) this amounts to removing the
contributions of sea quarks. This defines \index{Quenched approximation}
the quenched approximation mentioned in Sec.~\ref{sec:lattice-qcd}.

\chapter{Numerical Methods}
\label{sec:numerical-methods}
In this chapter, the numerical methods are introduced which are
required to simulate lattice gauge theories with and without dynamical
fermion contributions.

The properties of Monte-Carlo algorithms are introduced in
Sec.~\ref{sec:monte-carlo-algor}. They make use of Markov chains, as
it will be explained in Sec.~\ref{sec:markov-chains}.

Section~\ref{sec:autocorrelation} introduces into the subject of time
series analysis, in particular the analysis of autocorrelations in a
Monte-Carlo time series.

The measurement of hadronic masses is discussed in
Sec.~\ref{sec:meas-hadr-mass}.

The particular algorithms used for the Monte-Carlo integration scheme
are given in Secs.~\ref{sec:boson-sampl-algor}
and~\ref{sec:ferm-sampl-algor}. The former concentrates on the
algorithms required for scalar and gauge fields, while the latter
introduces algorithms applicable to simulations with dynamical
fermionic contributions. The most important bosonic algorithms are the
Metropolis algorithm (Sec.~\ref{sec:metropolis-algorithm}), the
heatbath algorithm (Sec.~\ref{sec:heatbath-algorithm}) and the
overrelaxation technique (Sec.~\ref{sec:overrelaxation}).

The algorithms for sampling contributions of dynamical fermions are
treated in Sec.~\ref{sec:ferm-sampl-algor}. First the general problems
one encounters when evaluating the determinant of the Wilson matrix
are introduced in Sec.~\ref{sec:sampling-with-wilson}. It will become
clear, that any algorithm dealing with the fermionic determinant
requires the inversion of a large matrix describing the contribution
of the discretized fermionic degrees of freedom. Then the most widely
used algorithm for the simulation of dynamical fermion flavors, the
\textit{hybrid Monte-Carlo} (HMC) algorithm, is reviewed in
Sec.~\ref{sec:hybrid-monte-carlo}.

In this thesis, however, a more advanced algorithm for this subject
will be used, namely a variant of the \textit{multiboson} (MB)
algorithms. This class of algorithms is discussed in
Sec.~\ref{sec:mult-algor}.  These algorithms are able to overcome
several limitations and shortcomings of the HMC, but at the cost of
far more complexity.

As has been mentioned above, matrix inversion is an essential tool for
the implementation of fermion algorithms. The tools required for the
implementation of matrix inversion algorithms are described in
Sec.~\ref{sec:matr-invers-algor}. The inversion algorithms presented
are static algorithms in Sec.~\ref{sec:stat-polyn-invers}, the
Conjugate-Gradient iteration (Sec.~\ref{sec:conj-grad-iter}), the
GMRES algorithm (Sec.~\ref{sec:gmres-algorithm}), and the BiCGStab
scheme (Sec.~\ref{sec:stab-bi-conj}).

Finally, the tools for the computation of eigenvalues of matrices are
shortly reviewed in Sec.~\ref{sec:eigenv-algor}. They are important
for the application of static matrix inversion schemes and thus for
the implementation of multiboson algorithms.

\section{Monte-Carlo Algorithms}
\label{sec:monte-carlo-algor}
The path integral definition introduced in
Sec.~\ref{sec:path-integral} allows for an evaluation using ensembles
of field configurations as discussed in Sec.~\ref{sec:ensembles}.
This definition, however, requires to perform an integration on an
infinite space of operator-valued distributions $\Ensemble{\phi}$ with
a given probability distribution $\EnsembleDens{\phi}$ and a measure
$\EnsembleMeas{\phi}$. An approach different to the reformulation in
terms of Gaussian integrals discussed in Sec.~\ref{sec:path-integral}
is the application of a numerical integration using a Monte-Carlo
scheme.  That such an endeavor can indeed yield physical results in
quantum field theories was first demonstrated in
\cite{Creutz:1979dw,Creutz:1980zw}. In this section it will be
demonstrated how an algorithm can be designed in such a way that it
generates a finite set of independent gauge field configurations which
can be used as an estimator to the ensemble averages and thus to the
path integral (\ref{eq:operator-def}).

A Monte-Carlo integration algorithm is an algorithm which computes a
finite set of mesh points and yields a statistical approximation
$\bar{A}\approx\langle\hat{A}\rangle$ to the given problem. The error
of the approximation is given by the \textit{statistical error} of the
integration scheme. To be specific, let's consider an algorithm which
generates a finite sequence $\MarkovChain{\phi_n}$, $n=1,\dots,N$, of
$N$ statistically independent configurations. These must be
distributed according to the probability density $\EnsembleDens{\phi}$
of the underlying ensemble. The finite sequence is called the
\index{Sample} \textit{sample of configurations}. If they have been
taken randomly from the ensemble (\ref{eq:random-ensemble}), the
sample average
\begin{equation}
  \label{eq:rand-sample-av}
  \bar{A} = \frac{1}{N}\sum_{n=1}^N e^{-S\left[\phi_n\right]}
  \hat{A}\left[\phi_n\right]
\end{equation}
is an estimator for the ensemble average $\langle\hat{A}\rangle$ with
an error given by the variance of the statistical estimate
(\ref{eq:operator-def}). Consequently, the error of the Monte-Carlo
integration behaves as $1/\sqrt{N}$. This is different from the
standard integration schemes like Simpson's rule \cite{Sokal:1989ea}
whose error behaves as $N^{-4/d}$, with $d$ being the dimension of the
underlying space. Obviously this method is better for low dimensions
(for $d<8$) and worse for higher dimensions ($d>8$). There are better
algorithms than Simpson's rule, but none is competitive with
Monte-Carlo integrations in very large dimensions. On the other hand,
no Monte-Carlo integration is competitive with deterministic
algorithms at lower dimensions.

It is obvious how to generalize Eq.~(\ref{eq:rand-sample-av}) if the
sample configurations have been drawn from the canonical ensemble
(\ref{eq:can-ensemble}). Then the estimator is given by the sample
average
\begin{equation}
  \label{eq:can-sample-av}
  \bar{A} = \frac{1}{N}\sum_{n=1}^N \hat{A}\left[\phi_n\right]\,.
\end{equation}
Since such an integrand may be peaked rather narrowly around its
average value, the sampling algorithm should generate only the
relevant contributions. Such a procedure is called \textit{importance
  sampling}\index{Importance sampling}.

In the following, the theoretical basis needed to design Monte-Carlo
algorithms from Markov chains is founded.

\subsection{Markov Chains}
\label{sec:markov-chains}
An important concept for the design of an algorithm yielding the
desired sample of configurations is the \textit{Markov chain}:
\begin{description}
\item[Markov chain:] A Markov chain
  $\left(\MarkovChain{\phi},\MarkovDens{\phi},\MarkovTrans\right)$
  consists of a set of states $\MarkovChain{\phi_n}$ defined on a base
  space. For the purposes in this thesis, this is the space of
  discretized fields, $\mathbb{Z}^4_\Omega$. A specific element
  $\phi_{i+1}$ is generated from the previous element $\phi_i$ by a
  stochastic process $\MarkovTrans$: \[ \phi_{i+1} =
  \MarkovTrans\phi_i\,. \] The associated transition probability is
  given by the matrix element $\MarkovProb{\phi_i}{\phi_{i+1}}$. It
  solely depends on the state $\phi_i$. The Markov density
  $\MarkovDens{\phi}$ is a unit vector in the state space spanned by
  all $\phi$, in which the matrix $\MarkovProb{\cdot}{\cdot}$ acts.
\end{description}
If the states $\MarkovChain{\phi_n}$ have the probability distribution
$\MarkovDens{\phi}$, applying $\MarkovTrans$ once to the end of the
chain may change the probability distribution. With the initial
distribution given by $\MarkovDensVar{\phi}{n}$, one obtains a new
distribution $\MarkovDensVar{\phi}{n+1}$ via \[
\MarkovTrans\MarkovDensVar{\phi}{n}=
\sum_{\lbrace\phi_i\rbrace}\MarkovProb{\phi_i}{\phi}
\MarkovDensVar{\phi_i}{n}=\MarkovDensVar{\phi}{n+1}\,. \] Using this
language one can define the following notions related to Markov
chains:
\begin{description}
\item[Irreducibility:] Denote $\phi_j=\MarkovTrans^{(M)}\phi_i$ for
  $M$ repeated applications of $\MarkovTrans$ on $\phi_i$, yielding
  $\phi_j$. A chain is called \textit{irreducible} if for any states
  $\xi, \zeta\in\mathbb{Z}^4_\Omega$, there exists an $M\geq 0$ such
  that
  \[ \zeta=\MarkovTrans^{(M)}\xi\,. \]
\item[Aperiodicity:] Define
  $p_{ij}^{(M)}=\MarkovProb{\phi_i}{\phi_{i+1}}\dots
  \MarkovProb{\phi_{j-1}} {\phi_j}$ to be the $M$-step transition
  probability to reach $\phi_j$ from the starting element $\phi_i$ in
  $M$ steps. A chain is called \textit{irreducible and aperiodic} if
  for each pair $\phi_i, \phi_j\in\mathbb{Z}^4_\Omega$ there exists an
  $M_0=M_0\left(\phi_i,\phi_j\right)$ such that $p_{ij}^{(M)}>0$ for
  all $M\geq M_0\left(\phi_i,\phi_j\right)$.
\item[Recurrence time:] Take a state $\xi\in\mathbb{Z}^4_\Omega$. Let
  $\left(\MarkovProb{\xi}{\xi}\right)^{(M)}$ be the probability to
  reach $\xi$ after $M$ applications of $\MarkovTrans$ on $\xi$. Then
  the \textit{mean recurrence time} $\tau_\xi$ is given by \[
  \tau_\xi=\sum_{M=1}^\infty
  M\left(\MarkovProb{\xi}{\xi}\right)^{(M)}\,.
  \]
\item[Positivity:] A state $\xi\in\mathbb{Z}^4_\Omega$ is called
  \textit{positive} iff $\tau_\xi$ is finite.
\item[Stationary distribution:] A probability distribution
  $\MarkovDens{\phi}$ is called \index{Stationary
    distribution|see{Equilibrium}} \textit{stationary distribution} of
  the Markov chain if it stays invariant under application of
  $\MarkovTrans$:
  \[ \MarkovDens{\phi} = \MarkovTrans \MarkovDens{\phi}\,. \]
\end{description}
A particularly important class of Markov chains is given by
the irreducible, aperiodic chains whose states are positive
\cite{Rothe:1992nt,Sokal:1989ea}. Indeed one can prove the following
theorem:
\begin{description}
\item[Existence and uniqueness of the stationary point:] Take an
  irreducible, aperiodic Markov chain with positive states
  $\MarkovChain{\phi_n}$ and transition function $\MarkovTrans$.
  Hence, the chain has the starting distribution
  $\MarkovDens{\phi_n}$. Then the limiting probability distribution
  $\MarkovDensEq{\phi}$, \[ \MarkovDensEq{\phi} =
  \MarkovTrans\MarkovDensEq{\phi} =
  \lim_{M\to\infty}\MarkovTrans^{(M)}\MarkovDens{\phi_n}\,, \] exists
  and is unique. It is thus a fixed point of $\MarkovTrans$.
\end{description}
From now on we will only consider Markov chain with this property.
The transition probability $\MarkovProb{\cdot}{\cdot}$ has to be
normalized, i.e.~for all $\xi\in\mathbb{Z}^4_\Omega$ the following
equation must hold:
\begin{equation}
  \label{eq:transprob-normal}
  \sum_{\lbrace\phi_i\rbrace} \MarkovProb{\xi}{\phi_i} = 1\,.
\end{equation}
Now we can generate the desired sample of field configurations
$\MarkovChain{\phi_n}$ as the states of a Markov chain by the repeated
application of $\MarkovTrans$ on the last state $\phi_{n}$ of the
sample thus generating new members of the sample and improving the
approximation of Eq.~(\ref{eq:rand-sample-av}). However, we must
ensure that the transition probability is designed in such a way that
the samples are taken from the desired ensemble of configurations,
i.e.~the density of the sample, $\MarkovDens{\phi}$, must equal the
density of the ensemble, $\EnsembleDens{\phi}$.

Since the application of $\MarkovTrans$ on a state $\phi$ may change
the probability density $\MarkovDens{\phi}$, we have to design the
process in such a way that the stationary distribution of the Markov
chain is given by the ensemble density of the ensemble under
consideration:
\begin{equation}
  \label{eq:markov-fixedp}
  \MarkovTrans\MarkovDensEq{\phi} = \MarkovDensEq{\phi} \equiv
  \EnsembleDens{\phi}\,.
\end{equation}
Knowing from above that the fixed point exists and that it is unique,
one can formulate the following sufficient (but not necessary)
condition for the transition probability
$\MarkovProb{\phi_n}{\phi_{n+1}}$ of the transition matrix
$\MarkovTrans$:
\begin{equation}
  \label{eq:detailed-balance}
  \MarkovProb{\phi_{n+1}}{\phi_n} \MarkovDens{\phi_n} =
  \MarkovProb{\phi_n}{\phi_{n+1}} \MarkovDens{\phi_{n+1}}\,.
\end{equation}
Summing on both sides over the complete state space $\phi_n$ and using
(\ref{eq:transprob-normal}) one arrives at
\[ \sum_{\lbrace\phi_n\rbrace} \MarkovProb{\phi_{n+1}}{\phi_n}
\MarkovDens{\phi_n} = \sum_{\lbrace\phi_n\rbrace}
\MarkovProb{\phi_n}{\phi_{n+1}} \MarkovDens{\phi_{n+1}} =
\MarkovDens{\phi_{n+1}}\,, \] which is identical to
Eq.~(\ref{eq:markov-fixedp}). Relation (\ref{eq:detailed-balance}) is
known as \index{Detailed balance} \textit{detailed balance}. It does
not determine the transition probability uniquely and thus one can
design different algorithms sampling the field configurations.
However, since Eq.~(\ref{eq:markov-fixedp}) is not a sufficient
condition, it may happen that the algorithm gets ``stuck'' in a local
maximum of the density. Such a situation is difficult to detect and
even more difficult to handle. The only way to proceed in such cases
is by using \index{Ensembles!multi-canonical} multicanonical sampling
(see Sec.~\ref{sec:ensembles}). If one manages to find an action
$\tilde{S}\left[\phi\right]$ which no longer has several distinct
local maxima, this problem is avoided. A typical situation where this
may happen is if the system is close to a first-order phase
transition, where the system has comparable probabilities to exist in
either one of two different phases \cite{Arnold:1998vn}.

For an infinitely long Markov chain, we define the mean value $\langle
\hat{A}\rangle$ by
\begin{equation}
  \label{eq:markov-mean}
  \bar{A} = \sum_i \MarkovDensEq{\phi_i}
  \hat{A}\left[\phi_i\right]\,,
\end{equation}
with $\MarkovDensEq{\phi}$ being the stationary distribution of the
Markov chain. Then the mean value $\bar{A}$ coincides with the
expectation value $\langle \hat{A}\rangle$ from
Eq.~(\ref{eq:operator-def}). For a \textit{finite sample}
$\MarkovChain{\phi_n}$, $n<\infty$, the estimator $\bar{A}$
approximates $\langle \hat{A}\rangle$ with an error of order
$\sqrt{N}$ as discussed above.

\section{Autocorrelation}
\label{sec:autocorrelation}
Although the Markov chain generates a new state only from the previous
one without any knowledge of older states, the new state may be rather
similar to the old one. Thus, the sample of configurations generated
as states of the Markov chain will in general not be statistically
independent. The correlation in the sequence of generated
configurations can be made mathematically precise using the
\textit{autocorrelation function} of a time series.  In the following
$A_i$ denotes the measurement of $\hat{A}\left[\phi_i\right]$ on a
configuration $\phi_i$. The time series then consists of the set of
$\lbrace A_i\rbrace$, $i=1,\dots,N$.

\subsection{Autocorrelation Function}
\label{sec:autoc-funct}
The autocorrelation function is defined by
\begin{equation}
  \label{eq:autocorr-def}
  C_{AA}(\tau) = \langle A_t A_{t+\tau}\rangle - \left(\langle
  A_t\rangle\right)^2\,,
\end{equation}
where the average of the infinite series is denoted as
$\langle\cdot\rangle$. The set of states underlying
Eq.~(\ref{eq:autocorr-def}) is infinite. However, as already noted
above, in practical calculations one deals with finite samples and
therefore is unable to compute the exact averages but only estimators.
The estimator based on a finite sample of length $N$ for
(\ref{eq:autocorr-def}) is given by
\begin{equation}
  \label{eq:autocorr-sample}
  \bar{C}_{AA}(\tau) = \frac{1}{N-\tau}\sum_{t=1}^{M-\tau} \left(
  A_t-\bar{A_0} \right)\left(
  A_{t+\tau}-\bar{A}_\tau\rangle\right)\,.
\end{equation}
The autocorrelation function with $\tau=0$ is the standard deviation
of the series. The normalized autocorrelation function
$\Gamma_{AA}(\tau)$ is defined by
\begin{equation}
  \label{eq:normed-ac-def}
  \Gamma_{AA}(\tau) = C_{AA}(\tau) / C_{AA}(0)\,.
\end{equation}

\subsection{Exponential Autocorrelation Time}
\label{sec:expon-autoc-time}
One important information the autocorrelation function yields is the
time the system needs to \index{Equilibrium} equilibrate, i.e.~the
time needed until the system goes from an arbitrary starting point
$\MarkovDens{\phi}$ to the stationary probability density
$\MarkovDensEq{\phi}$. To study this behavior, let
$\MarkovDensVar{\phi}{\mbox{\tiny inter}}$ be a given probability
measure on $\mathbb{Z}^4_\Omega$ at an arbitrary intermediate state
taken from the Markov chain and $\MarkovDensEq{\phi}$ the equilibrium
distribution of the Markov chain.  Let $l^2(\phi)$ denote the Banach
space of complex-valued functions $f(\MarkovDens{\phi})$ on the state
space $\mathbb{Z}^4_\Omega$ having finite norm
\begin{equation}
  \label{eq:banach-norm}
  \| f\|_{l^2(\phi)} = \left( \sum_{\phi\in\mathbb{Z}^4_\Omega}
    f(\MarkovDens{\phi})\MarkovTrans f(\MarkovDens{\phi}) \right) <
  \infty\,.
\end{equation}
The inner product in this space is given by
\begin{equation}
  \label{eq:banach-prod}
  \langle f\vert g\rangle = \sum_{\phi\in\mathbb{Z}^4_\Omega}
  f^*(\MarkovDens{\phi})\MarkovTrans g(\MarkovDens{\phi})\,.
\end{equation}
Then we define the deviation of $\MarkovDensVar{\phi}{\mbox{\tiny
    inter}}$ from $\MarkovDensEq{\phi}$ by
\cite{Sokal:1989:zz,Sokal:1989ea,Lippert:2001ha}:
\begin{eqnarray}
  \label{eq:distance-from-eq}
  d_2\left(\MarkovDensVar{\phi}{\mbox{\tiny inter}},
    \MarkovDensEq{\phi}\right) &=& \| \MarkovDensVar{\phi}{\mbox{\tiny
      inter}}-\MarkovDensEq{\phi} \| \nonumber \\
  &=& \sup_{\|
    f\|_{l^2\left(\phi\right)}\leq 1} \left\vert
    \sum_{\phi\in\mathbb{Z}^4_\Omega}\left[
      f(\MarkovDensVar{\phi}{\mbox{\tiny inter}}) -
      f(\MarkovDensEq{\phi}) \right] \right\vert\,.
\end{eqnarray}
In general, the transition $\MarkovTrans$ for an irreducible,
positive-recurrent Markov chain has the following properties:
\begin{description}
\item[Contraction:] The spectrum of $\MarkovTrans$ lies in the closed
  unit disc. Consequently, $\MarkovTrans$ is a contraction.
\item[Eigenvalues of the stationary distribution:] The eigenvalue $1$
  of $\MarkovTrans$ is simple. The operator $\MarkovTrans^*$ has the
  same properties.
\item[Uniqueness:] If the chain is aperiodic, then $1$ is the only
  eigenvalue of $\MarkovTrans$ (and of $\MarkovTrans^*$) on the unit
  circle. The eigenvector is the unit vector in $\mathbb{Z}^4_\Omega$.
\end{description}
If $\MarkovDensVar{\phi}{\mbox{\tiny inter}}$ has been obtained from a
starting distribution $\MarkovDensVar{\phi}{\mbox{\tiny start}}$ by a
single application of $\MarkovTrans$, it follows that
\begin{equation}
  \label{eq:distance-evol}
  d_2\left(\MarkovTrans\MarkovDensVar{\phi}{\mbox{\tiny start}},
    \MarkovDensVar{\phi}{\mbox{\tiny start}}\right)
  \leq \| \MarkovTrans\restriction\mathbf{1}^{\perp}\|
  d_2\left( \MarkovDensVar{\phi}{\mbox{\tiny start}},
    \MarkovDensEq{\phi} \right)\,.
\end{equation}
The spectral radius formula \cite{Sokal:1989:zz} yields:
\begin{equation}
  \label{eq:spectral-radius}
  \| \MarkovTrans\restriction\mathbf{1}^{\perp}\| \propto R :=
  \exp\left( \frac{-1}{\tau_{\mbox{\tiny exp}}}\right)\,.
\end{equation}
Thus, $R$ is the spectral radius of $\MarkovTrans$ on the orthogonal
complement of the identity, i.e.~the largest modulus of the
eigenvalues of $\MarkovTrans$ with $|\lambda|<1$. The definition of
$\tau_{\mbox{\tiny exp}}$, Eq.~(\ref{eq:spectral-radius}), maps the
spectral radius $R\in[0,1[$ onto $\tau_{\mbox{\tiny
    exp}}\in[0,\infty[$. Hence, a scale in the Markov chain has been
introduced. After $M$ applications of $\MarkovTrans$ one arrives at
\begin{equation}
  \label{eq:iterated-distance}
  d_2\left(\MarkovTrans^{(M)} \MarkovDensVar{\phi}{\mbox{\tiny
        start}},
    \MarkovDensEq{\phi} \right)
  \leq\exp\left(\frac{-M}{\tau_{\mbox{\tiny exp}}} \right)\,.
\end{equation}
The meaning of $\tau_{\mbox{\tiny exp}}$ is that of a relaxation
parameter. The number of steps required for the system to reach the
fixed point distribution starting from an arbitrary distribution is
characterized by this time scale. It may happen that
$\tau_{\mbox{\tiny exp}}$ even becomes infinite \cite{Sokal:1989ea}.
In such a case, one can never reach the equilibrium by starting from
an arbitrary configuration in finite time.

To actually compute $\tau_{\mbox{\tiny exp}}$ for a given algorithm,
one must find a good test function, i.e.~an appropriate observable in
(\ref{eq:normed-ac-def}) with sufficient overlap to the slowest mode
of the system. Thus, one can define $\tau_{\mbox{\tiny exp}}$ via
\begin{equation}
  \label{eq:tauexp-def}
  \tau_{\mbox{\tiny exp}} = \sup_{\lbrace\hat{A}\rbrace}
  \lim_{\tau\rightarrow\infty} \frac{\tau}{-\ln\Gamma_{AA}(\tau)}\,,
\end{equation}
where several different observables $\hat{A}$ must be considered. Of
course, in practice one can never be sure that the slowest mode of the
system is captured by the set of observables chosen.

In practical situations, however, one does not work with the total
density vector $\MarkovDens{\phi}$ of the system, but rather one
considers only the finite sample of configurations obtained by
repeated application of $\MarkovTrans$ to a single starting
configuration. The probability of this configuration in the
equilibrium density $\MarkovDensEq{\phi}$ may be rather small, but it
cannot be zero. The way to estimate a given density vector in the
state space of the Markov chain is then to histogram an observable and
examine its distribution. For a gauge theory on the lattice this could
e.g.~be the gluonic action. Unless the system hasn't thermalized, the
histogram will still change its shape when adding new configurations.

For the starting configuration it is common to either use a
homogeneous set of variables, the \index{Cold start} \textit{cold
  start}, or a set of random variables, the \index{Hot start}
\textit{hot start}.

\subsection{Integrated Autocorrelation Time}
\label{sec:integr-autoc-time}
Once the Markov chain has reached the equilibrium density, there is
still an autocorrelation between subsequent measurements. This
autocorrelation can be assessed by considering the \textit{integrated
  autocorrelation time}, $\tau_{\mbox{\tiny int}}$. For an observable
$\hat{A}$, the latter is defined via
\cite{Sokal:1989ea,Lippert:2001ha}
\begin{equation}
  \label{eq:autocorr-int-def}
  \tau^{\hat{A}}_{\mbox{\tiny int}} = \frac{1}{2}\sum_{\tau'=1}^\infty
  \Gamma_{AA}(\tau')\,.
\end{equation}
The factor of $1/2$ in (\ref{eq:autocorr-int-def}) is a matter of
convention. It ensures that $\tau^{\hat{A}}_{\mbox{\tiny int}}\approx
\tau^{\hat{A}}_{\mbox{\tiny exp}}$ if $\Gamma_{AA}(t)\simeq
\exp(-|t|/\tau)$ for $\tau\gg 1$. When applied to a finite sample of
lengths $N$, one obtains an estimate via
\begin{equation}
  \label{eq:tauint-estimate}
  \bar{\tau}^{\hat{A}}_{\mbox{\tiny int}} =
  \frac{1}{2}\sum_{\tau'=1}^N \Gamma_{AA}(\tau')\,.
\end{equation}
$\tau^{\hat{A}}_{\mbox{\tiny int}}$ characterizes the statistical
error of an observable $A$. This can be seen by considering the
variance $\sigma(\hat{A})$ of the mean (\ref{eq:rand-sample-av}):
\begin{eqnarray}
  \label{eq:variance-tauint}
  \sigma(\hat{A}) &=& \frac{1}{N^2} \sum_{i,j=1}^N C_{AA}(i-j)
  \nonumber \\
  &=& \frac{1}{N}\sum_{t=-(N-1)}^{N-1} \left( 1-\frac{t}{N} \right)
  C_{AA}(t) \nonumber \\
  &\buildrel{N\gg\tau}\over{\approx}& \frac{1}{N}
  (2\tau^{\hat{A}}_{\mbox{\tiny int}}) C_{AA}(0)\,.
\end{eqnarray}
Thus, the error in case of stochastically dependent configurations is
decreased by the factor $2\tau^{\hat{A}}_{\mbox{\tiny int}}$ if
autocorrelations are present. It is obvious, that the integrated
autocorrelation time will in general depend on the observable
$\hat{A}$, meaning that some quantities are harder to measure than
others from finite samples. This also depends on the algorithm
underlying the Markov chain, i.e.~on the choice of the transition
matrix $\MarkovTrans$.

As it is discussed in \cite{Lippert:2001ha}, the autocorrelation
function $C_{AA}(t)$ may be composed of several different
exponentials. The fast decaying modes lead to a decrease of the
contribution of the slower modes in the integral. Therefore,
observables with only a small overlap on the slowly decaying modes
will usually exhibit a smaller $\tau^{\hat{A}}_{\mbox{\tiny int}}$
than those dominated by the slower modes. Large spatial correlations
on the lattice may induce modes in the autocorrelation functions which
are also large (since the information has to propagate a larger
distance through the lattice along the Markov chain). This results to
the fact that large correlation lengths which one encounters for
smaller masses exhibit larger integrated autocorrelation times --- a
result which was clearly visible in the samples contained in
\cite{Lippert:2001ha}.

Recalling that $\tau_{\mbox{\tiny exp}}$ is associated with the
slowest mode in the system, one concludes that
$\tau^{\hat{A}}_{\mbox{\tiny int}}<\tau^{\hat{A}}_{\mbox{\tiny exp}}$
for any observables $\hat{A}$. This can also be shown by considering
again the spectrum of $\MarkovTrans$. If detailed balance holds,
$\MarkovTrans$ is self-adjoint on the space $l^2(\phi)$. Hence, the
spectrum is real and lies in an interval $[\lambda_{\mbox{\tiny
    min}},\lambda_{\mbox{\tiny max}}]\subseteq [-1,1]$ with
\begin{eqnarray}
  \label{eq:markovtrans-ev}
  \lambda_{\mbox{\tiny min}} &=& \inf
  \mbox{spec}\left(\MarkovTrans\restriction\mathbf{1}^{\perp}\right)\,,
  \nonumber \\
  \lambda_{\mbox{\tiny max}} &=& \sup
  \mbox{spec}\left(\MarkovTrans\restriction\mathbf{1}^{\perp}\right)\,.
\end{eqnarray}
Using the spectral radius formula (\ref{eq:spectral-radius}) again
yields \[ \tau_{\mbox{\tiny exp}} = \frac{1}{\ln\lambda_{\mbox{\tiny
      max}}}\,, \] where the slowest mode is associated with
$\lambda_{\mbox{\tiny max}}$. By considering an estimator
$\tau^{\hat{A}}_{\mbox{\tiny exp}}$ for $\tau_{\mbox{\tiny exp}}$, one
can write it in form of a spectral representation
\begin{equation}
  \label{eq:autocorr-spec}
  \Gamma_{AA}(\tau) = \int_{\lambda^A_{\mbox{\tiny
        min}}}^{\lambda^A_{\mbox{\tiny max}}} \lambda^{|\tau|}
  d\sigma_{\hat{A}}(\lambda)\,.
\end{equation}
The largest and slowest modes contributing to
$\tau^{\hat{A}}_{\mbox{\tiny exp}}$ have been denoted by
$\lambda^{\hat{A}}_{\mbox{\tiny min}}$ and
$\lambda^{\hat{A}}_{\mbox{\tiny max}}$.  They form a subinterval of
$[\lambda_{\mbox{\tiny min}},\lambda_{\mbox{\tiny max}}]$. Summing
(\ref{eq:autocorr-spec}) over $\tau$ one finally arrives at \[
\tau^{\hat{A}}_{\mbox{\tiny int}} = \frac{1}{2}
\int_{\lambda^{\hat{A}}_{\mbox{\tiny
      min}}}^{\lambda^{\hat{A}}_{\mbox{\tiny max}}}
\frac{1+\lambda}{1-\lambda} d\sigma_{\hat{A}}(\lambda) \leq
\frac{1}{2}\int_{\lambda^{\hat{A}}_{\mbox{\tiny
      min}}}^{\lambda^{\hat{A}}_{\mbox{\tiny max}}}
\frac{1+\lambda^{\hat{A}}_{\mbox{\tiny
      max}}}{1-\lambda^{\hat{A}}_{\mbox{\tiny max}}}
d\sigma_{\hat{A}}(\lambda)\,. \] This leads to \[
\tau^{\hat{A}}_{\mbox{\tiny int}}\leq\frac{1}{2}
\left(\frac{1+\exp(-1/\tau^{\hat{A}}_{\mbox{\tiny exp}})}
  {1-\exp(-1/\tau^{\hat{A}}_{\mbox{\tiny exp}})}\right) \approx
\tau^{\hat{A}}_{\mbox{\tiny exp}} \leq \tau_{\mbox{\tiny exp}}\,. \]

\subsection{Scaling Behavior}
\label{sec:scaling-behavior}
As has been discussed in Sec.~\ref{sec:eucl-field-theory}, a quantum
field theory usually will undergo a second order phase transition as
the continuum limit is approached. This implies that the correlation
length, $\xi$, associated with the system diverges. This divergence
claims an increase in the lattice size, $L$, and usually also means
that the autocorrelation time increases rapidly. This phenomenon is
known as \textit{critical slowing down}. In particular, the
autocorrelation time diverges as \cite{Sokal:1989ea}:
\begin{equation}
  \label{eq:tauint-diverge}
  \tau\propto \min\left(L,\xi\right)^z\,,
\end{equation}
which defines the \textit{dynamic critical exponent} $z$. Critical
slowing poses a problem for the numerical simulation of dynamical
systems since especially the critical points are points of major
physical interest.

\subsection{Short Time Series}
\label{sec:short-time-series}
When using Eq.~(\ref{eq:tauint-estimate}) to estimate
$\tau^{\hat{A}}_{\mbox{\tiny int}}$ for an observable on a finite time
series, one still needs a sufficient amount of measurements. The
particular problem is that large $\tau$ values of $C_{AA}(\tau)$ will
have large noise, but only small signals since the function does
approach zero while the errors don't \cite{Sokal:1989ea}. To be
specific the error can be computed using the approximation $\tau\ll
M\ll N$:
\begin{equation}
  \label{eq:tauint-error}
  \sigma(\bar{\tau}^{\hat{A}}_{\mbox{\tiny int}}) \approx
  \frac{2(2M+1)}{N}\left(\bar{\tau}^{\hat{A}}_{\mbox{\tiny
        int}}\right)^2\,.
\end{equation}
If the sum in (\ref{eq:tauint-estimate}) is cut off at a point $M<N$
(introducing a ``window'' of size $M$), one obtains
$\tilde{\tau}^{\hat{A}}_{\mbox{\tiny int}}(M)$ via
\begin{equation}
  \label{eq:tauint-window}
  \tilde{\tau}^{\hat{A}}_{\mbox{\tiny int}}(M) = \frac{1}{2}
  \sum_{\tau'=1}^M \Gamma_{AA}(\tau')\,.
\end{equation}
The trade-off is that by using (\ref{eq:tauint-window}), one
introduces a bias
\begin{equation}
  \label{eq:tauint-bias}
  \mbox{bias}(\bar{\tau}^{\hat{A}}_{\mbox{\tiny int}}) =
  -\frac{1}{2}\sum_{|\tau'|>M} \Gamma_{AA}(\tau') + {\cal
    O}\left(\frac{1}{N}\right)\,.
\end{equation}
Thus, the bias will only be a finite-length effect of the time series
which will vanish once the series is long enough.

The choice of $M$ should be guided by the desire to make
$\sigma(\bar{\tau}^{\hat{A}}_{\mbox{\tiny int}})$ small while on the
other hand still keeping the
$\mbox{bias}(\bar{\tau}^{\hat{A}}_{\mbox{\tiny int}})$ small.

\subsubsection{Windowing Procedure}
\label{sec:windowing-procedure}
One way to choose the window parameter $M$ is to apply the following
recipe \cite{Sokal:1989ea,Lippert:2001ha}: Find the smallest integer
$M$ such that \[ M>c\tilde{\tau}^{\hat{A}}_{\mbox{\tiny int}}(M)\,. \]
If $\Gamma_{AA}(\tau)$ was a pure exponential, then it would suffice
to take $c\approx 4$. This implies that $\Gamma_{AA}(\tau)$ would have
decayed by $98\%$ since $e^{-4}<2\%$. However, if $\Gamma_{AA}(\tau)$
does not show a clear exponential behavior, then one has to consider
$c\approx 6$ or still larger. For time series of the order of
$N\approx 1000\tau$ this algorithm works fine \cite{Sokal:1989ea},
however it is not clear how stable this procedure is for much smaller
samples. Sadly, in the numerical simulation of Euclidean field
theories, one usually only has $N\approx (100-200)\tau$ or even less,
so this method \textit{alone} is insufficient for obtaining a reliable
estimate of $\tau^{\hat{A}}_{\mbox{\tiny int}}$.

\subsubsection{Lag-Differencing Method}
\label{sec:lag-diff-meth}
A typical indicator of a systematic bias might be that the
autocorrelation function does not converge to zero but rather
approaches a constant before dropping to zero in a non-exponential
manner. It could also be that the autocorrelation function exhibits
linear behavior. Being conservative, one would conclude that in such a
situation the time series is simply too short to give answers and that
there is no way to extract further information from it. If one is more
practical, one may try to extract only the exponential modes from the
series and discard the linear behavior.  This is what
\textit{differencing} does.  In \cite{Lippert:2001ha}, this new method
for eliminating, or at least reducing the bias (\ref{eq:tauint-bias})
of the time series has been suggested by \textsc{Lippert}. The idea is
to apply a differencing prescription to the original series in order
to reveal the true autocorrelation behavior. This approach is
justified, because once the Markov density $\MarkovDens{\phi}$ becomes
stationary, the system will be unaffected by a shift in the time
origin.

Define the order-$k$-lag-$l$-differenced time series by
\begin{eqnarray}
  \label{eq:differencing-def}
  (D_l^{(k)} A)_i &=& (D_l^{(k-l)} A)_{i+l} - (D_l^{(k-l)} A)_i\,,
  \nonumber \\
  (D_l^{(1)} A)_i &=& A_{i+l} - A_i\,.
\end{eqnarray}
Examining the estimator for the average $\langle (D_l^{(1)}
A)_s\rangle$ shows, that the first-order-differenced series indeed
goes to zero: \[ \langle (D_l^{(1)} A)_s\rangle = \frac{1}{N-\tau-l}
\sum_{\tau'=1}^{N-\tau-l} (D_l^{(1)} A)_{\tau'+s}
\buildrel{N\rightarrow\infty}\over{\rightarrow} 0\,. \] One possible
way to apply definition (\ref{eq:differencing-def}) is to examine the
correlation between the original series $\lbrace A_i\rbrace$ and the
order-$1$ differenced series $\lbrace (D_l^{(1)} A)_i\rbrace$:
\begin{eqnarray}
  \label{eq:corr-diffser}
  C_{A,(D_l^{(1)} A)}(\tau) &=& \frac{1}{N-\tau-l}
  \sum_{\tau'=1}^{N-\tau-l} \left(A_t-\langle A_0\rangle\right) \left(
    (D_l^{(1)} A)_{\tau'+\tau} - \langle (D_l^{(1)}
    A)_{\tau}\rangle\right) \nonumber \\
  &&\buildrel{N\rightarrow\infty}\over{\rightarrow} C_{AA}(\tau) -
  C_{AA}(\tau+l)\,.
\end{eqnarray}
A constant bias will be removed for $l>\tau^{\hat{A}}_{\mbox{\tiny
    exp}}$, while the modes with scales below $l$ should not be
affected. However, when choosing $l$ too small,
Eq.~(\ref{eq:corr-diffser}) will destroy also exponential modes larger
than $l$. On the other hand, the procedure will be ineffective if $l$
is too large since the statistical quality of the sample will get
worse. For this reason, we also believe that higher order differencing
will not be useful for practical purposes.

In practice one has to examine the autocorrelation function for a
number of different lags. In the ideal case, a plateau should form
when plotting the estimated value for $\tau^{\hat{A}}_{\mbox{\tiny
    int}}$ from Eq.~(\ref{eq:corr-diffser}) vs.~the lag $l$. This
fortunate case is, however, only rarely given since one would not need
to apply the differencing procedure in the first place if the
statistics were good enough.

The recipe to apply this procedure which is used in this thesis
consists of the following steps: (i) Get a first rough estimate about
the autocorrelation time $\tilde{\tau}^{\hat{A}}_{\mbox{\tiny int}}$.
This may be obtained by comparison to different time series or by the
other methods for computing autocorrelation times. (ii) Vary the lag
$l$ and measure a the function $\tau^{\hat{A}}_{\mbox{\tiny int}}(l)$
for the different lags. (iii) If the function exhibits a plateau with
$l>\tau^{\hat{A}}_{\mbox{\tiny int}}$, the estimate for
$\tau^{\hat{A}}_{\mbox{\tiny int}}(l)$ at the plateau is taken. If no
plateau is formed even when going to
$l>2\tilde{\tau}^{\hat{A}}_{\mbox{\tiny int}}$, the method fails to
give any reasonable answer.

\subsubsection{Jackknife Method}
\label{sec:jackknife-method}
As an independent consistency check, one can also exploit relation
(\ref{eq:variance-tauint}) to obtain an estimate for
$\tau^{\hat{A}}_{\mbox{\tiny int}}$. The method discussed in the
following is called \index{Jackknife} \textit{Jackknife binning} and
allows to find the ``true'' variance of a sample. In addition, it
allows to estimate the variance of ``secondary quantities'', i.e.~a
function obtained from the average of the original sample. In the
context of quantum field theories, secondary quantities are given by
observables which are defined to be expectation values and thus
require an averaging over the ensemble.

Reference \cite{Efron:1982jk} contains an introduction to the
Jackknife procedure; for a complete discussion and further
applications consult Ref.~\cite{Orth:2002ph}.

The Jackknife method consists of the following steps:
\begin{enumerate}
\item Choose a block size $B\geq 1$ and partition the series in a
  number of blocks of size $B$. The total number of blocks is then
  given by $M=N/B$. In the following it will be assumed that all blocks
  have equal size (if $B$ is not a divisor of $N$, one can simply make
  the last block smaller; this has no practical influence).
\item Define the averages $\lbrace\bar{A}_j^{(B)}\rbrace$,
  $j=1,\dots,M$, by
  \begin{equation}
    \label{eq:jackknife-av-def}
    \bar{A}_j^{(B)} = \frac{1}{N-B}\left( \sum_{n=1}^{N_1-1} A_n +
      \sum_{n=N_2+1}^N A_n \right)\,,
  \end{equation}
  with $N_1=B(j-1)+1$ and $N_2=jB$. Thence, $\bar{A}_j^{(B)}$ is the
  average of the sample $\lbrace A_i\rbrace$ with the $j$th block of
  size $B$ (ranging from $N_1$ to $N_2$, included) being left out.
\item Then define the Jackknife estimator for the average and its
  variance for bin size $B$ by
  \begin{eqnarray}
    \label{eq:jackbin-av-var}
    \bar{A}^{(B)} &=& \nonumber\displaystyle\frac{1}{M} \sum_{n=1}^M
    \bar{A}_n^{(B)}\,, \\
    \sigma_B^2(\hat{A}) &=&
    \displaystyle\frac{M-1}{M}\sum_{n=1}^M\left(
      \bar{A}_n^{(B)} - \bar{A}^{(B)} \right)^2\,.
  \end{eqnarray}
\item Repeat the above procedure for different values of $B$ and take
  the limit $B\rightarrow\infty$. The corresponding value of
  $\sigma_{B\rightarrow\infty}(\hat{A})=\sigma(\hat{A})$ is the true
  variance of the sample. In practice, one has to plot the variance
  $\sigma_B$ vs.~the bin size $B$ until a plateau emerges. The
  resulting plateau will then give an estimate of the true variance.
  However, in general the resulting variances will fluctuate strongly,
  making a precise determination impossible. The best one can do is
  then to take the average value of the plateau as an estimate and the
  fluctuations as the errors on the variances.
\end{enumerate}
After knowing the true variance, the integrated autocorrelation time
can be estimated by
\begin{equation}
  \label{eq:tauint-jackknife}
  \bar{\tau}^{\hat{A}}_{\mbox{\tiny int}} =
  \frac{1}{2}\left(\frac{\sigma(\hat{A})}
  {\sigma_{B=1}(\hat{A})}\right)^2\,.
\end{equation}
This approach, however, only allows for a crude estimate of
$\tau^{\hat{A}}_{\mbox{\tiny int}}$, since one has no systematic
control of the error (see above). This has to be contrasted to the
autocorrelation function where one can use
Eq.~(\ref{eq:tauint-error}).

The generalization of the Jackknife method to secondary quantities,
i.e.~functions of the sample average, $f(\lbrace A_i\rbrace)$, is
straightforward. Starting from the averages defined in
(\ref{eq:jackknife-av-def}), one defines the functions
$\bar{f}_j^{(B)}$ of $\bar{A}_j^{(B)}$ and their variances analogously
to Eq.~(\ref{eq:jackbin-av-var}) by
\begin{eqnarray}
  \label{eq:jackbin-func-av-var}
  \bar{f}^{(B)} &=& \nonumber\displaystyle\frac{1}{M} \sum_{n=1}^M
  f\left(\bar{A}_n^{(B)}\right)\,, \\
  \sigma_B^2(f) &=& \displaystyle\frac{M-1}{M}\sum_{n=1}^M\left(
    f\left(\bar{A}_n^{(B)}\right) - \bar{f}^{(B)} \right)^2\,.
\end{eqnarray}
With the obtained variances, one can proceed as before and apply
(\ref{eq:tauint-jackknife}) to get the autocorrelation time of the
secondary quantity.

The Jackknife method is applied in this thesis both to obtain an
independent estimate of the autocorrelation time and to obtain the
true variance and thus the true error of both primary and secondary
quantities.

\section{Measuring Hadron Masses}
\label{sec:meas-hadr-mass}
\index{Hadrons} In order to measure hadronic masses on the lattice,
one needs to compute correlation functions of operators carrying the
same quantum numbers as the hadron under consideration. For general
reviews see
\cite{Creutz:1983bo,Montvay:1987wh,Rothe:1992nt,Montvay:1994cy} and
\cite{Gusken:1999hb}. On the lattice one has again a certain freedom
for the construction of these operators. In this thesis the simplest
operators are taken in accordance with \cite{Montvay:1987wh}. For
instance, in the case of the charged pion and rho-meson
(cf.~Eqs.~(\ref{eq:pseudoscal-mes-wavefunctions}) and
(\ref{eq:vect-mes-wavefunctions})), one obtains
\begin{eqnarray}
  \label{eq:meson-lattice-operators}
  \Phi_{\pi^+}(x)  &=& \sum_a \bar{d}^a(x) \gamma_5
  u^a(x)\,, \nonumber \\
  \Phi_{\rho^+}(x) &=& \sum_a \bar{d}^a(tx)
  \vec{\gamma} u^a(x)\,,
\end{eqnarray}
where $d^a(x)$ is the $d$-flavored quark field with color index $a$,
and $u^a(x)$ the $u$-flavored quark field, respectively.
$\vec{\gamma}$ means that summation over the three spatial
$\gamma_i$-matrices has to be performed.

As discussed in \cite{Montvay:1987wh}, one can use the
K{\"a}llen-Lehmann representation of two-point functions in the
Euclidean region to derive the mass formula. In the case of a scalar
field this is done via \[ \langle
\Omega|\mathscr{T}\left\lbrace\phi(x)
  \phi(y)\right\rbrace|\Omega\rangle = \int_{m_0^2}^\infty dm^2\;
\rho(m^2) \Delta_E(x-y;m^2)\,, \] where the spectral weight function
is positive and has the shape of a $\delta$-peak for single-particle
states. The Euclidean propagator, $\Delta_E(x-y;m^2)$, is given by \[
\Delta_E(x-y;m^2) = \int\frac{d^4k}{(2\pi)^4} \frac{\exp[\mbox{i}
  k^\mu(x_\mu-y_\mu)]} {m^2+k^\nu k_\nu}\,. \] Integrating over
three-space yields a single ``time-slice'', defining the correlation
function
\begin{eqnarray}
  \label{eq:corrfunc-deriv}
  \Gamma_\phi(t_1-t_2) &\equiv& \int d^3x
  \langle\Omega|\mathscr{T}
  \left\lbrace\phi(t_1,\vec{x})\phi(t_2,\vec{y})\right\rbrace
  |\Omega\rangle
  \nonumber \\
  &=& \int_{m_0}^\infty dm\; \rho(m^2) \exp[-m (t_1-t_2)]\,.
\end{eqnarray}
For large time separations, $t_1-t_2\rightarrow\infty$, the lowest
mass state $m_0$ dominates.

In order to extent this construction to fermionic correlation
functions, one needs the generalization of
(\ref{eq:grassmann-exp-int}) to arbitrary integrals of the Gaussian
type (see \cite{Montvay:1994cy} for a mathematical derivation):
{\setlength{\arraycolsep}{0pt}\begin{eqnarray}
    \label{eq:grassmann-exp-int-arb}
    &&\int\EnsembleMeas{\eta^\dagger}\EnsembleMeas{\eta} \exp\left[
      -\eta^\dagger A\eta\right] \eta_{j_1}\eta^\dagger_{i_1} \dots
    \eta_{j_M}\eta^\dagger_{j_M} \nonumber \\
    && \propto \det A \sum_{k_1\dots k_M} 
    \varepsilon^{k_1\dots k_M}_{j_1\dots j_M} (A^{-1})_{k_1i_1} 
    \dots (A^{-1})_{k_Mi_M}\,,
  \end{eqnarray}}
with \[ \varepsilon^{k_1\dots k_M}_{j_1\dots j_M} =
\left\lbrace\begin{array}{rl} 1, \qquad & \mbox{where $k_1\dots k_M$
      is an even permutation of $j_1\dots j_M$}, \\ -1, \qquad &
    \mbox{where $k_1\dots k_M$ is an odd permutation of $j_1\dots
      j_M$, and} \\ 0, \qquad & \mbox{where $k_1\dots k_M$ is no
      permutation of $j_1\dots j_M$.} \end{array} \right. \] The sign
factor from Eq.~(\ref{eq:grassmann-exp-int}) has been dropped. The
evaluation of a hadronic matrix element thus requires one to recourse
to the fermionic matrix, $Q(y,x)$. The bare mass which enters here is
related to the valence quark content of the hadron in question and is
therefore termed $\kappa_{\mbox{\tiny val}}$. It is therefore
possible, as already argued in Sec.~\ref{sec:lattice-qcd}, to choose
the valence quark mass appearing in the hadronic operators different
from the sea quark mass appearing in the measure which is used for the
sampling process, Eq.~(\ref{eq:yang-mills-lattice}). In fact, for
quenched simulations this is a necessity to derive hadronic masses.
See Sec.~\ref{sec:lattice-qcd} for a discussion of these methods.

A meson correlation function
\begin{equation}
  \label{eq:hadron-correlators}
  \Gamma_{\mbox{\tiny m}}(t) = \langle \Phi^\dagger(t) \Phi(t)\rangle
\end{equation}
for large Euclidean times will then yield the desired mass. However,
on a lattice with a finite extent, one has to take into account finite
size effects. Since periodic boundary conditions are usually
implemented, the lattice correlation function will be symmetric and
the lattice masses will have to be extracted using:
\begin{equation}
  \label{eq:lattice-meson-masses}
  \Gamma_{\mbox{\tiny m}}(t) = \exp[-t (am_{\mbox{\tiny m}})] +
  \exp[-(L_t-t) (am_{\mbox{\tiny m}})]\,,
\end{equation}
where the temporal lattice extension is taken to be $L_t$. The case of
baryons is more involved, however. See for the latest methods and
results \cite{Orth:2002ph}.

Combining Eqs.~(\ref{eq:meson-lattice-operators}),
(\ref{eq:grassmann-exp-int-arb}), and (\ref{eq:hadron-correlators}),
the correlation function for the pion is given by
\begin{equation}
  \label{eq:pion-correlator}
  \Gamma_{\pi^+}(t) = \sum_{\vec{x},\vec{y}} \langle\mbox{Tr}\,
  (\gamma_5 Q^{-1}((0,\vec{x}),(t,\vec{y}))\gamma_5 Q^{-1}((t,\vec{y}),
  (0,\vec{x}))\rangle\,.
\end{equation}
Fitting the resulting $\Gamma_{\pi^+}(t)$ to
(\ref{eq:lattice-meson-masses}) for large values of $t$ will then
yield the lattice pion mass, $(am_{\pi^+})$.

\section{Bosonic Sampling algorithms}
\label{sec:boson-sampl-algor}
The task of this section is to describe several algorithms realizing a
Markov chain for the field configurations
$\phi_i\in\mathbb{Z}^4_\Omega$. Any algorithm should thus generate a
new configuration $\phi_{i+1}$ from a given configuration satisfying
ergodicity and detailed balance Eq.~(\ref{eq:detailed-balance}). Once
the new configuration $\phi_{i+1}$ has been generated by updating all
degrees of freedom (d.o.f.), one denotes this procedure as a single
\textit{sweep}.

In general, one can divide the algorithms into two different classes:
\begin{description}
\item[Local algorithms:] \index{Algorithms!local} The \textit{local
    algorithms} consider a subset
  $\mathscr{I}\subset\mathbb{Z}^4_\Omega$ of sites --- usually only a
  single site at a time --- and change this point according to a
  certain prescription. Then a different subset will be considered
  until the whole space $\phi_i$ has been processed at least once.
  There is no global decision taking place on the lattice. Usually
  local algorithm are constructed such that they satisfy detailed
  balance and ergodicity locally, thus ensuring that the total sweep
  also satisfies these properties.
\item[Global algorithms:] \index{Algorithms!global} All sites are
  being updated at once according to a prescription \textit{not}
  depending on any sublattice or subset. These \textit{global update
    algorithms} usually induce larger autocorrelations than the local
  ones since the changes which can be applied to all sites at once
  will only be small compared to a change which can be applied at a
  single site only.
\end{description}
There are also several hybrid forms of algorithms. The multiboson
algorithms discussed in this thesis are usually a mixture of several
local sweeps combined with a global step. Furthermore, local forms of
the multicanonical algorithms \cite{Arnold:1998mt,Moschny:2000mt},
also may require the evaluation of the global action.

To estimate the dynamical critical exponent for a local algorithm, one
has to remember that in a single step the ``information'' is
transmitted from a single site to its neighbors \cite{Sokal:1989ea}.
Consequently, the information performs a random walk around the
lattice. In order to obtain a ``new'' configuration, the information
must travel at least a distance of $\xi$, the correlation length.
Therefore one would expect $\tau\propto\xi^2$ near criticality,
i.e.~$z=2$.

The potential advantage of global algorithms is that they may have a
critical scaling exponent smaller than for local algorithms. This can
be attributed to the fact that since all sites are update at once, the
information need not travel stepwise from one lattice site to its
neighbor, as it was the case for a local algorithm.

\subsection{Metropolis Algorithm}
\label{sec:metropolis-algorithm}
The Metropolis algorithm has been introduced in
\cite{Metropolis:1953am}. It can be implemented both locally and
globally and has the following general form which has been formulated
in \cite{Kennedy:1985pg,Kennedy:1988yy}): The transition probability
$\MarkovProb{\phi_i}{\phi_{i+1}}$ is the product of two probabilities
$\MarkovTrans=\MarkovTrans_A\cdot\MarkovTrans_C$, where
\begin{enumerate}
\item $\MarkovTrans_C(\MarkovChain{\phi_i}\rightarrow
  \MarkovChain{\phi_{i+1}})$ generates a given probability density for
  the proposed change of the configuration. A convenient choice may be
  that $\phi_{i+1}$ is taken from the random ensemble
  Eq.~(\ref{eq:random-ensemble}) independent of $\phi_i$.
\item The transition probability $\MarkovTrans_A$ is then given by
  \begin{equation}
    \label{eq:metropolis}
    \MarkovTrans_A(\MarkovChain{\phi_{i}}\rightarrow
    \MarkovChain{\phi_{i+1}}) \propto\min\left(1,
    \frac{\MarkovTrans_C(\MarkovChain{\phi_{i+1}}
    \rightarrow\MarkovChain{\phi_{i}})
    \MarkovDensEqElem{\phi_{i+1}}}
    {\MarkovTrans_C(\MarkovChain{\phi_{i}}
    \rightarrow\MarkovChain{\phi_{i+1}}) \MarkovDensEqElem{\phi_i}}
    \right)\,,
  \end{equation}
\end{enumerate}
where $\MarkovDensEqElem{\phi_i}$ is the probability of $\phi_i$ in
the equilibrium density of the Markov process, $\MarkovDensEq{\phi}$.

\subsubsection{Local Metropolis Update}
\label{sec:local-metr-update}
As an example we consider a lattice with field variables $\phi(x)$,
$x\in\mathbb{Z}^4_\Omega$, which can take on continuous variables from
the interval $[a,b]$, $a,b\in\mathbb{R}$. The task is to design a
Markov process which generates field configurations distributed
according to a canonical ensemble, Eq.~(\ref{eq:can-ensemble}),
i.e.~according to $\exp\left[-S[\phi]\right]$, where $S[\phi]$ is a
multiquadratic action as discussed in
App.~\ref{sec:local-forms-actions}. A simple algorithm which
implements the local Metropolis update sweep is designed as follows:
\begin{enumerate}
\item \label{item:loop1} For each lattice site $y$ compute the local
  staple $\Delta\tilde{S}[\Delta\phi(y)]$ corresponding to $\phi(y)$.
  For a definition and actual computations of such a staple see
  App.~\ref{sec:local-forms-actions}.
\item \label{item:loop2} Suggest a randomly chosen new field variable
  $\phi'(y)$ from $[a,b]$ with staple
  $\Delta\tilde{S}[\Delta\phi(y)]$, $\Delta\phi(y)=\phi'(y)-\phi(y)$.
  Accept the new variable $\phi'(y)$ with probability
  \begin{equation}
    \label{eq:metro-example}
    \min\left( 1,
      \exp[\Delta\tilde{S}[\Delta\phi(y)]-
      \Delta\tilde{S}[\Delta\phi'(y)]] \right)\,, 
  \end{equation}
  otherwise keep the old value $\phi(y)$.
\item Iterate step~\ref{item:loop2} a number of times.
\item Continue to next loop in item~\ref{item:loop1}.
\end{enumerate}
Afterwards, the entire lattice will have been updated. This algorithm
is obviously ergodic since \textit{any} configuration can be reached
due to the random proposal of $\phi'(y)$. Furthermore it satisfies
detailed balance (\ref{eq:detailed-balance}) by construction. This
form is the special case of the general algorithm, where
$\MarkovTrans_C(\cdot,\cdot)=1$ and thus $\MarkovTrans=\MarkovTrans_A$
alone.

The algorithm discussed above is applicable to almost any system with
multiquadratic action, but it may not be efficient. It may happen that
those values of $\phi'(y)$ which have a high chance of being accepted
are strongly peaked around a small subinterval and consequently most
suggestions are rejected. In such cases it is therefore preferable to
take $\phi'(y)$ from a non-uniform distribution which is very close
(or even identical) to the desired distribution. In this case, the
Metropolis decision will have to be modified accordingly. In the
latter case, if $\phi'(y)$ has already been taken from the correct
distribution, the test can even be skipped (since this situation would
correspond to the case $\MarkovTrans_A=1$ and consequently
$\MarkovTrans=\MarkovTrans_C$. This is the idea of the heatbath
algorithm which is discussed in Sec.~\ref{sec:heatbath-algorithm}.

\subsubsection{Global Metropolis Update}
\label{sec:glob-metr-update}
In contrast to the algorithm above, it is also possible to postpone
the Metropolis decision until all lattice sites have been processed.
This is the idea of the global Metropolis update. This may be
necessary in a situation where the action cannot be written in the
form of a local staple, or if this step is too costly. In general, the
Metropolis decision will take the following form
\begin{equation}
  \label{eq:global-metropolis}
  \MarkovTrans_A =
  \min\left(1,\exp\left[S[\phi']-S[\phi]\right]\right)\,.
\end{equation}
However, when choosing $\phi'$ to be a random configuration, the
action $S[\phi']$ will usually be widely different from $S[\phi]$, and
thus the exponential will become huge. To be specific, the probability
of acceptance is given by the $\Omega$th power of the single-site
acceptance rate, where $\Omega$ is the lattice volume. For any
reasonable lattice size, this number will consequently be
prohibitively small, if even the single-site acceptance was of the
order of ${\cal O}(99\%)$.

Therefore a global Metropolis step can only be applied in the
following situations:
\begin{itemize}
\item The distribution of $\phi'$ is close to the desired one. Hence,
  the sampling process was able to generate almost the ``correct''
  distribution and one merely has to correct a small residual error.
\item The proposal $\phi'$ is very close to the old configuration
  $\phi$. In such a case one has to make sure that ergodicity still
  holds and even if it does, the danger of running into
  metastabilities may be larger. Furthermore the autocorrelation times
  may not be very favorable in this situation since the evolution in
  phase space is rather slow. For the effort of processing all sites a
  much smaller path has been traversed than in the case of the local
  algorithms; this explains why e.g.~the HMC algorithm is not
  competitive to local algorithms when the local form of the action is
  available (see below).
\end{itemize}
The global form of the Metropolis algorithms therefore usually appears
in combination with some other algorithm (either of global or local
nature) which generates a suitable proposal $\phi'$ such that the
acceptance rate, Eq.~(\ref{eq:global-metropolis}), stays reasonably
large.

\subsection{Heatbath Algorithm}
\label{sec:heatbath-algorithm}
As has already been pointed out in the previous section, the heatbath
algorithm generates a sample from a distribution which is identical to
the equilibrium distribution $\MarkovDensEq{\phi}$. The name of the
algorithm expresses the procedure of bringing the system in contact
with an infinite heatbath. If there exists a global heatbath
algorithm, then it will immediately generate the new configuration
independent of the old one, thereby eliminating all autocorrelations.
This fortunate situation is only seldom given, however. In many
situations, it is possible to apply the heatbath at least locally,
i.e.~to generate a candidate $\phi'(y)$ at a site $y$ independent from
the old value $\phi(y)$ such that $\phi'(y)$ is distributed according
to
\begin{equation}
  \label{eq:single-site-hb}
  \MarkovDens{\phi'(y)}\propto
  \exp\left[-\Delta\tilde{S}[\phi'(y)]\right]\,.
\end{equation}
Since repeated application of the local Metropolis update prescription
generates a Markov chain for $\phi$ at lattice site $y$, which also
satisfies detailed balance, it will have a fixed point distribution
which is precisely given by Eq.~(\ref{eq:single-site-hb}). Thence,
repeating the local Metropolis an infinite number of times on a single
site is identical to the local heatbath algorithm.

Finding the distribution (\ref{eq:single-site-hb}) is possible once
its integral is known, i.e.~\cite{Montvay:1994cy}
\begin{equation}
  \label{eq:heatbath-gen}
  \exp\left[-\Delta\tilde{S}[\phi'(y)]\right] d\phi'(y) =
  dE_{\Delta\tilde{S}} (\phi'(y))\,.
\end{equation}
Then one can generate the distribution of $\phi'(y)$ from a random
number $r\in [a,b]$ by
\begin{equation}
  \label{eq:heatbath-gen-new}
  \phi'(y) = E_{\Delta\tilde{S}}^{-1}\left( E_{\Delta\tilde{S}}(a) +
  r\left( E_{\Delta\tilde{S}}(b)-E_{\Delta\tilde{S}}(a) \right)
  \right)\,.
\end{equation}
Often, it is not possible to directly generate the desired
distribution (\ref{eq:heatbath-gen-new}), but rather only an
approximation. Call this approximation $\MarkovDensVar{\phi(y)}{0}$
with its integral $E^0_{\Delta\tilde{S}}$. Now generate the new
variable $\phi'(y)$ and correct for the difference to the desired
distribution, $\MarkovDens{\phi(y)}$, with a Metropolis step with
probability \cite{Montvay:1994cy} \[ \MarkovTrans =
\frac{\MarkovDens{\phi'(y)}}{\MarkovDensVar{\phi'(y)}{0}}
\min_{a\leq\phi(y)\leq b}
\frac{\MarkovDensVar{\phi(y)}{0}}{\MarkovDens{\phi(y)}}\,. \] The
total transition probability matrix for this process is given by
\[ \MarkovTrans = \langle R\rangle \MarkovDens{\phi(y)} + (1-\langle
R\rangle) \mathbf{1}\,, \] with $\langle R\rangle$ being the average
acceptance rate which depends on the quality of the approximation of
$\MarkovDensElemVar{\phi(y)}{0}$. Iterating this step for $M$ times
yields the transition probability matrix \[ \MarkovTrans^{(M)} =
(1-(1-\langle R\rangle)^M)\MarkovDens{\phi(y)} + (1-\langle
R\rangle)^M\mathbf{1}\,.
\] In the limit $M\rightarrow\infty$ the desired distribution is
recovered. However, it is sufficient to just iterate this step $M$
times (where the optimal value of $M$ should be determined such that
the algorithm has the highest efficiency) since the stationary
distributions of $\MarkovTrans^{(M)}$ and $\MarkovTrans^{(\infty)}$
coincide by virtue of the properties of the Markov process.

One can also choose to iterate the transition $\MarkovTrans^{(1)}$ as
long as the proposed change is accepted, i.e.~stop the iteration once
one proposal has been rejected. This procedure will also have the same
stationary distribution. Again, considerations of numerical efficiency
should decide which choice is optimal.

In the following, several implementations of local heatbath algorithms
which are needed for the multiboson algorithm are presented.

\subsubsection{Heatbath for Gauge Fields}
\label{sec:heatb-gauge-fields}
First consider the case of the Wilson action,
Eq.~(\ref{eq:wilson-staple-def}) from
App.~\ref{sec:local-forms-actions}, for an SU$(2)$ gauge theory
\cite{Creutz:1980zw,Brown:1987rr}. The distribution to be generated
for a single gauge variable $U\equiv U_\mu(y)$ then takes the form
\begin{equation}
  \label{eq:gauge-hb}
  dE_{\Delta\tilde{S}}(U) \propto
  \exp\left[\frac{\beta}{2}\mbox{Re}\,\mbox{Tr}\,
    U\tilde{\mathscr{S}} \right] dU\,.
\end{equation}
The link variable $U\in\mbox{SU}(2)$ can be parameterized as
\[ U = a_0 + \mbox{i}\sum_{r=1}^3\sigma_r \mathbf{a}_r\,. \]
The unitarity condition implies \[ U^\dagger\cdot U = a_0^2 +
\sum_{r=1}^3 \mathbf{a}_r^2 = a^2 = 1,\quad a_0 =
z(1-|\mathbf{a}|^2)^{1/2}\,, \] where $z=\pm 1$, and $|\mathbf{a}| =
|\sum_{r=1}^3 \mathbf{a}_r^2|^{1/2}$. The \index{Measure!Haar} Haar
measure $dU$ in Eq.~(\ref{eq:gauge-hb}) can be parameterized as \[ dU
= \frac{1}{2\pi^2} \delta(a^2-1) d^4a\,. \] In the present form, all
parameters depend in a non-linear way on the distribution and the
precise form of the staple $\tilde{\mathscr{S}}$. Thus, it appears
that generating the desired distribution is a tough problem.  However,
it is possible to exploit the invariance of the Haar measure $dU$ on
the gauge group and to rotate the l.h.s.~of (\ref{eq:gauge-hb}) to a
distribution which only depends on $\det\tilde{\mathscr{S}}$. This
step significantly simplifies the problem; consider the
SU$(2)$-projection $\bar{U}\equiv
\tilde{\mathscr{S}}/\sqrt{\det{\tilde{\mathscr{S}}}} \equiv
\tilde{\mathscr{S}}/k$ (cf.~Eq.~(\ref{eq:su2-project}) in
App.~\ref{sec:su2-group}). Obviously, $\bar{U}\in\mbox{SU}(2)$ holds,
so the Haar measure stays invariant under right multiplication with
$\bar{U}^{-1}$,
\begin{eqnarray}
  \label{eq:rotate-unity}
  dE_{\Delta\tilde{S}}(U\cdot\bar{U}^{-1}) &\propto&
  \exp\left[\frac{\beta} {2} k\mbox{Re}\,\mbox{Tr}\,
    U\tilde{\mathscr{S}}\bar{U}^{-1}\right]
  dU \nonumber \\
  &=& \exp\left[\frac{\beta} {2} k\mbox{Re}\,\mbox{Tr}\,
    U\right] dU \nonumber \\
  &=& \exp\left[\beta ka_0\right] \frac{1}{2\pi^2}
  \delta(a^2-1)d^4a\,.
\end{eqnarray}
In this form, the distribution only depends on the determinant of the
staple, $k$, and the point $a_0$ has a non-trivial distribution alone.
Once it has been chosen, the remaining components, $\mathbf{a}'$, are
a random point on the unit sphere in three-dimensional space, $S^3$,
and can be chosen, for instance according to $d^2\Omega_a=d\phi
d(\cos\theta)$. The distribution for $a_0$ is given by (with
$a_0\in[-1,1]$):
\begin{equation}
  \label{eq:distrib-a0}
  \MarkovDens{a_0} \propto \sqrt{1-a_0^2}\;\exp(\beta k a_0)\,.
\end{equation}
By applying the transformation $y\equiv\exp(\beta ka_0)$ one obtains
\begin{equation}
  \label{eq:distrib-a0-new}
  \MarkovDens{a_0} \propto \left( 1 - \left(\frac{\log y}{\beta
  k}\right)^2 \right)^{1/2}\,.
\end{equation}
This distribution can be generated by the method from
Eq.~(\ref{eq:heatbath-gen-new}) by choosing the proposal for $a'_0$
from the interval $[\exp(-\beta k),\exp(\beta k)]$. An alternative
method has been introduced by \textsc{Kennedy} and \textsc{Pendleton}
in \cite{Kennedy:1985nu}. This method is superior if the distribution
for $a_0$ is peaked close to one, a situation which is typically
encountered in multiboson algorithms. While the method from
Eq.~(\ref{eq:distrib-a0-new}) becomes less efficient for sharply
peaked distributions, the latter choice will soon become superior.

Once the new $\lbrace a'_0,\mathbf{a}'\rbrace$ have been obtained in
this way, the new link proposal can be obtained by applying the
inverse rotation in Eq.~(\ref{eq:rotate-unity}) thus yielding
\begin{equation}
  \label{eq:rotate-inverse}
  U' = \left(a'_0\mathbf{1} + \mbox{i}\sum_{r=1}^3 \sigma_r
  \mathbf{a}'_r \right)\cdot\bar{U}\,.
\end{equation}
An extension of this procedure to the case of SU$(N)$ gauge theories
with $N>2$ is more difficult since they do not share the property that
any sum of group elements is proportional to a group element. A
possible generalization has been proposed in \cite{Cabibbo:1982zn}.
The basic idea is to decompose the whole SU$(N)$ group into an
appropriate set of SU$(2)$ subgroups such that no subgroup is left
invariant. Call this set $\lbrace a_k\rbrace$, $k=1,\dots,q$. A
possible choice is $q=N-1$ with \[ a_k = \left(\begin{array}{*{7}{c}}
    1 & & & & & & \\ & \ddots & & & & & \\ & & 1 & & & & \\ & & &
    \left(\alpha_k\right) & & & \\ & & & & 1 & & \\ & & & & & \ddots &
    \\ & & & & & & 1 \end{array}\right)\,, \quad
\alpha_k\in\mbox{SU}(2)\,. \] The new field variable $U'$ is finally
chosen to be \[ U' = a_q\cdot a_{q-1}\cdot\dots\cdot a_1\cdot U\,. \]
Defining \[ U^{(k)}\equiv a_k\cdot a_{k-1}\cdot\dots\cdot a_1 U\,,
\quad U^{(0)} = U\,, \] one obtains the recursion \[ U^{(k)} =
a_k\cdot U^{(k-1)}, \quad U^{(q)} = U'\,. \] Now each multiplication
with $a_k$ gives rise to a heatbath distribution of the SU$(2)$ group,
Eq.~(\ref{eq:gauge-hb}). Hence, one has to take
\begin{equation}
  \label{eq:cabibbo-subgroup}
  \frac{\beta}{N}\mbox{Re}\,\mbox{Tr}\,(a_k\cdot U^{(k-1)}
  \tilde{\mathscr{S}} + \dots) =
  \frac{\beta}{N}\mbox{Re}\,\mbox{Tr}\,(\alpha_k \rho_k) + \dots\;,
\end{equation}
where $\rho_k$ now takes over the role of the SU$(2)$-staple in
Eq.~(\ref{eq:gauge-hb}). For the proof that this procedure does indeed
generate the desired distribution consult
\cite{Montvay:1994cy,Cabibbo:1982zn}.

\subsubsection{Heatbath for Scalar Fields}
\label{sec:heatb-scal-fields}
In the case of scalar fields one encounters actions of the type
(\ref{eq:staples3}). One prominent example is the evaluation of the
fermion matrix in sampling algorithms (see
Sec.~\ref{sec:sampling-with-wilson}). Another case of major importance
is the evaluation of correlation functions like
Eq.~(\ref{eq:pion-correlator}). These systems allow for a rather
simple implementation of both local and global heatbath algorithms. In
fact, this is one of the few cases, where a global heatbath algorithm
exists. The application of the local algorithm is straightforward: For
each site $x\in\mathbb{Z}^4_\Omega$ generate a Gaussian random number
$\eta$ with width $1$, i.e. \[ \MarkovDens{\eta} \propto
\exp(-|\eta|^2)\,. \] The new field variable, $\phi'(x)$ is then given
by
\begin{equation}
  \label{eq:scalar-local-hb}
  \phi'(x) = \tilde{a}_1^{-1}\left(\eta - \frac{1}{2}\sum_{i=2}^M a_i
    \left[\phi\left(f_i^2(x)\right)
      \cdots \phi\left(f_i^{n_i}(x)\right) \right] \right)\,.
\end{equation}
There also exists a more powerful variant which is applicable if the
\textit{total} action admits the following form (as it is the case for
fermionic actions):
\begin{equation}
  \label{eq:scalar-global-hb-action}
  S = \sum_{xyz}\phi^\dagger(y)Q^\dagger(y,z) Q(z,x) \phi(x)\,.
\end{equation}
Similarly to the local case, the generation of the $\phi(x)$ proceeds
by taking a random Gaussian vector $\eta(x)$ with unit width, i.e. \[
\MarkovDens{\eta(x)} \propto \exp\left(-\sum_x
  \eta^*(x)\eta(x)\right)\,. \] Then solve the equation
\begin{equation}
  \label{eq:global-hb-invert}
  \sum_x Q(y,x) \phi(x) = \eta(y)\,.
\end{equation}
Thus, the global heatbath requires a matrix inversion for each new
sample $\phi(x)$. This is rather costly compared to the local variant;
however, the advantage is that there is no autocorrelation at all for
the whole sample of $\lbrace\phi(x)\rbrace$ generated.

Several methods how to perform the matrix inversion are discussed in
detail in Sec.~\ref{sec:matr-invers-algor}. All these methods provide
an approximation with a residual error $\epsilon$. The question
arises, how small this error should be made. Choosing the residual
error too large will result in a bias introducing systematic errors
beyond control. One could make the residual error extremely small,
i.e.~several orders of magnitude below the statistical error inherent
in the Monte Carlo integration. But this will waste computer time in
generating an inverse with too large accuracy. This question has been
addressed in several publications, see
\cite{Dong:1994pk,Viehoff:1998wi,Eicker:1996gk} and references
therein. An improvement to these standard methods has been suggested
in \cite{deForcrand:1998je}, which allows for a reduction of the
computer time required by a factor of about $2-3$ while still
generating the correct distribution. The idea is again to sample an
approximate distribution and apply a Metropolis correction step.
Consider a vector distributed according to
$\MarkovDens{\chi(x)}\propto \exp(-|\chi(x)-\sum_y Q(x,y)\chi(y)|^2)$.
Then consider the joint distribution \[ \MarkovDens{\phi(x),\eta(x)}
\propto \exp\left[ -\left\vert\sum_x Q(y,x)\eta(x)\right\vert
  -\left\vert\chi(y)-\sum_x Q(y,x) \phi(x)\right\vert\right]\,. \] By
virtue of
\begin{eqnarray*} &&
  \frac{1}{Z_\phi}\int\EnsembleMeas{\phi}
  \exp\left[-\left\vert\sum_x Q(y,x)\eta(x)\right\vert\right] \\
  &=& \frac{1}{Z_{\phi}Z_{\chi}}\int
  \EnsembleMeas{\phi}\EnsembleMeas{\chi} \exp\left[-\left\vert\sum_x
      Q(y,x)\eta(x)\right\vert -\left\vert\chi(y)-\sum_x Q(y,x)
      \phi(x)\right\vert\right]\,,
\end{eqnarray*}
the distribution of $\MarkovDens{\phi}$ is unchanged. Now one can
update $\chi(x)$ and $\phi(x)$ with the following alternate
prescription:
\begin{enumerate}
\item Perform a global heatbath on $\chi(x)$, \[ \chi(x) = \eta(x) +
  \sum_y Q(x,y)\phi(y)\,, \] where $\eta(x)$ is a random Gaussian
  vector with unit width.
\item Perform the reflection
  \begin{equation}
    \label{eq:scalar-global-hb-reflect}
    \phi'(x) = \sum_y Q^{-1}(x,y) \chi(x) - \phi(x)\,,
  \end{equation}
  which yields the new vector $\phi'(x)$.
\end{enumerate}
The second step conserves the probability distribution of $\phi$ but
is not ergodic. The first step ensures ergodicity. The matrix
inversion in (\ref{eq:scalar-global-hb-reflect}) now can be performed
with a finite accuracy $\epsilon$ yielding the approximate solution \[
\sum_y Q^{-1}(x,y)\zeta(y) = \chi(x) - r(x)\,, \] where $r(x)$ is the
residual. Now the second step can be considered as a proposal for
$\phi'(x)=\zeta(x) - \phi(x)$. It will be accepted in a Metropolis
step with probability (cf.~Eq.~(\ref{eq:metropolis}))
\begin{equation}
  \label{eq:scalar-global-hb-metroacc}
  P_{\mbox{\tiny acc}} (\phi(x)\rightarrow\phi'(x)) =
  \min(1,\exp(-\Delta S))\,,
\end{equation}
where
\begin{eqnarray}
  \label{eq:scalar-global-hb-metroexp}
  \Delta S &=& \left\vert \sum_y Q(x,y)\phi'(y)\right\vert^2 +
  \left\vert \chi(x) - \sum_y Q(x,y)\phi'(y)\right\vert^2 \nonumber \\
  && \qquad - \left\vert
    \sum_y Q(x,y) \phi(y)\right\vert^2 - \left\vert \chi(x) - \sum_y
    Q(x,y) \phi(y)\right\vert^2 \nonumber \\
  &=& 2\mbox{Re}\,\sum_x r^\dagger(x) \sum_{y}\left(Q(x,y)\phi(y) -
    Q(x,y)\phi'(y)\right)\,.
\end{eqnarray}
If the matrix inversion is solved exactly, i.e.~$|r(x)|=0$, then one
will recover the original global heatbath algorithm. It has been
discussed in \cite{deForcrand:1998je} that there exists an optimal
choice of $\epsilon\simeq 10^{-3}-10^{-4}$ which reduces the computer
time by a factor of $2-3$ over the older methods.

\subsection{Overrelaxation}
\label{sec:overrelaxation}
A particular method to improve the behavior of the system near
criticality consists of \textit{overrelaxation}. It is similar to the
technique of overrelaxation in differential equation algorithms
\cite{Adler:1981sn,Adler:1988ce}. The idea can also be generalized to
gauge theories \cite{Brown:1987rr,Creutz:1987xi}. An overrelaxation
step performs a reflection in the space of field elements, which keeps
the action invariant. When applying the local Metropolis decision,
Eq.~(\ref{eq:metro-example}), the change is thus always accepted.
Since the action does not change, the algorithm is non-ergodic and
generates the microcanonical ensemble,
Eq.~(\ref{eq:micro-can-ensemble}); it does, however, satisfy detailed
balance, Eq.~(\ref{eq:detailed-balance}). Consequently, it cannot be
used as the only updating scheme, but it can increase the motion of
the system in phase space if mixed with an ergodic algorithm. In this
way, the expected improvement may result in a dynamical critical
scaling exponent of about $z\simeq 1$, cf.~\cite{Montvay:1994cy}.

For a multiquadratic action of the form (\ref{eq:staples3}), a local
overrelaxation step may simply be implemented by choosing the new
field $\phi'(y)$ to be
\begin{equation}
  \label{eq:overrelax-scalar}
  \phi'(y) = -\phi(y)-\frac{1}{2}\tilde{a}_1^{-1} \sum_{i=2}^M a_i
  \left[\phi\left(f_i^2(y)\right) \cdots
  \phi\left(f_i^{n_i}(y)\right) \right]\,.
\end{equation}
For the Wilson action of the SU$(2)$ gauge theory, the overrelaxation
step can be performed by choosing the new element $U'_\mu(y)$ as
\begin{equation}
  \label{eq:wilson-overrelax}
  U'_\mu(y) = \tilde{\mathscr{S}}^\dagger_\mu(y)U^\dagger_\mu(y)
  \tilde{\mathscr{S}}^{-1}_\mu(y)\,,
\end{equation}
where $\tilde{\mathscr{S}}^{-1}_\mu(y)$ is given by
$\tilde{\mathscr{S}}^{-1}_\mu(y)= \tilde{\mathscr{S}}^\dagger_\mu(y)/
\det\tilde{\mathscr{S}}_\mu(y)$. This replacement leaves the action
invariant since (note that no summation over the index $\mu$ must take
place!): \[ \mbox{Re}\,\mbox{Tr}\,\left(
  U'_\mu(y)\tilde{\mathscr{S}}_\mu(y) \right) = \mbox{Re}\,\mbox{Tr}\,
\left( U_\mu(y)\tilde{\mathscr{S}}_\mu(y) \right)\,. \]
Eq.~(\ref{eq:wilson-overrelax}) is equivalent to the following
transformation:
\begin{equation}
  \label{eq:wilson-transform-or}
  U'_\mu(y) = U_0 U^{-1} U_0, \quad U_0 =
  \tilde{\mathscr{S}}^{-1}_\mu(y)
  \sqrt{\det\tilde{\mathscr{S}}_\mu(y)} =
  \tilde{\mathscr{S}}^\dagger_\mu(y)/
  \sqrt{\det\tilde{\mathscr{S}}_\mu(y)}\,.
\end{equation}
It is possible to generalize (\ref{eq:wilson-transform-or}) to the
case of SU$(N)$, $N>2$, with the same Cabibbo-Marinari decomposition
as discussed in Sec.~\ref{sec:heatbath-algorithm}.

\section{Fermionic Sampling Algorithms}
\label{sec:ferm-sampl-algor}
The algorithms discussed in the previous sections have for a long time
only been applicable to the case of theories without dynamical
fermions, i.e.~the quenched approximation. The typical cost one has to
pay if one includes dynamical fermion contributions is a factor of
about $100-1000$. It was not before the mid-90's when sufficient
computer power became available to treat also dynamical fermions
numerically. One further problem is that in a Yang-Mills theory
including dynamical fermion contributions,
Eq.~(\ref{eq:yang-mills-lattice}), the fermion determinant is a
non-local object. Therefore global algorithms like the HMC had to be
employed. A possible way to rewrite (\ref{eq:yang-mills-lattice}) to
obtain a purely local action has been put forward by
\textsc{L{\"u}scher} \cite{Luscher:1994xx}. This was the key to also
use local sampling algorithms for systems with dynamical fermions.

\subsection{Sampling with the Wilson Matrix}
\label{sec:sampling-with-wilson}
\index{Wilson matrix} The essential problem of lattice fermions is the
evaluation of the determinant from Eq.~(\ref{eq:state-sum-fermions}).
This can be achieved by using a Gaussian integral over boson fields
$\Phi^\dagger(x), \Phi(x)$ \cite{Weingarten:1981hx}
\begin{equation}
  \label{eq:boson-determinant}
  \int \EnsembleMeas{\Phi^\dagger}\EnsembleMeas{\Phi}
  \exp\left[-\sum_{xy}\Phi^\dagger(y) Q(y,x)\Phi(x)\right] \propto
  \frac{1}{\det {Q}}\,,
\end{equation}
where the field $\Phi(x)$ has the same indices as the Grassmann field
$\Psi(x)$. The prefactor from the integration in
Eq.~(\ref{eq:boson-determinant}) is a constant which cancels in any
observable and will hence be dropped from now on. The determinant can
be evaluated using a stochastic sampling process similar to the
measurement of observables using (\ref{eq:rand-sample-av}) for the
evaluation of (\ref{eq:operator-def}). Thus, the fermionic
contributions can also be written as a part of a measure. The
prefactor is a constant and will cancel for any observable. Therefore,
it will be disregarded in the following.

However, there are some problems with the application of
(\ref{eq:boson-determinant}) to the Wilson matrix,
Eq.~(\ref{eq:wilson-matrix}). The former is only defined for a
Hermitian and positive-definite matrix, a condition clearly not
fulfilled by the Wilson matrix. Nonetheless, the product
$Q^\dagger\cdot Q$ is Hermitian and positive definite, so
(\ref{eq:boson-determinant}) is applicable. This expression
corresponds to two dynamical, degenerate fermionic flavors.

A second problem regards the fact that the Wilson matrix will have
eigenvalues close to zero when describing sufficiently light fermions
(cf.~Sec.~\ref{sec:fermion-fields}). Since
(\ref{eq:boson-determinant}) computes the inverse determinant, a
single noisy estimate may oscillate over several magnitudes and in
sign, see e.g.~\cite{Montvay:1987wh}. Therefore, one instead tries to
compute the determinant instead of its inverse.

The above arguments result in the expression
\begin{equation}
  \label{eq:total-wilson-sampling}
  \int \EnsembleMeas{\Phi^\dagger}\EnsembleMeas{\Phi}
  \exp\left[-\sum_{xy}\Phi^\dagger(y) (Q^\dagger
    Q)^{-1}(y,x)\Phi(x)\right] = \det {Q}^2\,.
\end{equation}
When approximating the determinant in (\ref{eq:total-wilson-sampling})
with a finite sample of configurations, $\lbrace\Phi_i(x)\rbrace$, one
can use the global heatbath applied to the scalar boson fields
$\Phi(x)$ as discussed in Sec.~\ref{sec:heatb-scal-fields}. This
requires a matrix inversion. Algorithms to perform this inversion will
be discussed in Sec.~\ref{sec:matr-invers-algor}.

\subsection{Hybrid Monte-Carlo Algorithm}
\label{sec:hybrid-monte-carlo}
The idea behind the \textit{molecular dynamics}-based algorithms is
different from those discussed in the previous sections. The key
feature consists of using quantities obtained from averages of the
microcanonical ensemble (\ref{eq:micro-can-ensemble}) as an
approximation to the average as given in Eq.~(\ref{eq:operator-def})
obtained from the canonical ensemble. This identification works in the
thermodynamic limit, i.e.~in the case of large lattices. The first
time such an algorithm was used in the context of pure gauge field
theory was in \cite{Callaway:1982eb}.  This class of algorithms turned
out to be applicable to the case of dynamical fermions and became the
standard method for this type of systems. Closely related to this line
of thinking is the idea of stochastic quantization
\cite{Damgaard:1987rr}.

In order to simulate the pure gauge action
(\ref{eq:wilson-gauge-action}) using some classical Hamiltonian
formalism, consider the partition function (\ref{eq:gen-func-euclid})
for the random ensemble (\ref{eq:random-ensemble}), applied to
quenched action, \[ Z = \int\EnsembleMeas{U} \exp\left[-S_{\mbox{\tiny
      g}}[U(x)]\right]\,. \] Inserting a unit Gaussian integration
with a field $P_\mu(x)$ carrying the same indices as $U_\mu(x)$ into
the partition function introduces an overall constant which does not
change observables,
\begin{equation}
  \label{eq:hmc-partsum}
  Z' = \int\EnsembleMeas{U} \EnsembleMeas{P} \exp\left[
    -H[U,P]\right]\,,
\end{equation}
where $H[U,P]$ is given by
\begin{equation}
  \label{eq:hmc-hamilton-def}
  H[U,P] = \frac{1}{2}\sum_x \mbox{Tr}\, P^2_\mu(x) + S_{\mbox{\tiny
      g}}[U(x)]\,.
\end{equation}
The phase space has been enlarged by the introduction of the new
fields. Now the idea of the molecular dynamics methods is to simulate
a classical system, interpreting the function $H[U,P]$ in
Eq.~(\ref{eq:hmc-hamilton-def}) as the corresponding Hamiltonian and
thus the new fields $P_\mu(x)$ as the canonical conjugate momenta of
$U_\mu(x)$. This method, however, will only simulate the
microcanonical ensemble with the fixed ``energy'' $H[U,P]$. Since the
microcanonical ensemble can be used as an approximation to the
canonical ensemble as one approaches the thermodynamic limit, one can
take the samples from sufficiently long classical trajectories for
very large lattices to compute observables.

Since the canonical momenta appearing in (\ref{eq:hmc-hamilton-def})
have a Gaussian distribution independent of the fields, one can extend
the algorithm by not only considering a single classical trajectory,
but several of them, all starting with Gaussian distributed initial
momenta. This would clearly solve the problem of lacking ergodicity of
the purely microcanonical approach. If the momenta are refreshed
regularly during the molecular dynamics evolution, one arrives at the
\textit{Langevin} algorithms \cite{Parisi:1981ys}. The extreme case is
to refresh the momenta at each step which would imply that one
performs a random walk in phase space. The other extreme is the purely
molecular dynamics evolution which moves fastest without ever changing
direction by refreshing the momenta, but lacking ergodicity. A
combination of both approaches are the \textit{hybrid classical
  Langevin} algorithms \cite{Duane:1985ym}, where at each step a
random decision takes place whether to reshuffle the momenta or not.

The culmination point of the molecular dynamics algorithms is the
hybrid Monte-Carlo algorithm (see for the foundations
\cite{Scalettar:1986uy,Duane:1987de}, for a more detailed discussion
\cite{Creutz:1988wv} and for recent reviews
\cite{Kennedy:2000dm,Lippert:2001ha}).  The idea is again to simulate
the classical equations of motion along a trajectory of a certain
length; this is easily achieved by integrating the canonical equations
of motion, Eq.~(\ref{eq:canonical-motion}),
\begin{eqnarray}
  \label{eq:hmc-eq-of-motion}
  \dot{U}_\mu &=& \frac{\partial H[U,P]}{\partial P_\mu}, \nonumber \\
  \dot{P}_\mu &=& -\frac{\partial H[U,P]}{\partial U_\mu}\,.
\end{eqnarray}
The integration of the equations of motion can be done with various
algorithms available for molecular dynamics. Of particular interest
are the symplectic integration schemes, see
e.g.~\cite{McLachlan:1992ar,Yoshida:1993ar} for an introduction. A
scheme which is of second order and which requires only a single force
evaluation per step is the \textit{leap-frog} integration scheme. The
integration of the equations of motion proceeds with a finite step
length, $\Delta t$.  The leap-frog method has a systematic error or
the order of ${\cal O}(\Delta t^2)$, so the actual trajectory in the
simulation may differ from the exact solution of
(\ref{eq:hmc-eq-of-motion}). This deviation can be corrected for by a
global Metropolis step similar to Eq.~(\ref{eq:global-metropolis}),
but with the action $S[U]$ replaced by the ``Hamiltonian'' $H[U,P]$.
The integration of (\ref{eq:hmc-eq-of-motion}) is done for a certain
number of steps, $n_{\mbox{\tiny MD}}$, which is thus the length of an
HMC trajectory. After the Metropolis decision has taken place, a new
set of Gaussian random ``momenta'' is shuffled and the whole
integration is started again.

A crucial point for the application of the molecular dynamics
evolution is the reversibility of the trajectory, i.e.~replacing
$\Delta t$ by $-\Delta t$ should return to the system to exactly the
same point in parameter space, where it has started. This condition is
necessary for detailed balance, Eq.~(\ref{eq:detailed-balance}) to
hold.

The generalization to fermionic field theory was suggested in
\cite{Polonyi:1983tm}. It proceeds by considering the partition
function
\begin{equation}
  \label{eq:hmc-fermion-state-sum}
  Z'=\int\EnsembleMeas{U}\EnsembleMeas{\phi^\dagger}
  \EnsembleMeas{\phi}\EnsembleMeas{P}
  \EnsembleMeas{\pi^\dagger}\EnsembleMeas{\pi}
  \exp\left[-H[U,\phi^\dagger,\phi,P,\pi^\dagger,\pi]\right]\,,
\end{equation}
where the ``Hamiltonian'' $H[U,\phi^\dagger,\phi,P,\pi^\dagger,\pi]$
is now given by
\begin{eqnarray}
  \label{eq:hmc-fermion-hamiltonian}
  H[U,\phi^\dagger,\phi,P,\pi^\dagger,\pi] &=&
  \sum_x\biggl(\frac{1}{2}\mbox{Tr}\, P_\mu^2(x) +
  \mbox{Tr}\,\pi^\dagger(x)\pi(x) \nonumber \\
  && \quad + S_{\mbox{\tiny g}}[U] + \sum_y
  \phi(x) \left(Q^\dagger Q\right)^{-1}(x,y)\phi(y)
  \biggr)\,.
\end{eqnarray}
The explicit form of the resulting equations of motion can be found
e.g.~in \cite{Lippert:1993ph,Lippert:2001ha}.

By adjusting the step length, $\Delta t$, and the trajectory length,
$n_{\mbox{\tiny MD}}$ between two Metropolis decisions, one can tune
the acceptance rate and optimize the algorithm to achieve best
performance. In general, the larger the total trajectory length,
$\Delta t\cdot n_{\mbox{\tiny MD}}$, the lower the acceptance rate,
since a longer trajectory introduces larger numerical errors. This can
be compensated by making $\Delta t$ smaller (and consequently
$n_{\mbox{\tiny MD}}$ larger), but this will increase the required
computer time per trajectory by the same factor.

The suggestion by \textsc{Creutz} \cite{Creutz:1987id,Creutz:1989wt}
was to choose $\Delta t\cdot n_{\mbox{\tiny MD}}\simeq {\cal O}(1)$
and modify the two parameters such that the acceptance rate is about
$P_{\mbox{\tiny acc}}> 70\%$. This proposal has been tested
numerically in \cite{Gupta:1990ka}. A different investigation has been
performed in the case of compact QED by \textsc{Arnold} in
\cite{Arnold:1998mt}. Of particular interest is also the impact of
$32$-bit precision on the feasibility of the algorithm. As has been
shown in \cite{Conti:1997ba,Ritzenhofer:1997ph}, the systematic error
(i.e.~the non-reversibility of the HMC trajectory) in case of QCD with
two dynamical fermion flavors is of the order of $2\%$ on a
$\Omega=40\times 24^3$ lattice. This error should in any case be small
compared to the statistical error of the quantities under
consideration.

In conclusion, the advantages of the HMC are that it has only two
parameters namely $\Delta t$ and $n_{\mbox{\tiny MD}}$, that its
optimization and tuning is well understood and under control, and that
it is rather simple to implement even for more complicated systems. In
direct comparison to the local algorithms, it is, however, less
efficient. This is related to the fact that a local sweep changes each
variable by a greater amount than a global sweep, while it may still
have a similar computational cost. In the quenched case of QCD it soon
became clear that algorithms like the HMC are not competitive with
heatbath algorithms, in particular if they are used together with
overrelaxation techniques, see for a recent algorithmic review
e.g.~\cite{Peardon:2001pt}.

There exist extensions of molecular dynamics-based algorithms which
allow to handle also odd numbers of dynamical quark flavors. One
method is the $R$-algorithm, which has a residual systematic error
which has to be kept smaller than the statistical error of observables
\cite{Gottlieb:1987mq}. A method which is free of systematic errors
has been proposed by \textsc{Lippert} in \cite{Lippert:1999up}, but
its efficiency may be rather limited due to the presence of a nested
iteration\footnote{It is possible, that the quadratically optimized
  polynomials discussed in Sec.~\ref{sec:stat-polyn-invers} are able
  to handle this iteration in an efficient way}. A different approach
has been suggested in \cite{deForcrand:1997ck} and exploited in
\cite{Takaishi:2000rk,Takaishi:2000bt}.

One potential problem of the HMC scheme is related to the question of
ergodicity. Although the algorithm is exact and ergodic in the
asymptotic limit, for finite time series it may get ``stuck'' in
certain topological sectors. In particular, in a study of dense
adjoint matter, it has been shown in \cite{Hands:2000ei} that the HMC
method is not ergodic, while the MB algorithm retains ergodicity.
This question has also been raised by \textsc{Frezzotti} and
\textsc{Jansen} \cite{Frezzotti:1998sa} who introduced a variant of
the HMC algorithm \cite{Frezzotti:1998eu,Frezzotti:1998yp}, using a
static polynomial inversion similar to those discussed in
Sec.~\ref{sec:stat-polyn-invers} (see also \cite{deForcrand:1997ck}).
For a recent comparison of efficiencies of current algorithms see
\cite{Frezzotti:2000rp}.

\subsection{Multiboson Algorithms}
\label{sec:mult-algor}
As discussed in the previous subsection, the standard HMC allows to
simulate the situation with an even number of degenerate, dynamical
fermion flavors, at the expense of having a global algorithm.
Furthermore, since ergodicity is only ensured in an asymptotical
sense, one may ask whether it is possible to use a different approach
for the same problem. As has been shown by \textsc{L{\"u}scher}
\cite{Luscher:1994xx}, it is possible to rewrite the action
(\ref{eq:yang-mills-lattice}) in such a way that a purely local action
is obtained which can be treated by more efficient algorithms like
local heatbath and overrelaxation. The algorithms based on this idea
are called \textit{multiboson algorithms} (MB). For an overview of
recent investigations consult \cite{deForcrand:1999nw}. For
theoretical estimates of efficiency especially compared to the HMC
consider \cite{Borici:1996tk}.

Consider a similarity transformation
Eq.~(\ref{eq:wilson-diagonalize}), but applied to the non-Hermitian
Wilson matrix $\tilde{Q}(y,x)$. The resulting diagonal matrix,
$\mathbf{\tilde{Q}}(y,x)$ will have all eigenvalues of
$\tilde{Q}(y,x)$, $\lambda_i$, on its diagonal. Then consider a
polynomial of order $n$,
\begin{equation}
  \label{eq:polynomial}
  P_n(x) = c_n\prod_{j=1}^{n}(x-z_j)\,,
\end{equation}
which approximates the function $1/x$ over the whole spectrum of
$\tilde{Q}^2(y,x)$ with a certain accuracy, $\epsilon$. Applying this
polynomial to the matrix $\tilde{Q}^2(y,x)$ will yield an
approximation to $\tilde{Q}^{-2}(y,x)$ as can be seen by applying
(\ref{eq:polynomial}) to the diagonal matrix from
Eq.~(\ref{eq:wilson-diagonalize}), since the resulting matrix will
have the inverse eigenvalues, $1/\lambda_i$, on its diagonal. This
allows the fermionic action to be rewritten:
\begin{equation}
  \label{eq:eferm-wilson}
  S_{\mbox{\tiny f}} = \sum_j\sum_{xyz} \phi_j^\dagger(y)
  \left(\tilde{Q}(y,z)-\rho_j^*\right)
  \left(\tilde{Q}(z,x)-\rho_j\right) \phi_j(x)\,,
\end{equation}
where the $\rho_j$ are the roots of the $z_j$. The determinant is then
computed via \[ \det Q^2 \approx \frac{1}{\det P(\tilde{Q}^2)} =
\frac{1}{c_n}\int\EnsembleMeas{\phi_j^\dagger} \EnsembleMeas{\phi_j}
\exp\left[ -\phi_j^\dagger(\tilde{Q}-\rho_j^*)(\tilde{Q}
  -\rho_j)\phi_j\right]\,. \] This action has the form of
Eq.~(\ref{eq:multiquadratic}) and thus can be treated by local
heatbath and overrelaxation techniques, as they are discussed in
Sec.~\ref{sec:boson-sampl-algor}. The system now incorporates the
gauge fields $\lbrace U_\mu(x)\rbrace$ as before, but in addition also
$4N\times n$ scalar fields $\lbrace\phi_j(x)\rbrace$ since the
polynomial has $n$ roots and each field has the same indices as a
Dirac spinor times the Yang-Mills group number $N$. In the following
these fields will be referred to as ``boson fields''. Hence, it is
apparent that the system of (\ref{eq:eferm-wilson}) has both a huge
memory consumption and may have a relatively complicated phase space.
In any case, one will have to deal with $n$ additional fields and the
computational effort will still be enormous.

The central question now regards the optimal choice of the polynomial.
Clearly, its order $n$ should be kept as small as possible while still
maintaining a sufficiently good approximation. In any case, the
polynomial approximation in (\ref{eq:polynomial}) is a static
inversion (cf.~Sec.~\ref{sec:matr-invers-algor}). This means that once
the choice has been fixed, one cannot alter the polynomial during the
sampling process anymore. For an overview of the choices available,
see Sec.~\ref{sec:stat-polyn-invers}.

\subsubsection{Even-Odd Preconditioning for MB Algorithms}
\label{sec:even-odd-prec-mb}
\index{Even-odd preconditioning} It is possible to incorporate the
preconditioning technique introduced in Sec.~\ref{sec:even-odd-prec}
to the multiboson approximation (\ref{eq:eferm-wilson}). However,
since the matrix in (\ref{eq:wilson-eo-precond}) contains
next-to-nearest neighbor interactions, the square in
(\ref{eq:eferm-wilson}) would introduce an even more complicated
action which may have up to fourth-neighbor terms and thence would be
almost impossible to implement:
\begin{eqnarray}
  \label{eq:locmb-det-eo}
  \det\hat{Q}^2 &\approx& \displaystyle\nonumber \left(\det
    P\left(\hat{Q}^2\right) \right)^{-1} \\
  &=& \displaystyle \prod_j \left( \det\left( \hat{Q} - \rho_j^*
    \right) \left( \hat{Q}-\rho_j \right) \right)^{-1}\,.
\end{eqnarray}
This problem has been solved in \cite{Jegerlehner:1997px} by applying
the Schur decomposition from Eq.~(\ref{eq:schur-decomp}) again to the
preconditioned action:
\[ \det\left( \hat{Q} - \rho_j \right) \propto \det\left(
  \begin{array}{cc} \gamma_5 & - \gamma_5\kappa D_{eo} \\
    -\gamma_5\kappa D_{oe} & \gamma_5-\rho_j \end{array} \right) =
\det\left(\tilde{Q} - P_o\rho_j\right)\,, \]
where $P_o$ denotes the projector on ``odd'' sites ($P_o =
\mbox{diag}\left(0,\dots,0,1,\dots,1\right)$, which contains $0$ on the
first half diagonal and $1$ on the second half). The resulting
preconditioned action is then given by
\begin{equation}
  \label{eq:eferm-wilsoneo}
  S_{\mbox{\tiny f}} = \sum_j\sum_{xyz} \phi_j^\dagger(y)
  \left(\tilde{Q}(y,z)-P_o\rho_j^*\right)
  \left(\tilde{Q}(z,x)-P_o\rho_j\right) \phi_j(x)\,.
\end{equation}
This is the action which will be considered from here on.

\subsubsection{Exact Multiboson Algorithms}
\label{sec:exact-mult-algor}
\index{Correction step} The multiboson algorithm as discussed so far
only uses an approximate polynomial with a residual error $\epsilon$.
One could decide to stay with this error and try to minimize it by
increasing the order of the polynomial $n$. But this would indeed be a
bad idea since the computer time and memory requirement would become
enormous. Thus, different proposals have been made to get rid of the
residual error.  The original proposal \cite{Luscher:1994xx} was to
generate a sample of configurations using the action
(\ref{eq:eferm-wilson}) as an approximation to the ``real'' action in
the sense of (\ref{eq:multi-can-ensemble}). Then one performs a
\index{Reweighting} \index{Ensembles!multi-canonical} reweighting of
the observables using (\ref{eq:multi-can-operator}).  This procedure
is free of systematic errors but it may introduce additional noise in
the measurement of observables if the initial approximation of
(\ref{eq:polynomial}) is bad. Therefore, this approach has been
abandoned in practical simulations.

The method which is used in current simulations is to apply a
Metropolis step (\ref{eq:global-metropolis}) after a set of local
sweeps
\cite{Alexandrou:1995nd,Peardon:1995ji,Borici:1996bk,Borrelli:1996re}.
In this way the algorithm is free of any systematic error provided the
correction factor is computed with sufficient accuracy. The exact
acceptance probability is given by
\begin{equation}
  \label{eq:mb-correction-probability}
  P_{\mbox{\tiny acc}} = \min\left(1,\frac{\det{(\tilde{Q}^2[U']
  P_n(\tilde{Q}^2[U']))}} {\det{(\tilde{Q}^2[U]
  P_n(\tilde{Q}^2[U]))}}\right)\,,
\end{equation}
with $U'$ being the gauge field configuration \textit{after} the local
update sweeps and $U$ being the gauge field configuration
\textit{prior} to the sweeps.

Still the problem remains to actually compute the ratio of the
determinants in (\ref{eq:mb-correction-probability}). The
straightforward evaluation with a noisy estimate vector $\eta$ using a
global heatbath as discussed in Sec.~\ref{sec:heatbath-algorithm}
will result in a nested iteration of an inversion algorithm and the
polynomial $P_n(\tilde{Q}^2)$. In this sense, the polynomial will act
as a preconditioner.

Another approach has been suggested in \cite{Alexandrou:1995nd}: One
can obtain an estimate to the determinants by computing the low-lying
eigenvalues for which the chosen polynomial was only a bad
approximation. This allows to compute the correction factor directly.
For the smallest $L'$ eigenvalues $\lbrace\lambda_i\rbrace$,
$i=1,\dots,L'$, this yields
\begin{equation}
  \label{eq:mb-correction-approx}
  \det{(\tilde{Q}^2P_n(\tilde{Q}^2))} \approx
  \prod_{i=1}^{L'}\lambda_i P_n(\lambda_i)\,.
\end{equation}
This approximation is reasonable if the approximation $P_n(x)$ is
inaccurate only for small $x$. Nonetheless there is no way to limit
the systematic error if one doesn't want to determine $L'$
dynamically. Furthermore, this approach can be expected to scale badly
with the volume since the eigenvalue density is proportional to volume
$\Omega$ and the total effort will at best scale as $\Omega^2$.

For a discussion of the effect of the polynomial quality on the
acceptance factor, see \cite{Borrelli:1996re}.

Another suggestion lies at at the basis of the \textit{Two-Step
  Multiboson} (TSMB) algorithm proposed by \textsc{Montvay} in
\cite{Montvay:1996ea}. This is discussed below.

\subsubsection{Non-Hermitian Variant}
\label{sec:non-herm-vari}
One can also use the non-Hermitian Wilson matrix, $Q(y,x)$, instead of
$\tilde{Q}(y,x)$ for the construction of the polynomial approximation.
In this case, the action (\ref{eq:eferm-wilson}) takes on the
following form:
\begin{equation}
  \label{eq:eferm-wilson-nonherm}
  S_{\mbox{\tiny f}} = \sum_j\sum_{xyz} \phi_j^\dagger(y)
  \left(Q^{\dagger}(y,z)-\rho_j^*\right)
  \left(Q(z,x)-\rho_j\right) \phi_j(x)\,.
\end{equation}
This suggestion has first been put forward by \textsc{Bori\c{c}i} and
\textsc{de Forcrand} in \cite{Borici:1995am}. It is directly
applicable to the case of an even number of mass-degenerate fermion
flavors, just like the HMC\@. However, the approximation
(\ref{eq:polynomial}) fails once a real eigenvalue gets negative. This
problem is avoided as long as the fermion masses are still large. It
is unclear, however, what will happen if the masses get small enough,
so that fluctuations may eventually cause the smallest real eigenvalue
to cross the imaginary axis.

Since the effort of inverting the non-Hermitian matrix is lower than
in the Hermitian case, the algorithm is in principle more efficient,
whenever the aforementioned problem is avoided.

It is important to realize that also an algorithm based on the
expansion (\ref{eq:eferm-wilson-nonherm}) will be ``exact'' even if a
real eigenvalue gets negative, whenever it uses a correction step as
discussed above. The correction step will correct \textit{any} errors
in the polynomial approximation. However, the algorithm may become
inefficient since the acceptance rate would drop almost to zero once a
point in phase space is reached where the approximation becomes
invalid.

\subsubsection{TSMB Variant}
\label{sec:tsmb-variant}
An extension of multiboson algorithms which allows to handle
situations with an arbitrary number of fermion flavors has been
suggested by \textsc{Montvay} in \cite{Montvay:1996ea}. In particular,
supersymmetric Yang-Mills theory on the lattice has been examined (see
for early reviews \cite{Montvay:1998ak,Montvay:1997uq}). For the
physical results consult
\cite{Kirchner:1998nk,Koutsoumbas:1998de,Kirchner:1998mp,
  Feo:1999hw,Feo:1999hx,Farchioni:2000mp}.

This approach can immediately be generalized to the case of an
arbitrary number of dynamical fermions, in particular the physically
interesting case (cf.~Sec.~\ref{sec:lattice-qcd}) with three dynamical
quark flavors \cite{Montvay:1999gn}; this is done by choosing a
polynomial $P_{n_1}(x)$ (the reason for calling the polynomial order
$n_1$ instead of $n$ will become clear soon) which approximates
$x^{-\alpha}$, where $\alpha\neq 1$ is allowed. For $\alpha>1$ one
generally requires larger order $n$ to achieve the same accuracy while
for $\alpha<1$ one gets along with smaller $n$. The value of $\alpha$
determines the number of dynamical fermion flavors via $\alpha=N_f/2$
since the polynomial is still applied to $\tilde{Q}^2$. Thence, for
gluinos one has to choose $\alpha=1/4$ leading to $N_f=1/2$
\cite{Montvay:1996ea}. The case of three dynamical fermion flavors, as
discussed in Chapter~\ref{sec:expl-invest-param}, requires the choice
$\alpha=3/2$.

The central idea regards the computation of the correction factor
(\ref{eq:mb-correction-probability}). The generalized correction
factor for $\alpha\neq 1$ takes the form:
\begin{equation}
  \label{eq:tsmb-corr-prob}
  P_{\mbox{\tiny acc}} =
  \min\left(1,\frac{\det{(\tilde{Q}^{2\alpha}[U']
        P_n(\tilde{Q}^{2}[U']))}} {\det{(\tilde{Q}^{2\alpha}[U]
        P_n(\tilde{Q}^{2}[U]))}}\right)\,.
\end{equation}
The evaluation with a noisy estimate is highly difficult since now a
(possibly non-integer) power of the matrix $\tilde{Q}$ will have to be
inverted. The idea of Ref.~\cite{Montvay:1996ea} was to employ the
multicanonical sampling (cf.~Sec.~\ref{sec:ensembles}) to get an
approximate action
\begin{equation}
  \label{eq:tsmb-multican}
  \tilde{S}[U] = S_{\mbox{\tiny g}}[U] + \ln\frac{1}{\det
    P_{n_1}(\tilde{Q}^2) \det \tilde{P}_{n_2}(\tilde{Q}^2)}\,,
\end{equation}
where the polynomial $\tilde{P}_{n_2}(x)$ satisfies \[
\det\tilde{Q}^{2\alpha} \approx \frac{1}{\det P_{n_1}(\tilde{Q}^2)
  \det \tilde{P}_{n_2}(\tilde{Q}^2)}\,. \] This can be achieved by
replacing the TSMB noisy correction step (\ref{eq:tsmb-corr-prob}) by
\begin{equation}
  \label{eq:tsmb-approx-noisy}
  P_{\mbox{\tiny acc}} = \min\left(1, \frac{\det
      \tilde{P}_{n_2}(\tilde{Q}^2[U])} {\det
      \tilde{P}_{n_2}(\tilde{Q}^2[U'])}
  \right)\,.
\end{equation}
In order to compute this ratio using a noisy correction vector, one
uses the sampling prescription as discussed in
Sec.~\ref{sec:sampling-with-wilson}. This requires the application of
a global heatbath, as aforementioned in
Sec.~\ref{sec:heatb-scal-fields}, which is very expensive since it
would again require a nested inversion for the polynomial
$\tilde{P}_{n_2}(\cdot)$.  Therefore, the suggestion of
\cite{Montvay:1996ea} was to use a third polynomial $\hat{P}_{n_3}(x)$
with order $n_3$ which approximates the inverse square root of
$\tilde{P}_{n_2}(x)$,
\begin{equation}
  \label{eq:poly-invsqr-def}
  \hat{P}_{n_3}(x) \approx \left( \tilde{P}_{n_2}(x) \right)^{-1/2}\,.
\end{equation}
When applied to a matrix, one obtains
\begin{equation}
  \label{eq:invsqr-matrix}
  \hat{P}_{n_3}(\tilde{Q}[U]^2) \approx
  \left( \tilde{P}_{n_2}(\tilde{Q}[U]^2) \right)^{-1/2}\,.
\end{equation}
The reason for this procedure becomes clear, if one evaluates
(\ref{eq:tsmb-approx-noisy}) using a noisy estimate. In practice, a
single noisy vector is usually sufficient \cite{Montvay:1996ea}.  Then
the acceptance probability becomes
\begin{equation}
  \label{eq:noisy-acc-final}
  P_{\mbox{\tiny acc}} = \min\left( 1, \exp\left[
      -\eta^\dagger\left(\hat{P}_{n_3}(\tilde{Q}[U']^2)
        \tilde{P}_{n_2}(\tilde{Q}[U]^2)
        \hat{P}_{n_3}(\tilde{Q}[U']^2)-\mathbf{1}\right)\eta \right] 
        \right)\,,
\end{equation}
with $\eta(x)$ being a random Gaussian vector with unit width.

The approximation of $\hat{P}_{n_3}(\cdot)$ in
(\ref{eq:poly-invsqr-def}) determines the total residual error of the
algorithm. There is no way to correct for this error after the
sampling has taken place since the error appears in the correction
step and cannot be rewritten as an extra term in the action.  It is of
vital importance to keep this influence small. A precise investigation
of the effects associated with this residual error can be found in
Sec.~\ref{sec:tuning-quadr-optim}.

After $n_3$ has been chosen sufficiently large, the total systematic
error is governed by the second polynomial,
$\tilde{P}_{n_2}(\cdot)$. This systematic error, however, is present
in the action (\ref{eq:tsmb-multican}) and can therefore be corrected
by the measurement correction, Eq.~(\ref{eq:multi-can-operator}). As
shown in \cite{Campos:1999du}, this can be done by considering yet a
further polynomial, $\tilde{P}_{n_4}(x)$, defined by
\begin{equation}
  \label{eq:poly-rew-def}
  P_{n_1}(x) \tilde{P}_{n_2}(x) \tilde{P}_{n_4}(x) \approx
  x^{-\alpha}\,.
\end{equation}
The calculation of the expectation value of an operator
$\langle\hat{A}\rangle$ then proceeds by applying
Eq.~(\ref{eq:multi-can-operator}):
\begin{equation}
  \label{eq:poly4-rew}
  \langle\hat{A}\rangle = Z^{-1}
  \int\EnsembleMeas{\eta}\EnsembleMeas{U}\; \hat{A}[U] \exp\left[
    \eta^\dagger(1-\tilde{P}_{n_4}(\tilde{Q}[U]^2))\eta\right]\,,
\end{equation}
with \[ Z = \int\EnsembleMeas{U}\EnsembleMeas{\eta}\; \exp\left[
  \eta^\dagger(1-\tilde{P}_{n_4}(\tilde{Q}[U]^2))\eta\right]\,. \] The
interval of the polynomial approximation for $\tilde{P}_{n_4}(\cdot)$
must be sufficiently large to cover the entire eigenvalue spectrum of
$\tilde{Q}^2[U]$ for all gauge fields in the sample. Since this may be
problematic if exceptional configurations with extremely small
eigenvalues are present, one can combine the noisy estimation of the
correction factor in (\ref{eq:poly4-rew}) with an exact computation of
the corresponding factor for the smallest eigenvalues (see
\cite{Campos:1999du}):
\begin{equation}
  \label{eq:multican-ev-rew}
  \langle\hat{A}\rangle = Z^{-1}\int\EnsembleMeas{U}\; \hat{A}[U]
  \prod_j \left(\lambda_j\right)^\alpha
  P_{n_1}(\lambda_j) \tilde{P}_{n_2}(\lambda_j)\,,
\end{equation}
with $\lambda_j$ being the $j$th eigenvalue of the matrix
$\tilde{Q}[U]^2$.

This procedure can also act as a preconditioner to the computation of
$\tilde{P}_{n_4}(\tilde{Q}^2)$. The accuracy of
$\tilde{P}_{n_4}(\cdot)$ can be adjusted until the correction factor
has converged. However, as will be shown in Sec.~\ref{sec:rewe-corr},
it is in general not necessary to compute \textit{both} the smallest
eigenvalues \textit{and} the correction factor using the sampling in
(\ref{eq:poly4-rew}). Since the eigenvalue approximation of the
quadratically optimized polynomials converges extremely fast, it is
sufficient to approximate the correction factor using the smallest
eigenvalues only.


\section{Matrix Inversion Algorithms}
\label{sec:matr-invers-algor}
For the computation of the polynomials and the global heatbath in the
previous sections, an inversion of the fermion matrix is required. The
problem is to find a solution vector $\phi(x)$ which solves the
equation
\begin{equation}
  \label{eq:mat-invers}
  \sum_{x}Q(y,x) \phi(x) = \eta(y)\,,
\end{equation}
for a given matrix $Q(y,x)$ and a given vector $\eta(x)$. The
numerical effort of this problem depends cubically on the size of the
matrix \cite{numerical-recipes} and monotonically on the condition
number (see Sec.~\ref{sec:eigenv-algor}). If the inverse condition
number is of the order of or smaller than the machine precision, the
matrix is said to be ``ill-conditioned'', because the algorithms will
in general be unable to yield a stable solution, although the matrix
entries may not pose any direct problem themselves. The aim of
preconditioning techniques is thus to reduce the condition number of
the matrix $Q(y,x)$ without altering the solution. Often techniques
like those discussed in Sec.~\ref{sec:even-odd-prec} also go by the
name ``preconditioner'', although the even-odd preconditioned matrix
is \textit{different} from the original one.

For the case under consideration in the thesis, inversion of the
lattice Dirac matrices (\ref{eq:eferm-wilson}) or
(\ref{eq:eferm-wilsoneo}) is required.  These matrices typically have
sizes of the order $N=\left(12\cdot\Omega\right)$ which for a lattice
of size $\Omega=32\times 16^3$ is $N=1572864$. Storage of the complete
matrix would thus require about $18$ TBytes and is out of reach for
current computer technology. Consequently, for the inversion of
$Q(y,x)$ only an iterative solver may be considered. These solvers do
not require the whole matrix to be stored in memory, but rather
require the presence of a matrix-vector multiplication. This step
typically consumes most of the computer time of the algorithm.

From the repeated application of the matrix-vector multiplication, an
approximation $\phi^l(x)$ of order ${\cal O}\left(Q^{l}\right)$ to the
solution vector $\phi(x)$ is generated. Thus, these algorithms apply a
polynomial $P(\cdot)$ of order $l$ with the matrix $Q(y,x)$ as its
argument to the starting vector $\eta(x)$ yielding the solution
vector:
\begin{eqnarray}
  \label{eq:matinv-poly}
  \phi(y) &\approx& \phi^l(y) = \nonumber\displaystyle \sum_{x}
  P\left(Q\right)(y,x) \eta(x) \\
  &=& \nonumber\displaystyle \sum_{x}\left( p_0 +
    p_1 Q(y,x) + p_2 Q^2(y,x) + \dots + p_l Q^l(y,x)\right)\eta(x) \\
  &=& \displaystyle \sum_{x} \left( p_0 + Q \cdot \left( {p}_1 +
  Q\cdot \left( {p}_2 + Q\cdot\left( \dots + Q\cdot{p}_l\right)
  \right) \right) \right)(y,x) \eta(x)\,,
\end{eqnarray}
where the order of the polynomial is given by $l$ (which is thus the
number of iterations required).

Sometimes the iteration prescription can be cast in the form
\begin{equation}
  \label{eq:matinv-polstation}
  \phi^{k+1}(y) = \sum_x S(y,x) \phi^k(x) + c(y)\,,
\end{equation}
where the matrix $S(y,x)$ and the vector $c(y)$ are independent of the
iteration number $k$. Such methods are called ``stationary''. The
\textit{Jacobi method}, the \textit{Gauss-Seidel method} and the
\textit{(S)SOR methods} are examples of such cases
(cf.~\cite{netlib-templates,numerical-recipes}).

A measure of the quality of the approximation in equation
(\ref{eq:matinv-poly}) is given by the norm of the residual vector
\begin{equation}
  \label{eq:vec-norm}
  \| r^l\| = \frac{\sum_x {r^l}^{\dagger}(x)r^l(x)} {\sum_x
    \eta^{\dagger}(x)\eta(x)}\,, 
\end{equation}
where $r^l(x)$ is defined to be
\begin{equation}
  \label{eq:matinv-residual}
  r^l(y) = \sum_{x} Q(y,x) \phi^l(x) - \eta(y)\,,
\end{equation}
which should converge to zero as $l$ approaches infinity. In some
cases the exact solution is already found after a finite number of
steps. In most practical situations, however, the exact solution
cannot be found due to the limited accuracy of the machines and one is
interested only in finding the solution in as few steps $l$ as
possible up to a certain accuracy $\| r^l\| < \varepsilon$.

The solver determines the coefficients $\lbrace p_0, \dots,
p_l\rbrace$ of the polynomial $P(\cdot)$ in Eq.~(\ref{eq:matinv-poly})
or, in some cases, the recurrence coefficients of a recurrence
relation. The algorithms may be divided into two classes:
\begin{itemize}
\item The coefficients of the polynomial are fixed prior to the
  iteration and do not depend on the shape of the matrix $Q(y,x)$.
  This does not allow to exploit any knowledge gained by the algorithm
  during the iteration process and it does not allow to compensate for
  any rounding errors. Rather, the rounding errors will usually add up
  causing the iteration to saturate at some point where further
  iterations do not increase the accuracy of the solution. This class
  of solvers is called \textit{non-adaptive} and is of great
  importance for multiboson algorithms; generally they are important
  in those cases where an approximate inverse is required with a fixed
  series of coefficients. This is the case e.g.~for
  \index{Reweighting} reweighting purposes.
\item The coefficients are determined dynamically during the iteration
  itself. Thus, the solver may adapt to the specific form of the
  matrix $Q(y,x)$. These algorithms are called \textit{adaptive
    solvers} and are in general superior to the non-adaptive
  algorithms in terms of required matrix-vector operations.
  Furthermore they are able to compensate better for rounding errors
  so the accuracy which may be achieved is higher than for
  non-adaptive ones. The reaction on ill-conditioned matrices is
  consequently improved as well. These algorithms are the method of
  choice if the inverse up to a fixed accuracy is required.
\end{itemize}
For a complete overview of iterative solvers consult
\cite{numerical-recipes,netlib-templates}. The algorithms which have
been employed in this thesis are discussed in the following sections;
all of them are efficiently parallelizable both on \index{MIMD}
\index{SIMD|see{MIMD}} MIMD (Multiple Instruction, Multiple Data) and
on SIMD (Single Instruction, Multiple Data) machines. For an
explanation of the architectures see e.g.~\cite{Fosdick:1996bo}.

\subsection{Static Polynomial Inversion}
\label{sec:stat-polyn-invers}
The choice of the polynomial $P_n(x)$ in Eq.~(\ref{eq:polynomial}) is
crucial for the applicability of multiboson algorithms. The
construction of any polynomial requires one to know at least the
condition number of the Wilson matrix. Usually more information is
available regarding the spectrum, cf.~Sec.~\ref{sec:fermion-fields},
and also the spectral density plots in
Sec.~\ref{sec:optim-polyn-appr}. The original proposal of
\textsc{L{\"u}scher} \cite{Luscher:1994xx} is to use an approximation
build from \textit{Chebyshev polynomials} \cite{numerical-recipes}.
This approximation does not take care of the peculiarities of the
Wilson matrix and thus this choice is not the optimal one. It is,
however, a safe method which is applicable to any fermion
representation if only the condition number is known.

\subsubsection{Quadratically Optimized Polynomials}
\label{sec:quadr-optim-polyn}
The \textit{quadratically optimized polynomials} have been introduced
by \textsc{Montvay} \cite{Montvay:1996ea}. For a thorough discussion
and comparison to the Chebyshev polynomials see \cite{Montvay:1997vh}
and for further technical details
\cite{Montvay:1999ty,Montvay:1999kq}. The basic idea is to find the
polynomial $P_n(x)$ which approximates a function $x^{-\alpha}$ (with
$\alpha=N_f/2$, cf.~Sec.~\ref{sec:mult-algor}) in a given interval
$[\epsilon,\lambda]$ in such a way that the \textit{relative deviation
  norm} $\Delta$ defined via
\begin{equation}
  \label{eq:quadred-pol-def}
  \Delta = \left( (\lambda-\epsilon)^{-1}\int_{\epsilon}^{\lambda} dx
  \left( 1-x^\alpha P_n(x)\right)^2 \right)\,,
\end{equation}
is minimized. If $P_n(x)$ is expanded in coefficient form, \[ P_n(x) =
\sum_{\nu=0}^n c_\nu x^{n-\nu}\,, \] the coefficients $\lbrace
c_\nu\rbrace$ of the polynomial minimizing (\ref{eq:quadred-pol-def})
are given by \cite{Montvay:1997vh}
\begin{equation}
  \label{eq:quadred-min-coeffs}
  c_\nu = \sum_{\nu_1=0}^n M_{\nu\nu_1}^{-1} V_{\nu_1}\,,
\end{equation}
with
\begin{eqnarray*}
  V_\nu &=& \frac{\lambda^{1+\alpha+n-\nu}-\epsilon^{1+\alpha+n-\nu}}
  {(\lambda-\epsilon)(1+\alpha+n-\nu)}\,, \\
  M_{\nu_1\nu_2} &=& \frac{\lambda^{1+2\alpha+2n-\nu_1-\nu_2}
  -\epsilon^{1+2\alpha+2n-\nu_1-\nu_2}}
  {(\lambda-\epsilon)(1+2\alpha+2n-\nu_1-\nu_2)}\,.
\end{eqnarray*}
A straightforward computation of the $P_n(x)$ in terms of the
expansion coefficients (\ref{eq:quadred-min-coeffs}) is not practical,
however. The coefficients will soon become arbitrarily large and the
computation of larger polynomial orders is not feasible anymore, since
then typically orders of $n>100$ are required. Fortunately, the
polynomials can be computed in terms of a recurrence relation which is
stable even for orders of $n\approx 1000$ and beyond, at least if
$64$-bit precision is used.

Take a set of polynomials $\lbrace\Phi_\nu\rbrace$ (e.g.~Jacobi
polynomials are possible choices \cite{Montvay:1999ty}) satisfying the
orthogonality relation
\begin{equation}
  \label{eq:orthogon}
  \int_\epsilon^\lambda dx\;
  w(x)^2\Phi_\mu(x)\Phi_\nu(x)=\delta_{\mu\nu} q_\nu\,.
\end{equation}
The weight function $w(x)=1/x^{-\alpha}$ can be chosen. Then $P_n(x)$
can be expanded in terms of the $\lbrace\Phi_\nu\rbrace$ with
coefficients $d_{\nu}$,
\begin{equation}
  \label{eq:orthog-expand}
  P_n(x) = \sum_{\nu=0}^n d_\nu \Phi_\nu(x)\,.
\end{equation}
The coefficients $\lbrace d_\nu\rbrace$ are given by
\begin{equation}
  \label{eq:orthog-expand-coeffs}
  d_\nu = \frac{b_\nu}{q_\nu}\,, \qquad b_\nu=\int_\epsilon^\lambda
  dx\; w(x)^2 f(x) \Phi_\nu(x)\,.
\end{equation}
The polynomials $\lbrace\Phi_\nu\rbrace$ can be constructed by the
three-term recurrence relation (see \cite{Montvay:1997vh,Fox:1968bo})
\begin{equation}
  \label{eq:orthog-recurs}
  \Phi_{\mu+1}(x) = (x+\beta_\mu)\Phi_\mu(x) +
  \gamma_{\mu-1}\Phi_{\mu-1}(x)\,,
\end{equation}
with
\begin{equation}
  \label{eq:orthog-recurs-coeffs}
  \beta_\mu = -\frac{p_\mu}{q_\mu}, \quad \gamma_\mu =
  -\frac{q_{\mu+1}}{q_\mu}\,.
\end{equation}
The factors $\lbrace p_\mu\rbrace$ are given by \[ p_\mu =
\int_\epsilon^\lambda dx\; w(x)^2 \Phi_\mu(x)^2 x\,. \] The advantages
of the quadratically optimized polynomials are that they only require
the knowledge of the eigenvalue interval $[\epsilon,\lambda]$ of the
matrices whose inverse one is interested in. They provide a very good
approximation which is worse at the lower end of the interval where
the eigenvalue density is decreasing,
cf.~Sec.~\ref{sec:optim-polyn-appr}. Furthermore, the quadratically
optimized polynomials give a very simple way to control the number of
dynamical fermions to be simulated by a multiboson algorithm. This can
directly be done by by adjusting the value of $\alpha$. Of great value
is also the fact that they are very stable even for large orders.
Finally, they can efficiently be implemented on parallel computers
since they only require matrix-vector-multiplications and
vector-vector-additions.

The disadvantage is that they may not take into account all
information which is available about the matrix under consideration.
In particular, the eigenvalue density is also decreasing on the upper
end of the interval, although the quadratically optimized polynomials
have good accuracy at this point. In this sense, one might hope to
achieve better results by modifying the weight function $w(x)$. This
still leaves room for further improvement in the future.

\subsubsection{Ultra-Violet Filtering}
\label{sec:ultra-viol-filt}
An important preconditioning technique which has been introduced to
the field of multiboson algorithms by \textsc{de Forcrand}
\cite{deForcrand:1998sv} is known as \textit{UV-filtering}. It makes use
of the identity \[ e^{-\mbox{Tr}\, A}\det\, e^A = 1\,, \] so that
\begin{equation}
  \label{eq:uv-filter-def}
  \det(\mathbf{1}-\kappa D) = \exp\left[-\sum_{j=0}^M a_j\mbox{Tr}\;
  D^j\right] \det\left((\mathbf{1}-\kappa D)\exp\left[\sum_{j=0}^M
  a_j D^j \right] \right)\,.
\end{equation}
The order $M$ of the hopping parameter expansion can be adjusted to
minimize the total effort. The effect of UV-filtering on the order of
the polynomial approximation has been examined in
\cite{deForcrand:1998sv} and shown to be superior to standard HMC in
\cite{Alexandrou:1999ii,Alexandrou:1999vw}. It turns out that the
order $n$ can be reduced by a factor of about two.

In order to find the polynomial $P_n(x)$ one applies an adaptive
inverter (Ref.~\cite{Borrelli:1996re} uses the GMRES method for this
purpose) to a thermalized gauge field configuration. The polynomial
will then approximate
\begin{equation}
  \label{eq:uv-filter-apply}
  P_n(x)\approx (\mathbf{1}-\kappa D)^{-\alpha}\exp\left[ -\sum_j a_j
    D^j\right]\,.
\end{equation}
However, for larger orders $n$, the iterations used to fix the
coefficients of the polynomial become numerically unstable. This is
the reason why one needs the recursion form of the quadratically
optimized polynomials. The instability will thus limit the
applicability of the expansion (\ref{eq:uv-filter-def}).

Concluding, UV-filtering is a highly effective way to reduce the order
of the polynomial and thus to improve the algorithm to a large extend.
On the other hand, one needs a thermalized configuration (or even
several of them) at the physical point one is interested in. In this
respect, the method to fix the polynomial $P_n(x)$ discussed in
\cite{Alexandrou:1999vw} will only become optimal after a certain
run-time once thermalization is achieved.

\subsection{Conjugate-Gradient Iteration}
\label{sec:conj-grad-iter}
The simplest adaptive iterative inverter is the \textit{Conjugate
  Gradient} (CG) scheme, see e.g.~\cite{netlib-templates} for a
reference implementation. It is also the oldest and best-known method
for this problem. It requires that the matrix $Q(y,x)$ is Hermitian
and positive definite. The idea is to minimize the function
\begin{equation}
  \label{eq:cg-function}
  f\left(\phi(y)\right) = \frac{1}{2}\sum_{xy} \phi^\dagger(y) Q(y,x)
  \phi(x) - \sum_{y} \phi(y)^\dagger \eta(y)\,.
\end{equation}
This function is minimized when the gradient
\[ \partial_y f\left(\phi(y)\right) = \sum_{x} Q(y,x)\phi(x) -
\eta(y) \] vanishes which is simply equivalent to
Eq.~(\ref{eq:mat-invers}). The iteration prescription is to choose
orthogonal search directions $p^k(y)$ and minimize the function
(\ref{eq:cg-function}) along this direction in any iteration step:
\begin{equation}
  \label{eq:cg-iteration-phi}
  \phi^k(y) = \phi^{k-1}(y) + \alpha_k p^k(y)\,.
\end{equation}
Correspondingly, the residuals $r^k(y)$ are updated as
\begin{equation}
  \label{eq:cg-iteration-r}
  r^k(y) = r^{k-1}(y) - \alpha_k \sum_y Q(y,x) p^k(x)\,.
\end{equation}
The coefficients $\alpha_k$ are computed as to minimize the function
\[ \sum_{xy}\left(\phi^k-\phi\right)^\dagger(y) Q(y,x)
\left(\phi^k-\phi\right)(x) \] at each iteration step. Note that the
existence of this minimum requires $Q(y,x)$ to be positively definite
--- this is the reason why the CG algorithm only works for positively
definite matrices. The minimization is performed by choosing
\[ \alpha_k = \frac{\|{r^{k-1}(y)}\|} {\|
  {p^k}(y)\|_Q}\,, \] with $\| a(y)\|_Q$
denoting the following norm of a vector $a(y)$:
\[ \| a(y)\|_Q = \sum_{xy} a^\dagger(y) Q(y,x)
a(x)\,. \] The search directions are iterated via
\begin{eqnarray}
  \label{eq:cg-iteration-p}
  p^k(y)    &=& r^k(y) + \beta_{k-1} p^{k-1}(y)\,, \nonumber \\
  \beta_{k} &=& \displaystyle\frac{\|{r^k}(y)\|}
                {\|{r^{k-1}}(y)\|}\,.
\end{eqnarray}
This choice of $\beta_k$ makes it possible that $p^k$ is orthogonal to
\textit{all} previous $Ap^m$ and that $r^k$ is orthogonal to all
previous $r^m$ ($m<k$) (cf.~\cite{netlib-templates}). This is also the
reason why the algorithm is called CG, since it generates a series of
orthogonal (or ``conjugate'') vectors. The iterate $\phi^k(x)$ is
chosen from the $k$-dimensional subspace spanned by these vectors
which is known as the \index{Krylov space} ``Krylov'' subspace
$K_k\left(Q(y,x),\eta(y)\right)$
\begin{equation}
  \label{eq:krylov-space} K_k\left(Q(y,x),\eta(y)\right) =
  \mathop{\mbox{span}}\left\lbrace r^0(y),\sum_x
    Q(y,x)r^0(x),\dots,\sum_x Q^{k-1}(y,x)r^0(x)\right\rbrace\,.
\end{equation}
It can be shown \cite{golub-vanloan,paige-saunders} that for a
Hermitian matrix $Q(y,x)$ an orthogonal basis for the Krylov subspace
can be constructed using only a three-term recurrence relation. Thus,
such a recurrence is also sufficient for constructing the residuals.
In the CG algorithm this relation is replaced by two two-term
recurrences: one for the residuals $r^k(y)$ and one for the search
direction $p^k(y)$.

The starting points of the iterations are chosen to be
\[ \phi^0(y) = \eta(y), \qquad p_0(y) = \eta(y) - \sum_x Q(y,x)
\phi^0(x)\,. \] Of course it is possible to choose a different vector
as starting vector for $\phi^0(y)$, e.g.~a good guess if possible or a
random vector if all else fails.

The convergence of CG depends on the distribution of eigenvalues. With
$\kappa_2$ being the spectral condition number, an upper bound for the
effort can be given \cite{golub-vanloan}:
\[ \|\phi^k(x)-\phi(x)\|_Q \leq 2\frac{\sqrt{\kappa_2}-1}
{\sqrt{\kappa_2}+1} \|\phi^0-\phi(x)\|_Q\,. \] Thus, the number of
iterations to achieve a relative reduction of $\varepsilon$ in the
error is at most proportional to $\sqrt{\kappa_2}$. In the case of
well-separated eigenvalues, however, often a better convergence can be
observed. This can be explained by the fact that the CG tends to
optimize the solution in the direction of extremal eigenvalues first,
thereby reducing the effective condition number of the residual
subspace. For a discussion cf.~\cite{sluis-vorst}.

This method can also be extended to the case of non-Hermitian
matrices: if Eq.~(\ref{eq:mat-invers}) is multiplied from the left by
the conjugate matrix $Q^\dagger(y,x)$, the resulting equation becomes
\begin{equation}
  \label{eq:mat-invers-cg}
  \sum_{xz} Q^\dagger(y,z)Q(z,x)\phi(x) = \sum_x R(y,x)\phi(x) =
  \sum_x Q^\dagger(y,x)\eta(x)\,.
\end{equation}
In this form the iteration is done using the new matrix $R(y,x)\equiv
\sum_z Q^\dagger(y,z) Q(z,x)$. Thence, this method requires two matrix
multiplications per iteration. But the situation is even worse: Since
the new matrix $R(y,x)$ has a condition number
$\kappa_2(R)=\kappa_2^2(Q)$ exponentially larger than $Q(y,x)$, the
number of iterations required is increased by a factor of
$\kappa_{2}$. Consequently, the CG algorithm is much worse for these
applications and should only be considered as a last resort if other
methods fail.

\subsection{GMRES Algorithm}
\label{sec:gmres-algorithm}
In the case of a non-Hermitian matrix $Q(y,x)$ an orthogonal basis of
the Krylov space can no longer be constructed by a recurrence relation
among the residues $r^k$. Thus, the whole space has to be
orthogonalized; this can be done using the Gram-Schmidt construction:
\begin{eqnarray}
  \label{eq:gmres-arnoldi}
  v^0(y)       &=& r^0(y)\,, \nonumber \\
  w^{k,0}(y)   &=& \sum_x Q(y,x) v^k(x)\,, \nonumber \\
  w^{k,i+1}(y) &=& w^{k,i}(y) - \left(w^{k,i}(y),v^i(y)\right),
                   \qquad \left(i=\lbrace 1,\dots,k\rbrace\right)\,,
                   \nonumber \\
  v^{k+1}(y)   &=& w^{k,k}(y) / \| w^{k,k}(y)\|\,.
\end{eqnarray}
From the orthogonal basis of the Krylov space \[
K_l\left(Q(y,x),v^0(y)\right)=\mathop{\mbox{span}}\left\lbrace
  v^0(y),\dots,v^l(y)\right\rbrace\,, \] the iterate $\phi^l(y)$ can
be constructed via
\begin{equation}
  \label{eq:gmres-iterate}
  \phi^l(y) = \phi^0(y) + \sum_k y_k v^k(y)\,,
\end{equation}
where the coefficients minimize the residual norm
\begin{equation}
  \label{eq:gmres-function}
  f\left(\phi(y)\right) = \| \eta(y) - \sum_x
  Q(y,x)\phi(y)\|\,.
\end{equation}
This method is known as the \index{Arnoldi iteration} ``Arnoldi
method'' \cite{arnoldi51}. Thus, the \textit{Generalized Minimal
  Residual} (GMRES) algorithm minimizes the function
(\ref{eq:gmres-function}) instead of (\ref{eq:cg-function}) in case of
the CG iteration.

The advantages of this method are that it can be used to minimize
non-Hermitian functions and that the residual norms $\| r^k(y)\|$ can
be determined without computing the iterates $\phi^k(y)$. The major
disadvantage is its huge memory consumption if the iteration number
$l$ and the problem size $N$ are large. Although this method converges
exactly in $N$ steps, this point is out of reach in the cases of
interest in this thesis. Hence, only iterates up to a certain order
$l\simeq {\cal O}(100)$ can be formed. In case higher accuracy is
required, the method should be restarted several times discarding the
previous Krylov subspace. Furthermore, the convergence properties can
be improved by replacing the Gram-Schmidt orthogonalization by the
Householder method. Thus, greater computer time consumption can be
traded for higher stability.

This method has been proposed in \cite{deForcrand:1998sv} for the
generation of the polynomial $P_n(x)$ to be used in multiboson
algorithms, Eq.~(\ref{eq:polynomial}).

\subsection{Stabilized Bi-Conjugate Gradient Algorithm}
\label{sec:stab-bi-conj}
The \textit{Bi-Conjugate Gradient} (Bi-CG) method is an extension to
the CG algorithm which is also applicable to non-Hermitian matrices.
Unlike the proposal in Eq.~(\ref{eq:mat-invers-cg}) it does not square
the original matrix and thus does not worsen the condition number.
Instead it requires the computation of the Hermitian conjugate matrix
$Q^\dagger(y,x)$ to a conjugate set of residual and direction vectors,
doubling the memory requirements of the CG algorithm.

The updating prescription for the residuals then becomes
\begin{eqnarray}
  \label{eq:bicg-iteration-r}
  r^k(y)         &=& r^{k-1}(y) - \alpha_k \sum_x Q(y,x) p^k(x)\,,
                     \nonumber \\
 {\tilde r}^k(y) &=& {\tilde r}^{k-1}(y) - \alpha_k \sum_x
                     Q^\dagger(y,x) {\tilde p}^k(x)\,,
\end{eqnarray}
while for the search directions one gets
\begin{eqnarray}
  \label{eq:bicg-iteration-p}
  p^k(y)          &=& p^{k-1}(y) + \beta_{k-1} p^{k-1}(y)\,, \nonumber
  \\
  {\tilde p}^k(y) &=& {\tilde p}^{k-1}(y) + \beta_{k-1} {\tilde
    p}^{k-1}(y)\,.
\end{eqnarray}
Now the choices
\[ \alpha_k = \frac{\left( {\tilde r}^{k-1}(y), r^{k-1}(y)\right)}
{\left( {\tilde p}^k(y), \sum_x Q(y,x) p^k(x)\right)}\,, \qquad
\beta_k = \frac{\left( \tilde{r}^k(y),r^k(y)\right)}
{\left(\tilde{r}^{k-1}(y),r^{k-1}\right)} \] enforce the
bi-orthogonality relations
\begin{eqnarray*}
  \left( {\tilde r}^{k}(y),r^{l}(y) \right)               &=& 0\,, \\
  \left( {\tilde p}^{k}(y),\sum_x Q(y,x) p^{l}(x) \right) &=& 0\,,
  \qquad \mbox{for\ } k\neq l.
\end{eqnarray*}
This method allows inversion of non-Hermitian matrices but does not
show a stable convergence pattern in all cases. It may converge
irregularly or even fail completely. Therefore several modifications
have been proposed to make the convergence smoother (for an overview
see \cite{netlib-templates}). The method known as \textit{Stabilized
  Bi-Conjugate Gradient} (Bi-CGStab) as introduced by \textsc{van der
  Vorst} in \cite{vanderVorst:1992xx} does not require the Hermitian
conjugate matrix to be used, but has an overall cost similar to the
BiCG method just discussed.

\section{Eigenvalue Algorithms}
\label{sec:eigenv-algor}
An important ingredient of the application of multi-boson algorithms
as described in this thesis is the knowledge of how to tune the
polynomials to the eigenvalue spectrum of the matrix. Thus, it is of
great importance to have methods available to correctly compute at
least the borders of the eigenvalue spectrum. Another possible
application is the preconditioning of the matrix to make evaluation of
observables more simple. This approach has only been used in the
measurement of the correction factor (\ref{eq:mb-correction-approx})
in this thesis. But a different application also covers the
measurement of other observables as discussed in
Sec.~\ref{sec:meas-hadr-mass}. This approach has been examined in
\cite{Neff:2001ph}.

The matrix $Q(y,x)$ is said to have an eigenvector $\xi_i(y)\neq 0$
with corresponding eigenvalue $\lambda_i$ iff
\begin{equation}
  \label{eq:eigenval-def}
  \sum_x Q(y,x) \xi_i(x) = \lambda_i\xi_i(y)\,.
\end{equation}
A necessary condition for~(\ref{eq:eigenval-def}) is
\begin{equation}
  \label{eq:eigenval-cond}
  \det\vert Q(y,x) - \lambda\delta(y,x)\vert = 0\,,
\end{equation}
which translates to a polynomial of degree $N$ which has exactly $N$
complex roots. These roots need not be distinct. Furthermore,
Eq.~(\ref{eq:eigenval-cond}) implies that to every eigenvalue
$\lambda_i$ there corresponds an eigenvector since the matrix
${Q}(y,x)-\lambda_i\delta(y,x)$ is singular and thus has a kernel with
dimension $\geq 1$ \cite{numerical-recipes}. One important property is
that the eigenvalues may be shifted by some constant $\tau$ by adding
$\tau\xi_i(y)$ to both sides of Eq.~(\ref{eq:eigenval-def}).

Some matrices fulfill the \textit{normality condition}\footnote{As
  discussed in Sec.~\ref{sec:fermion-fields}, the Wilson matrix does
  not share this property}
\begin{equation}
  \label{eq:normal-def}
  \sum_{z} Q(y,z) Q^\dagger(z,x) = \sum_{z} Q^\dagger(y,z) Q(z,x)\,.
\end{equation}
The eigenvectors of a matrix fulfilling Eq.~(\ref{eq:normal-def}) span
the whole vector space $\mathbb{C}^N$. Applying Gram-Schmidt
orthogonalization to this set of eigenvalues yields an orthonormal
basis and thus a matrix fulfilling the unitarity condition
\begin{equation}
  \label{eq:unitary-def}
  \sum_{z} Q(y,z) Q^\dagger(z,x) = \sum_{z} Q^\dagger(y,z) Q(z,x) =
  \delta(y,x)\,.
\end{equation}
Although an arbitrary matrix has exactly $N$ eigenvalues and
consequently $N$ eigenvectors, these eigenvectors do not necessarily
span the whole vector space $\mathbb{C}^N$. In such a case the matrix
is said to be \textit{defective}.

The order-$k$ \index{Krylov space} Krylov subspace of a matrix
$Q(y,x)$ on a certain starting vector $\eta(y)$ defined in
Sec.~\ref{sec:conj-grad-iter} in Eq.~(\ref{eq:krylov-space}) can be
used for this purpose in the following way: Given the eigenvectors of
$Q(y,x)$, $\lbrace\xi_1(y),\dots,\xi_i(y)\rbrace$, which span a space
of dimension $\mbox{dim}\lbrace\xi_1(y),\dots, \xi_i(y)\rbrace = l\leq
i$, then choosing a vector
\[ \eta(y) = \sum_{j=1}^i c_i\xi_i(y)\,, \]
will result in a Krylov space whose dimension
$\mbox{dim}\,K\left(Q(y,x),\eta(y)\right)$ can be at most $l$.
Repeated application of $Q(y,x)$ on the starting vector will yield for
the $k$th element of the Krylov space
\[ \sum_{x} Q^k(y,x) \eta(x) = \sum_{j=1}^i c_i
\lambda_i^k\xi_i(y)\,. \] This recipe will increase the projection of
$\eta(y)$ on the eigenvector whose corresponding eigenvalue has the
largest magnitude $\lambda_{\mbox{\tiny max}}$. Thus, in the limit
$k\to\infty$, the iteration will converge to the largest eigenvector.

For a properly chosen starting vector $\eta(y)$ which has an overlap
will all eigenvectors, the Krylov iteration will consequently yield
the eigenvector corresponding to the eigenvalue with largest
magnitude. Repeated application of this procedure with
orthogonalization of the starting vector to previously found
eigenvectors allows in principle to restore the complete spectrum.

However, the straightforward application is quite cumbersome. In
practice it has turned out to be more economical to compute the
subspace of several eigenvalues from the border of the spectrum
together and afterwards to determine the largest eigenvector from
these iterates. One of the methods achieving this goal is called
\index{Arnoldi iteration} \textit{Arnoldi iteration}
\cite{golub-vanloan}\footnote{This algorithm is already
  coded in the \textbf{ARPACK} package and can be found in\\
  \texttt{http://www.caam.rice.edu/software/ARPACK/}}. If more than a
single eigenvalue/-vector are required, this method is the most
efficient way to determine the spectrum of a matrix.

Once the eigenvalue(s) closest to the origin are known, one can also
use this knowledge to simplify the inversion of a matrix using any of
the algorithms discussed in Sec.~\ref{sec:matr-invers-algor}.  With
$N$ eigenvectors $\lbrace\xi_i(y)\rbrace$ and their corresponding eigenvalues
$\lbrace\lambda_i\rbrace$ known, one can compute
\begin{equation}
  \label{eq:ev-precond}
  \eta'(y) = \eta(y) - \sum_{i=1}^N \frac{\sum_{x} \xi_i(x)^*
    \eta(x)}{\sum_{x} \xi_i(x)^*\xi_i(x)} \xi_i(y)\,,
\end{equation}
and then use $\eta'(y)$ as a starting point for the inversion. The
resulting inverse $\phi(y)$ is then given by
\begin{equation}
  \label{eq:ev-precond-inv}
  \phi(y) = \sum_{x}Q^{-1}(y,x)\eta'(x) + \sum_{i=1}^N
  \frac{1}{\lambda_i} \frac{\sum_{x} \xi_i^*(x)
    \eta(x)}{\sum_{x} \xi_i^*(x)\xi_i(x)} \xi_i(y)\,.
\end{equation}
The problem to compute $\sum_{x}Q^{-1}(y,x)\eta'(x)$ may now have a
significantly reduced condition number since the $N$ smallest
eigenvalues have been removed. For a highly singular matrix $Q(y,x)$,
the cost to compute the eigenvalues (which is independent of the
condition number) may be lower than the cost for the complete
inversion. However, it has been shown in \cite{Neff:2001ph} that for
the condition numbers used in the \SESAM\ project
\cite{Eicker:1996gk,Lippert:2001ha,Eicker:2001ph,Orth:2002ph} (which
are similar or even higher than those considered in this thesis), this
is not yet the case.

\chapter{Tuning of Multiboson Algorithms}
\label{sec:tuning-mult-algor}
The main focus of this chapter is the optimization and tuning of
multiboson algorithms with an emphasis on the TSMB algorithm
introduced by \textsc{Montvay} \cite{Montvay:1996ea}. The details of
the algorithm have been discussed in Sec.~\ref{sec:mult-algor}.
Throughout this chapter, the focus lies mainly on the survey of QCD
with two degenerate, dynamical fermion flavors on various lattice
sizes with fixed physical parameters given in Tab.~\ref{tab:phys-par}
(for a precise measurement see \cite{Orth:2002ph}, also
cf.~\cite{Eicker:1998sy,Eicker:2001ph}). These numbers are an excerpt
from Tab.~\ref{tab:phys-par-complete}.
\begin{table}[hb]
  \begin{center}
    \begin{tabular}[c]{*{6}{c|}c}
      \hline\hline
      \multicolumn{3}{c}{\textbf{Bare parameters}} & 
      \multicolumn{4}{|c}{\textbf{Physical parameters}} \\ \hline
      $\mathbf{N_f^{\mbox{\tiny sea}}}$ & $\mathbf{\beta}$ &
      $\mathbf{\kappa}$ & $\mathbf{(am_\pi)}$ & $\mathbf{(am_\rho)}$ &
      $\mathbf{m_\pi/m_\rho}$ & \textbf{$\mathbf{a}$/fm} \\ \hline
      $2$ & $5.5$ & $0.159$ & $0.4406(33)$ & $0.5507(59)$ &
      $0.8001(104)$ & $0.141$ \\
      \hline\hline
    \end{tabular}
    \caption{Bare and physical parameters for most runs
      presented.}
    \label{tab:phys-par}
  \end{center}
\end{table}

In Sec.~\ref{sec:optim-polyn-appr}, the static aspects of the
polynomial approximations are discussed. The question to be answered
is how to choose the approximation of an inverse power of the Wilson
matrix in the most efficient way if one recourses to a static
approximation (cf.~Sec.~\ref{sec:stat-polyn-invers}).

Section~\ref{sec:tuning-dynam-param} investigates the tuning of the
dynamical aspects of multiboson algorithms. After a detailed
presentation of the tools used for the efficiency analysis
in~\ref{sec:pract-determ-autoc}, the practical application to an
aspect of major importance, namely the dependence of the performance
on the order $n_1$ of the polynomial (\ref{eq:polynomial}) is
investigated in~\ref{sec:acceptance-rates-vs}. The results presented
here should be independent of the particular implementation of the
algorithm and thus apply to other variants of MB algorithms apart from
TSMB as well. The impact of reweighting is analyzed in
Sec.~\ref{sec:dynam-rewe-fact}, and finally the updating strategy is
discussed in Sec.~\ref{sec:updating-strategy}. The updating strategy
consists of the proper combination of local updating sweeps which make
up a single \textit{trajectory}\index{Trajectory}. A trajectory is
then the logical partition after which an iteration of update sweeps
restarts.

The practical implementations of multiboson algorithms are discussed
in Sec.~\ref{sec:impl-syst}. The two major platforms, where the
multiboson algorithm has been implemented are compared and performance
measurements are presented.

Section~\ref{sec:summary} summarizes the results from this chapter.

\section{Optimizing the Polynomial Approximation}
\label{sec:optim-polyn-appr}
In order to find the required approximations for the TSMB algorithm,
one has to focus first on the behavior of the polynomial approximation
in the static case. This regards the application of the inversion to a
single gauge field configuration with known condition number and
eigenvalue distribution.

\begin{figure}[htb]
  \begin{center}
    \includegraphics[scale=0.3,clip=true]{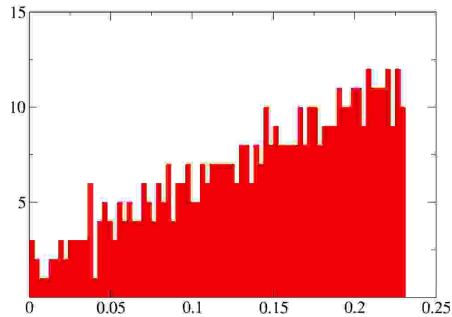}
    \caption{Histogram of the $512$ smallest eigenvalues of
      $\tilde{Q}^2(y,x)$.}
    \label{fig:low_EV_Histo}
  \end{center}
\end{figure}

\begin{figure}[htb]
  \begin{center}
    \includegraphics[scale=0.3,clip=true]{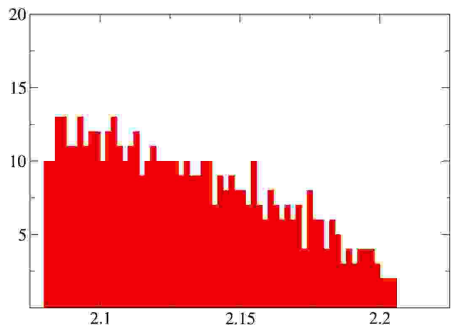}
    \caption{Histogram of the $512$ largest eigenvalues of
      $\tilde{Q}^2(y,x)$.}
    \label{fig:big_EV_Histo}
  \end{center}
\end{figure}

In the following, a particular thermalized gauge field configuration
at the physical point given in Tab.~\ref{tab:phys-par} on an
$\Omega=8^4$ lattice will be considered. The extremal eigenvalues and
the condition number of the Wilson matrix $\tilde{Q}^2(y,x)$ for this
gauge field configuration are given in Tab.~\ref{tab:extremal-evs}.
\begin{table}[ht]
  \begin{center}
    \begin{tabular}[c]{*{2}{c|}c}
      \hline\hline
      \textbf{\strut$\lambda_{\mbox{\tiny min}}$} &
      \textbf{$\lambda_{\mbox{\tiny max}}$} &
      \textbf{$\lambda_{\mbox{\tiny max}}/\lambda_{\mbox{\tiny min}}$}
      \\ \hline
      $\Exp{5.4157}{-4}$ & $2.2052$ & $4071.9$ \\ \hline\hline
    \end{tabular}
    \caption{Extremal eigenvalues and the condition number of
      $\tilde{Q}^2(y,x)$.}
    \label{tab:extremal-evs}
  \end{center}
\end{table}
A histogram of the lowest $512$ eigenvalues is shown in
Fig.~\ref{fig:low_EV_Histo}.  Figure~\ref{fig:big_EV_Histo} shows the
corresponding histogram of the largest eigenvalues. As it is evident
from these plots, the eigenvalue density is small at the lower and
upper ends of the interval and increases towards the middle.

\subsection{Tuning the Quadratically Optimized Polynomials}
\label{sec:tuning-quadr-optim}
\begin{figure}[htb]
  \begin{center}
    \includegraphics[scale=0.3,clip=true]{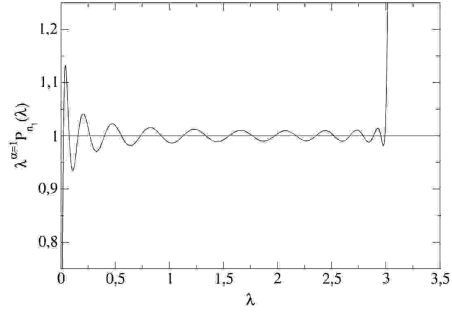}
    \caption{Test function $\lambda^{\alpha=1} P_{n_1=20}(\lambda)$
      for a quadratically optimized polynomial.}
    \label{fig:pn1_quality}
  \end{center}
\end{figure}

The quality of the approximation provided by the polynomial
(\ref{eq:polynomial}) does not only depend on its order, but also on
the choice of the interval, where it should approximate the function
under consideration.  Now the optimal choice of the approximation
interval, $[\epsilon,\lambda]$, will be determined for a quadratically
optimized polynomial introduced in Sec.~\ref{sec:stat-polyn-invers}.
Figure~\ref{fig:pn1_quality} displays the function
\[ \lambda^{\alpha}P_{n_1}(\lambda) \] of a quadratically optimized
polynomial with $n_1=20$, $\alpha=1$ and
$[\epsilon,\lambda]=[\Exp{7.5}{-4},3]$. The quality of the
approximation is best at the upper end of the interval, while already
slightly above the upper limit it will soon become useless.  At the
lower end of the interval the approximation is worse, but the limit is
not as stringent as in the former case.

These observations fix the strategy for finding the optimal interval:
The upper limit must be chosen very conservatively --- large enough
that during the simulation runs an eigenvalue never leaves this
interval. In the following, the choice $\lambda=3$ will be adopted
unless otherwise stated. The lower end may be chosen more freely, in
particular it may be chosen larger than the smallest eigenvalue since
the eigenvalue density is largest in the middle of the interval.
Raising the lower limit will make the approximation for the smallest
eigenvalues worse, but will increase the quality of the polynomial in
the middle, where the majority of eigenvalues is located.

\subsubsection{Measures of Accuracy}
\label{sec:measure-accuracy}
\index{Matrix norm} To find a measure for the quality of the
polynomial approximation for a particular matrix (in this case the
square of the Hermitian Wilson matrix, $\tilde{Q}^2(y,x)$, for the
gauge field configuration discussed above), the following two
definitions of matrix norms will be adopted: Consider the matrix
$R_n(\cdot)$ defined by
\begin{equation}
  \label{eq:mat-resid}
  R_{n}(\tilde{Q}^2) = \mathbf{1}-\tilde{Q}^{2\alpha}
  P_{n}(\tilde{Q}^2)\,.
\end{equation}
Then the following two definitions of matrix norms will be used:
\begin{enumerate}
\item Measure the vector norm of $|\xi(x)|$ defined by 
  \begin{equation}
    \label{eq:vector-norm}
    \xi(y) = \sum_x R_{n}(\tilde{Q}^2)(y,x) \eta(x)\,,
  \end{equation}
  where $\eta(x)$ is a Gaussian random vector with unit width. The
  average vector norm $|\xi(x)|\equiv |\sum_x \xi(x)|$ for a sample of
  $\lbrace\eta\rbrace$ will be denoted by $|R_{n}(\tilde{Q}^2)|$.
\item Measure the expectation value
  \begin{equation}
    \label{eq:determinant-norm}
    \langle R_{n}(\tilde{Q}^2)\rangle = \sum_{\eta(x)} \sum_{xy}
    \eta^\dagger(y)R_{n}(\tilde{Q}^2)(y,x)\eta(x)\,,
  \end{equation}
  where $\eta(x)$ is again a Gaussian random vector with width one.
  This quantity is not a norm, however. Since it is not positive
  definite the absolute value of $\langle R_{n}(\tilde{Q}^2)\rangle$
  will be used in the following and will be denoted by
  $\|R_{n}(\tilde{Q}^2)\|$.
\end{enumerate}
These definitions can also be applied to the case of the inverse
square root defined in Eq.~(\ref{eq:poly-invsqr-def}). This is done by
replacing $R_n(\cdot)$ by $\hat{R}_{n_3}^{n_2}(\cdot)$, which is
defined by
\begin{equation}
  \label{eq:r3-determinant-norm}
  \hat{R}_{n_3}^{n_2}(\tilde{Q}^2) =
  \mathbf{1}-\hat{P}_{n_3}(\tilde{Q}^2)
  (\tilde{P}_{n_2}(\tilde{Q}^2))^2\,.
\end{equation}
In particular, $\|\hat{R}_{n_3}^{n_2}(\tilde{Q}^2)\|$ is the
exponential factor in the noisy correction step of the TSMB algorithm,
Eq.~(\ref{eq:noisy-acc-final}), if the old configuration is chosen
equal to the new one.

When computing the matrix norms of $R_{n}$ for too small orders $n$,
the fluctuations of the norms will be large. In particular, the matrix
norm (\ref{eq:determinant-norm}) at small $n$ will be close to
(\ref{eq:boson-determinant}), i.e.~the inverse of the determinant of
$\tilde{Q}$. The opposite limit $n\rightarrow\infty$ will correspond
to the determinant itself. As has been discussed in
Sec.~\ref{sec:sampling-with-wilson}, the fluctuations of
(\ref{eq:boson-determinant}) are huge, while those of its inverse are
small. Thus, the fluctuations will decrease for increasing values of
$n$. Therefore, the optimization of the static approximation should be
performed for comparatively large orders.

\subsubsection{Fixing the Lower Limit}
\label{sec:fixing-lower-limit}
\begin{figure}[htb]
  \begin{center}
    \includegraphics[scale=0.3,clip=true]{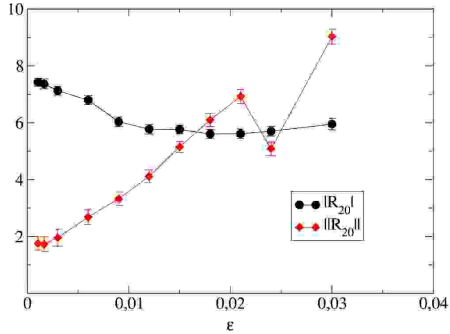}
    \caption{Norms $|R_{20}|$ and $\|R_{20}\|$ vs.~the lower interval
      limit $\epsilon$.}
    \label{fig:smalln1-var-epsilon}
  \end{center}
\end{figure}

As has been argued, it is of importance to have a recipe for fixing
the lower limit of a quadratically optimized polynomial for a given
order.  First consider the choice $n_1=20$ for which the two matrix
norms together with their standard errors are displayed in
Fig.~\ref{fig:smalln1-var-epsilon} for varying values of $\epsilon$.
For each point a sample of $100$ Gaussian vectors has been considered.
While $\|R_{20}\|$ displays a minimum at the lower end of the interval
(where the smallest eigenvalue is located), $|R_{20}|$ stays more or
less constant over a range of more than one order of magnitude. Thus,
for small orders, one cannot rule out that a choice
$\epsilon\gg\lambda_{\mbox{\tiny min}}$ is practical.

\begin{figure}[htb]
  \begin{center}
    \includegraphics[scale=0.3,clip=true]{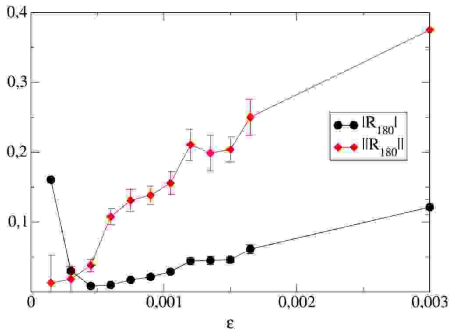}
    \caption{Norms $|R_{180}|$ and $\|R_{180}\|$ as defined in
      Eq.~(\ref{eq:vector-norm}) vs.~the lower interval limit
      $\epsilon$.}
    \label{fig:hugen1-var-epsilon}
  \end{center}
\end{figure}

Next consider the case $n_1=180$ which should already provide a very
good approximation to the inverse function.
Figure~\ref{fig:hugen1-var-epsilon} again shows the two matrix norms
for varying values of $\epsilon$. The curve of $|R_{180}|$ clearly
displays a minimum at $\epsilon_{\mbox{\tiny opt}}=\Exp{4.5}{-4}$,
which is about $20\%$ smaller than $\lambda_{\mbox{\tiny min}}$. The
curve of $\|R_{180}\|$ shows a more or less continuous increase with
larger errors.

Finally the situation regarding the third polynomial must be
clarified. In general, the systematic error of a simulation run should
be bounded to be much smaller than the statistical error of any
quantity measured. The magnitude of the error can be estimated by
considering a noisy estimate for the determinant,
Eq.~(\ref{eq:noisy-acc-final}), with the old configuration being equal
to the new one, i.e. \[ U'=U\,. \] If the approximation was exact the
acceptance probability would be equal to one. However, any deviation
in the exponential could cause spurious acceptances or rejections.
Since any negative value in the exponential in
(\ref{eq:noisy-acc-final}) would cause the configuration to be
accepted in any case, the case of large negative values of the
exponential factor can be completely disregarded. On the other hand,
for large positive values of the argument, the influence of any error
on the acceptance rate will be minor due to the flat tail of the
exponential function. Thence, the largest influence is to be expected
for values around zero.

To quantify the influence of this systematic error one can consider
the following model for the exponential correction factor:
\begin{equation}
  \label{eq:exp-corr-model}
  E(\sigma,b,x) = \frac{1}{\sqrt{2\pi}\sigma} \exp\left[
  -\frac{1}{2\sigma^2} \left(b-x\right)^2\right]\,.
\end{equation}
The resulting acceptance rate can be computed to yield
\begin{eqnarray}
  \label{eq:acc-rate-model}
  P_{\mbox{\tiny acc}}(\sigma,b) &=& \int_{-\infty}^0 dx\;
  E(\sigma,b,x) + \int_0^{\infty} dx\; E(\sigma,b,x) \exp(-x)
  \nonumber \\
  &=& \frac{1}{2}\left(1-\mbox{Erf}\left[\frac{b}{\sqrt{2}\sigma}
    \right]\right) + \frac{1}{2}\left(1-\mbox{Erf}\left[\frac{
        \sigma^2-b}{ \sqrt{2}\sigma}\right]\right)
  \exp\left[-b+\frac{1}{2}\sigma^2\right]\,.
\end{eqnarray}
Using Eq.~(\ref{eq:acc-rate-model}), one can compute the actual
systematic error by measuring $\sigma$, $b$ and
$\hat{R}^{n_2}_{n_3}(\tilde{Q}^2)$ in a given run and considering the
resulting change in acceptance rates
\begin{equation}
  \label{eq:systematic-error-est}
  \Delta P_{\mbox{\tiny acc}}(\sigma,b,\Delta
  b=\|\hat{R}^{n_2}_{n_3}(\tilde{Q}^2)\|) = | P_{\mbox{\tiny
      acc}}(\sigma,b) - P_{\mbox{\tiny
      acc}}(\sigma,b-\|\hat{R}^{n_2}_{n_3}(\tilde{Q}^2)\|) |\,.
\end{equation}
The resulting number is the systematic error for a single trajectory.
As a rule of thumb one should not allow $\Delta P_{\mbox{\tiny
    acc}}(\sigma,b,\Delta b)$ to exceed values of $\Exp{1}{-3}$. In
most situations, however, it is possible to make it as small as
$\Exp{1}{-5}$. Any systematic error will then be negligible.

\begin{figure}[htb]
  \begin{center}
    \includegraphics[scale=0.3,clip=true]{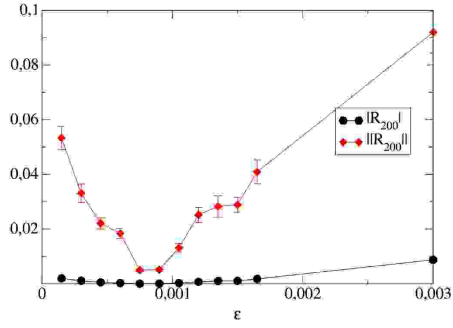}
    \caption{Residual norms for $\hat{R}^{160}_{200}(\tilde{Q}^2)$
      vs.~the lower interval limit $\epsilon$.}
    \label{fig:n3-var-epsilon}
  \end{center}
\end{figure}

Figure~\ref{fig:n3-var-epsilon} shows plots of the two norms with
$n_3=200$ (the value $n_2=160$ has been chosen compatible to the
situation discussed above) vs.~the lower limit of the interval,
$\epsilon$. Both norms obtain their minimal values at lower interval
limits of $\epsilon\approx\Exp{7.5}{-4}\approx
1.5\times\lambda_{\mbox{\tiny min}}$. However, the important norm in
this case is $\|\hat{R}_{n_3}(\tilde{Q}^2)\|$ --- already if
$\epsilon$ is varied by a factor of $2$, one approaches a region where
the systematic error may become significant. One has to keep in mind
that in a dynamical simulation fluctuations may cause the smallest
eigenvalue to become smaller than in the present case. Therefore, the
interval for the third polynomial has to be chosen far more
conservatively than for the other polynomials and the residual norm
must be adjusted by increasing the order $n_3$.
In some cases, there is a different problem related to this strategy,
cf.~Sec.~\ref{sec:accur-cons-test} below.

One can conclude that the choice of the approximation interval for
quadratically optimized polynomials can have a large impact on the
quality of the approximation. While in the case of the first
polynomial (where one deals with a comparatively small order), the
choice of the lower limit only has a small impact, the situation
changes as the order is increased. For orders as large as the second
polynomial in the two-step approximation, the optimal choice for the
lower limit is slightly smaller than the smallest eigenvalue of the
matrix, while for the third polynomial, the choice of the lower limit
should be made extremely conservative. The interval should always
cover every single eigenvalue and ensure that $\Delta P_{\mbox{\tiny
    acc}}$ in Eq.~(\ref{eq:systematic-error-est}) is sufficiently
small for the algorithm to be free of systematic errors.

In the case of a dynamical simulation, the choice of the interval
should be guided by the average values of the smallest eigenvalue.
However, this information is only rarely available prior to the run.
It may therefore be necessary to readjust the polynomials during a run
as more information becomes available. In this way one can reduce the
total runtime, but at the price of more effort and logistics.
Appendix~\ref{sec:logist-runn-large} gives a framework for handling
this type of runs.

In any case, one has to make sure that the third polynomial is
sufficiently good by making a very conservative decision regarding the
lower limit and making the order sufficiently large as to keep the
systematic error bounded to at most a few percent.

\subsection{Algorithm for Polynomials}
\label{sec:algor-polyn}
As has been discussed in Sec.~\ref{sec:mult-algor}, different methods
are possible for finding the polynomial, Eq.~(\ref{eq:polynomial}),
which approximates the inverse of the Wilson matrix.  The
quadratically optimized polynomials do not require explicit knowledge
of the eigenvalue density in the approximation interval. On the other
hand, the method proposed by \textsc{de Forcrand} in
\cite{deForcrand:1998sv} requires a thermalized gauge field
configuration to be available. As has been noted in the latter
publication, taking a single gauge field configuration may already be
sufficient since the results from several configurations are similar.

\begin{figure}[htb]
  \begin{center}
    \includegraphics[scale=0.3,clip=true]{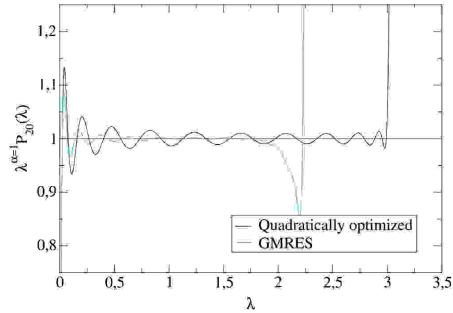}
    \caption{Polynomial $\lambda^{\alpha=1} P_{20}(\lambda)$ which has
      been obtained by applying the GMRES algorithm to a thermalized
      gauge field configuration together with the corresponding
      quadratically optimized polynomial.}
    \label{fig:pGMRES_quality}
  \end{center}
\end{figure}

Figure~\ref{fig:pn1_quality} already showed the deviation of a
quadratically optimized polynomial. The GMRES algorithm discussed in
Sec.~\ref{sec:gmres-algorithm} will now be used to construct the
polynomial dynamically on a (different) thermalized configuration. The
resulting plot of $\lambda^{\alpha=1} P_{n_1=20}(\lambda)$ is
displayed in Fig.~\ref{fig:pGMRES_quality}. For comparison the
quadratically optimized polynomial from Fig.~\ref{fig:pn1_quality} is
also shown.

It is apparent that the quadratically optimized polynomial performs
worse in the middle of the spectrum, where most eigenvalues are
located. In contrast, the GMRES polynomial respects the spectral
density of the Wilson matrix and thus results in a better
approximation. The disadvantage is that the underlying algorithm will
become unstable when computing the coefficients of the polynomials for
larger orders if it is only run on a machine with $64$-bit
\index{Numerical precision} precision. This instability will already
become apparent for orders slightly beyond $n_1=20$. A further problem
is that usually one does not have a thermalized gauge field
configuration for a particular physical point \textit{prior} to the
calculation. It is therefore necessary to perform the optimization
process dynamically during the sampling and readjust the polynomials
after a certain number of trajectories.  Similar to the case of
quadratically optimized polynomials, this requires more effort.

\begin{figure}[htb]
  \begin{center}
    \includegraphics[scale=0.3,clip=true]{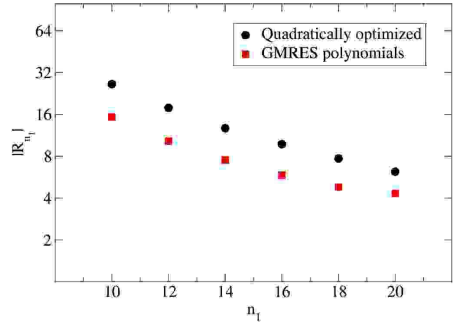}
    \caption{Residual vector norm for both the GMRES and the
      quadratically optimized polynomials.}
    \label{fig:Polynomial_comparison}
  \end{center}
\end{figure}

The influence of the qualitative difference is displayed in
Fig.~\ref{fig:Polynomial_comparison}. The GMRES results are compared
to the quadratically optimized polynomials for a number of different
orders. The polynomial interval for the quadratically optimized
polynomials has been chosen to be
$[\epsilon,\lambda]=[\Exp{6}{-4},2.5]$. The former choice clearly
exhibits smaller residuals and is thus superior. However, since the
computation has only been performed with $64$-bit
\index{Numerical precision} precision, the numerical instabilities are
already visible at $n_1=20$.

This scheme can easily be extended to the case of any rational number
$\alpha$, i.e.~any rational number of fermion flavors. However, the
instability of this method will also grow as the number of
multiplications required increases. Therefore, this method has not
been applied in the following. This is a place where further research
is in demand. One way to solve this problem would be to implement the
polynomial algorithm using very high \index{Numerical precision}
precision arithmetics, similar to what has been done in
\cite{Bergmann:1999ud}. Another way could consist of using a different
scheme which does not recourse directly to the expansion coefficients
in Krylov space.

\subsection{Computing the Reweighting Correction}
\label{sec:rewe-corr}
When using the TSMB algorithm for the correction step as discussed in
Sec.~\ref{sec:mult-algor}, one will perform a noisy estimate for the
inverse determinant using a static inversion algorithm with the
polynomial $\tilde{P}_{n_2}(x)$. Any residual systematic error in the
ensemble of gauge field configurations generated will then have to be
repaired with a multicanonical reweighting. An observable will be
computed using either (\ref{eq:poly4-rew}) or
(\ref{eq:multican-ev-rew}).

\begin{figure}[htb]
  \begin{center}
    \includegraphics[scale=0.3,clip=true]{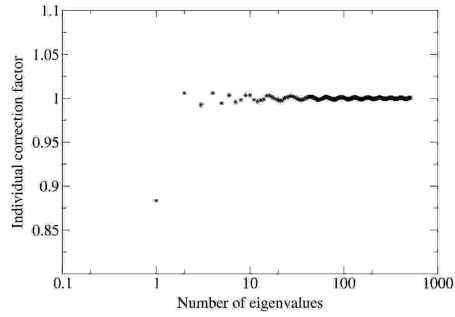}
    \caption{Individual correction factors as computed using the
      $512$ lowest eigenvalues of $\tilde{Q}^2(y,x)$ for the
      quadratically optimized polynomial $P_{180}(\tilde{Q}^2)$.}
    \label{fig:low_EV_indivReweight}
  \end{center}
\end{figure}

In order to find an efficient way to perform this reweighting, it is
assumed that the configuration under consideration has been computed
using the TSMB algorithm with action (\ref{eq:tsmb-multican}), where
the quadratically optimized polynomial
(cf.~Sec.~\ref{sec:tuning-quadr-optim})
$P_{n_1}(x)\tilde{P}_{n_2}(x)\simeq P_{n_1+n_2=180}(x)$ has been
employed with the interval $[\epsilon,\lambda]=[\Exp{7.5}{-4},3]$. The
overall systematic error from the simulation run is thus determined
from the polynomial $P_{180}(x)$ alone. The observable under
consideration is now $\hat{A}=1$, i.e.~the correction alone is being
measured. The correction factor from the individual $512$ lowest
eigenvalues of $\tilde{Q}^2$ is plotted in
Fig.~\ref{fig:low_EV_indivReweight}. Obviously, the correction is
mostly related to the lowest eigenvalue alone\footnote{The situation
  changes if the GMRES polynomials had been used since they also
  perform a worse approximation on the upper end of the interval,
  cf.~Fig.~\ref{fig:pGMRES_quality}}.
\begin{figure}[htb]
  \begin{center}
    \includegraphics[scale=0.3,clip=true]{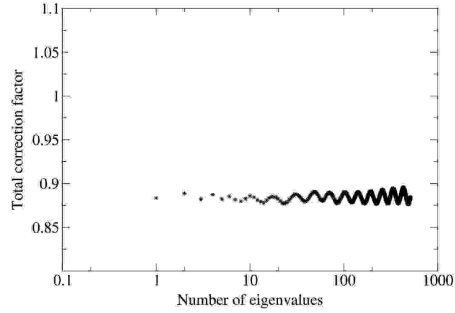}
    \caption{Similar to Fig.~\ref{fig:low_EV_indivReweight}, but with
      the cumulative correction factor from the $512$ lowest
      eigenvalues.}
    \label{fig:low_EV_cumulReweight}
  \end{center}
\end{figure}

This finding is confirmed when examining the convergence behavior of
the correction factor with respect to the number of eigenvalues
computed. The cumulative factor in Eq.~(\ref{eq:multican-ev-rew}) as a
function of the number of eigenvalues taken into account is shown in
Fig.~\ref{fig:low_EV_cumulReweight} and gives an average value of
$0.885$. Although larger eigenvalues still introduce fluctuations, the
major impact comes from the smallest eigenvalue alone.

The alternative way to compute the correction factor is provided by
the evaluation of (\ref{eq:poly4-rew}). From $100$ noisy vectors one
observes that the approximation already has converged at order
$n_4=500$. The total correction factor from this method is \[ \langle
1\rangle_{P_{n_4=500}} = 0.8783\pm 0.0113\,, \] which is completely
consistent with the value obtained from
Fig.~\ref{fig:low_EV_cumulReweight}.

In conclusion, when estimating the correction factor on the basis of
eigenvectors on an $\Omega=8^4$ lattice alone, it makes sense to use
only a small fraction (definitely less than $32$) of the lowest
eigenvalues. The fluctuations introduced from the larger eigenvalues
do not have a significant influence on the total result. The
evaluation of the correction factor using a fourth polynomial is a
practical alternative and avoids having to compute a fraction of all
eigenvalues. The only potential problem is that the smallest
eigenvalue of $\tilde{Q}^2(y,x)$ may lie outside the interval of
$P_{n_4}(\cdot)$, which could result in incorrect results. This
problem can again be controlled by choosing an extremely conservative
lower limit $\epsilon$ or even $\epsilon=0$\footnote{As has been
  discussed in \cite{Montvay:1999kq}, the convergence will no longer
  be exponential in this case. Since the total runtime of the
  correction step is negligible compared to the whole run, this
  approach still appears to be justified}. Once one considers larger
lattices up to $\Omega=32\times 16^3$ and beyond, the eigenvalue
approach may become too costly since the eigenvalue density increases
linearly with the volume and consequently a larger number of
eigenvalues need to be computed to cover an equivalent fraction of the
spectrum.

\section{Tuning the Dynamical Parameters}
\label{sec:tuning-dynam-param}
After the optimal matrix inversion using a non-adaptive polynomial for
a single (or a limited set of) gauge field configurations has been
found, there remains the task to examine the dynamical behavior of
these approximations. This is a question of major interest for the
practical implementation of any multiboson algorithm, since after all
one is interested in using the approximations in a dynamical updating
process. It may happen, that fluctuations of the eigenvalue density
may temporarily cause eigenvalues to run out of the approximation
interval. This can have a dramatic impact on the performance of the
algorithm. It is therefore of considerable importance to assess the
size of these fluctuations and what impact they could have on a
simulation run. It is important to notice that these aspects may still
be explored on rather small lattices since they will exhibit larger
fluctuations and will thence show a larger sensitivity to these
vulnerabilities.

\subsection{Practical Determination of Autocorrelations}
\label{sec:pract-determ-autoc}
Before proceeding further, the tools must be prepared to compute the
primary measure of efficiency in the dynamical case, namely the
autocorrelation time of a time series. Since the aim of any simulation
algorithm is to generate statistically independent gauge field
configurations with minimal effort, the autocorrelation time is the
key monitor for the cost determination of a particular algorithm. The
theoretical bases of methods to compute autocorrelations of time
series have been laid in Sec.~\ref{sec:autocorrelation}. The purpose
of this section is to apply them to two different time series obtained
from actual simulation runs.

In the first case, the series has a low fluctuation and is
sufficiently long for the autocorrelation time to be measured. The
second situation is less suitable: the time series exhibits large
fluctuations and a rather large autocorrelation time. Furthermore, it
shows a contamination of a very long mode which introduces
fluctuations on a time scale comparable to the length of the series
itself. This mode appears to be separated from the other modes
contained in the series. Given the fact that the total lattice size is
given by $L\approx 1.128$ fm (cf.~Tab.~\ref{tab:phys-par}), one may
suspect that the simulation is already very close to the shielding
transition, see also Sec.~\ref{sec:non-zero-temperature}. This could
explain the observed behavior and the presence of the long-ranged
mode. This mode contaminates the results and unless it is possible to
perform simulations on a series at least two orders of magnitude
longer, no statement can be made about its length. In this case it
will become evident, that the lag-differencing method as discussed in
Sec.~\ref{sec:short-time-series} is still able to extract information
from the series although the other methods fail.

\subsubsection{Case I: Low Fluctuations}
\label{sec:case-i-low}
This time series has been taken from a simulation run using the
physical parameters displayed in Tab.~\ref{tab:phys-par} on an
$\Omega=32\times 16^3$-lattice. The algorithm employed is the HMC
algorithm with SSOR-preconditioning. The molecular dynamics
integration algorithm is the leap-frog scheme with a time step of
$\Delta t=\Exp{1}{-2}$ and a trajectory length of $n_{\mbox{\tiny
    MD}}=100\pm 20$. The resulting acceptance rate is $71.6\%$.

\begin{figure}[htb]
  \begin{center}
    \includegraphics[scale=0.3,clip=true]%
    {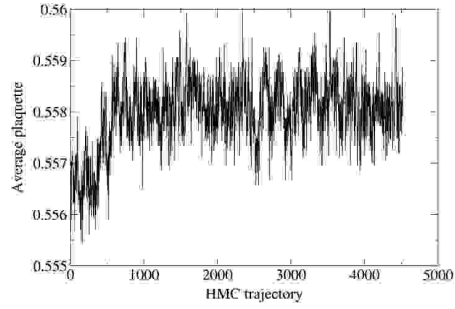}
    \caption{Plaquette history of HMC run.}
    \label{fig:hmc-series}
  \end{center}
\end{figure}

The total size of the sample consists of $4518$ trajectories, from
which the leading $1000$ trajectories have been discarded. The
complete time series is given in Fig.~\ref{fig:hmc-series}.

\begin{figure}[htb]
  \begin{center}
    \includegraphics[scale=0.3,clip=true]%
    {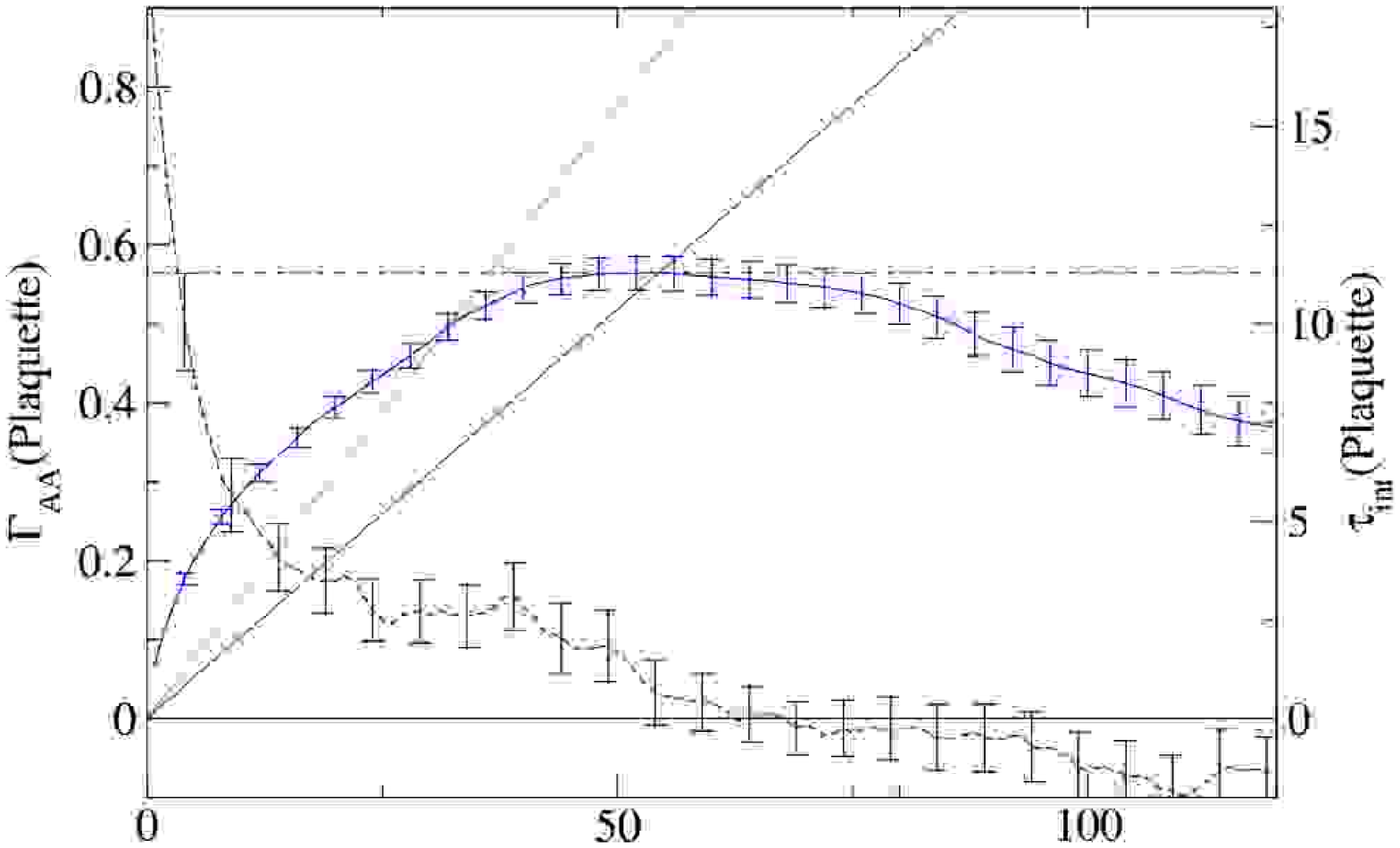}
    \caption{Normalized autocorrelation function and integrated
      autocorrelation time vs.~the cutoff for the HMC plaquette
      history.}
    \label{fig:hmc-autocorrs}
  \end{center}
\end{figure}

Figure~\ref{fig:hmc-autocorrs} shows the normalized autocorrelation
function (dotted curve), together with the integrated autocorrelation
time (blue curve) as a function of the cutoff. The windowing
procedure discussed in Sec.~\ref{sec:short-time-series} has been
applied with $c=4,6$ resulting in the green and red lines,
respectively. From the $c=6$ line, one can read off an
integrated autocorrelation time of $\tau_{\mbox{\tiny int}}=11.28\pm
0.43$. The dashed-dotted line displays the maximum of the curve, which
is clearly compatible with the $c=6$ window.
A particular problem is already visible in the behavior of the
normalized autocorrelation function. It does not approach zero
exponentially (as one would expect), but appears to reach a plateau
above zero, before it suddenly drops. The curve is not compatible with
zero at this point since it is almost two standard deviations too
high. This is a typical case of a linear bias mentioned in
Sec.~\ref{sec:short-time-series}. The lag-differencing method which
will be applied below is able to handle this situation.

\begin{figure}[htb]
  \begin{center}
    \includegraphics[scale=0.3,clip=true]%
    {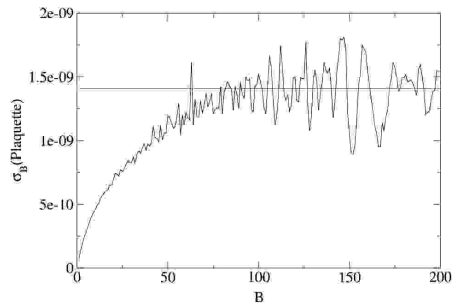}
    \caption{Variance $\sigma_B(\mbox{Plaquette})$ vs.~the bin
      size $B$ using the Jackknife method for the HMC plaquette
      history.}
    \label{fig:hmc-jackknife}
  \end{center}
\end{figure}

Next, the variance is estimated using the Jackknife method
(cf.~Sec.~\ref{sec:jackknife-method}). Figure~\ref{fig:hmc-jackknife}
shows the variance $\sigma_B(\mbox{Plaquette})$ as a function of the
bin size $B$. The variance reaches a plateau (red line) at
$\sigma(\mbox{Plaquette})\approx\Exp{(1.41\pm 0.53)}{-9}$ which yields
the true variance of the plaquette. The result for $B=1$ is given by
$\sigma_{B=1}(\mbox{Plaquette}) = \Exp{7.53}{-11}$.  Applying
Eq.~(\ref{eq:tauint-jackknife}) now yields $\tau_{\mbox{\tiny
    int}}\approx 9.36\pm 3.52$, which is slightly below the result
from the previous methods, but with a much larger uncertainty. It must
be realized that this procedure should only give a rough estimate of
the ``true'' value of $\tau_{\mbox{\tiny int}}$, see
\cite{Orth:2002ph} for a thorough discussion.

\begin{figure}[htb]
  \begin{center}
    \includegraphics[scale=0.3,clip=true]%
    {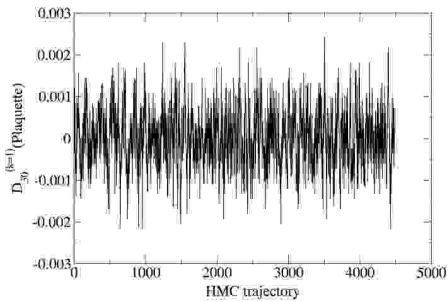}
    \caption{Order-$1$-lag-$30$ differenced series of the
      HMC plaquette history.}
    \label{fig:hmc-differenced-series}
  \end{center}
\end{figure}

\begin{figure}[htb]
  \begin{center}
    \includegraphics[scale=0.3,clip=true]%
    {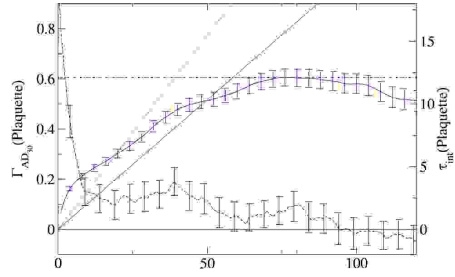}
    \caption{Correlation function $\Gamma_{AD_{30}^{(1)} A)}(t)$
      together with its integral vs.~the cutoff.}
    \label{fig:hmc-diffser-corr}
  \end{center}
\end{figure}

Finally the lag-differencing method
(cf.~Sec.~\ref{sec:short-time-series}) is applied to the time series.
As a first step, the order-$1$-lag-$30$-differenced series,
$D_{l=30}^{(k=1)}(\mbox{Plaquette})$, is computed using the definition
(\ref{eq:differencing-def}). It is displayed in
Fig.~\ref{fig:hmc-differenced-series}. As the next step, the
correlation between the plaquette and $D_{30}^{(1)}$ is being
computed, cf.~(\ref{eq:corr-diffser}). The normalized correlation
function, $\Gamma_{A,(D_{30}^{(1)} A)}(t)$, together with its integral
as a function of the cut-off is shown in
Fig.~\ref{fig:hmc-diffser-corr}. The former is given by the dotted
curve, while the latter is visualized by the blue line. The windowing
method proposes the values $\tau_{\mbox{\tiny int}}=5.94\pm 0.34$
(green line) and $\tau_{\mbox{\tiny int}}=10.26\pm 0.55$ (red line)
for windows of $c=4$ and $c=6$, respectively.  The maximum of the
autocorrelation function, however, is reached at $\tau_{\mbox{\tiny
    int}}=12.11\pm 0.48$. It is obvious that there is no significant
improvement from the differencing prescription and that the resulting
function $\Gamma_{A,(D_{30}^{(1)} A)}(t)$ is not compatible with a
single exponential mode, just like the original function
$\Gamma_{AA}(t)$ was not. Furthermore, the maximum of the curve has
shifted to the left and the windowing prescription does no longer give
the maximum at $c=6$. In fact, the differencing prescription impairs
the statistics if the lag is large compared to the ``true''
autocorrelation time.

To obtain a better result from the lag-differencing method, one has to
repeat the procedure leading to the estimate for $\tau_{\mbox{\tiny
    int}}$ using a series of different lags and look for the stability
of results. As long as the lag $l$ stays above the autocorrelation
time, no physical modes should get lost. Once the autocorrelation time
obtained becomes as large as or larger than the lag $l$, one may cut
off physical modes. Thus --- in accordance with the discussion in
Sec.~\ref{sec:short-time-series} --- one would look for a plateau at
some intermediate values of $l$, where the autocorrelation function
should exhibit an exponential behavior.

\begin{figure}[htb]
  \begin{center}
    \includegraphics[scale=0.3,clip=true]%
    {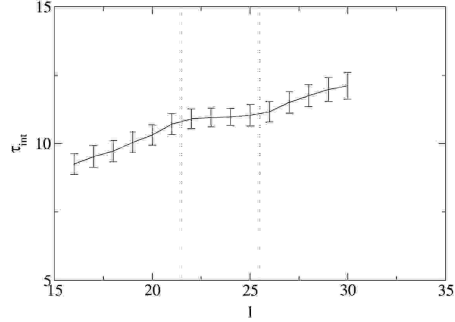}
    \caption{Integrated autocorrelation time as obtained from the
      lag-differencing method for varying lags $l$.}
    \label{fig:hmc-diffser-tauint}
  \end{center}
\end{figure}
\begin{figure}[htb]
  \begin{center}
    \includegraphics[scale=0.3,clip=true]%
    {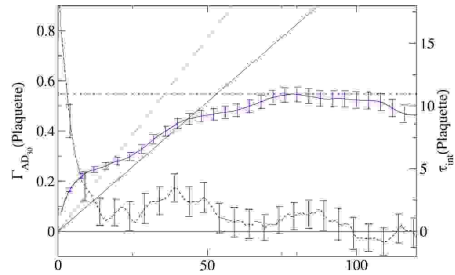}
    \caption{Similar to Fig.~\ref{fig:hmc-diffser-corr}, but with a
      differencing lag $l=23$.}
    \label{fig:hmc-diffser-corr2}
  \end{center}
\end{figure}

The results from this analysis are displayed graphically in
Fig.~\ref{fig:hmc-diffser-tauint}. Indeed, one finds a plateau
reaching from $l\approx 22$ up to $l\approx 26$ giving rise to
$\tau_{\mbox{\tiny int}}=10.96\pm 0.39$. The self-consistency
criterion $\tau_{\mbox{\tiny int}}>l$ is clearly met. The question
arises, whether the differencing prescription does indeed result in a
correlation function where the linear bias is suppressed. To address
this question, Fig.~\ref{fig:hmc-diffser-corr2} shows the correlation
function for the case $l=23$. Now the function indeed decays to zero
for already a short value of the cutoff, but still increases later on.
This may be no exponential mode, but a polynomial mode giving rise to
a higher order bias. Although in theory one could get rid of this bias
by considering a higher-order differencing scheme, the impact of this
procedure on the quality of statistics would invalidate this approach
pretty soon. In particular, if the quality of the series was good
enough to allow for higher-order differencing, the impact of the bias
would be significantly smaller in the first place.

The lesson from this investigation is that the linear bias can be
removed by applying the lag-differencing prescription and the result
obtained in this way is consistent with the one obtained from the
original autocorrelation function and from the Jackknife method. The
analysis shows that for the HMC with dynamical fermions one has to use
a time series with a length of at least $4000$ trajectories to gain
accurate information about the true autocorrelation behavior.

\subsubsection{Case II: Large Fluctuations}
\label{sec:case-ii:-large}
The second series has been obtained from the history of the average
plaquettes using the TSMB algorithm discussed in
Sec.~\ref{sec:mult-algor}. The simulation has been performed using the
same physical parameters as in the previous case except for the volume
and the algorithmic parameters given in
Tab.~\ref{tab:tsmb-multi-genpars}.  The first polynomial order was
$n_1=20$. This run is part of the tuning series discussed in
Sec.~\ref{sec:acceptance-rates-vs} below.  Reweighting of the
observables has been neglected, since this would have introduced
another source of autocorrelation effects, see \cite{Janke:1994se} for
a discussion. The total length of the series was $51196$ trajectories,
where the thermalization phase has already been subtracted.
\begin{figure}[htb]
  \begin{center}
    \includegraphics[scale=0.3,clip=true]%
    {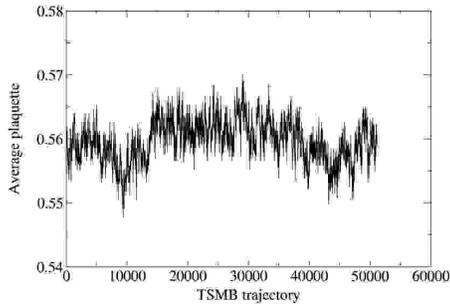}
    \caption{Time series of the average plaquette from the TSMB
      algorithm.}
    \label{fig:tsmb-ser}
  \end{center}
\end{figure}

As has already been pointed out, a very long mode possibly related to
the shielding transition is present in the series which cannot be
examined in a time series of such a length. However, since this mode
does not appear to have any connection to the other, short-range
fluctuations, it may be questioned if it has any significance for the
efficiency considerations of the multiboson algorithms.

\begin{figure}[htb]
  \begin{center}
    \includegraphics[scale=0.3,clip=true]%
    {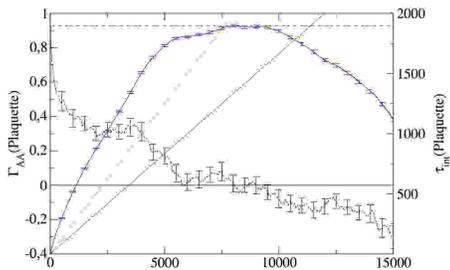}
    \caption{Autocorrelation function and the corresponding integral
      as a function of the cutoff for the plaquette history from the
      TSMB run.}
    \label{fig:tsmb-autocorr}
  \end{center}
\end{figure}

As in the previous case, first the autocorrelation function is
visualized together with the corresponding integral in
Fig.~\ref{fig:tsmb-autocorr}. The estimated autocorrelation time is
$\tau_{\mbox{\tiny int}}=1897\pm 135$. Again, the problem is that the
autocorrelation function does not go to zero exponentially and that it
appears to reach a plateau, before it drops to zero and then decreases
linearly. This behavior may indicate a linear bias, which could be
removed by the lag-differencing method.

\begin{figure}[htb]
  \begin{center}
    \includegraphics[scale=0.3,clip=true]%
    {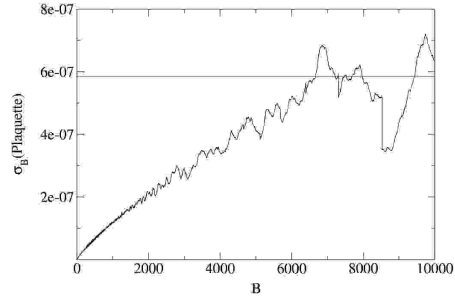}
    \caption{Jackknife variance calculated for different bin sizes for
      the time series from the TSMB run.}
    \label{fig:tsmb-jackknife}
  \end{center}
\end{figure}

The variance obtained from the Jackknife analysis for different bin
sizes is shown in Figure~\ref{fig:tsmb-jackknife}. The plateau can be
estimated to lie at about
$\sigma(\mbox{Plaquette})\approx\Exp{(5.855\pm 2.422)}{-7}$. Together
with $\sigma_{B=1}(\mbox{Plaquette})=\Exp{2.032}{-10}$ one obtains an
estimate of $\tau_{\mbox{\tiny int}}\approx 1441\pm 596$. This number
is compatible with the previous estimate, however it should not be
trusted since the original time series was contaminated with the mode
too long to be reliably examined.

\begin{figure}[htb]
  \begin{center}
    \includegraphics[scale=0.3,clip=true]{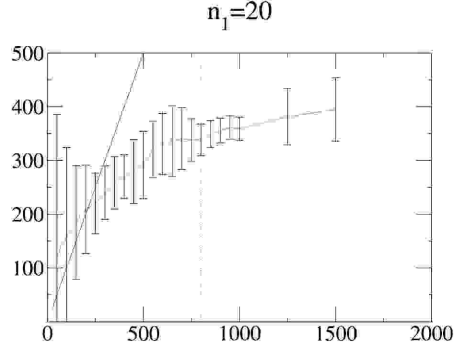}
    \caption{Integrated autocorrelation time vs.~the differencing
      lag for the TSMB run.}
    \label{fig:tsmb-diff-tauint}
  \end{center}
\end{figure}

Finally, the lag-differencing method has to shed some light on the
behavior of the autocorrelation time.
Figure~\ref{fig:tsmb-diff-tauint} displays the results from measuring
the integrated autocorrelation time with various lags. If the
long-range mode is indeed separated from the other modes, one should
be able to see a plateau from the other modes \textit{after} the
long-range mode has been cut out. There is a clear signal for the
formation of such a plateau at lags between $l=600$ and $l=800$. Using
the error bars from the single points and making a linear fit yields
$\tau_{\mbox{\tiny int}} = 334.3 \pm 65.6$.  This result is about a
factor of four below the previous estimates.

In conclusion, the lag-differencing method allows to get rid of a
linear bias and thus enables the evaluation of a series with large
fluctuations. The stability criterion is met, i.e.~the estimated
autocorrelation time exhibits a plateau when plotted as a function of
the lag $l$. The self-consistency criterion is also met, i.e.~the
resulting value for $\tau_{\mbox{\tiny int}}$ is larger than the
current value of $l$ used. A large autocorrelation mode associated
with a fluctuation that is expected to vanish with increasing volume,
has successfully been cut off. In the following section, a series of
these runs will be presented, which are all identical except that a
single parameter has been changed during all runs.

\subsection{Acceptance Rates vs.~Polynomial Approximation Quality}
\label{sec:acceptance-rates-vs}
It has already been argued that the number of boson fields enters
linearly into the autocorrelation time of a multiboson algorithm. On
the other hand, one can expect that a small number of boson fields
gives rise to a small acceptance rate and thence to an increase in the
autocorrelation time again for small numbers of fields. The number of
boson fields and thus the first polynomial order is of critical
importance for a multiboson algorithm. Until today, however, no
systematic analysis of this effect has been performed and the impact
of this choice on practical simulations is unclear. This is certainly
related to the fact that any systematic analysis is exacerbated by the
requirement to measure autocorrelation times with a reasonable
accuracy. Therefore, we base our study on very long runs. Beyond that,
we employ the efficient tools described in detail in the previous
section.

The algorithmic parameters shared by all runs are displayed in
Tab.~\ref{tab:tsmb-multi-genpars}. Only the order of the first
polynomial, $n_1$, has been varied.
\begin{table}[htb]
  \begin{center}
    \begin{tabular}[c]{*4{c|}c}
      \hline\hline
      $\mathbf{n_1}$ & $\mathbf{n_2}$ & $\mathbf{n_3}$ &
      $\mathbf{[\epsilon,\lambda]}$ & \textbf{Updates/Trajectory} \\
      \hline
      var. & $160$ & $200$ & $[\Exp{7.5}{-4},3]$ & $1$ boson HB, $3$
      boson OR, $2$ gauge Metropolis, $1$ noisy corr. \\ \hline
      \multicolumn{5}{c}{Volume $\Omega=8^4$} \\
      \hline\hline
    \end{tabular}
    \caption{General algorithmic parameters for high-statistics TSMB
      runs.}
    \label{tab:tsmb-multi-genpars}
  \end{center}
\end{table}

Table~\ref{tab:alice-runs} shows the statistics generated together
with the acceptance rates of the noisy correction step and the cost of
a single trajectory. These runs have been performed on the \ALiCE\ 
computer cluster installed at Wuppertal University\footnote{See
  \texttt{http://www.theorie.physik.uni-wuppertal.de/Computerlabor/ALiCE.phtml}
  for technical details and further information}. The machine
configurations were both a single node configuration (with no
parallelization) and a four-node partition with the lattice
parallelized in $z$- and $t$-direction. The local lattice size was
consequently $\Omega_{\mbox{\tiny loc}} = 4\times 8\times 8\times 4$.
The numerical efforts are given for the latter situation.
\begin{table}[htb]
  \begin{center}
    \begin{tabular}[c]{r|*{2}{c|}c}
      \hline\hline
      \textbf{$n_1$} & \textbf{Number of confs.} & \textbf{Acceptance
        rates} & \textbf{Numerical effort/MV-Mults} \\ \hline
      $12$  & $101111$ &  $8.00\%$&  $463.0$ \\
      $18$  & $62462$  & $39.61\%$&  $499.6$ \\
      $20$  & $61196$  & $51.51\%$&  $511.8$ \\
      $22$  & $42248$  & $57.94\%$&  $524.0$ \\
      $24$  & $49704$  & $64.56\%$&  $536.2$ \\
      $26$  & $50684$  & $69.45\%$&  $548.4$ \\
      $28$  & $50412$  & $74.46\%$&  $560.6$ \\
      $32$  & $50238$  & $80.84\%$&  $585.0$ \\
      \hline\hline
    \end{tabular}
    \caption{Runs for the parameter tuning of the TSMB algorithm.}
    \label{tab:alice-runs}
  \end{center}
\end{table}
It is specified in terms of a multiplication by the preconditioned
fermion matrix $\tilde{Q}^2(y,x)$ with an arbitrary colorspinor
$\eta(x)$. Since the TSMB algorithm uses non-adaptive polynomials in
the noisy correction step, the number of explicit matrix-vector
multiplications is straightforwardly given by $n_2+n_3$. In the case
under consideration we thus have $n_2+n_3 = 360$. To estimate the
total effort we assume that the efficiency of the implementation for
the local algorithms is roughly equivalent to the efficiency of the
matrix-vector multiplication routine\footnote{This assumption is only
  roughly valid leading to a machine- and compiler-dependence of the
  effort defined in this way. See Sec.~\ref{sec:arch-effic} for a
  thorough discussion}. Thencefrom, we measure the time needed for a
complete trajectory\index{Trajectory}, $t_{\mbox{\tiny traj}}$, and
the time needed for the noisy correction alone, $t_{\mbox{\tiny
    noisy}}$. Using these times, we can define the total effort
$E_{\mbox{\tiny MV-mults}}$ as\footnote{Actually the total time may
  fluctuate due to the load of the whole communication. Therefore, the
  results in Tab.~\ref{tab:alice-runs} have been averaged}
\begin{equation}
  \label{eq:numerical-effort}
  E_{\mbox{\tiny MV-mults}} = \left(n_2+n_3\right)
  \frac{t_{\mbox{\tiny traj}}} {t_{\mbox{\tiny noisy}}}\,.
\end{equation}

\subsubsection{Behavior of the Correction Factor}
\label{sec:behav-corr-fact}
\begin{figure}[htb]
  \begin{center}
    \includegraphics[scale=0.3,clip=true]%
    {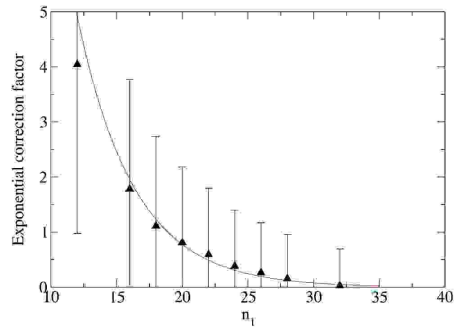}
    \caption{Dependence of the exponential correction factor on
      the number of boson fields, $n_1$.}
    \label{fig:exp-correction}
  \end{center}
\end{figure}

As a first step, the dependence of the acceptance rate on the
magnitude of the exponential correction factor
$\exp\left(-C_{12}\lbrace U_{\mbox{\tiny old}},U_{\mbox{\tiny
      new}}\rbrace\right)$ should be clarified.
Figure~\ref{fig:exp-correction} shows the exponential correction
factor together with its standard deviation. It depends exponentially
on the order $n_1$.

\begin{figure}[htb]
  \begin{center}
    \includegraphics[scale=0.3,clip=true]%
    {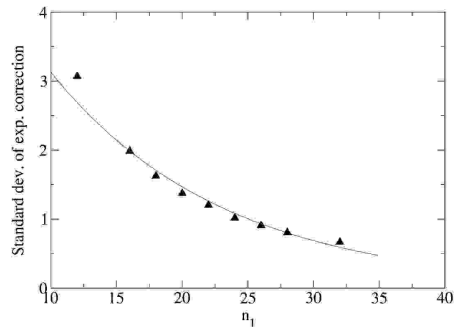}
    \caption{Dependence of the standard deviation of the
      exponential correction factor on the number of boson fields,
      $n_1$.}
    \label{fig:exp-correction-stddev}
  \end{center}
\end{figure}

The function approximating the average value is given by
\begin{eqnarray}
  \label{eq:exp-correction-av}
  \left[\exp\left(-C_{12}\right)\right]\left(n_1\right) &=&
  A\cdot\exp\left(-B\cdot n_1\right)\,,\nonumber \\
  && A = 78.657\,,\; B = 0.23092\,.
\end{eqnarray}
This behavior is in line with the expectations that the convergence of
the first polynomial is exponential. On the other hand, the standard
deviation of the exponential correction does not follow a precise
exponential dependence; this is shown in
Fig.~\ref{fig:exp-correction-stddev}.

Restricted to the intermediate regime, nevertheless an exponential
function yields a good fit to the data points:
\begin{eqnarray}
  \label{eq:exp-correction-stddev}
  \langle\left[\exp\left(-C_{12}\right)\right]\rangle\left(n_1\right)
  &=& A^\prime\cdot\exp\left(-B^\prime\cdot n_1\right)\,,\nonumber \\
  && A^\prime = 6.6765\,,\; B^\prime = 0.075757\,.
\end{eqnarray}
Finally, inserting the models (\ref{eq:exp-correction-av}) and
(\ref{eq:exp-correction-stddev}) into Eq.~(\ref{eq:acc-rate-model})
allows to check their validity by comparing them to the measured
acceptance rates from Tab.~\ref{tab:alice-runs}.
Table~\ref{tab:accrate-comp} compares the predicted and the measured
acceptance rates. The numbers are obviously in perfect agreement.
\begin{table}[htb]
  \begin{center}
    \begin{tabular}[c]{c|c|c}
      \hline\hline
      $\mathbf{n_1}$ & \textbf{Model} & \textbf{Measured} \\ \hline
      $12$  & $8.64\%$  &  $8.00\%$  \\
      $18$  & $43.84\%$ & $39.61\%$  \\
      $20$  & $53.34\%$ & $51.51\%$  \\
      $22$  & $60.88\%$ & $57.94\%$  \\
      $24$  & $66.84\%$ & $64.56\%$  \\
      $26$  & $71.59\%$ & $69.45\%$  \\
      $28$  & $75.45\%$ & $74.46\%$  \\
      $32$  & $81.37\%$ & $80.84\%$  \\
      \hline\hline
    \end{tabular}
    \caption{Comparison of acceptance rates as a function of $n_1$
      from the predictions of Eq.~(\ref{eq:acc-rate-model}) and from
      the actuals runs in Tab.~\ref{tab:alice-runs}.}
    \label{tab:accrate-comp}
  \end{center}
\end{table}

In conclusion, the exponential correction factor shows an exponential
dependence on the order $n_1$. Its standard deviation also
approximately follows an exponential decay. Therefore, the acceptance
rate can be predicted as a function of the polynomial order $n_1$ once
at least two points have been determined which allow to fit the
functions (\ref{eq:exp-correction-av}) and
(\ref{eq:exp-correction-stddev}).

\subsubsection{Fermionic Energy}
\label{sec:fermionic-energy}
\begin{figure}[htb]
  \begin{center}
    \includegraphics[scale=0.3,clip=true]{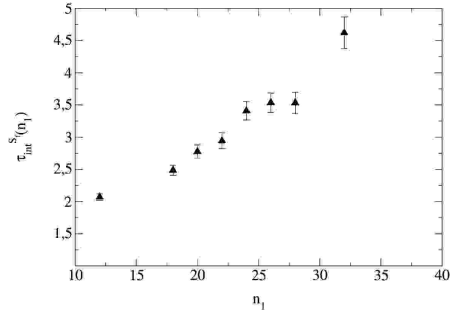}
    \caption{Integrated autocorrelation time $\tau_{\mbox{\tiny
          int}}\left(S_{\mbox{\tiny f}}\right)$ vs.~the number of
      boson fields $n_1$.}
    \label{fig:tau-int-sferm}
  \end{center}
\end{figure}

First consider the fermionic action $S_{\mbox{\tiny f}}$. This
quantity is not affected by the correction step, since in the
trajectory in Tab.~\ref{tab:tsmb-multi-genpars} the correction step
may only reject an update of the gauge field. However, it is still
linked to the full dynamics of the system by its coupling to the gauge
field. Thus, it is expected to display a linear dependence on the
number of boson fields, $n_1$. Figure~\ref{fig:tau-int-sferm} shows
the integrated autocorrelation times $\tau_{\mbox{\tiny
    int}}\left(S_{\mbox{\tiny f}}\right)$ versus the polynomial order,
$n_1$. The autocorrelation times have been measured using the
windowing prescription from Sec.~\ref{sec:short-time-series} on the
integrated autocorrelation function.

One finds a linear dependence on the number of boson fields.
Certainly, the small absolute values of the fermionic autocorrelation
times help to make the measurement very precise. However, the
fermionic energy does not directly give rise to any useful physical
information. Rather quantities computed directly from the gauge fields
(like the plaquette) give rise to physical information about the
system. Since the ``fermionic force'' on the gauge field is directly
related to the boson fields, one can nonetheless expect the influence
on the autocorrelation time of gauge-field related quantities to be
linear in $n_1$. Unfortunately, the situation is far more involved in
that case.

\subsubsection{Gauge Field Plaquette}
\label{sec:gauge-field-plaq}
Since the plaquette is a purely gauge-field dependent quantity, its
autocorrelation time will be affected by the acceptance rate. It can
be expected to increase at too small $n_1$ because of the correlations
caused by identical configurations. Furthermore, as already noted
above, the plaquette contains a strong noise and is thus very
difficult to be measured. The desired behavior will therefore be
embedded in huge fluctuations. A standard analysis of the effect is
therefore bound to fail. It is in this situation, where the
lag-differencing method becomes important and provides a useful source
of information.

\begin{figure}[ht]
  \begin{center}
    \begin{tabular}[c]{l|r}
      \includegraphics[scale=0.22,bb=0 0 27cm 22cm,angle=0,clip=true]%
      {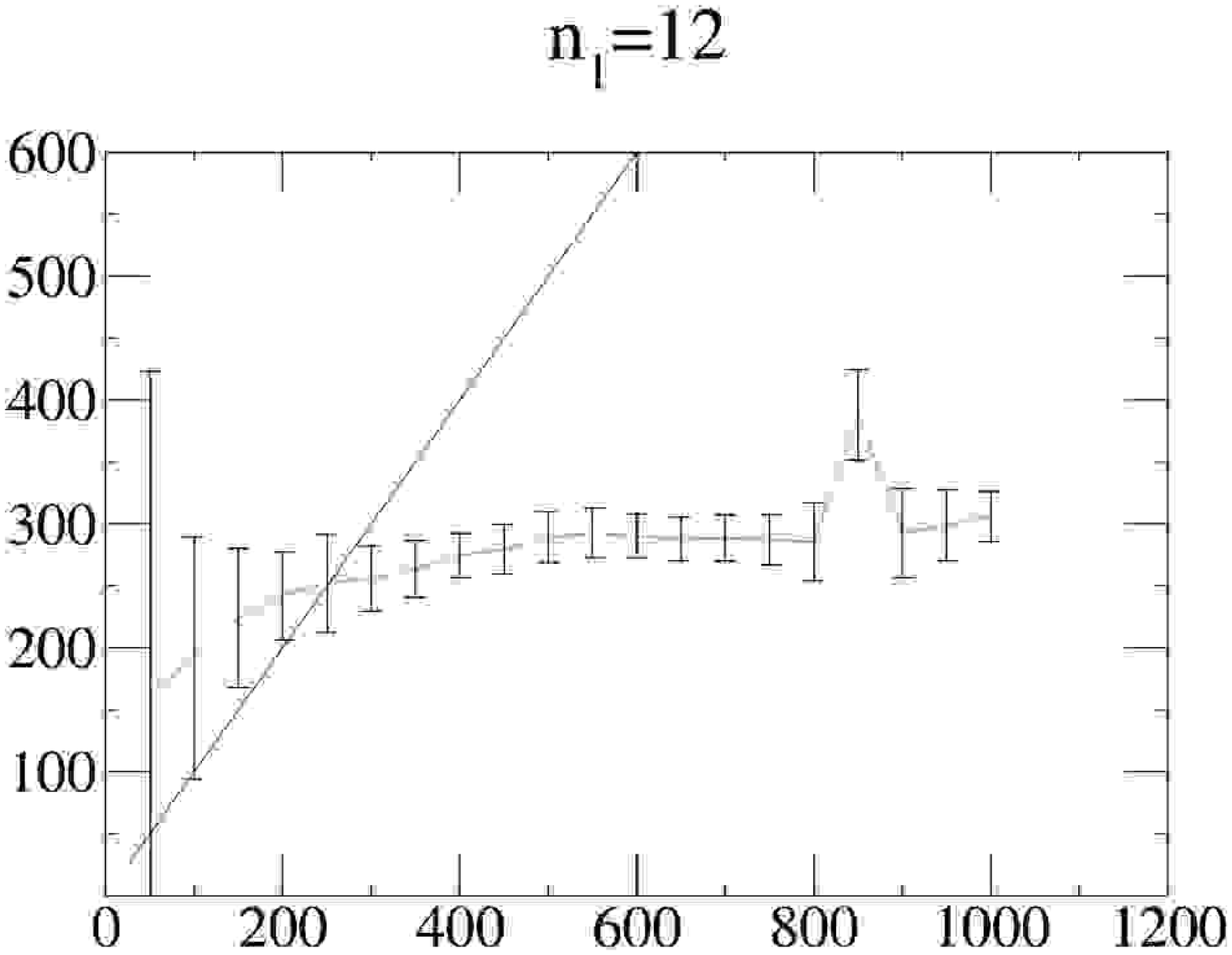} &
      \includegraphics[scale=0.22,bb=0 0 27cm 22cm,angle=0,clip=true]%
      {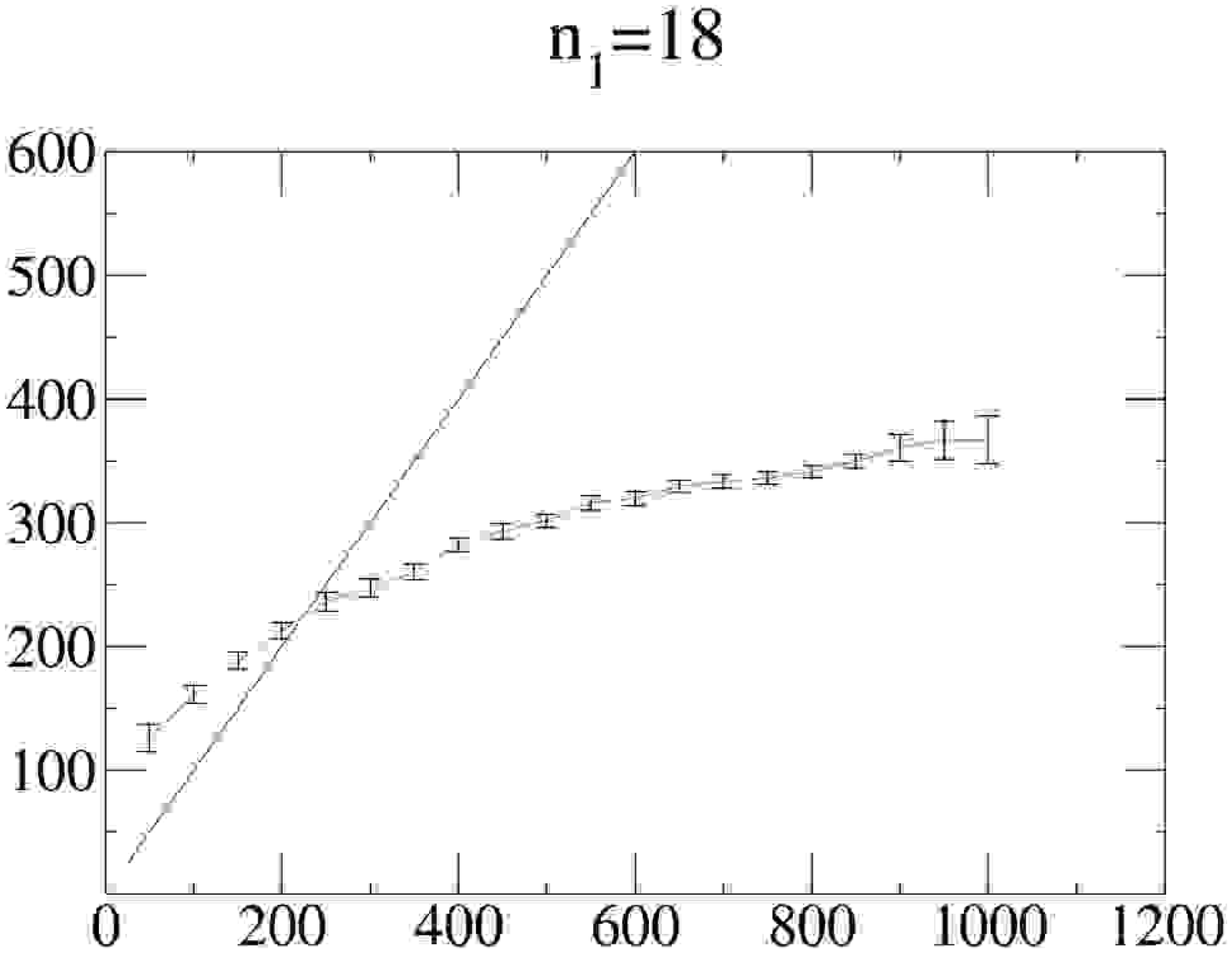} \\ \hline                
      \includegraphics[scale=0.22,bb=0 0 27cm 22cm,angle=0,clip=true]%
      {ACT_vs_Lag_n1_20.eps} &
      \includegraphics[scale=0.22,bb=0 0 27cm 22cm,angle=0,clip=true]%
      {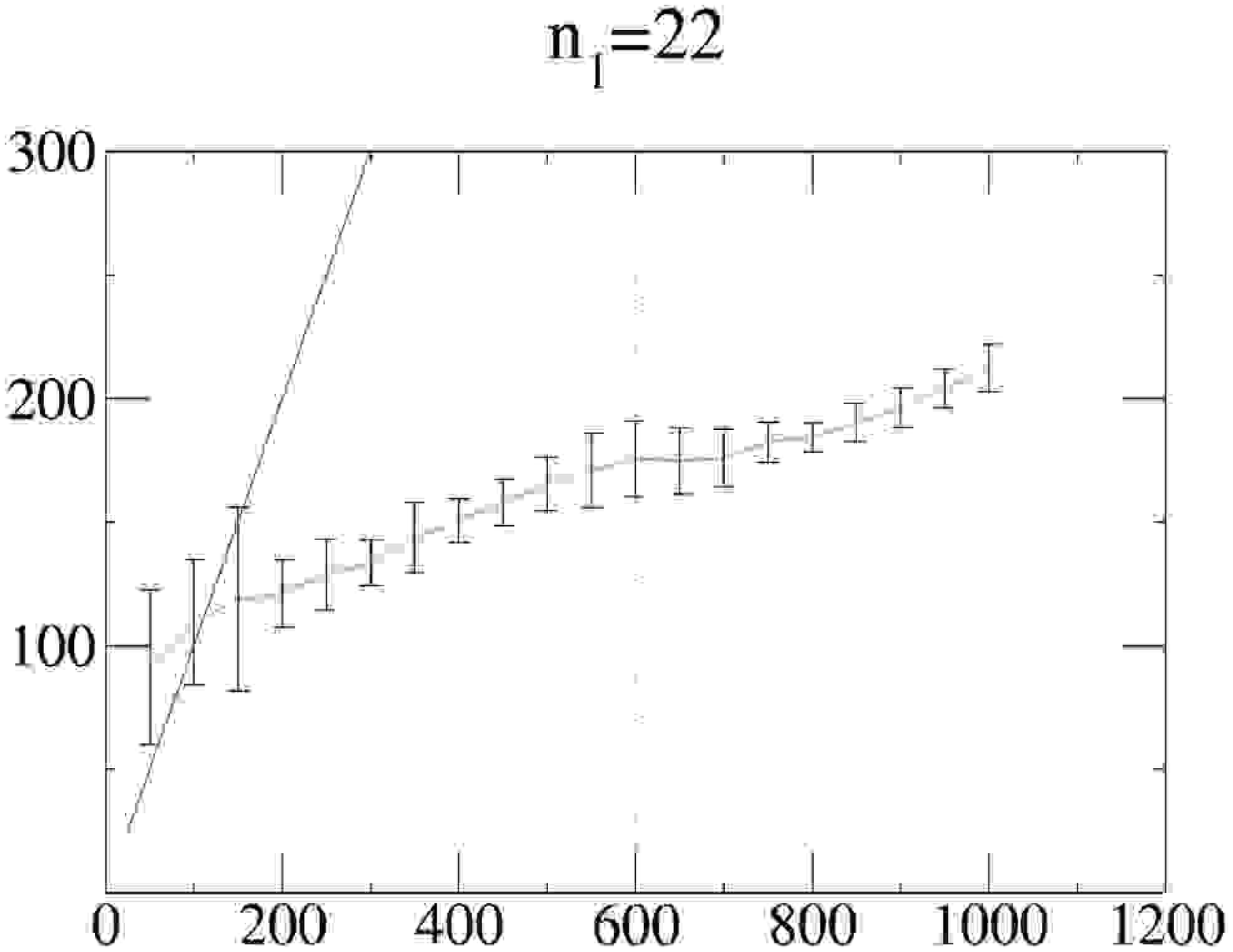} \\ \hline
      \includegraphics[scale=0.22,bb=0 0 27cm 22cm,angle=0,clip=true]%
      {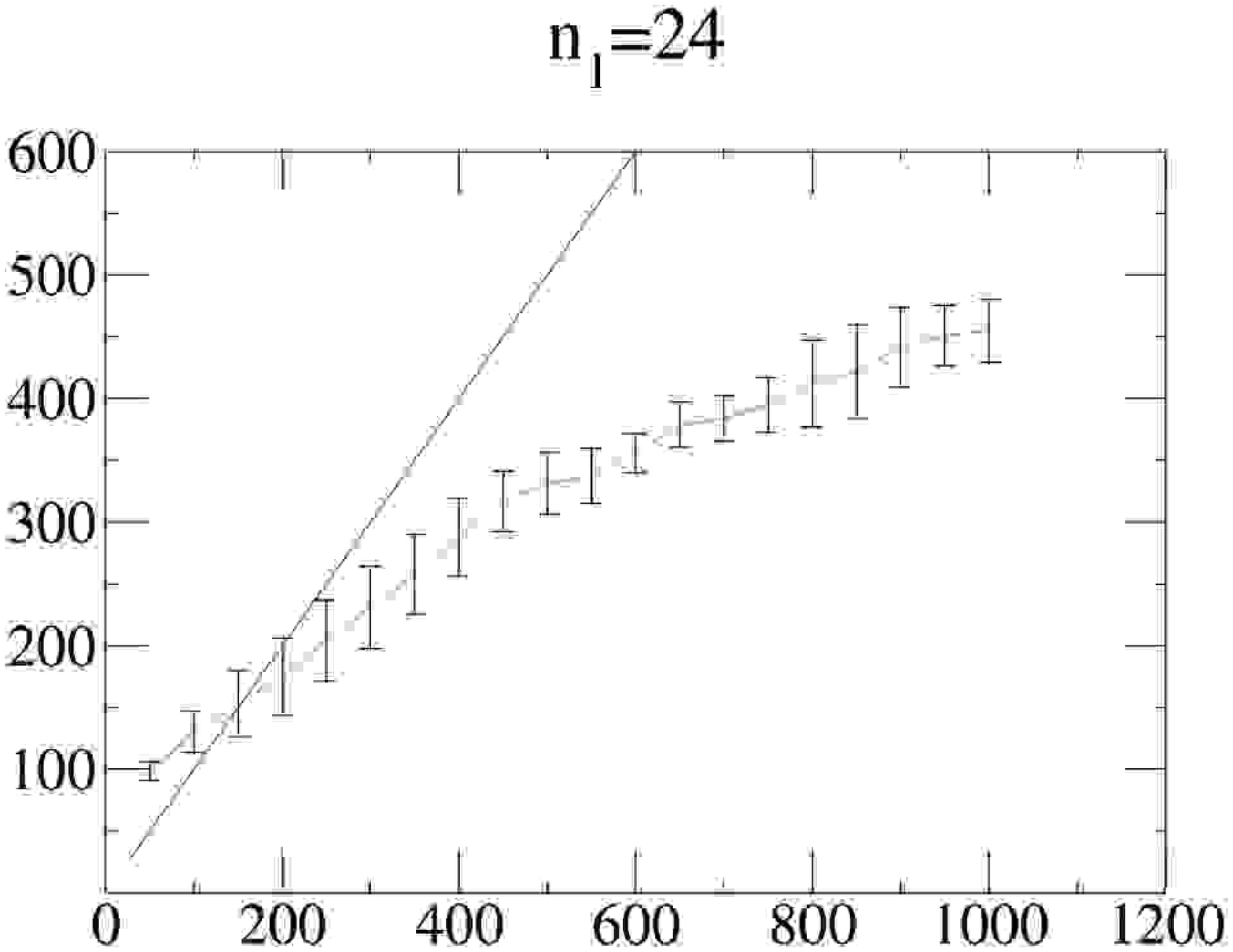} &
      \includegraphics[scale=0.22,bb=0 0 27cm 22cm,angle=0,clip=true]%
      {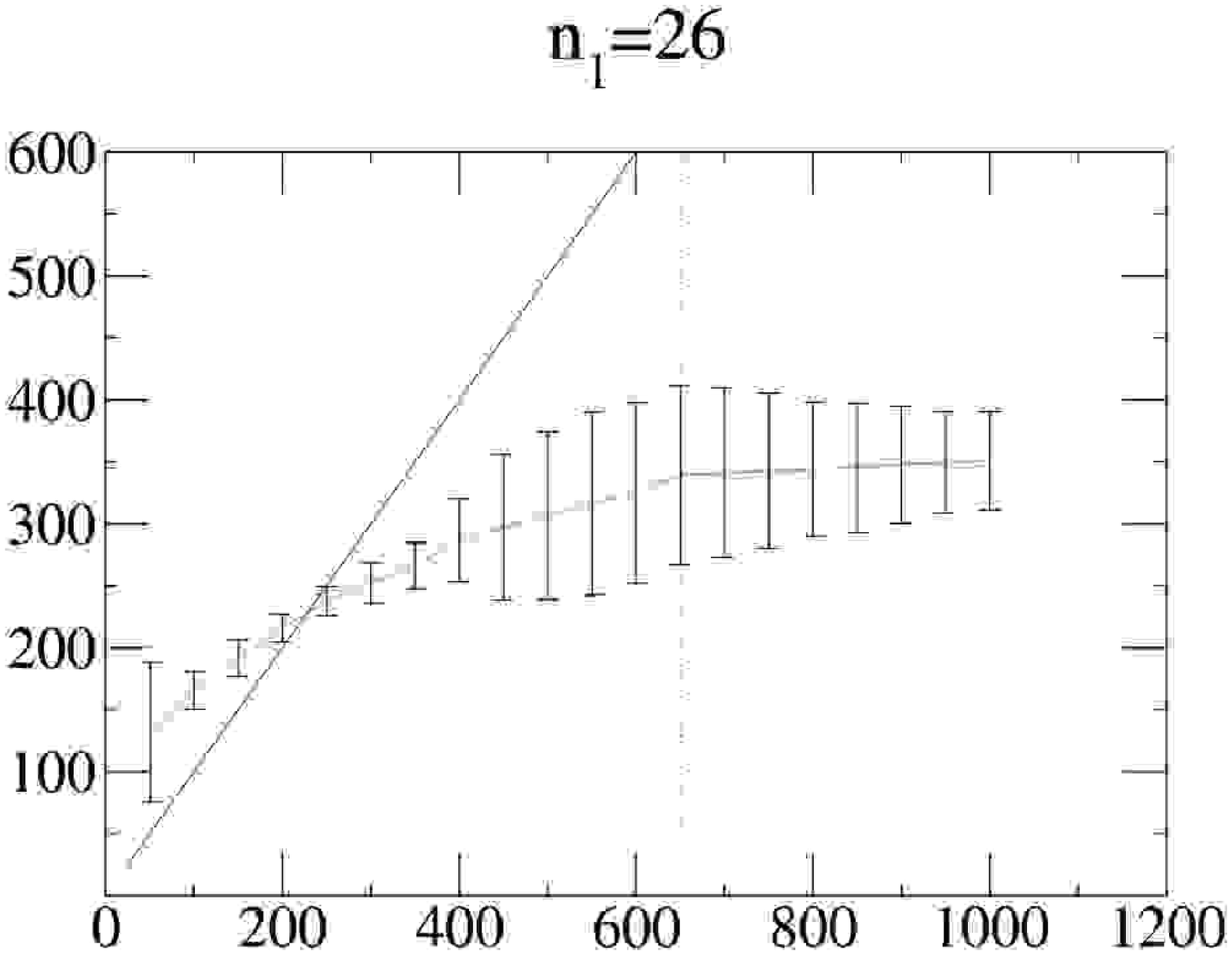} \\ \hline           
      \includegraphics[scale=0.22,bb=0 0 27cm 22cm,angle=0,clip=true]%
      {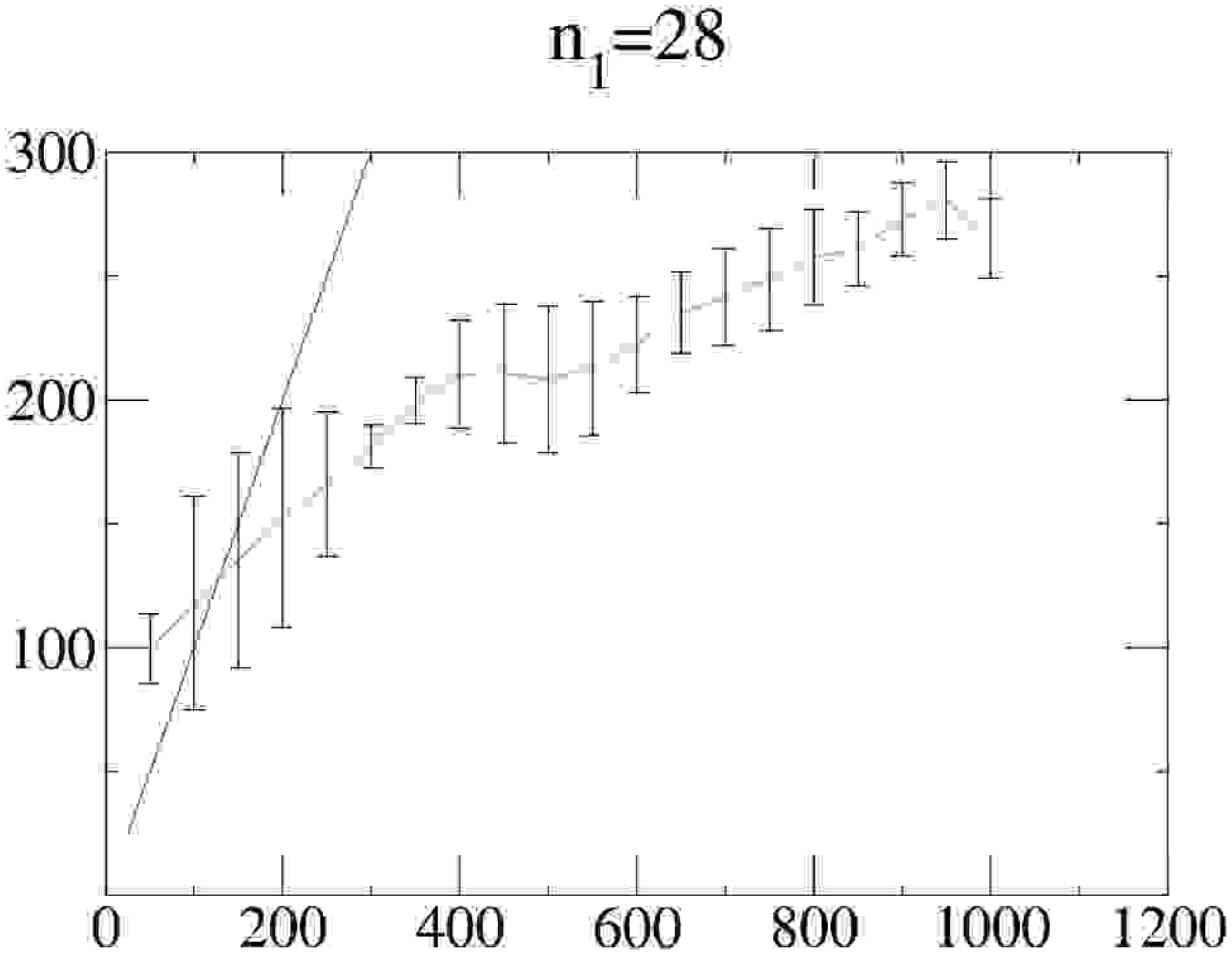} &
      \includegraphics[scale=0.22,bb=0 0 27cm 22cm,angle=0,clip=true]%
      {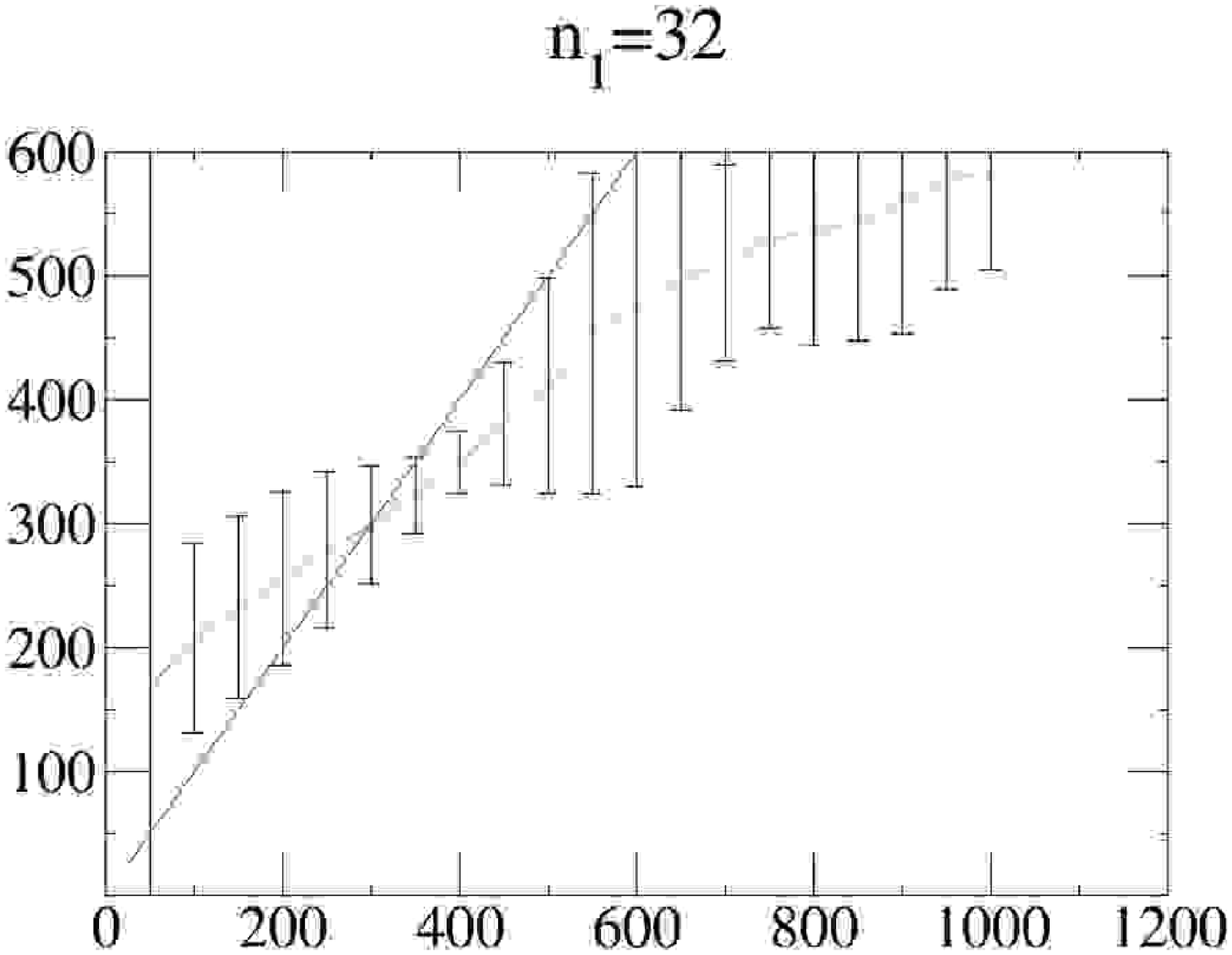}
    \end{tabular}
    \caption{Integrated autocorrelation times of the plaquette
      together with their standard errors as a function of the
      differencing lag for various $n_1$.}
    \label{fig:tau-int-Plaq-lags}
  \end{center}
\end{figure}

The integrated autocorrelation time as a function of the differencing
lag is shown in Fig.~\ref{fig:tau-int-Plaq-lags} for all available
values of $n_1$. The errors are larger than in the case of
$S_{\mbox{\tiny f}}$ since the autocorrelation times are now larger
and thus the statistics for this observable is worse. The left diagram
in the second row has already been discussed in
Sec.~\ref{sec:pract-determ-autoc}, Fig.~\ref{fig:tsmb-diff-tauint}.

In the case $n_1=12$ a plateau is clearly visible, indicating that the
lag-differencing yields a stable solution. In the case $n_1=18$ the
situation is less clear. A pseudo-plateau may be suspected around
$l=600$, but in general the method is unstable and the result should
be disregarded. The case $n_1=20$ has been discussed in
Sec.~\ref{sec:pract-determ-autoc}, while $n_1=22$ is again very stable
with a plateau determined around $l=600$. Absolutely nothing can be
learned from $n_1=24$; there is no plateau and obviously any attempt
to find one is futile. For $n_1=26$ a clear plateau is again visible,
starting at about $l=650$. Regarding $n_1=28$ a plateau can be found
from $l=400$ to $l=600$. For $n_1=32$, again no result can be found.

\begin{figure}[ht]
  \begin{center}
    \begin{tabular}[c]{c|c}
      \includegraphics[scale=0.22,bb=0 0 29cm 22cm,angle=0,clip=true]%
      {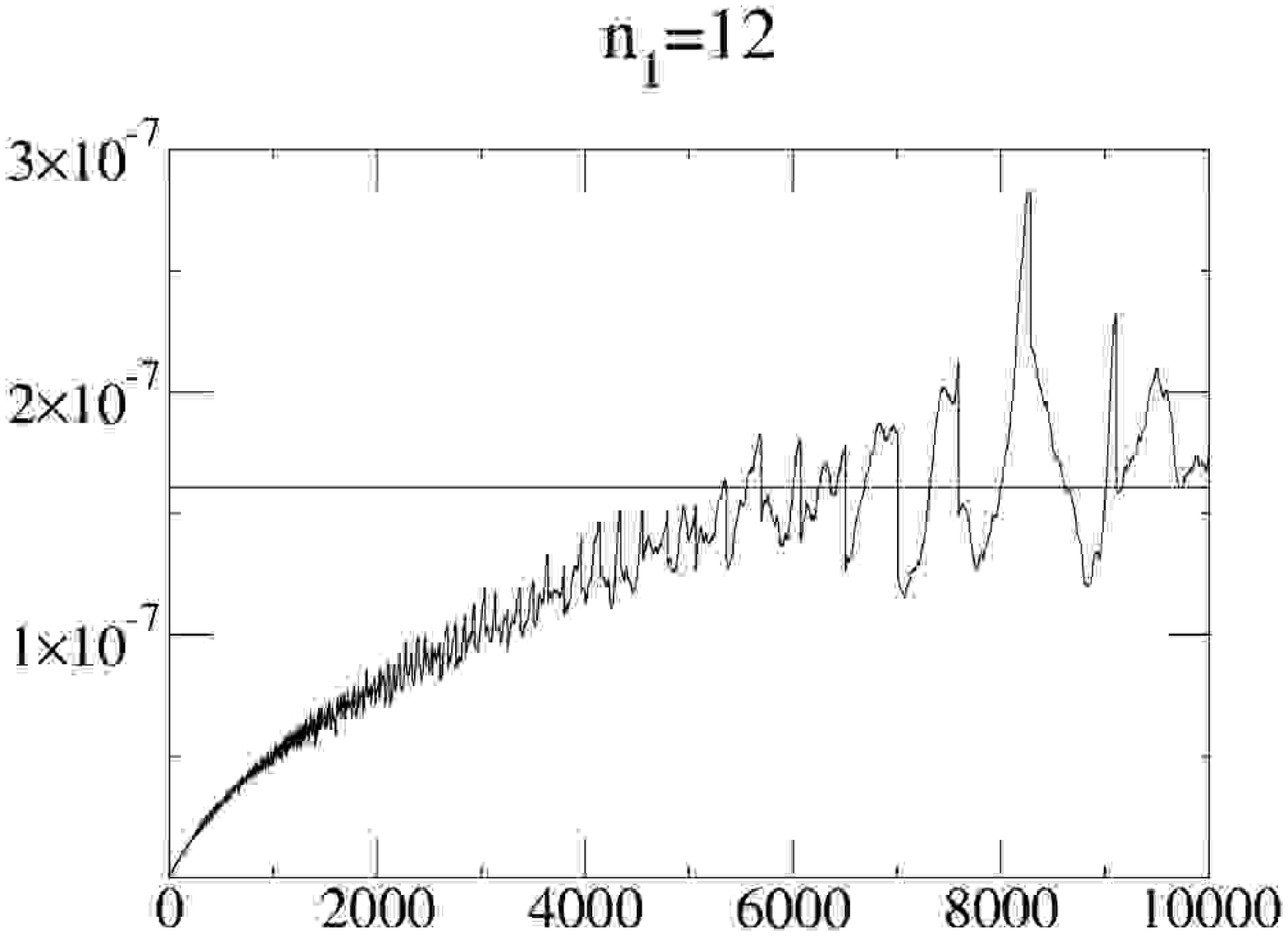} &
      \includegraphics[scale=0.22,bb=0 0 29cm 22cm,angle=0,clip=true]%
      {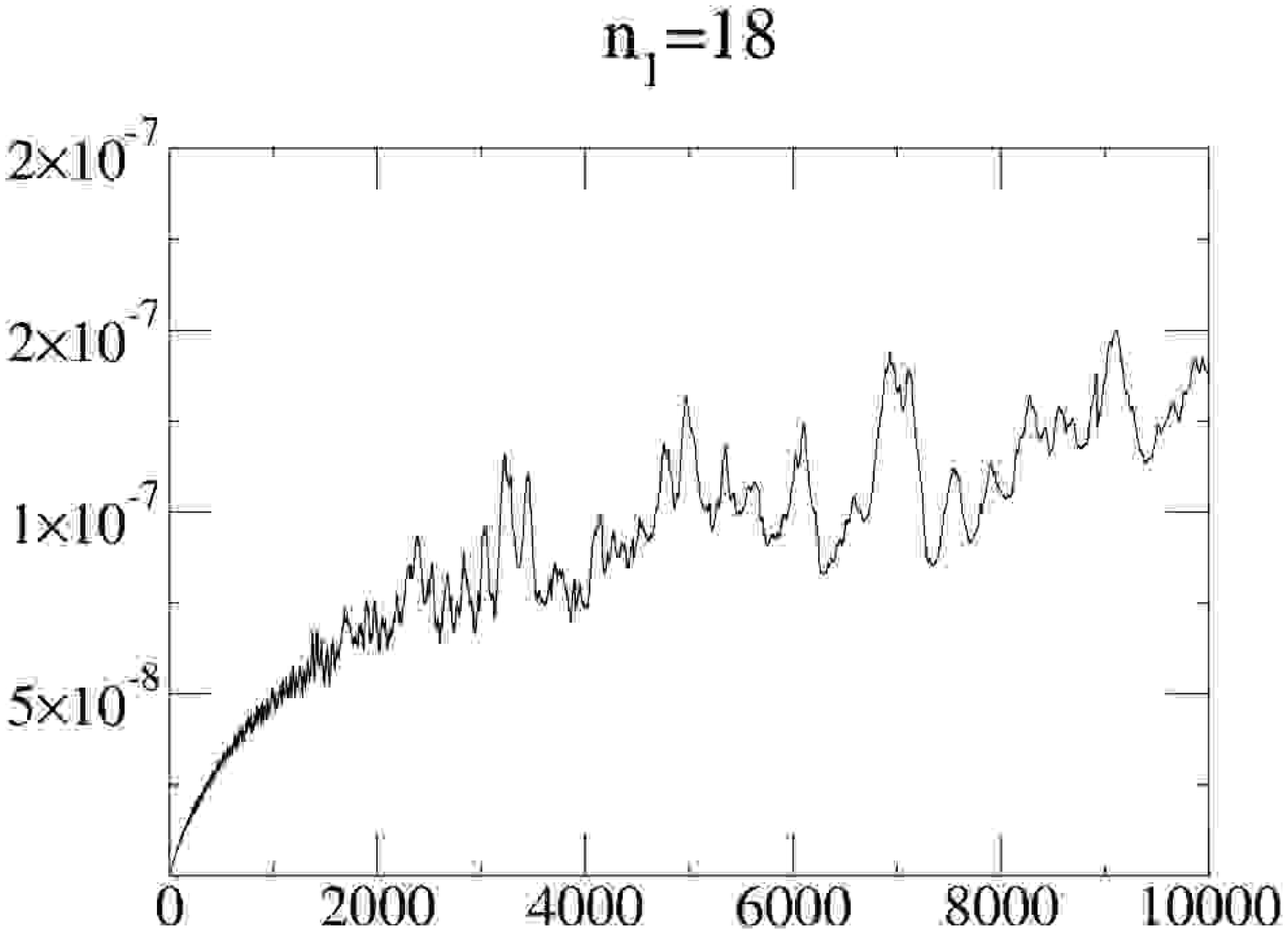} \\ \hline                  
      \includegraphics[scale=0.22,bb=0 0 29cm 22cm,angle=0,clip=true]%
      {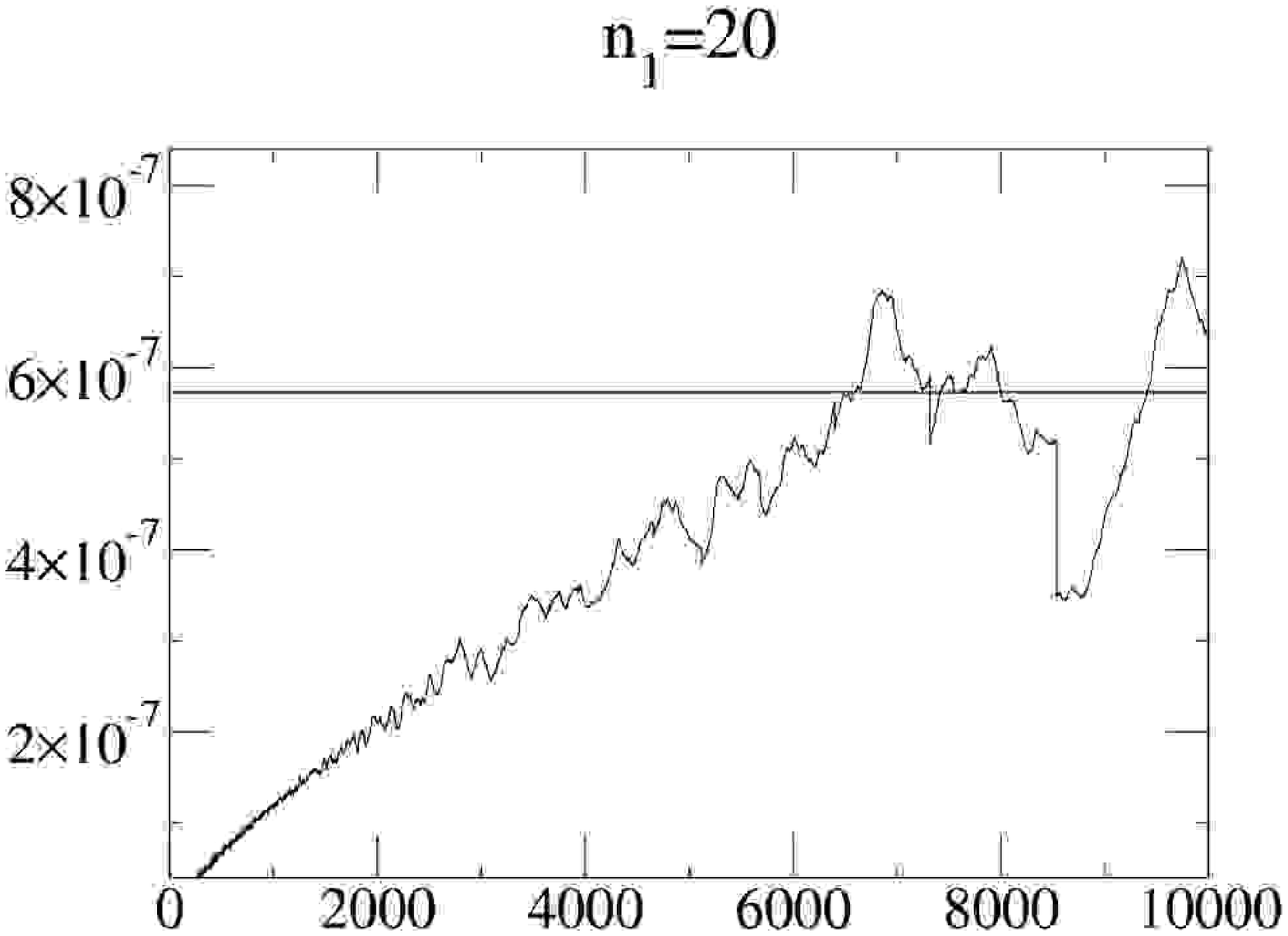} &
      \includegraphics[scale=0.22,bb=0 0 29cm 22cm,angle=0,clip=true]%
      {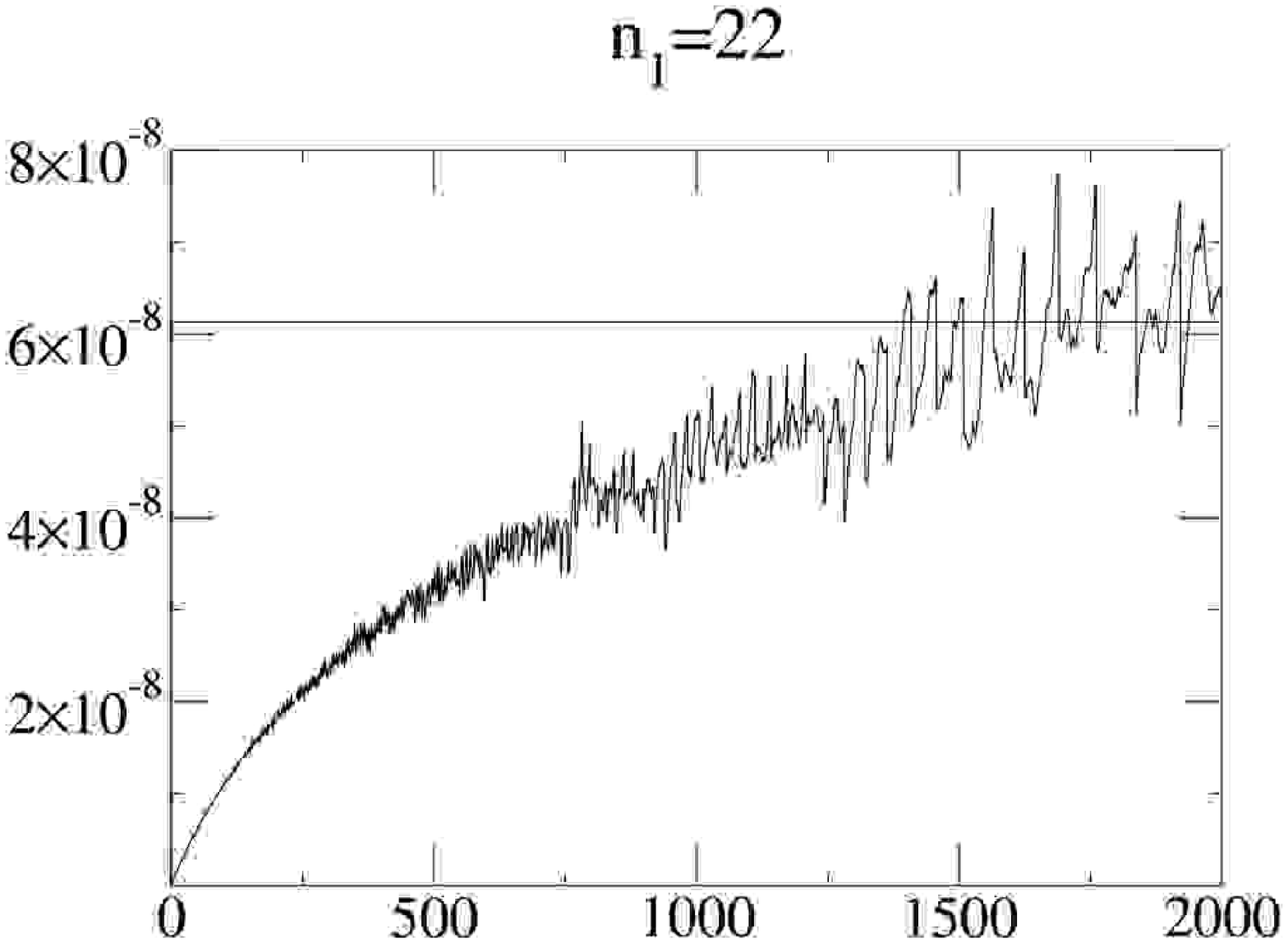} \\ \hline                  
      \includegraphics[scale=0.22,bb=0 0 29cm 22cm,angle=0,clip=true]%
      {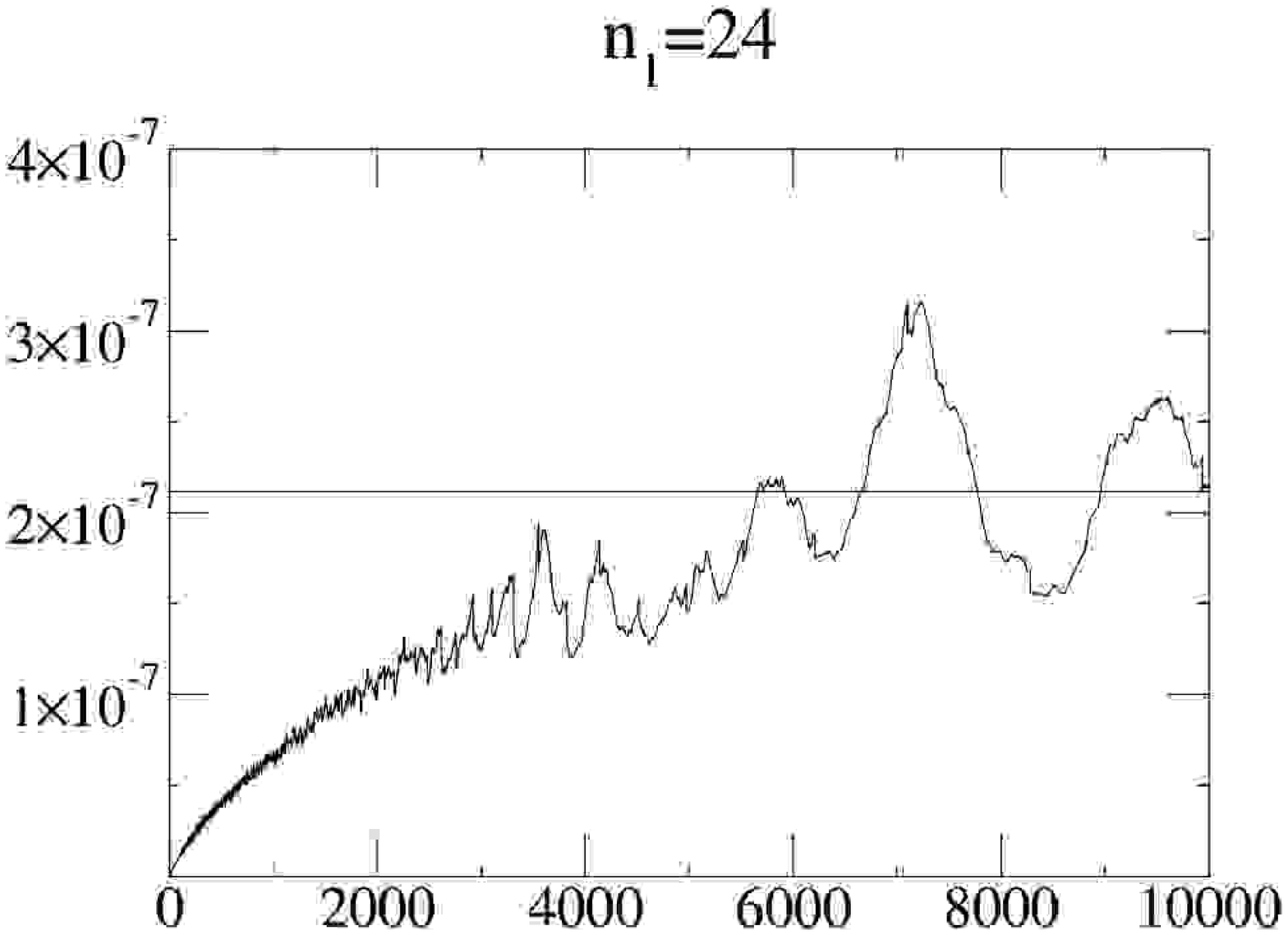} &
      \includegraphics[scale=0.22,bb=0 0 29cm 22cm,angle=0,clip=true]%
      {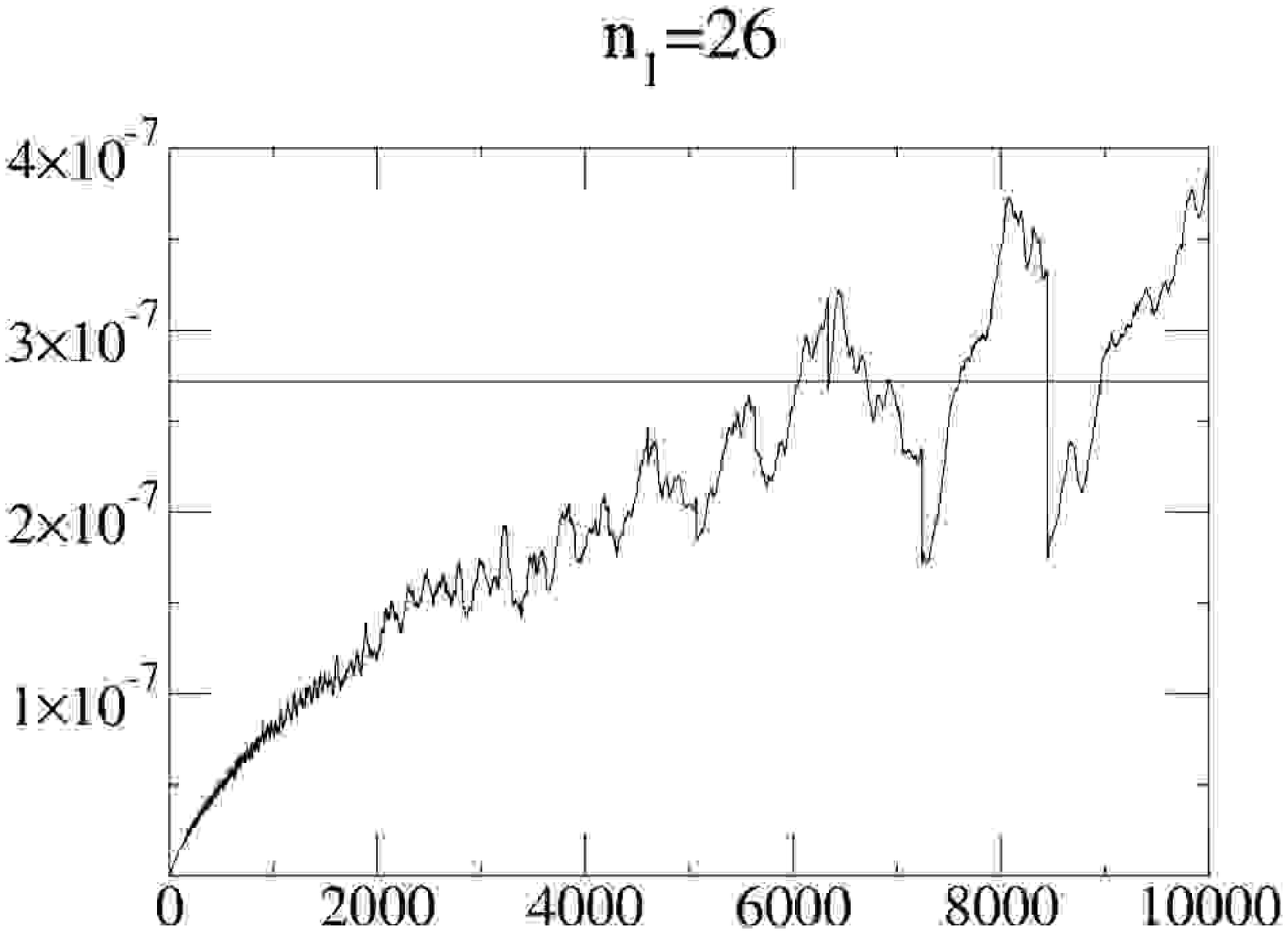} \\ \hline                  
      \includegraphics[scale=0.22,bb=0 0 29cm 22cm,angle=0,clip=true]%
      {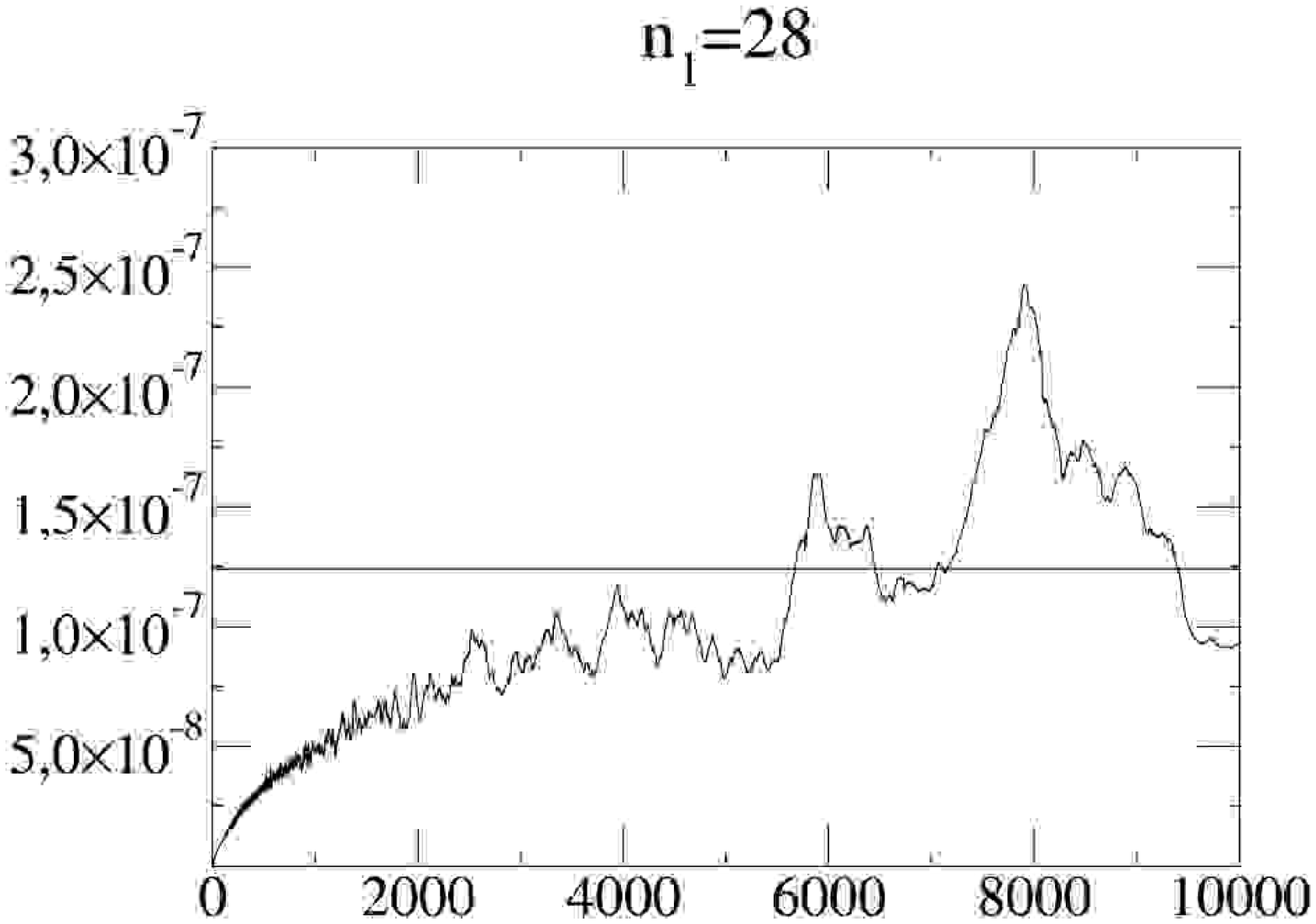} &
      \includegraphics[scale=0.22,bb=0 0 29cm 22cm,angle=0,clip=true]%
      {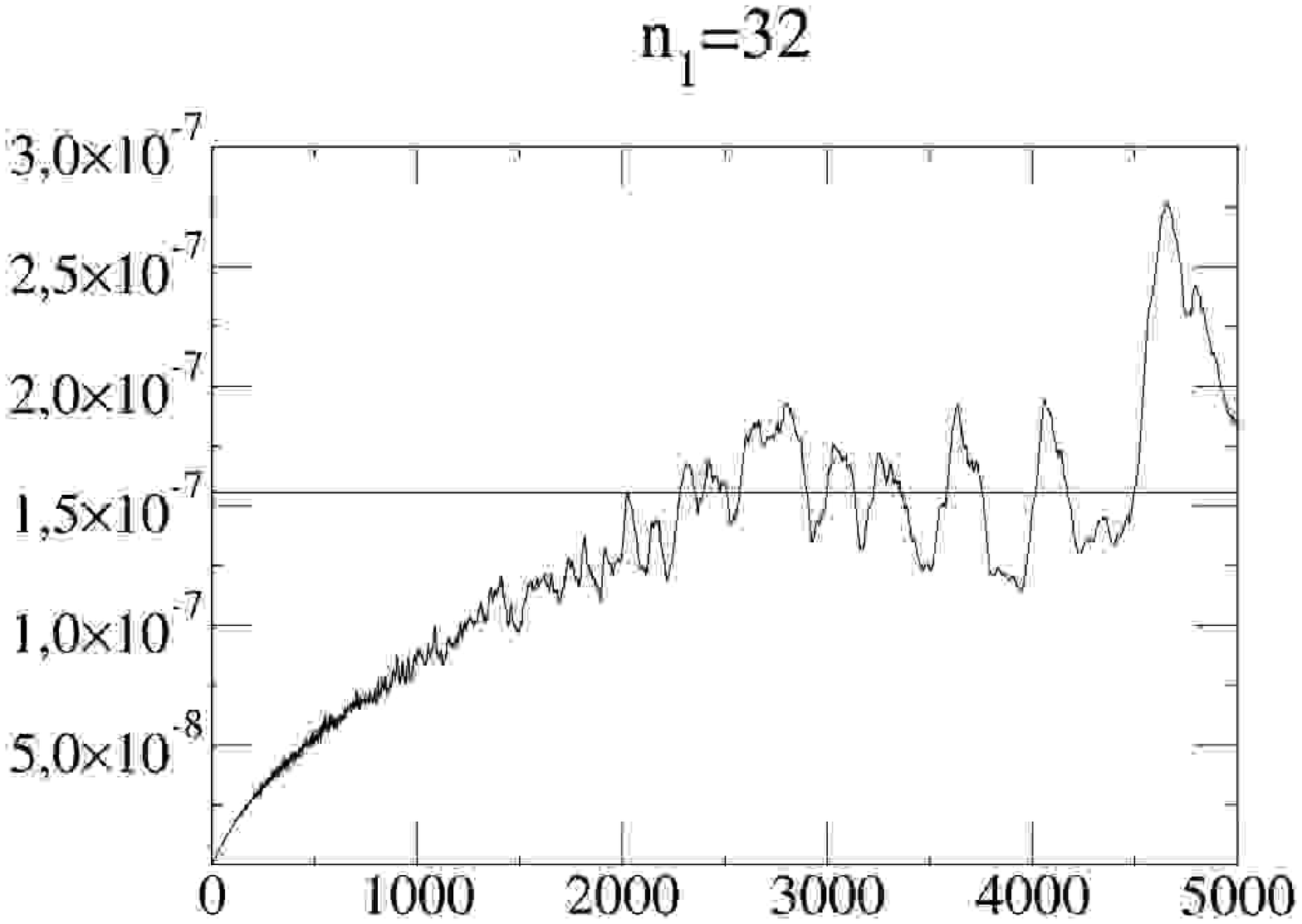}
    \end{tabular}
    \caption{Jackknife variances as a function of the bin size for
      various values of $n_1$.}
    \label{fig:tau-int-Plaq-jack}
  \end{center}
\end{figure}
For comparison the Jackknife procedure as discussed in
Sec.~\ref{sec:short-time-series} has been applied to all samples. The
variances vs.~the bin sizes are displayed in
Fig.~\ref{fig:tau-int-Plaq-jack}. The straight lines give the plateau
values. Again, the first graph in the second row is identical to
Fig.~\ref{fig:tsmb-jackknife} in Sec.~\ref{sec:pract-determ-autoc}.

\begin{figure}[htb]
  \begin{center}
    \includegraphics[scale=0.3,clip=true]{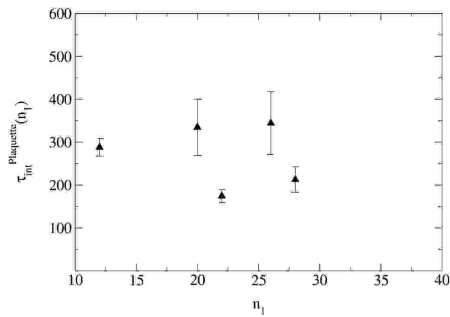}
    \caption{Integrated autocorrelation times of the plaquette
      vs.~$n_1$.}
    \label{fig:tau-int-Plaq}
  \end{center}
\end{figure}

The resulting integrated autocorrelation times read off from
Figs.~\ref{fig:tau-int-Plaq-lags} and~\ref{fig:tau-int-Plaq-jack} are
given in Tab.~\ref{tab:tau-Plaq-result}. Figure~\ref{fig:tau-int-Plaq}
displays these results graphically.

The conclusion is that there is no measurable increase in the
autocorrelation time as $n_1$ is increased if one relies solely on the
lag-differencing method. The increased acceptance rate from larger
$n_1$ compensates for the loss of mobility in phase space. From this
point of view it appears reasonable to simulate at comparatively small
acceptance rates.

The Jackknife method exhibits a similar behavior, but it is compatible
with a decrease of the autocorrelation time with increasing $n_1$.
There is no indication, however, that this decrease exceeds a factor
of two, see the cases $n_1=12$ and $n_1=32$. In fact, it is very
likely that there is a non-trivial dependence (which results in
differences of the order of a factor of two, but not much larger)
which will become visible if one performs the same runs with far
larger statistics. This is impossible with current computer
technology, and would not be worth the effort given the fact that its
influence is so small. The conclusion is therefore unchanged.

\begin{table}[htb]
  \begin{center}
    \begin{tabular}[ht]{r|*{2}{c|}c}
      \hline\hline
      \textbf{$n_1$} & \textbf{$\tau_{\mbox{\tiny int}}$ from LDM} &
      \textbf{Lag $l$} &
      \textbf{$\tau_{\mbox{\tiny int}}$ from Jackknife}   \\
      \hline
      $12$ & $288.0\pm 20.5$ & $400-800$  &  $885\pm 679$ \\
      $18$ & $320.1$ (?)     & $600$ (?)  &  -            \\
      $20$ & $334.3\pm 65.6$ & $600-800$  & $1441\pm 596$ \\
      $22$ & $174.3\pm 15.3$ & $600-700$  &  $203\pm  54$ \\
      $24$ & -               & -          &  $680\pm 333$ \\
      $26$ & $344.3\pm 73.7$ & $650-1000$ &  $779\pm 335$ \\
      $28$ & $212.9\pm 29.6$ & $400-600$  &  $409\pm 390$ \\
      $32$ & -               & -          &  $386\pm  89$ \\
      \hline\hline
    \end{tabular}
    \caption{Integrated autocorrelation times of the plaquette
      together with their standard errors as read off from
      Fig.~\ref{fig:tau-int-Plaq-lags}.}
    \label{tab:tau-Plaq-result}
  \end{center}
\end{table}
\clearpage

\subsection{Dynamical Reweighting Factor}
\label{sec:dynam-rewe-fact}
\begin{figure}[htb]
  \begin{center}
    \includegraphics[scale=0.3,clip=true]{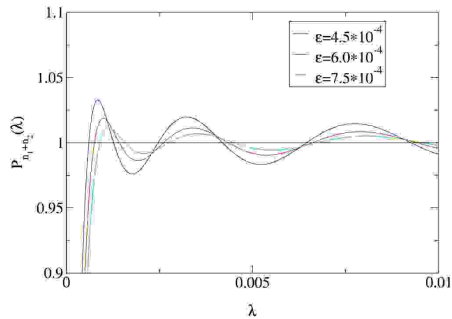}
    \caption{Polynomials $P_{n_1}(\lambda)\cdot P_{n_2}(\lambda)\simeq
      P_{n_1+n_2}(\lambda)$ for the three different values of the
      lower limit.}
    \label{fig:polys-var-eps}
  \end{center}
\end{figure}

\index{Reweighting} \index{Ensembles!multi-canonical} Of particular
importance is the stability of the polynomial approximation with
respect to the interval $\left[\epsilon,\lambda\right]$ chosen for the
TSMB algorithm. To access this problem, three runs have been performed
with different choices for the lowest limit $\epsilon$. The physical
parameters were again chosen according to Tab.~\ref{tab:phys-par} on
an $\Omega=8^4$ lattice using the TSMB algorithm with polynomial order
$n_1=20$ and parameters as in Tab.~\ref{tab:tsmb-multi-genpars} apart
from the value of $\epsilon$. The lower limit has been varied to be
$\epsilon=\Exp{4.5}{-4}$ in the first, $\epsilon=\Exp{6.0}{-4}$ in the
second, and $\epsilon=\Exp{7.5}{-4}$ in the final case. These
polynomials are visualized in Fig.~\ref{fig:polys-var-eps} for the
lower end of the approximation intervals.

\begin{figure}[htb]
  \begin{center}
    \begin{tabular}[c]{l}
      \includegraphics[scale=0.3,clip=true]%
      {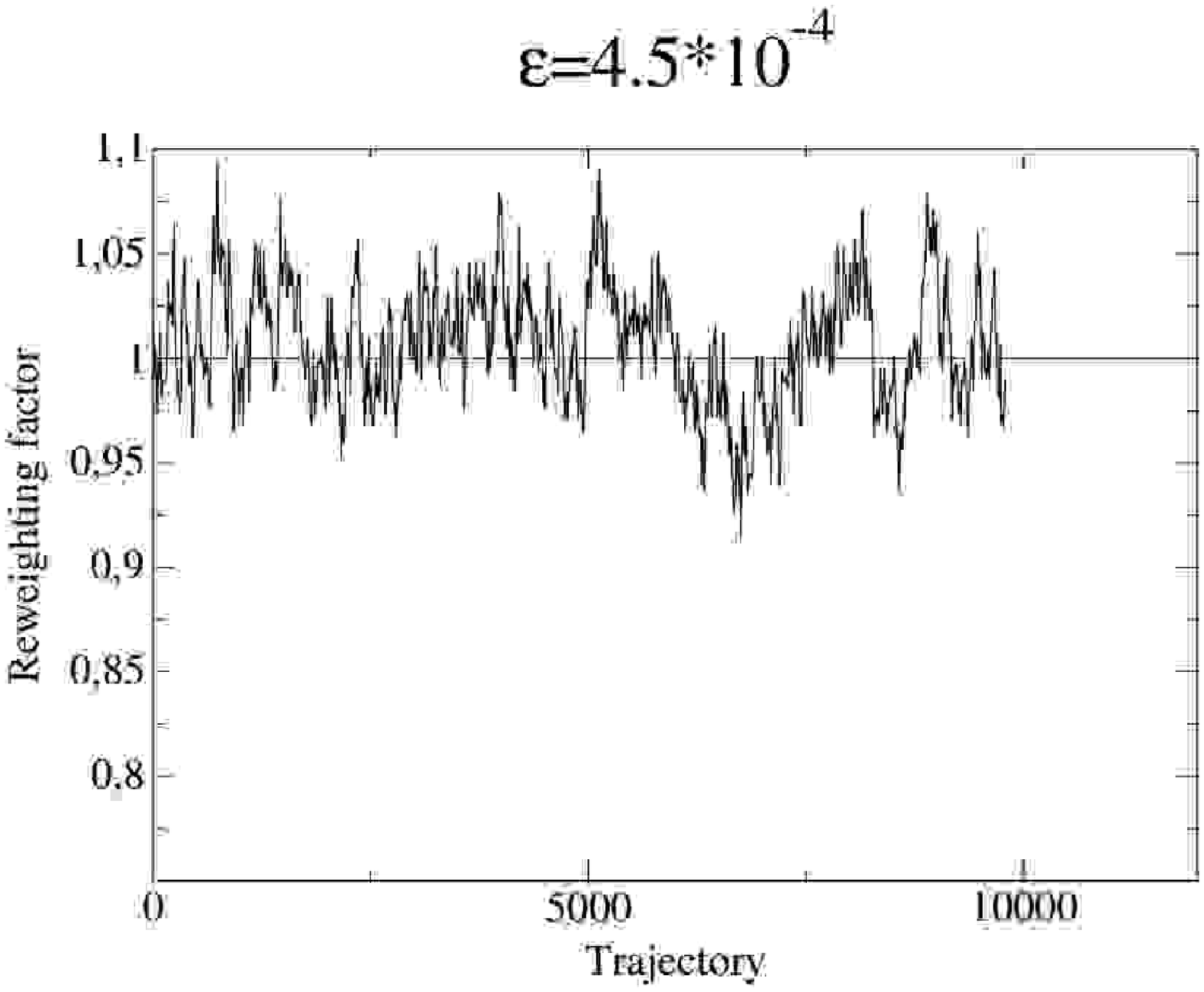} \\
      \includegraphics[scale=0.3,clip=true]%
      {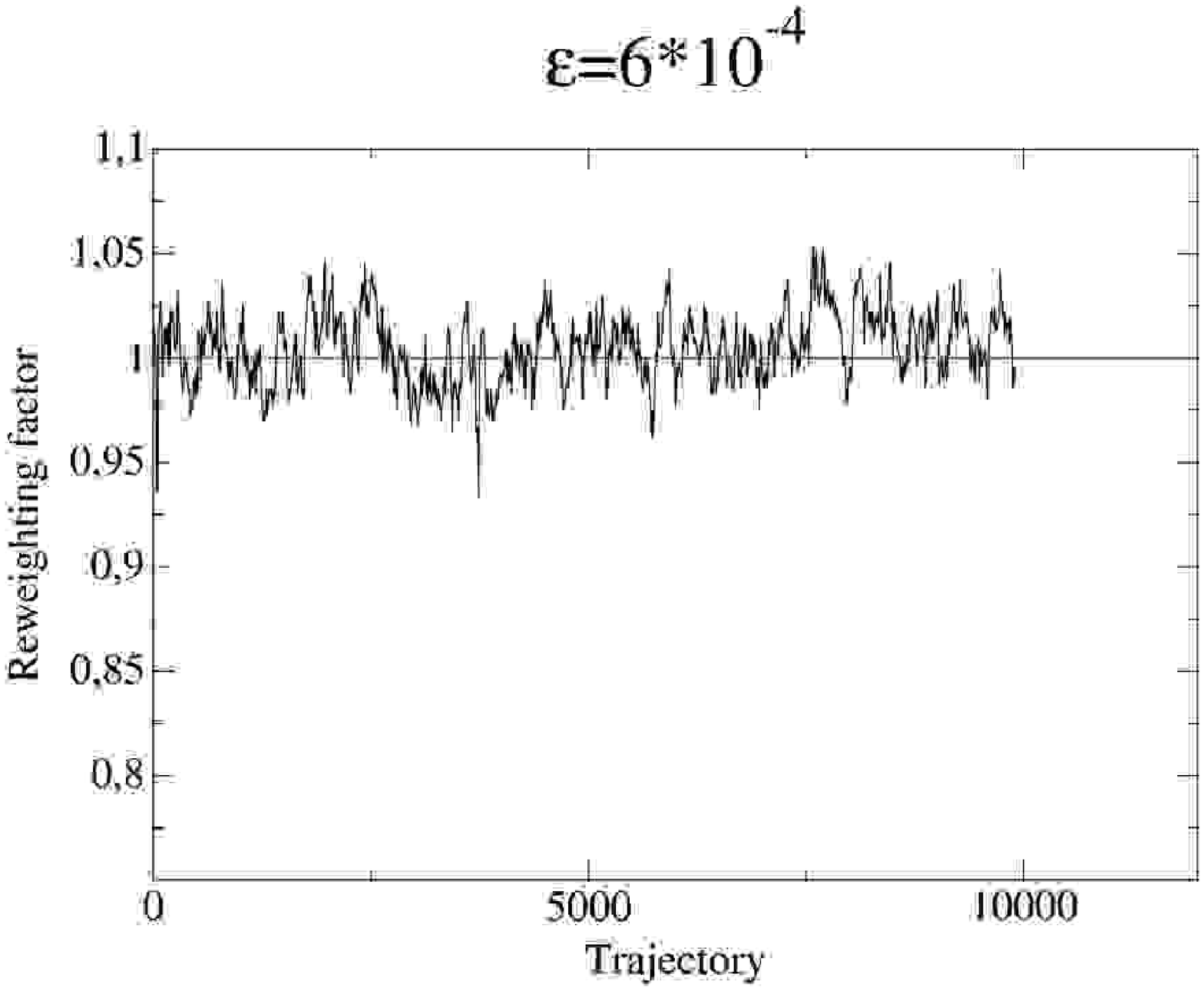}  \\
      \includegraphics[scale=0.3,clip=true]%
      {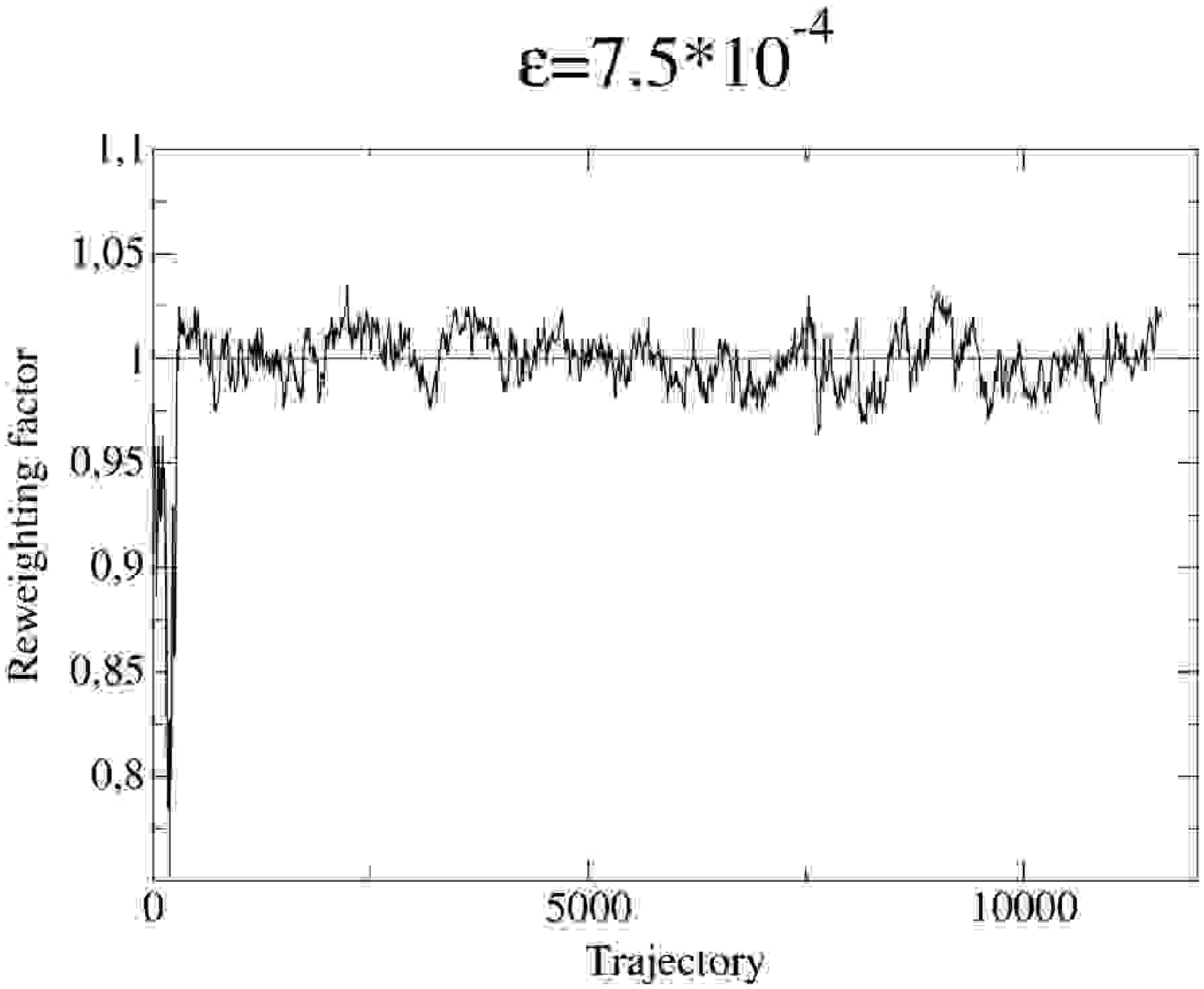} \\
    \end{tabular}
    \caption{Reweighting factors for the three values of $\epsilon$.}
    \label{fig:evreweight-intervalvar}
  \end{center}
\end{figure}

The histories of the reweighting factors measured during the runs are
displayed in Fig.~\ref{fig:evreweight-intervalvar}. They have been
computed from the lowest $32$ eigenvalues
(cf.~Sec.~\ref{sec:rewe-corr}).
\begin{figure}[htb]
  \begin{center}
    \begin{tabular}[c]{l}
      \includegraphics[scale=0.3,clip=true]%
      {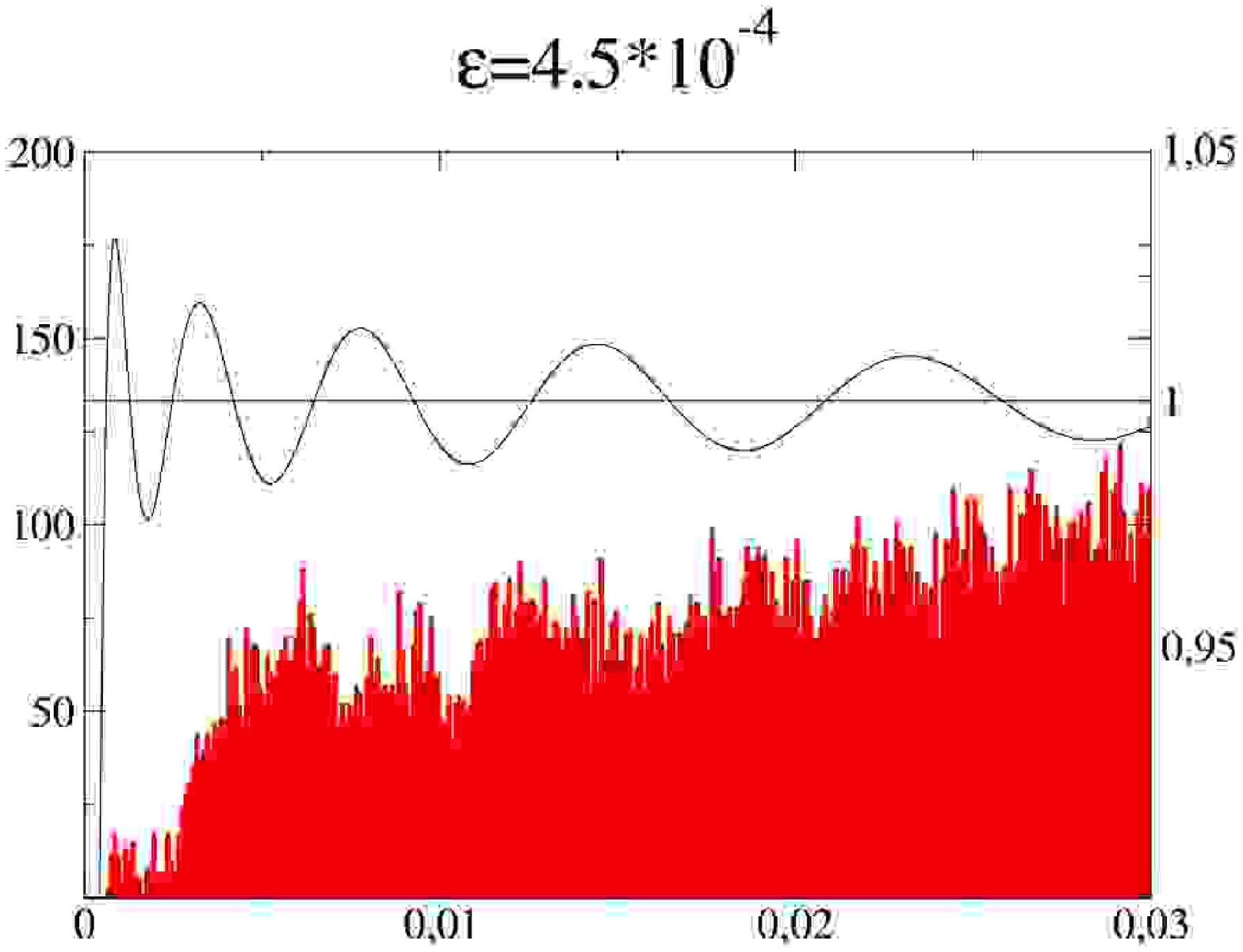} \\
      \includegraphics[scale=0.3,clip=true]%
      {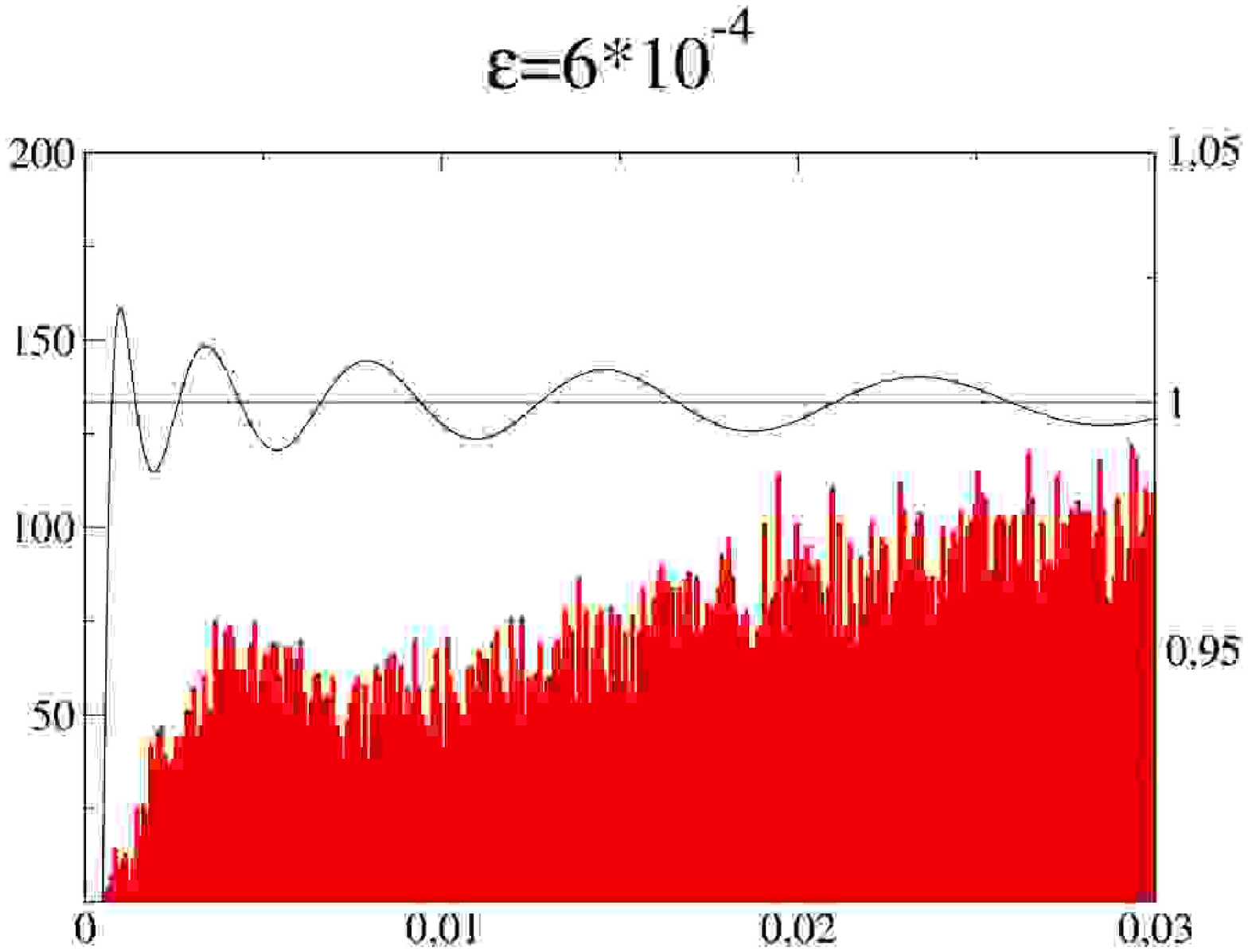}  \\
      \includegraphics[scale=0.3,clip=true]%
      {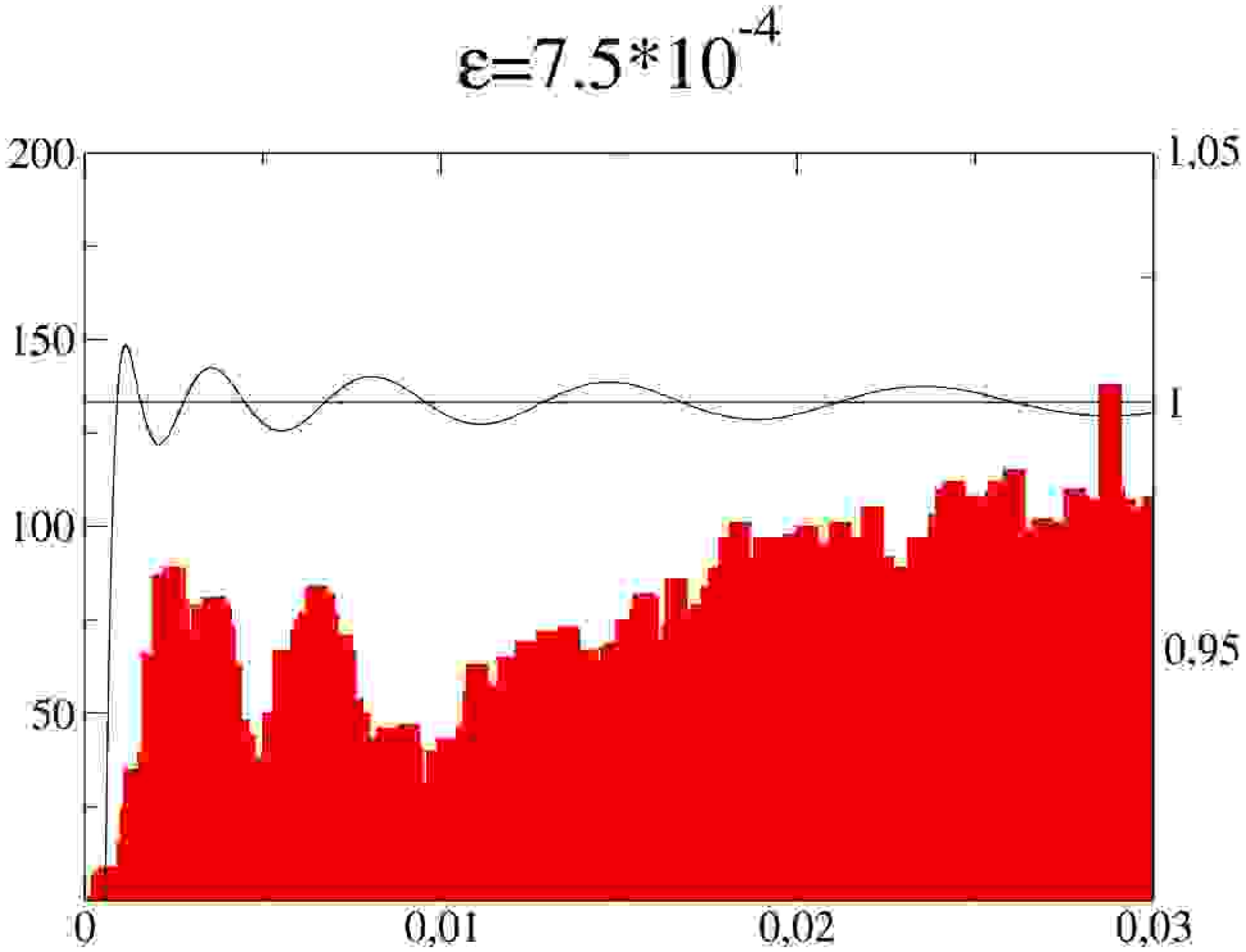} \\
    \end{tabular}
    \caption{Accumulated eigenvalue histograms for the $32$ lowest
      eigenvalues of $\tilde{Q}^2$ for the three runs.}
    \label{fig:ev-histograms}
  \end{center}
\end{figure}
\clearpage

In the first case, $\epsilon_1=\Exp{4.5}{-4}$, the correction factor
shows the largest fluctuation of all cases: $\Delta_1=1.006990\pm
0.028267$. For the second case, $\epsilon_2=\Exp{6.0}{-4}$, one
obtains a correction factor of $\Delta_2=1.005144\pm 0.016978$. In the
third case with $\epsilon_3=\Exp{7.5}{-4}$ the correction factor drops
down to $0.75$ meaning that the lowest eigenvalue left the region
where the polynomial approximation is good enough; although this
appears to be a problem, it only happens at a single place and never
repeats. If this outlier is not disregarded, the correction factor is
found to be $\Delta_3=0.997461\pm 0.021883$. Dropping the first $500$
trajectories from the sample --- which is certainly justified, since
it makes up only $5\%$ of the total runtime --- one arrives at
$\Delta_3=0.999619\pm 0.012296$. When computing any observable,
however, the configurations must not be disregarded, since the
algorithm would otherwise not be ergodic. If it is included in the
measurement via Eq.~(\ref{eq:poly4-rew}), the algorithm will be both
exact and ergodic.

As it is apparent from Fig.~\ref{fig:polys-var-eps} (see also
Fig.~\ref{fig:pn1_quality}), the correction factor oscillates for the
contributions from the smallest eigenvalues when using the static
inversion of the TSMB algorithm. Consequently, an eigenvalue which is
located in a ``valley'' will contribute with a smaller factor than an
eigenvalue located on a local maximum. The static inversion of the
TSMB algorithm may prefer to accumulate the eigenvalues in the local
minima compared to the distribution sampled by an adaptive inverter.
Since the distributions from the adaptive inversion (or from the
reweighted static inversion, respectively) do not know about the
existence of the oscillations, one may suspect that this distortion of
phase space introduces notable \textit{stochastic
  forces}\index{Stochastic forces}. If these forces were present, they
could result in larger autocorrelation times since motion of the
eigenvalues between separate valleys would be suppressed.
Figure~\ref{fig:ev-histograms} shows the histograms of the $32$ lowest
eigenvalues from the three histories of the runs, computed every $100$
trajectories.

There does not appear to be any correlation between the eigenvalue
fluctuations and the shape of the polynomials. As a result, one can
say that the static matrix inversion employed in the TSMB algorithm
does not result in significant stochastic forces. Furthermore, it is
possible to make a choice of $\epsilon$ guided by the optimal lower
limits found in Sec.~\ref{sec:tuning-quadr-optim}. If the lowest
eigenvalue ever leaves this interval, the correction factor will
suppress these configurations. On the average, it will be smaller
(with all other parameters fixed) than any choice with smaller
$\epsilon$. The magnitude of the reweighting factor is then controlled
by the choice of $n_2$. Since a large fluctuation of the reweighting
factor impairs the statistics, the value of $n_2$ should be chosen
such that the reweighting factor is always below the statistical
error. This choice ensures that its influence is so small that the
statistical quality of the sample is not perturbed too much.
Alternatively, one can also compute the determinant norm
$\|R(\tilde{Q}^2)\|$ given by Eq.~(\ref{eq:determinant-norm}) for the
second polynomial and keep its error below the statistical one. The
latter procedure is simpler and still gives a good handle of the
quality attained.

On the other hand, one should refrain from making $n_2$ far too large
and the lower limit $\epsilon$ too small --- this choice will simply
increase the computer time required and make the algorithm
inefficient.

\subsection{Updating Strategy}
\label{sec:updating-strategy}
After the recipe for choosing the polynomials and their orders are
fixed, finally the focus is placed on the tuning of the updating
algorithms for sampling a new configuration (the transition matrix
$\MarkovTrans$ in the Markov process). The available algorithms have
all been discussed in detail in Sec.~\ref{sec:boson-sampl-algor}.

\subsubsection{Boson Field Updates}
\label{sec:boson-field-updates}
The boson field updates should be performed using a global heatbath
(cf.~Sec.~\ref{sec:heatbath-algorithm}) to achieve optimal
decorrelation. This fact is already known theoretically from
\cite{Borici:1996tk} and has been confirmed numerically in
\cite{Alexandrou:1999ii}. The global updates may, however, be
accompanied by local updates, see e.g.~\cite{Jegerlehner:1997px}.

\begin{figure}[htb]
  \begin{center}
    \includegraphics[scale=0.3,clip=true]%
    {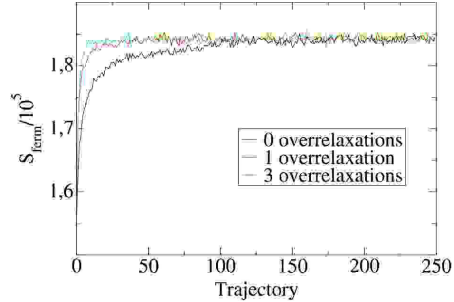}
    \caption{Boson field thermalization using local boson field
      updating algorithms.}
    \label{fig:scafi-therm}
  \end{center}
\end{figure}

Prior to a simulation run, the boson fields should be thermalized.
This can best be achieved by holding the gauge field configuration
fixed and updating the boson fields only. Observing the efficiency of
thermalization methods also allows to shed some light on the best
combination of local updating algorithms.
Figure~\ref{fig:scafi-therm} shows a history of the fermionic energy,
$S_{\mbox{\tiny f}}$, starting from a random boson field
configuration. They display the effect of local overrelaxation sweeps
to the thermalization. The runs have been performed using an
$\Omega=4^4$ lattice. The physical parameters have been chosen as in
Tab.~\ref{tab:phys-par} and the polynomial order has been chosen to be
$n_1=60$. This makes no sense for a production run on this lattice
size, but since the local boson field updates factorize, there should
be no dependence on the number $n_1$ chosen.  The interval of the
polynomial has been chosen to be
$[\epsilon,\lambda]=[\Exp{7.5}{-4},3]$. A trajectory always consisted
of a local boson heatbath update and either $0$, $1$ or $3$ local
boson field overrelaxations. It is obvious, that local boson
overrelaxations improve the thermalization rate and thus should also
be expected to decrease the exponential autocorrelation time.

Of course, in practical implementations, one should not use the local
updating algorithms for the thermalization, but instead directly use
the global boson heatbath.

\subsubsection{Gauge Field Updates}
\label{sec:gauge-field-updates}
For the gauge field updates one can either use a local Metropolis
algorithm (Sec.~\ref{sec:local-metr-update}), a local heatbath (see
Sec.~\ref{sec:heatbath-algorithm}), or local overrelaxations
(Sec.~\ref{sec:overrelaxation}). It turns out that after a certain
number of gauge field updates has been applied the acceptance rate of
the noisy correction step stays essentially constant.

To demonstrate this behavior, a simulation run at the physical
parameters given in Tab.~\ref{tab:gfu-physpar} on a lattice with
$\Omega=8^4$ has been performed. This run is a part of the
investigations performed in Sec.~\ref{sec:case-beta5.3}.
Table~\ref{tab:gfu-genpar} shows the algorithmic parameters in this
study. The only parameter varied is the number of gauge field
Metropolis sweeps, where a Metropolis algorithm with eight hits per
single link has been used.
\begin{table}[htb]
  \begin{center}
    \begin{tabular}[c]{c|c|c}
      \hline\hline
      $\mathbf{N_f^{\mbox{\tiny\textbf{sea}}}}$ & $\mathbf{\beta}$ &
      $\mathbf{\kappa}$ \\ \hline
      $3$ & $5.3$ & $0.150$ \\
      \hline\hline
    \end{tabular}
    \caption{Physical parameters for the investigation of the gauge
      field updating sequence.}
    \label{tab:gfu-physpar}
  \end{center}
\end{table}
\begin{table}[htb]
  \begin{center}
    \begin{tabular}[c]{*{4}{c|}c}
      \hline\hline
      $\mathbf{n_1}$ & $\mathbf{n_2}$ & $\mathbf{n_3}$ &
      $\mathbf{\left[\epsilon,\lambda\right]}$ &
      \textbf{Updates/Trajectory} \\ \hline
      $24$ & $100$ & $140$ & $[\Exp{1}{-2},3]$ & $2$ boson HB, $6$
      boson OR, var.~gauge Metropolis, $1$ noisy corr. \\ \hline
      \multicolumn{5}{c}{Volume: $\Omega=8^4$} \\
      \hline\hline
    \end{tabular}
    \caption{Algorithmic parameters for the investigation of the gauge
      field updating sequence.}
    \label{tab:gfu-genpar}
  \end{center}
\end{table}

The resulting values of the exponential correction together with their
standard deviations and the corresponding acceptance rates are shown
in Tab.~\ref{tab:gfu-varupdate}.
\begin{table}[htb]
  \begin{center}
    \begin{tabular}[c]{*{3}{c|}c}
      \hline\hline
      \textbf{Metropolis sweeps} & $\mathbf{\exp(-C_{12})}$ &
      \textbf{Std.~dev.~of $\mathbf{\exp(-C_{12})}$} &
      \textbf{Acceptance rate} \\ \hline
      $2$      & $0.663$ & $1.125$ & $59\%$ \\
      $4$      & $1.000$ & $1.421$ & $48\%$ \\
      $6$      & $1.350$ & $1.371$ & $41\%$ \\
      $8$      & $1.188$ & $1.523$ & $52\%$ \\
      $12$     & $1.120$ & $1.429$ & $48\%$ \\
      \hline\hline
    \end{tabular}
    \caption{Exponential correction with standard deviation and the
      corresponding acceptance rates as a function of the number of
      gauge field Metropolis sweeps.}
    \label{tab:gfu-varupdate}
  \end{center}
\end{table}

The average exponential correction increases slightly with an
increasing number of sweeps between the noisy corrections. However,
this increase is accompanied by a slight increase in the standard
deviation. Already after six sweeps have been performed, the changes
can be attributed to fluctuations. The net effect is that the
acceptance rate does not vary more than $10\%$. Hence, one finds that
indeed the acceptance rate will saturate once a certain amount of
updates has been applied to the gauge field.

Therefore it is possible to choose a rather large number of gauge
field updates between the noisy corrections.

\subsubsection{Choice of Updating Sequence}
\label{sec:choice-updat-sequ}
During a single trajectory, all d.o.f.~of the system need to be
updated. Therefore, a trajectory always consists of a certain number
of boson field updates and a certain number of gauge field updates
followed by a noisy correction step. The optimal sequence, however,
might also depend on the architecture used for the configuration
sampling. The reason is that the ratio of local to global update
sweeps itself depends on the architecture used, see
Sec.~\ref{sec:impl-syst} below for details.

The acceptance rate is only slightly influenced by the number of local
gauge field sweeps, as has been shown in
Sec.~\ref{sec:acceptance-rates-vs}. Hence, it is obvious that the
noisy correction does not contribute to any reduction of the
autocorrelation time while the update sweeps do. Consequently, one
should always keep the number of local updates in a trajectory as
large as possible, such as to minimize the contribution of the noisy
correction step to the total runtime.

Now two different kinds of trajectories\index{Trajectory} are
proposed:
\begin{itemize}
\item Perform a number of boson field updates, followed by a number of
  gauge field updates and a correction step.
\item Perform an alternating sequence of gauge and boson field updates
  prior to a correction step.
\end{itemize}
Since the correction step only depends on the gauge field
configuration, but not on the boson fields, a rejection in the first
proposal does not imply the necessity of restoring the boson field
configuration since it has always been obtained in the background of a
``valid'' gauge field ``background'' configuration. Conversely, the
second scheme will have to restore both the old boson and gauge field
configurations in case the update is rejected. Hence, the memory
requirements of the second proposal will be significantly larger than
in the first case.

These considerations do not yet fix the optimal mixture of gauge and
boson field updates. A very simple proposal is to use a single local
gauge overrelaxation sweep followed by a single local boson
overrelaxation sweep. The overrelaxation sweeps allow for a very fast
movement through phase space, but do not ensure ergodicity, see
Sec.~\ref{sec:overrelaxation}. So this sequence has to be complemented
by at least one ergodic heatbath and/or Metropolis sweep. This simple
sequence turns out to be quite efficient in practice, see
Sec.~\ref{sec:setting-up-mbor} for an application.

However, given the fact that the fermionic energy decorrelates much
faster than the gauge field plaquette (see
Sec.~\ref{sec:acceptance-rates-vs} above), one might hope that
subsequent updates of the gauge field alone might still decrease the
plaquette autocorrelation, although essentially only a small subset of
all d.o.f.~is being updated. The reason why this could be efficient
is, as will be shown below in Sec.~\ref{sec:arch-effic}, that the
caching of the boson field contributions as proposed in
App.~\ref{sec:local-forms-actions} in
Eqs.~(\ref{eq:ferm-staple-cache}) and (\ref{eq:cache-def}) allows for
a larger number of subsequent gauge-field-only update sweeps. On the
other hand it is clear that this effect will very soon lead to a
saturation since the gauge field will thermalize with the fixed boson
field background.

The question when this saturation occurs can only be answered in a
practical simulation. The physical parameters have again been chosen
from Tab.~\ref{tab:phys-par}, and the algorithmic parameters are given
in Tab.~\ref{tab:diff-trajs-polys}. They are identical to the
parameters given in Tab.~\ref{tab:tsmb-multi-genpars}.
\begin{table}[htb]
  \begin{center}
    \begin{tabular}[c]{*{3}{c|}c}
      \hline\hline
      $\mathbf{n_1}$ & $\mathbf{n_2}$ & $\mathbf{n_3}$ &
      $\mathbf{[\epsilon,\lambda]}$ \\ \hline
      $20$ & $160$ & $200$ & $[\Exp{7.5}{-4},3]$ \\ \hline
      \multicolumn{4}{c}{Volume: $\Omega=8^4$} \\
      \hline\hline
    \end{tabular}
    \caption{Polynomial and lattice volume for runs with different
      updating sequences.}
    \label{tab:diff-trajs-polys}
  \end{center}
\end{table}

Table~\ref{tab:diff-trajs-setup} shows the different sequences used
for a number of simulations and the number of trajectories computed
together with the total cost in MV-Mults for a single trajectory. The
machine configuration used for all runs was an eight-node partition of
the \ALiCE\ computer cluster; parallelization was used in the $z$- and
$t$-direction resulting in local lattices of $\Omega_{\mbox{\tiny
    loc}}=2\times 8\times 8\times 4$.
\begin{table}[htb]
  \begin{center}
    \begin{tabular}[c]{l|c|c|c}
      \hline\hline
      & \textbf{Updating strategy} & \textbf{Trajectories} &
      \textbf{Cost/MV-Mults} \\ \hline
      Sequence I   & $2$ boson HB, $6$ boson OR,
                   & $71400$ & $1208.15$            \\
                   & $8$ gauge OR, $1$ noisy corr.& \\
      Sequence II  & $1$ boson HB, $3$ boson OR,
                   & $68400$ & $1403.77$            \\
                   & $16$ gauge OR, $1$ noisy corr.&\\
      Sequence III & $2$ boson HB, $16$ boson OR,
                   & $31400$ & $1631.09$            \\
                   & $8$ gauge OR, $1$ noisy corr.& \\
      Sequence IV  & $1$ boson HB, $3$ boson OR,
                   & $40300$ & $1123.14$            \\
                   & $2$ gauge Metro, $1$ noisy corr.,&\\
                   & $2$ gauge OR, $1$ noisy corr.& \\
      \hline\hline
    \end{tabular}
    \caption{Different updating sequences used for a single
      trajectory.}
    \label{tab:diff-trajs-setup}
  \end{center}
\end{table}

Sequence I corresponds to an intermediate number of boson sweeps and
gauge field sweeps. For the gauge field local overrelaxations have
been used. Hence, ergodicity is ensured by the boson field heatbath
only. Sequence II applies a small number of boson field sweeps, but a
large number of gauge field sweeps. Sequence III applies a large
number of boson field updates and an intermediate number of gauge
field updates. Sequence IV consists of a small number of boson field
updates and only an intermediate number of gauge field updates. The
latter run makes direct contact with the run in
Sec.~\ref{sec:acceptance-rates-vs} at $n_1=20$. It will be denoted by
``Sequence 0''.

In the following, we will not only consider the autocorrelation times
in terms of trajectories, $\tau_{\mbox{\tiny int}}$, but also the
efficiencies of the algorithms, $E_{\mbox{\tiny indep}}$. These
efficiencies are defined as the number of MV-Mults required per
statistically independent gauge field configuration,
\begin{equation}
  \label{eq:eff-def}
  E_{\mbox{\tiny indep}} = 2 E_{\mbox{\tiny MV-mults}}
  \tau_{\mbox{\tiny int}}\,,
\end{equation}
where $E_{\mbox{\tiny MV-mults}}$ has been defined in
Eq.~(\ref{eq:numerical-effort}) and $\tau_{\mbox{\tiny int}}$ is the
integrated autocorrelation time of the observable under consideration,
in this case the plaquette. Thus, the quantity $E_{\mbox{\tiny
    indep}}$ is a measure for the total cost which is, to a large
extent, independent of the technical details of the underlying
algorithm.

\begin{figure}[ht]
  \begin{center}
    \begin{tabular}[c]{l|r}
      \includegraphics[scale=0.22,bb=0 0 27cm 22cm,angle=0,clip=true]%
      {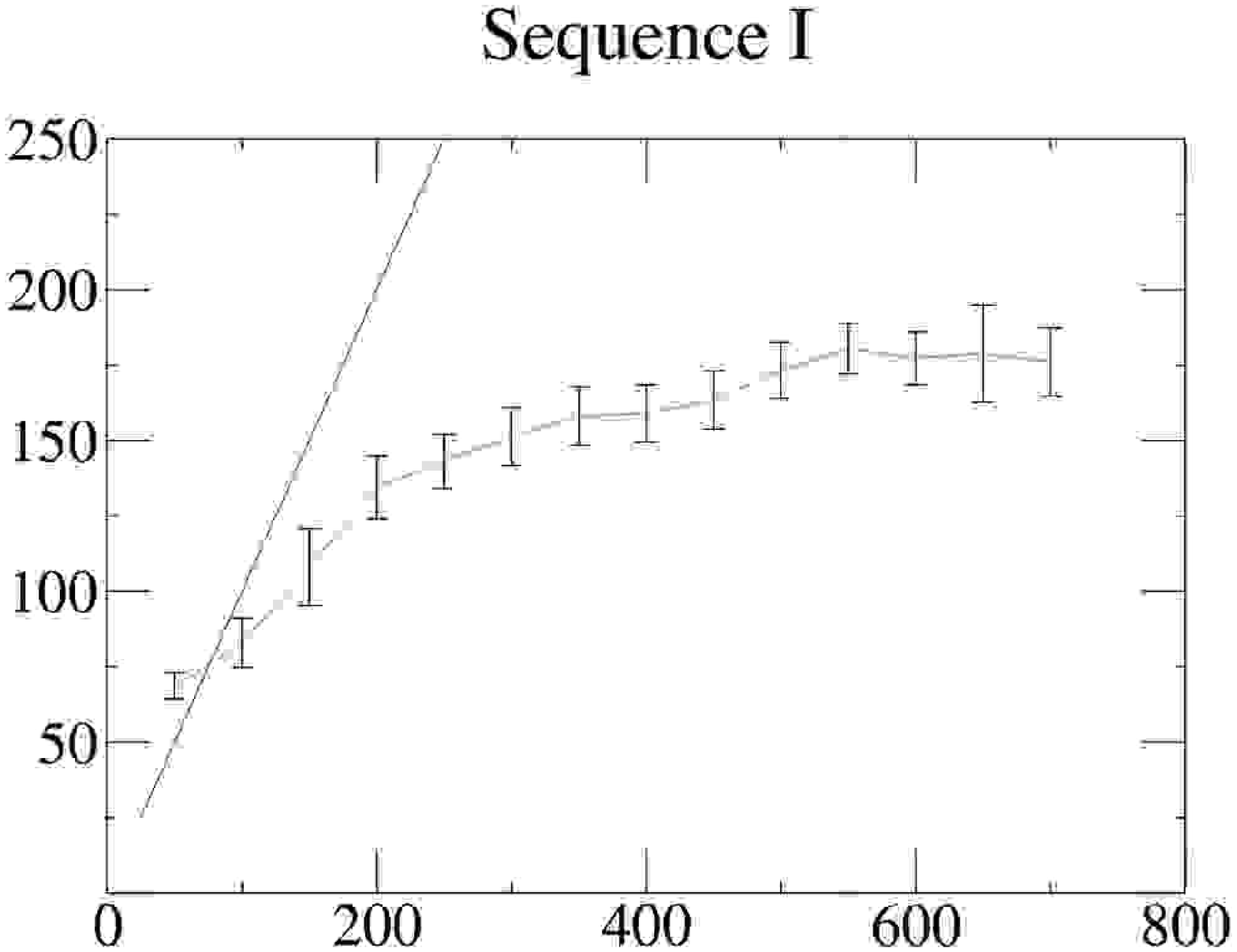} &
      \includegraphics[scale=0.22,bb=0 0 27cm 22cm,angle=0,clip=true]%
      {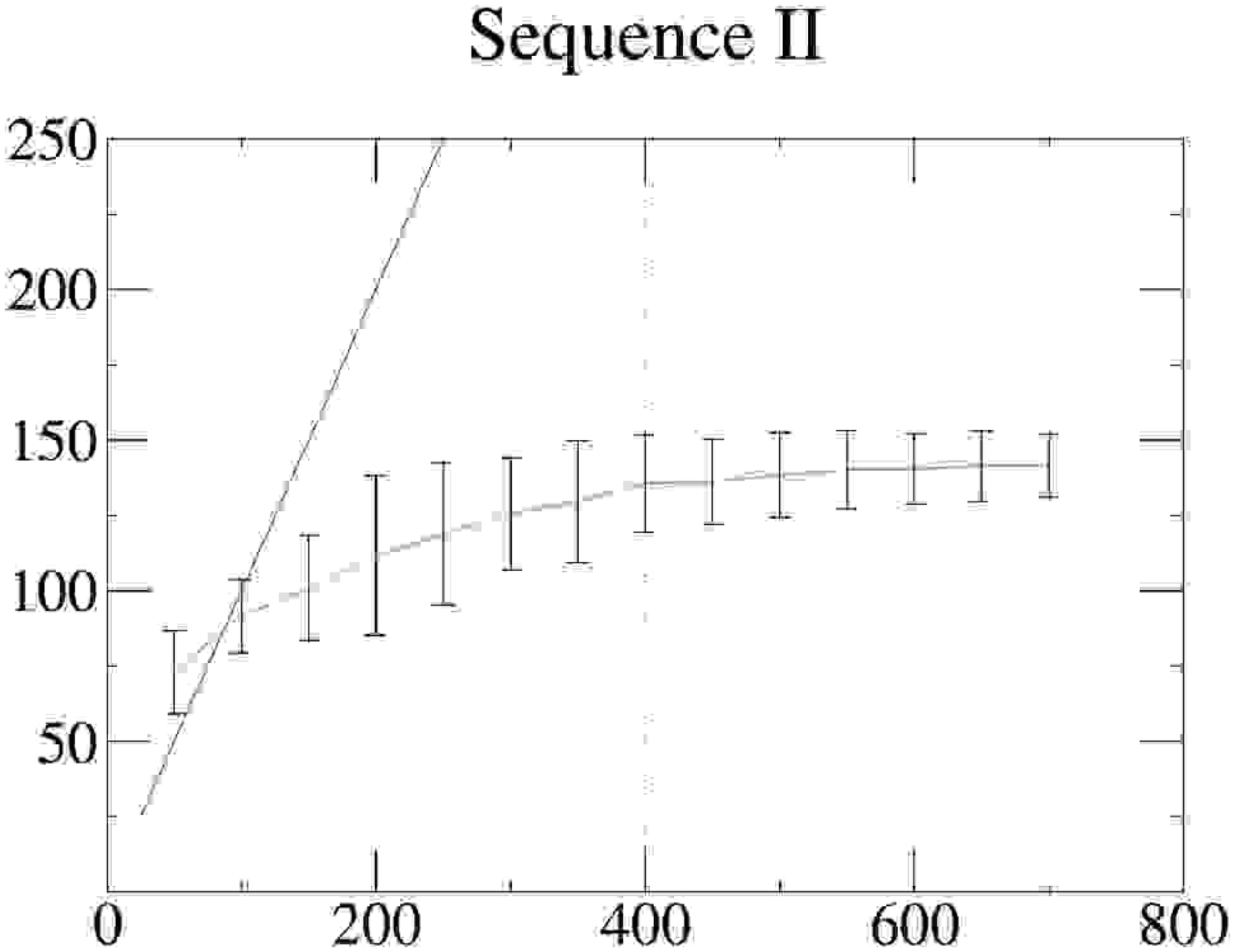} \\ \hline                
      \includegraphics[scale=0.22,bb=0 0 27cm 22cm,angle=0,clip=true]%
      {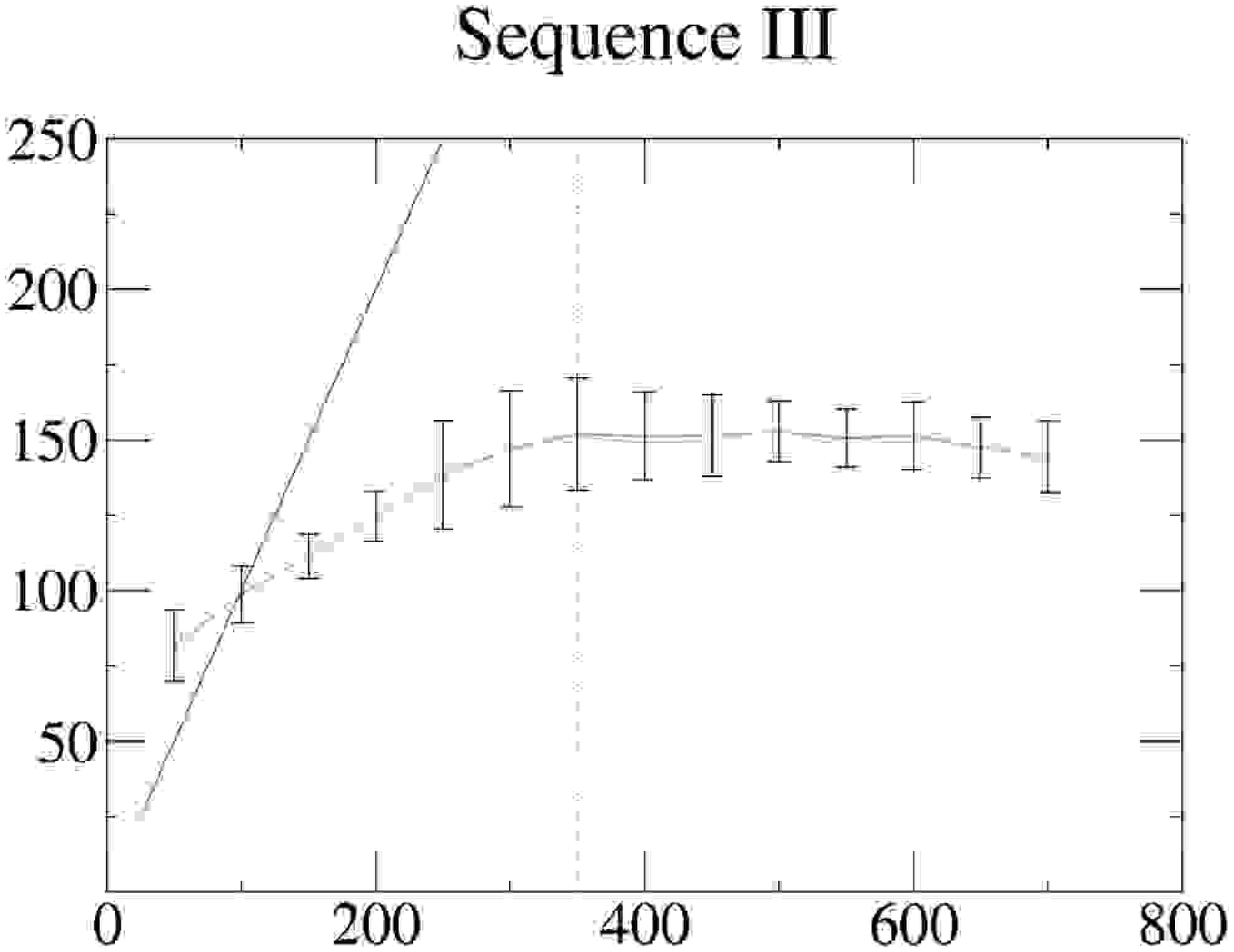} &
      \includegraphics[scale=0.22,bb=0 0 27cm 22cm,angle=0,clip=true]%
      {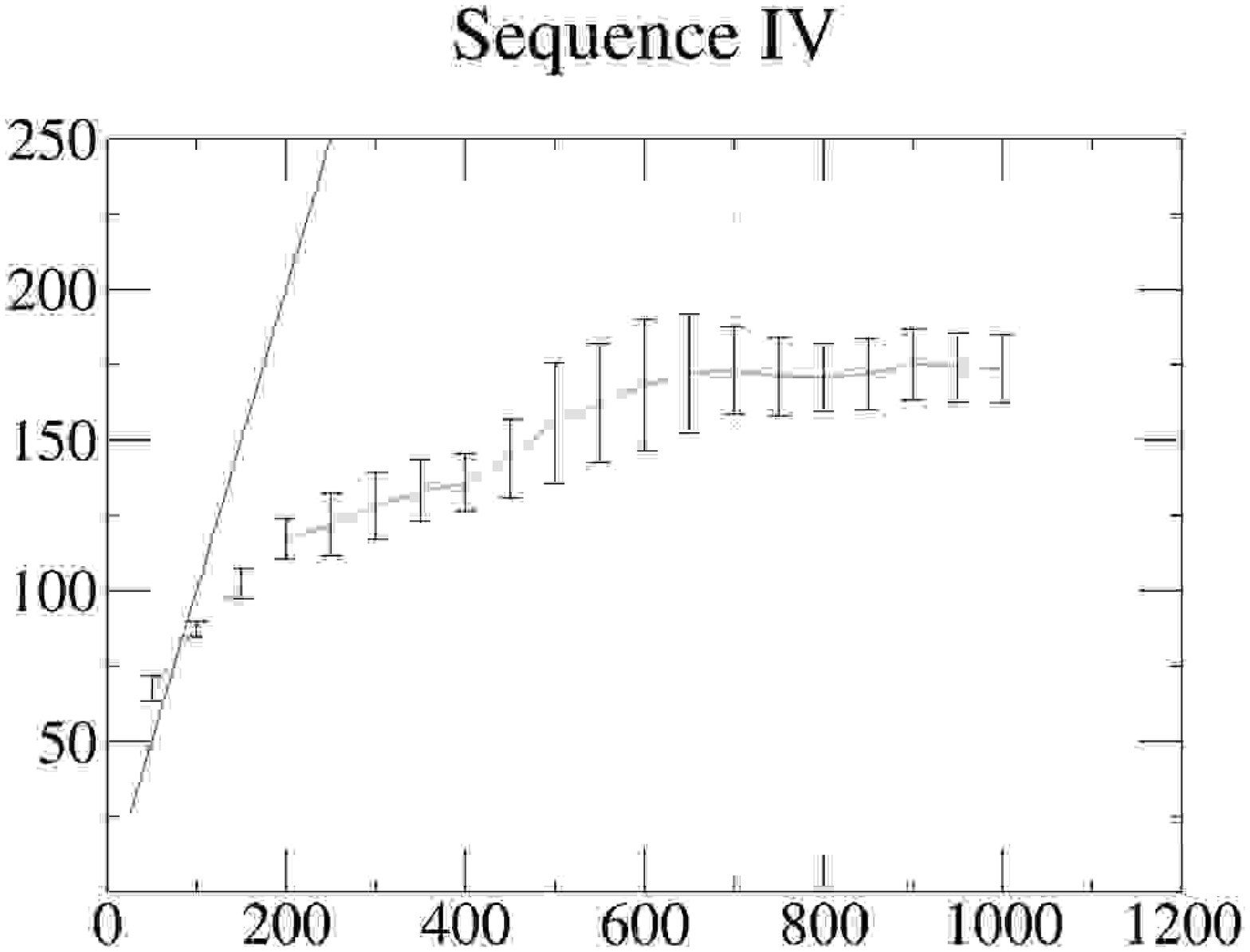}
    \end{tabular}
    \caption{Integrated autocorrelation times of the plaquette
      together with their standard errors as a function of the
      differencing lag for the different update sequenced.}
    \label{fig:diff-trajs-tau-int-Plaq-lags}
  \end{center}
\end{figure}

The resulting plaquette autocorrelation times have been computed using
both the lag-differencing and the Jackknife method.
Figure~\ref{fig:diff-trajs-tau-int-Plaq-lags} shows the results for
the former method. For Sequence I one finds that a plateau emerges
beyond $l=500$, while for Sequence II it starts already around
$l=400$. Sequence III exhibits a stable region between $l=350$ and
$l=600$, while Sequence IV becomes steady beyond $l=700$. Hence, in
all cases, the differencing method was able to yield conclusive
results.

\begin{figure}[ht]
  \begin{center}
    \begin{tabular}[c]{c|c}
      \includegraphics[scale=0.22,bb=0 0 29cm 22cm,angle=0,clip=true]%
      {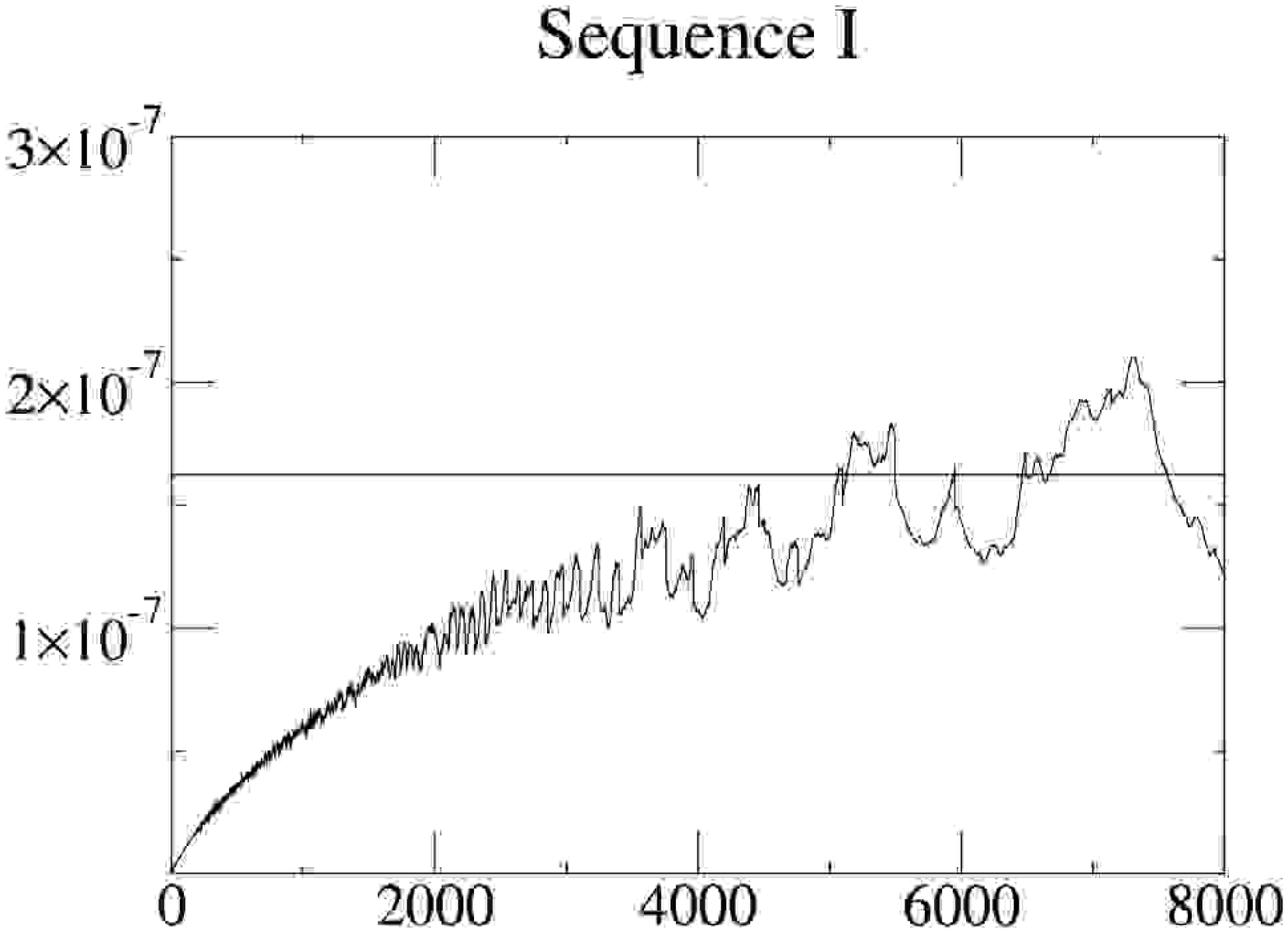} &
      \includegraphics[scale=0.22,bb=0 0 29cm 22cm,angle=0,clip=true]%
      {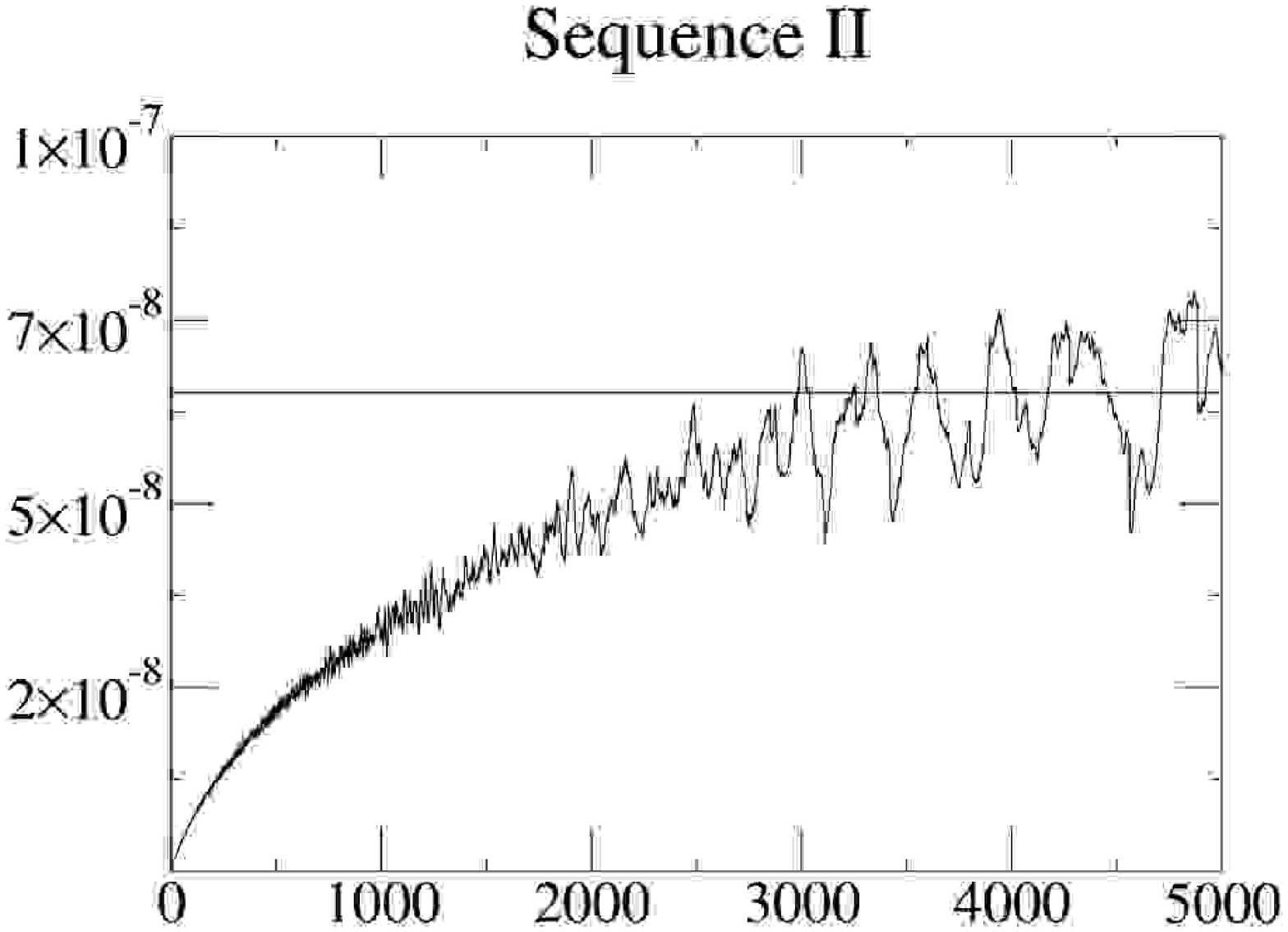} \\ \hline                  
      \includegraphics[scale=0.22,bb=0 0 29cm 22cm,angle=0,clip=true]%
      {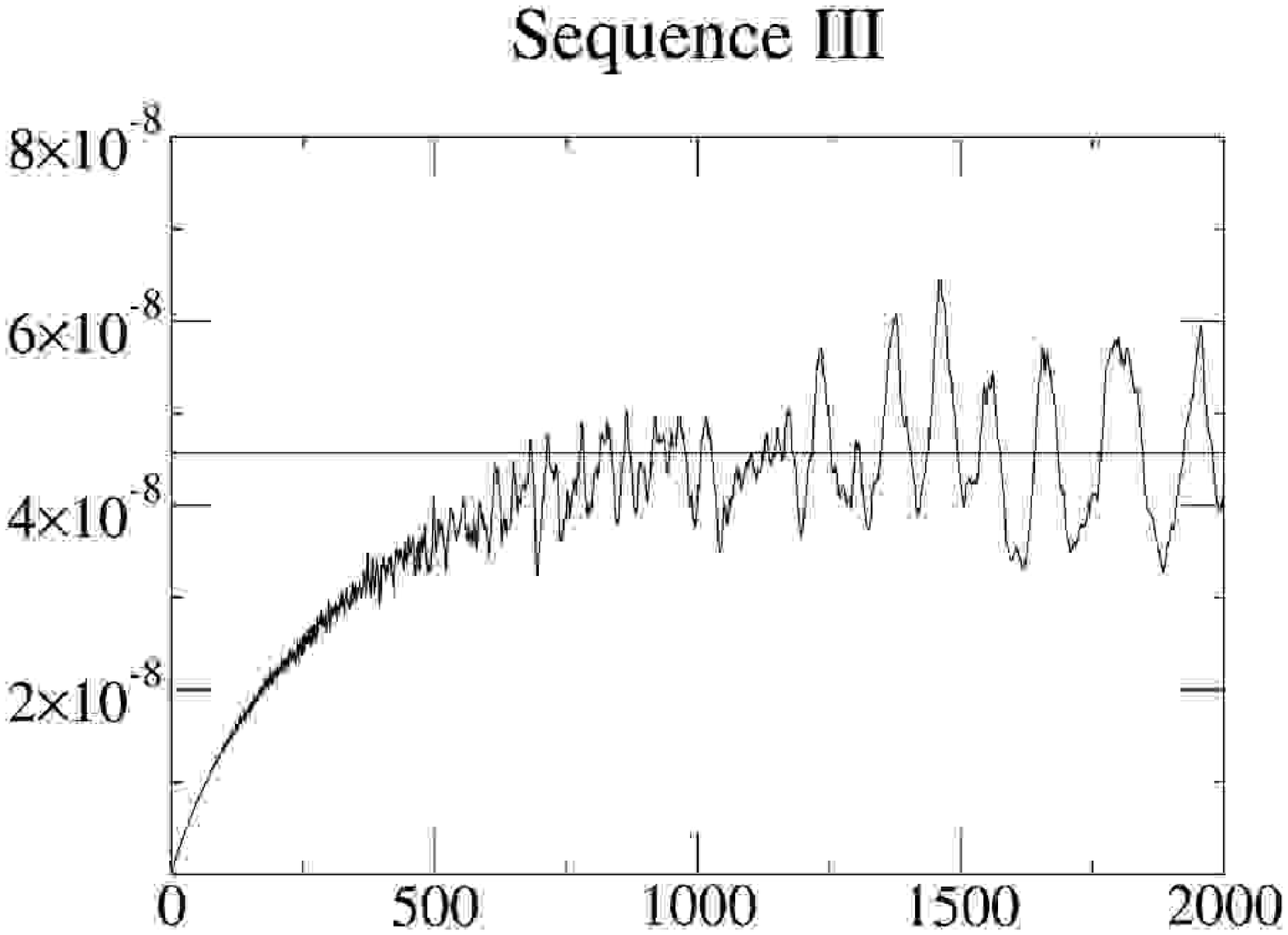} &
      \includegraphics[scale=0.22,bb=0 0 29cm 22cm,angle=0,clip=true]%
      {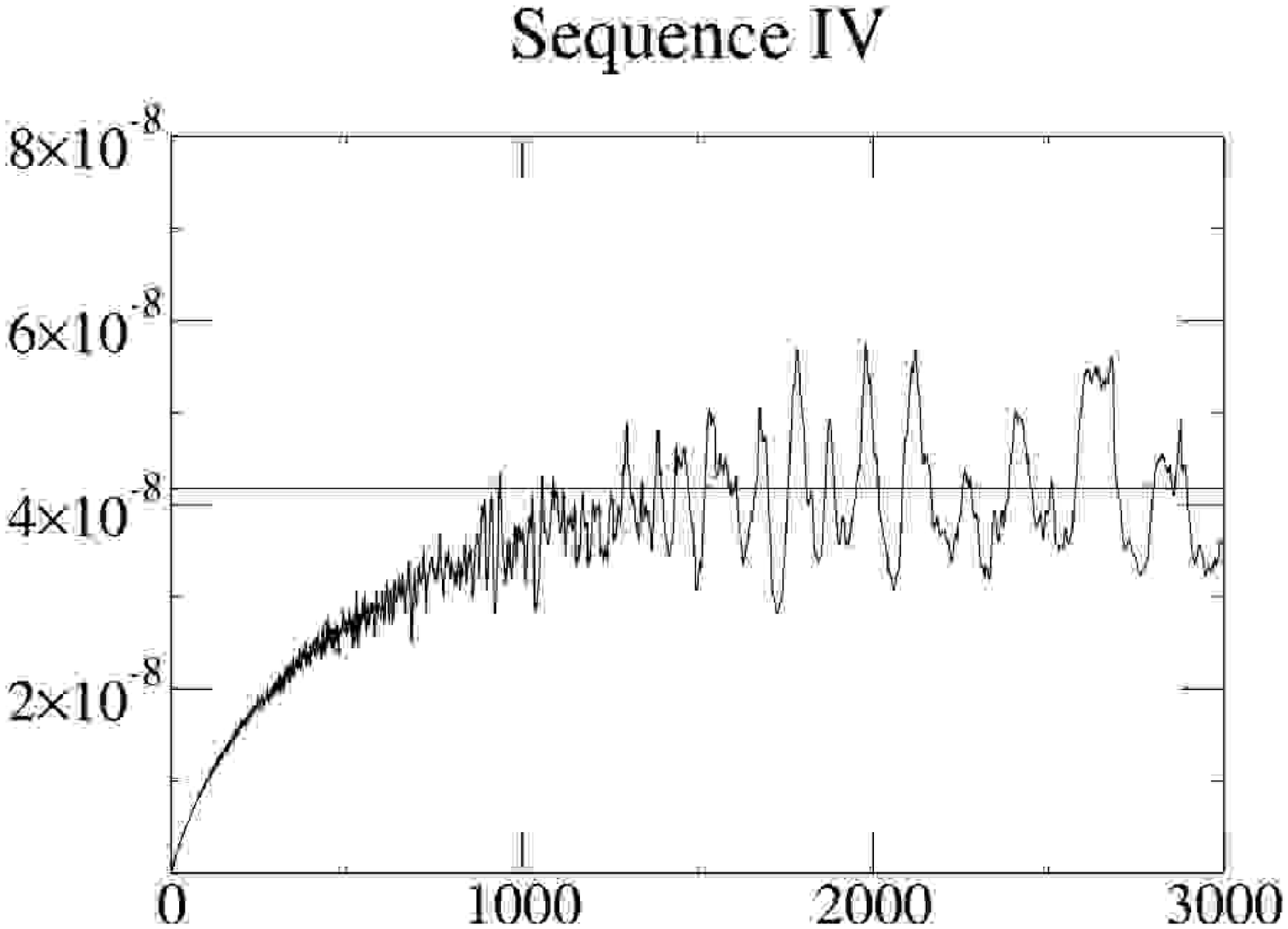}
    \end{tabular}
    \caption{Jackknife variances as a function of the bin size for
      the different updating sequences.}
    \label{fig:diff-trajs-tau-int-Plaq-jack}
  \end{center}
\end{figure}

The Jackknife variances as a function of the bin size are displayed in
Fig.~\ref{fig:diff-trajs-tau-int-Plaq-jack}. In all cases, plateaus
can be identified. Note again, that there is no systematic control of
the errors when using this method.

The resulting autocorrelation times and the total numerical efforts
(as computed from the lag-differ\-en\-cing method) are summarized in
Tab.~\ref{tab:diff-trajs-results}. The data for Sequence 0 has been
taken from Tab.~\ref{tab:tau-Plaq-result}.
\begin{table}[htb]
  \begin{center}
    \begin{tabular}[c]{l|c|c|c|c}
      \hline\hline
      & $\mathbf{\tau_{\mbox{\tiny int}}}$ \textbf{from LDM} &
      \textbf{Lag $\mathbf{l}$} & $\mathbf{\tau_{\mbox{\tiny int}}}$
      \textbf{from Jackknife} & \textbf{Total effort/MV-Mults} \\
      \hline
      Sequence 0   & $334.3\pm 65.6$ & -         & $1441$  & $342189$ \\
      Sequence I   & $180.5\pm 8.3$  & $> 550$   & $611.8$ & $436070$ \\
      Sequence II  & $141.4\pm 10.5$ & $> 400$   & $318.0$ & $397098$ \\
      Sequence III & $152.8\pm 10.1$ & $350-600$ & $109.8$ & $498461$ \\
      Sequence IV  & $175.2\pm 11.7$ & $> 700$   & $129.7$ & $393503$ \\
      \hline\hline
    \end{tabular}
    \caption{Integrated plaquette autocorrelation times from the
      lag-differencing and the Jackknife methods together with the
      total costs for a statistically independent configuration.}
    \label{tab:diff-trajs-results}
  \end{center}
\end{table}

When examining the integrated autocorrelation times one finds that
indeed some gain can be achieved by increasing the number of local
gauge field sweeps. However, the effect clearly saturates already
after as few as four consecutive sweeps have been performed in
Sequence IV\@. On the other hand, increasing the number of boson field
sweeps in Sequence III did not produce any practical gain.

In contrast, the picture is different if the total efforts are
considered. Apparently a decrease in the autocorrelation time is
accompanied by an increase in the effort for a single trajectory. From
this finding one can conclude that one should better mix the local
update sweeps. There is, however, one subtlety which is not reflected
in Tab.~\ref{tab:diff-trajs-results}: As it will be shown later in
Tab.~\ref{tab:performances-ape}, on the \APE-100 architecture a
caching of the boson fields allows for an efficient implementation of
subsequent gauge field sweeps. This would reduce the total cost in
Sequence IV compared to Sequence 0 by the amount of three gauge field
sweeps. The net effect of such an implementation is that the cost
would only slightly increase for the execution of subsequent gauge
field sweeps. Hence, one would find that for the \APE-implementation
an updating sequence like Sequence IV is superior to the sequence used
previously.

In conclusion, the optimal updating sequence consists of a mixture of
local gauge and boson field sweeps. One can update the gauge fields
for several sweeps while holding the boson field background fixed. It
does not appear to be efficient, however, to perform more than four
gauge field sweeps in this way. In case the caching discussed in
App.~\ref{sec:local-forms-actions} is available, this method is indeed
effective in reducing the total effort for a single trajectory,
$E_{\mbox{\tiny MV-mults}}$, and hence also the total cost for a
statistically independent configuration, $E_{\mbox{\tiny indep}}$.

\section{Implementation Systems}
\label{sec:impl-syst}
The discussion so far has been limited to the machine-independent part
of multiboson algorithms. In practical simulations, however, a
particular architecture for the large-scale simulations has to be
selected. This choice will have a considerable impact on the project
since the complexity of a multiboson algorithm is huge compared to
other algorithms in use today and the program is expected to run for
several months.

For the purposes of this thesis, two platforms have been given major
focus: The first platform was the \APE-100 platform
\cite{Battista:1993qc} which is also compatible with the \APE-1000
\cite{Aglietti:1997nx} architecture\footnote{Further material on these
  machines can also be found on the web under
  \texttt{http://chimera.roma1.infn.it/ape.html}}. The machines used
are installed at \DESY/Zeuthen and at the \textit{Forschungszentrum
  J{\"u}lich} in J{\"u}lich, Germany. The second target platform for
the implementation was the \ALiCE\ computer cluster
\cite{Alice:2001ar} installed at Wuppertal University. In the
following, the machines together with their specific merits and
drawbacks will be presented. The properties of the implementations are
discussed and the influence of the numerical precision on the
calculations is examined.

\subsection{\APE\ Platform}
\label{sec:ape-platform}
The \APE\ is a \index{MIMD} SIMD machine which executes a program in
parallel on a number of nodes arranged in a three-dimensional mesh.
The smallest configuration of nodes is a $2\times 2\times
2$-partition, the largest configuration available is an $8\times
8\times 8$-machine.

The \APE-100 architecture can only execute single-precision floating
point numbers efficiently in parallel.  However, for special
operations like global sums, a library for double-precision addition
is available \cite{Simma:1995fs}.  Given that a global sum only
consumes a fraction of the total runtime of a program, there is no
performance degradation to be expected. In contrast to floating point
operations, integer calculations are being done globally on a single
CPU (with no parallelization possible) with less efficiency.
Especially integer operations on array indices should be kept at a
minimum for the program to run efficiently. One further obstacle is
the fact that the complete multi-boson program would be too large to
fit into the memory of the machine --- thus only a portion of the
complete code can be written on the \APE\ machines and the remaining
parts must be run on conventional parallel computers.

One further problem is the bad I/O-performance of the machine. A
save/restore of the complete machine state requires about $1$-$2$
hours of time (for a typical lattice of size $\Omega=32\times 16^3$
and $n_1=30-60$) which means that about $5-10\%$ of the whole runtime
of a job (which is usually about $20-30$ hours) would be wasted for
I/O operations. This problem can be overcome, however, by not storing
the boson fields on disc, but rather performing a global heatbath
thermalization sweep (see Sec.~\ref{sec:updating-strategy}) to
initialize them prior to a run. This strategy is more efficient (and
also is more effective in terms of autocorrelation times of
observables) than saving and restoring the complete machine state each
time.

The compiler and optimizer technologies lag behind the industry
standards of conservative parallel computers --- the CPUs have no
caches (only the registers of the floating point processors serve as a
kind of $1$st level cache). Most optimization strategies (like loop
unrolling, prefetching etc.) have to be implemented manually using the
high-level language of the platform. This language is called \TAO\ 
\cite{TAO_Manual} and is a language build on \Zz, which is a compiler
construction language.  However, due to the fact that \Zz\ is still
accessible (to extend the features of \TAO\ and to implement manual
loop-unrolling etc.), the system is effectively using a dynamic
grammar, which is known to bear a lot of responsibility on the
implementor. The drawback is that more complicated programs developed
on the machine cannot easily be ported to different architectures and
thus the maintenance costs will soon become a reasonable factor. This
is no concern for trivial algorithms like the HMC, but will become a
serious problem once a larger source code base is to be established on
the machine.

One particular problem is that the sources of the compilers are not
publicly available, meaning that bugs are hard to locate and fix
compared to e.g.~the \GNU\ compilers\footnote{Further information and
  resources related to this system can be found under\\
  \texttt{http://www.gnu.org/software/gcc/gcc.html}}. This implies
that the development tools could not be run on modern and fast
machines --- the typical compile times during the early phases of the
project were of the order of $20-30$ minutes resulting in turn-around
times of more than half an hour.

The advantages of the \APE\ architecture are that a lot of computer
time is available and it has proven to be the ideal platform for
simple algorithms like the HMC, which essentially rely only on the
implementation of an efficient matrix-vector multiplication.
Furthermore the platform scales very efficiently since the
communicational overhead is minimal. This results in a rather small
latency and is thus a counterpoint to the workstations clusters
available today, see \cite{Schroers:2000ap} for a different
application of workstation clusters which demonstrates the same
properties. Several of these shortcomings have improved with the
advent of the \APE-1000 architecture \cite{Aglietti:1997nx}, but
experience is still too sparse to include major results in this
thesis. The \APE-1000 architecture still has problems regarding the
maximum machine size and the fact that double precision calculations
will introduce a performance hit of a factor of four in the peak
performance.

The first implementation of the TSMB algorithm used in this thesis has
been written on the \texttt{Q4open} machine located at the \NIC\ in
J\"ulich, Germany\footnote{See \texttt{http://www.fz-juelich.de/nic/}
  for further information on the \textit{John von Neumann ---
    Institut f{\"u}r Computing}}. The machine had a configuration of
$2\times 4\times 4$ nodes and served as the major development platform
until Spring 2000. Sadly, it went out of service due to a defect
board.

\subsection{\ALiCE\ Cluster}
\label{sec:alice-cluster}
At a later stage of this project, development was shifted to the
\ALiCE\ computer cluster installed at Wuppertal University\footnote{A
  large contribution to this program has been provided by
  Prof.~I.~Montvay, \DESY, Hamburg}, where modern compilers and
development tools are available. Most results have in fact been
obtained on this machine.  The cluster consists of $128$ Compaq
\texttt{DS 10} workstations, each equipped with a 21264 Alpha
processor running at $616$ MHz. The size of the second level cache is
$2$ Mbyte. The network is based on a \texttt{Myrinet} network with a
peak performance of $1.28$ Gbit/$s$.

The coding done on this platform was immediately usable on other
parallel machines, like the \CRAY\ T3E located at the \ZAM,
J{\"u}lich\footnote{The official homepage of the \textit{Central
    Institute for Applied Mathematics} can be found at\\
  \texttt{http://www.fz-juelich.de/zam/}} and the \Nicse-cluster which
is also located at the \NIC\ institute.  The program has been proven
to run also on a cluster of standard, Intel-based workstations
installed at Wuppertal University. The network of these machines is
based on standard Ethernet which made the installation not competitive
from a performance point of view, but very attractive for development
and debugging purposes. This illustrates the particular advantage of
standard tools over proprietary solutions: although the hardware costs
might be smaller (the situation might be less clear once development
costs are included, however), the \textit{Total Cost of Ownership}
(TCO) may outweigh the former price. In fact, the total costs for
maintenance and software may become larger than the pure hardware
costs.

\subsection{Accuracy Considerations and Test Suites}
\label{sec:accur-cons-test}
The complexity of code for the multiboson algorithm is high compared
to the case of other algorithms in use today like the HMC\@. The
multiboson code on the \APE\ machine (together with the production
environment) amounted to more than $11000$ lines of code; the program
on the \ALiCE\ cluster consisted of $17000$ lines of code (for the
single-node and the parallel version) and the administrative software
required another $17000$ lines. For the measurement of hadronic
masses, a program with a size of $29000$ lines of code was required.
This clearly asks for having efficient test suites available to track
down possible sources of errors.

\subsubsection{Local Fermionic Action}
\label{sec:local-ferm-acti}
An important part of the program consists of the implementation of the
fermionic action. Explicit forms of the different expressions required
for the local action are given in App.~\ref{sec:local-forms-actions}.
The local forms have to be consistent with the implemented
matrix-vector multiplication\footnote{Strictly speaking, this is not a
  necessity for the program to be correct. One can implement the
  global matrix-vector multiplication $Q(y,x)$ with another convention
  than that used for the local actions. However, in this case the
  tests suggested here will fail. Thus, it appears to be a good idea
  to keep the actions consistent and proceed as discussed}. Then one
can alter a single link at an arbitrary site and compare the results
of Eqs.~(\ref{eq:eferm-wilsoneo}), (\ref{eq:eferm-eo-wilson-expand})
and (\ref{eq:ferm-eo-staple2}). The second test consists of changing a
single color-spinor with an arbitrary index $j$, $1<j\leq n_1$, and
again comparing the results of (\ref{eq:eferm-wilsoneo}) and
(\ref{eq:eferm-eo-wilson-expand}). The residual error should only be
limited by the \index{Numerical precision} machine precision. This can
also act as a test on whether the single precision of the \APE\ 
machines is a real limitation.

\begin{figure}[htb]
  \begin{center}
    \includegraphics[scale=0.3,clip=true]%
    {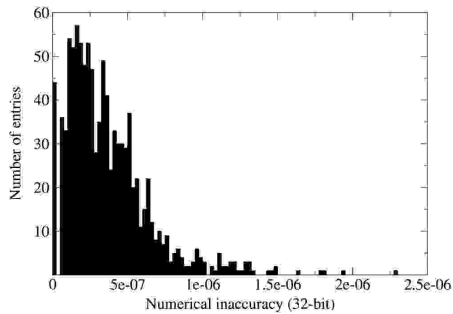}
    \caption{Maximum numerical error between the different
      implementations of the fermionic actions for local changes of
      the gauge field.}
    \label{fig:local-gauge-numerical-error}
  \end{center}
\end{figure}

In fact, for the application of (\ref{eq:ferm-eo-staple2}) one has to
sum up $n_1$ terms in single precision to get a complex $3\times 3$
matrix which may introduce already difficulties at moderate values of
$n_1$. To examine the errors as they occur in practical computations,
one can already get along with a very small lattice since the major
source of numerical errors occurs in the \textit{local} update part.
Therefore, a simulation has been performed using a thermalized
configuration on a $\Omega=4^4$ lattice at the physical parameters
given in Tab.~\ref{tab:phys-par} on the QH1-board at \DESY/Zeuthen.
The polynomial in question has been chosen to be $n_1=32$ with
$[\epsilon,\lambda]=[\Exp{3}{-3},3]$. The maximum numerical error in
the three expressions is displayed in
Figure~\ref{fig:local-gauge-numerical-error}, where the distribution
of the inaccuracies are shown. They have been obtained by considering
separately each site on a single node during a gauge field updating
sweep. The error obviously is bounded from above and only scarcely
exceeds $\Exp{1}{-6}$. In the development phase such a plot turned out
to be extremely useful since identifying the sites which give a huge
numerical error can help to track down program bugs rather easily.

The second message to be learned from
Fig.~\ref{fig:local-gauge-numerical-error} is that the $32$-bit
precision used is not an obstacle in actual simulations: The
systematical error introduced by the local gauge field updates is
obviously under control.

\begin{figure}[htb]
  \begin{center}
    \includegraphics[scale=0.3,clip=true]%
    {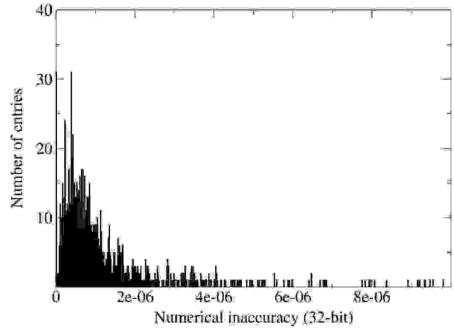}
    \caption{Maximum numerical error between the implemented actions
      for local changes of the boson field.}
    \label{fig:local-scafi-numerical-error}
  \end{center}
\end{figure}

The same can be done for local changes of the boson field and
considering the expressions Eqs.~(\ref{eq:eferm-wilsoneo}) and
(\ref{eq:eferm-eo-wilson-expand}). The corresponding results are
displayed in Fig.~\ref{fig:local-scafi-numerical-error}. Apparently,
the same can be said about the boson field case as has been stated
before in the gauge field case.

\subsubsection{The Inverse Square-root}
\label{sec:inverse-square-root}
Another problem is posed by the residual error of the inverse square
root required for the TSMB algorithm, Eq.~(\ref{eq:poly-invsqr-def}).
As has already been discussed in Sec.~\ref{sec:tuning-quadr-optim},
the systematic error can be computed via
(\ref{eq:r3-determinant-norm}) and must be less than one percent.

\begin{figure}[htb]
  \begin{center}
    \includegraphics[scale=0.3,clip=true]%
    {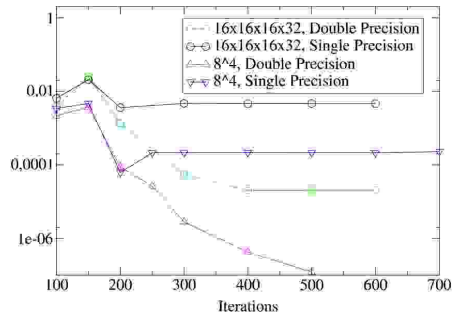}
    \caption{Residual error of the noisy correction
      $\|R_{n_3}^{n_2=160}(\tilde{Q}^2)\|$ vs.~the number of
      iterations $n_3$ for two lattice sizes. Both single ($32$ bit)
      and double precision ($64$ bit) have been used.}
    \label{fig:n3-accur}
  \end{center}
\end{figure}

The question arises, how accurate the approximation can be \textit{at
  best}. Given the fact that in case of the HMC algorithm one has a
residual error of $2\%$ if one uses $32$-bit floating point numbers on
an $\Omega=40\times 24^3$ lattice (see
Sec.~\ref{sec:hybrid-monte-carlo}), the question arises, how large the
lattice may be in the TSMB case if one only has access to
\index{Numerical precision} single precision on a particular
architecture. This question is answered by Fig.~\ref{fig:n3-accur},
where the residual error $\|\hat{R}_{n_3}^{n_2}(\tilde{Q}^2)\|$ of the
noisy correction step is plotted vs.~the order of the third
polynomial, $n_3$ (the polynomial $n_2$ has been chosen to be
$n_2=160$). The calculation has been performed using both $32$-bit
(single precision) and $64$-bit (double precision) algebra. The
lattice sizes which have been considered were $\Omega=8^4$ and
$\Omega=32\times 16^3$.

On the $\Omega=8^4$ lattice the results coincide up to orders of about
$n_3=200$. Beyond this point, the single precision result deviates
from the double precision curve. Finally, the single precision numbers
saturate at $n_3\geq 250$. The accuracy which can be reached is still
satisfactory since it is bounded by
$\|\hat{R}_{n_3}^{n_2}(\tilde{Q}^2)\|\approx\Exp{2.1}{-4}$. Hence, one
can conclude that single precision is adequate on an $\Omega=8^4$
lattice.

On the $\Omega=32\times 16^3$ lattice, the single precision result
saturates already at
$\|\hat{R}_{n_3}^{n_2}(\tilde{Q}^2)\|\approx\Exp{4.6}{-3}$. This is
more than on the $\Omega=8^4$ lattice, but it is still sufficiently
small. The double precision curve also shows a saturation, but not
before $n_3=400$ and an accuracy of
$\|\hat{R}_{n_3}^{n_2}(\tilde{Q}^2)\|\approx\Exp{2.1}{-5}$ has been
achieved. Again, this is completely acceptable.

The conclusion to be drawn from this test is that single precision
arithmetic is not at all a problem on an $\Omega=8^4$ lattice, and is
still acceptable on lattices as large as $\Omega=32\times 16^3$. On
larger lattices, however, single precision may no longer be feasible
and one should refrain from using $32$-bit arithmetics.

\subsection{Architectures and Efficiency}
\label{sec:arch-effic}
All architectures discussed so far have certain advantages and
disadvantages. In this section, the efficiencies of the
implementations are compared against each other.
Table~\ref{tab:performances} shows the execution times of the
different parts of the multiboson implementations for a selection of
architectures. The physical parameters of the runs are given in
Tab.~\ref{tab:phys-par}, while the algorithmic parameters are listed
in Tab.~\ref{tab:tsmb-multi-genpars} with $n_1=20$. The lattice size
was again $\Omega=8^4$.

In case of the \ALiCE\ cluster and the \CRAY\ T3E, an eight-node
partition with parallelization in $z$- and $t$-direction has been
employed which results in local lattices of $\Omega_{\mbox{\tiny
    loc}}=2\times 8\times 8\times 4$. Table~\ref{tab:performances}
displays the execution times of the different algorithms employed.The
last line shows the ratio of local update sweeps to the global
matrix-vector multiplications taking place in the noisy correction
step. This is the key quantity of interest, where the influence of a
particular architecture is most clearly exposed.
\begin{table}[htb]
  \begin{center}
    \begin{tabular}[c]{l|c|c}
      \hline\hline
      \textbf{Algorithm} & \multicolumn{1}{c|}{\textbf{Time on
      \ALiCE}} & \multicolumn{1}{c}{\textbf{Time on \CRAY\ T3E}} \\
      \hline
      Mat-Vect.~mult     & $0.04938 s$ & $0.06542 s$  \\
      Noisy correction   & $17.777  s$ & $23.551  s$  \\
      Boson heatbath     & $1.10352 s$ & $4.02112 s$  \\
      Boson overrel.     & $1.06641 s$ & $3.79800 s$  \\
      Gauge heatbath     & $1.17969 s$ & $4.12083 s$  \\
      \hline                                         
      Ratio local/global & $0.3748$    & $1.0045$     \\
      \hline\hline
    \end{tabular}
    \caption{Execution times of several parts of the TSMB algorithm on
      the \ALiCE\ cluster and the \CRAY\ T3E.}
    \label{tab:performances}
  \end{center}
\end{table}

For the \APE-100, an eight-node \texttt{Q1} board has been used with a
lattice size of $\Omega=4^4$. The local lattices were
$\Omega_{\mbox{\tiny loc}}=2\times 2\times 2\times 4$ per node. The
resulting execution times of these algorithms are quoted in
Tab.~\ref{tab:performances-ape}. In contrast to the above
implementation, the \APE\ program uses an efficient caching strategy,
where the contributions of the boson fields to the local staples,
Eq.~(\ref{eq:ferm-eo-staple2}), which are unchanged by a gauge field
sweep are held in memory, see App.~\ref{sec:local-forms-actions} for
details. This allows to reduce the computational costs for
\textit{repeated} local gauge field sweeps. The time required for the
initialization of the gauge field sweep is given in the fifth line of
Tab.~\ref{tab:performances-ape}. This strategy has only been
implemented on the \APE, but in principle this caching scheme is
machine-independent and could also be implemented in the other
program.
\begin{table}[htb]
  \begin{center}
    \begin{tabular}[c]{l|c}
      \hline\hline
      \textbf{Algorithm} & \textbf{Time} \\ \hline
      Mat-Vect.~mult     & $0.01354 s$   \\
      Noisy correction   & $4.876   s$   \\
      Boson heatbath     & $1.634   s$   \\
      Boson overrel.     & $1.482   s$   \\
      Init gauge sweep   & $1.695   s$   \\
      Gauge heatbath     & $0.124   s$   \\ \hline
      Ratio local/global & $1.6503$      \\
      \hline\hline
    \end{tabular}
    \caption{Execution times of the parts of the TSMB algorithm on an
      \APE\ board.}
    \label{tab:performances-ape}
  \end{center}
\end{table}

When comparing the \ALiCE\ and the \CRAY\ T3E, one realizes that the
\CRAY\ T3E architecture has a more efficient network, but lacks the
large second-level caches of the \ALiCE\ nodes. This explains why the
communication-intensive performance of the matrix-vector
multiplication is very efficient on the \CRAY\ T3E. On the other hand,
the local update sweeps are very cache intensive and communication
between nodes only plays a minor role. Therefore, the local update
sweeps contribute a much smaller fraction to the total runtime on the
\ALiCE\ cluster, while they account for about $50\%$ of the total
sweep time on the T3E.

The \APE-100 architecture shows an even more prominent dominance of
the local update sweeps, which consume about $2/3$ of the total
runtime. This can be attributed to the fact that the CPUs have no
second-level cache at all and the data has to be fetched from memory
each time. Therefore the local update sweeps are rather inefficient,
while the global matrix-vector multiplications are very efficient.

\section{Summary}
\label{sec:summary}
The TSMB algorithm requires three approximations to an inverse power
of the Hermitian Wilson matrix. All approximations are being performed
with static inversion algorithms as discussed in
Sec.~\ref{sec:stat-polyn-invers}. The first is a crude approximation
to an inverse power of $\tilde{Q}^{2}$ with order $n_1$. The second is
a refined approximation to the same function with order $n_2$. The
third one approximates the inverse square root of the polynomial with
which the second approximation is performed.

The best method to find the first polynomial consists of applying the
GMRES algorithm (see Sec.~\ref{sec:gmres-algorithm}) to one or more
thermalized gauge field configurations at the physical point one is
interested in. This method is limited by the numerical precision of
the architecture used to generate the polynomial. The other method
consists of using a quadratically optimized polynomial. The latter
choice only requires rough knowledge of the spectral bounds.

The second and third polynomials have to be quadratically optimized
polynomials. The value of $\epsilon$ for the product of the first and
the second polynomial should be chosen such that it is slightly
smaller than the average smallest eigenvalue. Its order has to be
adjusted such that the reweighting factor is not fluctuating more than
a few percent. Since the convergence is exponential, this can be
achieved without too much effort.

The third polynomial must have a sufficiently high order such that its
corresponding determinant norm, Eq.~(\ref{eq:r3-determinant-norm}), or
rather the resulting systematic error,
Eq.~(\ref{eq:systematic-error-est}), is never exceeding values of
$\simeq 10^{-3}$.

The order of the first polynomial influences the acceptance rate of
the correction step. It appears to be safe to make the acceptance rate
somewhat smaller than $50\%$. The motion in phase space depends
linearly on the number of boson fields. The decreased acceptance rate
counteracts the increased mobility in phase space and the resulting
efficiency does not appear to depend on the acceptance rate.  Since
the numerical cost of a single trajectory is proportional to the
number of boson fields, the total cost for a statistically independent
gauge field configuration, Eq.~(\ref{eq:eff-def}), is then given by
\begin{equation}
  \label{eq:indep-cost}
  E_{\mbox{\tiny indep}} \propto n_1^2\,.
\end{equation}
This formula should apply for runs at different parameters and
different orders $n_1$ if the acceptance rates are held constant and
if the field updates dominate the time needed for a single trajectory.

A trajectory is given by the updating sequence, i.e.~the transition
function of the Markov chain. It consists of a number of update sweeps
for the boson fields and local update sweeps of the gauge field. It
has turned out that the boson field updates should be mixed with the
gauge field updates, but a number of subsequent gauge field updates
allows a similarly efficient decorrelation. Each trajectory is
completed by a noisy correction step. In view of the fact that the
acceptance rate is rather independent from the local gauge field
updates, the optimal efficiency can be achieved by performing a larger
number of local gauge field updates between two noisy corrections.
Thus, the sequence should be arranged in such a way that the field
updates dominate the total runtime.

Furthermore, it is important to identify the machine architecture
which meets the specific demands of MB algorithms. While a
conventional massive-parallel machine with a network similar to the
\CRAY\ architecture, but small caches on the nodes does not perform
well with respect to the local update sweeps, its efficient network
allows for a rather efficient matrix-vector multiplication. However,
as it has been discussed in Sec.~\ref{sec:updating-strategy}, one can
(and, in fact, one should) always arrange the updating sequence in
such a manner that the local updating sweeps dominate the total
runtime of the code. Hence, the machine best suited for TSMB
calculations is found to be a cluster of workstations with large cache
and standard programming tools.

The \APE-100 system does not perform very well in the local
update sweeps and suffers from the problem that the coding of the
algorithm can until now not be used on any other machines. The former
problem may be overcome with the advent of the \APE-1000 architecture
which may (due to the large number of CPU registers) reduce the number
of memory accesses required. The latter problem can not be expected to
be solved before the advent of the \APE-Next platform
\cite{Alfieri:2001qf}. Due to the complexity of the algorithm it
appears reasonable to implement first a reference implementation in a
standard language on a different architecture before starting with the
coding on the \APE\ platform.

There is still room for further improvements, however. In particular,
improving the approximation scheme of the third polynomial would allow
to overcome the limitations of the current implementations and should
make the algorithm applicable to larger lattices even with single
precision arithmetics. A different place where further improvements
are in order is the first polynomial. Although the optimal way to find
its coefficients has been identified, this method still requires high
numerical accuracy of the implementation system.

\chapter{Comparison of Dynamical Fermion Algorithms}
\label{sec:comp-dynam-ferm}
In the first section, Sec.~\ref{sec:simulation-runs-at}, the variant
of the multiboson algorithm with TSMB correction step, which has been
studied in detail in Chapter~\ref{sec:tuning-mult-algor}, is applied
at different physical locations in parameter space. At all points,
results from the HMC method are available. This allows for a direct
comparative study.

Section~\ref{sec:effic-mult-algor} directly compares the efficiencies
of MB algorithms with that of the HMC\@. This investigation is not
limited to the MB-variant discussed so far, but also covers the
non-Hermitian variant with UV-filtering and gauge field overrelaxation
as proposed by \textsc{de Forcrand} in \cite{deForcrand:1998sv}
(cf.~Sec.~\ref{sec:mult-algor}). These results might be of importance
for future simulations of gauge theories with dynamical fermions, see
e.g.~\cite{Orth:2001mk}. The study is carried out on equal lattice
volumes and with identical physical parameters. Hence, it will allow
to probe the scaling of the algorithms as the continuum limit is
approached.

\section{Simulation Runs at Different Parameters}
\label{sec:simulation-runs-at}
The TSMB variant of the multiboson algorithm is applied to situations
at different physical points in parameter space with two dynamical
fermion flavors. A direct comparison with the HMC algorithm is given.
The latter acts as a benchmark for the alternative proposals.

The HMC simulations have all been carried out on volumes
$\Omega=32\times 16^3$. The physical parameters of the various runs
performed here and in the rest of this chapter are compiled in
Tab.~\ref{tab:phys-par-complete} and have been taken from
\cite{Orth:2002ph}, see also \cite{Eicker:1998sy,Eicker:2001ph}. The
second line is identical to Tab.~\ref{tab:phys-par} in
Chapter~\ref{sec:tuning-mult-algor}.
\begin{table}[htb]
  \begin{center}
    \begin{tabular}[c]{*{6}{c|}c}
      \hline\hline
      \multicolumn{3}{c}{\textbf{Bare parameters}} & 
      \multicolumn{4}{|c}{\textbf{Physical parameters}} \\ \hline
      $\mathbf{N_f^{\mbox{\tiny sea}}}$ & $\mathbf{\beta}$ &
      $\mathbf{\kappa}$ & $\mathbf{(am_\pi)}$ & $\mathbf{(am_\rho)}$ &
      $\mathbf{m_\pi/m_\rho}$ & \textbf{$\mathbf{a}$/fm} \\ \hline
      $2$ & $5.5$ & $0.158$ &
      $0.5528(40)$ & $0.6487(55)$ & $0.8522(95)$  & $0.166$ \\
      $2$ & $5.5$ & $0.159$ & 
      $0.4406(33)$ & $0.5507(59)$ & $0.8001(104)$ & $0.141$ \\
      $2$ & $5.5$ & $0.160$ &
      $0.3041(36)$ & $0.4542(78)$ & $0.6695(138)$ & $0.117$ \\
      $2$ & $5.6$ & $0.156$ & 
      $0.4464(27)$ & $0.5353(42)$ & $0.8339(66)$  & $0.137$ \\
      \hline\hline
    \end{tabular}
    \caption{Bare and physical parameters for the comparison between
      HMC and TSMB\@. Also cf.~Sec.~\ref{tab:phys-par}.}
    \label{tab:phys-par-complete}
  \end{center}
\end{table}

The multiboson algorithm has been operated on rather small lattices
with volumes $\Omega=8^4$ and $\Omega=16\times 8^3$. Measurements of
the physical masses can be expected to differ from those on the larger
volumes due to finite-size effects \cite{Orth:2002ph}. Hence, runs on
such small lattice sizes can only be preliminary studies which need to
be supplemented later by runs on larger volumes. The quantity under
consideration here is the average plaquette. This observable will
exhibit only a weak dependence on the lattice volume. These runs
corroborate the tests performed in Sec.~\ref{sec:accur-cons-test} of
Chapter~\ref{sec:tuning-mult-algor}.

Table~\ref{tab:var-phys-runs} gives the total statistics in numbers of
trajectories entering the analysis. The complete data set as given in
Tab.~\ref{tab:alice-runs} has been exploited, therefore the statistics
in this particular case is enormous. Furthermore, the machines on
which the data have been sampled are shown.
\begin{table}[htb]
  \begin{center}
    \begin{tabular}[c]{*3{c|}*2{l|}r}
      \hline\hline
      \multicolumn{2}{c|}{\textbf{Bare parameters}} \\ \cline{1-2}
      $\mathbf{\beta}$ & $\mathbf{\kappa}$ & \textbf{Volume
        $\mathbf{\Omega}$} & \textbf{Machine} &
      \textbf{Algorithm} &
      \textbf{Trajectories} \\
      \hline
      $5.5$ & $0.158$ & $8^4$ & \ALiCE\ cluster & TSMB & $17703$  \\
      $5.5$ & $0.158$ & $32\times 16^3$ & \APE-100 \textbf{QH4} & HMC
      & $3042$ \\ \hline
      $5.5$ & $0.159$ & $8^4$ & \ALiCE\ cluster & TSMB & $448055$ \\
      $5.5$ & $0.159$ & $32\times 16^3$ & \APE-100 \textbf{QH4} & HMC
      & $4518$ \\ \hline
      $5.6$ & $0.156$ & $16\times 8^3$  & \APE-100 \textbf{Q4}  & TSMB
      & $4424$ \\
      $5.6$ & $0.156$ & $32\times 16^3$ & \APE-100 \textbf{QH2} & HMC
      & $4697$ \\
      \hline\hline
    \end{tabular}
    \caption{Machines and statistics for the test runs at
      different physical parameters.}
    \label{tab:var-phys-runs}
  \end{center}
\end{table}

Table~\ref{tab:var-phys-plaqs} lists the resulting values for the
average plaquette together with their standard errors for the
different algorithms used. The standard errors have been obtained
using the Jackknife method on the plaquette time series.
\begin{table}[htb]
  \begin{center}
    \begin{tabular}[c]{*{3}{c|}c}
      \hline\hline
      \multicolumn{2}{c|}{\textbf{Bare parameters}} \\ \cline{1-2}
      $\mathbf{\beta}$ & $\mathbf{\kappa}$ &
      \textbf{Plaquette/HMC} & \textbf{Plaquette/TSMB} \\ \hline
      $5.5$ & $0.158$ & $0.55546(6)$ & $0.55380(41)$ \\
      $5.5$ & $0.159$ & $0.55816(4)$ & $0.55909(27)$ \\
      $5.6$ & $0.156$ & $0.56988(2)$ & $0.56865(86)$ \\
      \hline\hline
    \end{tabular}
    \caption{Average plaquette values with their standard errors
      obtained from the different samples.}
    \label{tab:var-phys-plaqs}
  \end{center}
\end{table}

The plaquette values obtained with the HMC coincide with those
generated by the TSMB algorithm up to three digits\footnote{The
  residual deviation is caused by finite-size effects. However, it may
  also indicate that the autocorrelation times are underestimated and
  the actual errors of the plaquettes are still larger}.

In conclusion, the TSMB implementation indeed produces identical
plaquettes. This comparison demonstrates the correct implementation
and the correct execution of the MB algorithm.

\section{Efficiency of Multiboson Algorithms}
\label{sec:effic-mult-algor}
In this section, three different algorithms for the simulation of
Lattice QCD with two dynamical fermion flavors are compared. The
physical parameters are the ones given in the second line of
Tab.~\ref{tab:phys-par-complete}. In all cases, the lattice volume has
been chosen to be $\Omega=32\times 16^3$. This allows for a
measurement of hadronic masses and opens the stage for a direct
comparison of both algorithms.

The HMC algorithm has been introduced in the previous section, see
Tab.~\ref{tab:var-phys-runs}. The two variants of MB algorithms used
are the implementation with quadratically optimized polynomials
discussed in the previous section Sec.~\ref{sec:simulation-runs-at}
and an implementation based on the UV-filtered non-Hermitian
approximation. The latter code has been written by \textsc{M.~D'Elia}
and \textsc{Ph.~de Forcrand}. Both programs have been implemented on
the \APE-100 \textbf{QH4} installed at \DESY/Zeuthen. While the former
variant uses heatbath sweeps for the gauge field updates (it will be
called ``MB-HB'' in the following) and the TSMB correction step, the
latter uses overrelaxation sweeps (in the following abridged with
``MB-OR'') for the gauge field and an exact correction step. To
reflect the different updating strategies used, the algorithms are
named after the corresponding local gauge field updates.

\subsection{Tuning the MB-HB Algorithm}
\label{sec:setting-up-mbhb}
For the MB-HB algorithm, the question arises how the acceptance rate
of the correction step changes with the volume and what the
consequences for the polynomial orders are.
Table~\ref{tab:tsmb-accrates-largevol} shows two different choices of
parameters and the corresponding acceptance rates. Thus, the choice
$n_1=60$ gives an acceptance rate of about $50\%$, while for smaller
values of $n_1$ the acceptance rate is decreasing rather fast. We
observe a significant volume dependence since in the case $\Omega=8^4$
one only needs $n_1=20$ to get similar acceptance rates (consult
Tab.~\ref{tab:alice-runs}). For the simulation run to be presented
below the parameters from the second line have been taken.
\begin{table}[htb]
  \begin{center}
    \begin{tabular}[c]{*{4}{c|}r}
      \hline\hline
      \multicolumn{4}{c|}{\textbf{Algorithmic parameters}}  \\
      \cline{1-4}
      $\mathbf{n_1}$ & $\mathbf{n_2}$ & $\mathbf{n_3}$ &
      $\mathbf{[\epsilon,\lambda]}$ & \textbf{Acc.~rate}    \\ \hline
      $42$ & $160$ & $250$ & $[\Exp{7.5}{-4},3]$ & $15.2\%$ \\
      $60$ & $150$ & $250$ & $[\Exp{6}{-4},3]$   & $49.4\%$ \\
      \hline\hline
    \end{tabular}
    \caption{Different polynomial parameters and the resulting
      acceptance rates for the MB-HB algorithm.}
    \label{tab:tsmb-accrates-largevol}
  \end{center}
\end{table}

The updating strategy is shown in Tab.~\ref{tab:mbhb-algpars} and is
chosen similar to Tab.~\ref{tab:tsmb-multi-genpars} in the previous
chapter. Note, as has been discussed in
Sec.~\ref{sec:updating-strategy}, the boson fields do not have to be
restored if the correction step rejects a proposed gauge field
configuration.
\begin{table}[htb]
  \begin{center}
    \begin{tabular}[c]{c}
      \hline\hline
      \textbf{Updates/Trajectory} \\ \hline
      $1$ boson HB, $3$ boson OR, $2$ gauge heatbath, $1$ noisy 
      corr. \\
      \hline\hline
    \end{tabular}
    \caption{Updating sequence for the MB-HB algorithm.}
    \label{tab:mbhb-algpars}
  \end{center}
\end{table}

The heatbath algorithm (cf.~Sec.~\ref{sec:heatbath-algorithm}) has
been employed for the gauge field updates. It is important to notice
that a single-hit heatbath algorithm is sufficient if the scheme from
\cite{Kennedy:1985nu} is used. In fact, acceptance rates exceeding
$99\%$ have been observed when generating the distribution for $a_0$
(consult Sec.~\ref{sec:heatbath-algorithm} for the notation).

In the version of the program employed, the contribution of the
unit-submatrix from the even points in the noisy vectors has been
included in the noisy correction step. This resulted in a systematic
error of about $2.48\%$ when applying
Eq.~(\ref{eq:systematic-error-est}). If this had not been done, one
could have reduced the polynomial orders $n_2$ and $n_3$. We do expect
this to influence neither the stochastic averages nor the
autocorrelation times in terms of sweeps, however.

\subsection{Tuning the MB-OR Algorithm}
\label{sec:setting-up-mbor}
In contrast to the former variant, the other MB algorithm does not
make use of a polynomial approximation in the correction step but
makes an adaptive inversion. As has been discussed in
Sec.~\ref{sec:mult-algor}, this requires a nested iteration of an
adaptive inversion and the polynomial $P_{n_1}$, which acts as a
preconditioner. This approach has the great practical advantage that
no multicanonical reweighting for the measurement of observables is
necessary, but has the shortcoming that one has an increased effort
once configurations with exceptionally small eigenvalues are
encountered. In the present case, however, we do not expect this to
have a major influence.

The power of the GMRES polynomials in conjunction with UV-filtering
for the non-Hermitian Wilson matrix is demonstrated if one considers
the order of the polynomial $P_{n_1}(\cdot)$ required to arrive at an
acceptance rate of $60.3\%$. The polynomial needed in this case has an
order of only $n_1 = 24$.

Thus, the number of boson fields could have been reduced by a factor
of $2.5$ (and even more if one aims for an acceptance rate of about
$50\%$). This number takes into account the combined effect of using
the non-Hermitian Wilson matrix, employing the expansion of
Eq.~(\ref{eq:uv-filter-apply}), and using the GMRES algorithm instead
of quadratically optimized polynomials. To actually find this
polynomial, however, a thermalized gauge field configuration had to be
provided from the HMC run. Had this configuration not been available
prior to the run, the run would have had to be performed with a
non-optimal polynomial instead for thermalization. This would have
increased the total investment into the algorithm.

\begin{table}[htb]
  \begin{center}
    \begin{tabular}[c]{c}
      \hline\hline
      \textbf{Updates/Trajectory}       \\ \hline
      $1$ gauge OR, $5\times\left(\right.$ $1$ boson
      OR, $1$ gauge OR $\left.\right)$, \\
      $1$ boson field global quasi-HB,  \\
      $5\times\left(\right.$ $1$ gauge OR, $1$ boson
      OR $\left.\right)$, $1$ gauge OR, \\
      $1$ noisy corr.                   \\
      \hline\hline
    \end{tabular}
    \caption{General algorithmic parameters for MB-OR run.}
    \label{tab:mbor-genpars}
  \end{center}
\end{table}
The precise update sequence for a single trajectory is given in
Tab.~\ref{tab:mbor-genpars}. In contrast to the former multiboson
implementation discussed in Sec.~\ref{sec:setting-up-mbhb}, only
overrelaxation sweeps (see Sec.~\ref{sec:overrelaxation}) have been
used for the gauge field. Although this algorithm alone is
non-ergodic, ergodicity is ensured by the boson field global
quasi-heatbath (this method has been discussed in
Sec.~\ref{sec:heatbath-algorithm}). In particular, instead of only two
gauge field updates between a correction step, in total $12$ gauge
field updates are being run. However, the mixing of gauge and boson
field updates requires to restore both kinds of fields in case the
correction step rejects the current configuration. This results in
much larger memory requirements. As has been argued in
Sec.~\ref{sec:updating-strategy}, one can expect that this updating
sequence results in a faster decorrelation than the updating sequence
in Tab.~\ref{tab:mbhb-algpars}.

In conclusion one can expect that the MB-OR implementation may perform
better since both the number of boson fields is reduced significantly
and the updating sequence ensures a faster decorrelation.

\subsection{Direct Algorithmic Comparison}
\label{sec:direct-algor-comp}
The observables under consideration were the average plaquette, the
(non-singlet) pseudoscalar meson mass (denoted as pion $\pi$) and the
(non-singlet) vector meson mass (denoted as rho-meson $\rho$). Their
expectation values (for the three different algorithms) together with
their standard errors are shown in Tab.~\ref{tab:multalgo-comp}. The
hadronic masses have been taken from \textsc{Orth} \cite{Orth:2002ph}.
\begin{table}[htb]
  \begin{center}
    \begin{tabular}[c]{l|c|c|c|c|c}
      \hline\hline
      \textbf{Algorithm} & \textbf{Trajectories} &
      \textbf{Configurations} &
      \textbf{Plaquette} & $\mathbf{(am_\pi)}$ &
      $\mathbf{(am_\rho)}$ \\ \hline
      HMC   & $3521$ & $140$ & $0.55816(4)$ & $0.4406(33)$ & $0.5507(59)$ \\
      MB-HB & $5807$ & $108$ & $0.55819(6)$ & $0.448(10)$  & $0.578(17)$  \\
      MB-OR & $6217$ & $177$ & $0.55804(7)$ & $0.4488(37)$ & $0.5635(83)$ \\
      \hline\hline
    \end{tabular}
    \caption{Average plaquette and hadronic masses for the three
      different sampling algorithms for Lattice QCD used.}
    \label{tab:multalgo-comp}
  \end{center}
\end{table}

The plaquette values agree within errors, while the meson masses agree
within at most two standard deviations. The statistics for the MB-HB
algorithm is worse than in the other cases.

Table~\ref{tab:eff-results-mesons} shows the resulting total efforts
as defined in Eq.~(\ref{eq:eff-def}) for the three algorithms
employed. The quantities under consideration are the meson masses. The
efforts have been computed by \textsc{Orth} in \cite{Orth:2002ph}
using the Jackknife method. See also \cite{Schroers:2001if} for the
latest results.
\begin{table}[htb]
  \begin{center}
    \begin{tabular}[c]{l|c|c}
      \hline\hline
      & \multicolumn{2}{c}{\textbf{Efforts for meson masses}} \\
      \cline{2-3}
      \textbf{Algorithm} & $\mathbf{E_{\mbox{\tiny
            indep}}(m_{\pi})/\mbox{MV-mults}}$
      & $\mathbf{E_{\mbox{\tiny indep}}(m_{\rho})/\mbox{MV-mults}}$ \\
      \hline
      HMC   & $<810000$  & $<810000$  \\
      MB-HB & $>2000000$ & $>2000000$ \\
      MB-OR & $264000$   & $352000$   \\
      \hline\hline
    \end{tabular}
    \caption{Numerical efforts for meson masses obtained by employing
      three different sampling algorithms. Courtesy \textsc{B.~Orth}.}
    \label{tab:eff-results-mesons}
  \end{center}
\end{table}

\begin{figure}[htb]
  \begin{center}
    \includegraphics[scale=0.3,clip=true]%
    {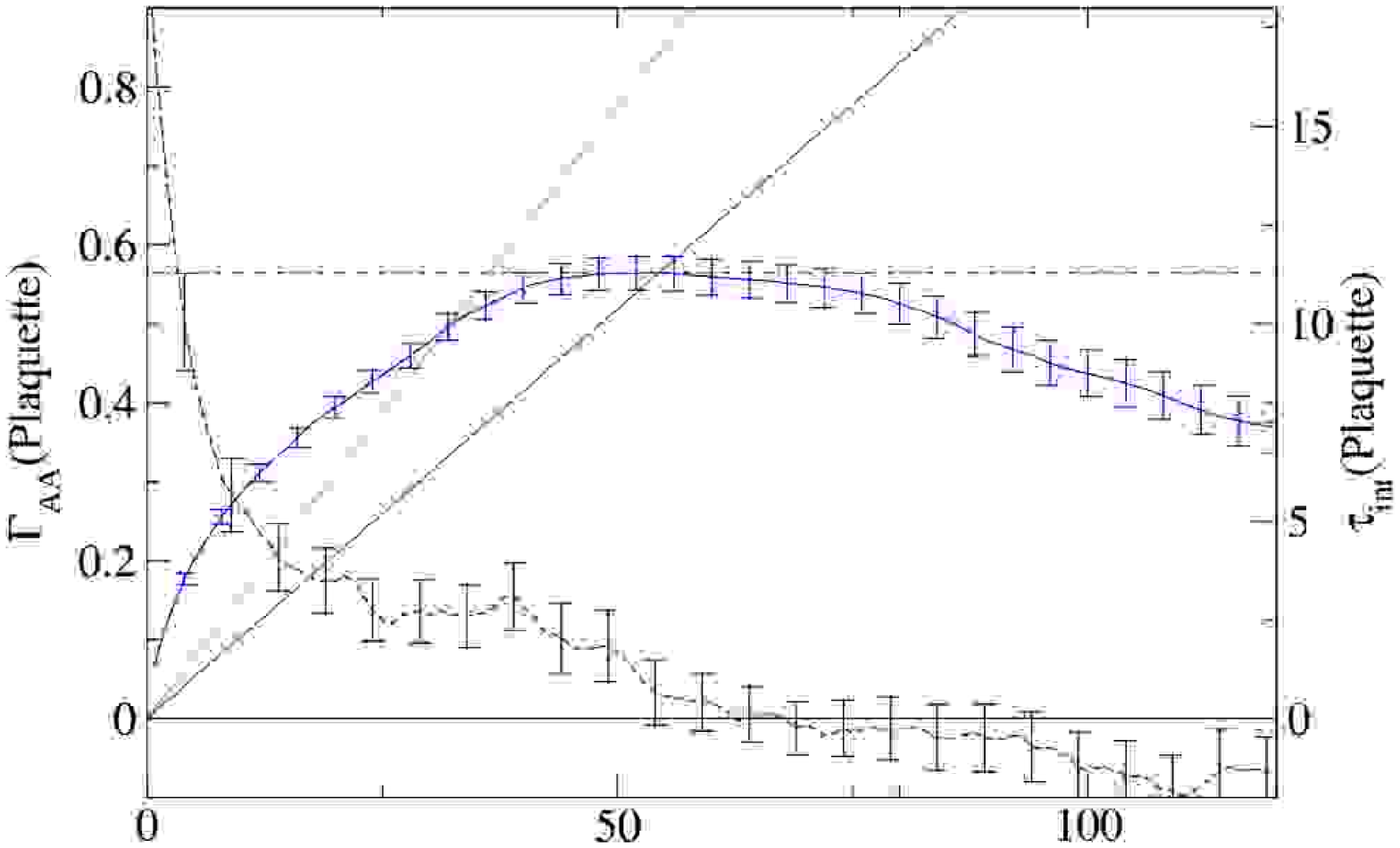}
    \caption{Autocorrelation function and the corresponding integral
      as a function of the cutoff for the plaquette history from the
      HMC run. This figure is identical to
      Fig.~\ref{fig:hmc-autocorrs}.}
    \label{fig:hmc-autocorr-chap5}
  \end{center}
\end{figure}

\begin{figure}[htb]
  \begin{center}
    \includegraphics[scale=0.3,clip=true]%
    {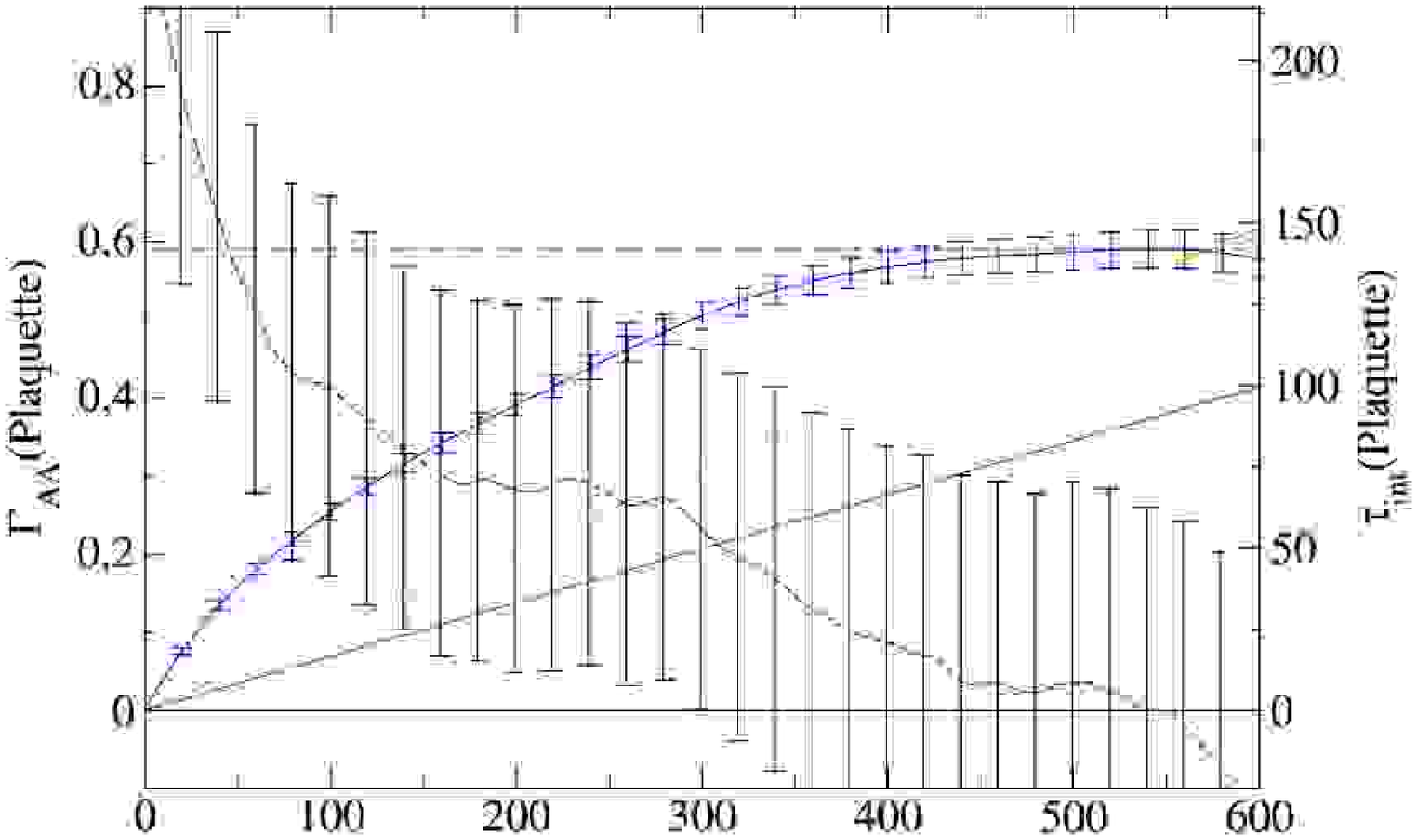}
    \caption{Autocorrelation function and integrated autocorrelation
      time for the plaquette histories of the MB-HB algorithm.}
    \label{fig:kappa159-mbhb-ac}
  \end{center}
\end{figure}

\begin{figure}[htb]
  \begin{center}
    \includegraphics[scale=0.3,clip=true]%
    {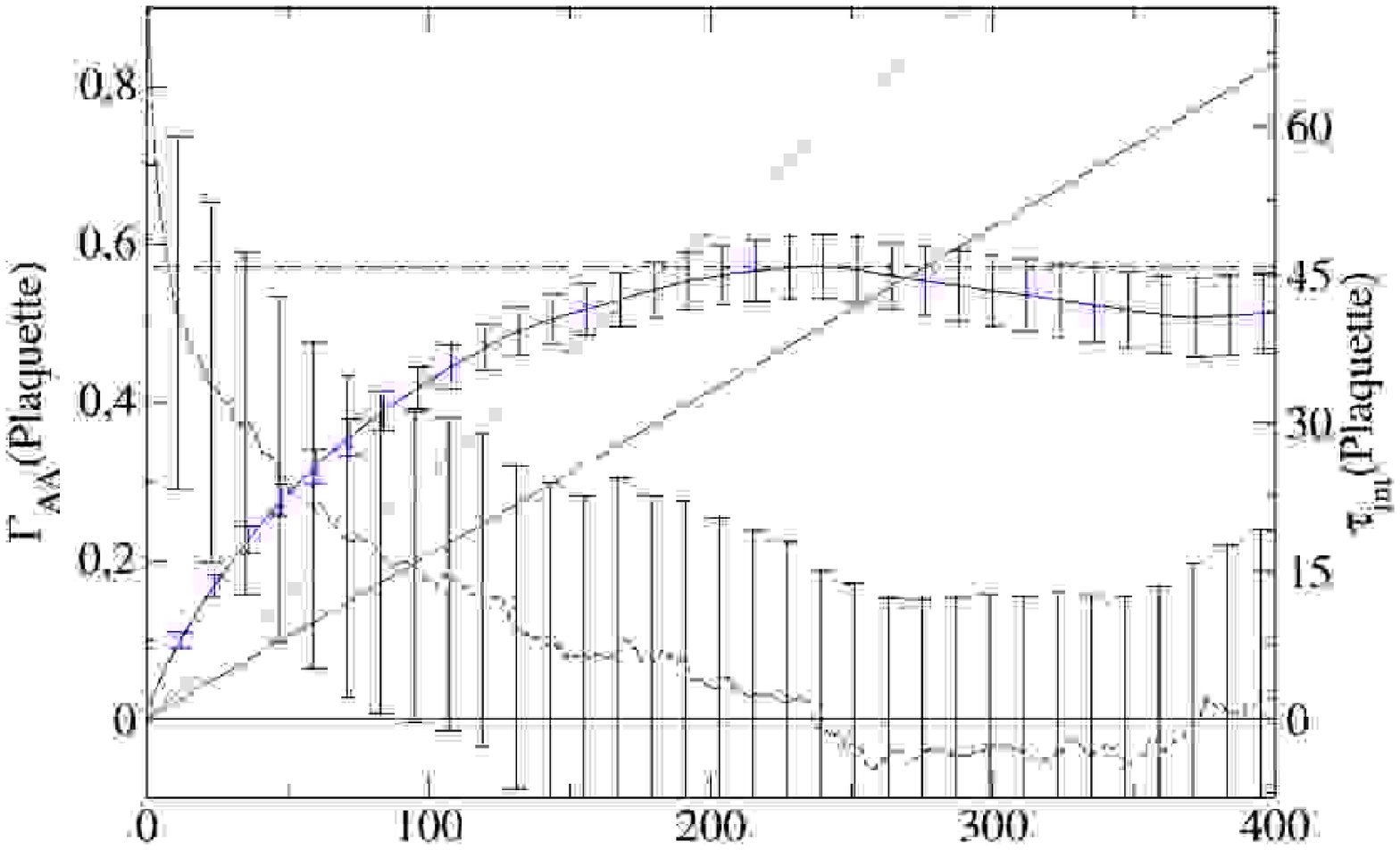}
    \caption{Autocorrelation function and integrated autocorrelation
      time for the plaquette histories of the MB-OR algorithm.}
    \label{fig:kappa159-mbor-ac}
  \end{center}
\end{figure}

Finally, the plaquette is investigated. The time series for the HMC
method at these physical parameters has already been examined in
Sec.~\ref{sec:pract-determ-autoc}.
Figures~\ref{fig:hmc-autocorr-chap5} (this figure is identical to
Fig.~\ref{fig:hmc-autocorrs}),~\ref{fig:kappa159-mbhb-ac}
and~\ref{fig:kappa159-mbor-ac} show the autocorrelation functions and
the corresponding autocorrelation times computed for the plaquette
histories from the HMC, the MB-HB and the MB-OR algorithms
respectively.

The efforts for each single trajectory, the corresponding
autocorrelation times, and the total efforts to obtain one
statistically independent plaquette measurement are listed in
Tab.~\ref{tab:eff-results-plaq}. Note that --- as has been pointed out
in Sec.~\ref{sec:setting-up-mbhb} --- the effort for a single
trajectory could have been reduced in the case of the MB-HB algorithm.
The integrated autocorrelation times have been determined using the
windowing procedure which has been discussed in
Sec.~\ref{sec:short-time-series}.
\begin{table}[htb]
  \begin{center}
    \begin{tabular}[c]{l|c|c|c}
      \hline\hline
      \textbf{Algorithm} & \textbf{Effort/Trajectory} &
      $\mathbf{\tau_{\mbox{\tiny int}}(\mbox{\textbf{Plaquette}})}$ &
      $\mathbf{E_{\mbox{\tiny
            indep}}(\mbox{\textbf{Plaquette}})/\mbox{MV-mults}}$ \\
      \hline
      HMC   & $16200$ &  $11.0\pm 0.4$  & $356400\pm 12960$  \\
      MB-HB &  $2000$ & $141.8\pm 32.7$ & $567200\pm 130800$ \\
      MB-OR &  $4400$ &  $44.7\pm 3.4$  & $393360\pm 29920$  \\
      \hline\hline
    \end{tabular}
    \caption{Autocorrelation times and efforts for independent
      plaquette measurements for the three different algorithms at
      $\beta=5.5$ and $\kappa=0.159$.}
    \label{tab:eff-results-plaq}
  \end{center}
\end{table}

The time series from the HMC algorithm contains $320$ autocorrelation
times which is sufficient to obtain a reliable estimate for the
autocorrelation time. The MB-OR algorithm was run for $140$
autocorrelation times, which should be enough for a good estimate. The
MB-HB algorithm, however, has only accumulated of the order of ${\cal
  O}(40)$ autocorrelation times if the value of $\tau_{\mbox{\tiny
    int}}$ is correct. This is too short for a safe determination of
$\tau_{\mbox{\tiny int}}$, therefore, these numbers have to be taken
with a grain of salt. One cannot be sure that already the longest mode
has been measured in the time series, but one can consider the
autocorrelation mode giving rise to this value as a lower limit of the
true autocorrelation time.

As it has already been anticipated, the MB-HB algorithm can not
compete with the MB-OR algorithm at this point in parameter space. The
observed autocorrelation time for the plaquette in terms of
trajectories is a factor of about $3.2$ larger than for the MB-OR
algorithm. However, the statistics which went into the MB-HB run is
not yet sufficient. Given the large difference in the number of boson
fields and the small number of gauge field updates between the noisy
corrections during each trajectory compared to the MB-OR algorithm,
the efficiency may consequently be even worse than what is expressed
in Tab.~\ref{tab:eff-results-plaq}. The results for the meson masses
(cf.~Tab.~\ref{tab:eff-results-mesons}) are compatible with the these
findings. Again, the numbers should only be considered to be lower
limits and may not capture the longest mode of the time series in
question.

When comparing the MB-OR and the HMC algorithms regarding the
plaquette autocorrelation times, one finds that the algorithms are
similarly efficient. In the case of the meson masses, the problem
occurs that for the HMC only the configuration at every $25$th
trajectory has been analyzed. There is no residual autocorrelation in
the sample, therefore the actual autocorrelation times may be even
smaller than the numbers given in Tab.~\ref{tab:eff-results-mesons}.

In conclusion, at the physical point given in Tab.~\ref{tab:phys-par},
the MB-OR algorithm performs for the decorrelation of the hadronic
masses at least as good as the HMC\@. For the measurement of hadronic
masses, the results are similar. However, one finds that the tuning of
MB algorithms is crucial for their performance.

\subsection{Scaling Behavior of Algorithms}
\label{sec:scal-behav-algor}
The ultimate goal of Lattice QCD simulations has been formulated in
\cite{Sharpe:2000bc} (cf.~Sec.~\ref{sec:lattice-qcd}), namely the
demand to simulate with three light fermionic flavors at quark masses
of about $1/4m_{s}$. For this goal to be reached, an algorithm is
required which has a sufficiently weak critical scaling exponent when
approaching the chiral regime (see Eq.~(\ref{eq:tauint-diverge})). The
challenge is now to apply the algorithms from the previous comparison
to a point in phase space with lighter fermion masses. The point has
been chosen from the third line in Tab.~\ref{tab:phys-par-complete},
i.e.~$\beta=5.5$ and $\kappa=0.160$. It corresponds to lighter quark
masses and should allow to shed some light on the scaling behavior of
the algorithms under consideration.

The updating sequence of the multiboson algorithm has been chosen
identical to the previous run, see Tab.~\ref{tab:mbor-genpars}. The
number of boson fields had to be increased, however, and is now
$n_1=42$. This results in an acceptance rate of $65.85\%$.

As has been found in Eq.~(\ref{eq:indep-cost}), the total cost for a
single trajectory should depend quadratically on the number of boson
fields, $n_1$. From the cost obtained in
Tab.~\ref{tab:eff-results-plaq} for $n_1=24$ we read off that
an estimate for the cost with $n_1=42$ is given by
\begin{equation}
  \label{eq:cost-estimate}
  E_{\mbox{\tiny indep}} \simeq \left(\frac{42}{24}\right)^2 \times
  393360\pm 29920 = 1204665\pm 91630\,.
\end{equation}
This estimate neglects the non-quadratic contribution of the
correction step to the trajectory, but should still be a good
approximation given the fact that the updating sweeps dominate the
total cost.

The number of trajectories performed in each case together with the
average plaquette is listed in Tab.~\ref{tab:light-compare-pars}. The
plaquettes coincide within their standard errors.
\begin{table}[htb]
  \begin{center}
    \begin{tabular}[c]{l|c|c}
      \hline\hline
      \textbf{Algorithm} & \textbf{Trajectories} & \textbf{Plaquette}
      \\ \hline
      HMC   & $5003$ & $0.56077(6)$ \\
      MB-OR & $9910$ & $0.56067(5)$ \\
      \hline\hline
    \end{tabular}
    \caption{Statistics and average plaquette for the HMC and the MB-OR
      algorithms at $\beta=5.5$ and $\kappa=0.160$.}
    \label{tab:light-compare-pars}
  \end{center}
\end{table}

\begin{figure}[htb]
  \begin{center}
    \includegraphics[scale=0.3,clip=true]%
    {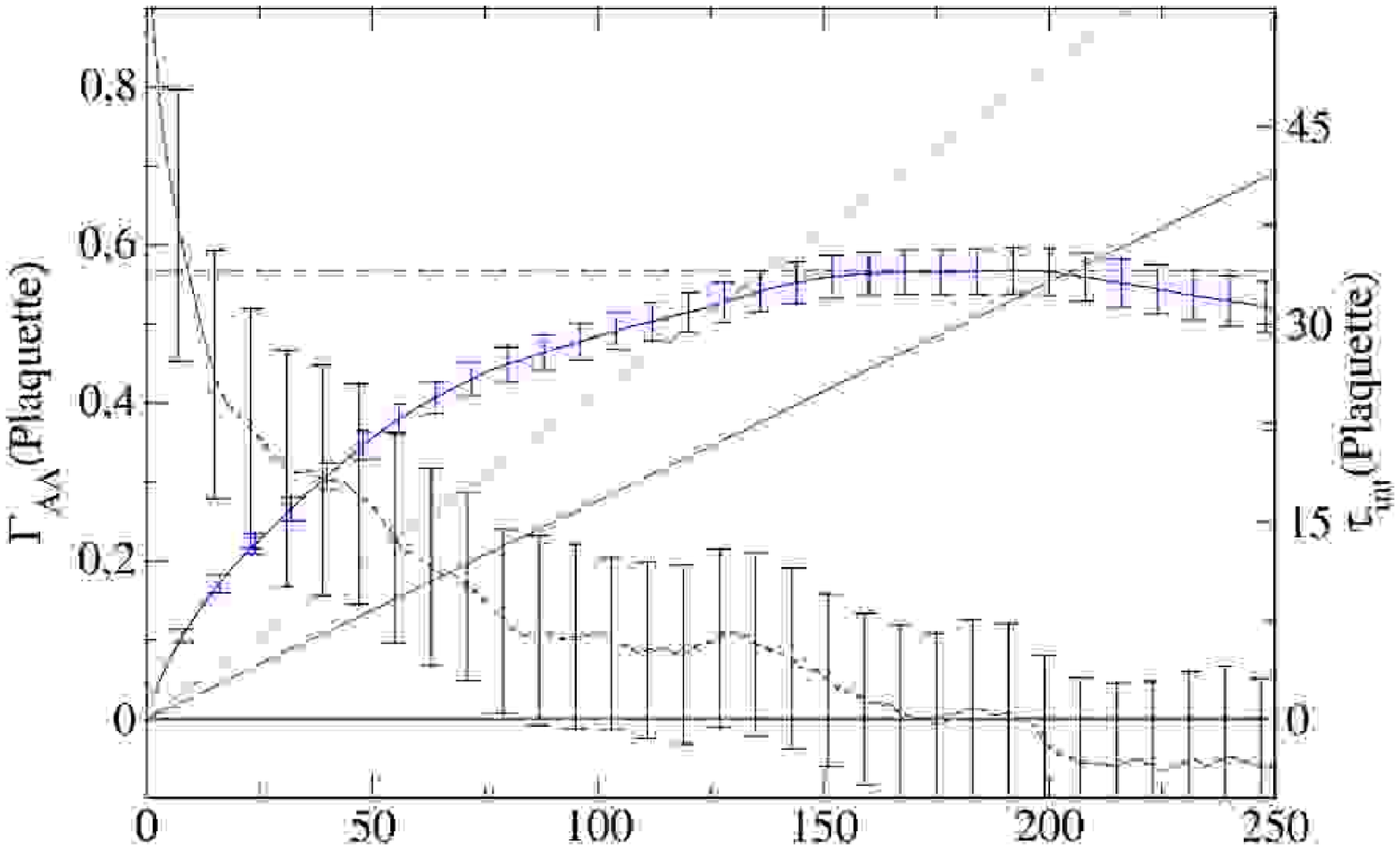}
    \caption{Autocorrelation function and integrated autocorrelation
      time for the plaquette histories of the HMC algorithm.}
    \label{fig:kappa160-hmc-ac}
  \end{center}
\end{figure}

\begin{figure}[htb]
  \begin{center}
    \includegraphics[scale=0.3,clip=true]%
    {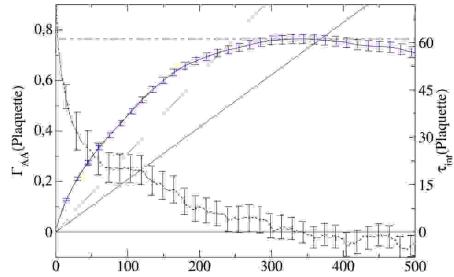}
    \caption{Autocorrelation function and integrated autocorrelation
      time for the plaquette histories of the MB-OR algorithm.}
    \label{fig:kappa160-mbor-ac}
  \end{center}
\end{figure}

The autocorrelation functions corresponding to the plaquette histories
are displayed in Figs.~\ref{fig:kappa160-hmc-ac}
and~\ref{fig:kappa160-mbor-ac}.

The corresponding efforts for a single trajectory, the integrated
autocorrelation times for the plaquettes and the resulting efforts are
given in Tab.~\ref{tab:eff2-results-plaq}. The statistics for the HMC
algorithm are now more than $140$ autocorrelation times, while the
MB-OR has generated about $160$ autocorrelation times. These numbers
should allow for a reliable estimate of the efficiencies in both
cases. In addition, the cost estimate from
Eq.~(\ref{eq:cost-estimate}) is in excellent agreement with the
measured effort $E_{\mbox{\tiny indep}}$ in
Tab.~\ref{tab:eff2-results-plaq}.
\begin{table}[htb]
  \begin{center}
    \begin{tabular}[c]{l|c|c|c}
      \hline\hline
      \textbf{Algorithm} & \textbf{Effort/Trajectory} &
      $\mathbf{\tau_{\mbox{\tiny int}}(\mbox{\textbf{Plaquette}})}$ &
      $\mathbf{E_{\mbox{\tiny indep}}(\mbox{\textbf{Plaquette}})}$ \\
      \hline
      HMC   & $42000$ & $34.1\pm 3.1$ & $2864400\pm 260400$ \\
      MB-OR &  $8800$ & $61.1\pm 4.1$ & $1075360\pm 72160$ \\
      \hline\hline
    \end{tabular}
    \caption{Autocorrelation times and efforts for independent
      plaquette measurements for the two algorithms at $\beta=5.5$ and
      $\kappa=0.160$.}
    \label{tab:eff2-results-plaq}
  \end{center}
\end{table}

In light of these results, it is clear that the MB-OR gained a lot of
ground in comparison to the HMC\@. For the former, the total effort to
generate one statistically independent configuration only increased by
a factor of about $2.7$, while for the latter the effort has increased
by a factor of $8.0$. The MB algorithm has become an overall factor of
almost three more effective than the HMC\@. For the simulation of two
light, degenerate fermion flavors, the algorithm of choice is
therefore definitely a multiboson algorithm.

\section{Summary}
\label{sec:summary-5}
It has been shown that all implementations of MB algorithms considered
indeed produce the same physical results as the HMC algorithm.
However, MB algorithms are more complicated to operate and tune and it
has turned out that a suboptimal choice can easily lead to a
degradation in performance. The efficiency of MB algorithms depends
strongly on the polynomial and the updating sequence. The optimal
setup for the polynomial at the chosen working points has been
identified in Sec.~\ref{sec:algor-polyn}. Furthermore, it has been
discussed in Sec.~\ref{sec:updating-strategy} that one should apply
sufficiently many gauge field updates between the correction steps to
ensure a fast decorrelation. In this way, the field updates will
dominate the runtime of the algorithm and lead to an optimal
exploitation of resources.

For intermediate quark masses the HMC is able to perform equivalently
to a well-tuned multiboson algorithm. When going to lighter quark
masses, however, the MB will pretty soon outrival the HMC\@. It still
remains to be seen, to what extend a non-Hermitian polynomial
approximation is a viable candidate for further simulations in the
deep chiral regime as they are planned in \cite{Orth:2001mk}. One may
have to switch to a Hermitian approximation after one starts to
encounter ``exceptional'' configurations with negative real
eigenvalues to get reasonable acceptance rates. This step might be
accompanied with an increase of $n_1$. However, first indications
regarding the behavior of the smallest real eigenvalues in simulations
in the deep chiral regime are given in \cite{Farchioni:2001di} and
references therein. These preliminary results hint that in actual
simulations the sign problem may be absent unless one gets extremely
close to the chiral limit.

The optimal tuning of the MB algorithm can only be found after a
certain runtime has already been invested since the best polynomial
approximating the fermionic contribution to the action can only be
gained from one or more thermalized gauge field configurations. This
additional effort requires more logistics and should also be
considered when estimating the efficiencies.

Due to the price in complexity one has to pay, the HMC can
consequently still be the preferred choice whenever it can be expected
to be comparable or only slightly inferior to MB algorithms.
Nonetheless, for simulations at very light quark masses close to the
physical regime, it cannot be expected that the HMC is competitive
anymore. An excellent candidate for future simulations at such masses
is therefore the MB algorithm.

\chapter{Exploring the Parameter Space with Three Degenerate Dynamical
  Flavors}
\label{sec:expl-invest-param}
Up to this point, the emphasis has been put on the simulation of two
degenerate dynamical fermion flavors. However, as has been argued in
Sec.~\ref{sec:lattice-qcd}, realistic numerical simulations of Lattice
QCD require a simulation with \textit{three} dynamical fermionic
degrees of freedom. Reference \cite{Sharpe:2000bc} shows that it is
sufficient to concentrate first on the case of three mass-degenerate
dynamical fermion flavors with dynamical quark masses of the order of
$1/4m_{s}$. One possible goal is to obtain the Gasser-Leutwyler
coefficients from those runs. However, such an endeavor requires
lattice sizes and Wilson-matrix condition numbers beyond what we are
capable of handling today.

In this section, a first step in such type of program will be taken,
namely the application of a multiboson algorithm with TSMB correction
step to this physically interesting situation.

In order to prepare the stage, we will work on $\Omega=8^4$ and
$\Omega=16\times 8^3$ lattices. This will help to acquire some insight
onto the chances of doing more realistic simulations on
$\Omega=32\times 16^3$ lattices, as previously carried out for $N_f=2$
in the \SESAM-project \cite{Eicker:1998sy,Lippert:2001ha}. So the
question is whether, in the $N_f=3$ scenario, we can establish an
operational window to achieve a reasonably large pion correlation
length without hitting the shielding transition that has been found in
$N_f=2$ at finite volumes and fixed $\beta$, as $\kappa$ was increased
towards $\kappa_{\mbox{\tiny crit}}$.

Section~\ref{sec:non-zero-temperature} gives a short overview of the
determination of the non-zero temperature crossover and the shielding
transition. It is important to avoid this region in parameter space
since the physical properties of the non-hadronized region are
different from the zero-temperature phase of QCD\@. In particular, no
hadrons are expected to exist and consequently one cannot extract
useful information on their masses.

The physically interesting point in parameter space in an infinite
lattice volume $\Omega\rightarrow\infty$ is the critical point where
the Wilson matrix describes massless fermions. This property has been
discussed in Sec.~\ref{sec:fermion-fields}. The practical ways to find
this chiral limit are reviewed in Sec.~\ref{sec:chiral-limit}.

The application to two different values of $\beta$ is discussed in
Sec.~\ref{sec:explorative-studies}. These runs have been performed
with the TSMB algorithm and might allow to identify a potential
working point for future simulations.

At this stage we would like to mention some previous algorithmic work
on $N_f=3$ physics, which was mainly carried out at finite
temperatures. Reference \cite{Iwasaki:1996zt} presents a detailed
study of the thermodynamical properties of three flavor QCD\@. It
employs the $R$-algorithm for the numerical simulations.

Note that, algorithmically, extensions of the HMC can also be used for
these kinds of simulations
\cite{Lippert:1999up,Takaishi:2000rk,Takaishi:2000bt,Aoki:2001yj}.

\section{The Non-Zero Temperature Crossover}
\label{sec:non-zero-temperature}
\index{Finite-temperature} It is expected that the phase space of QCD
contains a ``deconfined phase'', where chiral symmetry is restored and
the quarks and gluons form a plasma with color-charges being
Debye-screened. This transition takes place at some critical
temperature. For general introductions to this topic consult
\cite{Montvay:1994cy,Rothe:1992nt}. This phase is interesting for the
description of hadronic matter at high temperatures and densities.
However, when performing simulations relevant for the low-temperature
phase of QCD --- where the phenomenology is dominated by hadronized
particles --- this phase should be avoided.

This transition is accompanied by a jump in the free energy of the
system. An order parameter is given by the Polyakov loop, which is
defined to be \cite{Karsch:1998sc}:
\begin{equation}
  \label{eq:def-polyloop}
  P(\vec{x}) =
  \frac{1}{3}\frac{1}{L_s^3}\;\mbox{Tr}\,\prod_{x_0}^{L_0}
  U_0\left((x_0,\vec{x})\right)\,,
\end{equation}
with $\vec{x}$ being a point in three-space, and $L_s$ and $L_0$ the
spatial and temporal lattice sizes, respectively. The physical picture
of $P(\vec{x})$ is the description of the average world line of a
static quark. Information about the free energy of a static
quark-antiquark pair can be obtained from the correlation of two such
loops having opposite direction
\begin{equation}
  \label{eq:poly-corr}
  \Gamma(\vec{x},\vec{y}) = \langle L(\vec{x}) L^\dagger(\vec{y})
  \rangle\,.
\end{equation}
One can show \cite{Rothe:1992nt} that this quantity is related to the
free energy $F_{q\bar{q}}(\vec{x},\vec{y})$ of a static
quark-antiquark pair via
\begin{equation}
  \label{eq:poly-wilson}
  \Gamma(\vec{x},\vec{y}) = \exp\left[-\beta
    F_{q\bar{q}}(\vec{x},\vec{y})\right]\,.
\end{equation}
Assuming that $\Gamma(\vec{x},\vec{y})$ satisfies clustering, one
finds
\begin{equation}
  \label{eq:poly-cluster}
  \Gamma(\vec{x},\vec{y}) = \langle L(\vec{x}) L^\dagger(\vec{y})
  \rangle \buildrel{|\vec{x}-\vec{y}|\rightarrow\infty}\over
  {\rightarrow} |\langle L\rangle|^2\,.
\end{equation}
Hence, one obtains that if $\langle L\rangle=0$, the free energy
increases for large $|\vec{x}-\vec{y}|$ with the separation of the
quarks. This is a signal for the hadronization phase.

Therefore the order parameter indicates the phase of the system via
\begin{equation}
  \label{eq:poly-crit}
  \langle P\rangle = \left\lbrace\begin{array}{rl}
      = 0    & \mbox{hadronization}\;, \\
      \neq 0 & \mbox{finite-temperature phase}\;.
    \end{array}\right.
\end{equation}
This argumentation so far is only valid in the absence of dynamical
quarks. It may, however, also be extended to the case of dynamical
quarks with finite mass, see \cite{Rothe:1992nt}. In this case, the
Polyakov loop might similarly indicate the non-zero temperature
crossover.

Up to this point, the discussion has always considered the case where
the temporal lattice extension is smaller than the spatial one,
$L_0<L_s$. In actual simulations, the situation can also arise that a
transition similar to the non-zero temperature crossover occurs for
too small lengths $L_s$, even if $L_0$ is sufficiently large. In this
case, one is similarly unable to measure hadronic masses properly.
This phenomenon is called the \textit{shielding transition}%
\index{Shielding transition}.

\section{The Chiral Limit}
\label{sec:chiral-limit}
\index{Critical point} Of particular importance for any simulation of
QCD is the critical line in parameter space, where the mass of the
pion vanishes. The vicinity of this point allows for a treatment using
$\chi$PT, as it has been argued in Sec.~\ref{sec:lattice-qcd}. As
explained in Sec.~\ref{sec:fermion-fields}, the Wilson matrix then
contains a zero-mode.

The critical line can be found by varying the hopping parameter
$\kappa$ appearing in the action Eq.~(\ref{eq:yang-mills-lattice}) at
a fixed value of the gauge bare parameter $\beta$. Then one has to
find the critical value $\kappa_{\mbox{\tiny crit}}$, where the
fermionic contribution to the action describes massless fermions.
Repeating this procedure for several values of $\beta$ yields the
critical line in parameter space.  This procedure is impeded once the
shielding transition sets in.

\begin{figure}[htb]
  \begin{center}
    \includegraphics[scale=1.0,clip=true]{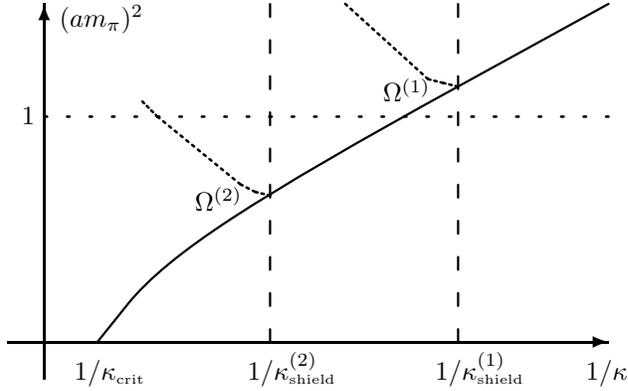}
    \caption{Sketch of the FSE-induced shielding transition. The
      squared pseudoscalar meson mass, $(am_\pi)^2$, is plotted
      vs.~the inverse hopping parameter $1/\kappa$.}
    \label{fig:shielding-trans}
  \end{center}
\end{figure}

A qualitative illustration of the shielding transition and the
critical behavior is given in Fig.~\ref{fig:shielding-trans}. The
figure shows the squared pseudoscalar meson mass, $(am_\pi)^2$, at a
fixed value of $\beta$, as a function of $1/\kappa$. The solid curve
shows the mass in the infinite volume limit,
$\Omega\rightarrow\infty$. The dotted line corresponds to a
correlation length $\xi_\pi^2=1/(am_\pi)^2=1$.

\index{Shielding transition}%
\index{Finite-size effects|see{Shielding transition}} As has already
been pointed out, finite-size-effects (FSE) will induce the shielding
transition which might inhibit a reliable extraction of
zero-temperature physics. We illustrate this scenario by sketching the
FSE for two different volumes, $\Omega^{(1)}<\Omega^{(2)}$, with
lengths $L_s^{(1)}< L_s^{(2)}$. When measuring the mass on the smaller
lattice volume, $\Omega^{(1)}$, one finds that the curve can be
followed reliably up to the point $\kappa_{\mbox{\tiny
    shield}}^{(1)}$. Beyond this point, the shielding transition sets
in and the mass can no longer be measured correctly on the smaller
volume. The larger lattice volume, $\Omega^{(2)}$, allows to go closer
to the critical point, but will still run into finite-size-effects at
some higher value, $\kappa_{\mbox{\tiny shield}}^{(2)}$. The ``true''
value of $\kappa_{\mbox{\tiny crit}}$ (as defined in the limit
$\Omega\rightarrow\infty$) can be estimated the better the larger the
available volume.

Lattice results become meaningful, once the pseudoscalar correlation
length, stays larger than unity, $\xi_\pi\gg 1$. This condition is
impossible to fulfill on the lattice $\Omega^{(1)}$ --- the shielding
transition sets in before the desired parameter region is reached. On
the lattice $\Omega^{(2)}$, however, it is in fact possible to go
beyond $\xi_\pi>1$ before shielding is observed. Hence, for a given
set of parameters one has to increase the lattice volume until one
reaches a ``window'', where the FSE are under control while the mass
already became sufficiently small.

But how to estimate $\kappa_{\mbox{\tiny crit}}$? On a given large
enough lattice, one can use the following recipes (see also
Sec.~\ref{sec:fermion-fields})
\begin{enumerate}
\item \label{lab:critkap-diverge} the point in $\kappa$-space where
  the condition number of the Hermitian Wilson matrix $\tilde{Q}$
  diverges,
\item \label{lab:critkap-spectrum} the point where the smallest real
  eigenvalue of the non-Hermitian Wilson matrix $Q$ reaches the
  imaginary axis,
\item \label{lab:critkap-pion} the point where the pseudoscalar meson
  mass $(am_\pi)$ vanishes. This is the physical definition of the
  chiral limit.
\end{enumerate}
\textbf{Comment:} From a physical point of view, the last criterion is
the approach of choice for estimating the critical point. The first
two definitions will coincide and give identical results since, in
both cases, the matrix contains a zero-mode. Furthermore, the mass of
the pseudoscalar mesons is strongly dominated by the smallest
eigenvalues and this dominance becomes more pronounced as the chiral
limit is approached, cf.~\cite{Neff:2001ph}. Hence, the results from
all these methods will coincide sufficiently close to the chiral
limit\footnote{However, it is extremely difficult to actually work
  ``sufficiently close'' to the chiral limit}. For larger masses,
however, one can expect that the results from the methods differ in
practical simulations.

\index{Chiral perturbation theory} To properly apply the physical
definition, one can use an extrapolation inspired by $\chi$PT\@. To be
specific, one employs (see e.g.~\cite{Eicker:1998sy})
\begin{equation}
  \label{eq:pion-vs-kappa}
  \frac{1}{2}\left(\frac{1}{\kappa_{\mbox{\tiny sea}}} -
    \frac{1}{\kappa_{\mbox{\tiny crit}}}\right) \propto
  (am_{\pi})^2\,.
\end{equation}
Strictly speaking, $\chi$PT only applies in the continuum. However, it
is customary to nonetheless use such type of fitting function at a
fixed value of $\beta$, see again \cite{Eicker:1998sy} and also
\cite{Eicker:2001ph} for latest results. Furthermore, this relation
might have to be modified by logarithmic corrections which could cause
the linear behavior predicted by Eq.~(\ref{eq:pion-vs-kappa}) to be
inaccessible in current simulations \cite{Sharpe:2000bc}. For the
moments of structure functions it has indeed been shown in
\cite{Detmold:2001jb} that a logarithmically modified extrapolation
formula appears to yield best agreement with experimental data.
Therefore, one should be careful when interpreting all predictions
obtained by linear fits only.

\section{Explorative Studies}
\label{sec:explorative-studies}
In this section, results from simulations at two different values of
$\beta$ are presented, namely at $\beta=5.3$
(Sec.~\ref{sec:case-beta5.3}) and $\beta=5.2$
(Sec.~\ref{sec:case-beta5.2}). For several values of
$\kappa_{\mbox{\tiny sea}}$, the average plaquette is determined. In
both cases, the critical value, $\kappa_{\mbox{\tiny crit}}$, is
measured. In the latter case, both methods discussed in
Sec.~\ref{sec:chiral-limit} are applied, while in the former case only
a single method is used.

A discussion about prospects for future simulations concludes these
investigations.

\subsection{The Case $\beta=5.3$}
\label{sec:case-beta5.3}
The simulations discussed here have been run at a value of $\beta=5.3$
on lattices with volume $\Omega=8^4$ and varying values of
$\kappa_{\mbox{\tiny sea}}$.
\begin{table}[ht]
  \begin{center}
    \begin{tabular}[c]{c|c|c|c}
      \hline\hline
      $\mathbf{\kappa_{\mbox{\tiny sea}}}$ & 
      \textbf{Number of confs.} & \textbf{Plaquette} & 
      $\mathbf{\tau_{\mbox{\tiny\bf int}}}$   \\ \hline
      $0.125$ & $3750$  & $0.4627(6)$ & $75$  \\
      $0.135$ & $15710$ & $0.4717(7)$ & $281$ \\
      $0.145$ & $15700$ & $0.4840(6)$ & $297$ \\
      $0.150$ & $11600$ & $0.4956(8)$ & $265$ \\
      $0.155$ & $9400$  & $0.5118(8)$ & $305$ \\
      $0.160$ & $6100$  & $0.5498(18)$ & -    \\
      $0.161$ & $6200$  & $0.5533(3)$ & -     \\
      $0.162$ & $5600$  & $0.5564(5)$ & -     \\
      $0.163$ & $5500$  & $0.5595(5)$ & -     \\
      \hline\hline
    \end{tabular}
    \caption{Hopping parameter, $\kappa_{\mbox{\tiny sea}}$, number of
      trajectories, and plaquette values with resulting
      autocorrelation times for the runs with $N_f=3$ and
      $\beta=5.3$.}
    \label{tab:beta53-plaqs}
  \end{center}
\end{table}
The different values of $\kappa_{\mbox{\tiny sea}}$, the number of
trajectories after thermalization, and the resulting average plaquette
values are listed in Tab.~\ref{tab:beta53-plaqs} together with an
estimate for the integrated autocorrelation time of the average
plaquette. The standard errors on the plaquettes together with the
estimate for $\tau_{\mbox{\tiny int}}$ have been determined using the
Jackknife method. This data has been obtained from runs on both the
\Nicse\ and the \ALiCE\ clusters, see Sec.~\ref{sec:alice-cluster} for
further details. The algorithmic parameters have been varied in the
runs.  Table~\ref{tab:beta53-algpars} shows the algorithmic parameters
together with the resulting acceptance rates.

\begin{table}[htb]
  \begin{center}
    \begin{tabular}[c]{c||*{5}{c|}l}
      \hline\hline
      $\mathbf{\kappa_{\mbox{\tiny sea}}}$ & $\mathbf{n_1}$ & 
      $\mathbf{n_2}$ & $\mathbf{n_3}$ & 
      $\mathbf{[\epsilon,\lambda]}$ & \textbf{Updates/Trajectory} &
      $\mathbf{P_{\mbox{\tiny\bf acc}}}$ \\ \hline
      $0.125$ & $8$ & $60$ & $80$ & $[0.1,3]$ & $1$ boson HB, $3$
      boson OR, & \\
      &&&&& $3\times(2$ gauge Metropolis, $1$ noisy corr.$)$ &
      $28.2\%$ \\
      $0.135$ & $24$ & $100$ & $140$ & $[0.01,3]$ & $2$ boson HB, $6$
      boson OR, & \\
      &&&&& $8$ gauge Metropolis, $1$ noisy corr. & $45.9\%$ \\
      $0.145$ & $24$ & $100$ & $140$ & $[0.01,3]$ & identical to
      $\kappa_{\mbox{\tiny sea}}=0.135$ & $53.9\%$ \\
      $0.150$ & $24$ & $100$ & $140$ & $[0.01,3]$ & First $6000$
      trajs: identical to $\kappa_{\mbox{\tiny sea}}=0.125$ & $56.3\%$
      \\
      &&&&& Remaining: identical to $\kappa_{\mbox{\tiny sea}}=0.135$
      & $58.4\%$ \\
      $0.155$ & $24$ & $100$ & $140$ & $[0.01,3]$ & First $5000$
      trajs: identical to $\kappa_{\mbox{\tiny sea}}=0.125$ & $56.0\%$
      \\
      &&&&& Remaining: identical to $\kappa_{\mbox{\tiny sea}}=0.135$
      & $52.7\%$ \\
      $0.160$ & $32$ & $300$ & $400$ & $[\Exp{7.5}{-3},3]$ & identical
      to $\kappa_{\mbox{\tiny sea}}=0.135$ & $59.2\%$ \\
      $0.161$ & $32$ & $300$ & $400$ & $[\Exp{7.5}{-3},3]$ & identical
      to $\kappa_{\mbox{\tiny sea}}=0.135$ & $58.7\%$ \\
      $0.162$ & $32$ & $300$ & $400$ & $[\Exp{7.5}{-3},3]$ & identical
      to $\kappa_{\mbox{\tiny sea}}=0.135$ & $58.8\%$ \\
      $0.163$ & $32$ & $300$ & $400$ & $[\Exp{7.5}{-3},3]$ & identical
      to $\kappa_{\mbox{\tiny sea}}=0.135$ & $54.9\%$ \\ \hline
      \multicolumn{7}{c}{Volume: $\Omega=8^4$\strut} \\
      \hline\hline
    \end{tabular}
    \caption{Algorithmic parameters for the runs with three dynamical
      quark flavors at $\beta=5.3$.}
    \label{tab:beta53-algpars}
  \end{center}
\end{table}

\begin{figure}[htb]
  \begin{center}
    \includegraphics[scale=0.3,clip=true]{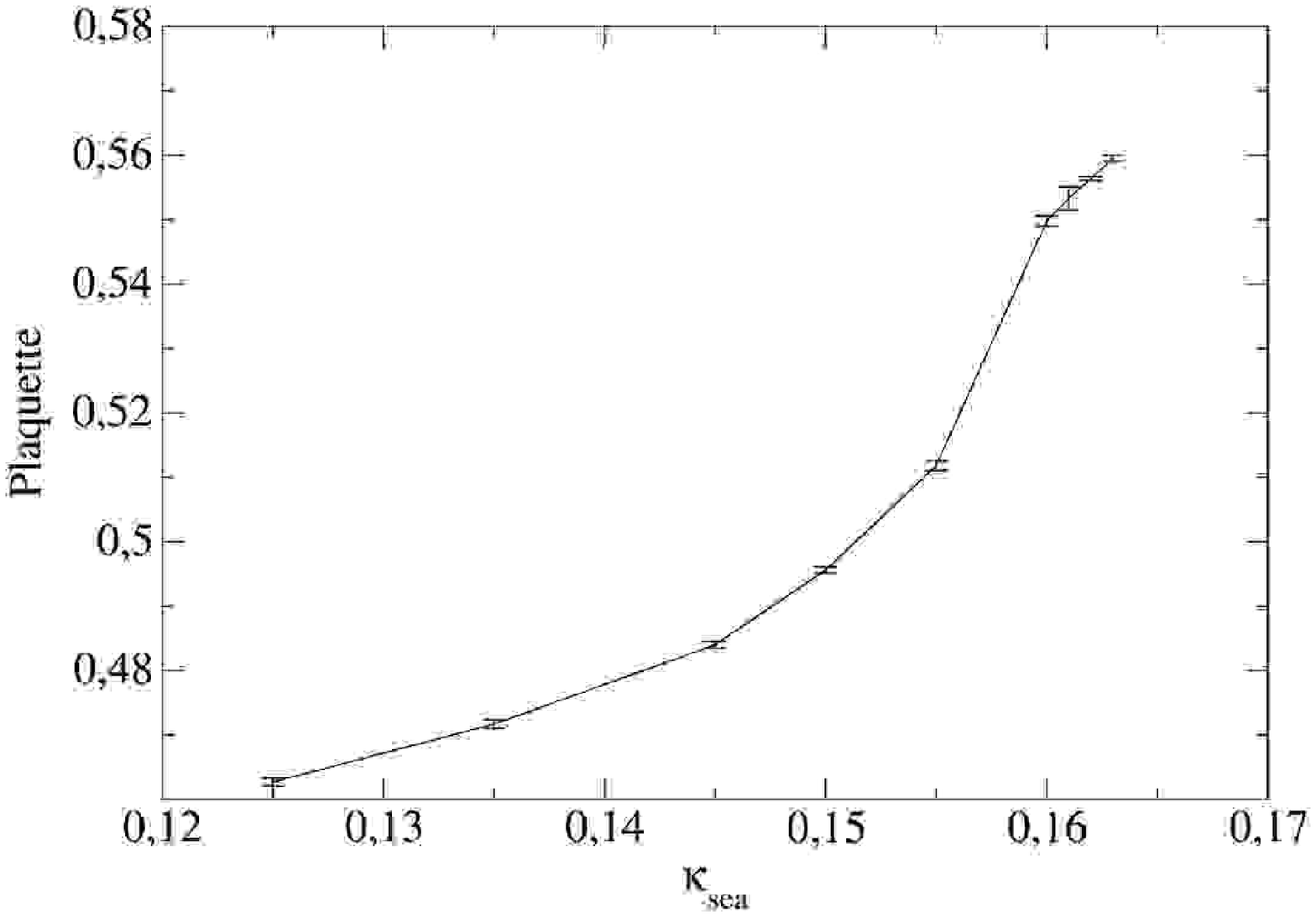}
    \caption{Average plaquettes for runs with three dynamical flavors
      at $\beta=5.3$.}
    \label{fig:b53-plaqs}
  \end{center}
\end{figure}

The average plaquette is visualized in Fig.~\ref{fig:b53-plaqs}.
Between $\kappa_{\mbox{\tiny sea}}=0.150$ and $\kappa_{\mbox{\tiny
    sea}}=0.160$ a large jump in the plaquette occurs which indicates
the presence of the shielding transition\index{Shielding transition}.
The values beyond this transition are therefore not particularly
interesting and hence less statistics has been generated.  An estimate
for the autocorrelation time has not been obtained here.  Therefore,
the statistical error may be underestimated.

For the determination of the critical value, $\kappa_{\mbox{\tiny
    crit}}$, the first method from Sec.~\ref{sec:chiral-limit} is
adopted. Table~\ref{tab:beta53-eigvals} shows the average smallest and
largest eigenvalues of $\tilde{Q}^2$. The eigenvalues have been
computed every $100$ trajectories, and the errors have again been
estimated using the Jackknife method.
\begin{table}[htb]
  \begin{center}
    \begin{tabular}[c]{c|c|c|c}
      \hline\hline
      $\mathbf{\kappa}$ &
      $\mathbf{\lambda_{\mbox{\tiny\bf min}}}$ &
      $\mathbf{\lambda_{\mbox{\tiny\bf max}}}$ &
      $\mathbf{\lambda_{\mbox{\tiny\bf max}}/\lambda_{\mbox{\tiny\bf
            min}}}$ \\ \hline
      $0.125$ & $0.2055(6)$   & $1.7034(3)$  & $8.288(12)$   \\
      $0.135$ & $0.1331(7)$   & $1.8426(5)$  & $13.844(49)$  \\
      $0.145$ & $0.07298(52)$ & $1.9997(5)$  & $27.40(19)$   \\
      $0.150$ & $0.04539(76)$ & $2.0836(10)$ & $45.91(73)$   \\
      $0.155$ & $0.02251(58)$ & $2.1678(5)$  & $96.3\pm 2.3$ \\
      $0.160$ & $0.00764(65)$ & $2.2367(18)$ & $292.9\pm 25.1$ \\
      $0.161$ & $0.00894(75)$ & $2.2545(6)$  & $252.2\pm 4.6$ \\
      $0.162$ & $0.00633(51)$ & $2.2722(8)$  & $359.2\pm 28.9$ \\
      $0.163$ & $0.00885(66)$ & $2.2908(12)$ & $258.9\pm 19.3$ \\
      \hline\hline
    \end{tabular}
    \caption{Average extremal eigenvalues and condition numbers for
      runs with three dynamical flavors at $\beta=5.3$.}
    \label{tab:beta53-eigvals}
  \end{center}
\end{table}

\begin{figure}[htb]
  \begin{center}
    \includegraphics[scale=0.3,clip=true]{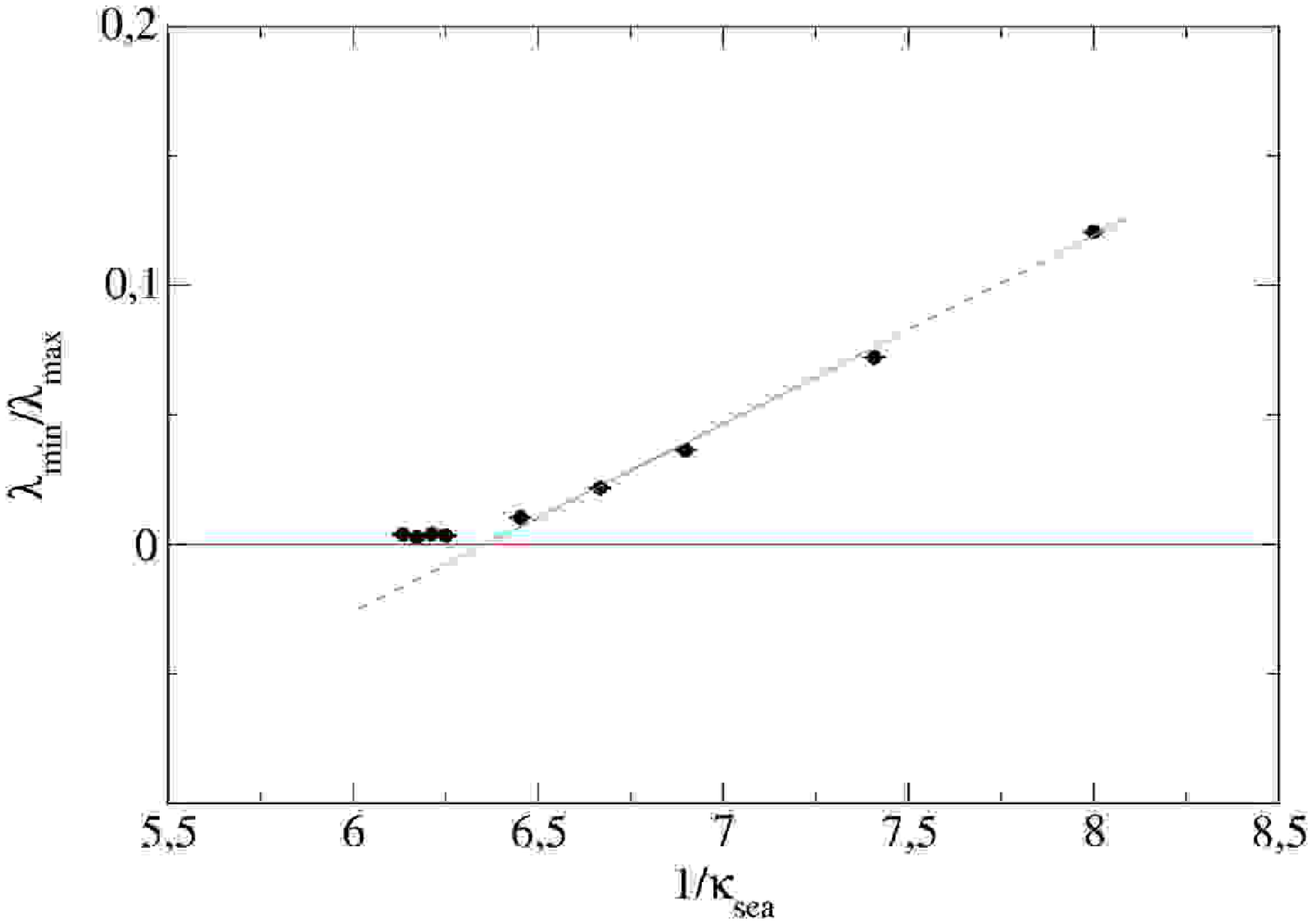}
    \caption{Inverse condition number of $\tilde{Q}^2$
      vs.~$1/\kappa$ for three dynamical fermions at $\beta=5.3$.}
    \label{fig:b53-condnum}
  \end{center}
\end{figure}

Figure~\ref{fig:b53-condnum} shows the resulting plot of $1/\kappa$
vs.~the inverse condition number $\lambda_{\mbox{\tiny
    min}}/\lambda_{\mbox{\tiny max}}$ of $\tilde{Q}^2$.

The straight line is a fit to the points between $1/\kappa=6.452$ and
$1/\kappa=8.0$ which is parameterized by
\begin{equation}
  \label{eq:b53-kappafit}
  \lambda_{\mbox{\tiny min}}/\lambda_{\mbox{\tiny max}} = -0.4610(13)
  + 0.07252(17) /\kappa\,.
\end{equation}
From the point where $\lambda_{\mbox{\tiny max}}/\lambda_{\mbox{\tiny
    min}}$ diverges (and thus $\kappa\rightarrow \kappa_{\mbox{\tiny
    crit}}$) one finds
\begin{equation}
  \label{eq:b53-kappacrit-result}
  \kappa_{\mbox{\tiny crit}} = 0.1573(4)\,.
\end{equation}
This method requires little effort and has a rather small error on the
critical value of $\kappa$. However, the estimate
(\ref{eq:b53-kappacrit-result}) still contains a systematic
uncertainty due to the fact that one is still rather far from the
chiral regime.

\subsection{The Case $\beta=5.2$}
\label{sec:case-beta5.2}
The point considered in the previous section already showed signs of
the shielding transition as the condition number of $\tilde{Q}^2$
still was below $100$. Hence, this value of $\beta$ does not allow to
probe the chiral regime further if one is limited to such small
lattices. It can, however, be considered as a working point for future
studies on larger lattices. As a different starting point, the focus
will now be placed on the point $\beta=5.2$ with lattice sizes of
$\Omega=16\times 8^3$ instead. This lattice size might already allow
for a measurement of the ratio $m_{\pi}/m_{\rho}$ for degenerate sea
and valence quark masses and hence for an independent estimate of the
chiral transition. Again, the finite-temperature phase of QCD has to
be avoided.

For the actual simulation, again several values for
$\kappa_{\mbox{\tiny sea}}$ have been chosen. The polynomial
parameters are given in Tab.~\ref{tab:large-latt-polpars}. The runs
have been performed on the \ALiCE-cluster with a partition of eight
nodes for each run.
\begin{table}[htb]
  \begin{center}
    \begin{tabular}[c]{*{4}{c|}c}
      \hline\hline
      $\mathbf{n_1}$ & $\mathbf{n_2}$ & $\mathbf{n_3}$ &
      $\mathbf{[\epsilon,\lambda]}$ & \textbf{Updates/Configuration} \\
      \hline
      $24$ & $300$ & $450$ & $[\Exp{7.5}{-4},3]$ & $1$ boson HB, $5$
      boson OR, \\ \cline{1-4} \multicolumn{4}{c|}{} &
      $2$ gauge Metropolis, $1$ noisy corr. \\ \hline
      \multicolumn{5}{c}{Volume: $\Omega=16\times 8^3$} \\
      \hline\hline    
    \end{tabular}
    \caption{Algorithmic parameters for each configuration for the
      runs at $\beta=5.2$ with $N_f=3$.}
    \label{tab:large-latt-polpars}
  \end{center}
\end{table}

In general, one can expect that the polynomial orders and intervals
are chosen somewhat conservatively and one could achieve some gain by
adapting them manually with respect to the spectrum of $\tilde{Q}^2$
obtained during the production. Despite the lengths of the runs, it
might still make sense to improve the statistics further.

The working points chosen are listed in
Tab.~\ref{tab:large-latt-three-nf} together with the acceptance rate
of the noisy correction step, the number of performed trajectories,
and the average plaquette with the error determined from the Jackknife
method. From the Jackknife estimate, the plaquette autocorrelation
time has been determined. Finally the correction factor with its
standard deviation is shown.
\begin{table}[htb]
  \begin{center}
    \begin{tabular}[c]{l|c|c|c|c|c}
      \hline\hline
      $\mathbf{\kappa_{\mbox{\tiny sea}}}$ &
      \textbf{Number of confs.} & $\mathbf{P_{\mbox{\tiny\bf acc}}}$ &
      \textbf{Plaquette} & $\mathbf{\tau_{\mbox{\tiny\bf int}}}$ &
      \textbf{Rew.~factor}                                    \\
      \hline
      $0.156$ & $16700$ & $30.30\%$ & $0.4794(9)$  & $1135$ &
      $1.0000(1)$ \\
      $0.158$ & $19980$ & $27.27\%$ & $0.4860(11)$ & $1359$ &
      $1.0001(2)$ \\
      $0.160$ & $20100$ & $30.35\%$ & $0.4923(4)$  &  $359$ &
      $1.0001(2)$ \\
      $0.162$ & $34710$ & $24.62\%$ & $0.5105(?)$  &      - &
      $1.0000(2)$ \\
      $0.163$ & $9200$  & $17.92\%$ & $0.5356(8)$  &  $735$ &
      $0.9998(6)$ \\
      $0.164$ & $24490$ & $18.99\%$ & $0.5443(6)$  & $1062$ &
      $0.97(10)$  \\
      $0.165$ & $11510$ & $25.54\%$ & $0.5478(3)$  &  $286$ &
      $1.0000(2)$ \\
      $0.166$ & $10900$ & $26.51\%$ & $0.5508(2)$  &  $116$ &
      $1.0000(2)$ \\
      \hline\hline
    \end{tabular}
    \caption{Simulation runs using three dynamical fermion
      flavors at $\beta=5.2$.}
    \label{tab:large-latt-three-nf}
  \end{center}
\end{table}

\begin{figure}[htb]
  \begin{center}
    \includegraphics[scale=0.3,clip=true]{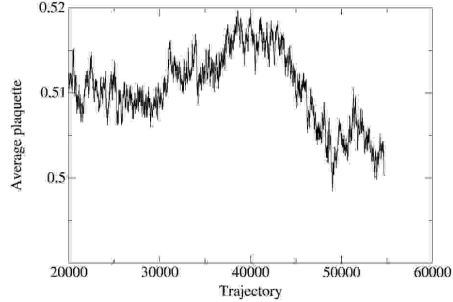}
    \caption{Plaquette history of the run at $\beta=5.2$ and
      $\kappa_{\mbox{\tiny sea}}=0.162$ with $N_f=3$.}
    \label{fig:b52-k162-plaq}
  \end{center}
\end{figure}

The plaquette for the run at $\kappa_{\mbox{\tiny sea}}=0.162$ showed
a fluctuation between two different points and is plotted in
Fig.~\ref{fig:b52-k162-plaq}.  This is an indication that the
shielding transition takes place around this point. Since the series
is too short to make any statement about this fluctuation, the
standard error is not shown here.

In the cases $\kappa_{\mbox{\tiny sea}}=0.156$ and
$\kappa_{\mbox{\tiny sea}}=0.158$ the autocorrelation time appears to
be very large. Hence, the statistics are still comparatively small at
these working points.

\begin{figure}[htb]
  \begin{center}
    \includegraphics[scale=0.3,clip=true]{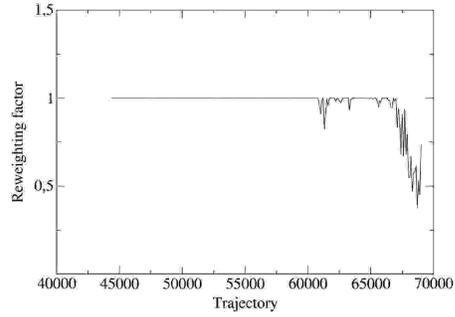}
    \caption{History of reweighting factors for the run at $\beta=5.2$
      and $\kappa_{\mbox{\tiny sea}}=0.164$ with $N_f=3$.}
    \label{fig:k0164-reweight-hist}
  \end{center}
\end{figure}

The magnitude of the reweighting factors in
Tab.~\ref{tab:large-latt-three-nf} confirms the expectation that the
polynomial has been chosen very conservatively in most cases.
However, the run at $\kappa_{\mbox{\tiny sea}}=0.164$ has a large
fluctuation in the reweighting factor, which means that the smallest
eigenvalue went off the polynomial interval. The precise situation is
displayed in Fig.~\ref{fig:k0164-reweight-hist} after the
thermalization phase has been subtracted. If this run was to be
continued, one may consider to use polynomials with a smaller value of
the lower limit for the approximation interval. The properly
reweighted values may still be used for this analysis, but the
statistics may be worse for this case.  For the other simulation runs,
one can conclude that reweighting can safely be disregarded.

\begin{figure}[htb]
  \begin{center}
    \includegraphics[scale=0.3,clip=true]{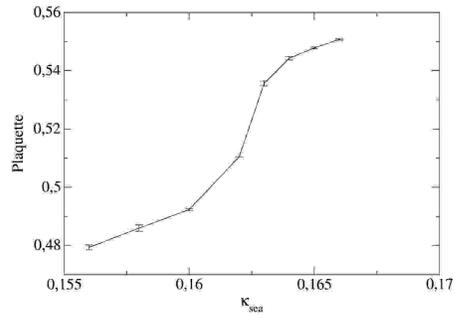}
    \caption{Average plaquettes for runs with three dynamical flavors
      at $\beta=5.2$.}
    \label{fig:b52-plaqs}
  \end{center}
\end{figure}

Figure~\ref{fig:b52-plaqs} shows the resulting values of the average
plaquette as a function of the hopping parameter $\kappa_{\mbox{\tiny
    sea}}$. This plot corroborates that the shielding transition is
located around $\kappa_{\mbox{\tiny sea}}=0.162$.
\clearpage

\begin{figure}[ht]
  \begin{center}
    \begin{tabular}[c]{l|r}
      \includegraphics[scale=0.22,bb=0 0 27cm 22cm,angle=0,clip=true]%
      {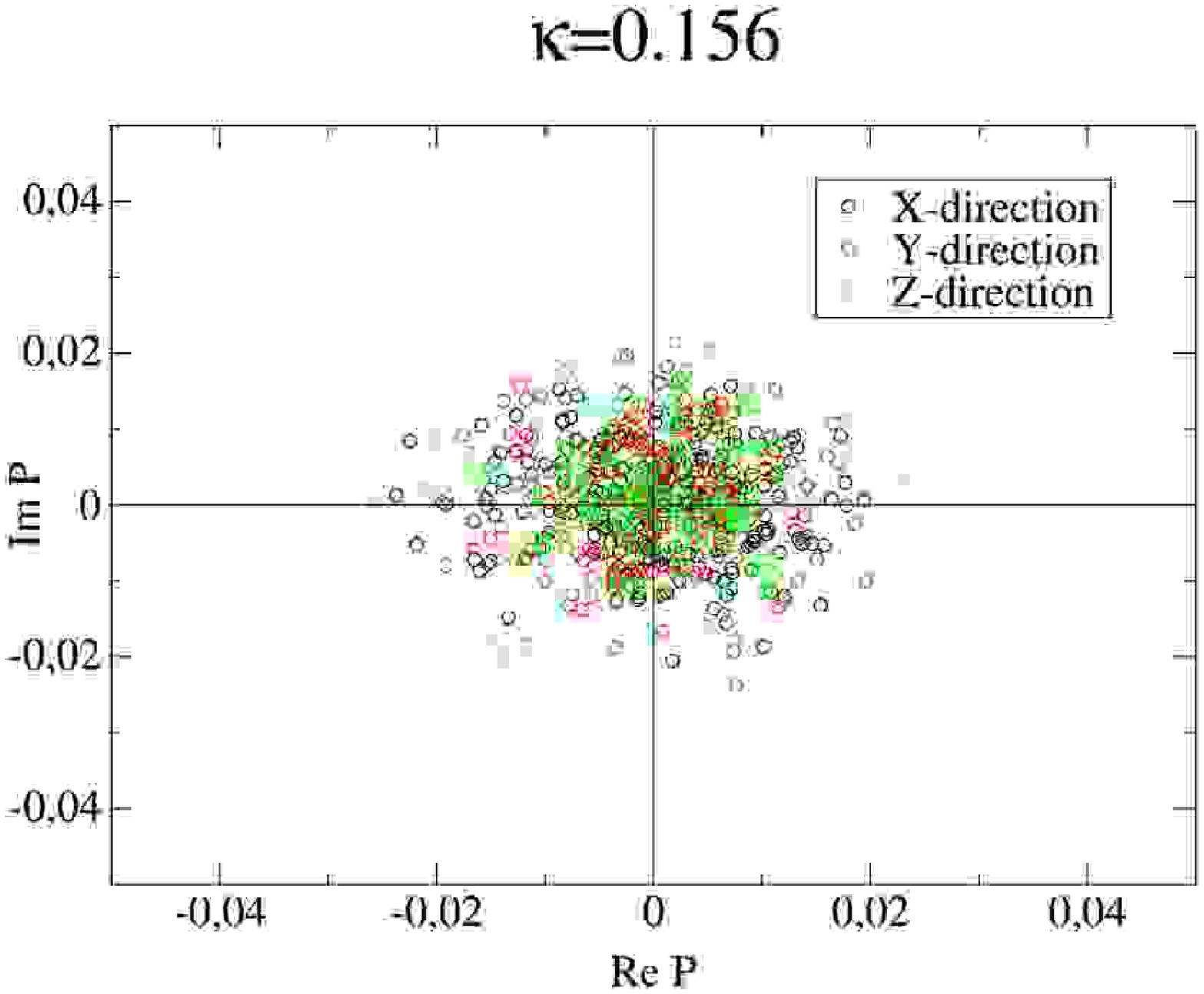} &
      \includegraphics[scale=0.22,bb=0 0 27cm 22cm,angle=0,clip=true]%
      {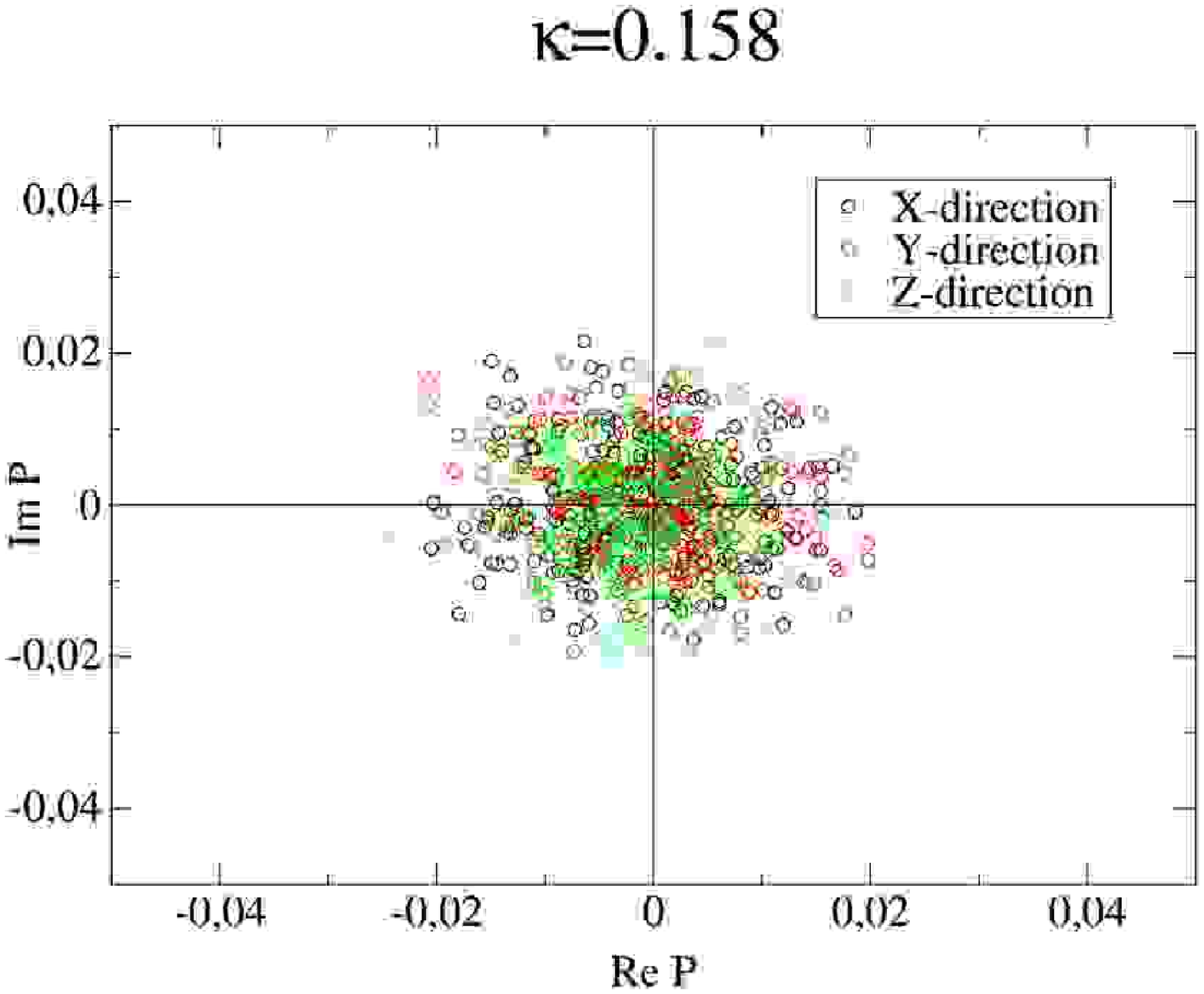} \\ \hline                
      \includegraphics[scale=0.22,bb=0 0 27cm 22cm,angle=0,clip=true]%
      {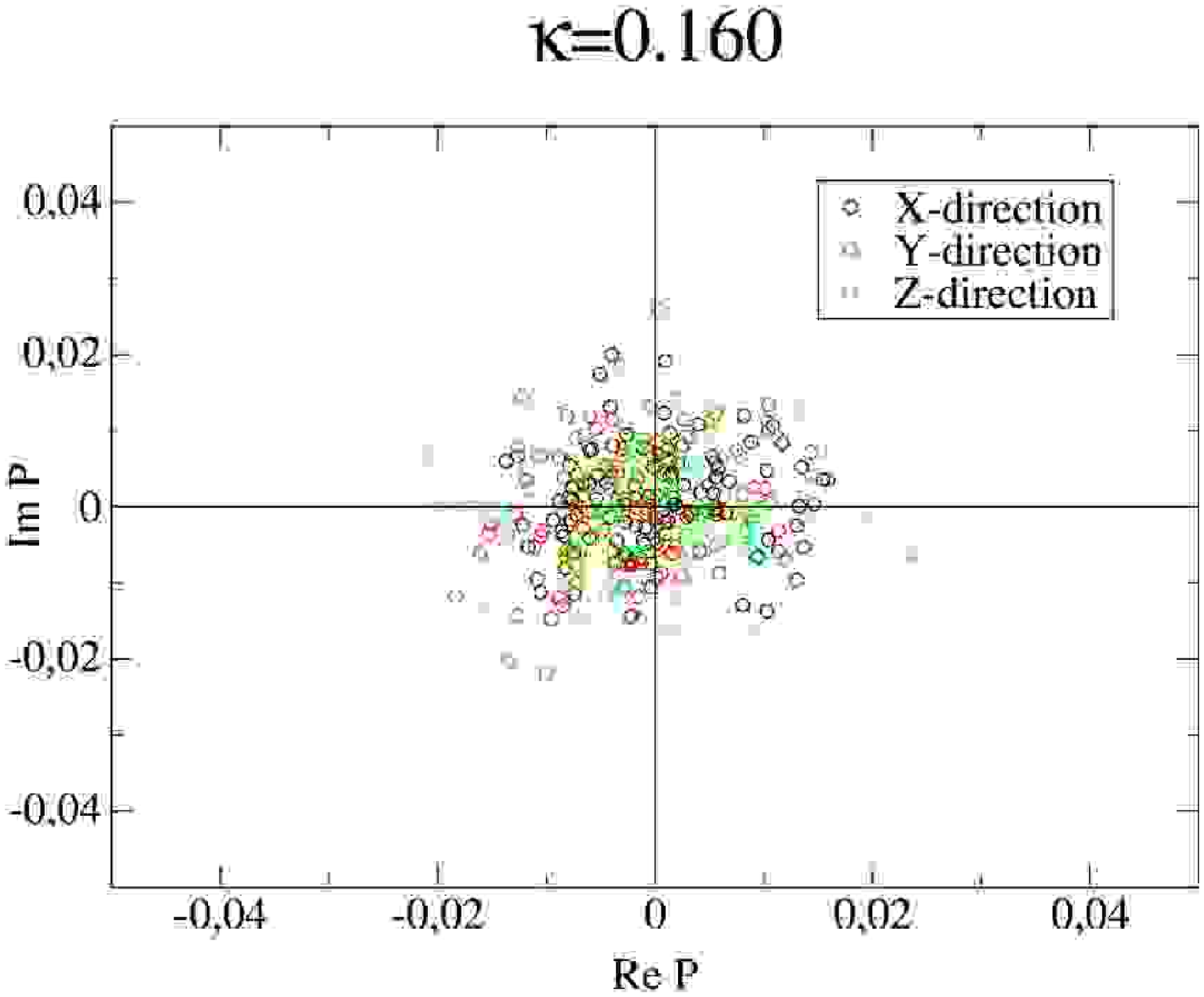} &
      \includegraphics[scale=0.22,bb=0 0 27cm 22cm,angle=0,clip=true]%
      {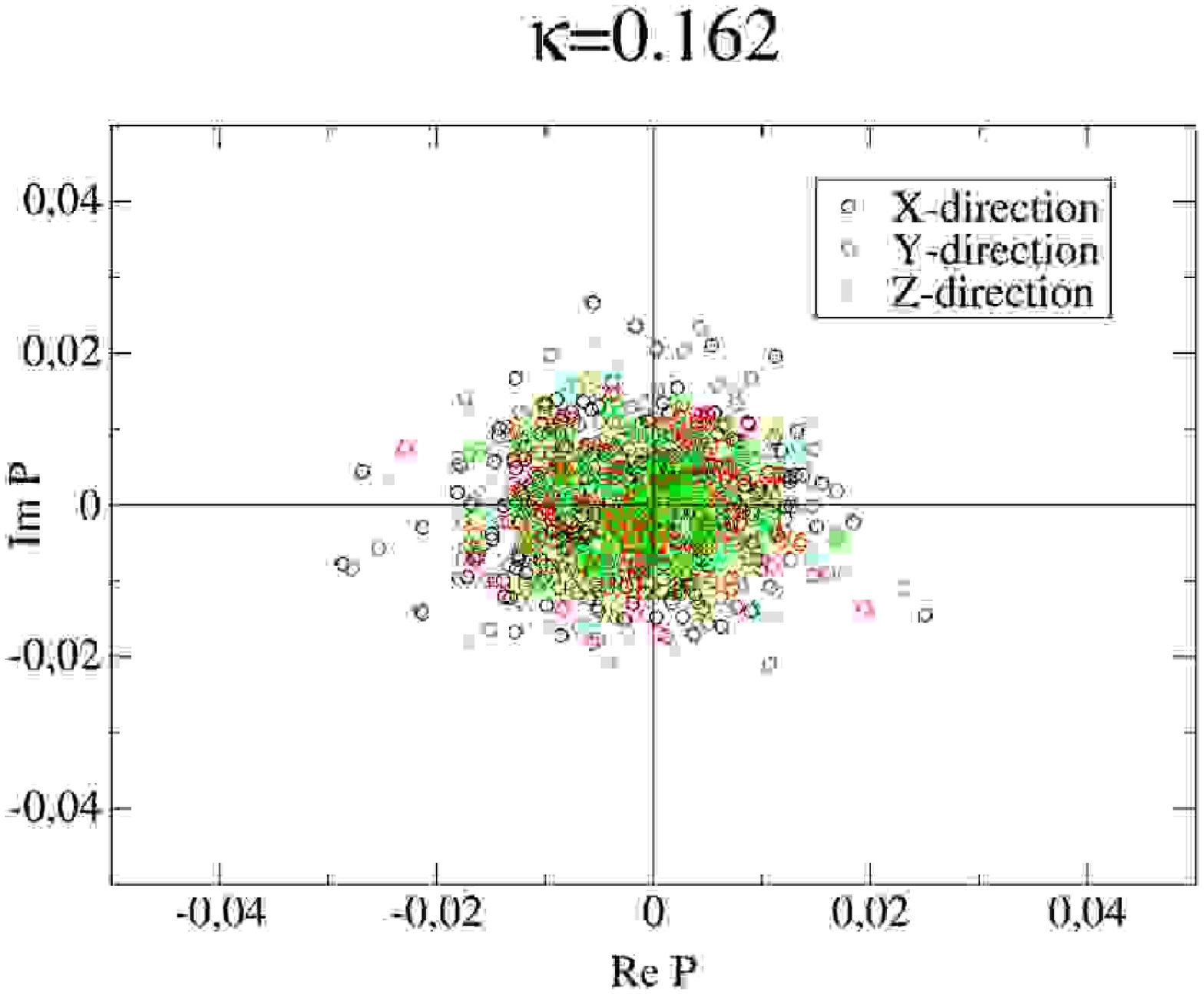} \\ \hline
      \includegraphics[scale=0.22,bb=0 0 27cm 22cm,angle=0,clip=true]%
      {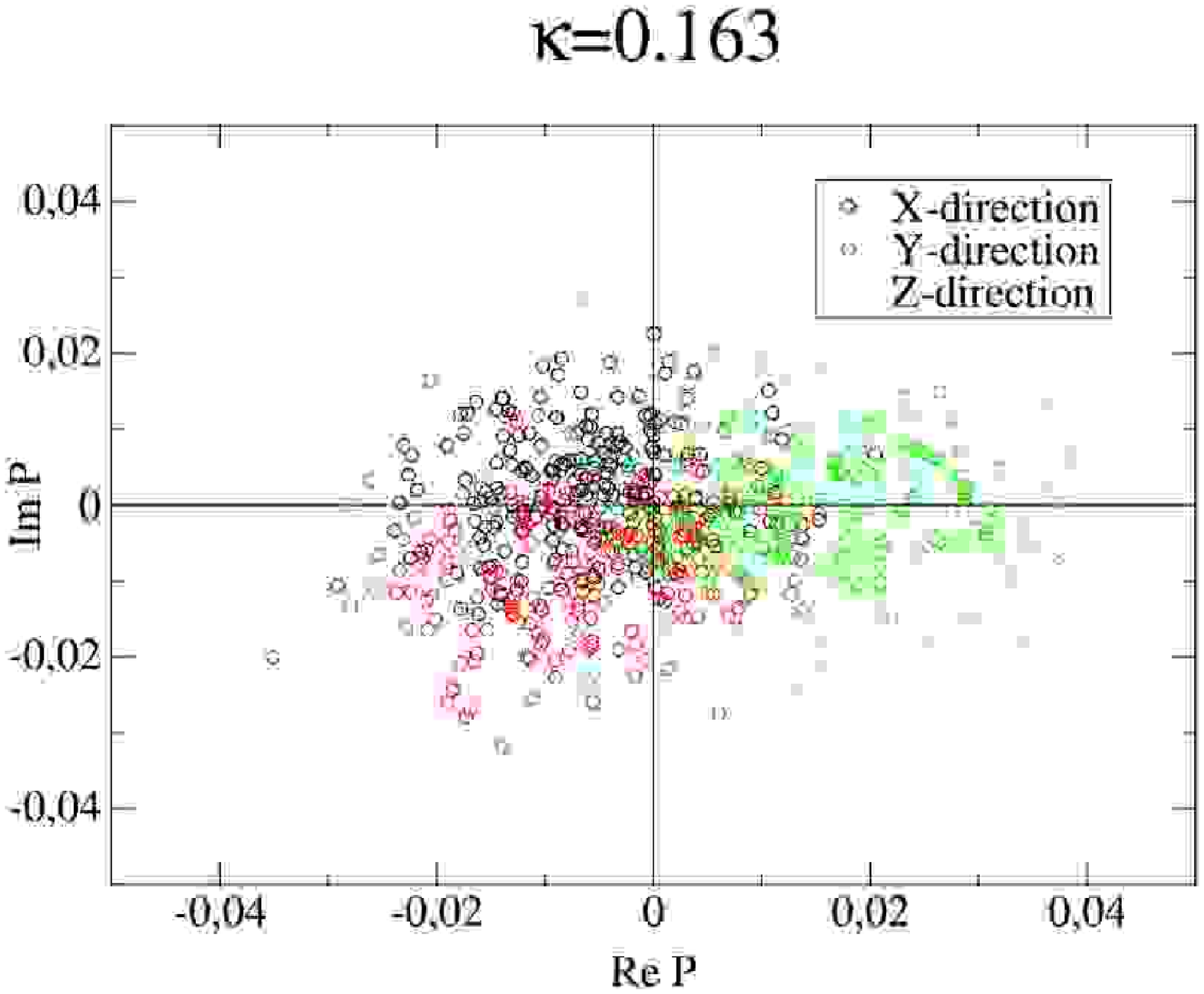} &
      \includegraphics[scale=0.22,bb=0 0 27cm 22cm,angle=0,clip=true]%
      {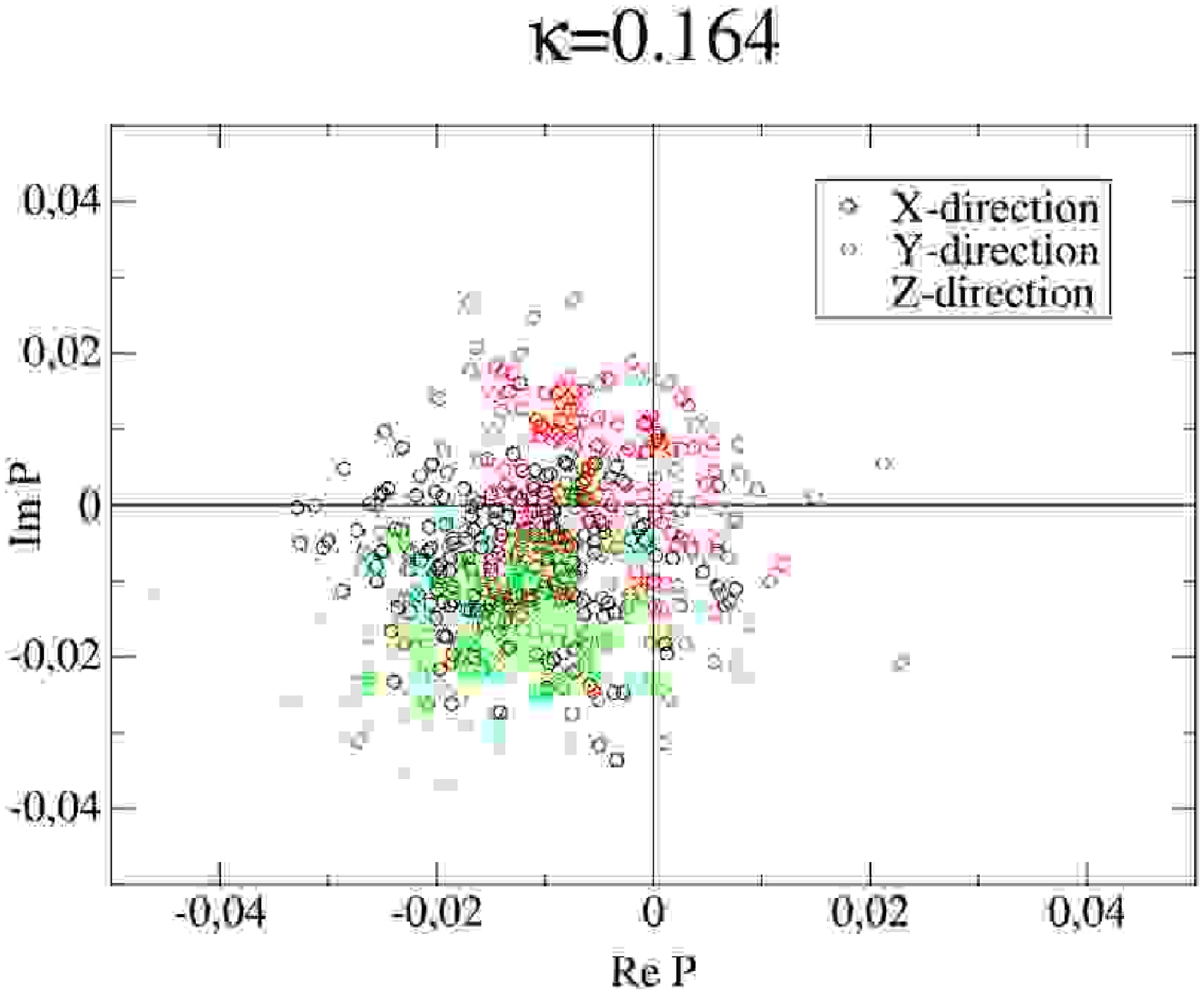} \\ \hline           
      \includegraphics[scale=0.22,bb=0 0 27cm 22cm,angle=0,clip=true]%
      {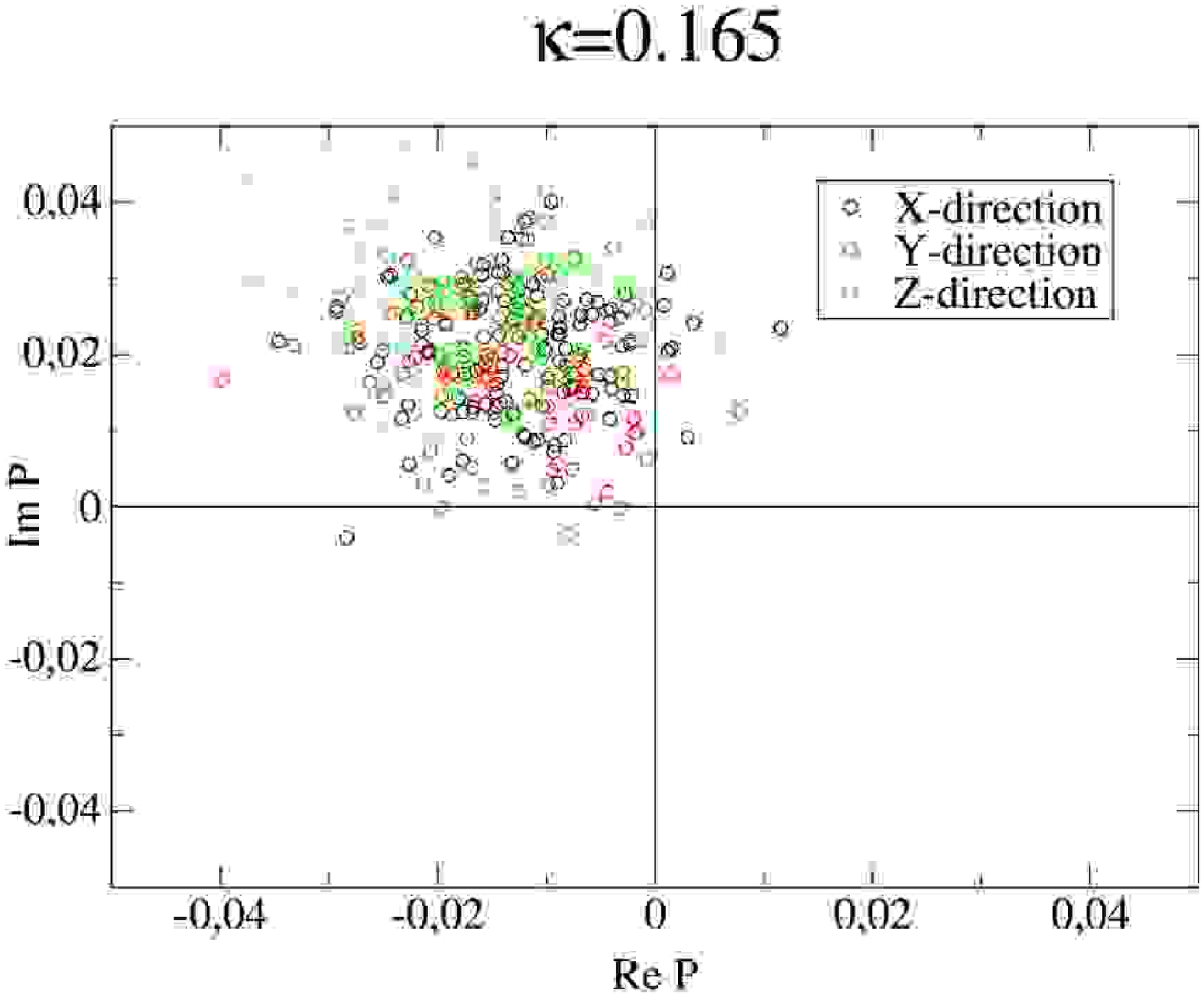} &
      \includegraphics[scale=0.22,bb=0 0 27cm 22cm,angle=0,clip=true]%
      {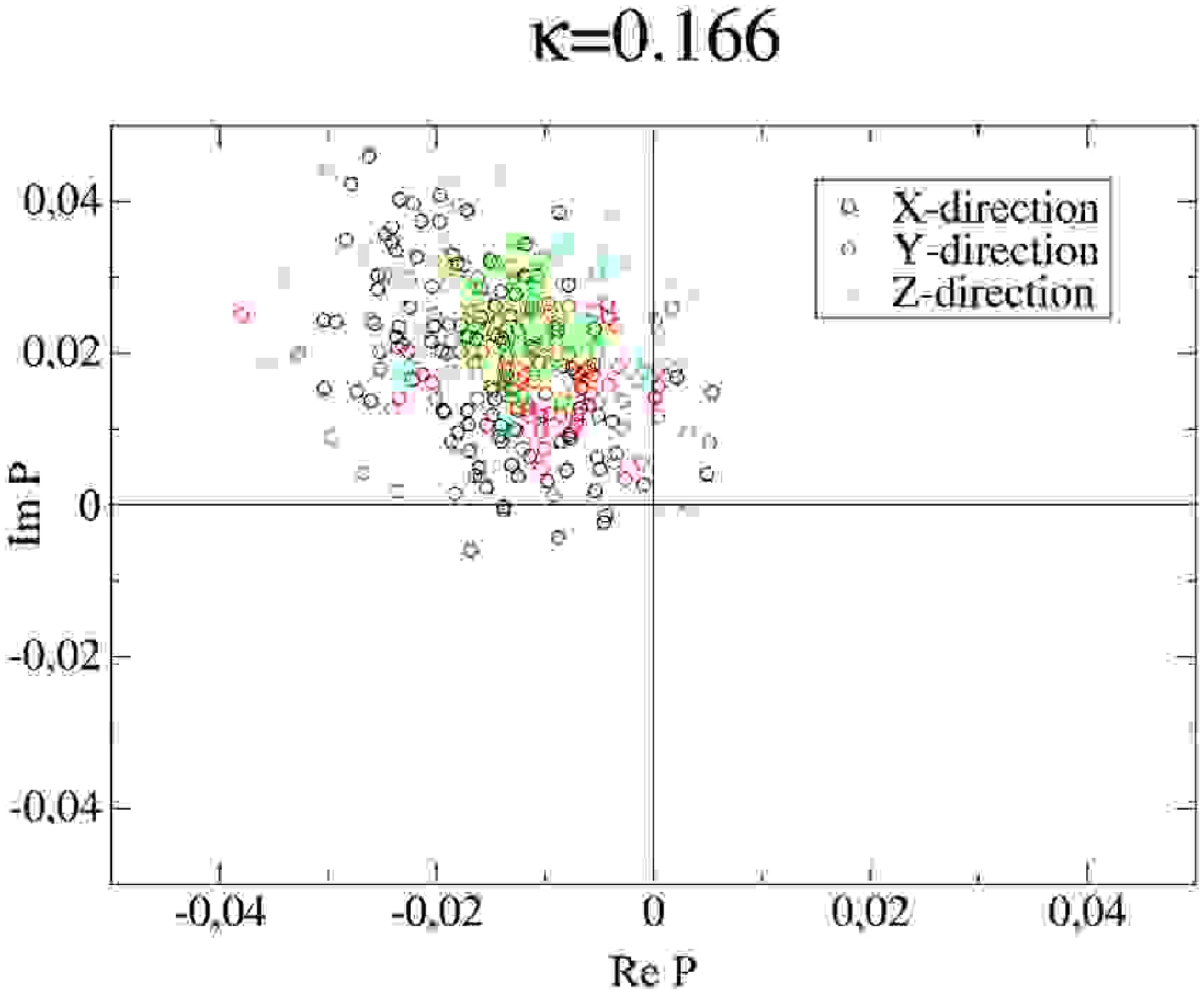}
    \end{tabular}
    \caption{Polyakov loops along the shortest length, $L_s=8$, for
      the simulation of three dynamical fermion flavors at
      $\beta=5.2$.}
    \label{fig:b52-polyakov}
  \end{center}
\end{figure}
\clearpage
\subsubsection{Locating the Shielding Transition}
\label{sec:locating-non-zero}
As a first guideline of where the crossover to the shielded phase
takes place, the plaquette fluctuation in Fig.~\ref{fig:b52-k162-plaq}
and the jump in the average plaquette in Fig.~\ref{fig:b52-plaqs} have
been considered. To gain further insight one can investigate the
behavior of the average Polyakov loop, see
Sec.~\ref{sec:non-zero-temperature}. However, in this case it should
now be measured in spatial (i.e.~$x$, $y$, and $z$) direction since
the $t$-direction is now the longest. The Polyakov line in that
direction can be expected to show no sign of the finite-temperature
phase\index{Shielding transition}.

The Polyakov loops have been measured every $100$ trajectories. The
resulting values are shown in Fig.~\ref{fig:b52-polyakov}. Starting
with $\kappa_{\mbox{\tiny sea}}=0.163$, one clearly sees a clustering
in one of the three sectors. It is surprising that despite the rather
long runs, in each case the values are clustered in only one sector.
This indicates that the samples are not decorrelated with respect to
this observable. At $\kappa_{\mbox{\tiny sea}}=0.162$ the shielding
transition is not yet apparent in the Polyakov loop. However, when
considering the previous indications, it appears safer to disregard
the latter run from the following analysis.

\subsubsection{Computing $m_\pi/m_\rho$}
\label{sec:computing-m_pim_rho}
The details for the measurement of hadronic masses have been given in
Sec.~\ref{sec:meas-hadr-mass}. As has been discussed above, only the
points $\kappa_{\mbox{\tiny sea}}\leq 0.160$ should be considered for
this analysis. The reweighting factor has been included, although it
had no practical influence in these productions.

\begin{figure}[ht]
  \begin{center}
    \includegraphics[scale=0.3,clip=true]{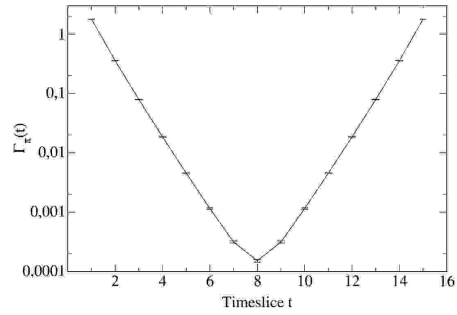}
    \caption{Symmetrized correlation function for the pseudoscalar
      meson with three dynamical fermi\-ons at $\beta=5.2$ and
      $\kappa=0.156$.}
    \label{fig:b52-k156-pionprop}
  \end{center}
\end{figure}

\begin{figure}[ht]
  \begin{center}
    \includegraphics[scale=0.3,clip=true]{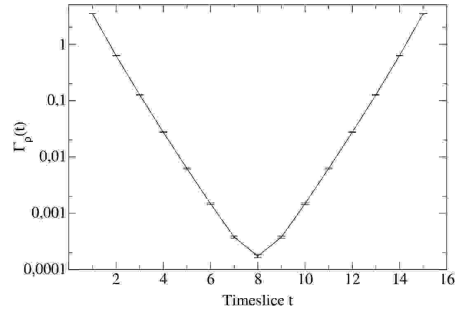}
    \caption{Symmetrized correlation function for the vector
      meson with three dynamical fermi\-ons at $\beta=5.2$ and
      $\kappa=0.156$.}
    \label{fig:b52-k156-rhoprop}
  \end{center}
\end{figure}

In the run with $\kappa=0.156$ the correlation functions for the
(non-singlet) pseudoscalar and the vector mesons are visualized in
Figs.~\ref{fig:b52-k156-pionprop} and~\ref{fig:b52-k156-rhoprop}.
These functions have already been symmetrized, i.e.~the plot shows
(cf.~Eq.~(\ref{eq:hadron-correlators})) \[ \Gamma_{\mbox{\tiny
    $\pi,\rho$}}^{\mbox{\tiny sym}}(t) = \frac{1}{2}\left(
  \Gamma_{\mbox{\tiny $\pi,\rho$}}(t) + \Gamma_{\mbox{\tiny
      $\pi,\rho$}}(L_0-t) \right)\,, \] with $L_0$ being the lattice
extension in $t$-direction.

These functions should follow the behavior given in
Eq.~(\ref{eq:lattice-meson-masses}). However, for small values of $t$,
one expects the results to be too large (due to the contamination with
higher modes, cf.~Sec.~\ref{sec:meas-hadr-mass}), while for larger
values of $t$, larger autocorrelations of the greater lengths may
result in worse statistics.

\begin{figure}[ht]
  \begin{center}
    \includegraphics[scale=0.3,clip=true]{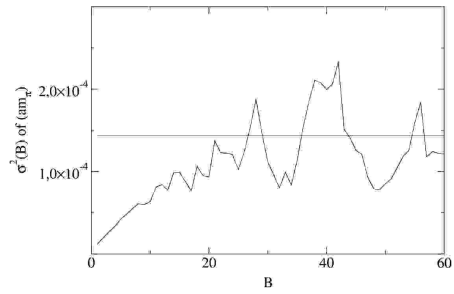}
    \caption{Jackknife variances for different bin sizes for the
      mass of the pseudoscalar meson with three dynamical fermions at
      $\beta=5.2$ and $\kappa=0.156$. The mass is obtained from a fit
      to an interval $[t_{\mbox{\tiny min}}=5, t_{\mbox{\tiny
          max}}=7]$.}
    \label{fig:b52-k156-pionerr}
  \end{center}
\end{figure}

\begin{figure}[ht]
  \begin{center}
    \includegraphics[scale=0.3,clip=true]{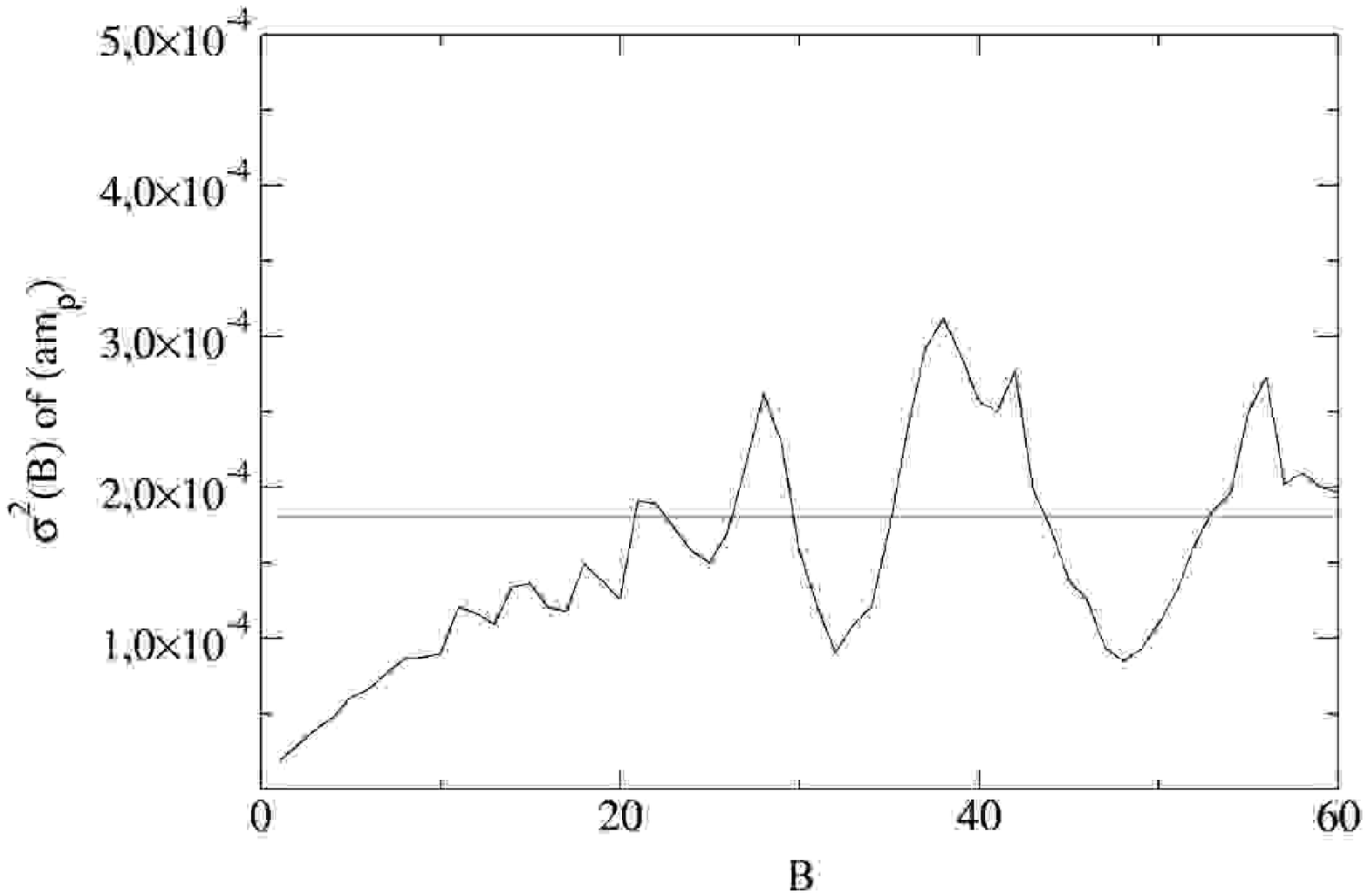}
    \caption{Jackknife variances for different bin sizes for the
      mass of the vector meson with three dynamical fermions at
      $\beta=5.2$ and $\kappa=0.156$. The mass is obtained from a fit
      to an interval $[t_{\mbox{\tiny min}}=5, t_{\mbox{\tiny
          max}}=7]$.}
    \label{fig:b52-k156-rhoerr}
  \end{center}
\end{figure}

To obtain an estimate for the autocorrelation time of these masses,
the Jackknife method has again been employed. The case which is
considered in detail is the run at $\kappa=0.156$.
Figures~\ref{fig:b52-k156-pionerr} and~\ref{fig:b52-k156-rhoerr} show
the variances of the masses for a fit interval from timeslice $t=5$ to
timeslice $t=7$.

\begin{table}[th]
  \begin{center}
    \begin{tabular}[c]{l|c|c|c|r}
      \hline\hline
      \textbf{Quantity} & \textbf{Expect.~value} & \textbf{Variance}
      $\mathbf{\sigma^2}$ & $\mathbf{\sigma^2(B=1)}$ &
      $\tau_{\mbox{\tiny int}}$ \\ \hline
      $(am_\pi)$  & $1.374(12)$ & $\Exp{1.436}{-3}$ &
      $\Exp{1.163}{-5}$ & $6.18$ \\
      $(am_\rho)$ & $1.440(13)$ & $\Exp{1.803}{-3}$ &
      $\Exp{1.804}{-5}$ & $5.00$ \\ \hline\hline
    \end{tabular}
    \caption{Jackknife variances together with the corresponding
      variances for bin size $B=1$ for the simulation run at
      $\beta=5.2$ and $\kappa=0.156$. From there, an estimate for the
      integrated autocorrelation time is obtained.}
    \label{tab:b52-k162-errors}
  \end{center}
\end{table}
The resulting values are given in Tab.~\ref{tab:b52-k162-errors}
together with the variance estimate for bin size $B=1$. By exploiting
Eq.~(\ref{eq:tauint-jackknife}), one can as usual obtain an estimate
for the autocorrelation time of the quantity under consideration.
\clearpage

Since the correlators have been computed every $100$ trajectories, the
results imply that the $\pi$- and $\rho$-mesons have autocorrelation
times of $\tau_{\mbox{\tiny int}}^{\pi}\approx 618$ and
$\tau_{\mbox{\tiny int}}^{\rho}\approx 500$ trajectories,
respectively. These numbers are slightly better than what the
plaquette has indicated, albeit still large. A source of this problem
is the choice of the first polynomial. If the GMRES method had been
used instead, one might have achieved a faster decorrelation by
reducing the polynomial order, $n_1$, see Sec.~\ref{sec:algor-polyn}
and also Sec.~\ref{sec:direct-algor-comp}.

\begin{figure}[htb]
  \begin{center}
    \begin{tabular}[c]{c|c}
      \includegraphics[scale=0.22,bb=0 0 27cm 22cm,clip=true]
      {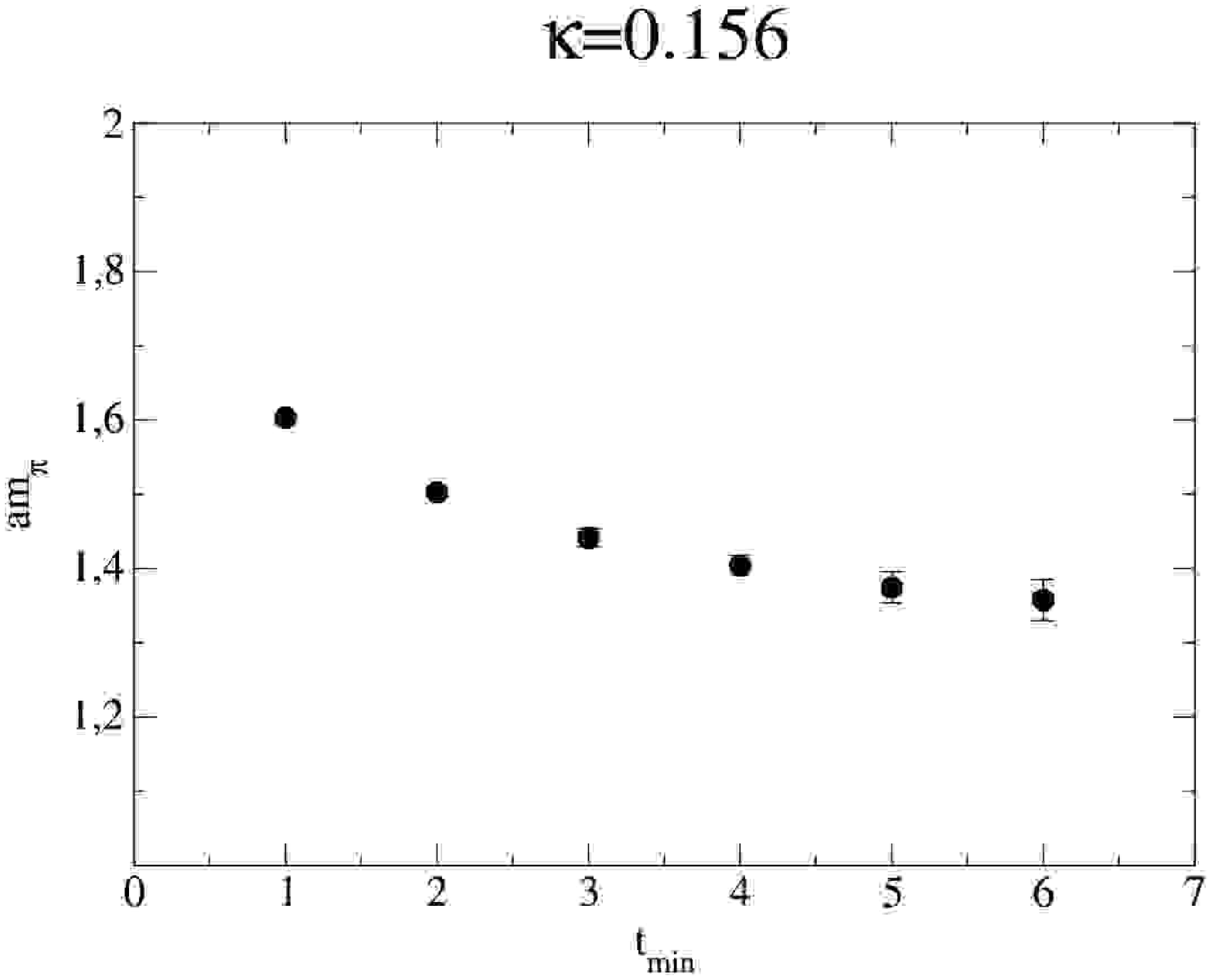} &
      \includegraphics[scale=0.22,bb=0 0 27cm 22cm,clip=true]
      {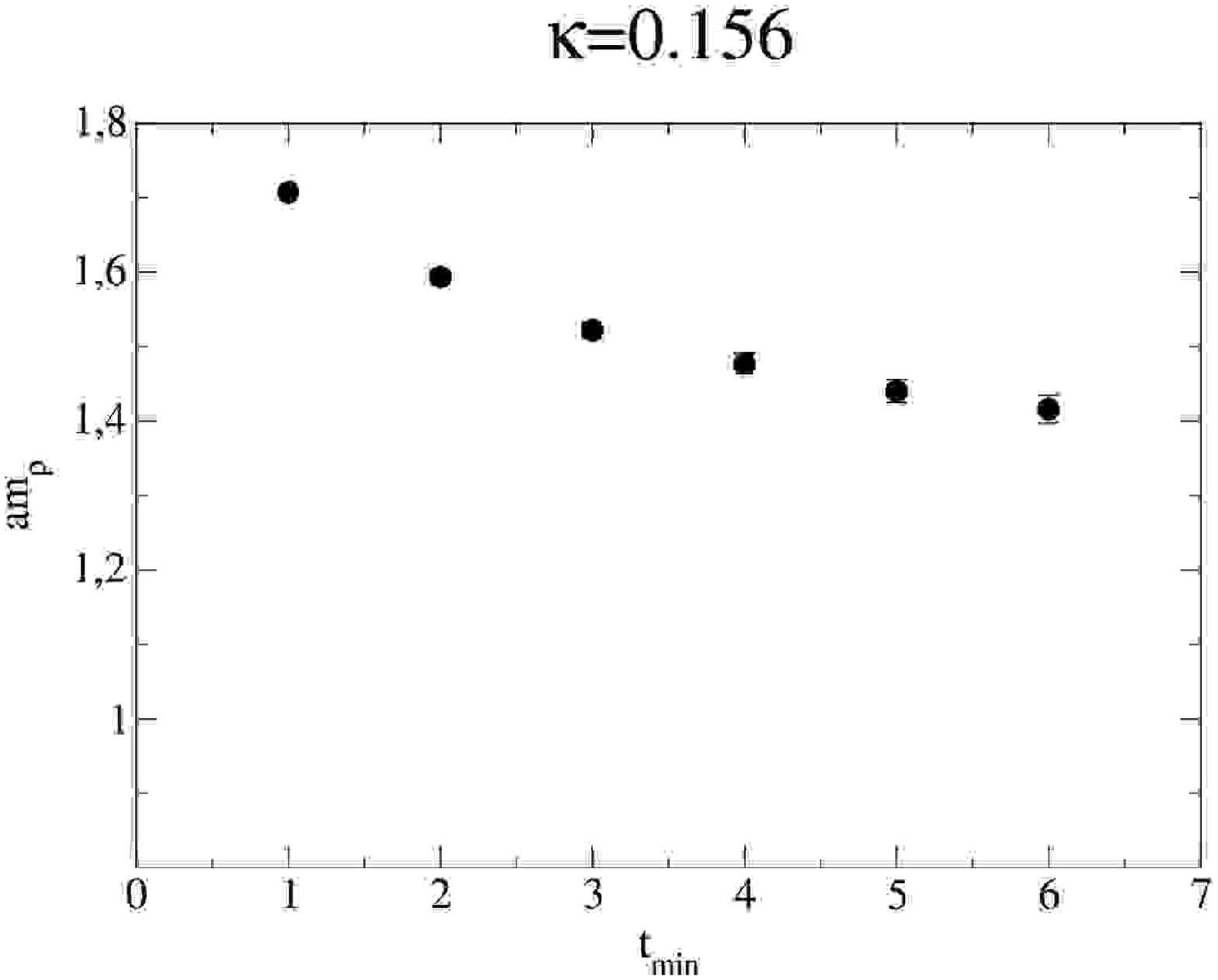} \\ \hline
      \includegraphics[scale=0.22,bb=0 0 27cm 22cm,clip=true]
      {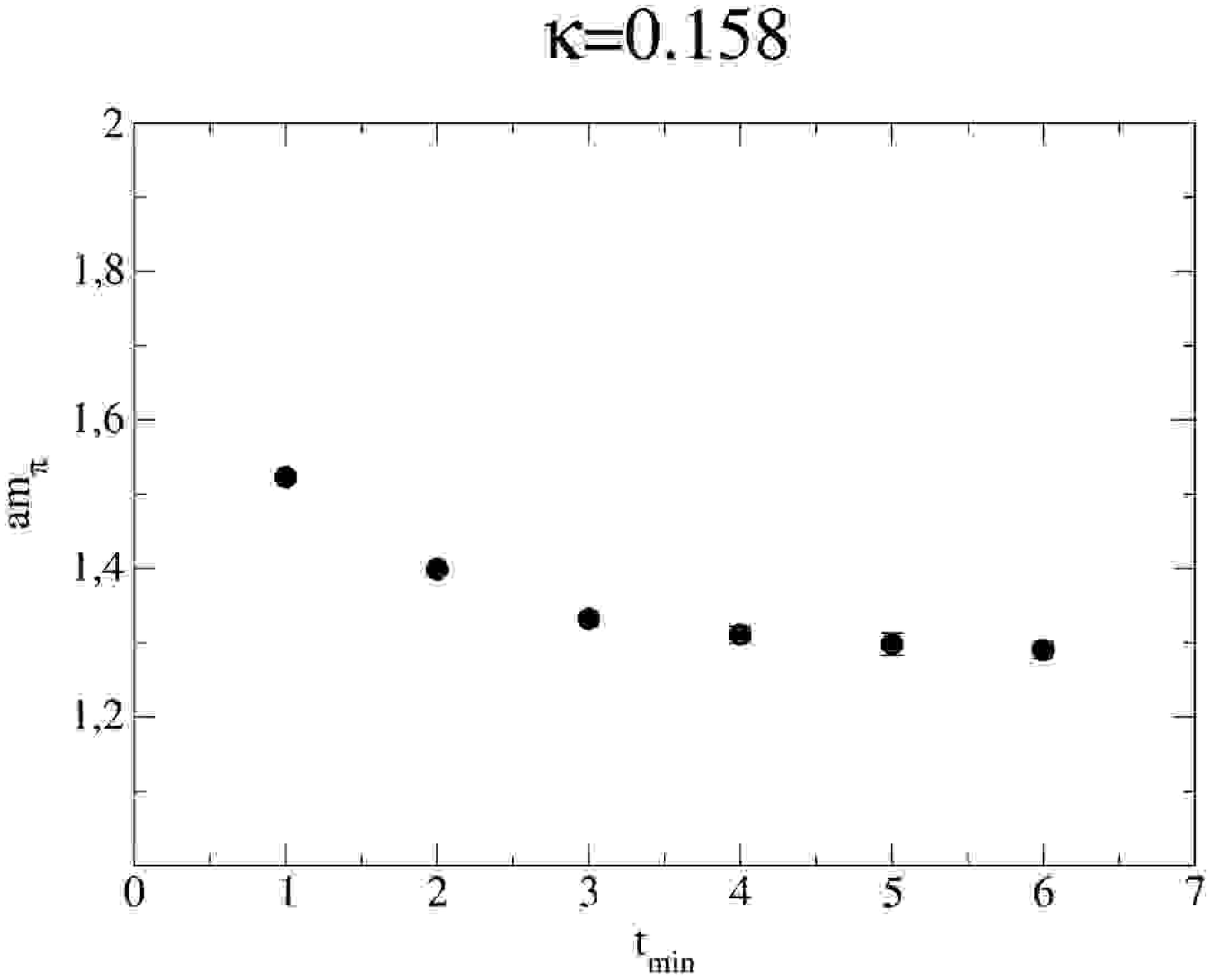} &
      \includegraphics[scale=0.22,bb=0 0 27cm 22cm,clip=true]
      {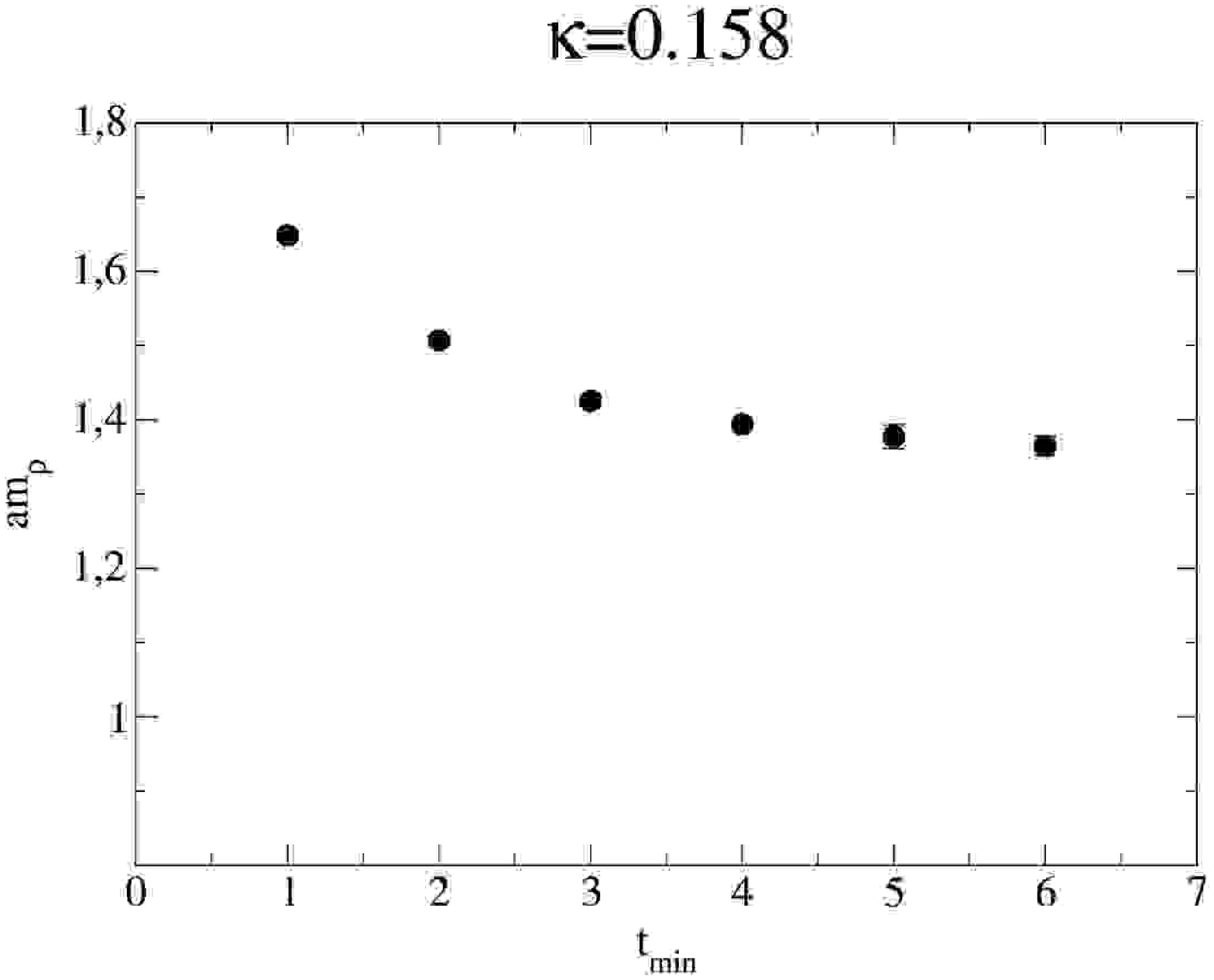} \\ \hline
      \includegraphics[scale=0.22,bb=0 0 27cm 22cm,clip=true]
      {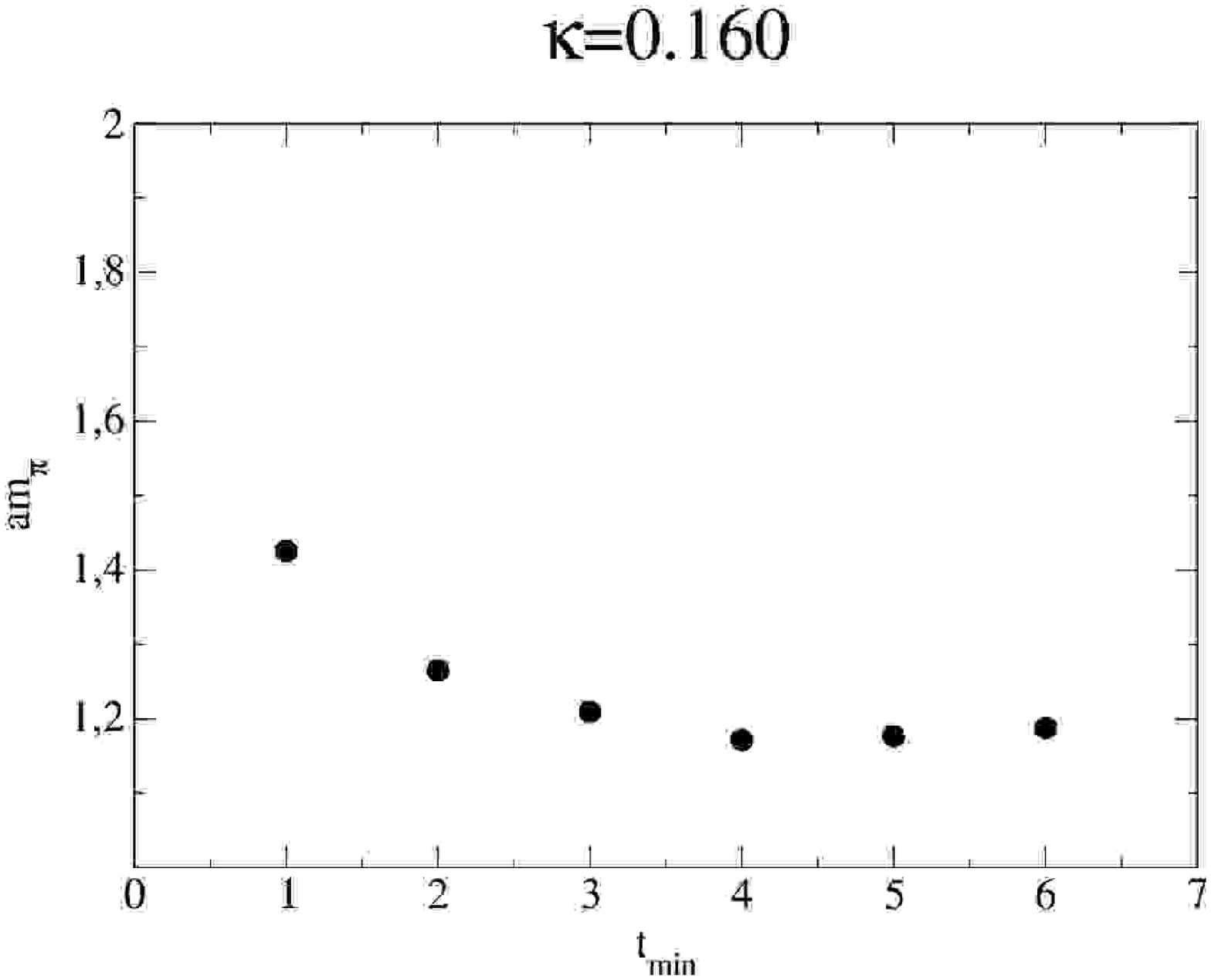} &
      \includegraphics[scale=0.22,bb=0 0 27cm 22cm,clip=true]
      {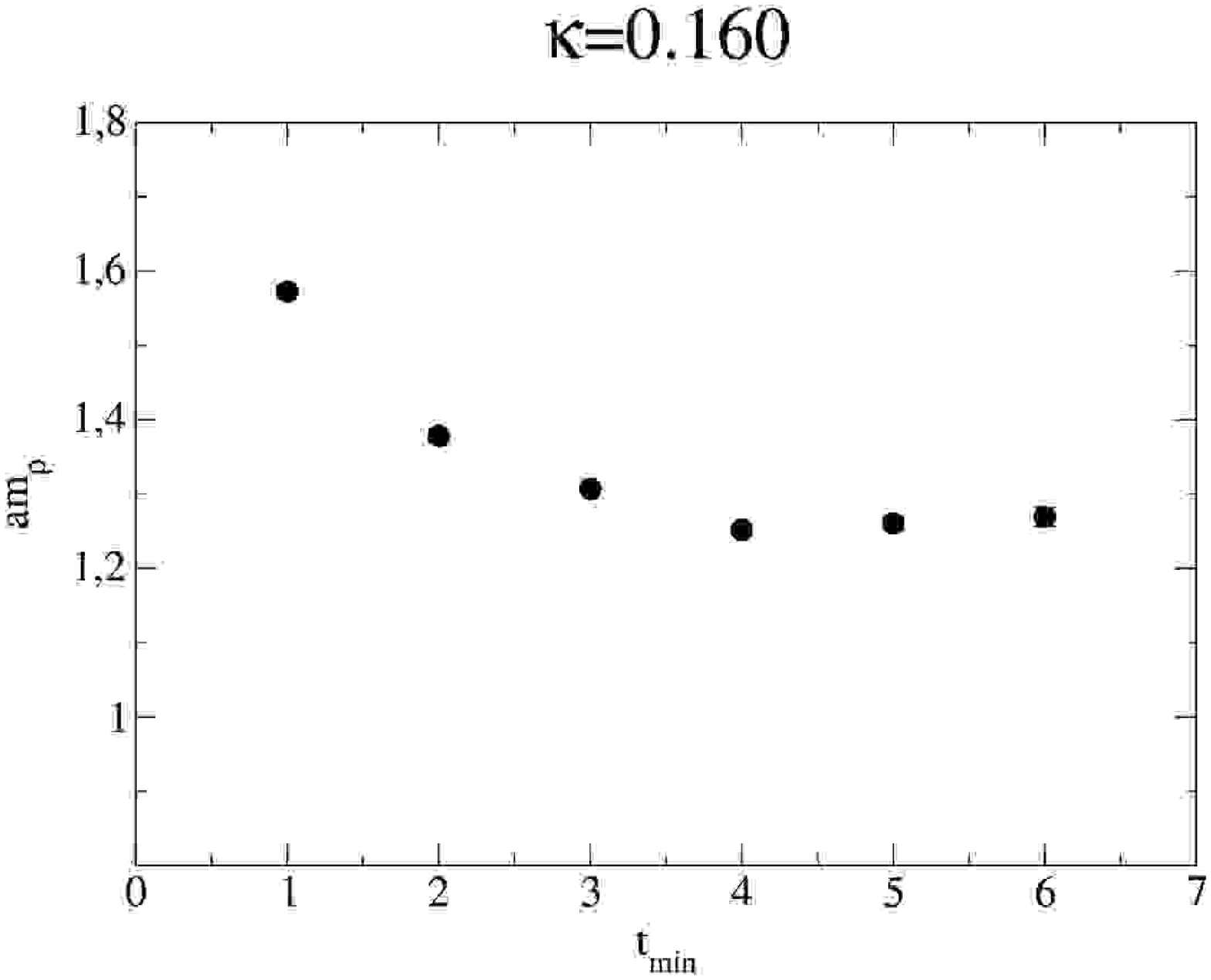} \\
    \end{tabular}
    \caption{Meson masses in lattice units as a function of the
      fitting interval for three dynamical fermions at $\beta=5.2\,$.}
    \label{fig:b52-masses}
  \end{center}
\end{figure}

Figure~\ref{fig:b52-masses} shows the resulting values for masses in
lattice units. The lower limit of the fit is given by the timeslice
$t$, while for the upper limit, always the next-to-last limit has been
used, i.e.~$L_0^{\mbox{\tiny max}}=7$.  The error is again taken to be
the standard error, which has been computed using the Jackknife
procedure as above in Tab.~\ref{tab:b52-k162-errors}. The method
follows the results discussed in
\cite{Gusken:1999hb,Orth:2002ph,Eicker:2001ph,Struckmann:2000ts}.

The plateaus in Fig.~\ref{fig:b52-masses} are reached at $t=5$.
Therefore, the values obtained at this point will be used in the
following. Table~\ref{tab:b52-masses} summarizes all results together
with the autocorrelation times determined using the Jackknife scheme.
In the case $\kappa=0.160$, no plateau could have been identified and
the results are compatible with an integrated autocorrelation time
below $100$ trajectories.
\clearpage

\clearpage
\clearpage
\begin{table}[ht]
  \begin{center}
    \begin{tabular}[c]{l|*{5}{c|}c}
      \hline\hline
      $\mathbf{\kappa}$ & $\mathbf{(am_\pi)}$ &
      $\mathbf{(am_\rho)}$   & $\mathbf{m_\pi/m_\rho}$ & 
      $\mathbf{a}$ & $\mathbf{\tau_{\mbox{\tiny int}}^\pi}$ &
      $\mathbf{\tau_{\mbox{\tiny int}}^\rho}$ \\ \hline
      $0.156$ & $1.374(12)$  & $1.440(13)$  & $0.954(18)$  &
      $0.368(3)$ & $618$ & $508$ \\
      $0.158$ & $1.298(15)$  & $1.377(16)$  & $0.943(16)$  &
      $0.352(4)$ & $620$ & $565$ \\
      $0.160$ & $1.1767(52)$ & $1.2604(84)$ & $0.9336(75)$ &
      $0.323(2)$ & $<100$ & $<100$ \\
      \hline\hline
    \end{tabular}
    \caption{Masses and their autocorrelation times for the
      determination of the ratio $m_\pi/m_\rho$ for three dynamical
      fermions at $\beta=5.2$.}
    \label{tab:b52-masses}
  \end{center}
\end{table}

\subsubsection{Locating the Critical Point}
\label{sec:locat-crit-point}
To locate the critical point, again the smallest and largest
eigenvalues have been computed and the condition numbers have been
determined for the runs at $\beta=5.2$. The eigenvalues have been
computed every $100$ trajectories. Table~\ref{tab:beta52-eigvals}
summarizes the findings.
\begin{table}[htb]
  \begin{center}
    \begin{tabular}[c]{c|c|c|c}
      \hline\hline
      $\mathbf{\kappa}$ &
      $\mathbf{\lambda_{\mbox{\tiny\bf min}}}$ &
      $\mathbf{\lambda_{\mbox{\tiny\bf max}}}$ &
      $\mathbf{\lambda_{\mbox{\tiny\bf max}}/\lambda_{\mbox{\tiny\bf
            min}}}$\\ \hline
      $0.156$ & $0.0330(6)$ & $2.2137(5)$  & $67.1\pm 1.1$  \\
      $0.158$ & $0.0249(3)$ & $2.2503(9)$  & $90.4\pm 3.2$  \\
      $0.160$ & $0.0178(2)$ & $2.2879(5)$  & $128.5\pm 3.3$ \\
      \hline\hline
    \end{tabular}
    \caption{Average extremal eigenvalues and condition numbers for
      runs with three dynamical flavors at $\beta=5.2$.}
    \label{tab:beta52-eigvals}
  \end{center}
\end{table}

\begin{figure}[htb]
  \begin{center}
    \includegraphics[scale=0.3,clip=true]{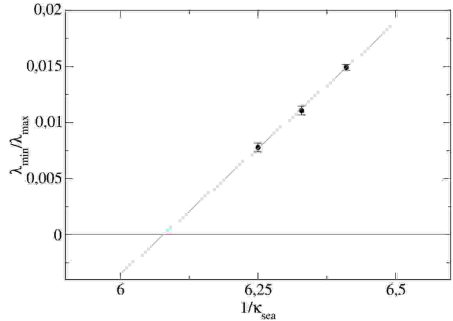}
    \caption{Inverse condition number of $\tilde{Q}^2$
      vs.~$1/\kappa$ for three dynamical fermions at $\beta=5.2$.}
    \label{fig:b52-condnum}
  \end{center}
\end{figure}

The inverse condition number is plotted vs.~the inverse quark mass in
Fig.~\ref{fig:b52-condnum}.

The fit to all three data points yields
\begin{equation}
  \label{eq:b52-kappafit}
  \lambda_{\mbox{\tiny min}}/\lambda_{\mbox{\tiny max}} = -0.272(18) 
  + 0.0448(28) /\kappa\,.
\end{equation}
The zero of the line gives
\begin{equation}
  \label{eq:b52-kappacrit-result}
  \kappa_{\mbox{\tiny crit}} = 0.1645(29)\,.
\end{equation}

\begin{figure}[htb]
  \begin{center}
    \includegraphics[scale=0.4,clip=true]{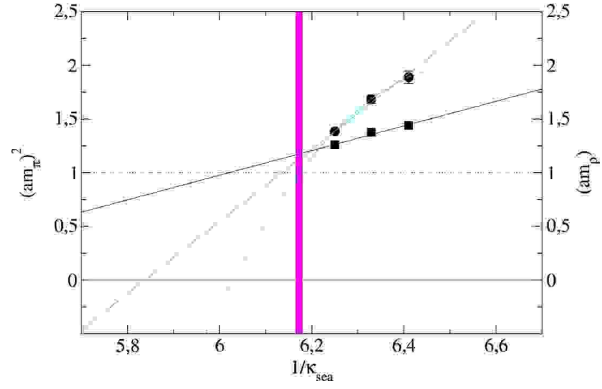}
    \caption{Square of the pion mass, $(am_\pi)^2$, (black circles)
      and the rho mass, $(am_\rho)$, (black squares) for $\beta=5.2$
      with $N_f=3$.}
    \label{fig:b52-chiral}
  \end{center}
\end{figure}

Finally, the fitting function from Eq.~(\ref{eq:pion-vs-kappa}) is
applied to the situation at hand with the pion masses given by
Tab.~\ref{tab:b52-masses}. In Fig.~\ref{fig:b52-chiral} the inverse
value of the quark mass, $1/\kappa$, is plotted versus the square of
the pion mass, $(am_\pi)^2$. In addition, the rho mass, $(am_\rho)$,
is also included in this plot. The former is visualized as circles,
while the latter is pictured by squares. The shielding transition is
shown as a magenta bar.

The linear fit to $(am_\pi)^2$ is given by the solid green line in
Fig.~\ref{fig:b52-chiral}. The curve is parameterized by
\begin{equation}
  \label{eq:b52-kappafit2}
  (am_\pi)^2 = -(19.60\pm 1.94) + (3.359\pm 0.309)
  /\kappa\,.
\end{equation}
The critical value of $\kappa_{\mbox{\tiny crit}}$ is then found to be
\begin{equation}
  \label{eq:b52-kappacrit-result2}
  \kappa_{\mbox{\tiny crit}} = 0.1713(67)\,.
\end{equation}
The result from Eq.~(\ref{eq:b52-kappacrit-result2}) agrees within the
errors with the previous result from
Eq.~(\ref{eq:b52-kappacrit-result}).

Furthermore, a quadratic curve has been drawn through the values for
$(am_\pi)^2$, given by the green dashed line. It is parameterized by
\begin{equation}
  \label{eq:b52-kappafit-square}
  (am_\pi)^2 = -340.81 + 105.06/\kappa
  -8.05/\kappa^2\,.
\end{equation}
When using this curve, the resulting value for $\kappa_{\mbox{\tiny
    crit}}$ is found to be
\begin{equation}
  \label{eq:b52-kappacrit-result3}
  \kappa_{\mbox{\tiny crit}} = 0.1659\,.
\end{equation}
The linear fit to $(am_\rho)$ is given by the blue line in
Fig.~\ref{fig:b52-chiral}. The curve is parameterized by
\begin{equation}
  \label{eq:b52-kappafitrho}
  (am_\rho) = -(5.91\pm 0.60) + (1.147\pm 0.095)/\kappa\,.
\end{equation}
Obviously, it is not possible to reach values of $\xi_\pi>1$ before
the shielding transition sets in on the current lattice size,
cf.~Sec.~\ref{sec:chiral-limit}. Therefore, the linear extrapolation
in Eq.~(\ref{eq:b52-kappafit2}) may be biased with an uncontrolled
systematic uncertainty. To estimate this effect, one may compare the
resulting critical point, Eq.~(\ref{eq:b52-kappacrit-result2}), with
the result obtained from the quadratic fit,
Eq.~(\ref{eq:b52-kappacrit-result3}). This uncertainty makes further
investigations closer to the chiral point necessary and consequently
implies the need to go to larger lattices.

\subsubsection{Prospects for Future Simulations}
\label{sec:working-point-future}
Up to this point, one could only achieve ratios of $m_\pi/m_\rho>0.9$
with $\xi_\pi<1$. When going to larger lattices, the shielding
transition will set in at larger values of $\kappa_{\mbox{\tiny
    sea}}$, allowing to probe lighter quark masses.

A procedure for continuing along this line of research consists of
going to $\Omega=24\times 12^3$ lattices, starting from
$\kappa_{\mbox{\tiny sea}}\geq 0.160$ until the shielding transition
for the new lattice sets in. In light of the fact that the shielding
transition in Fig.~\ref{fig:b52-chiral} is located shortly before one
arrives at $\xi_\pi\geq 1$, a lattice size of $L_s=12$ might already
be sufficiently large to obtain a set of data points all fulfilling
the requirement $\xi_\pi>1$. In such a situation, one could obtain an
extrapolation to the chiral point, $\kappa_{\mbox{\tiny crit}}$, with
reduced systematic uncertainty.

\begin{table}[bht]
  \begin{center}
    \begin{tabular}[c]{*{6}{c|}c}
      \hline\hline
      \multicolumn{3}{c}{\textbf{Bare parameters}} & 
      \multicolumn{4}{|c}{\textbf{Physical parameters}} \\ \hline
      $\mathbf{N_f^{\mbox{\tiny sea}}}$ & $\mathbf{\beta^{\mbox{\tiny
            spec}}}$ & $\mathbf{\kappa}^{\mbox{\tiny spec}}$ &
      $\mathbf{(am_\pi)}$ & $\mathbf{(am_\rho)}$ &
      $\mathbf{m_\pi/m_\rho}$ & \textbf{$\mathbf{a}$/fm} \\ \hline
      $3$ & $5.2$ & $0.169(23)$ & $0.5$ & $0.882$ &
      $0.567$ & $0.226$ \\
      \hline\hline
    \end{tabular}
    \caption{The suggested working point for future spectroscopic
      studies on lattices with $\Omega=32\times 16^3$ and beyond.}
    \label{tab:spec-point}
  \end{center}
\end{table}
With the available information we can, however, still try to locate a
working point at this particular value of $\beta$ in the
$\beta$-$\kappa$-plane with properties similar to the point chosen for
the \SESAM-project \cite{Lippert:2001ha}.

This working point will now be denoted $\left(\beta^{\mbox{\tiny
      spec}},\kappa^{\mbox{\tiny spec}}\right)$. With the
uncertainties discussed above in mind, we imposes the following
constraints:
\begin{eqnarray}
  \label{eq:finite-size-param}
  z &\equiv& \xi_\pi/L < 1/4\,, \nonumber \\
  \xi_\pi &\geq& 2\,.
\end{eqnarray}
The actual parameters
can be identified from the extrapolations
Eqs.~(\ref{eq:b52-kappafit2}) and (\ref{eq:b52-kappafitrho}). First,
from setting $\xi_\pi=2$, we obtain
\begin{equation}
  \label{eq:kappa-sea-sesam}
  \kappa^{\mbox{\tiny spec}} = 0.169(23)\,.
\end{equation}
From the requirement (\ref{eq:finite-size-param}) that the value of
the finite-size parameter should be $z<1/4$, one finds, in accordance
with the \SESAM-data from \cite{Eicker:2001ph}, that one has to go to
lattices with at least $L_s=16$ if one wants to explore this region in
parameter space.

The parameters for this working point are summarized in
Tab.~\ref{tab:spec-point}. The estimated values for $(am_\rho)$,
$m_\pi/m_\rho$, and $a$ have been computed from the fit
(\ref{eq:b52-kappafitrho}). The total physical lattice size $L_s$
would then be $L_s=3.616\; \mbox{fm}$. A possible criticism against
this working point might be that this lattice spacing is rather
coarse. To actually increase the resolution, one would need to go to
higher values of $\beta$, thus moving closer to the continuum limit.

Finally, the question arises how large the total effort might be for
such a project. For the case of quadratically optimized polynomials,
it has been argued in \cite{Montvay:1999kq}, that the required
increase in $n_1$ when going from $N_f=2$ to $N_f=3$ is only of the
order of about $30\%$. Reference \cite{Farchioni:2001di} confirms this
finding by stating that going from $N_f=1/2$ to $N_f=3$ will only
increase $n_1$ by about $50\%$. Taking --- as a very conservative
estimate --- the latter number to be applicable also to the
simulations performed for $N_f=2$ in
chapter~\ref{sec:comp-dynam-ferm}, we find by applying
Eq.~(\ref{eq:indep-cost}) that the total cost for an independent
configuration (with respect to the plaquette) is about
\begin{equation}
  \label{eq:effort-final}
  E_{\mbox{\tiny indep}} \simeq (1.5)^2\cdot (1075360\pm 72160)\approx
  (2420000\pm 162000)\,,
\end{equation}
when considering the lightest quark mass, where $m_\pi/m_\rho=0.6695$.
Hence, this estimate marks the upper limit for the effort required in
a simulation similar to the \SESAM-project, provided one decides to
take recourse to an MB algorithm. The total cost quoted in
Eq.~(\ref{eq:effort-final}) is still smaller than the corresponding
cost for the HMC run with $N_f=2$. Therefore, one can expect the
simulations at the lighter quark masses to be even cheaper than they
were in the case of the \SESAM-project.

\section{Summary and Outlook}
\label{sec:summary-outlook}
A first step towards the simulation of QCD with three degenerate
dynamical quark flavors has been taken. The TSMB algorithm been
applied successfully to this physically interesting situation. The
simulations have yielded first results for the shielding transitions
on the current lattice size with $L_s=8$ and the critical points at
two values of $\beta$. It has become clear that there is no window for
doing spectroscopy at the parameters chosen.

Prospects for future simulations have been given and a potential
working point has been estimated, although with large systematic
uncertainty. It has been argued that a study with physical masses
similar to the \SESAM-project is feasible today and might even cost
slightly less than the HMC-based program.

In an ongoing research project, such type of simulations will be
performed on larger lattices and closer to the chiral limit. For the
current status of the comprehensive project see
\cite{Farchioni:2001di}.

A potential obstacle for future simulations with three dynamical
fermion flavors may still be posed by the fermionic sign problem. As
has been noted in Sec.~\ref{sec:fermion-fields}, the fermionic
determinant will change its sign if an odd number of real eigenvalues
becomes negative. The polynomial approximations in
Sec.~\ref{sec:stat-polyn-invers}, however, are applied to the square
of the Hermitian Wilson matrix. Hence, they will always yield a
positive sign. Consequently, the sign would have to be included into
the measurement of observables which may eventually spoil the
statistical quality of the sample. A similar problem is known to occur
in the simulation of gauge theories with a non-zero chemical
potential, see \cite{Hands:2000ei}.

Such a problem does not show up for an even number of degenerate
fermion flavors since in that case squaring the Wilson matrix will
always yield a positive sign. The only known way to overcome this
obstacle directly in a sampling process has been found for some
quantum spin systems (cf.~Sec.~\ref{sec:d-theory}). It is yet unclear,
if any quantum spin system similar to gauge theories with dynamical
fermion flavors can be simulated efficiently in such a manner.
However, as has already pointed out in Sec.~\ref{sec:summary-5}, it
may be that this sign problem is not significant in actual simulations
of QCD\@.

\chapter{Summary and Outlook}
\label{sec:summary-outlook-1}
\begin{flushright}
  \includegraphics[scale=1.0,clip=true]%
  {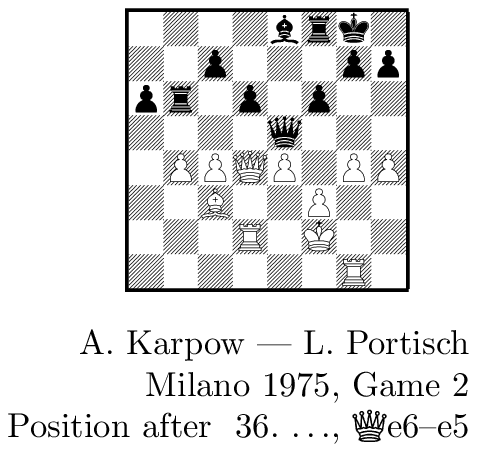}                    \\
  A foresighted strategy can help to find \\
  a winning move in a superior position.
\end{flushright}
In this thesis algorithms for the simulation of quantum field theories
with dynamical fermionic degrees of freedom have been presented.
Special emphasis has been put on a new class of algorithms, namely the
multiboson algorithms which represent the fermionic determinant by a
number of boson fields. They allow for the implementation of local
updating algorithms, which are known to be superior to global schemes
applicable to gauge theories so far.

A particular variant of these algorithms, the TSMB method, has been
implemented on several machines. This scheme relies on the computation
of powers of matrices using static polynomials. The parameters which
fix a given polynomial are the order and the interval of the
approximation. Beyond that, it has turned out that the choice of the
updating scheme is important.

The optimal settings for these parameters have been determined and the
sensitivity of the system to sub-optimal tuning has been analyzed.
Furthermore, different updating schemes have been examined with
respect to their efficiency and recommendations for the
implementations of multiboson algorithms in general have been given.

Due to the complexity of MB schemes, however, there is still room for
improvement. MB algorithms remain open for refinements in the future,
but can already be used for large-scale simulations today.

Major emphasis has been put on how multiboson algorithms compare to
their competitors in the field of dynamical fermion simulations. We
have shown that, with sufficient tuning, MB algorithms appear to be
superior to the HMC algorithm in the case of light quark masses. We
would expect that further improvements in MB algorithms will be found
with growing experience in future simulations. This might help the MB
scheme to replace the HMC method as the standard algorithm in Lattice
QCD.

The final part of this thesis has considered the application of the
TSMB algorithm to the case of three dynamical fermionic flavors, a
situation which is of great importance for realistic simulations of
QCD\@. Based on this experience, a proposal for future simulations has
been formulated. In fact, one can be optimistic to perform a project
similar to \SESAM\ at reasonable cost.  A working point for such type
of simulations on $\Omega=32\times 16^3$ lattices has been estimated,
where semi-realistic simulations with good statistics should be run.
This would provide an assessment of an operating window in the $N_f=3$
scenario.

In conclusion, we find that multiboson algorithms provide a great leap
forward in the simulation of Lattice QCD and give us the means to
perform simulations in realistic scenarios.


\begin{appendix}

\chapter{Notations and Conventions}
\label{sec:notat-conv}
Unless otherwise explicitly stated, natural units have been adopted
throughout this thesis by setting
\begin{equation}
  \label{eq:natural-units}
  \hbar = c = k_B = 1\,.
\end{equation}
The four-dimensional Minkowski-space is denoted by $\mathbb{M}^4$ and
has the canonical flat-space metric
\begin{equation}
  \label{eq:minkowski-metric}
  g_{\mu\nu} = \left(\begin{array}{*{4}{p{0.35cm}}}
      1 &  0 &  0 &  0 \\
      0 & -1 &  0 &  0 \\
      0 &  0 & -1 &  0 \\
      0 &  0 &  0 & -1 \end{array}\right)\,.
\end{equation}
The \index{Wick rotation} Wick rotation \eqcite{eq:Wick-rotation}
\begin{equation}
  x \mapsto x^{\prime}: x^\prime =
  \left(x^\prime_0,\mathbf{x}^\prime\right) = \left(-\mbox{i}x_0,
  \mathbf{x}\right)
\end{equation}
\fineqcite transforms vectors from Minkowski-space to the Euclidean
space $\mathbb{R}^4$ with the metric given by the Kronecker symbol
$\delta_{\mu\nu}$:
\begin{equation}
  \label{eq:euclidean-metric}
  \delta_{\mu\nu} = \left\lbrace\begin{array}{l}
      1,\qquad\mbox{for }\mu=\nu, \\
      0,\qquad\mbox{for }\mu\neq\nu.
    \end{array}\right.
\end{equation}
The discrete space of the lattice theory is denoted by $\mathbb{Z}^4$.
Vectors in space are always denoted by $x$. Unit vectors are only used
in $\mathbb{Z}^4$, where they are written as $\hat{\mu}$ with
$\mu=0,\dots,3$. If the lattice volume is finite, the lengths $L_\mu$
in direction $\mu$ are also denoted by $L_t=L_0$ and
$L_s=L_1=L_2=L_3$, where $L_t$ is the lattice size in ``time'' and
$L_s$ in ``space'' direction. The total volume is denoted by
$\Omega=\prod_\mu L_\mu=L_t\times L_s^3$ and the corresponding space
is $\mathbb{Z}^4_\Omega$. For different lengths $L_1$, $L_2$, and
$L_3$, the notation $\Omega=L_0\times L_1\times L_2\times L_3$ will be
used. For a bosonic field, $\phi(x)\in\mathbb{Z}^4_\Omega$, periodic
boundary conditions are imposed:
\[ \phi\left(x+\hat{\mu}L_\mu\right) = \phi\left(x\right)\,. \] Here
$x+\hat{\mu}$ denotes the point adjacent to $x$ in direction $\mu$.

The totally antisymmetric $4$-tensor $\epsilon^{\mu\nu\rho\sigma}$
obeys
\[
  \epsilon^{\left[\mu\nu\rho\sigma\right]} =
  \left\lbrace\begin{array}{ll}
      -1, \qquad\mbox{for $\left[\mu\nu\rho\sigma\right]$ being an odd
        permutation of $0123$}, \\
      +1, \qquad\mbox{for $\left[\mu\nu\rho\sigma\right]$ being an
        even permutation of $0123$}, \\
      \phantom{+}0, \qquad\mbox{otherwise}.
    \end{array}\right.
\]
The commutator of two objects $A,B$ for which multiplication and
addition are defined is denoted by \[ \left[ A, B\right] = A\cdot B -
B\cdot A\,. \] The anti-commutator is denoted by \[ \left\lbrace A,
  B\right\rbrace = A\cdot B + B\cdot A\,. \]

\section{Dirac Matrices}
\label{sec:dirac-matrices}
The Dirac matrices $\tilde{\gamma}_\mu$, $\mu=0,\dots,3$, in Minkowski
space are defined by
\begin{equation}
  \label{eq:dirac-mink-def}
  \left\lbrace\tilde{\gamma}_\mu,\tilde{\gamma}_\nu\right\rbrace = 2
  g_{\mu\nu}\,.
\end{equation}
The matrix $\tilde{\gamma}_5$ is defined by
\begin{equation}
  \label{eq:gamma5-def}
  \tilde{\gamma}_5 = \tilde{\gamma}^5 =
  \mbox{i}\tilde{\gamma}_0\tilde{\gamma}_1\tilde{\gamma}_2\tilde{\gamma}_3
  = -\frac{\mbox{i}}{4!} \epsilon_{\mu\nu\rho\sigma}
  \tilde{\gamma}^\mu\tilde{\gamma}^\nu\tilde{\gamma}^\rho\tilde{\gamma}^\sigma
  = \tilde{\gamma}_5^\dagger\,.
\end{equation}
It satisfies $\tilde{\gamma}_5^2 = \mathbf{1}$, and
$\lbrace\tilde{\gamma}_5,\tilde{\gamma}_\mu\rbrace = 0$. The chirality
projectors $P_{R,L}$ are given by
\begin{equation}
  \label{eq:chirality-proj-def}
  P_{R,L} = \frac{1}{2}\left(1\pm\tilde{\gamma}_5\right)\,.
\end{equation}
When performing the Wick rotation, the Euclidean Dirac matrices are
given by
\begin{equation}
  \label{eq:dirac-eucl-def}
  \left\lbrace\gamma_\mu,\gamma_\nu\right\rbrace = 2 \delta_{\mu\nu}\,.
\end{equation}
They are related via \cite{Montvay:1994cy} \[ \gamma_{1,2,3} =
-\mbox{i}\tilde{\gamma}_{1,2,3}, \qquad \gamma_0 = -\tilde{\gamma}_0\,.
\] The Euclidean $\gamma_5$-matrix is given by
$\gamma_5=\gamma_1\gamma_2\gamma_3\gamma_0$.

The matrices with smallest dimension satisfying
(\ref{eq:dirac-mink-def}) are $4\times 4$ matrices
\cite{Nachtmann:1990ta}. The representation of the
$\gamma_\mu$-matrices employed in this thesis has been chosen to be
the ``chiral'' one\footnote{It should be remarked that this is
  \textit{different} from the representation which has been employed
  in the standard \TAO-libraries \cite{TAO_Manual}, where the Dirac
  form has been used}, where the Minkowski-space matrices are given by
{\setlength{\arraycolsep}{0pt}
  \begin{eqnarray}
    && \begin{array}{ll}
      \tilde{\gamma}_0 = \left(\matrix{ 0 & 0 & -1 & 0 \\ 0 & 0 & 0 &
          -1 \\ -1 & 0 & 0 & 0 \\ 0 & -1 & 0 & 0 \\
      }\right),\qquad &
      \tilde{\gamma_1} = \left(\matrix{ 0 & 0 & 0 & 1 \\ 0 & 0 & 1 & 0
          \\ 0 & -1 & 0 & 0 \\ -1 & 0 & 0 & 0 \\  }\right), \\ & \\
      \tilde{\gamma}_2 = \left(\matrix{ 0 & 0 & 0 & -\mbox{i} \\ 0 & 0
          & \mbox{i} & 0 \\ 0 & \mbox{i} & 0 & 0 \\ -\mbox{i} & 0 & 0 &
          0 \\ }\right),\qquad &
      \tilde{\gamma}_3 = \left(\matrix{ 0 & 0 & 1 & 0 \\ 0 & 0 & 0 & -1
          \\ 1 & 0 & 0 & 0 \\ 0 & -1 & 0 & 0 \\  }\right), \\ &
    \end{array} \nonumber \\
    && \begin{array}{l}
      \tilde{\gamma}_5 = \left( \matrix{ 1 & 0 & 0 & 0 \\ 0 & 1 & 0 & 0
          \\ 0 & 0 & -1 & 0 \\ 0 & 0 & 0 & -1 \\  }
      \right)\,.
    \end{array}
    \label{eq:gamma-chiral-minkowski}
  \end{eqnarray}}
The Euclidean $\gamma_\mu$ matrices are then given by
{\setlength{\arraycolsep}{0pt}
  \begin{eqnarray}
    && \begin{array}{ll}
      \gamma_0 = \left(\matrix{ 0 & 0 & 1 & 0 \\ 0 & 0 & 0 & 1 \\
          1 & 0 & 0 & 0 \\ 0 & 1 & 0 & 0 \\  }\right),\qquad &
      \gamma_1 = \left(\matrix{ 0 & 0 & 0 & -\mbox{i}
          \\ 0 & 0 & -\mbox{i} & 0 \\ 0 & \mbox{i} & 0 & 0 \\
          \mbox{i} & 0 & 0 & 0 \\  }\right), \\ & \\
      \gamma_2 = \left(\matrix{ 0 & 0 & 0 & -1 \\ 0 & 0 & 1 & 0 \\
          0 & 1 & 0 & 0 \\ -1 & 0 & 0 & 0 \\  }\right),\qquad &
      \gamma_3 = \left(\matrix{ 0 & 0 & -\mbox{i} & 0 \\ 0 & 0 & 0
      & \mbox{i} \\ \mbox{i} & 0 & 0 & 0 \\ 0 & -\mbox{i} & 0 & 0
      \\  }\right), \\ &
    \end{array} \nonumber \\
    && \begin{array}{l}
      \gamma_5 = \left(\matrix{ 1 & 0 & 0 & 0 \\ 0 & 1 & 0 & 0
          \\ 0 & 0 & -1 & 0 \\ 0 & 0 & 0 & -1 \\  }\right)\,.
    \end{array}
    \label{eq:gamma-chiral-euclid}
  \end{eqnarray}}
The contraction of a four-vector $A^\mu$ with the Dirac matrices is
denoted by
\begin{equation}
  \label{eq:feynman-slash-def}
  \slash\!\!\!\! A = \sum_{\mu=0}^3 A^\mu \gamma_\mu.
\end{equation}

\chapter{Groups and Algebras}
\label{sec:groups-algebras}
One of the central concepts of particle physics is the notion of
\textit{symmetry groups}.  All particles transform according to a
symmetry of space-time and the symmetries of the Lagrangian. In the
relativistic case this will mean that they transform as
representations of the proper orthochroneous Poincar\'{e}-group, see
below.

\section{Groups and Representations}
\label{sec:groups-repr}
A group is a pair $\left(G,\cdot\right)$ of a set $G$ and a relation
$\cdot$, satisfying the following axioms
\begin{enumerate}
\item The operation $\cdot$ is associative, i.e.~$\forall\ x,y,z\in G$
  holds $\left(x\cdot y\right)\cdot z = x\cdot\left(y\cdot z\right)$.
\item There is a unit element $e\in G$ satisfying: $\forall\ x\in G$
  holds $x\cdot e=e\cdot x=x$.
\item For every $x\in G\,\; \exists\ x^{-1}\in G$ such that $x\cdot
  x^{-1}=x^{-1}\cdot x=e$.
\end{enumerate}
The group is called \textit{Abelian} (or \textit{commutative}) if
additionally $\left[x,y\right]=x\cdot y - y\cdot x, \forall\ x,y\in G$
holds. It follows immediately that the unit element is uniquely
determined. Furthermore it follows that the inverse element $x^{-1}$
for each $x$ is unique.

A representation $\mathscr{R}(G)$ of the group $G$ is a group
homomorphism from $G$ to the group of vector space endomorphisms of a
representation space $V$, $\mathscr{R}(G):g\mapsto M(g)$, $g\in G$ and
$M(g)\in V$ with the following properties:
\begin{enumerate}
\item $M(g)\cdot M(h) = M(g\cdot h)$, i.e.~the representation respects
  the group multiplication of $G$,
\item $M(\mathbf{1}) = \mathbf{1}$, i.e.~the image of the unit element
  in $G$ is the identity in $V$,
\item $M(g^{-1}) = M^{-1}(g)$, i.e.~the image of the inverse element
  is the inverse of the group element.
\end{enumerate}
A representation is called \textit{irreducible} if it can not be
written as the direct sum of other representations. Thus, there are no
invariant subspaces under the action of the $M(g)$ for all $g\in G$.
In the following, a matrix in $V$ (with an appropriate basis) with the
above properties will be called a representation of $G$.

Particles, as observed in nature, should certainly be independent of
the way we choose our coordinate system, i.e.~how we choose the basis
for the representation space $V$ (this requirement parallels the
requirement of the theories to be coordinate invariant). Thus, they
should always be classified by irreducible representations of a group.
These irreducible representations also go under the name
\textit{multiplet}.

Of particular interest to physics are the \index{Lie groups}
\textit{Lie groups}, see for a textbook \cite{Carter:1995bo}. A Lie
group is a group for which the multiplication law and taking the
inverse are smooth functions. Thus, the group space must be a manifold
and one can form the tangent space on any point in the group. The
tangent space on the unit element is called the \textit{Lie algebra}
of the group. A basis of the Lie algebra is called the set of
\index{Group generators} \textit{generators} of the group.
Accordingly, an element $a$ of a Lie group $a\in L$ can be written in
terms of the generators $\lbrace g_i\rbrace$, $i=1,\dots,N$ as:
\begin{equation}
  \label{eq:lie-generators}
  a = \exp\left[ \sum_{i=1}^N \omega_i g_i \right]\,,
\end{equation}
where the element $a$ is parameterized using the $\lbrace
\omega_i\rbrace$ as coordinates. The \textit{dimension} of a Lie group
is thus the dimension of the underlying manifold of the group space. A
Lie algebra can be specified by the \index{Structure constants}
\textit{structure constants} $f_{abc}$, which are defined via
\begin{equation}
  \label{eq:lie-structure-consts}
  \left[ g_a, g_b \right] = f_{abc} g_c\,.
\end{equation}

For the integration over the group space, there exists a unique
measure on $G$ called the \index{Measure!Haar} \textit{Haar measure},
$dU$, which obeys:
\begin{enumerate}
\item Consider a function $f:G\rightarrow\mathbb{C}$. Then $dU$ obeys
  for all $V\in G$
  \[ \int_G\; dU f(U) = \int_G\; dU f(V\cdot U) = \int_G\; dU f(U\cdot
  V)\,. \]
\item The integral is normalized, i.e.~$\int_G dU = 1$.
\end{enumerate}
It satisfies \[ \int_G dU\; f(U) = \int_G dU\; f(U^{-1})\,. \]

The \textit{rank} of a group is the number of generators that
simultaneously commute among themselves. It is thus the maximum number
of generators which can simultaneously be diagonalized.

A Lie algebra is called \textit{semisimple}, if for some $z\in G$,
there are $x,y\in G$ with $z=\left[x,y\right]$. It can be shown, that
for any compact Lie group, the algebra can always be written as the
direct sum of a semisimple Lie algebra and an Abelian one. The
semisimple Lie algebras can be decomposed into a set of groups which
are called \textit{simple}. The latter cannot be written as sums of
anything else. The simple groups fall into the following
categories\footnote{A compact and readable introduction to the subject
  of simple, finite groups and this classification can also be found
  in \texttt{http://math.ucr.edu/home/baez/week63.html},
  \texttt{http://math.ucr.edu/home/baez/week64.html}, and\newline
  \texttt{http://math.ucr.edu/home/baez/week66.html}}:
\begin{enumerate}
\item The algebra sl$_{N}(\mathbb{C})$, the $N\times N$ complex
  matrices with vanishing trace. The compact real form of
  sl$_{N}(\mathbb{C})$ is su$(N)$ and the corresponding Lie group is
  SU$(N)$, the $N\times N$ unitary matrices with unit determinant. In
  the case $N=1$, we speak of the group U$(1)$, which consists of the
  complex numbers on the unit circle.
\item \label{it:soNodd} The Lie algebra so$_{2N+1}(\mathbb{C})$, the
  $(2N+1)\times(2N+1)$ skew-symmetric complex matrices with vanishing
  trace. The compact real form is so$_{N}$, and the Lie group
  generated is SO$(N)$, the $N\times N$ real, orthogonal matrices with
  determinant one. They form the rotation group in $N$-dimensional
  Euclidean space. The rotation group in Minkowski space whose metric
  changes sign on the diagonal is usually denoted with SO$(3,1)$, but
  still belongs to this category.
\item The Lie algebra sp$_{N}(\mathbb{C})$, the $2N\times 2N$ complex
  matrices of the form \[ \left(\begin{array}{cc} A & B \\ C & D
    \end{array}\right)\,, \] where $B$ and $C$ are symmetric, and $D$ is the
  negative transpose of $A$.  The compact real form is sp$(N)$ and the
  Lie group is Sp$(N)$, which forms the group of $N\times N$
  quaternionic matrices which preserve the inner product on the space
  $H^N$ of $N$-tuples of quaternions.
\item The Lie algebra so$_{2N}(\mathbb{C})$, the $(2N)\times(2N)$
  skew-symmetric complex matrices with vanishing trace. This is the
  even-dimensional analog of item~\ref{it:soNodd} for even $N$. These
  groups are to be distinguished, since the physics in these cases may
  differ.
\end{enumerate}
Apart from these classical algebras, there are also the groups
$G_{4}$, $F_{2}$, $E_{6}$, $E_{7}$, and $E_{8}$. Some of these also
have applications in physics, however, so far they are not considered
to play any role for the purposes of this thesis.

The dimensions and ranks of the three important kinds of semi-simple
groups in this thesis are shown in Tab.~\ref{tab:group-dim-rank}.

\begin{table}[htb]
  \begin{center}
    \begin{tabular}[c]{l|l|l}
      \hline\hline
      \textbf{Group} & \textbf{Dimension} & \textbf{Rank} \\
      \hline
      \highstrut SO$(N)$, $N$ even & $\frac{1}{2}N(N-1)$ & $N/2$ \\
      \highstrut SO$(N)$, $N$ odd  & $\frac{1}{2}N(N-1)$ & $(N-1)/2$ \\
      \highstrut SU$(N)$           & $N^2-1$             & $N-1$ \\
      \hline\hline
    \end{tabular}
    \caption{Most important semi-simple groups together with their
      dimension and rank. The table is taken from \cite{Kaku:1993bo}.}
    \label{tab:group-dim-rank}
  \end{center}
\end{table}

\section{The U$(1)$ Group}
\label{sec:u1-group}
The U$(1)$ group is a special case of the SU$(N)$ groups. It consists
of the group of complex numbers on the unit circle. It is a
commutative group since the complex numbers commute under
multiplications.

\section{The SU$(N)$ Groups}
\label{sec:sun-groups}
The SU$(N)$ groups consist of elements isomorphic to the $N\times N$
unitary matrices with unit determinant:
\begin{equation}
  \label{eq:su_n-def}
  U\cdot U^\dagger = U^\dagger\cdot U = \mathbf{1}, \qquad
  \mbox{det}\,U = 1\,.
\end{equation}
Obviously the matrices $U$ in (\ref{eq:su_n-def}) form already the
fundamental representation of the SU$(N)$ group. In this thesis the
groups SU$(2)$ and SU$(3)$ play a central role. The generators chosen
for the specific realizations used in this thesis are listed in the
following sections.

\subsection{The SU$(2)$ Group}
\label{sec:su2-group}
The standard choice for the generators of the SU$(2)$ group are the
\textit{Pauli matrices}:
\begin{equation}
  \label{eq:pauli-matrices}
  \sigma_1 = \sumat{0}{1}{1}{0}\,,\quad \sigma_2 =
  \sumat{0}{-\mbox{i}}{\mbox{i}}{0}\,, \quad \sigma_3 =
  \sumat{1}{0}{0}{-1}\,.
\end{equation}
The peculiarity of SU$(2)$ is that these matrices together with the
unit matrix, \[ \mathbf{1} = \sumat{1}{0}{0}{1}\,, \] form a basis of
the complex $2\times 2$ matrices. The expansion coefficients form a
hypersurface in the space of complex $2\times 2$ matrices, where the
expansion coefficients are real. A consequence of this observation is
that any sum of SU$(2)$ matrices is again proportional to an SU$(2)$
matrix. This property only exists in the case $N=2$. The
proportionality factor can be computed by considering the inverse of a
matrix $A$, \[ A = a_0 + \mbox{i} \sum_{r=1}^3 \sigma_r \mathbf{a}_r =
\sumat{a_0+\mbox{i}a_3}{a_2+\mbox{i}a1}{-a_2+\mbox{i}a_1}{a_0-\mbox{i}a_3}\,.
\] Then the inverse is given by \[ A^{-1} = \frac{1}{\det A} A^\dagger
= \frac{1}{\det A} \sumat{a_0-\mbox{i}a_3}{-a_2-\mbox{i}a_1}
{a_2-\mbox{i}a_1}{a_0+\mbox{i}a_3 }\,. \] Consequently, the
proportionality factor is given by $k=\sqrt{\det A}$, i.e.~the matrix
\begin{equation}
  \label{eq:su2-project}
  B = A/\sqrt{\det A}
\end{equation}
is an SU$(2)$ matrix.

From Tab.~\ref{tab:group-dim-rank} it follows that the group has rank
one, thus the representations correspond to the eigenvalues of a
single operator. Usually, the eigenvalues of $\sigma_3$ are taken to
classify the multiplets. SU$(2)$ is locally isomorphic to SO$(3)$
\cite{Scheck:1992me}, which means that the algebras of the two groups
are identical, although this does not hold for their global topology.
To be specific, SU$(2)$ is the double-cover of SO$(3)$.  While the
latter is \textit{not} simply connected, the former is.

\subsection{The SU$(3)$ Group}
\label{sec:su3-group}
In this thesis the Gell-Man matrices $\lbrace\lambda_i\rbrace$,
$i=1,\dots,8$, as defined in \cite{Montvay:1994cy} have been chosen as
the generators of the SU$(3)$ group: {\setlength{\arraycolsep}{0pt}
  \begin{eqnarray}
    && \begin{array}{lll}
      \lambda_1 = \pmatrix{0 & 1 & 0 \\ 1 & 0 & 0 \\ 0 & 0 & 0 \\
        }\,,\qquad &
      \lambda_2 = \pmatrix{0 & -\mbox{i} & 0 \\ \mbox{i} & 0 & 0 \\
        0 & 0 & 0 \\ }\,,\qquad &
      \lambda_3 = \pmatrix{1 & 0 & 0 \\ 0 & -1 & 0 \\ 0 & 0 & 0 \\
        }\,, \\ &\\
      \lambda_4 = \pmatrix{0 & 0 & 1 \\ 0 & 0 & 0 \\ 1 & 0 & 0 \\
        }\,,\qquad &
      \lambda_5 = \pmatrix{0 & 0 & -\mbox{i} \\ 0 & 0 & 0 \\
        \mbox{i} & 0 & 0 }\,,\qquad &
      \lambda_6 = \pmatrix{0 & 0 & 0 \\ 0 & 0 & 1 \\ 0 & 1 & 0
        }\,, \\ &\\
      \lambda_7 = \pmatrix{0 & 0 & 0 \\ 0 & 0 & -\mbox{i} \\ 0 &
        \mbox{i} & 0 }, \qquad & \multicolumn{2}{l}{
        \lambda_8 = \pmatrix{1/\sqrt{3} & 0 & 0 \\ 0 & 1/\sqrt{3} &
          0 \\ 0 & 0 & -2/\sqrt{3} }\,.}
    \end{array}
  \end{eqnarray}}

\section{The Poincar\'{e} Group}
\label{sec:poincare-group}
The space-time manifold underlying the physical theories discussed in
this thesis is given by the Minkowski-space. The metric is
pseudo-Euclidean and can be transformed globally to the form
\eqcite{eq:minkowski-metric}\begin{equation}
  g_{\mu\nu} = \left(\begin{array}{*{4}{p{0.35cm}}}
      1 &  0 &  0 &  0 \\
      0 & -1 &  0 &  0 \\
      0 &  0 & -1 &  0 \\
      0 &  0 &  0 & -1 \end{array}\right)\,.
\end{equation}\fineqcite
An event is associated with a point in space-time, and the distance
between two events is defined as
\begin{equation}
  \label{eq:distance-minkowski}
  (x-y)^2 = (x-y)^\mu(x-y)^\nu g_{\mu\nu}\,.
\end{equation}
Here and in the following the Einstein summation convention that
identical indices are to be summed over is understood. The quantity
$a^\mu a^\nu g_{\mu\nu}\equiv a^\mu a_\nu$ is called the \textit{norm}
of $a^\mu$. This norm, however, is not positive definite. Depending on
the sign of $a^2$, one defines the following classes of vectors:
\begin{description}
\item[Timelike region:] If $a^2=(x-y)^2>0$, the distance is called
  \textit{timelike}. In such a case, the two events at $x^\mu$ and
  $y^\mu$ may have a causal influence on each other and there exists a
  unique Lorentz transformation which reduces the spatial components
  of $a^\mu$ to zero. However, there is no transformation which
  rotates the $a^0$ component to $0$.
\item[Spacelike region:] If $a^2=(x-y)^2<0$, the distance is
  \textit{spacelike}. In this case, two events at $x^\mu$ and $y^\mu$
  cannot be causally related. This requirement is equivalent to the
  colloquial saying that ``no information can travel faster than the
  speed of light''. There is a unique Lorentz transformation which
  rotates the $a^0$ component to $0$, but there is none which reduces
  the spatial components of $a^\mu$ to zero.
\item[Likelight region:] If $a^2=(x-y)^2=0$, the distance between
  $x^\mu$ and $y^\mu$ is \textit{lightlike}. The two events can be
  causally related if the interaction happens by exchanging
  information using massless particles traveling at the velocity $c$.
\end{description}
The Poincar\'{e} group consists of the four-dimensional rotations in
Minkowski space, the group SO$(3,1)$, and the translation group. An
element of the Poincar\'{e} group is denoted by
$\left(\Lambda^\mu_{\phantom{\mu} \nu},a^\mu\right)$ and transforms a
four-vector $x^\mu$ in the following manner:
\begin{equation}
  \label{eq:lorentz-transform}
  x'^\mu=\Lambda^\mu_{\phantom{\mu} \nu}x^\nu+a^\mu\,.
\end{equation}
The inverse transformation of $\left(\Lambda,a\right)$ is given by
$\left(\Lambda^{-1},\Lambda^{-1}a\right)$. The multiplication law is
given by \[ \left(\Lambda_1,a_1\right)\cdot\left(\Lambda_2,a_2\right)
= \left(\Lambda_1\Lambda_2,a_1+\Lambda_1a_2\right)\,. \] Thus, the set
of Poincar\'{e}-transformations form a non-Abelian group.

According to the postulates of special relativity, the coordinate
transformations are linear and real. When changing the frame of
reference, the distance of two events will be unchanged, which implies
that the norm of a vector is conserved. This means that for
$a^\mu=\mathbf{0}$ (this subgroup is called the \textit{homogeneous}
Poincar\'{e}-group):
\begin{eqnarray}
  \label{eq:lorentz-properties}
  \Lambda^\mu_{\phantom{\mu} \nu} &=& \Lambda^{*\mu}_{\phantom{*\mu}
    \nu}\,, \nonumber \\
  \Lambda^\mu_{\phantom{\mu} \alpha} \Lambda^\alpha_{\phantom{\alpha}
    \nu} &=& \delta^{\mu}_{\phantom{\mu} \nu}\,.
\end{eqnarray}
Consequently, one finds \[ \det\,\Lambda^\mu_{\phantom{\mu} \nu} = \pm
1\,, \] and one can distinguish four kinds of transformations as
displayed in Tab.~\ref{tab:transform-kinds}. From the four subsets,
only the proper, orthochroneous set contains the unit element and is
therefore the only subgroup. This subgroup is connected, while the
entire homogeneous Poincar\'{e} group is not connected.

\begin{table}[htb]
  \begin{center}
    \begin{tabular}[c]{*{4}{l|}*{2}{r|}c}
      \hline\hline
      \multicolumn{1}{c|}{\strut\textbf{Group}} &
      \multicolumn{1}{c|}{$\mathbf{|
      \Lambda^{\mu}_{\phantom{\mu}\nu}|}$} &
      \multicolumn{1}{c|}{$\mathbf{\Lambda^0_{\phantom{0}0}}$} &
      \multicolumn{4}{c}{\textbf{Category}} \\ \hline
      \strut$\mathcal{L}^\uparrow_+$ & $+1$ & $\geq +1$ & proper &
      \multicolumn{1}{c}{} & \\ \cline{1-4}
      \strut$\mathcal{L}^\uparrow_-$ & $-1$ & $\geq +1$ &
      \multicolumn{3}{r|}{orthochroneous} & \\ \cline{1-6}
      \strut$\mathcal{L}^\downarrow_+$ & $+1$ & $\leq -1$ &
      \multicolumn{1}{c}{} & \multicolumn{3}{r}{homogeneous} \\
      \cline{1-3}
      \strut$\mathcal{L}^\downarrow_-$ & $-1$ & $\leq -1$ \\
      \hline\hline
    \end{tabular}
    \caption{All four kinds of homogeneous Poincar\'{e}-transformations
      compatible with (\ref{eq:lorentz-properties}).}
    \label{tab:transform-kinds}
  \end{center}
\end{table}

The Poincar\'{e} group has six generators for rotations in the
$\mu-\nu$-plane, $L_{\mu\nu}$ (which are antisymmetric,
$L_{\mu\nu}=-L_{\nu\mu}$), and four generators $P_\mu$ for
translations. Their commutation relations give rise to the
Poincar\'{e} algebra \cite{Kaku:1993bo}:
\begin{eqnarray}
  \label{eq:poincare-algebra}
  \left[ L_{\mu\nu}, L_{\rho\sigma}\right] &=& \mbox{i}\left(
  g_{\nu\rho} L_{\mu\sigma} - g_{\mu\rho} L_{\nu\sigma} -
  g_{\nu\sigma}L-{\mu\rho} + g_{\mu\sigma}L_{\nu\rho}\right)\,,
  \nonumber \\
  \left[ L_{\mu\nu}, P_{\rho}\right] &=& \mbox{i}\left(
    -g_{\mu\rho}P_\nu + g_{\nu\rho}P_\mu\right)\,, \nonumber \\
  \left[ P_\mu, P_\nu\right] &=& 0\,.
\end{eqnarray}
The algebra admits a representation in Minkowski space in terms of
differential operators:
\begin{eqnarray}
  \label{eq:poincare-representation}
  L_{\mu\nu} &=& \mbox{i}\left( x_\mu\partial_\nu -
    x_\nu\partial_\mu\right)\,, \nonumber \\
  P_\mu &=& \mbox{i}\partial_\mu\,.
\end{eqnarray}
Defining the Pauli-Lubanski tensor by
\begin{equation}
  \label{eq:pauli-lub-tensor}
  W^\mu = \frac{1}{2}\epsilon^{\mu\nu\rho\sigma}P_\nu
  L_{\rho\sigma}\,,
\end{equation}
one finds \cite{Kaku:1993bo} that the Poincar\'{e} group has two
Casimir operators: $P^\mu P_\mu$, and $W^\mu W_\mu$. This allows to
classify all irreducible representations \cite{Haag:1992hx}:
\begin{description}
\item[$P^\mu P_\mu\equiv m^2>0$, $P_0> 0$:] The energy states lie
  on the hyperboloid in the forward light cone. This describes massive
  particles with spin $s$, $| m,s\rangle$, $s=0,1/2,1,3/2,\dots$,
\item[$P^\mu P_\mu=0$, $P_0\geq 0$:] The energy states lie on the
  forward cone. This describes massless particles with helicities $h$,
  $| h\rangle$, $h=\pm s$, s counts as above,
\item[$P_\mu=0$:] This is the single point at the origin.
\item[$P^\mu P_\mu=0$, $P_0\leq0$:] The energy states lie on the
  surface of the backward light cone. The quantum number $s$ is
  continuous.
\item[$P^\mu P_\mu\equiv m^2>0$, $P_0< 0$:] The energy states lie on
  the hyperboloid in the backward light cone.
\item[$P^\mu P_\mu\equiv -\kappa^2<0 (\kappa\in\mathbb{R})$:] The
  particles lie on a spacelike hyperboloid. This would describe
  tachyonic particles with velocities greater than $c$.
\end{description}
Only the first two classes are realized for observable particles in
nature. If there was no lower bound to the energy of a particle as it
would be the case if the last class did correspond to any physical
particle, an arbitrary amount of energy could spontaneously be created
from any point in spacetime. According to the rules of quantum
mechanics this would happen with finite probability. Thus, this
possibility seems to be incompatible with the formulations of quantum
field theories known so far.

\section{Spin-Statistics Theorem}
\label{sec:spin-stat-theor}
An important relation between the particles of different spins is the
\textit{spin-statistics theorem} \cite{Haag:1992hx,Weinberg:1995mt}.
For a relativistic quantum field theory the observable particles
(i.e.~the physical states) must have the following properties if one
requires that causality holds: Taking $(x-y)^2<0$ to be a spacelike
distance in Minkowski space, the fields $\Phi(x)$ must satisfy the
following (anti-) commutativity relations:
\begin{description}
\item[Bose fields:] $\left[\Phi(x),\Phi^\dagger(y)\right] = 0$, if the
  fields $\Phi(x)$ transform according to a particle with even spin.
  The particles described by $\Phi(x)$ are called \textit{bosons}.
  Consequently, a single state may be occupied by an arbitrary amount
  of bosons.
\item[Fermi fields:] $\left\lbrace\Phi(x),\Phi^\dagger(y)\right\rbrace
  = 0$, if the fields described by $\Phi(x)$ transform as a
  representation with odd spin. The corresponding particles are called
  \textit{fermions} and a single state may only be occupied by a
  single or none fermion.
\end{description}

\section{Grassmann Algebras}
\label{sec:grassmann-algebras}
As has been noted in Sec.~\ref{sec:spin-stat-theor}, the fields
describing fermions anticommute for spacelike distances. The
anticommutativity is an essential property of Grassmann fields. Thus,
Grassmann algebras are an important ingredient for the description of
fermionic degrees of freedom. The discussion follows
Ref.~\cite{Roepstorff:1991xb}.

\subsection{Definitions}
\label{sec:definitions}
Consider a map from $p$ coordinates in $\mathbb{C}^N$, $\lbrace
u_i\rbrace$, $i=1,\dots,p$, onto the complex numbers,
\[ S:\underbrace{\mathbb{C}^N\otimes
  \cdots\otimes\mathbb{C}^N}_{p} \rightarrow \mathbb{C}: (u_1, \dots,
u_p) \mapsto \mathbb{C}\,. \] $S(u_1,\dots,u_p)$ is called $p$-linear
if $S$ is separately linear in each argument. It is called
\textit{antisymmetric} if, for any permutation $\pi\lbrace
1,\dots,p\rbrace$, we have \[ S(u_{\pi(1)},\dots,u_{\pi(p)}) =
\mbox{sgn}(\pi)\, S(u_1,\dots,u_p)\,, \] where $\mbox{sgn}(\pi)$
denotes the signature of the permutation $\pi$.

Now we consider the space $A^p(\mathbb{C}^N)$ of $p$-linear
antisymmetric functions on $\mathbb{C}^N$. By definition, we set
$A^0(\mathbb{C}^N)=\mathbb{C}$. For $p\geq 1$, one finds
\begin{eqnarray*}
  \dim A^p(\mathbb{C}^N) &=& {N \choose p}\,, \qquad 0\leq
  p\leq N\,, \\
  A^p(\mathbb{C}^N) &=& 0, \qquad p>N\,.
\end{eqnarray*}
The \textit{Grassmann product map} assigns to any two vectors $S\in
A^p$ and $T\in A^q$ a vector $S\wedge T\in A^p\otimes A^q = A^{p+q}$ via
{\setlength{\arraycolsep}{0pt}\begin{eqnarray}
    \label{eq:grassmann-prod-def}
    && S\wedge T(u_1,\dots, u_{p+q}) = \nonumber \\ && \qquad
    \frac{1}{p!q!} \sum_{\pi}
    \mbox{sgn}(\pi)\, S(u_{\pi(1)}, \dots, u_{\pi(p)})
    T(u_{\pi(p+1)},\dots, u_{\pi(p+q)}).
  \end{eqnarray}}
The Grassmann product is associative, \[ R\wedge(S\wedge T) = (R\wedge
S)\wedge T\,, \] and the commutation law becomes 
\begin{equation}
  \label{eq:grassmann-prod-mult}
  S\wedge T = (-1)^{pq}T\wedge S\,.
\end{equation}
The direct sum of vector spaces, \[ A(\mathbb{C}^N) =
\bigoplus_{p=0}^N A^p(\mathbb{C}^N)\,, \] together with the Grassmann
product Eq.~(\ref{eq:grassmann-prod-def}) form a graded algebra,
called the \textit{Grassmann algebra over $\mathbb{C}^N$}. An element
of $A(\mathbb{C}^N)$ can always be written as a sum
$S_0+S_1+\dots+S_N$ such that $S_p\in A^p(\mathbb{C}^N)$. The
dimension of the algebra is given by \[ \dim A(\mathbb{C}^N) =
2^{\dim\mathbb{C}^N} = 2^N\,. \] $A$ may be decomposed into an even
and an odd part,
\begin{eqnarray}
  \label{eq:grassmann-decomp}
  A   &=& A_+ \oplus A_-\,, \nonumber \\
  A_+ &=& A^0 \oplus A^2 \oplus \dots\;, \qquad\mbox{(even subspace)},
  \nonumber \\
  A_- &=& A^1 \oplus A^3 \oplus \dots\;, \qquad\mbox{(odd subspace)}.
\end{eqnarray}
Using the decomposition (\ref{eq:grassmann-decomp}) allows to write
the product rule (\ref{eq:grassmann-prod-mult}) as follows:
\begin{equation}
  \label{eq:grassmann-product-new}
  S\wedge T = \left\lbrace\begin{array}{ll}
      T\wedge  S & \mbox{if $S\in A_+$ or $T\in A_+$,} \\
      -T\wedge S & \mbox{if both $S,T\in A_-$.} \\
    \end{array}\right.
\end{equation}
Thus, the even part $A_+$ is a commutative subalgebra.

Let $\lbrace \hat{e}_i\rbrace$, $i=1,\dots,N$, be a basis of
$\mathbb{C}^N$. Any vector $u\in\mathbb{C}^N$ has then the coordinates
$\lbrace u^i\rbrace$. Then define special elements $\eta^i\in A^1$ via
\[ \eta^i(u) = u^i\,. \] The following properties then express the fact
that the $\lbrace\eta^i\rbrace$ generate the Grassmann algebra:
\begin{enumerate}
\item The $\lbrace\eta^i\rbrace$ anticommute:
  $\lbrace\eta^i,\eta^j\rbrace = 0$.
\item Each vector $S\in A^p$ may be represented as \[ S=\frac{1}{p!}
  s_{i_1\dots i_p} \eta^{i_1}\dots\eta^{i_p}\,, \] where $s_{i_1\dots
    i_p}$ are complex expansion coefficients which are antisymmetric
  with respect to permutations of their indices.
\end{enumerate}
The above definitions still make sense when the limit
$N\rightarrow\infty$ is considered. This is the interesting situation
when applying Grassmann variables to continuum field theories.
However, when constructing the Schwinger functions $\schwinger_N$ of
Grassmann fields on the lattice, cf.~Sec.~\ref{sec:fermion-fields},
the behavior of the fermionic degrees of freedom in the continuum
limit will also matter.

\subsection{Derivatives}
\label{sec:derivatives}
The \textit{derivative}, $d_u$, is a map \[ d_u:
A^p(\mathbb{C}^)\rightarrow A^{p-1}(\mathbb{C}^N)\,, \qquad p>0\,, \]
which is given by
\begin{equation}
  \label{eq:grassmann-diff-def}
  d_u = \sum_{i=1}^N u^i\frac{\partial}{\partial\eta^i}\,,
\end{equation}
with respect to the basis $\lbrace\eta^i\rbrace$. It obeys the
following rules
\begin{enumerate}
\item $\frac{\partial}{\partial\eta_i}(\alpha S+\beta T) =
  \alpha\frac{\partial}{\partial\eta_i} S +
  \beta\frac{\partial}{\partial\eta_i} T\,, \qquad
  \alpha,\beta\in\mathbb{C}\,$,
\item $\frac{\partial}{\partial\eta^i} 1 = 0\,,$
\item $\frac{\partial}{\partial\eta^i} (\eta^k S) = \delta_i^k S -
  \eta^k \frac{\partial}{\partial\eta^i} S\,$.
\end{enumerate}

\subsection{Integration}
\label{sec:integration}
The integral $\int\EnsembleMeas{\eta}:A^p\rightarrow\mathbb{C}$ has
the following properties:
\begin{enumerate}
\item The integral $\int\EnsembleMeas{\eta}\, S$ is a complex number
  and the map $S\mapsto\int\EnsembleMeas{\eta}\, S$ is linear,
\item $\int\EnsembleMeas{\eta}\,\frac{\partial}{\partial\eta^i} S =
  0$, $i=1,\dots,N\,$,
\item $\int\EnsembleMeas{\eta}\, \eta^1\dots\eta^N = 1\,$.
\end{enumerate}
It is straightforward to proof the following rules:
\begin{enumerate}
\item The relation between integration and differentiation is given by
  \[ \int\EnsembleMeas{\eta} S = \frac{\partial}{\partial\eta^N}\dots
  \frac{\partial}{\partial\eta^1} S\,. \]
\item Integration by parts is performed via ($S\in A^p$, $T\in A$)
  \[ \int\EnsembleMeas{\eta} \left(\frac{\partial}{\partial\eta^i}
    S\right) \wedge T = (-1)^{(p+1)} \int\EnsembleMeas{\eta}\, S\wedge
  \frac{\partial}{\partial\eta^i} T\,. \]
\item Consider a linear transformation
  $a:\mathbb{C}^N\mapsto\mathbb{C}^N$ of the coordinates $\lbrace
  u_i\rbrace$ of $S$. Then the following rule holds:
  \[ \int\EnsembleMeas{\eta}\, S(au) = \det a\int\EnsembleMeas{\eta}\,
  S(u)\,. \] This integral is the counterpart of the corresponding
  integral in a real vector space, $x\in\mathbb{R}^N$,
  $a\in\mathbb{R}^N\otimes\mathbb{R}^N$, \[ \int dx\, f(ax) = |\det
  a|^{-1} \int dx\, f(x)\,. \]
\item The exponential integral of the linear transformation
  $a:\mathbb{C}^N\mapsto\mathbb{C}^N$ is given by
  \begin{equation}
    \label{eq:grassmann-exp-int}
    \int\EnsembleMeas{\eta}\EnsembleMeas{\zeta} \exp\left\lbrace
      -\sum_{ik=1}^N a_{ik} \eta^i\zeta^k \right\rbrace =
    (-1)^{\mbox{\tiny${\setlength{\arraycolsep}{0pt}\left(\begin{array}{c}N
            \\ 2\end{array}\right)}$}} \det (-a)\,.
  \end{equation}
  This rule is again the counterpart of the exponential integral in a
  real vector space.  However, in the latter case, the integral only
  exists for a positive definite transformation $a$, while the former
  exists for any $a$.
\end{enumerate}
In fact, the generating functional (\ref{eq:gen-func-euclid}) for
bosonic fields can be generalized to an integral over Grassmann fields
$\lbrace\eta^i\rbrace$ if fermions are considered. Then
Eq.~(\ref{eq:grassmann-exp-int-arb}) is the central tool for
evaluating the path integral on a finite lattice
$\mathbb{Z}^4_\Omega$. It should be pointed out that the sign-factor
in Eq.~(\ref{eq:grassmann-exp-int}) drops out in the case of Dirac
fermions since a Dirac spinor is composed of two Weyl spinors which
are separately described by Grassmann variables. This in turn implies
that $N$ will always be even in case of Dirac fermions. Thus, the
overall sign is $+1$.

\chapter{Local Forms of Actions Used}
\label{sec:local-forms-actions}
For the local updating algorithms on the lattice discussed in
Sec.~\ref{sec:boson-sampl-algor} the lattice actions have to be cast
into a form where the contribution of a single site factorizes from
the contributions of the other sites. This is not possible for all
actions, but in many cases it is possible to find an approximative
action which fulfills the above condition and which has sufficient
overlap with the original action under consideration. This idea is in
fact the basis of L\"uscher's original proposal for a multiboson
algorithm \cite{Luscher:1994xx}. After the action has been rearranged
in the form above, the local ``staples'' can be used for the local
updating algorithms.

\section{General Expressions}
\label{sec:general-expressions}
Consider a lattice action of the following general form
\begin{equation}
  \label{eq:multiquadratic}
  \tilde{S} = \sum_{i=1}^M\sum_{x} a_{i}
  \left(\phi(f_{i}^1(x))\phi(f_{i}^2(x))\cdots
  \phi(f_{i}^{n_{i}}(x))\right)\,,
\end{equation}
i.e.~on a given space $\Omega$ with coordinate vectors denoted by
$x\in\Omega$ we have a discretized field $\lbrace\phi(x)\rbrace$. The
action is given by a sum of $M$ terms containing products of the field
$\lbrace\phi(x)\rbrace$ such that each coordinate appears in the
action only once; i.e.~the functions of the coordinates $\lbrace
f_k^r(x)\rbrace$ (with $k=1\dots M$, and $r=1\dots n_{k}$,
$f_k^r:\Omega\mapsto \Omega$) must be distinct:
\begin{equation}
  \label{eq:coordrestriction}
  f_k^i(x) \neq f_k^j(x) \quad\forall\ i\neq j,\ x\in\Omega; \quad
  i, j = 1\dots n_k\,.
\end{equation}
Furthermore the functions $\lbrace f_k^r(x)\rbrace$ must be
invertible.

Then we can choose the functions $\lbrace f_k^r(x)\rbrace$ such that
$f_k^1(x)=x$ without loss of generality. If the action contains $N$
different fields $\phi_k(x)$, $k=1,\dots,N$, each field-type
$\phi_j(x)$ has to be considered separately in
Eq.~(\ref{eq:multiquadratic}). The other fields $\phi_{k\neq j}(x)$
are then contained in the constants $a_i$.

From Eq.~(\ref{eq:multiquadratic}) we can compute the staples of the
action, i.e.~the change $\Delta \tilde{S}$ in the action $\tilde{S}$
if we vary the field $\lbrace\phi(x)\rbrace$ at a single point $y$
about $\Delta\phi(y)$, with the following formula:
\begin{eqnarray}
  \label{eq:staples1}
  \Delta\tilde{S}\left[ \Delta\phi(y)\right] &=& \displaystyle
  \Delta\phi(y)\sum_{i=1}^M a_{i}
  \Biggl[ \phi(f_i^2(y))\cdots\phi(f_i^{n_{i}}) +
  \sum_{p=2}^{n_{i}} \phi\left(\left(f_i^p(y)\right)^{-1}\right)
  \nonumber\\
  && \displaystyle\nonumber \times
  \phi\left(f_i^p\left(f_i^2(y)\right)^{-1}\right)\cdots 
  \underbrace{\phi\left(f_i^p\left(f_i^p(y)\right)^{-1}
  \right)}_{\mbox{omitted}}\cdots
  \phi\left(f_i^p\left(f_i^{n_{i}}(y)\right)^{-1}\right)\Biggr]\,. \\
\end{eqnarray}
The above form may also be generalized to the case where $\phi(x)$
denotes a field with several components, e.g.~a complex $3\times 3$
matrix in the case of gluon fields $U_\mu(x)$. The action
(\ref{eq:multiquadratic}) will then be the trace over the resulting
matrix; however, equation (\ref{eq:staples1}) will have to be modified
to account for the non-commutativity of the fields. Since the trace is
not invariant under commutation, but under cyclic permutations, the
expression reads
\begin{eqnarray}
  \label{eq:staples2}
  \Delta\tilde{S}\left[ \Delta\phi(y)\right] &=& \displaystyle
  \Delta\phi(y)\sum_{i=1}^M a_{i} \Biggl[
  \phi(f_i^2(y))\cdots\phi(f_i^{n_{i}}) \nonumber \\
  &&\displaystyle\quad + \sum_{p=2}^{n_{i}}
  \phi\left(\left(f_i^p\left(f_i^{p+1}(x)\right)\right)^{-1}
  \right)\cdots \phi\left(f_i^p\left(f_i^{n_i}(x)\right)^{-1} \right)
  \nonumber \\
  && \displaystyle\qquad \times \phi\left(f_i^p(x)^{-1}\right)
  \phi\left( f_i^p\left(f_i^2(x)\right)^{-1} \right)\cdots\phi\left(
    f_i^p\left(f_i^{p-1}(x)\right)^{-1}\right) \Biggr]\,. \nonumber \\
\end{eqnarray}
There is another important situation where the field
$\lbrace\phi(x)\rbrace$ at site $x$ appears quadratically in the
lattice action. In this case the action can be rewritten as a Gaussian
and the heatbath algorithm discussed in
Sec.~\ref{sec:heatb-scal-fields} can immediately be applied. Such an
action will have the following form:
\begin{eqnarray}
  \label{eq:staples3}
  \tilde{S}\left[ \phi_l(x)\right] &=&
  \displaystyle\sum_{x}\Biggl\lbrace a_1 \left(\phi(x)^2
    \phi\left(f_1^2(x)\right)\cdots \phi\left(f_1^{n_1}(x)\right)
  \right) \nonumber\\
  &&\displaystyle\qquad + \sum_{i=2}^M a_i \left(\phi(x)
    \phi\left(f_i^2(x)\right)\cdots \phi\left(f_i^{n_i}(x)\right)
  \right) \nonumber\Biggr\rbrace\\
  &=&\displaystyle\nonumber \sum_x \tilde{a}_1 \left\lbrace \phi(x) +
    \frac{1}{2}\tilde{a}_1^{-1} \sum_{i=2}^M a_i
  \left(\phi\left(f_i^2(x)\right)
      \cdots \phi\left(f_i^{n_i}(x)\right) \right)
  \right\rbrace^2\nonumber\\
  &&\displaystyle\qquad\quad - \left(\mbox{\strut independent of
      $\phi(x)$}\right)\,,
\end{eqnarray}
where $\tilde{a}_1=a_1\left( \phi\left(f_1^2(x)\right)\cdots
  \phi\left(f_1^{n_1}(x)\right)\right)$. The remaining terms
independent of $\phi(x)$ are of no importance for the updating
algorithm and their precise form does not matter. The case where
$\phi(x)$ is a complex field or an $n$-component field (where the
trace has to be taken to compute the action) is straightforward.
However, the matrix $a_1$ must be invertible for this method to work.

\section{Local Forms of Various Actions}
\label{sec:local-forms-varactions}
To implement the local algorithms for gauge fields in
Sec.~\ref{sec:boson-sampl-algor}, one has to find the plaquette
staples $\tilde{\mathscr{S}}_\mu(x)$ for a given action. The local
action then takes the form
\begin{equation}
  \label{eq:wilson-staple-def}
  \Delta\tilde{S}\left[\Delta U_\mu(x)\right] = -\frac{\beta}{N}
  \mbox{Re}\,\mbox{Tr}\, \Delta U_\mu(x)
  \tilde{\mathscr{S}}_\mu(x)\,.
\end{equation}
In the following subsections, this form will be examined for the cases
needed in this thesis. Please note that in the following \textit{no}
implicit summation over the external index must be performed.

\subsection{Pure Gauge Fields}
\label{sec:pure-gauge-fields}
\index{Action!Wilson} As a first example, consider the pure gauge
action given by Eq.~(\ref{eq:wilson-gauge-action}),
\eqcite{eq:wilson-gauge-action}
\begin{equation}
  S[U(x)] = \beta\sum_x\sum_{\mu\nu} \left(1-\frac{1}{N}\mbox{Re}\,
  \mbox{Tr}\, U_{\mu\nu}(x)\right)\,,
\end{equation}
\fineqcite with the plaquette $U_{\mu\nu}(x)$ given by
\eqcite{eq:new-plaquette}
\begin{equation}
  U_{\mu\nu}(x) = U_\mu(x) U_\nu(x+\hat{\mu})
  U^\dagger_\mu(x+\hat{\nu}) U^\dagger_\nu(x)\,.
\end{equation}
\fineqcite Then one immediately finds for the local staple form of the
action:
\begin{eqnarray}
  \label{eq:wilson-gauge-staple}
  \Delta\tilde{S}\left[\Delta U_\mu(x)\right] &=&
  -\beta\frac{1}{N}\sum_{\nu\neq\mu} \mbox{Re}\,\mbox{Tr}\,\Delta
  U_\mu(x) \biggl( U_\nu(x+\hat{\mu}) U^\dagger_\mu(x+\hat{\nu})
  U^\dagger_\nu(x) \nonumber \\
  && \qquad\qquad\qquad + U^\dagger_\nu(x+\hat{\mu}-\hat{\nu})
  U^\dagger_\mu(x-\hat{\nu}) U_\nu(x-\hat{\nu}) \biggr)\,.
\end{eqnarray}

\subsection{Lattice Fermion Fields}
\label{sec:latt-ferm-fields}
\index{Wilson matrix} The Wilson matrix $Q(y,x)$ describing a single,
massive fermion flavor is given by the expression
(\ref{eq:wilson-matrix}): \eqcite{eq:wilson-matrix}
\begin{eqnarray}
  Q(y,x) &=& \nonumber\displaystyle\delta(y,x) -
  \kappa\sum_{\rho=0}^3\Bigl( 
  U_\rho\left(y-\hat{\rho}\right) \left(1+\gamma_\rho\right)
  \delta\left(y, x+\hat{\rho}\right) \\
  && \displaystyle\phantom{\delta(y,x) -
    \kappa\sum} + U_\rho^\dagger(y)
  \left(1-\gamma_\rho\right) \delta\left(y, x-\hat{\rho}\right)
  \Bigr)\,,
\end{eqnarray}
\fineqcite where $\kappa < \kappa_{\mbox{\tiny crit}}$. Up to now the
boundary conditions have been chosen implicitly to be periodic in the
lattice $1$-, $2$- and $3$-directions and anti-periodic in the lattice
$0$-direction. For the actual implementation of the local action, it
is more convenient to impose periodic boundary conditions in all four
lattice directions, and consequently have a symmetric treatment of the
lattice volume $\Omega$. Respecting the anti-periodicity can be done
by introducing an explicit factor which implements the anti-periodic
boundary conditions in the lattice $0$-direction (also called
$T$-direction). We define the fermionic sign function to be:
\begin{equation}
  \label{eq:fermsign}
  \theta_\mu(x) = 1 - 2\,\delta(\mu,0)\,
  \delta\left(x_0,T_{\mbox{\tiny max}}\right)\,,
\end{equation}
i.e.~the function $\theta(x)$ is equal to $-1$ on the hyperslice with
$x_0=T_{\mbox{\tiny max}}$ for $\mu=0$ only and $+1$ everywhere else.
With this convention the Wilson matrix takes the form
\begin{eqnarray}
  \label{eq:wilson-matrix-antiper}
  Q(y,x) &=& \displaystyle\nonumber \delta(y,x) -
  \kappa\sum_{\rho=0}^3\Bigl( U_\rho\left(y-\hat{\rho}\right)
  \left(1+\gamma_\rho\right) \delta\left(y, x+\hat{\rho}\right)
  \theta_\rho\left( y-\hat{\rho}\right) \\
  && \displaystyle\phantom{\delta(y,x) -
    \kappa\sum} + U_\rho^\dagger(y) 
  \left(1-\gamma_\rho\right) \delta\left(y, x-\hat{\rho}\right)
  \theta_\rho\left( y\right) \Bigr)\,.
\end{eqnarray}
This staple can be used directly for the implementation on a computer.

\subsubsection{Wilson Fermions (Hermitian)}
\label{sec:wils-ferm-herm}
Using the Hermitian fermion matrix the fermionic energy is given by
\eqcite{eq:eferm-wilson}
\begin{equation}
  S_{\mbox{\tiny f}} = \sum_j\sum_{xyz} \phi_j^\dagger(y)
  \left(\tilde{Q}(y,z)-\rho_j^*\right)
  \left(\tilde{Q}(z,x)-\rho_j\right) \phi_j(x)\,,
\end{equation}
\fineqcite where the $\rho_j$ are the roots of the polynomial in
(\ref{eq:polynomial}). If one uses even-odd preconditioning, the
fermionic energy is given by Eq.~(\ref{eq:eferm-wilsoneo}):
\eqcite{eq:eferm-wilsoneo}
\begin{equation}
  S_{\mbox{\tiny f}} = \sum_j\sum_{xyz} \phi_j^\dagger(y)
  \left(\tilde{Q}(y,z)-P_o\rho_j^*\right)
  \left(\tilde{Q}(z,x)-P_o\rho_j\right) \phi_j(x)\,.
\end{equation}
\fineqcite Inserting the Wilson matrix (\ref{eq:wilson-matrix}) into
Eq.~(\ref{eq:eferm-wilson}) one gets the action in the form of
Eq.~(\ref{eq:multiquadratic}):
\begin{eqnarray}
  \label{eq:eferm-wilson-expand}
  \lefteqn{\begin{array}{rcrl}
      S_{\mbox{\tiny f}} &=&
      \lefteqn{\displaystyle\nonumber\sum_j\sum_{xy}
        \phi_j^\dagger(y)\left(
          \tilde{Q}^2(y,x)-\left(\rho^*_j+\rho_j\right)
          \tilde{Q}(y,x) +
          \rho^*_j\rho_j\delta(y,x)\right)\phi_j(x)}\hfill \\
      &=& \displaystyle\sum_j\sum_{xyz}
      \phi_j^\dagger(y)\Biggl\lbrace \gamma_5\biggl[ \delta(y,z)-
      \kappa\sum_\rho\Bigl( & \hspace{-1.5\arraycolsep}
      U_\rho\left(y-\hat{\rho}\right) 
      \left(1+\gamma_\rho\right) \delta(y,z+\hat{\rho})
      \theta_\rho(y-\hat{\rho}) \\
      && + & \displaystyle \hspace{-1.5\arraycolsep}
      U^\dagger_\rho(y)\left(1-\gamma_\rho\right) 
      \delta(y,z-\hat{\rho})\theta_\rho(y) \Bigr)\biggr] \\
      && \displaystyle \times\gamma_5\biggl[ \delta(z,x)-
      \kappa\sum_\sigma\Bigl( & \hspace{-1.5\arraycolsep}
      U_\sigma\left(z-\hat{\sigma}\right)\left(1+\gamma_\sigma\right)
      \delta(z,x+\hat{\sigma}) \theta_\sigma(z-\hat{\sigma}) \\
      && + & \displaystyle \hspace{-1.5\arraycolsep}
      U^\dagger_\sigma(z)\left(1-\gamma_\sigma\right)
      \delta(z,x-\hat{\sigma})\theta_\sigma(z) \Bigr)\biggr] \\
      && \displaystyle -\left(\rho^*_j+\rho_j\right)\gamma_5\biggl[
      \delta(y,x)-\kappa\sum_\rho\Bigl( &
      \displaystyle\hspace{-1.5\arraycolsep} 
      U_\rho(y-\hat{\rho})\left(1+\gamma_\rho\right)
      \delta(y,x+\hat{\rho})\theta_\rho(y-\hat{\rho}) \\
      && + & \displaystyle \hspace{-1.5\arraycolsep}
      U^\dagger_\rho(y)\left(1-\gamma_\rho\right)
      \delta(y,x-\hat{\rho}) \theta_\rho(y)\Bigr) \biggr]
    \end{array}} \\
  \phantom{S_{\mbox{\tiny f}}} & \phantom{=} & \displaystyle
  \qquad\qquad + \rho^*_j\rho_j\delta(y,x)\Biggr\rbrace\phi_j(x)
  \nonumber \\
  &=& \displaystyle\nonumber \sum_j \sum_{y}
  \phi_j^\dagger(y)\Biggl\lbrace
  \left(1+16\kappa^2+\rho_j^*\rho_j
    - (\rho_j^*+\rho_j)\gamma_5 \right) \phi_j(y) \\ \nonumber
  \lefteqn{\begin{array}{rcrl}
      \phantom{S_{\mbox{\tiny f}}} & \phantom{=} &
      \displaystyle \qquad + \kappa\sum_\rho\biggl[
      &\hspace{-1.5\arraycolsep}
      \left((\rho_j^*+\rho_j)\gamma_5 (1+\gamma_\rho)-2\right)
      U_\rho(y-\hat{\rho}) \phi_j(y-\hat{\rho})
      \theta_\rho(y-\hat{\rho}) \\
      &&& \displaystyle\hspace{-1.5\arraycolsep} +
      \left((\rho_j^*+\rho_j)\gamma_5(1-\gamma_\rho)-2\right)
      U_\rho^\dagger(y) \phi_j(y+\hat{\rho}) \theta_\rho(y)\biggr] \\
      && \displaystyle +
      \kappa^2\sum_{\rho_1\neq\rho_2}
      \biggl[ &\hspace{-1.5\arraycolsep} U_{\rho_1}(y-\hat{\rho}_1)
        U_{\rho_2}(y-\hat{\rho}_1-\hat{\rho}_2) (1-\gamma_{\rho_1})
        (1+\gamma_{\rho_2}) \\
        &&& \phantom{\biggl[}\qquad\times
        \phi_j(y-\hat{\rho}_1-\hat{\rho}_2)
        \theta_{\rho_1}(y-\hat{\rho}_1)
        \theta_{\rho_2}(y-\hat{\rho}_1-\hat{\rho}_2) \\
        &&&\phantom{\biggl[}\hspace{-1.5\arraycolsep} +
        U_{\rho_1}(y-\hat{\rho}_1)
        U^\dagger_{\rho_2}(y-\hat{\rho}_1) (1-\gamma_{\rho_1})
        (1-\gamma_{\rho_2}) \\
        &&& \phantom{\biggl[}\qquad\times
        \phi_j(y-\hat{\rho}_1+\hat{\rho}_2)
        \theta_{\rho_1}(y-\hat{\rho}_1)
        \theta_{\rho_2}(y-\hat{\rho}_1) \\
        &&&\phantom{\biggl[}\hspace{-1.5\arraycolsep} +
        U^\dagger_{\rho_1}(y)
        U_{\rho_2}(y+\hat{\rho}_1-\hat{\rho}_2) (1+\gamma_{\rho_1})
        (1+\gamma_{\rho_2}) \\
        &&& \phantom{\biggl[}\qquad\times
        \phi_j(y+\hat{\rho}_1-\hat{\rho}_2)
        \theta_{\rho_1}(y)
        \theta_{\rho_2}(y+\hat{\rho}_1-\hat{\rho}_2) \\
        &&&\phantom{\biggl[}\hspace{-1.5\arraycolsep} +
        U^\dagger_{\rho_1}(y)
        U_{\rho_2}(y+\hat{\rho}_1) (1+\gamma_{\rho_1})
        (1-\gamma_{\rho_2}) \\
        &&& \phantom{\biggl[}\qquad\times
        \phi_j(y+\hat{\rho}_1+\hat{\rho}_2)
        \theta_{\rho_1}(y)
        \theta_{\rho_2}(y+\hat{\rho}_1) \biggr]\Biggr\rbrace\,. \\
      \end{array}} \\
\end{eqnarray}
This expression can be cast into the form (\ref{eq:staples3}) to yield
\begin{eqnarray}
  \label{eq:phi-staple3}
  S_{\mbox{\tiny f}} &=& \displaystyle\nonumber \sum_j\sum_y
  \mbox{Re}\,\left(
    \phi^\dagger_j(y) A \phi_j(y) + \phi^\dagger_j(y) V_j(y)\right) \\
  &=& \displaystyle\nonumber \sum_j\sum_y \left( \phi^\dagger_j(y) A
    \phi_j(y) + \frac{1}{2}\left( \phi^\dagger_j(y) V_j(y) +
      V^\dagger_j(y) \phi_j(y) \right) \right) \\
  &=& \displaystyle\nonumber \sum_j\sum_y \left( \phi^\dagger_j(y) +
    \frac{1}{2}V^\dagger_j(y) A^{-1}\right) A \left( \phi_j(y) +
    \frac{1}{2}A^{-1} V_j(y) \right) + \mbox{indep.~of $y\;$.}
\end{eqnarray}
With the chiral representation of the $\lbrace\gamma\rbrace$-matrices,
cf.~Eq.~(\ref{eq:gamma-chiral-euclid}), the matrix $A^{-1}$ takes a
very simple form
\begin{eqnarray*}
  A &=& \left(1 + 16\kappa^2 + \rho^*_j\rho_j\right) - \left(\rho^*_j
    + \rho_j\right)\gamma_5 \\
  &\equiv& f_1 + f_2 \gamma_5\,, \\
  A^{-1} &=& \frac{f_1}{f_1^2-f_2^2} +
  \frac{-f_2}{f_1^2-f_2^2}\gamma_5\,.
\end{eqnarray*}
A local boson field heatbath is then computed by (see also
Eq.~(\ref{eq:scalar-local-hb}))
\begin{equation}
  \label{eq:scalar-local-hb-gen}
  \phi'_j(y) = \Omega_j(y) - A^{-1} V_j(y)\,,
\end{equation}
with $\Omega_j(y)$ being a random number taken from a Gaussian
distribution with unit width. A local boson field overrelaxation is
performed by
\begin{equation}
  \label{eq:scalar-local-or}
  \phi'_j(y) = -\phi_j(y) - 2 A^{-1} V_j(y)\,.
\end{equation}
In both cases, the order of the sites being updated matters.

For the local gauge field updates, expression
(\ref{eq:eferm-wilson-expand}) has to be cast into the form
(\ref{eq:staples2}). Then $\Delta S_{\mbox{\tiny f}}[\Delta U_\mu(y)]$
takes the form
\begin{eqnarray}
  \label{eq:ferm-staple2}
  \Delta S_{\mbox{\tiny f}}[\Delta U_\mu(y)] &=& \mbox{Re}\,\mbox{Tr}\,
  \displaystyle\Delta U_\mu(y)\theta_\mu(y) \sum_j\Biggl\lbrace
  -2\kappa\phi_j^\dagger(y+\hat{\mu}) \left(
    2-(\rho^*_j+\rho_j)\gamma_5 (1+\gamma_\mu)\right) \phi_j(y)
  \nonumber\\ \nonumber\lefteqn{
    \begin{array}{rl}
      \displaystyle \qquad + 2\kappa^2\sum_{\rho\neq\mu}\biggl[ &
      \displaystyle\hspace{-1.5\arraycolsep}\phi^\dagger_j(y+\hat{\mu})
      (1-\gamma_\mu)(1+\gamma_\rho) \phi_j(y-\hat{\rho})
      U_\rho(y-\hat{\rho})\theta_\rho(y-\hat{\rho}) \\
      & \displaystyle\phantom{\biggl[} \hspace{-5.5\arraycolsep} +
      U_\rho(y+\hat{\mu})\phi^\dagger_j(y+\hat{\mu}+\hat{\rho})
      (1-\gamma_\rho)(1+\gamma_\mu) \phi_j(y) \theta_\rho(y+\hat{\mu})
      \\
      & \displaystyle\phantom{\biggl[} \hspace{-5.5\arraycolsep} +
      \phi^\dagger_j(y+\hat{\mu}) (1-\gamma_\mu)(1-\gamma_\rho)
      \phi_j(y+\hat{\rho}) U^\dagger_\rho(y)\theta_\rho(y) \\
      & \displaystyle\phantom{\biggl[} \hspace{-5.5\arraycolsep} +
      U^\dagger_\rho(y+\hat{\mu}-\hat{\rho})
      \phi^\dagger_j(y+\hat{\mu}-\hat{\rho}) (1+\gamma_\rho)
      (1+\gamma_\mu) \phi_j(y) \theta_\rho(y+\hat{\mu}-\hat{\rho})
      \biggr]\Biggr\rbrace\,.
    \end{array}} \\
\end{eqnarray}
This expression can be implemented efficiently for the case of
\textit{repeated} local gauge field sweeps, as has already been noted
in Sec.~\ref{sec:arch-effic}. Equation (\ref{eq:ferm-staple2}) admits
a representation in the following form
\begin{eqnarray}
  \label{eq:ferm-staple-cache}
  \Delta S_{\mbox{\tiny f}}[\Delta U_\mu(y)] &=&
  \mbox{Re}\,\mbox{Tr}\, \displaystyle\Delta U_\mu(y)
  \Biggl\lbrace C^1_\mu(y) + \sum_{\rho\neq\mu}\biggl[
  C^2_{\mu\rho}(y) U_\rho(y-\hat{\rho}) +
  U_\rho(y+\hat{\mu}) C^3_{\mu\rho}(y)  \nonumber \\
  &&\displaystyle\qquad\qquad\qquad\qquad\qquad\quad +
  C^4_{\mu\rho}(y) U^\dagger_\rho(y)    +
  U^\dagger_\rho(y+\hat{\mu}-\hat{\rho})
  C^5_{\mu\rho}(y) \biggr]\Biggr\rbrace\,,
\end{eqnarray}
with the cache fields $\lbrace C^1_\mu, C^2_{\mu\rho}, C^3_{\mu\rho},
C^4_{\mu\rho}, C^5_{\mu\rho}\rbrace$ given by
\begin{eqnarray}
  \label{eq:cache-def}
  C^1_\mu(y) &=& \displaystyle\nonumber
  -2\kappa\theta_\mu(y)\theta_\rho(y-\hat{\rho})
  \sum_j\phi_j^\dagger(y+\hat{\mu}) \left(
    2-(\rho^*_j+\rho_j)\gamma_5 (1+\gamma_\mu)\right)\phi_j(y)\,, \\
  C^2_{\mu\rho}(y) &=& \displaystyle\nonumber
  2\kappa^2\theta_\mu(y)\theta_\rho(y-\hat{\rho})
  \sum_j\phi^\dagger_j(y+\hat{\mu})(1-\gamma_\mu)(1+\gamma_\rho)
  \phi_j(y-\hat{\rho})\,, \\
  C^3_{\mu\rho}(y) &=& \displaystyle\nonumber
  2\kappa^2\theta_\mu(y)\theta_\rho(y+\hat{\mu})
  \sum_j\phi^\dagger_j(y+\hat{\mu}+\hat{\rho})
  (1-\gamma_\rho)(1+\gamma_\mu) \phi_j(y)\,, \\
  C^4_{\mu\rho}(y) &=& \displaystyle\nonumber
  2\kappa^2\theta_\mu(y)\theta_\rho(y)
  \sum_j\phi^\dagger_j(y+\hat{\mu})
  (1-\gamma_\mu)(1-\gamma_\rho)\phi_j(y+\hat{\rho})\,, \\
  C^5_{\mu\rho}(y) &=& \displaystyle
  2\kappa^2\theta_\mu(y)\theta_\rho(y+\hat{\mu}-\hat{\rho})
  \sum_j\phi^\dagger_j(y+\hat{\mu}-\hat{\rho})
  (1+\gamma_\rho)(1+\gamma_\mu) \phi_j(y)\,.
\end{eqnarray}
Any expression similar to (\ref{eq:ferm-staple2}) can be written in
the form (\ref{eq:ferm-staple-cache}).

By inserting the matrix (\ref{eq:wilson-matrix}) into
Eq.~(\ref{eq:eferm-wilsoneo}), one arrives at the corresponding
expressions in the preconditioned case. $P_o$ designates the projector
to odd, and $P_e$ the projector to even sites:
\begin{eqnarray}
  \label{eq:eferm-eo-wilson-expand}
  S_{\mbox{\tiny f}} &=& \displaystyle\nonumber \sum_j \sum_{y}
  \phi_j^\dagger(y)\Biggl\lbrace
  \left(1+16\kappa^2+P_o\rho_j^*\rho_j
    - (P_e \rho_j^*+P_o \rho_j)\gamma_5 \right) \phi_j(y) \\ \nonumber
  \lefteqn{\begin{array}{rcrl}
      \phantom{S_{\mbox{\tiny f}}} & \phantom{=} &
      \displaystyle \qquad + \kappa\sum_\rho\biggl[
      &\hspace{-1.5\arraycolsep}
      \left((P_e \rho_j^*+P_o \rho_j)\gamma_5 (1+\gamma_\rho)-2\right)
      U_\rho(y-\hat{\rho}) \phi_j(y-\hat{\rho})
      \theta_\rho(y-\hat{\rho}) \\
      &&& \displaystyle\hspace{-1.5\arraycolsep} +
      \left((P_e \rho_j^*+P_o \rho_j)\gamma_5(1-\gamma_\rho)-2\right)
      U_\rho^\dagger(y) \phi_j(y+\hat{\rho}) \theta_\rho(y)\biggr] \\
      && \displaystyle +
      \kappa^2\sum_{\rho_1\neq\rho_2}
      \biggl[ &\hspace{-1.5\arraycolsep} U_{\rho_1}(y-\hat{\rho}_1)
      U_{\rho_2}(y-\hat{\rho}_1-\hat{\rho}_2) (1-\gamma_{\rho_1})
      (1+\gamma_{\rho_2}) \\
      &&& \phantom{\biggl[}\qquad\times
      \phi_j(y-\hat{\rho}_1-\hat{\rho}_2)
      \theta_{\rho_1}(y-\hat{\rho}_1)
      \theta_{\rho_2}(y-\hat{\rho}_1-\hat{\rho}_2) \\
      &&&\phantom{\biggl[}\hspace{-1.5\arraycolsep} +
      U_{\rho_1}(y-\hat{\rho}_1)
      U^\dagger_{\rho_2}(y-\hat{\rho}_1) (1-\gamma_{\rho_1})
      (1-\gamma_{\rho_2}) \\
      &&& \phantom{\biggl[}\qquad\times
      \phi_j(y-\hat{\rho}_1+\hat{\rho}_2)
      \theta_{\rho_1}(y-\hat{\rho}_1)
      \theta_{\rho_2}(y-\hat{\rho}_1) \\
      &&&\phantom{\biggl[}\hspace{-1.5\arraycolsep} +
      U^\dagger_{\rho_1}(y)
      U_{\rho_2}(y+\hat{\rho}_1-\hat{\rho}_2) (1+\gamma_{\rho_1})
      (1+\gamma_{\rho_2}) \\
      &&& \phantom{\biggl[}\qquad\times
      \phi_j(y+\hat{\rho}_1-\hat{\rho}_2)
      \theta_{\rho_1}(y)
      \theta_{\rho_2}(y+\hat{\rho}_1-\hat{\rho}_2) \\
      &&&\phantom{\biggl[}\hspace{-1.5\arraycolsep} +
      U^\dagger_{\rho_1}(y)
      U_{\rho_2}(y+\hat{\rho}_1) (1+\gamma_{\rho_1})
      (1-\gamma_{\rho_2}) \\
      &&& \phantom{\biggl[}\qquad\times
      \phi_j(y+\hat{\rho}_1+\hat{\rho}_2)
      \theta_{\rho_1}(y)
      \theta_{\rho_2}(y+\hat{\rho}_1) \biggr]\Biggr\rbrace\,. \\
    \end{array}} \\
\end{eqnarray}
The corresponding staple $\Delta S_{\mbox{\tiny f}}[\Delta U_\mu(y)]$
takes the form
\begin{eqnarray}
  \label{eq:ferm-eo-staple2}
  \Delta S_{\mbox{\tiny f}}[\Delta U_\mu(y)] &=&
  \mbox{Re}\,\mbox{Tr}\,
  \displaystyle\Delta U_\mu(y)\theta_\mu(y) \sum_j\Biggl\lbrace
  -2\kappa\phi_j^\dagger(y+\hat{\mu}) \left(
    2-(P_e \rho^*_j+P_o \rho_j)\gamma_5 (1+\gamma_\mu)\right)
  \phi_j(y)
  \nonumber\\ \nonumber\lefteqn{
    \begin{array}{rl}
      \displaystyle \qquad + 2\kappa^2\sum_{\rho\neq\mu}\biggl[ &
      \displaystyle\hspace{-1.5\arraycolsep}\phi^\dagger_j(y+\hat{\mu})
      (1-\gamma_\mu)(1+\gamma_\rho) \phi_j(y-\hat{\rho})
      U_\rho(y-\hat{\rho})\theta_\rho(y-\hat{\rho}) \\
      & \displaystyle\phantom{\biggl[} \hspace{-5.5\arraycolsep} +
      U_\rho(y+\hat{\mu})\phi^\dagger_j(y+\hat{\mu}+\hat{\rho})
      (1-\gamma_\rho)(1+\gamma_\mu) \phi_j(y) \theta_\rho(y+\hat{\mu})
      \\
      & \displaystyle\phantom{\biggl[} \hspace{-5.5\arraycolsep} +
      \phi^\dagger_j(y+\hat{\mu}) (1-\gamma_\mu)(1-\gamma_\rho)
      \phi_j(y+\hat{\rho}) U^\dagger_\rho(y)\theta_\rho(y) \\
      & \displaystyle\phantom{\biggl[} \hspace{-5.5\arraycolsep} +
      U^\dagger_\rho(y+\hat{\mu}-\hat{\rho})
      \phi^\dagger_j(y+\hat{\mu}-\hat{\rho}) (1+\gamma_\rho)
      (1+\gamma_\mu) \phi_j(y) \theta_\rho(y+\hat{\mu}-\hat{\rho})
      \biggr]\Biggr\rbrace\,.
    \end{array}} \\
\end{eqnarray}

\subsubsection{Wilson Fermions (non-Hermitian)}
\label{sec:wilson-fermions-non}
The fermionic action in terms of the non-Hermitian Wilson fermions is
obtained by replacing $\tilde{Q}(y,x)$ with $Q(y,x)$ in
(\ref{eq:eferm-wilson}). Without even-odd preconditioning one arrives
then at\eqcite{eq:eferm-wilson-nonherm}
\begin{equation}
  S_{\mbox{\tiny f}} = \sum_j\sum_{xyz} \phi_j^\dagger(y)
  \left(Q^{\dagger}(y,z)-\rho_j^*\right)
  \left(Q(z,x)-\rho_j\right) \phi_j(x)\,,
\end{equation}\fineqcite
The local fermionic action becomes
\begin{eqnarray}
  \label{eq:eferm-noherm-wilson-expand}
  S_{\mbox{\tiny f}} &=& \displaystyle\nonumber \sum_j \sum_{y}
  \phi_j^\dagger(y)\Biggl\lbrace
  \left(1+16\kappa^2+\rho_j^*\rho_j
    - (\rho_j^*+\rho_j) \right) \phi_j(y) \\ \nonumber
  \lefteqn{\begin{array}{rcrl}
      \phantom{S_{\mbox{\tiny f}}} & \phantom{=} &
      \displaystyle \qquad - \kappa\left(1+\rho_j^*+\rho_j\right)
      \sum_\rho\biggl[
      &\hspace{-1.5\arraycolsep}
      U_\rho(y-\hat{\rho}) \left(1+\gamma_\rho\right)
      \phi_j(y-\hat{\rho}) \theta_\rho(y-\hat{\rho}) \\
      &&& \displaystyle\hspace{-1.5\arraycolsep} +
      U_\rho^\dagger(y) \left(1-\gamma_\rho\right)
      \phi_j(y+\hat{\rho}) \theta_\rho(y)\biggr] \\
      && \displaystyle +
      \kappa^2\sum_{\rho_1\neq\rho_2}
      \biggl[ &\hspace{-1.5\arraycolsep} U_{\rho_1}(y-\hat{\rho}_1)
      U_{\rho_2}(y-\hat{\rho}_1-\hat{\rho}_2) (1+\gamma_{\rho_1})
      (1+\gamma_{\rho_2}) \\
      &&& \phantom{\biggl[}\qquad\times
      \phi_j(y-\hat{\rho}_1-\hat{\rho}_2)
      \theta_{\rho_1}(y-\hat{\rho}_1)
      \theta_{\rho_2}(y-\hat{\rho}_1-\hat{\rho}_2) \\
      &&&\phantom{\biggl[}\hspace{-1.5\arraycolsep} +
      U_{\rho_1}(y-\hat{\rho}_1)
      U^\dagger_{\rho_2}(y-\hat{\rho}_1) (1+\gamma_{\rho_1})
      (1-\gamma_{\rho_2}) \\
      &&& \phantom{\biggl[}\qquad\times
      \phi_j(y-\hat{\rho}_1+\hat{\rho}_2)
      \theta_{\rho_1}(y-\hat{\rho}_1)
      \theta_{\rho_2}(y-\hat{\rho}_1) \\
      &&&\phantom{\biggl[}\hspace{-1.5\arraycolsep} +
      U^\dagger_{\rho_1}(y)
      U_{\rho_2}(y+\hat{\rho}_1-\hat{\rho}_2) (1-\gamma_{\rho_1})
      (1+\gamma_{\rho_2}) \\
      &&& \phantom{\biggl[}\qquad\times
      \phi_j(y+\hat{\rho}_1-\hat{\rho}_2)
      \theta_{\rho_1}(y)
      \theta_{\rho_2}(y+\hat{\rho}_1-\hat{\rho}_2) \\
      &&&\phantom{\biggl[}\hspace{-1.5\arraycolsep} +
      U^\dagger_{\rho_1}(y)
      U_{\rho_2}(y+\hat{\rho}_1) (1-\gamma_{\rho_1})
      (1-\gamma_{\rho_2}) \\
      &&& \phantom{\biggl[}\qquad\times
      \phi_j(y+\hat{\rho}_1+\hat{\rho}_2)
      \theta_{\rho_1}(y)
      \theta_{\rho_2}(y+\hat{\rho}_1) \biggr]\Biggr\rbrace\,. \\
    \end{array}} \\
\end{eqnarray}
The gauge action staples become
\begin{eqnarray}
  \label{eq:ferm-nonherm-staple2}
  \Delta S_{\mbox{\tiny f}}[\Delta U_\mu(y)] &=&
  \mbox{Re}\,\mbox{Tr}\,
  \displaystyle\Delta U_\mu(y)\theta_\mu(y) \sum_j\Biggl\lbrace
  -2\kappa\phi_j^\dagger(y+\hat{\mu}) \left(1+\rho_j^*+\rho_j\right)
  (1+\gamma_\mu) \phi_j(y)
  \nonumber\\ \nonumber\lefteqn{
    \begin{array}{rl}
      \displaystyle \qquad + 2\kappa^2\sum_{\rho\neq\mu}\biggl[ &
      \displaystyle\hspace{-1.5\arraycolsep}\phi^\dagger_j(y+\hat{\mu})
      (1+\gamma_\mu)(1+\gamma_\rho) \phi_j(y-\hat{\rho})
      U_\rho(y-\hat{\rho})\theta_\rho(y-\hat{\rho}) \\
      & \displaystyle\phantom{\biggl[} \hspace{-5.5\arraycolsep} +
      U_\rho(y+\hat{\mu})\phi^\dagger_j(y+\hat{\mu}+\hat{\rho})
      (1+\gamma_\rho)(1+\gamma_\mu) \phi_j(y) \theta_\rho(y+\hat{\mu})
      \\
      & \displaystyle\phantom{\biggl[} \hspace{-5.5\arraycolsep} +
      \phi^\dagger_j(y+\hat{\mu}) (1+\gamma_\mu)(1-\gamma_\rho)
      \phi_j(y+\hat{\rho}) U^\dagger_\rho(y)\theta_\rho(y) \\
      & \displaystyle\phantom{\biggl[} \hspace{-5.5\arraycolsep} +
      U^\dagger_\rho(y+\hat{\mu}-\hat{\rho})
      \phi^\dagger_j(y+\hat{\mu}-\hat{\rho}) (1-\gamma_\rho)
      (1+\gamma_\mu) \phi_j(y) \theta_\rho(y+\hat{\mu}-\hat{\rho})
      \biggr]\Biggr\rbrace\,.
    \end{array}} \\
\end{eqnarray}
Finally the even-odd preconditioned form of
(\ref{eq:eferm-wilson-nonherm}) is given by
\begin{equation}
  \label{eq:eferm-eo-nonherm}
  S_{\mbox{\tiny f}} = \sum_j\sum_{xyz} \phi_j^\dagger(y)
  \left(Q^{\dagger}(y,z)-P_o \rho_j^*\right)
  \left(Q(z,x)-P_o \rho_j\right) \phi_j(x)\,.
\end{equation}
This leads to the local fermionic action
\begin{eqnarray}
  \label{eq:eferm-noherm-eo-wilson-expand}
  S_{\mbox{\tiny f}} &=& \displaystyle\nonumber \sum_j \sum_{y}
  \phi_j^\dagger(y)\Biggl\lbrace
  \left(1+16\kappa^2+P_o \rho_j^*\rho_j
    - (P_e \rho_j^*+P_o \rho_j) \right) \phi_j(y) \\ \nonumber
  \lefteqn{\begin{array}{rcrl}
      \phantom{S_{\mbox{\tiny f}}} & \phantom{=} &
      \displaystyle \qquad - \kappa\left(1+P_e \rho_j^*+P_o
        \rho_j\right)
      \sum_\rho\biggl[
      &\hspace{-1.5\arraycolsep}
      U_\rho(y-\hat{\rho}) \left(1+\gamma_\rho\right)
      \phi_j(y-\hat{\rho}) \theta_\rho(y-\hat{\rho}) \\
      &&& \displaystyle\hspace{-1.5\arraycolsep} +
      U_\rho^\dagger(y) \left(1-\gamma_\rho\right)
      \phi_j(y+\hat{\rho}) \theta_\rho(y)\biggr] \\
      && \displaystyle +
      \kappa^2\sum_{\rho_1\neq\rho_2}
      \biggl[ &\hspace{-1.5\arraycolsep} U_{\rho_1}(y-\hat{\rho}_1)
      U_{\rho_2}(y-\hat{\rho}_1-\hat{\rho}_2) (1+\gamma_{\rho_1})
      (1+\gamma_{\rho_2}) \\
      &&& \phantom{\biggl[}\qquad\times
      \phi_j(y-\hat{\rho}_1-\hat{\rho}_2)
      \theta_{\rho_1}(y-\hat{\rho}_1)
      \theta_{\rho_2}(y-\hat{\rho}_1-\hat{\rho}_2) \\
      &&&\phantom{\biggl[}\hspace{-1.5\arraycolsep} +
      U_{\rho_1}(y-\hat{\rho}_1)
      U^\dagger_{\rho_2}(y-\hat{\rho}_1) (1+\gamma_{\rho_1})
      (1-\gamma_{\rho_2}) \\
      &&& \phantom{\biggl[}\qquad\times
      \phi_j(y-\hat{\rho}_1+\hat{\rho}_2)
      \theta_{\rho_1}(y-\hat{\rho}_1)
      \theta_{\rho_2}(y-\hat{\rho}_1) \\
      &&&\phantom{\biggl[}\hspace{-1.5\arraycolsep} +
      U^\dagger_{\rho_1}(y)
      U_{\rho_2}(y+\hat{\rho}_1-\hat{\rho}_2) (1-\gamma_{\rho_1})
      (1+\gamma_{\rho_2}) \\
      &&& \phantom{\biggl[}\qquad\times
      \phi_j(y+\hat{\rho}_1-\hat{\rho}_2)
      \theta_{\rho_1}(y)
      \theta_{\rho_2}(y+\hat{\rho}_1-\hat{\rho}_2) \\
      &&&\phantom{\biggl[}\hspace{-1.5\arraycolsep} +
      U^\dagger_{\rho_1}(y)
      U_{\rho_2}(y+\hat{\rho}_1) (1-\gamma_{\rho_1})
      (1-\gamma_{\rho_2}) \\
      &&& \phantom{\biggl[}\qquad\times
      \phi_j(y+\hat{\rho}_1+\hat{\rho}_2)
      \theta_{\rho_1}(y)
      \theta_{\rho_2}(y+\hat{\rho}_1) \biggr]\Biggr\rbrace\,. \\
    \end{array}} \\
\end{eqnarray}
The fermionic contribution to the gauge field staple for non-Hermitian
even-odd preconditioned Wilson fermions is then given by
\begin{eqnarray}
  \label{eq:ferm-nonherm-eo-staple2}
  \Delta S_{\mbox{\tiny f}}[\Delta U_\mu(y)] &=&
  \mbox{Re}\,\mbox{Tr}\,
  \displaystyle\Delta U_\mu(y)\theta_\mu(y) \sum_j\Biggl\lbrace
  -2\kappa\phi_j^\dagger(y+\hat{\mu}) \left(1+P_e \rho_j^*+P_o
    \rho_j\right)
  (1+\gamma_\mu) \phi_j(y)
  \nonumber\\ \nonumber\lefteqn{
    \begin{array}{rl}
      \displaystyle \qquad + 2\kappa^2\sum_{\rho\neq\mu}\biggl[ &
      \displaystyle\hspace{-1.5\arraycolsep}\phi^\dagger_j(y+\hat{\mu})
      (1+\gamma_\mu)(1+\gamma_\rho) \phi_j(y-\hat{\rho})
      U_\rho(y-\hat{\rho})\theta_\rho(y-\hat{\rho}) \\
      & \displaystyle\phantom{\biggl[} \hspace{-5.5\arraycolsep} +
      U_\rho(y+\hat{\mu})\phi^\dagger_j(y+\hat{\mu}+\hat{\rho})
      (1+\gamma_\rho)(1+\gamma_\mu) \phi_j(y) \theta_\rho(y+\hat{\mu})
      \\
      & \displaystyle\phantom{\biggl[} \hspace{-5.5\arraycolsep} +
      \phi^\dagger_j(y+\hat{\mu}) (1+\gamma_\mu)(1-\gamma_\rho)
      \phi_j(y+\hat{\rho}) U^\dagger_\rho(y)\theta_\rho(y) \\
      & \displaystyle\phantom{\biggl[} \hspace{-5.5\arraycolsep} +
      U^\dagger_\rho(y+\hat{\mu}-\hat{\rho})
      \phi^\dagger_j(y+\hat{\mu}-\hat{\rho}) (1-\gamma_\rho)
      (1+\gamma_\mu) \phi_j(y) \theta_\rho(y+\hat{\mu}-\hat{\rho})
      \biggr]\Biggr\rbrace\,.
    \end{array}} \\
\end{eqnarray}
This concludes the discussion of local forms of the actions used.

\chapter{Logistics for Running Large Numerical Productions}
\label{sec:logist-runn-large}
It has become clear in the discussion of the TSMB algorithm in
Chapter~\ref{sec:tuning-mult-algor}, that the effort for maintaining
and running a production run to generate a sufficiently large sample
of field configurations is enormous. A large amount of data is being
generated. But already in the simpler case of the HMC, a large number
of gauge field configuration is generated which will have to be stored
in a large file-server. In particular, for the \SESAM/\TXL-projects
\cite{Eicker:1996gk,Lippert:2001ha}, several TBytes of data have been
accumulated.

In the case of a large-scale multiboson production, one may in
addition want to change the polynomials during the production run and
thus end up with a selection of sub-samples all distributed with a
different multicanonical action. Therefore it is inevitable to have a
powerful machinery available which allows to maintain and use a
TByte-sized archive over several years and conserve the data for
potential later use by other groups\footnote{Already the \SESAM/\TXL\ 
  groups have realized that an efficient, standardized system for the
  storage and handling of their configurations was in demand. The
  contributions discussed in the following originally were developed
  as a solution to their problems}.

To meet these goals, an \texttt{SQL}-based database system has been
devised. The design has been a part of the TSMB development in this
thesis and it has turned out to be very useful for practical
applications. In particular, the following components have been
developed:
\begin{enumerate}
\item A library to read and write gauge field configurations on
  variable lattice sizes in the standardized \texttt{Gauge Connection}
  format\footnote{See for a definition and description
    \texttt{http://qcd.nersc.gov}}. The library is usable both from
  \texttt{C} and \texttt{Fortran} and allows to access the gauge
  fields in the form of a comfortable data structure. Furthermore, a
  selection of different and proprietary formats is supported which is
  used mainly for data exchange with the \APE-machines. The majority
  of gauge field configurations generated on these machines is still
  available in this format only\footnote{The program can be
    downloaded from \\
    \texttt{http://www.theorie.physik.uni-wuppertal.de/\~\ $\!$%
      wolfram/publications/downloads/unic.tar.gz}}.
\item A database programmed in \texttt{SQL} which employs the fast and
  efficient \texttt{MySQL}-database engine\footnote{The database can
    be found at \texttt{http://www.mysql.com/}}. Albeit its lack of
  certain features of modern databases, it is very suitable for the
  purpose of storing information from numerical simulations. The
  reason is that write accesses (which usually consist of adding a new
  configuration) only take place once every minutes or even hours
  during a production run and almost never concurrently. The same is
  valid for queries: queries are used to request information for
  measurements and are unlikely to happen concurrently. Thus, usage of
  the \texttt{MySQL} engine appears to be perfectly justified for the
  purposes of lattice field theory simulations.
\item Programs to support adding configurations to the database and to
  support specific types of queries. The database can be accessed
  using a high-level language via their corresponding interfaces. This
  allows a direct combination with the conversion library discussed
  above. A further alternative is the access to the database using
  script languages like shell scripts or \texttt{Perl} scripts.
\end{enumerate}

\clearpage
\section{Design of the Database}
\label{sec:design-database}
A number of text books is available which describe the design process
of a database in detail, see e.g.~\cite{Sauer:1994bo}. The basic
structure of a database is characterized by a set of
\textit{entities}, their corresponding \textit{properties}, and
\textit{relations} between the entities. Important design goals are
\begin{itemize}
\item avoidance of UPDATE-anomalies,
\item elimination of redundancies,
\item the creation of an understandable model,
\item and the minimization of restructuring the relations for the
  introduction of new data types. This should prolong the life
  expectancy of the applications.
\end{itemize}
The above points can be satisfied, if the underlying database is
\textit{normalized}. There exist a number of properties the relations
need to satisfy for the database to be normalized. The most important
are ones are given by the first five normal forms.

\begin{figure}[htb]
  \begin{center}
    \includegraphics[scale=0.3,clip=true]{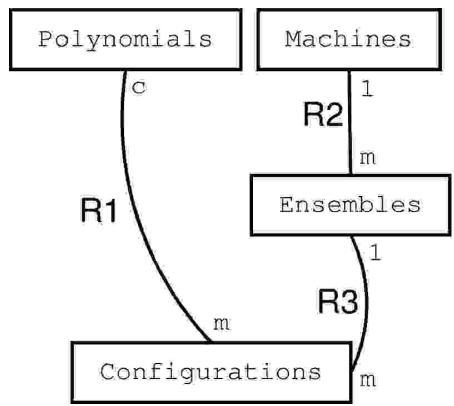}
    \caption{Entity-relation diagram for the configuration database.}
    \label{fig:er-database}
  \end{center}
\end{figure}

Figure~\ref{fig:er-database} shows the entities together with the
relations between them. These ingredients will now be discussed in
detail.

The entities in the database are given by
\begin{description}
\item[Configurations:] Any single gauge field configuration needs to
  be stored separately. Several pieces of information are required for
  the configurations to be reproduced correctly. The \texttt{Gauge
    Connection} format stores all necessary information as a part of
  the file in the header. The \textit{Configurations} entity therefore
  needs to have similar properties.
  Table~\ref{tab:prop-configurations} lists all attributes of this
  entity.
  \begin{table}[htb]
    \begin{center}
      \begin{tabular}[c]{l|l|l}
        \hline\hline
        \textbf{Attribute} & \textbf{Type (SQL)} & \textbf{Content} \\
        \hline
        \textsf{CONFID} & \sqlint & Configuration identification,
        primary key \\
        \textsf{Format} & \sqlenm & One of (\texttt{Gauge Connection},
        \\ && \texttt{Q1}, \texttt{Q4}, \texttt{Q4open}, \texttt{QH1},
        \texttt{QH2}, \texttt{QH4}) \\
        \textsf{Ordering} & \sqlenm & One of (\texttt{Gauge
          Connection}, \\
        && \texttt{GRAL\_TSMB}, \texttt{SESAM\_SSOR},
        \texttt{SESAM\_EO}) \\
        \textsf{Dimension\_1} & \sqlint & Lattice size in
        $x$-direction, $L_1$ \\
        \textsf{Dimension\_2} & \sqlint & Lattice size in
        $y$-direction, $L_2$ \\
        \textsf{Dimension\_3} & \sqlint & Lattice size in
        $z$-direction, $L_3$ \\
        \textsf{Dimension\_4} & \sqlint & Lattice size in
        $t$-direction, $L_0$ \\
        \textsf{Trajectory} & \sqlint & Trajectory or sweep number \\
        \textsf{Link\_Trace} & \sqldou & Sum of traces over all links
        \\
        \textsf{Plaquette} & \sqldou & Sum over all plaquettes \\
        \textsf{Creation\_date} & \sqldat & Sampling date and time \\
        \textsf{Archive\_date} & \sqldat & Archive date and time
        \\
        \textsf{EnsembleID} & \textsf{ENSID} & Foreign key, references
        \textit{Ensembles} \\
        \textsf{PolynomialID} & \textsf{POLID} & Foreign key,
        references \textit{Polynomials} \\
        \textsf{Location} & \sqlvch & Complete path to the
        \texttt{Gauge Connection} file \\
        \textsf{Comment} & \sqltxt & (Optional) comment \\
        \hline\hline
      \end{tabular}
      \caption{Attributes of the \textit{Configurations} entity.}
      \label{tab:prop-configurations}
    \end{center}
  \end{table}
\item[Polynomials:] The TSMB algorithm (see Sec.~\ref{sec:mult-algor})
  requires a multicanonical reweighting with a correction factor
  depending on the choice of the polynomial used (see
  Sec.~\ref{sec:rewe-corr}). Hence, it is important to know the
  polynomial the configuration has been sampled with. Therefore, the
  \textit{Polynomials} entity will contain all necessary information
  about up to three polynomials used. However, if reweighting is not
  required --- if either the configuration has been sampled using an
  algorithm like the HMC or with a multiboson algorithm using an exact
  correction step --- no polynomial will be associated with the
  configurations. The relation \textsf{\textbf{R1}} between the
  \textit{Polynomials} and the \textit{Configurations} entity is thus
  $c:m$. The attributes implemented for \textit{Polynomials} are
  displayed in Tab.~\ref{tab:prop-polynoms}.
  \begin{table}[htb]
    \begin{center}
      \begin{tabular}[c]{l|l|l}
        \hline\hline
        \textbf{Attribute} & \textbf{Type (SQL)} & \textbf{Content} \\
        \hline
        \textsf{POLID} & \sqlint & Polynomial-identification, primary
        key \\
        $\mathsf{\alpha}$ & \sqldou & Power of the polynomial
        (cf.~Eq.~(\ref{eq:tsmb-corr-prob})) \\
        $\mathsf{\epsilon}$ & \sqldou & Lower end of polynomial
        approximation interval \\
        $\mathsf{\lambda}$ & \sqldou & Upper end of polynomial
        approximation interval \\
        $\mathsf{n_1}$ & \sqlint & Order of first polynomial \\
        $\mathsf{n_2}$ & \sqlint & Order of second polynomial \\
        $\mathsf{n_3}$ & \sqlint & Order of third polynomial  \\
        \textsf{Location} & \sqlvch & Complete path to polynomial
        input file \\
        \textsf{Comment} & \sqltxt & (Optional) comment         \\
        \hline\hline
      \end{tabular}
      \caption{Attributes of the \textit{Polynomials} entity.}
      \label{tab:prop-polynoms}
    \end{center}
  \end{table}
\item[Ensembles:] For the Monte-Carlo integration schemes as discussed
  in Sec.~\ref{sec:monte-carlo-algor} one has to compute a sample of
  gauge field configurations which can then be used to measure a
  physical quantity with a certain statistical error. For this
  procedure it is important to categorize all configuration in the
  database into distinct classes according their physical parameters,
  the people who contributed to them etc. This classification is
  implemented using the \textit{Ensembles} entity. It is important to
  realize that this entity need not classify the configurations only
  by their physical properties, but can also categorize the
  configurations by certain ``organizational'' considerations,
  i.e.~the origin of the configurations, the projects they are
  intended for etc. The relation \textsf{\textbf{R3}} between
  \textit{Ensembles} and \textit{Configurations} is $1:m$, i.e.~each
  configuration must be part of one and only one ensemble, but each
  ensemble can contain several configurations. The corresponding
  attributes are shown in Tab.~\ref{tab:prop-ensembles}.
  \begin{table}[htb]
    \begin{center}
      \begin{tabular}[c]{l|l|l}
        \hline\hline
        \textbf{Attribute} & \textbf{Type (SQL)} & \textbf{Content} \\
        \hline
        \textsf{ENSID} & \sqlint & Ensemble identification, primary
        key \\
        \textsf{Ensemble} & \sqltxt & Description of ensemble
        (physical \& organizational) \\
        \textsf{HWID} & \textsf{MACHID} & Foreign key, references
        \textit{Machines} \\
        \textsf{Comment} & \sqltxt & (Optional) comment \\
        \hline\hline
      \end{tabular}
      \caption{Attributes of the \textit{Ensembles} entity.}
      \label{tab:prop-ensembles}
    \end{center}
  \end{table}
\item[Machines:] It is useful to know on which particular machine a
  certain ensemble has been sampled. This is one example of the
  categorization of the \textit{Ensembles} entity, and the only
  example which has been implemented in this thesis. The practical use
  of this information is the evaluation of efficiency analysis, where
  one usually performs simulations at equivalent physical parameters,
  but on different implementation systems
  (cf.~Sec.~\ref{sec:impl-syst}).  This is again an $1:m$ relation
  (see relation \textsf{\textbf{R2}}), since each ensemble has to be
  created on a particular implementation system, but each
  implementation can give rise to several ensembles.  Attributes
  relating to the \textit{Machines} entity are given in
  Tab.~\ref{tab:prop-machines}.
  \begin{table}[htb]
    \begin{center}
      \begin{tabular}[c]{l|l|l}
        \hline\hline
        \textbf{Attribute} & \textbf{Type (SQL)} & \textbf{Content} \\
        \hline
        \textsf{MACHID} & \sqlint & Machine identification, primary
        key \\
        \textsf{Hardware} & \sqlvch & Description of hardware \\
        \hline\hline
      \end{tabular}
      \caption{Attributes of the \textit{Machines} entity.}
      \label{tab:prop-machines}
    \end{center}
  \end{table}
\end{description}
Beyond what has been done here, it is possible to introduce further
entities to categorize the \textit{Ensembles} further, like different
research groups or different projects where the configurations are to
be used. This topic has so far been outside the scope of this thesis
and has therefore not been implemented.

In the practical implementation, the gauge field configurations cannot
be stored in the database itself. In fact, the storage requirements
are enormous --- a typical configuration on an $\Omega=32\times 16^3$
lattice will use about $24$ MB of RAM, and a typical sample consists
of several thousands of these. It is clear that a dedicated storage
device is required. The solution was to store the configurations on a
tape archive installed at the \textit{Forschungszentrum J{\"u}lich},
Germany. The database contains only the path to the configurations in
the archive.  If the configurations are in \texttt{Gauge Connection}
format, they will contain redundant information about their physical
and logical affiliation. This redundancy ensures that the archive can
also be used independently from the database. For the same reason, the
information about the lattice volume are stored in the
\textit{Configurations} table and \textit{not} in the
\textit{Ensembles} table, in contrast to what one would expect from a
normalized relation.

With the extended definition of the \textit{Configurations} entity
which also includes the \textsf{Format} and \textsf{Ordering}
properties in Tab.~\ref{tab:prop-configurations}, one is also able to
store configurations in formats different from the \texttt{Gauge
  Connection} scheme. In particular, all other structures used by the
\SESAM/\TXL-collaboration are supported by the current design. In this
case, the information in the table is \textit{not} redundant and is
required to successfully access a particular configuration.
Furthermore the \textsf{Link\_Trace} and \textsf{Plaquette} properties
are simple and efficient checksum implementations for these
applications.

There is an important subtlety regarding the approximation interval
for quadratically optimized polynomials discussed in
Sec.~\ref{sec:stat-polyn-invers} as used in
Tab.~\ref{tab:prop-polynoms}: the interval applies to the first and
second polynomials and it is assumed that these intervals are
identical. If this is not the case, one will have to store two sets of
$[\epsilon,\lambda]$ values for the two polynomials. The corresponding
information about the third polynomial is not required since it is not
used for reweighting purposes. The information about $n_3$ can
therefore also be considered optional.

In all cases, the \textsf{Location} entry should contain sufficient
information to uniquely locate a file in the archive. Therefore, the
format \texttt{user@host:/complete-path-to-file} has been used, which
allows the file to be accessed directly using the \texttt{scp}
program\footnote{The program and documentation can be obtained from
  \texttt{http://www.openssh.com/}}.

In conclusion, the configuration database allows to store all
necessary information about gauge field configurations. It supports
different formats and allows to salvage all data from the \SESAM/\TXL\ 
projects. It uses a modern database design which can be accessed from
a diversity of different implementation systems. The configurations in
\texttt{Gauge Connection} format can also be accessed independently
from the database.

\end{appendix}


\bibliography{phd_thesis}


\clearpage\newpage
\include{index}


\end{document}